\newcounter{myctr}
\def\myitem{\refstepcounter{myctr}\bibfont\noindent\ifnum\themyctr>9\else\phantom{0}\fi\hangindent17pt\themyctr.\enskip}

\documentclass{ws-ijqi}
\usepackage{hyperref}
\usepackage[super,sort,compress]{cite}
\usepackage{url}
\usepackage{bbm}
\usepackage{mathrsfs} 
\usepackage{arydshln} 

\def\Id{{\mathbbm 1}}
\DeclareMathOperator{\Tr}{Tr}
\def\dag{{^{\dagger}}}
\def\erf{{\rm erf}}

\usepackage{xparse}
\NewDocumentCommand{\ceil}{s O{} m}{%
  \IfBooleanTF{#1} 
    {\left\lceil#3\right\rceil} 
    {#2\lceil#3#2\rceil} 
}

\newcounter{postulate}
\newtheorem{postulate}{Postulate}

\begin{document}

\catchline{}{}{}{}{}

\title{Quantum communications in continuous variable systems\footnote{This is a PhD thesis, and parts of it have been published in M.~N. Notarnicola, M.~G.~A. Paris, and S. Olivares, J.~Opt.~Soc.~Am.~B {\bf 40} (2023) 705--714; M.~N. Notarnicola and S. Olivares, Phys.~Rev.~A {\bf 108} (2023) 022404; M.~N. Notarnicola, M. Jarzyna, S. Olivares, and K. Banaszek, New~J.~Phys. {\bf 25} (2023) 103014; M.~N. Notarnicola and S. Olivares, Phys. Rev.~A {\bf 108}  (2023) 042619; M.~N. Notarnicola, S. Olivares, E. Forestieri, E. Parente, L. Potì, and M. Secondini, IEEE Trans.~Commun. {\bf 72} (2024) 375--386; M.~N. Notarnicola, F. Cieciuch, and M. Jarzyna, New J. Phys. {\bf 26} (2024) 043015; M.~N. Notarnicola and S. Olivares, Int.~J.~Quantum~Inf. {\bf 22} (2024) 2450008.}}

\author{Michele N. Notarnicola}
\address{Department of Optics, Palack\'y University, \\
17 Listopadu 12, 779 00 Olomouc (Czech Republic) \\[1ex]
Dipartimento di Fisica ``Aldo Pontremoli'', \\
Universit\`a degli Studi di Milano, I-20133 Milano, (Italy) \\[1ex]
michelenicola.notarnicola@upol.cz}

\maketitle

\begin{history}
\received{\today}
\end{history}

\begin{abstract}
Nowadays, quantum communications provide a vast field of research in rapid expansion, with a huge potential impact on the future developments of quantum technologies. In particular, continuous variable systems, employing coherent-state encoding and quadrature measurements, represent a suitable platform, due to their compatibility with both the modulation and detection systems currently employed in standard fiber-optical communications. In this work, we address some relevant aspects of the field, and provide innovative results being also experimentally oriented. In particular, we focus on two relevant paradigms: quantum decision theory and continuous variable quantum key distribution (CVQKD). 
In the former case, we address the problem of coherent state discrimination and design new hybrid receivers for binary phase-shift-keying discrimination, obtaining a quantum advantage over conventional detection schemes, being also robust against typical experimental imperfections. In the latter scenario, we proceed in two different directions. On the one hand, we design new CVQKD protocols employing discrete modulation of coherent states, being a feasible solution compatible with the state of the art in optical communications technologies; on the other hand, we address the more fundamental problem of performing channel losses mitigation to enhance existing protocols, and investigate the role of optical amplifiers for the task. Finally, we make a first step towards a fully non-Gaussian CVQKD scheme by proposing, for the first time, the adoption of an optimized state discrimination receiver, commonly adopted for quantum decision theory, within the context of CVQKD, obtaining a genuine quantum enhancement over conventional protocols.
\end{abstract}


\tableofcontents

\markboth{M.~N.~Notarnicola}
{Quantum communications in continuous variable systems}


\section{Introduction}

\subsection{Motivation}

The appearance of quantum mechanics in the mid 1920's yielded a milestone progress in modern science, providing a fundamental theory to describe natural phenomena at the microscopic and sub-microscopic scale, being untenable within the framework of classical physics.
However, if on the one hand it establishes a consistent description of highly non-classical phenomena, involving atoms and molecules, semi- and super-conductors, \ldots; on the other hand, in time, physicists questioned themselves about the non-intuitive aspects of the theory, namely, non-commutativity of observables, the Einstein-Podolsky-Rosen paradox, quantum non-locality and the von Neumann reduction induced by quantum measurements \cite{Zeilinger1998}.
Starting from the 1950's, the perspective changed, and the emerging results in the quantum mechanics foundations gave birth to a new field: {\it quantum information science}.
While quantum mechanics limits itself to explain the natural phenomena at the microscopic level, quantum information science starts from the Landauer's observation that information is a physical entity, being dependent of the physical laws used to store and processes it \cite{Landauer1991, Landauer1996, Landauer1999}, and focuses on its transmission and processing by means of the quantum features of a physical system.

Historically, the first developments of quantum information arose from the field of {\it quantum communications}. 
Indeed, in the 1960's, Gordon \cite{Gordon1962}, Stratonovich \cite{Stratonovich1965:1,Stratonovich1965:2}, and Helstrom \cite{Helstrom1969, Helstrom1976} firstly proposed a formulation of optical communications in the quantum regime, with the intent of establishing the fundamental limits posed by quantum mechanics in the transmission of classical information over optical communication links.
In particular, Gordon and Stratonovich addressed the information capacity of optical channels, while Helstrom focused himself on quantifying the error probability in a decision strategy when a finite set of classical symbols is encoded onto non-orthogonal quantum states of radiation.
These results provided the first step for all the subsequent investigations, assessing the ultimate limits of quantum protocols and operations ranging from channel parameter estimation \cite{QCRB}, information capacity \cite{Holevo1979CAP}, optical amplification \cite{Caves1982}, and so on.
Moreover, thanks to the recent technological progresses, these limits not only provide useful theoretical results, but have also been experimentally demonstrated and, nowadays, are commonly encountered in several contexts, from near- and deep-space communications to loss mitigation of realistic metropolitan fiber channels.
 
On the other hand, the 1980's brought a remarkable change in perspective.
Instead of merely considering quantum properties as {\it passive} features that pose limitations on the performance of classical protocols, scientists realized that they could be exploited as {\it active} resources to design completely new protocols and paradigms that outperform  the existing classical schemes. In other words, since information is physical, we can employ quantum mechanical effects, e.g. Heisenberg's uncertainty, superposition, \ldots, to transmit it in novel and more powerful fashion.
The main result in this direction has been achieved by quantum key distribution (QKD), firstly introduced by Bennett and Brassard in 1984 \cite{BB84}, that allows two distant parties to distill a random key via the exchange of quantum states, with unconditional security guaranteed by the quantum mechanics laws.
In time, QKD has become one of the milestone aspects of quantum information, and its application has been extended from discrete variable to continuous variable systems \cite{Grosshans2002}.
Following this philosophy, in more recent years, a big enhancement has been obtained also regarding the information transmission over quantum channels, by addressing the transmission of genuine quantum information, i.e. quantum states (possibly entangled), instead of the simple encoding of classical symbols onto a quantum optical carrier field \cite{Holevo2011}.

\subsection{Thesis overview}

Given the motivations described above, nowadays quantum communications provide a vast field of research in rapid expansion, with a huge potential impact on the future developments of quantum technologies.
The scope of this Thesis is then to address some relevant aspects of the field, and provide innovative results being also experimentally oriented. In particular, here we focus on two relevant paradigms, namely quantum state discrimination and continuous variable (CV) QKD, with particular reference to optical platforms. 
In the former case, we design new hybrid receivers for discrimination of phase-shift-keyed coherent states, obtaining a quantum advantage over conventional detection schemes.
In the latter scenario, we proceed in two different directions. On the one hand, we design new CVQKD protocols employing discrete modulation of coherent states, being a more feasible solution compatible with the state of the art in optical communications technologies; on the other hand, we address the more fundamental problem of performing channel losses mitigation to enhance the key generation rate (KGR), by considering both feasible conventional optical amplifiers and more sophisticated schemes like probabilistic noiseless linear amplifiers.
Finally, we make a first step towards a fully non-Gaussian CVQKD scheme by proposing, for the first time, the adoption of an optimized state discrimination receiver, commonly adopted for quantum decision theory, within the context of CVQKD, obtaining a genuine quantum enhancement over conventional protocols in particular ranges of transmission distance.  

Here below, the structure and the main original results of the Thesis are summarized in more detail. 

\subsubsection{Organizational note}
The present PhD Thesis consists of three Parts, followed by Conclusions and a list of Appendices, for a total of ten Sections.
In particular, Part~I provides a preliminary part that introduces the notation and the fundamental tools employed throughout the rest of the Thesis, and it is divided into three Sections.
Part~II is devoted to quantum state discrimination theory, presenting a comprehensive analysis of binary and $M$-ary discrimination protocols, and it is composed of two Sections. Finally, Part~III deals with QKD in continuous variable systems, addressing the different existing approaches to assess security, and it consists of four Sections.

\begin{itemize}
\item In Sec.~\ref{chap:QMech}, we summarize the modern tools of quantum mechanics, presenting the postulates of the theory regarding quantum states, quantum evolution and quantum measurements.

\item In Sec.~\ref{chap:QOpt}, we present the basic elements of quantum optics, with relevant examples of quantum states of radiation, quantum maps and quantum measurements. In particular, we give a detailed presentation of the Gaussian state formalism, that will be widely exploited in the main Parts of the Thesis.

\item In Sec.~\ref{chap:QComm}, we introduce the main aspects of realistic quantum communication systems and information theory, highlighting the difference between the classical and quantum description.\vspace{0.5cm}

\item In Sec.~\ref{chap:GeneralFeatures}, we outline the framework of quantum state discrimination theory: we introduce the decision error probability as the main figure of merit, and address the fundamental case of binary discrimination of quantum states. In particular, we focus on coherent-state discrimination and present a comprehensive analysis of the quantum receivers proposed in literature, assessing their performance also in the presence of realistic inefficiencies, e.g. non-unit quantum efficiency, dark counts, visibility reduction and phase noise. Results have been published in~\cite{Notarnicola2023:HYNORE, Notarnicola2023:FF, Notarnicola2023:PhN}.

\item In Sec.~\ref{chap:MaryDisc}, we extend the analysis to multiple-state discrimination. We present a deep review of all the fundamental theoretical results of the theory and, eventually, specialize it to quadrature phase-shift-keying discrimination of coherent states, discussing the functioning and the limits of the most relevant quantum receivers. Results have been published in~\cite{Notarnicola2023:KB}. \vspace{0.5cm}

\item In Sec.~\ref{chap:CVQKD}, we address QKD: at first, we provide a basic overview of its  main aspects and, thereafter, we focus on CVQKD, where coherent states are employed as information carrier. We present three different approaches to assess the security of the protocols, namely unconditional security, trusted-device scenario and wiretap channel assumption, and introduce the KGR as the fundamental figure of merit. Subsequently, we focus on the unconditional scenario and provide security proofs by the ``optimality of Gaussian attacks" theorem. Then, we apply the obtained theoretical results to both Gaussian modulation and discrete modulation protocols. Results have been published in~\cite{Notarnicola2024:SEC}.

\item In Sec.~\ref{chap:RESTREAV}, we study the two other security frameworks previously introduced, that represent examples of restricted eavesdropping. At first, we address the trusted-device scenario and extend the validity of the optimality of Gaussian attacks, accounting also for the limitations of the eavesdropper. Thereafter, we study the security of a wiretap channel, in which the eavesdropping attack is assumed to be known, comparing the resulting KGR with the unconditional security case. Results have been published in~\cite{Notarnicola2024:SEC}.

\item In Sec.~\ref{chap:Amplifiers}, we discuss the potentiality of optical amplifiers to perform mitigation of the transmission losses and enhance CVQKD. We address the problem of optical amplification at the quantum limit, introducing both conventional amplifiers, i.e. phase-insensitive and phase-sensitive amplifiers, and probabilistic noiseless linear amplifiers. Then, we study their application in CVQKD schemes, providing security analysis under different security frameworks. Results have been published in~\cite{Notarnicola2023:NLA, Notarnicola2023:MJ}.

\item In Sec.~\ref{chap:nonGauss}, we proceed beyond the standard CVQKD protocols, employing Gaussian measurements, and investigate the potentiality of non-Gaussian detection schemes to increase the KGR. In particular, we resort to $M$-ary quantum state discrimination theory, discussed in Sec.~\ref{chap:MaryDisc}, and design an optimized quantum receiver maximizing the KGR, comparing its performance with respect to standard Gaussian receivers under the wiretap channel assumption. Results have been published in~\cite{Notarnicola2023:KB}.
\end{itemize}

\subsubsection{Main results}

\begin{list}{}{\leftmargin 15pt \itemsep 0pt \topsep 3pt}
\item {\bf Binary discrimination of coherent states (see Sec.s~\ref{sec4:Hybrid},~\ref{sec4:IneffDiscr}, and~\ref{sec4:PhNDiscr}):} Starting from the state-of-the-art quantum receivers, namely the homodyne and displacement receivers (e.g. the Kennedy one), we propose a new hybrid scheme, the hybrid near-optimum receiver (HYNORE), that combines both the homodyne-like and displacement-photon counting setups via feed-forward operations to obtain an enhanced discrimination strategy. The receiver not only outperforms the Kennedy, but also provide a fascinating proposal for experimental implementations, as it only relies on photon-number-resolving (PNR) detectors and electro-optic modulators (EOMs) to implement conditional displacements. Thereafter, we also extend the scheme to a multi-copy receiver, to design a second hybrid receiver, the hybrid feed-forward receiver (HFFRE), where further reduction of the decision error probability is obtained by splitting the encoded signal into many rescaled copies and performing subsequent feed-forward operations. Finally, we detailedly address the robustness of the two proposed hybrid receivers against the typical experimental imperfections, that is non-unit quantum efficiency, dark counts, reduced interference visibility, and phase diffusion noise.

\item {\bf $\boldsymbol M$-ary state discrimination (see Sec.s~\ref{sec:M=GammaA} and~\ref{sec:OptwithGUS}):}
The problem of the optimal decision becomes highly nontrivial in the presence of a constellation of $M>2$ quantum states. Unlike the binary case, where Helstrom's theory provides an explicit derivation of the optimum receiver and the corresponding minimum error probability, for $M$-ary state discrimination, the decision problem is recast into a convex optimization problem. The optimum receiver is only indirectly characterized by Yuen's theorem, presenting necessary and sufficient conditions to be fulfilled, while, in general, calculation of the corresponding positive-operator valued measure (POVM) and decision error probability should be handled numerically. Given this limitation, suboptimal POVMs with simpler construction have been proposed, among which the paradigmatic example is provided by the pretty good measurement (PGM), being proved to yield the optimum receiver in the particular case of pure-state constellations satisfying the geometrically uniform symmetry (GUS).
In this Thesis, we present a new and simpler derivation of the optimum receiver in this latter scenario. In particular, we prove that, in the presence of GUS, every pure-state discrimination receiver is ultimately identified by a set of $M-1$ phases, which may be properly chosen to minimize the decision error probability, thus transforming a convex functional optimization into optimization of a real function with $M-1$ real variables. Remarkably, this result provide a huge simplification, and leads to straightforward construction of the optimum receiver that does not refer to suboptimal methods, obtained by setting all the free phases equal to $0$.

\item {\bf CVQKD with discrete modulation (see Sec.~\ref{subsec:QAMproto}):} Given the practical difficulty to implement Gaussian modulation of coherent states in the current CVQKD demonstrations, we propose a new CVQKD protocol employing quadrature amplitude modulation (QAM) of the coherent pulses, assisted by probabilistic amplitude shaping to obtain a non-uniform sampling probability distribution that approximates a Gaussian distribution. We prove QAM to both outperform the conventional discrete modulation formats based on phase-shift keying (PSK), and to close the gap with respect to Gaussian modulation, thus providing a solution to achieve high values of KGR with a feasible experimental setup.

\item {\bf Security of CVQKD in the trusted-device scenario (see Sec.~\ref{sec: Trusted}):} Usual security proofs for CVQKD are obtained in the unconditional security framework, where the whole channel connecting sender and receiver is considered to be untrusted, thus assuming the presence of an omnipotent eavesdropper. However, this represent an excessive assumption, that can be relaxed in practical conditions. To this aim, we consider the trusted-device scenario, in which some of the channel components (e.g. detection losses and noise) are trusted. We provide for the first time a security proof for discrete modulation protocols, by extending the ``optimality of Gaussian attacks" theorem, commonly adopted in the unconditional framework, to this scenario, providing a general tool to assess security also in the presence of restricted eavesdropping.

\item {\bf Long-distance CVQKD with optical amplifiers (see Sec.s~\ref{sec9-PIAPSAQKD} and~\ref{sec:CVQKD-NLA}):} We address the problem of channel losses mitigation in CVQKD by adopting different kinds of optical amplifiers. At first, we consider conventional amplifiers, namely phase-insensitive (PIAs) and phase-sensitive amplifiers (PSAs), arranged in a multi-span configuration, where the quantum channel is composed of many regenerative stations interspersed with lossy links. We address security under both the unconditional and the trust-device frameworks. In the former case, we prove that the KGR is improved with respect to the standard no-amplifier protocol only for PSA links where the de-amplified quadrature is measured. In the latter, we assume all amplifiers
and spans except one are trusted, and show that the position of the untrusted span greatly affects the
potential enhancement offered by amplification.
Thereafter, we study CVQKD assisted by heralded noiseless linear amplifiers (NLAs), being probabilistic operations that amplify signals without additional noise, provided that a particular outcome is retrieved from the measurement of some ancillary modes. For the sake of simplicity, we only address unconditional security, and prove that, remarkably, it is possible to distill a secure key at arbitrary large distances if the amplifier gain is properly optimized.

\item {\bf CVQKD with state-discrimination receivers (see Sec.~\ref{chap:nonGauss}):} Ultimately, we make a first step towards the design of fully non-Gaussian CVQKD protocols. By drawing inspiration on our original results for $M$-ary quantum state discrimination, we propose an innovative optimized state-discrimination receiver for the quadrature PSK (QPSK) protocol, referred to as the key-rate optimized receiver (KOR). For the sake of simplicity, we analyze security under a pure-loss wiretap channel, and obtain an enhancement with respect to the conventional QPSK protocol in the metropolitan-network distance regime. We also consider the performance of a feasible displacement feed-forward receiver, in which case we have an increase in the KGR with respect to Gaussian detection for short transmission distances.

\end{list}



\section*{Part I: Preliminaries}\label{part1}
\addcontentsline{toc}{section}{Part I : Preliminaries}


\def\rrangle{\rangle\!\rangle}
\def\llangle{\langle\!\langle}

\section{Mathematical tools of quantum mechanics}\label{chap:QMech}

This thesis deals with the analysis of communication protocol at the quantum limit; therefore this preliminary Part is devoted to a basic introduction to the main features that will be exploited throughout the work.

To begin with, this Section presents the fundamental tools of quantum mechanics from a modern perspective, by following a quantum information approach \cite{Paris2012}. The Section is organized as follows.
In Sec.~\ref{sec:1-ToolsQM}, we first introduce the standard prescriptions to establish a statistical theory, i.e. states, evolution and measurements. Then, we address the case of quantum mechanics, presenting the postulates of the theory in the standard fashion. Finally, we proceed beyond and generalize the postulates, accounting for the presence of an open quantum system.
In particular, in Sec.~\ref{sec:1-States} we present the generalized description of quantum states in terms of statistical operators, in Sec.~\ref{sec:1-Dyn} we introduce the concept of quantum completely positive maps as an extension of the usual unitary dynamics, and in Sec.~\ref{sec:1-Meas} we define positive-operator-valued measures, providing the most general description for quantum measurements.

\subsection{Mathematical tools of quantum mechanics}\label{sec:1-ToolsQM}
Generally speaking, any physical theory that provide a complete description of a physical system\footnote{With the term physical system we refer to a single given degree of freedom, e.g. position, spin, polarization, angular momentum,...} should be structured by a minimum set of prescriptions, establishing the connection between the observed phenomena and the adopted mathematical framework.
The typical fundamental building blocks are three. Firstly, we identify what is the {\it state} of the system, that is the mathematical object describing its preparation. Secondly, we introduce the {\it dynamics}, that is the evolution law of a physical state of the system. Finally, we
define {\it measurements}, specifying the objects that describe observables. 
As an example, in the case of classical mechanics, the state of a system is identified by a set of canonical variables, i.e. positions and momenta $\{x_j , p_j\}_j$, the evolution is provided by a suitable differential equation in the canonical variables, namely the Hamilton equation, while probability measures determine the rule that predicts the measurement outcomes.

Beside, in the case of quantum mechanics, these three prescriptions are defined through the postulates of the theory, representing the minimal requirements to describe the behaviour of a system in the presence of genuine quantum effects.
In their original formulation, the standard postulates of quantum mechanics can be summarized as follows \cite{Paris2012, Ercolessi2015}.
\begin{itemize}
\item \textbf{Postulate 1. (States)} The quantum state of a system is described by a normalized vector of a separable  Hilbert space $\mathcal{H}$, $|\psi \rangle \in \mathcal{H}$, $\langle \psi | \psi \rangle = 1$. For multipartite systems, the global Hilbert space is the tensor product of the Hilbert spaces associated with each subsystem. Given this vector space description, the superposition principle holds: if $|\psi_{1}\rangle$ and $|\psi_{2}\rangle$ are two physical states, then every (normalized) linear combination $\alpha |\psi_{1}\rangle + \beta |\psi_{2}\rangle$, $\alpha,\beta\in \mathbb{C}$, $|\alpha|^2+ |\beta|^2=1$, is also a possible physical state of the system.
\item \textbf{Postulate 2. (Dynamics)} Given the initial state $|\psi_0\rangle$ at time $t_0$, its time evolution is obtained through a unitary operator $U(t,t_0)$, that is $U^\dagger(t,t_0)U(t,t_0)=\Id$, $\Id$ being the identity operator over $\cal H$. The state at time $t$ then reads $|\psi(t)\rangle = U(t,t_0) |\psi_0\rangle$.
\item \textbf{Postulate 3. (Measurements)} Observables are Hermitian operators acting on the Hilbert space $\mathcal{H}$, $X \in {\cal B}(\mathcal{H})$. By the spectral theorem, $X= \sum_x x \mathbb{P}_x$, where $x\in \mathbb{R}$ and $\mathbb{P}_x=|x\rangle \langle x|$ are a complete set of orthogonal projectors $\mathbb{P}_x \mathbb{P}_{x'}= \delta_{x x'} \mathbb{P}_x$. The eigenvalues of $X$ represent all possible outcomes of the measurement, and the probability of retrieving $x$ given state $|\psi\rangle$ is provided by the Born rule: 
\begin{align}
p_x= |\langle x | \psi \rangle|^2 \, .
\end{align}
Finally, when the measurement brings the value $x$, the conditional state of the system after detection is the (normalized) projection of the probed state onto the eigenspace associated with $x$, namely
\begin{align}
|\psi_x\rangle= \frac{1}{\sqrt{p_x}} \mathbb{P}_x |\psi\rangle \, ,
\end{align}
referred to as Von Neumann reduction.
\end{itemize}

Actually, all these postulates presuppose a ``Postulate 0", that is to consider a {\it closed and isolated system}. 
On the other hand, in practical situations we mostly deal with systems interacting with the rest of the universe, either during their dynamical evolution, or when subjected to measurement. In turn, in the following we develop a suitable modification of the three prescriptions, referred to as the generalized postulates, that hold also in the presence of open quantum systems.

\subsubsection{Postulate 1: Quantum states}\label{sec:1-States}

The first generalized postulate takes into account that the preparation of a quantum system, in general, may not be completely under control. Thus, a probabilistic description must be considered, in terms of a given statistical ensemble $\{p_k, |\psi_k\rangle \}_k$, meaning that state $ |\psi_k\rangle$ is prepared with probability $p_k$. In this case, the description of the system is obtained by the statistical (density) operator $\rho=\sum_k p_k |\psi_k\rangle \langle \psi_k|$. By suitably characterizing this mathematical object, we rephrase Postulate 1 as follows.

\begin{postulate}\label{eq:Postulate1}
{\bf (States)} The quantum state of a system is described by a {\it statistical operator}, that is a positive operator of unit trace, $\rho \in {\cal L}({\cal H})$, $\rho\geq0$, $\Tr[\rho]=1$.
\end{postulate}
In the former expression, ${\cal L}({\cal H})$ refers to the set of all linear operator acting on he Hilbert space $\cal H$.
As a consequence, given state $\rho$, if we consider a traditional Hermitian observable $X$, the Born rule and the von Neumann reduction become:
\begin{align}\label{eq:BornRUle}
p_x= \Tr[\rho \, \mathbb{P}_x] \qquad \mbox{and} \qquad \rho_x = \frac{1}{p_x} \, \mathbb{P}_x \rho \, \mathbb{P}_x \, ,
\end{align}
respectively.
In the presence of a pure state $|\psi\rangle \in {\cal H}$, we have a $1$-rank density operator $\rho=|\psi\rangle\langle \psi|$, while, in the more general case, we have $1 \le {\rm rank}(\rho) \le d$, with $d={\rm dim}({\cal H})$, and the state is said to be mixed. 

The new generalized postulate can be reconciled with the standard Postulate 1 by observing that, for an isolated bipartite system $AB$, described by the vector state $|\psi \rangle\! \rangle_{AB}$, the state describing only subsystem $A$ is obtained as the partial trace of the global state:
\begin{align}
\rho_A = \Tr_B\Big[|\psi \rangle\! \rangle_{AB} \langle\! \langle\psi| \Big] \, ,
\end{align}
and the same undergoes for $\rho_B$. 
Conversely, any density operator on $\cal H$ can be viewed as the partial trace of a state vector on a larger Hilbert space.
In fact, given a quantum state $\rho$ it is also possible to construct its {\it purification} as follows.
We start from the spectral decomposition $\rho=\sum_k \lambda_k |\phi_k\rangle \langle \phi_k|$; now, we introduce another Hilbert space $\cal K$ with dimension ${\rm dim}({\cal K}) \ge {\rm rank}(\rho)$ and an orthonormal basis $\{|\theta_k\rangle\}_k$ in $\cal K$. Then, the vector $|\Psi\rrangle \in {\cal H} \otimes {\cal K}$, equal to:
\begin{align}\label{eq:SchmidtSDC}
|\Psi\rrangle = \sum_k \sqrt{\lambda_k} \, |\phi_k\rangle |\theta_k\rangle \, ,
\end{align}
provides a purification of $\rho$, namely a pure state such that $\Tr_{\cal K} [|\Psi\rrangle \llangle \Psi|]= \rho$.
We also note that the expression~(\ref{eq:SchmidtSDC}) represents the Schmidt decomposition of $|\Psi\rrangle$ \cite{Friedberg2014, Breuer2002}.
Given this argument, we underline that there exist infinitely many purifications of a density operator. However, thanks to the  ``freedom-in-purifications theorem" \cite{NielsenChuang}, all purifications are unitarily equivalent. That is, if $|\Psi_1\rrangle \in {\cal H} \otimes {\cal K}_1 $ and $|\Psi_2\rrangle \in {\cal H} \otimes {\cal K}_2$ are two purifications of $\rho$, there exists a unitary operation $U: {\cal K}_1 \to {\cal K}_2$ which transforms $|\Psi_1\rrangle$ into $|\Psi_2\rrangle$, namely:
\begin{align}
|\Psi_2\rrangle= \left( \hat{\Id}_{\cal H} \otimes U \right) |\Psi_1\rrangle \, .
\end{align}

Finally, to quantify the degree of mixedness of a quantum state $\rho$, that is how far a density operator is from a pure state, we introduce two typical measures:
\begin{itemize}
\item the {\it purity}, defined as the trace of the square density operator:
\begin{align}
\mu[\rho] = \Tr[\rho^2] \, ,
\end{align}
such that $1/d \le \mu[\rho] \le 1$, $d$ being the dimension of the Hilbert space. Pure states, satisfying $\rho^2=\rho$, have $\mu=1$, while for any $\mu<1$ we have a mixed state;

\item the {\it von Neumann entropy}:
\begin{align}
{\sf S}[\rho] = -\Tr[\rho \log \rho]= -\sum_n \lambda_n \log\lambda_n \, ,
\end{align}
where $\{\lambda_n\}_n$ are the eigenvalues of $\rho$, and the logarithm is typically taken in basis $2$. The von Neumann entropy captures the intuitive idea that, if a system is prepared in a pure state, we have 
the maximum possible information, and ${\sf S}[\rho]=0$; while mixed states are obtaining by tracing out some degrees of freedom of a larger system, thus ignoring the information encoded in the correlations between the portion under investigation and the rest of the universe. In this latter case, we have $0 <{\sf S}[\rho] \le \log d$.
\end{itemize}

\subsubsection{Postulate 2: Quantum dynamics}\label{sec:1-Dyn}
The second generalized postulate characterizes the dynamics of quantum states in the presence of open systems, in which case the unitary description is untenable.
Within this framework, the evolution of system is described in terms of a map $\cal E: {\cal L}({\cal H}) \rightarrow {\cal L}({\cal H})$ that transforms a quantum state $\rho$ into another quantum state ${\cal E}(\rho)$.
The fundamental requirements that should be fulfilled by $\cal E$ to describe a physical operation are the following \cite{Paris2012}:
\begin{itemize}\label{items}
\item[(a)] The map $\mathcal{E}$ is {\it positive and trace preserving}: if $\rho\geq0$, then $\mathcal{E} (\rho) \geq0$ and $\mathrm{Tr}[\mathcal{E} (\rho)] = \mathrm{Tr}[\rho]$, guaranteeing that the output state is a genuine quantum state satisfying Postulate~\ref{eq:Postulate1}.

\item[(b)] The map $\mathcal{E}$ is {\it linear}:  $\mathcal{E} \bigl(\sum_k p_k \rho_k\bigr)= \sum_k p_k \mathcal{E} (\rho_k)$, such that the state obtained 
by applying the map to the ensemble $\{p_k, \rho_k\}_k$ is the ensemble $\{p_k, {\cal E}(\rho_k)\}_k$.

\item[(c)] The map $\mathcal{E}$ is {\it completely positive} (CP): besides satisfying positivity, the map $\mathcal{E} \otimes \hat{\Id}_n : {\cal L}(\mathcal{H}\otimes \mathbb{C}^n) \rightarrow {\cal L}(\mathcal{H}\otimes \mathbb{C}^n)$ is also positive $\forall n \in \mathbb{N}$. This property captures the idea that the map should be preserve its physical meaning when applied to subsystems of a composite system.
\end{itemize}
Then, we have:
\begin{postulate}\label{eq:Postulate2}
{\bf (Dynamics)} The evolution of a quantum system is described by a map $\cal E: {\cal L}({\cal H}) \rightarrow {\cal L}({\cal H})$ satisfying $\rm (a)-(c)$, usually referred to with the terms {\it quantum CP map}, {\it quantum operation}, or {\it quantum channel}.
\end{postulate}

The reconciliation with the usual closed system dynamics is guaranteed by Kraus theorem \cite{Kraus1983, Breuer2002}.
\begin{theorem}[\textbf{Kraus, 1971}]
Let $\mathcal{E}: {\cal L}({\cal H}) \rightarrow {\cal L}({\cal H})$ be a linear map. Then, the following are equivalent:
\begin{itemize}
\item $\mathcal{E}$ is a quantum CP map,
\item there exists a group of operators $\{M_k\}_k \subset {\cal L}({\cal H})$ (named Kraus operators) such that
\begin{align}\label{eq:Kraus}
\mathcal{E}(\rho) = \sum_k M_k \rho M_k^\dagger \, , \hspace{1.cm} \sum_k M_k^\dagger M_k = \hat{\Id} \, ,
\end{align}
\item there exist a Hilbert space $\mathcal{H}_B$, a preparation $|\omega\rangle_B \in \mathcal{H}_B$ and a unitary operation $U \in {\cal L}(\mathcal{H}\otimes \mathcal{H}_B)$ such that
\begin{align}\label{eq:unitary}
\mathcal{E}(\rho) =  \mathrm{Tr}_B \Big[ U \ \rho \otimes |\omega\rangle_B\langle \omega| \ U^\dagger \Big].
\end{align}
\end{itemize}
\end{theorem}
We note that the theorem yields two equivalent representations for the map $\mathcal{E}$. Eq.~\ref{eq:Kraus} constitutes a ``local representation", the so-called Kraus representation, providing the generalization of unitary evolution in terms of a set of Kraus operators. In this framework, unitary maps are retrieved as CP maps associated with a single Kraus operator $M_k$.
On the contrary, Eq.~\ref{eq:unitary} is a ``microscopic representation", that constructs a {\it unitary dilation} of $\mathcal{E}$ as the partial trace of a unitary evolution on a larger system. In particular, this construction follows from the Stinespring dilation theorem \cite{Stinespring1955, Breuer2002}.


\subsubsection{Postulate 3: Quantum measurements}\label{sec:1-Meas}
Finally, we deal with quantum measurements. As introduced before, a quantum measurement in a closed and isolated system is characterized by a complete set of orthogonal projectors $\{\mathbb{P}_x\}_x$ associated with the self-adjoint operator $X=\sum_x x \mathbb{P}_x$. 
Analogously to quantum maps, we now relax these requirements and highlight the minimum prescriptions that should be satisfied by quantum observables, leading to the definition of generalized measurements.
In fact, we note that the Born rule in~(\ref{eq:BornRUle}) can be rewritten as:
\begin{align}
p_x= \Tr[\rho \, \mathbb{P}_x]= \Tr[\rho \, \mathbb{P}_x^2] \, .
\end{align}
In turn, the only request to let $p_x$ be a faithful probability distribution is that $\mathbb{P}_x^2$ should be a positive operator $\Pi_x \ge 0$, satisfying the normalization condition $\sum_x \Pi_x = \hat{\Id}$.
This condition defines a generalized quantum measurement as a {\it positive operator-valued measure} (POVM) $\{\Pi_x\}_x$, such that $\Pi_x \ge 0$ for all $x$, and $\sum_x \Pi_x = \hat{\Id}$. Moreover, due to positivity, the POVM elements $\Pi_x$ can be expressed in terms of {\it detection operators} $M_x$ as $\Pi_x= M_x^\dagger M_x$ \cite{Paris2012}.
In summary:
\begin{postulate}\label{eq:Postulate3}
{\bf (Measurements)} A generalized quantum measurement is described by a POVM, i.e. a collection $\{\Pi_x\}_x$ of positive operators $\Pi_x= M_x^\dagger M_x\ge 0$, such that $\sum_x \Pi_x = \hat{\Id}$.
Then, the Born rule and the Von Neumann reduction become:
\begin{align}
p_x= \mathrm{Tr}[\rho \, \Pi_x] \qquad \mbox{and} \qquad \rho_x = \frac{1}{p_x} M_x \rho M_x^\dagger \, ,
\end{align}
respectively.
\end{postulate}
Formally, the detection operators are obtained as $M_x=\sqrt{\Pi_x}$, thus representing the counterpart of the projectors $\mathbb{P}_x$ in the conventional projection-valued measures (PVMs). However, unlike PVMs, the orthogonality condition between the $\{M_x\}_x$ is not required, therefore the
number of the POVM elements need not to coincide with the dimension of the Hilbert space, and can also be larger.

Similarly to Kraus theorem, one can prove that any POVM $\{\Pi_x\}$ can be brought back to a suitable PVM performed on a larger Hilbert space. This provides conciliation with the standard postulate, guaranteed by Naimark theorem \cite{Paulsen2002, Breuer2002}.
\begin{theorem}[\textbf{Naimark, 1943}]
Given a generalized measurement $\{ \Pi_x\}_x$ on a Hilbert space $\mathcal{H}_A$, there exist a Hilbert space $\mathcal{H}_B$, a preparation $|\omega\rangle_B \in \mathcal{H}_B$, a unitary operation $U_{AB} \in {\cal L}(\mathcal{H}_A\otimes \mathcal{H}_B)$ and a projective measurement $\{\mathbb{P}_x\}_x$ on $\mathcal{H}_B$ such that:
\begin{subequations}
\begin{align}
p_x&= \mathrm{Tr}_A[\rho \Pi_x] = \mathrm{Tr}_{AB}\Big[U_{AB} \  \rho \otimes |\omega\rangle_B \langle \omega| \ U_{AB}^\dagger \ (\hat{\Id}_A \otimes \mathbb{P}_x) \Big] \, , \\[1ex]
\rho_x &= \frac{1}{p_x} M_x \rho M_x^\dagger = \frac{1}{p_x} \mathrm{Tr}_B \Big[ U_{AB} \ \rho \otimes |\omega\rangle_B \langle \omega| \ U_{AB}^\dagger \ (\hat{\Id}_A \otimes \mathbb{P}_x) \Big] \, .
\end{align}
\end{subequations}
\end{theorem} 

As for the case of quantum maps, Naimark theorem provides a ``microscopic representation" of a generalized quantum measurement. In fact, in practical contexts, performing a quantum measurement implies the presence of a detection apparatus, prepared into an ancillary state interacting with the quantum state of the system and, thereafter, being measured. As a matter of fact, within this description, the measured quantity may be always described by a PVM on the global Hilbert space of both the system and the apparatus. Then, an equivalent description in terms of POVMs is obtained when tracing out the degrees of freedom of the apparatus. Conversely, any conceivable POVM is associated with a {\it Naimark extension}, providing a physical implementation via ancilla-based standard measurement.
We also note that the possible Naimark extensions are infinite, corresponding to the intuitive idea that exists infinitely many apparatuses, with an arbitrary number of ancillary systems, to measure a physical quantity: indeed, for a given POVM element $\Pi_x$ there are infinite possible detection operators $M_x$. For this reason, the construction reported in the statement of Naimark theorem is usually referred to as the {\it canonical extension} of the POVM, which does not necessarily coincide with a feasible actual implementation. 

\def\bmalpha{\boldsymbol\alpha}
\def\bmbeta{\boldsymbol\beta}
\def\Re{{\rm Re \, }}
\def\Im{{\rm Im \,}}
\def\G{{\rm G}}
\def\IN{{\rm in}}
\def\OUT{{\rm out}}
\def\bmsigma{\boldsymbol\sigma}
\def\RMeas{  {{\bf r}_{\rm m}}  }
\def\sigmaMeas{  {\boldsymbol\sigma_{\rm m}}  }
\def\WFH{{\rm WH}}
\def\HL{{\rm HL}}

\section{Basics of quantum optics}\label{chap:QOpt}

In this Section, we present a general overview of continuous variable systems, with particular attention to quantum optical systems, that will provide the main platform for the quantum communication protocols discussed throughout the thesis.
We establish the quantum description of optical electromagnetic fields, both in the Hilbert space and the quantum phase space representations, and, in particular, we focus on the Gaussian state formalism, that describes most of the optical elements being commonly exploited in the state-of-art technologies.

The structure of the Section is the following.
In Sec.~\ref{sec:2-CV} we introduce continuous variable systems, namely physical systems described by position- and momentum-like operators that obey the canonical commutation rule. Then, Sec.~\ref{sec2:QuantumStates} specializes the analysis to quantum optical fields, presenting the description of quantum states in both the Hilbert space and the quantum phase space. Thereafter, Gaussian states are introduced and characterized, and some relevant examples are reported. Instead, in Sec.~\ref{sec2:EvoQO} we discuss the quantum evolution of quantum optical states, focusing on the case of Gaussian dynamics, transforming input Gaussian states into output Gaussian states. Finally, in Sec.~\ref{sec2:MeasQO}, we study quantum measurements of optical systems. At first, we provide a complete description of Gaussian measurement, yielding Gaussian statistics and Gaussian conditional states when performed on a Gaussian probe. Then, we present some relevant examples of non-Gaussian measurements that will be discussed in the rest of the thesis, i.e. photon-number resolving detection and weak-field homodyne detection.

\subsection{Introduction to continuous variable systems}\label{sec:2-CV}

In this thesis, we deal with continuous variable systems. With this terminology, we refer to a non-relativistic degrees of freedom, each one being quantized by a pair of Hermitian position-like and momentum-like operators $q$ and $p$ that satisfy the canonical commutation relation (CCR):
\begin{align}\label{eq:qpcomm}
[q,p]= 2 i \sigma_0^2\, ,
\end{align}
where $\sigma_0\ge 0$ is a proper multiplicative constant, whose physical meaning will be discussed thereafter.
These systems require the adoption of an infinite dimensional Hilbert space, hence the terminology ``quantum continuous variables”. In fact, operators $q$ and $p$ have continuous spectrum, $q=\int_\mathbb{R} q \, |q\rangle \langle q|$ and $p=\int_\mathbb{R} p \, |p\rangle \langle p|$, with eigenvalues over the whole real line, and improper eigenstates $\{|q\rangle \}$ and $\{|p\rangle \}$ such that $\langle q_1|q_2\rangle= \delta(q_1-q_2)$, $\langle p_1|p_2\rangle= \delta(p_1-p_2)$, and $\langle q|p\rangle= \exp(i q p)/\sqrt{2\pi}$ \cite{Serafini2017}.

Equivalently, for bosonic systems we also introduce the creation and annihilation operators:
\begin{align}
a= \frac{q+i p}{2\sigma_0} \qquad \mbox{and} \qquad a^\dagger=\frac{q-ip}{2\sigma_0} \, ,
\end{align}
effecting the creation and annihilation of energy quanta, e.g. photons, phonons, \ldots, of an harmonic oscillator, that obey $[a,a^\dagger]= 1$ \cite{GerryKnight}.
Furthermore, the extension of the CCR to $n$ pairs of canonical variables $\{q_j, p_j \}$, $j=1,\ldots, n$, is straightforward, and reads $[q_j,p_k]= 2 i \sigma_0^2 \delta_{jk}$, or, equivalently, $[a_j,a_k]=[a_j^\dagger,a_k^\dagger]=0$ and $[a_j, a_k^\dagger]=\delta_{jk}$, $\delta_{jk}$ being the Kronecker delta.
To get a more compact notation, it is useful to introduce a vector representation of the canonical operators:
\begin{align}
\hat{\textbf{r}}= (q_1,p_1, ..., q_n,p_n)^{\sf T} \, ,
\end{align}
and re-express the CCRs as:
\begin{align}
[\hat{{\bf r}},\hat{{\bf r}}^{\sf T}]= 2 i \sigma_0^2 \, \Omega^{(n)} \, ,
\end{align}
where $[{\bf A},{\bf B}] =  {\bf A B} - ({\bf A B})^{\sf T}$ and $\Omega^{(n)}= \oplus_{j=1}^{n} \Omega$ is the $n$-mode symplectic form, with
\begin{align}
\Omega= \begin{pmatrix} 0 & 1 \\ -1 & 0 \end{pmatrix} \, ,
\end{align}
and $\oplus$ denoting direct sum.

\subsection{Quantum states of radiation}\label{sec2:QuantumStates}

The previous description provides theoretical modeling of a wide range of physical bosonic platforms, such as mechanical harmonic oscillators, electromagnetic field, trapped ions, \ldots. For the purposes of this thesis, we will only focus on quantum optical platforms, in which case the canonical operators $q$ and $p$ represent the quantized {\it quadratures} of a single mode optical field at given angular frequency $\omega$, associated with the Hamiltonian:
\begin{align}\label{eq:Hamw}
H= \hbar \omega \left(\hat{n} + \frac12\right) \, ,
\end{align}
where $\hbar$ is the reduced Planck's constant and $\hat{n}=a^\dagger a$ is the photon-number operator, revealing the number of excitation quanta (photons) of a given quantum state \cite{GerryKnight, Serafini2017, Ferraro2005}. Moreover, the corresponding electric field operator $\hat{{\bf E}}({\bf r})$ at position ${\bf r}$, is obtained as:
\begin{align}
\hat{{\bf E}}({\bf r})&= \mathscr{E}_0 \Big[a \, e^{i {\bf k} \cdot {\bf r}}  +a^\dagger \, e^{-i {\bf k} \cdot {\bf r}} \Big] = \frac{\mathscr{E}_0}{\sigma_0} \Big[q \, \cos ({\bf k} \cdot {\bf r})  - p \, \sin ({\bf k} \cdot {\bf r}) \Big] \, ,
\end{align}
where $\bf k$ is the wave-vector associated with the carrier frequency, with modulus $k=|{\bf k}|=\omega/c$, and $\mathscr{E}_0=\sqrt{\hbar \omega/(2\epsilon_0 V)}$ is the so-called electric field ``per single photon", in which $c$, $\epsilon_0$ and $V$ are the speed of light, the vacuum electric permittivity, and the quantization volume, respectively.
If the number of modes of the field is $n > 1$, the Hamiltonian becomes $H=\sum_{j=1}^{n} \hbar \omega_j (\hat{n}_j + 1/2)$, $\hat{n}_j $ being the photon-number operator associated with the bosonic mode $a_j$  \cite{GerryKnight}.

Given these considerations, we present some paradigmatic examples of quantum states for optical fields.

\subparagraph{Fock states.} They are the eigenstates of the Hamiltonian~(\ref{eq:Hamw}), corresponding to the eigenstates of the number operator, $\{|n\rangle\}$, $n\in \mathbb{N}$, such that $\hat{n} |n\rangle = n |n\rangle$. By the spectral theorem, they form a complete orthonormal system. In particular, among Fock states, we find the {\it vacuum state} $|0\rangle$ such that $\hat{n} |0\rangle=0$, for which the quadrature mean values and variances read:
\begin{align}
\langle q \rangle = \langle p \rangle = 0 \qquad \mbox{and} \qquad \Delta^2q = \Delta^2p=\sigma_0^2 \, ,
\end{align}
revealing the physical meaning of constant $\sigma_0^2$ introduced in~(\ref{eq:qpcomm}), namely the {\it zero-point fluctuations} of  the field quadratures, also referred to as {\it vacuum fluctuations} or {\it shot noise variance}.
In turn, the shot noise $\sigma_0^2$ provides a measurement scale to evaluate the quadrature variances of all quantum states; thus, in principle, its value can be arbitrarily chosen. Typically, there are two conventional choices: either fixing $\sigma_0^2=1/2$, the so-called canonical representation, or $\sigma_0^2= 1$, corresponding to shot-noise units (SNU).

\subparagraph{Coherent states.} They are the eigenstates of the annihilation operator $a$, that is $a |\alpha\rangle = \alpha |\alpha\rangle$, $\alpha \in \mathbb{C}$, expanded in the Fock basis as:
\begin{align}
|\alpha\rangle= e^{-\frac{|\alpha|^2}{2}} \sum_{\mathrm{n}=0}^{\infty} \frac{\alpha^n}{\sqrt{n!}} |n\rangle = D(\alpha) |0\rangle \, ,
\end{align}
where $D(\alpha) = \exp(\alpha a^\dagger - \alpha^* a)$ is the \textit{displacement operator}.
Coherent states are not orthogonal with one another, since eigenstates of a non-Hermitian operator,
\begin{align}
\langle \beta|\alpha\rangle= e^{-|\alpha-\beta|^2/2} e^{(\alpha \beta^*-\alpha^* \beta)/2} \, , \quad \alpha,\beta \in \mathbb{C} \, ,
\end{align}
but, nevertheless, they form an overcomplete set, as:
\begin{align}
\int_{\mathbb{C}} \frac{d^2 \alpha}{\pi} |\alpha \rangle \langle \alpha |= \hat{\Id} \, .
\end{align}
Moreover, these states are usually considered as {\it quasi} classical states, namely the quantum states that describe classical optical fields in the absence of noise. Indeed, given a coherent state with amplitude $\alpha= |\alpha| e^{i \phi}$, $0\le \phi<2\pi$, the expectations values of quadratures read $\langle q \rangle = 2 \sigma_0 |\alpha| \cos\phi$ and $\langle p \rangle = 2 \sigma_0 |\alpha| \sin\phi$, reproducing the behaviour of a classical harmonic oscillator, albeit exhibiting shot noise fluctuations $\Delta^2q = \Delta^2p=\sigma_0^2$, that become negligible in the classical limit of high-intensity fields, $|\alpha|^2 \gg \sigma_0^2$.

\subparagraph{Thermal state.} It is the mixed state describing radiation emitted by thermal sources, namely:
\begin{align}
\nu^{\mathrm{th}}(\bar{n}) = \frac{1}{1+ \bar{n}} \sum_{\mathrm{n}=0}^{\infty} \left(\frac{\bar{n}}{1+ \bar{n}}\right)^n |n \rangle \langle n| \, ,
\end{align}
$\bar{n}$ being the mean number of photons. It has null mean values of quadratures and variances larger than vacuum fluctuations, as $\Delta^2q = \Delta^2p=\sigma_0^2 (1+2 \bar{n})$.

\subparagraph{Single-mode squeezed vacuum state.} It is the state of radiation obtained from second order interaction of light within a non linear crystal of nonzero second order susceptibility, expressed as:
\begin{align}
|r\rangle = S(r) |0\rangle \,,
\end{align} 
where $S(r)= \exp[r(a^{\dagger 2} - a^2)/2]$ is the {\it single-mode squeezing operator}, $r\in\mathbb{C}$.
If $r>0$, we compute the expectation values of quadratures as $\langle q \rangle = \langle p \rangle = 0$ and:
\begin{align}
\Delta^2q = e^{2r} \sigma_0^2 \qquad \mbox{and} \qquad \Delta^2p=e^{-2r} \sigma_0^2 \, .
\end{align}
The name \textit{squeezing} derives from the fact that fluctuations on $p$ are reduced below the vacuum, at the expense of enlarging those on $p$ above the shot-noise limit. In other words, quadrature $p$ is \textit{squeezed}, i.e. de-amplified, while quadrature $q$ is \textit{anti-squeezed}, that is amplified.

\subparagraph{Two-mode squeezed vacuum state (TMSV).} It is the paradigmatic example of a two-mode entangled state, defined as:
\begin{align}
|{\rm TMSV}\rrangle = S_2(r) |0\rangle|0\rangle = \sqrt{1-\lambda^2} \sum_{n=0}^{\infty} \lambda^n |n\rangle |n\rangle \, ,
\end{align} 
where $S_2(r)= \exp[r(a^{\dagger} b^\dagger - a b)]$ is the {\it two-mode squeezing operator}, acting on two modes $a$ and $b$, $[a,b]=0$, and $\lambda= \tanh r$. In the previous expression, we assumed $r\ge 0$ for the sake of simplicity. The TMSV represents the maximally entangled state at fixed mean energy, as, after performing partial trace over one mode, we retrieve a thermal state with energy $\bar{n}= \lambda^2/(1-\lambda^2)$ \cite{Ferraro2005}.

\subsubsection{The quantum phase space description}\label{subsec2:PhaseSpaceMethods}
Beside the Hilbert space description, obtained by expansion over a suitable basis, an equivalent representation of optical quantum states is obtained in the so-called {\it quantum phase space}. 
It is introduced as a bijective mapping between $n$-mode density operators in the Hilbert space and complex scalar function defined in $\mathbb{R}^{2n}$.

Let us consider a $n$-mode bosonic system, described by the bosonic operators ${\bf a}= (a_1,a_2, \ldots, a_n)^\mathsf{T}$, arranged in vector notation. Then, according to Glauber's formula, also referred to as Fourier-Weyl relation, any quantum state $\rho$ can be expressed as \cite{Serafini2017, Ferraro2005,OlivaresGauss}:
\begin{align}\label{eq:Glauber}
\rho = \int_{\mathbb{C}^n} \frac{d^2 \bmalpha}{\pi^n} \chi(\bmalpha) D_{\bf a}(\bmalpha)\dag \, ,
\end{align}
where $\bmalpha= (\alpha_1,\alpha_2, \ldots, \alpha_n)^\mathsf{T} \in \mathbb{C}^n$ and 
\begin{align}
D_{\bf a}(\bmalpha) = \bigotimes_{k=1}^{n} D_{a_k}(\alpha_k) \, ,
\end{align}
where $D_{a_k}(\alpha_k)= \exp(\alpha_k a_k\dag - \alpha_k^* a_k) $ is the displacement operator acting on mode $a_k$. 
Some useful properties of the displacement operator are reported below:
\begin{subequations}\label{eq: propertiesDisplacementOperation}
\begin{align}
&D_{\bf a}(\bmalpha_1) D_{\bf a}(\bmalpha_2) = D_{\bf a}(\bmalpha_1+\bmalpha_2) \, , \quad \bmalpha_1,\bmalpha_2 \in\mathbb{C}^n \,,  \\
&D_{\xi \bf a}(\bmalpha)= D_{\bf a}(\xi \bmalpha)\, , \quad \xi \in\mathbb{R} \,,  \\
&\Tr\big[D_{\bf a}(\bmalpha)\big] = \pi^n \delta^{(n)} (\bmalpha) \, ,
\end{align}
\end{subequations}
$\delta^{(n)} (\bmalpha)$ being the complex $n$-mode Dirac delta distribution.
Moreover, for any pair of generic operators $O_1$ and $O_2$ acting on the Hilbert space $\cal H$ of $n$ modes the \textit{trace rule} holds:
\begin{align}
\Tr[O_1 O_2] = \int_{\mathbb{C}^n} \frac{d^2\bmalpha}{\pi^n} \chi[O_1](\bmalpha) \chi[O_2](-\bmalpha) \, ,
\end{align}
$\chi[O_{1(2)}](\bmalpha)$ being the characteristic function of $O_{1(2)}$, respectively.
As an example, for a single radiation mode $a$, we choose $O_1=D(\alpha)$, with $\alpha=x+i y$, and $O_2=q^2=\sigma_0^2 (a+a\dag)^2$ and obtain \cite{Ghalaii}:
\begin{align}\label{eq: CovDisp}
\Tr\big[D(\alpha) q^2\big]= \sigma_0^2 \,  e^{-(x^2+y^2)/2} \Bigg[\pi \delta^{(2)} (\alpha)+ 2 \pi y \delta(x) \frac{d}{d y} \delta(y) - \pi \delta(x) \frac{d^2}{d y^2} \delta(y)\Bigg] \, ,
\end{align}
where $\delta(x)$ is the Dirac delta distribution.

Remarkably, we note that Eq.~(\ref{eq:Glauber}) formally represents an expansion on the set of displacement operators, which constitutes a complete basis set on the Hilbert space ${\cal L}({\cal H})$ of the linear operators acting on $\cal H$.
In turn, the coefficient:
\begin{align}
\chi(\bmalpha) = \Tr\big[\rho D_{\bf a}(\bmalpha)\big] \, ,
\end{align}
referred to as the {\it characteristic function} associated with $\rho$, contains all information about the quantum state. We conclude that, as anticipated before, we can study and analyze the properties of $\rho$ by a function of $n$ complex variables rather than a density operator in the Hilbert space.

Starting from $\chi(\bmalpha)$ we also introduce the generalized characteristic function, or $s$-ordered characteristic function \cite{Serafini2017, Ferraro2005,OlivaresGauss}:
\begin{align}
\chi_s(\bmalpha)=\Tr\big[\rho D_{\bf a}(\bmalpha)\big]  \, e^{s \bmalpha^\dagger \bmalpha/2} \, ,
\end{align}
that satisfy the following properties:
\begin{itemize}
\item[i)] $\chi_s({\bf 0})= \Tr[\rho ]= 1$;
\item[ii)] for all $j=1,\ldots, n$ and $s=-1,0,1$, we have:
\begin{align}
\left(\frac{\partial}{\partial \alpha_j} \right)^m \left(-\frac{\partial}{\partial \alpha^*_j} \right)^k \chi_s(\bmalpha) \Bigg|_{\bmalpha={\bf 0}} = \left\langle \big(a_j^{\dagger}\big)^m a^k \right\rangle_s \, ,
\end{align}
where $\langle\cdot \rangle_s$ is the $s$-ordered expectation value, corresponding to anti-normal order ($s=-1$), symmetric order ($s=0$) and normal order ($s=1$). That is, the generalized characteristic function is the moment generating function for the state $\rho$.
\end{itemize}

Finally, by passing to the Fourier transform, we obtain the \textit{quasi-probability distribution} \cite{Serafini2017, Ferraro2005,OlivaresGauss}:
\begin{align}
W_s(\bmalpha) = \int_{\mathbb{C}^n} \frac{d^2\bmbeta }{\pi^{2n}}  \, e^{(\bmbeta^\dagger \bmalpha- \bmalpha^\dagger \bmbeta)/2} \, \chi_s(\bmbeta) \, ,
\end{align}
such that:
\begin{itemize}
\item[i)] $\int_{\mathbb{C}^n} d^2 \bmalpha \, W_s(\bmalpha)= \chi_s({\bf 0})= 1$;
\item[ii)] for all $j=1,\ldots, n$ and $s=-1,0,1$, we have:
\begin{align}
\left(\frac{\partial}{\partial \alpha_j} \right)^m \left(-\frac{\partial}{\partial \alpha^*_j} \right)^k \chi_s(\bmalpha) \Bigg|_{\bmalpha={\bf 0}} = \int_{\mathbb{C}^n} d^2\bmalpha \, \big(\alpha_j^*\big)^m \alpha_j^k \, W_s(\bmalpha) \, ,
\end{align}
and, by integration, we retrieve all statistical moments of state $\rho$;
\item[iii)] for $s=0$ the trace rule becomes:
\begin{align}\label{eq:TrRuleWf}
\Tr[O_1 O_2] = \pi^n \int_{\mathbb{C}^n} d^2\bmalpha \, W[O_1](\bmalpha) W[O_2](\bmalpha) \, ,
\end{align}
where we omitted the pedix $0$ for convenience.
\end{itemize}

We also remark that, by performing a suitable change of variables, we can re-express all the previously introduced functions in terms of $2n$ real variables in the domain $\mathbb{R}^{2n}$, rather than $n$ complex ones in $\mathbb{C}^n$.
To this aim, we introduce a set of Cartesian variables ${\bf r}=(x_1,y_1;x_2,y_2;\ldots; x_n,y_n)^{\sf T}= 2\sigma_0 (\Re \alpha_1, \Im \alpha_1;\Re \alpha_2, \Im \alpha_2; \ldots; \Re \alpha_n, \Im \alpha_n)^{\sf T} \in \mathbb{R}^{2n}$ and set:
\begin{align}
\chi_s({\bf r})=\chi_s(\bmalpha({\bf r})) \qquad \mbox{and} \qquad W_s({\bf r})=\frac{1}{(4\sigma_0^2)^n} W_s(\bmalpha({\bf r})) \, , 
\end{align}
where $\bmalpha({\bf r})$ denotes the re-parametrization of the complex vector $\bmalpha$ in terms of the new variables, i.e. $\alpha_j=(x_j+iy_j)/2\sigma_0$, $j=1,\ldots, n$, and the pre-factor $1/(4\sigma_0^2)^n$ in the quasi-probability distribution $W_s$ has been introduced to preserve its normalization.

Moreover, quasi-probability distributions provide useful tools to assess some relevant quantum features of the state $\rho$. In particular, the most common distributions refer to the values $s=-1,0,1$ and are presented below for the single mode case, $n=1$.

\subparagraph{$\boldsymbol{P}$-function} ($s=1$). If $s=1$, the function $P(\alpha)= W_1(\alpha)$ is called $P$- function, or Glauber-Sudarshan function, such that for each state $\rho$ we have:
\begin{align}
\rho= \int_{\mathbb{C}} d^2\alpha \ P(\alpha) \ |\alpha \rangle \langle \alpha| \, ,
\end{align}
$|\alpha\rangle$ being a coherent state.
We note that, in general, $P(\alpha)$ is not a regular function. In particular, if we deal with a coherent state $\rho=|\beta\rangle\langle \beta|$, $\beta \in \mathbb{C}$, we have $P(\alpha)= \delta^{(2)}(\alpha-\beta)$; otherwise, if the state cannot be expressed as a statistical mixture of coherent states, the corresponding $P(\alpha)$ is a pathological function, being more singular than a Dirac delta. In turn, the singularity of the $P$-function provides a sufficient condition to assess non-classicality of quantum states.

\subparagraph{Wigner function} ($s=0$). The function $W(\alpha)= W_0(\alpha)$ is called Wigner function, being the Fourier transform of the characteristic function $\chi(\alpha)$.
It represents the conventional choice to describe quantum states in the phase space, and can be also computed as \cite{OlivaresGauss, Ferraro2005}:
\begin{align}
W(\alpha) = \frac{2}{\pi} \sum_{n=0}^{\infty} \, (-1)^n \, \langle n | D\dag(\alpha) \, \rho \, D(\alpha) | n\rangle \, .
\end{align}
$\{|n\rangle\}_n$ being the Fock basis. Moreover, in the Cartesian notation, performing integration over one of the two variables yields:
\begin{align}
\int_{\mathbb{R}} dp \, W(q,p) = \langle q|\rho|q\rangle  \qquad \mbox{and} \qquad \int_{\mathbb{R}} dq \,W(q,p) = \langle p|\rho|p\rangle \, ,
\end{align}
where $|q\rangle$ and $|p\rangle$, $q,p\in\mathbb{R}$, are the improper eigenstates of the canonical operators $q$ and $p$, respectively.
Therefore the marginal distributions of the $W$-function yield the probability distributions associated with quadrature detection \cite{OlivaresGauss, Ferraro2005}.
Unlike the $P$-function, the Wigner function is non-singular, $W(\alpha) <\infty$, but it can get negative values. If so, the width of the Wigner negativity region in the phase space provides a measure of non-classicality \cite{Genoni2013}.

\subparagraph{$\boldsymbol{Q}$-function} ($s=-1$). Finally, the function $Q(\alpha)= W_{-1}(\alpha)$ is called $Q$-function (or Husimi function). Its fundamental property is that:
\begin{align}
Q(\alpha) = \frac{\langle \alpha | \rho | \alpha \rangle}{\pi} \ge 0 \, ,
\end{align}
$|\alpha\rangle$ being a coherent state. In turn, $Q(\alpha)$ is regular for all quantum states. 

\subsubsection{Gaussian states}\label{subsec:Gaussianstates}
Within the phase space representation, a quantum state $\rho_\G$ is a {\it Gaussian state} if its associated Wigner function (or, analogously, its characteristic function) is Gaussian, namely:
\begin{align}\label{eq:Wcart}
   W(\mathbf{r}) =  \frac{1}{(2\pi)^n \sqrt{\det(\boldsymbol{\sigma})}} \,  \exp \left[ 
   - \frac12 (\mathbf{r}-\mathbf{R})^\mathsf{T} \, \boldsymbol{\sigma}^{-1} \, (\mathbf{r}-\mathbf{R})
   \right]
\end{align}
where $\mathbf{r}= (x_1, y_1, \ldots ,x_n, y_n)^\mathsf{T} \in \mathbb{R}^{2n}$, and
\begin{align}
    {\bf R}= \Tr[\rho_\G \, \hat{{\bf r}}]
\end{align}
is the first moment vector (FM) and
\begin{align}
    \boldsymbol\sigma &= \frac12 \Tr \bigg[ \rho_\G \, \big\{(\hat{{\bf r}}-{\bf R}),(\hat{{\bf r}}-{\bf R})^\mathsf{T}\big\} \bigg] \nonumber \\[1ex]
&= \frac12 \Tr \bigg[ \rho_\G \, \big\{\hat{{\bf r}},\hat{{\bf r}}^\mathsf{T}\big\} \bigg] -{\bf R}{\bf R}^{\sf T} \, 
\end{align}
is the $2n\times 2n$ covariance matrix (CM), where $\{{\bf A}, \bf{B}\}={\bf AB}+(\bf {AB})^{\sf T}$ is the anti-commutator of $\bf A$ and $\bf B$. Thus, a Gaussian state is completely characterized by its FM and its CM.
Equivalently, a Gaussian state can be expressed as a Gibbs-state of a (at most) quadratic Hamiltonian $\hat{H}= \hat{\textbf{r}}^{\sf T}  \mathbb{H} \, \hat{\textbf{r}}/2 + {\bf a}^{\sf T}\hat{\textbf{r}}$, where $\mathbb{H}$ is a $2n \times 2n$ symmetric matrix and ${\bf a}\in \mathbb{R}^{2n}$ is a ``displacement" vector. That is:
\begin{align}\label{eq:rho_G}
\rho_\G = \frac{e^{-\beta \hat{H}}}{{\cal Z}} \, ,
\end{align}
${\cal Z}= \mathrm{Tr}[e^{-\beta \hat{H}}]$, where $\beta$ is a free parameter.

The expansion of a Gaussian state $\rho_\G$ onto the Fock basis has been recently derived in \cite{Quesada2019}. To this aim, we first re-express Eq.~(\ref{eq:Wcart}) as:
\begin{align}\label{eq:Wcomplex}
   W(\boldsymbol{\alpha}) =\frac{1}{\pi^n \sqrt{\det(\widetilde{\boldsymbol{\sigma}})}} \,  \exp \left[ 
   - \frac12 (\boldsymbol{\alpha}-\boldsymbol{\beta})^{\dagger} \,\, \widetilde{\boldsymbol{\sigma}}^{-1} \, (\boldsymbol{\alpha}-\boldsymbol{\beta})
   \right]
\end{align}
with $\boldsymbol{\alpha}= (\alpha_1, \alpha^*_1, \ldots ,\alpha_n, \alpha^*_n)^\mathsf{T} \in \mathbb{C}^{2n}$, and
\begin{align}
\boldsymbol{\beta}= \mathbb{U} \, \mathbf{x} \quad \mbox{and} \quad \widetilde{\boldsymbol{\sigma}} = \mathbb{U} \,\boldsymbol{\sigma} \,\mathbb{U}^{\dagger}
\end{align}
where $\mathbb{U}= \oplus_{k=1}^{n} \mathbb{U}_1$ and
\begin{align}
\mathbb{U}_1=
\frac{1}{2\sigma_0} \, 
\begin{pmatrix}
1 & i \\
1 & -i 
\end{pmatrix}
 \, .
\end{align}
Thereafter, we introduce the matrices
\begin{subequations}
\begin{align}
\boldsymbol{\sigma}_Q&= \widetilde{\boldsymbol{\sigma}} + \Id_{2n}/2  \\
\boldsymbol{A}&= \boldsymbol{X} (\Id_{2n}-\boldsymbol{\sigma}_Q^{-1})  \\
\boldsymbol{\gamma}^{\mathsf T} &= \boldsymbol{\beta}^{\dagger} \boldsymbol{\sigma}_Q^{-1}
\end{align}
\end{subequations}
where $\Id_{2n}$ is the $2n\times 2n$ identity matrix and $\boldsymbol{X}= \oplus_{s=1}^n \boldsymbol{\sigma}_x$, $\boldsymbol{\sigma}_x$ being the Pauli $x$-matrix.
Then, the matrix element in the Fock basis $\rho_{\boldsymbol{mk}}= \langle \boldsymbol{m}|\rho |\boldsymbol{k}\rangle$, $|\boldsymbol{k}\rangle=|k_1 \, k_2 \, \ldots k_n\rangle$ and $|\boldsymbol{m}\rangle=|m_1 \, m_2 \, \ldots m_n\rangle$, reads:
\begin{align}
\rho_{\boldsymbol{mk}} = T_{\boldsymbol{mk}}  \, \prod_{s=1}^{n} \left(\frac{\partial}{\partial\alpha_s}\right)^{k_s} \left(\frac{\partial}{\partial\alpha^*_s}\right)^{m_s} \,
 \exp \left(\frac12 \boldsymbol{\alpha}^{\mathsf T} \boldsymbol{A} \boldsymbol{\alpha} + \boldsymbol{\gamma}^{\mathsf T} \boldsymbol{\alpha}\right)\Big|_{\boldsymbol{\alpha}=0}
\end{align}
where
\begin{align}
T_{\boldsymbol{mk}} =\frac{1}{\sqrt{\det(\boldsymbol{\sigma}_Q) \prod_{s=1}^{n} k_s! m_s!}}\exp \left(-\frac12 \boldsymbol{\beta}^{\dagger} \boldsymbol{\sigma}_Q^{-1} \boldsymbol{\beta}\right) \, .
\end{align}

We now list some fundamental properties of Gaussian states that will be helpful throughout the thesis:

\begin{itemize}
\item {\it Tensor products of Gaussian states are Gaussian.} That is, given two Gaussian states $\rho_A(B)$ with FM $\textbf{R}_{A(B)}$ and CM $\bmsigma_{A(B)}$, respectively, the bipartite state $\rho= \rho_A \otimes \rho_B$ is Gaussian with FM $\textbf{R}= \textbf{R}_A \oplus \textbf{R}_B$ and CM $\boldsymbol \sigma= \boldsymbol \sigma_A \oplus\boldsymbol \sigma_B$. 

\item {\it The partial trace of a Gaussian state is Gaussian.} We consider a generic bipartite Gaussian state $\rho_{AB}$, with FM and CM:
\begin{align}\label{eq:BlockForm}
\textbf{R}= \begin{pmatrix} \textbf{R}_A \\ \textbf{R}_B \end{pmatrix}, \qquad \boldsymbol\sigma =  \begin{pmatrix} \boldsymbol\sigma_A & \boldsymbol\sigma_{AB} \\  \boldsymbol\sigma_{AB} & \boldsymbol\sigma_B \end{pmatrix} \, .
\end{align}
Then, $\rho_{A(B)}= \mathrm{Tr}_{B(A)}[\rho_{AB}]$ is still Gaussian with FM $\textbf{R}_{A(B)}$ and covariance $\boldsymbol\sigma_{A(B)}$, that is the sub-blocks associated with subsystem ${A(B)}$ in $\textbf{R}$ and $\boldsymbol\sigma$. 

\item {\it The purity and the von Neumann entropy of a Gaussian state only depend on its CM.} The purity of a Gaussian state $\rho_\G$, with CM $\bmsigma$, is retrieved as:
\begin{align}
\mu[\rho_\G]= \Tr[\rho_\G^2] = \frac{(\sigma_0^2)^n}{ \sqrt{\det(\bmsigma)}} \, ,
\end{align}
whereas its von Neumann entropy $S[\rho_\G]=-\Tr[\rho_\G \log_2\rho_\G]$ reads:
\begin{align}\label{eq:EntropyGS}
S\left[\rho_\G\right]= \sum_{j=1}^{n} h\left(\frac{{\rm d}_j/\sigma_0^2-1}{2} \right) \, ,
\end{align}
where $h(x)=(x+1) \log_2 (x+1) - x \log_2 x$, and $\{{\rm d}_j\}_j$ are the $n$ symplectic eigenvalues of $\bmsigma$, i.e. the positive eigenvalues of the $2n\times 2n$ matrix $i \Omega^{(n)} \bmsigma$ \cite{Ferraro2005, Serafini2017}.
Closed expressions for the symplectic eigenvalues are available for systems of $n=1,2$ modes. 
In particular, for $n=1$, we have ${\rm d}_1=\sqrt{\det(\bmsigma)}$, whereas for $n=2$ we express the CM in block form as in Eq.~(\ref{eq:BlockForm}) and obtain:
\begin{align}
    {\rm d}_{1(2)}= \sqrt{\frac{\Delta \pm \sqrt{\Delta^2-4 I_4}}{2}} \, ,
\end{align}
with $I_{1(2)}= \det(\bmsigma_{A(B)})$, $I_3= \det(\bmsigma_{AB})$, $I_4= \det(\bmsigma)$ and $\Delta= I_1+I_2+2I_3$. 

\end{itemize}

\begin{figure}
\centerline{\includegraphics[width=0.7\columnwidth]{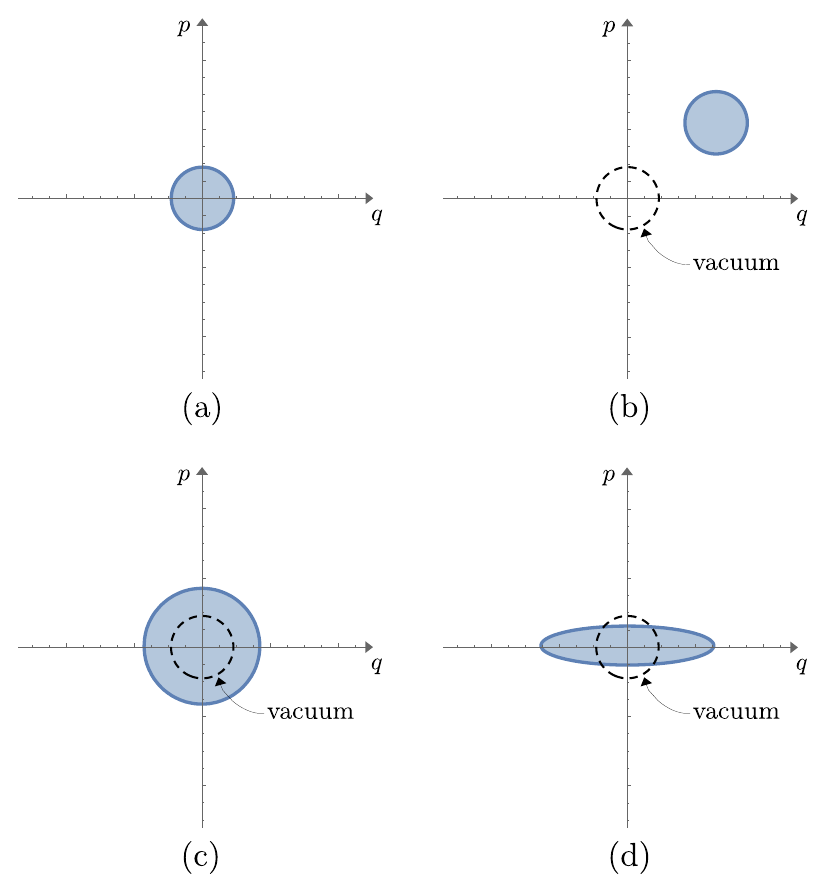}}
\centering
\caption{Phase space representation of Gaussian states: the vacuum $|0\rangle$ (a), a coherent state $|\alpha\rangle$ (b), $\alpha\in\mathbb{C}$, a thermal state $\nu^{\rm th} (\bar{n})$ (c), and a single-mode squeezed vacuum state $|r\rangle$, $r> 0$ (d).}\label{fig:01:sec2.2.2-PHSP}
\end{figure}

We now present some common examples of Gaussian states.
Moreover, for single-mode states it is also possible to provide a graphical phase space representation, depicted in Fig.~\ref{fig:01:sec2.2.2-PHSP}, obtained as the contour plot of the Wigner function $W(q,p)$ at the height of variances.

First of all, the vacuum $|0\rangle$ is a Gaussian state with $\textbf{R}=\textbf{0}$ and $\boldsymbol\sigma= \sigma_0^2 \Id_2$, being represented in the phase space as a circle with radius $\sigma_0$ centered in the origin, see Fig.~\ref{fig:01:sec2.2.2-PHSP}(a). Also coherent states $|\alpha\rangle = D(\alpha)|0\rangle$ are Gaussian, with $\textbf{R}=2 \sigma_0 (\Re\alpha,\Im \alpha)$ and $\boldsymbol\sigma= \sigma_0^2 \Id_2$; thus the displacement operator acts as a translation in the phase space, see Fig.~\ref{fig:01:sec2.2.2-PHSP}(b).
Thermal states are Gaussian, with null FM and $\boldsymbol\sigma= \sigma_0^2 (1+2\bar{n})$, see Fig.~\ref{fig:01:sec2.2.2-PHSP}(c), being, therefore, represented as a circle with a bigger radius than the vacuum. Finally, the single-mode squeezed state is Gaussian, with null FM and CM
\begin{align}
\boldsymbol\sigma= \sigma_0^2 \begin{pmatrix} e^{2r} & 0 \\ 0 & e^{-2r} \end{pmatrix} \, ,
\end{align}
where we assume $r\ge 0$: that is, the squeezing operator acts as a dilatation of the $q$-axis and a compression of the $p$ axis, deforming the vacuum circle into an ellipse of bigger semi-axis equals to $e^{r} \sigma_0$ and smaller semi-axis $e^{-r} \sigma_0$, see Fig.~\ref{fig:01:sec2.2.2-PHSP}(d). In the more general case $r= |r| e^{i \psi}$, the ellipse is rotated by an angle $\psi/2$.
We also note that, starting from these basic examples, it is possible to reconstruct all single-mode Gaussian state. In fact, by the former definition of Gaussian states of Eq. \ref{eq:rho_G}, it follows that a generic single mode Gaussian state can always be written as a displaced squeezed thermal state $\rho_\G = D(\alpha) S(r) \nu^{\mathrm{th}}(\bar{n}) S^\dagger(r) D^\dagger(\alpha)$,
for some parameters $\alpha,r \in \mathbb{C}$ and $\bar{n}\ge 0$.

Finally, within the class two-mode states, the typical example of Gaussian state is the TMSV, having null FM and CM:
\begin{align}
\boldsymbol\sigma= \sigma_0^2 \begin{pmatrix} V \, \Id_2 & Z \, \bmsigma_z \\  Z \, \bmsigma_z & V \, \Id_2 \end{pmatrix} \, ,
\end{align}
where $V=\cosh (2r)= (1+\lambda^2)/(1-\lambda^2)$ is the so-called quadrature variance, $Z=\sinh (2r)=\sqrt{V^2-1}$ is the correlation term, and $\bmsigma_z$ is the Pauli $z$-matrix.

\subsection{Quantum evolution of optical states}\label{sec2:EvoQO}

The dynamics of a quantum optical system follows the prescriptions outlined in the previous Section. That is, in the presence of a closed system, we have a unitary dynamics governed by a self-adjoint Hamiltonian operator $\hat{H}$ through the Schr\"odinger evolution $d\rho/dt= - i [\hat{H},\rho]$, where we adopted the natural units, $\hbar=1$.
Otherwise, open quantum dynamics are modeled by a suitable quantum master equation for the density operator $\rho$, obtained from a Hamiltonian dynamics over a larger system, that includes an environment, being, ultimately, traced out \cite{Serafini2017,Breuer2002}. Under proper dynamical assumptions, we retrieve a time-local master equation in the Lindblad form:
\begin{align}
\frac{d\rho}{dt}= \gamma \sum_j {\cal L}[A_j] \, \rho \, ,
\end{align}
where $\gamma$ is a decoherence rate and ${\cal L}[A] \rho= A \rho A^\dagger - \{A^\dagger A,\rho\}/2$, ${\cal L}[A]$ being the so-called Lindblad superoperator \cite{Breuer2002}.

In the following, we mainly focus on the special case of Gaussian dynamics, in which situation the theoretical description is rather simplified.

\subsubsection{Gaussian dynamics}\label{subsec2:GaussianDyn}

The class of Gaussian dynamics refers to evolutions of quantum states that preserve Gaussianity, namely mapping Gaussian states into Gaussian states. This corresponds either to unitary evolutions of closed systems associated with linear or bilinear Hamiltonian operators, or suitable master equations modeling quadratic interaction of an open system with a Gaussian environment \cite{Serafini2017, Ferraro2005}.
Since Gaussian states are completely characterized by first and second moments, we wonder to describe the dynamics in the phase space by introducing suitable transformation laws for the FM and the CM.
In more detail, if $\rho_{\IN(\OUT)}$ is the initial (evolved) state, with FM ${\bf R}_{\IN(\OUT)}$ and CM $\bmsigma_{\IN(\OUT)}$, the task is to determine transformation rules mapping $\textbf{R}_{\mathrm{in}} \to {\bf R}_\OUT$ and $\boldsymbol \sigma_{\mathrm{in}} \to \bmsigma_\OUT$. We consider both unitary and CP evolutions.

In the case of closed systems, a Gaussian evolution is a unitary evolution generated by linear or quadratic Hamiltonian $\hat{H}$. If $\hat{H}$ is linear, $\hat{H}=  {\bf a}^{\sf T} \hat{\textbf{r}}$, then the modes evolution reads \cite{Serafini2017, Ferraro2005}:
\begin{align}
\hat{\textbf{r}}_\OUT = \hat{\textbf{r}}_\IN+ {\bf d}\, ,
\end{align}
with ${\bf d}= \Omega^{(n)} \boldsymbol {\bf a} \, t$, $t$ being the duration of the time evolution. That is, the unitary operator $\hat{U}=\exp(-i \hat{H} t)$ takes the form of a displacement operator, and, accordingly:
\begin{align}
\textbf{R}_{\mathrm{out}}= \textbf{R}_{\mathrm{in}} + {\bf d} \qquad \mbox{and} \qquad \boldsymbol \sigma_{\mathrm{out}}= \boldsymbol \sigma_{\mathrm{in}} \, .
\end{align} 
Othwerwise, if the Hamiltonian $\hat{H}$ is quadratic, namely in the form $\hat{H}= \hat{\textbf{r}}^{\sf T} \ \mathbb{H} \ \hat{\textbf{r}}/2$, then:
\begin{align}
\hat{\textbf{r}}_\OUT = e^{i \hat{H} t} \ \hat{\textbf{r}}_\IN \ e^{-i \hat{H} t} = S \,\hat{\textbf{r}}_\IN \, ,
\end{align}
where $S=\exp[\Omega^{(n)} \mathbb{H} \, t]$ is a $2n \times 2n$ symplectic matrix, satisfying the fundamental property \cite{Serafini2017, Ferraro2005}:
\begin{align}
S \, \Omega^{(n)} \,S^{\sf T}= \Omega^{(n)} \, .
\end{align}
Then, the output first and second moments read:
\begin{align}
\textbf{R}_{\mathrm{out}}= S \, \textbf{R}_{\mathrm{in}} \qquad \mbox{and} \qquad  \boldsymbol \sigma_{\mathrm{out}}= S \ \boldsymbol \sigma_{\mathrm{in}} \ S^{\sf T} \, ,
\end{align} 
and the unitary evolution associated with $\hat{H}$ corresponds to a symplectic evolution of FM and CM under matrix $S$.

Finally,  we present some examples of symplectic matrices associated with commonly encountered transformations.
\begin{itemize}
\item \textit{Phase shift.} A phase shift is a single mode evolution associated with the unitary operator $U_\theta= \exp[-i \theta a^\dagger a]$. Its associated symplectic matrix is a rotation matrix of angle $\theta$:
\begin{align}
S= \begin{pmatrix} \cos \theta & \sin\theta \\ -\sin\theta & \cos\theta \end{pmatrix} \, .
\end{align}
\item \textit{Single-mode squeezing.} As introduced before, the single-mode squeezing evolution corresponds to the operator $S(r)= \exp[r(a^{\dagger 2}- a^2)/2]$. If $r>0$, the associated symplectic matrix is:
\begin{align}
S= \begin{pmatrix} e^r & 0 \\ 0 & e^{-r} \end{pmatrix} \, .
\end{align}
\item \textit{Beam splitter.} The beam splitter acts as a two-mode mixing evolution, mapping two input modes $a$ and $b$, into two output ports, associated with modes $c$ and $d$. It corresponds to the unitary operator $U_{\mathrm{BS}}=\exp[ \theta \ (a^\dagger b - a b^\dagger ) ]$, such that $\cos^2 \theta=T\le 1$ is the beam splitter trasmissivity. The symplectic matrix is the $4 \times 4$ matrix:
\begin{align}
S= \begin{pmatrix} \sqrt{T}\, \Id_2 & \sqrt{1-T}\, \Id_2 \\ -\sqrt{1-T}\, \Id_2 & \sqrt{T}\, \Id_2 \end{pmatrix} \, .
\end{align}
Accordingly, the input-output relations read:
\begin{subequations}
\begin{align}
c&= U_{\mathrm{BS}}^\dagger \ a \ U_{\mathrm{BS}} = \sqrt{T} \ a + \sqrt{1-T} \ b \, , \\[1ex]
d&= U_{\mathrm{BS}}^\dagger \ b \ U_{\mathrm{BS}} = \sqrt{T} \ b - \sqrt{1-T} \ a \,. 
\end{align}
\end{subequations}

\item \textit{Two-mode squeezing.} The two-mode squeezing operator $S_2(r)=\exp[r(a^\dagger b^\dagger - a b)]$ generates a Gaussian dynamics associated with the $4 \times 4$ symplectic matrix:
\begin{align}
S= \begin{pmatrix} \cosh(r) \, \Id_2 & \sinh (r)\, \bmsigma_z \\ \sinh(r)\, \bmsigma_z & \cosh (r) \, \Id_2 \end{pmatrix} \, ,
\end{align}
where we assumed $r\ge 0$. The input-output relations are the following:
\begin{subequations}
\begin{align}
c&= S_2(r)^\dagger \ a \ S_2(r) = \cosh(r) \ a + \sinh (r) \ b^\dagger \, , \\[1ex]
d&= S_2(r)^\dagger \ b \ S_2(r) = \cosh(r) \ b +\sinh(r) \ a^\dagger \,. 
\end{align}
\end{subequations}
\end{itemize}

In the opposite scenario of an open system, the Gaussian evolution of quantum states is provided by a suitable CP map, describing quadratic interaction with an environment prepared in a Gaussian state, and thereafter being traced out.
Accordingly, following Kraus theorem, any open Gaussian dynamics can be retrieved as a symplectic joint evolution of system and environment.
In more detail, if $\hat{\textbf{r}}_\IN$ and $\hat{\textbf{r}}_\IN^{(E)}$ are the input quadrature operator vectors of system and environment, respectively, and the environment is prepared in a Gaussian state with null FM and CM $\bmsigma_\IN^{(E)}$, there exists a symplectic matrix $S$, in the form:
\begin{align}
S= \begin{pmatrix} X & B \\ C & D \end{pmatrix} \, ,
\end{align} 
such that
\begin{align}
\begin{pmatrix} \hat{\textbf{r}}_\OUT \\ \hat{\textbf{r}}_\OUT^{(E)} \end{pmatrix} =
S \, \begin{pmatrix} \hat{\textbf{r}}_\IN \\ \hat{\textbf{r}}_\IN^{(E)} \end{pmatrix} \, .
\end{align} 
In turn, the input output relation for the system becomes:
\begin{align}
\hat{\textbf{r}}_\OUT = X\, \hat{\textbf{r}}_\IN  + B \, \hat{\textbf{r}}_\IN^{(E)} \, ,
\end{align}
and we obtain the FM and CM transformation rule as:
\begin{align}
\textbf{R}_{\mathrm{out}} = X \textbf{R}_{\mathrm{in}} \qquad \mbox{and} \qquad \boldsymbol\sigma_{\mathrm{out}} = X \boldsymbol\sigma_{\mathrm{in}} X^{\sf T} + Y \, ,
\end{align}
with $Y=B \, \bmsigma_\IN^{(E)} \, B^{\sf T}$.

In summary, a Gaussian CP map is defined by two $2n \times 2n$ matrices $X$ and $Y$ such that:
\begin{itemize}
\item[i)] $\textbf{R}_{\mathrm{out}} = X \textbf{R}_{\mathrm{in}} $ ,
\item[ii)] $\boldsymbol\sigma_{\mathrm{out}} = X \boldsymbol\sigma_{\mathrm{in}} X^{\sf T} + Y$,
\item[iii)] $Y+ i \sigma_0^2\Omega^{(n)} \geq i \sigma_0^2 X \Omega^{(n)} X^{\sf T}$,
\end{itemize}
where condition iii) guarantees that the additive noise matrix $Y$preserves Heisenberg's uncertainty relations for the output state.
In the particular case of $X$ being a symplectic matrix and $Y=0$, we regain the usual unitary evolution.

A paradigmatic example of a Gaussian CP map is the {\it thermal-loss channel}, corresponding to the loss master equation in the presence of a thermal bath:
\begin{align}
\frac{d\rho}{dt}= \gamma (\bar{n}+1) \, {\cal L}[a] \, \rho + \gamma \bar{n}  \,{\cal L}[a^\dagger]  \, \rho \, ,
\end{align}
wehre $\gamma$ is the system-bath coupling rate and $\bar{n}$ is the mean number of photons of the bath. The resulting dynamics is equivalent to a Gaussian CP map associated with the matrices \cite{Serafini2017}:
\begin{align}
X = \sqrt{T} \, \Id_2 \quad \mbox{and} \quad Y = (1-T)(1+2\bar{n}) \Id_2 \,,
\end{align}  
$\Id_2$ being the $2\times 2$ identity matrix, in which $T=\exp(-\gamma t) \le1$ is the channel transmissivity.



\subsection{Quantum measurements of optical states}\label{sec2:MeasQO}

In this section, we conclude the presentation of the fundamental aspects of quantum optics by discussing the main examples of quantum measurements, i.e. POVMs, that can be implemented. First of all, we provide characterization of Gaussian measurements, such that, if performed on a Gaussian state, yield a Gaussian probability distribution and a Gaussian conditional state, and, subsequently, we present some relevant non-Gaussian schemes, based on photo-detection.

\subsubsection{Gaussian measurements}\label{subsec2:Gauss}

Formally, with the term Gaussian measurement, we refer to a continuous-valued POVM $\{\Pi_\RMeas\}$, $\RMeas\in \mathbb{R}^{2n}$, whose elements $\Pi_\RMeas$ are associated with a Gaussian Wigner function:
\begin{align}
   W[\Pi_\RMeas](\mathbf{r}) =  \frac{1}{(4\pi\sigma_0^2)^n} \frac{1}{(2\pi)^{n} \sqrt{\det(\sigmaMeas)} } \,  \exp \left[ 
   - \frac12 (\mathbf{r}-\RMeas)^\mathsf{T} \, \sigmaMeas^{-1} \, (\mathbf{r}-\RMeas)
   \right] \, ,
\end{align}
where $\RMeas$ is the outcome of the measurement and $\sigmaMeas$ is the so-called ``covariance matrix of the measurement", being characteristic of the particular measurement performed.
We also note that, unlike quantum states, the function $W[\Pi_\RMeas](\mathbf{r})$ is not normalized, as the POVM elements constitute a resolution of the identity, $\int d\RMeas \Pi_\RMeas= \hat{\Id}$. In fact:
\begin{align}
&\int_{\mathbb{R}^{2n}} d\RMeas \, W[\Pi_\RMeas]({\bf r}) =   W \left[ \int_{\mathbb{R}^{2n}} d\RMeas \Pi_\RMeas \right]({\bf r}) \nonumber \\
&= W[\hat{\Id}]({\bf r})= \frac{1}{(4\pi\sigma_0^2)^n} \, ,
\end{align}
where we exploited the linearity of Wigner functions.
Then, by applying the trace rule~(\ref{eq:TrRuleWf}), when a Gaussian measurement is performed on a Gaussian state $\rho_\G$ with FM $\textbf{R}$ and CM $\bmsigma$, the probability of retrieving outcome $\RMeas$ reads:
\begin{align}\label{eq:generaldyneprob}
p(\RMeas) &= \Tr[\rho_\G \ \Pi_\RMeas] =(4\pi\sigma_0^2)^n \int_{\mathbb{R}^{2n}} d{\bf r} \, W[\rho_\G]({\bf r}) W[\Pi_\RMeas]({\bf r}) \\[1ex]
&= \frac{1}{(2\pi)^n \sqrt{\det (\boldsymbol\sigma + \sigmaMeas)}} \times \nonumber \\[1ex]
&\hspace{2cm}  \exp\left[-\frac{1}{2} (\RMeas- \textbf{R})^{\sf T} (\boldsymbol\sigma + \sigmaMeas)^{-1} (\RMeas- \textbf{R}) \right] \, .
\end{align}
That is, the outcomes of the Gaussian measurement are Gaussian distributed, with average value $\textbf{R}$ and covariance $\boldsymbol\sigma + \sigmaMeas$.

Finally, we also discuss the case of conditional measurements, that will be often encountered throughout the thesis. We consider a bipartite system $AB$, where subsystems $A$ and $B$ are composed of $n_{A}$ an $n_B$ modes, respectively. In the vector notation we have ${\bf a}= ({\bf a}_A,{\bf a}_B)$. We consider a bipartite quantum state $\rho_{AB}$ with characteristic functions $\chi_{AB}(\bmalpha)= \chi_{AB}(\bmalpha_A, \bmalpha_B)$.
We now perform a quantum measurement on subsystem $B$, described my means of the positive-operator-valued measurement (POVM) $\{\Pi_{\RMeas}\}_{\RMeas}$, whose elements are associated with the characteristic function $\chi_{\RMeas}(\bmalpha_B)$.
By applying the trace rule, the conditional state on $A$ reads:
\begin{align}
\rho_{A|\RMeas} &= \frac{1}{p(\RMeas)} \Tr_B\big[\rho_{AB} \big(\Id_A \otimes \Pi_{\RMeas} \big)\big] \notag \\
&= \frac{1}{p(\RMeas)} \int \frac{d^2\bmalpha_A} {\pi^{n_A}}\, \chi_{A|\RMeas}(\bmalpha_A)\,  D_{{\bf a}_A} (\bmalpha_A)\dag\, ,
\end{align}
where:
\begin{align}
\chi_{A|\RMeas}(\bmalpha_A) = 
\int \frac{d^2\bmalpha_B} {\pi^{n_B}}\,  \chi_{AB}(\bmalpha_A,\bmalpha_B)\, \chi_{\RMeas}(-\bmalpha_B) \, ,
\end{align}
and $p(\RMeas)$ is the detection probability:
\begin{align}\label{eq: SuccP}
p(\RMeas) &= \Tr_{AB} \big[\rho_{AB} \big(\Id_A \otimes \Pi_{\RMeas} \big)\big] \notag \\
&= \Tr_{A} \Bigg[\int \frac{d^2\bmalpha_A} {\pi^{n_A}}\, \chi_{A|\RMeas}(\bmalpha_A)\,  D_{{\bf a}_A} \Bigg]  
= \chi_{A|\RMeas}({\bf 0}) \, .
\end{align}
\par
An interesting result is obtained for Gaussian states and Gaussian measurements. We now assume $\rho_{AB}$ to be a Gaussian state with FM $\mathbf{R}=(\mathbf{R}_A,\mathbf{R}_B)$ and CM (written in block form)
\begin{align}
\boldsymbol\sigma = \begin{pmatrix} \boldsymbol\sigma_A & \boldsymbol\sigma_{Z} \\[1ex]\boldsymbol\sigma_{Z}^\mathsf{T} & \boldsymbol\sigma_B \end{pmatrix} \, .
\end{align}
Moreover, we consider a Gaussian POVM $\{\Pi_{\RMeas}\}_{\RMeas}$, with CM $\sigmaMeas$.
Then, the conditional state $\rho_{A|\RMeas}$ is still a Gaussian state with CM $\boldsymbol\sigma_{A|\RMeas}$ and FM $\mathbf{R}_{A|\RMeas}$ given by \cite{OlivaresGauss, Ferraro2005, Serafini2017}:
\begin{align}
    \boldsymbol\sigma_{A|\RMeas} = \boldsymbol\sigma_A - \boldsymbol\sigma_{Z} (\boldsymbol\sigma_B+\sigmaMeas)^{-1} \boldsymbol\sigma_{Z}^\mathsf{T}\,,
\end{align}
and
\begin{align}
    \mathbf{R}_{A|\RMeas} = \mathbf{R}_A + \boldsymbol\sigma_{Z} (\boldsymbol\sigma_B+\sigmaMeas)^{-1}(\RMeas-\mathbf{R}_B)\,,
\end{align}
respectively.

\paragraph{Homodyne and double-homodyne detection.}

\begin{figure}
\centerline{\includegraphics[width=0.95\columnwidth]{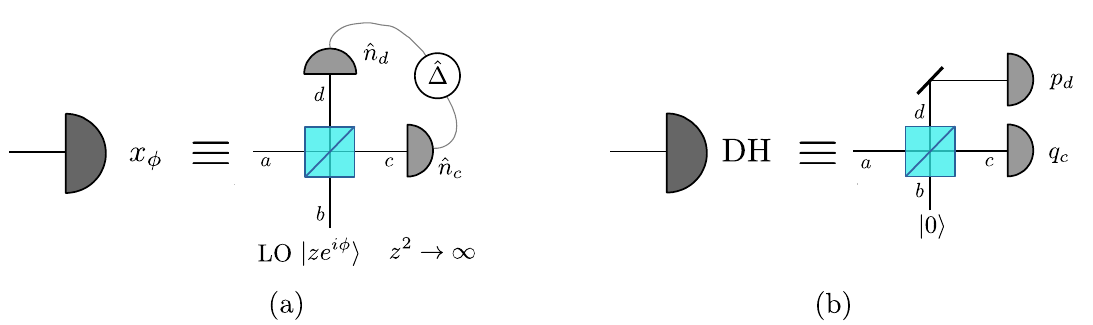}}
\centering
\caption{(a) Setup of homodyne detection of quadrature $x_\phi = \cos\phi \, q + \sin\phi \, p$. The signal interferes with a high-intensity LO $|z e^{i \phi}\rangle$ at a balanced beam splitter, with $z^2 \to \infty$; thereafter, photon-number detection is performed on both branches and the difference photocurrent $\hat{\Delta}$ is considered. (b) Setup of double homodyne (DH) detection. The
incoming signal is divided into two parts at a balanced beam splitter, on which
joint homodyne measurement of both quadratures $q_c$ and $p_d$ is performed. }\label{fig:01:sec2.41-HD}
\end{figure}

Typical examples of Gaussian measurements are the so-called {\it coherent detection schemes}, corresponding to homodyne and double-homodyne measurements. The name ``coherent" arises from the fact that, in experimental implementations, a strong local oscillator is let impinge with the incoming optical signal, which implies the two fields to be perfectly phase-matched in both the spatial and temporal domain.

To begin with, we introduce the {\it homodyne measurement}. With this term, we refer to measurement of one of the field quadratures:
\begin{align}
x_\phi = \cos\phi \, q + \sin\phi \, p = \sigma_0 \left(a e^{-i\phi}+a^\dagger e^{i \phi} \right) \, ,
\end{align}
$0\le \phi<\pi$. In turn, homodyne detection is described as a conventional projective measurement over the eigenstates of $x_\phi$, namely $\Pi_{x_\phi}=|x_\phi\rangle\langle x_\phi|$, and the probability of obtaining outcome $x$ from state $\rho$ reads $P(x)= \langle x_\phi | \rho | x_\phi \rangle$.

Remarkably, homodyne is a Gaussian measurement, associated with the CM:
\begin{align}
\sigmaMeas = \sigma_0^2 \,  \lim_{z\rightarrow 0} \left\{\mathcal{R}_\phi \begin{pmatrix} z & 0 \\ 0 & \frac{1}{z} \end{pmatrix} \mathcal{R}_\phi^{\sf T} \right\}\, ,
\end{align}
where $\mathcal{R}_\phi$ is a rotation matrix of angle $\phi$:
\begin{align}
\mathcal{R}_\phi= \begin{pmatrix} \cos \phi & \sin\phi \\ -\sin\phi & \cos\phi \end{pmatrix}.
\end{align}

The practical implementation is achieved by the scheme reported in Fig.~\ref{fig:01:sec2.41-HD}(a).
At first, the incoming signal is mixed at a balanced beam splitter, with transmissivity $T=1/2$, with a local oscillator (LO) excited in the coherent state $|z e^{i \phi}\rangle$, $z>0$; thereafter photon-number detection is performed on both the output beams \cite{Banaszek2020}. 
At the quantum limit, we describe the input optical modes associated with signal and LO by two bosonic operator $a$ and $b$, respectively, while
performing photon-number detection on the output modes $c$ and $d$ corresponds to measure the two Hermitian operators $\hat{n}_c= c^\dagger c$ and $\hat{n}_d=d^\dagger d$.
Ultimately, we evaluate the difference photocurrent $\hat{\Delta} = \hat{n}_{c}-\hat{n}_{d}$.
In the Heisenberg picture, namely by applying the modes transformations at a balanced beam splitter $a \to c = (a+b)/\sqrt{2}$ and $b \to d = (b-a)/\sqrt{2}$, we have \cite{Olivares2021}:
\begin{eqnarray}
\hat{\Delta} =  a b^\dagger + a^\dagger b \,  .\label{eq:DiffPhoto}
\end{eqnarray}
Therefore, considering the state $|z e^{i \phi}\rangle$ of the LO as fixed, we have:
 \begin{align}\label{eq:DiffP}
\langle z e^{i \phi}| \hat{\Delta} |z e^{i \phi}\rangle = \frac{z}{\sigma_0}  \left(a e^{-i \phi} + a^\dagger e^{i\phi} \right) = \frac{z}{\sigma_0} \, x_\phi \, .
\end{align}
In turn, measuring the rescaled photocurrent $\hat{\Delta}/(z\sigma_0)$ may provide an indirect measurement of the quadratures of the input field. If we compute, however,  the expectation value of $\hat{\Delta}^n$ we have:
\begin{eqnarray}\label{delta:moment}
\frac{\langle z e^{i \phi}| \hat\Delta^n | z e^{i \phi}\rangle}{(z/\sigma_0)^n}
= x_\phi^n + \gamma_{\phi}^{(n)}(a,a^\dagger;z) \, ,
\end{eqnarray}
where $\gamma_{\phi}^{(n)}(a,a^\dagger;z)$ is a function of both the field operators $a$ and $a^\dagger$ and the LO intensity, to be explicitly calculated; for instance, $\gamma_{\phi}^{(1)}(a,a^\dagger;z)=0$, $\gamma_{\phi}^{(2)}(a,a^\dagger;z)=a^\dagger a/(z/\sigma_0)^2$, $\gamma_{\phi}^{(3)}(a,a^\dagger;z)=(3 a^\dagger x_\phi\, a + x_\phi)/(z/\sigma_0)^2$ and
\begin{eqnarray}\label{delta:moment:4}
\gamma_{\phi}^{(4)}(a,a^\dagger)=\frac{3 (a^\dagger)^2 a^2+ a^\dagger a}{(z/\sigma_0)^4} + \frac{6 a^\dagger x_\phi^2\, a + 4(x_\phi^2-1)+1}{(z/\sigma_0)^2} \, .
\end{eqnarray}
Therefore, it is clear that the actual measurement of the quadrature is achieved only in the limit $z^2 \to \infty$ in which the measurement of the moments $(\hat\Delta/z)^n$ corresponds to measure $x_\phi^n$ \cite{Olivares2020}.
Moreover, in practical realizations, in the presence of high LO, photon-number measurement is implemented by p-i-n photodiodes, namely photodetectors generating macroscopic photocurrents proportional to the intensity of the incoming light.

The homodyne measurement allows us to retrieve information on a single field quadrature, raising the question to design a scheme for the joint detection of the two canonical quadratures $q$ and $p$. This task is not straightforward, as $[q,p]=2i\sigma_0^2$, therefore $q$ and $p$ cannot be jointly measured with maximum precision, according to Heisenberg's uncertainty principle. However, this kind of measurement can be realized in terms of a POVM, referred to as {\it double-homodyne} (DH) measurement.

Formally, DH detection is described as a $1$-rank (non-orthogonal) projection on coherent states, with associated POVM:
\begin{align}\label{eq:HETPOVM}
\Pi_{q,p}= \frac{|\zeta_{q,p}\rangle \langle \zeta_{q,p}|}{2\pi\sigma_0^2} \, ,
\end{align}
where $|\zeta_{q,p}\rangle$ is a coherent state with amplitude $\zeta_{q,p}=(q+ip)/2\sigma_0$, such that $\int dqdp \, \Pi_{q,p}= \hat{\Id}$.
The corresponding probability distribution is $P(q,p) = \langle \zeta_{q,p} | \rho | \zeta_{q,p} \rangle / 2\pi\sigma_0^2 = Q(q,p)$, that is the Husimi $Q$-function; whilst the CM of the measurement is $\sigmaMeas=\sigma_0^2 \Id_2$.
In turn, measurement of both quadratures introduces a ineludible excess noise, equal to the shot noise variance, that makes the quadrature variances larger than the usual homodyne detection.

The practical implementation of this scheme is reported in Fig.~\ref{fig:01:sec2.41-HD}(b). Now we split the incoming signal in two parts thanks to a balanced beam splitter, in whose second port we have the vacuum state $|0\rangle$. Then, we implement joint homodyne detection of quadrature $q_c=(q_a+q_b)/\sqrt{2}$ (corresponding to $\phi=0$) and $p_d=(p_b-p_a)/\sqrt{2}$ ($\phi=\pi/2$) on the transmitted and reflected branch $c$ and $d$, respectively.

\subsubsection{Some relevant examples of non-Gaussian measurements}

Beyond the Gaussian realm, the paradigmatic example of a non-Gaussian measurement is provided by photo-detection, namely {\it photon-number measurement}, described as projection onto the Fock states, $\Pi_n=|n\rangle\langle n|$. Usually, photo-detection is also referred to as {\it incoherent detection} or {\it direct detection}, as it does not require the presence of auxiliary fields nor an interferometric scheme, as for homodyne detection.
Nevertheless, from a practical point of view, resolving individual photons is a highly nontrivial task. In fact, all the existing photo-detectors actually implement a destructive measurement based on the photoelectric effect, and indirectly probe the particle-like properties of radiation by measuring the electron flow, i.e. the current, generated inside the detector by the absorption of photons.
In light of this, some examples of commonly employed photo-detectors are the following:
\begin{itemize}
\item {\it p-i-n photodiodes}, namely proportional detectors that generate a macroscopic current proportional to the incoming number of photons. They provide an efficient solution for detection of semi-classical fields, e.g. homodyne detection, whilst not being able to resolve photons;
\item {\it on-off detectors}, that is click-no click detectors that can only resolve the vacuum state, distinguishing the absence or presence of photons. Accordingly, they are described by the binary POVM $\{\Pi_{\rm off}, \Pi_{\rm on} \}$, with $\Pi_{\rm off}=|0\rangle \langle 0|$ and $\Pi_{\rm on} = \hat{\Id} - \Pi_{\rm off}$;
\item {\it photon-number resolving (PNR) detectors}, capable of resolving the lowest Fock states $n=0,1,\ldots, M$ up to a maximum number $M< \infty$. 
\end{itemize}
Within these classes, PNR detectors provide the more advanced solution from the technological point of view, and its functioning will be discussed in detail in the following.

\subsubsection{Photon-number resolving detection}\label{subsec2:PNR}

As mentioned above, PNR detectors are able to resolve any number of photons $n$ up to a maximum number $M< \infty$, hence referred to as the photon number resolution; to highlight this features, throughout this thesis we will refer to them as PNR($M$) detectors. For instance, PNR(3) refers to a detector that has only four possible outcomes $n \in \{0,1,2,\geq\!3\}$, where ``$\geq\!3$'' means 3 or more photons. 
Clearly, PNR(1) detectors correspond to on-off detectors, whereas ideal photo-detection requires PNR($M$) with $M=\infty$ \cite{Notarnicola2023:HYNORE}.
Accordingly, we describe PNR($M$) detection by a finite-valued POVM with $M+1$ possible outcomes, $\{\Pi_0 , \Pi_1 ,\ldots, \Pi_M \}$, with:
\begin{align}
\Pi_n = 
\left\{\begin{array}{ll} 
 | n \rangle\langle n| &\mbox{if } n=0,\ldots, M-1 \, , \\[2ex]
 \displaystyle \hat{\Id} - \sum_{j=0}^{M-1} | j \rangle\langle j| &\mbox{if } n=M .
\end{array}
\right.
\end{align}
As a consequence, if we are performing PNR($M$) measurement on a generic coherent state $|\alpha\rangle$, $\alpha \in \mathbb{C}$, the probability of detecting the outcome $n=0,\ldots,M$ reads:
\begin{align}\label{eq:qnMU}
    p_n(N) = \langle \alpha| \Pi_n |\alpha \rangle=
    \left\{\begin{array}{l l}
    {\displaystyle e^{-N} \ \frac{N^{n}}{n!}}  &\mbox{if } n=0,\ldots, M-1 \ , \\[2ex]
    {\displaystyle 1- e^{-N} \sum_{j=0}^{M-1} \frac{N^{j}}{j!}} &\mbox{if } n = M \ ,
    \end{array}
    \right.
\end{align}
with a mean photon number $N=|\alpha|^2$, namely a truncated Poisson distribution.

Good candidates as PNR detectors are the hybrid photodetectors, which are endowed with partial photon-number resolution and a linear response up to 100 photons \cite{Olivares2019PNR}, though with a quantum efficiency of about $50\%$ in the green spectral region. Very high quantum efficiencies are obtained with transition-edge sensors (TES), but their dynamic range falls to approximatively 10 photons \cite{Thekkadath2020,Nehra2020}. More recently, also Silicon photomultipliers (SiPMs) has been investigated as
PNR detectors, since they are more compact and with higher dynamic range \cite{Chesi2019, Cassina2021, Sanvito2024}.

\subsubsection{Weak-field homodyne detection}\label{subsec2:HL}

\begin{figure}
\centerline{\includegraphics[width=0.65\columnwidth]{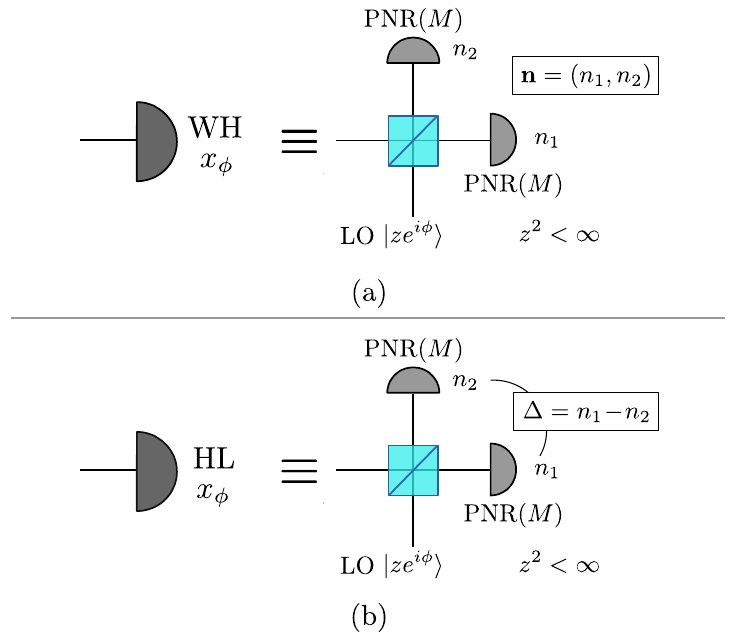}}
\centering
\caption{(a) Scheme of weak-field homodyne ($\WFH$) detection employing a low LO, $z^2 < \infty$, and PNR($M$) detection. The outcome is a pair of integer values ${\bf n}=(n_1,n_2)$, $n_{1(2)}=0,\ldots, M$. (b) Homodyne-like ($\HL$) detection, namely a $\WFH$ scheme where we only consider the difference photocurrent $\Delta=-M,\ldots, M$ at the output.}\label{fig:01:sec2.4.2-WH}
\end{figure}

Given the previous considerations, an intriguing question arises: that is, to investigate the performance of the homodyne scheme, depicted in Fig.~\ref{fig:01:sec2.41-HD}(a), beyond the high-LO limit, when $z^2 < \infty$, in which case the setup does not implement quadrature detection anymore.

We can identify two possible scenarios. The first one is referred to as weak-field homodyne ($\WFH$) detection, whose scheme is reported in Fig.~\ref{fig:01:sec2.4.2-WH}(a): the conventional p-i-n photodiodes of the standard homodyne detection scheme are replaced by PNR($M$) detectors \cite{Donati2014, Thekkadath2020}. As a matter of fact, realistic PNR detectors have a finite resolution $M<\infty$, hence a low-intensity LO is required. The use of PNR detectors gives access to the two local photon-number statistics and allows
jointly probing both the wake- and particle-like properties of the field.
Accordingly, $\WFH$ detection returns a pair of integer outcomes ${\bf n}=(n_1,n_2)$, $n_{1(2)}=0,\ldots, M$. Given an input coherent state $|\alpha\rangle$, $\alpha \in \mathbb{C}$, the resulting probability distribution reads:
\begin{align}
P^{(\phi)}_{\bf n}(\alpha)=  p_{n_1}\big(\mu_{+}(\alpha;\phi)\big) p_{n_2}\big(\mu_{-}(\alpha;\phi)\big) \,,
\end{align}
where 
\begin{align}
\mu_{\pm}(\alpha;\phi)= \frac{\left|\alpha \pm z e^{i\phi}\right|^2}{2}\, ,
\end{align}
is the mean energy on the two output branches, respectively, and $p_{n}(\mu)$ is the PNR($M$) distribution reported in Eq.~(\ref{eq:qnMU}).

In the second scenario we consider the $\WFH$ scheme where we evaluate the difference $\Delta$ between the number of photons measured at output beams. Due to the clear analogy with the standard homodyne detection, we refer to this configuration as homodyne-like ($\HL$) detection, depicted in Fig.~\ref{fig:01:sec2.4.2-WH}(b) \cite{Allevi2017, Olivares2019PNR, Chesi2019}. The probability of obtaining the value $\Delta=n_1-n_2$, with $-M \le \Delta \le M$, becomes:
\begin{align}\label{eq:probDelta}
{\cal S}^{(\phi)}_\Delta(\alpha)= \sum_{n_1,n_2=0}^{M} p_{n_1}\big(\mu_{+}(\alpha;\phi)\big) p_{n_2}\big(\mu_{-}(\alpha;\phi)\big) \, \delta_{n_1-n_2,\Delta} \,,
\end{align}
$\delta_{jk}$ being the Kronecker delta. In particular, when $M=\infty$, $\Delta$ represents the difference between two Poisson variables, thus Eq.~(\ref{eq:probDelta}) approaches a Skellam distribution \cite{Notarnicola2023:HYNORE}.

\def\ns{n_{\rm S} }
\def\nn{n_{\rm N} }

\section{Fundamentals of quantum communication systems}\label{chap:QComm}

In this last Section of Part~I, we discuss the fundamental aspects of quantum communication systems, whose general goal is to transmit information between a sender and a receiver located at a certain distance.
Given this scenario, different protocols can be designed, according to both the type of information to be shared, either classical information (e.g. a classical message) or quantum information (i.e. a quantum state), and the receiver's task. In particular, in the rest of the thesis we will deal with two different classes of protocols involving transmission of classical information carried by quantum states over a quantum channel, that is quantum state discrimination and continuous variable quantum key distribution.
Instead, in this Section we maintain a general approach, and provide a general description, being valid for all the subsequent case studies.

The Section is organized as follows. At first, in Sec.~\ref{sec3:GeneralschemeQC}, we present a general scheme of a telecommunication system, discussing the role and the composition of the sender and receiver stations. Then, we focus on two relevant aspects. In Sec.~\ref{sec3:OptSig}, we discuss in detail the methods adopted for encoding of classical information onto optical signals, both at the classical and quantum limit. Thereafter, in Sec.~\ref{sec3:InfoTh}, we briefly review the basics of information theory, firstly developed by Shannon \cite{Shannon1949}, that provide the theoretical description for all the considered protocols.

\subsection{A general scheme of telecommunication systems}\label{sec3:GeneralschemeQC}

Generally speaking, any communication system, operating both at the classical and the quantum limit, involves two distant parties, the sender (also referred to as transmitter or modulator) and the receiver (or demodulator), usually renamed as Alice and Bob, respectively.

Alice handles a classical source that emits a message to be reliably transmitted to Bob via a communication channel.
In more detail, the message $m$ is a sequence of classical symbols $m=(x_1,\ldots, x_n)$, for some $n\ge 1$, where the $\{x_j\}$ are outcomes of a random variable $X$, whose values $x$ are drawn from a given alphabet ${\cal A}$ and distributed according to the probability $\{p(x)\}$.
In classical information theory, the amount of information contained in each of these sequences is established by the first Shannon coding theorem, being equal to the minimum number of binary digits (bits) needed to determine its binary representation \cite{Shannon1949, Holevo2011}. In the aymptotic limit $n\gg 1$, this quantity is approximately equal to $n H(X)$, where
\begin{align}
H(X)= {\sf H}[p(x)] \equiv -\sum_{x \in {\cal A}} p(x) \log_2 p(x)
\end{align}
is the Shannon entropy of the probability distribution $\{p(x)\}$.
To convey this information to Bob, Alice encodes the message into optical signals, e.g. laser pulses, being injected into a communication channel, being, in general, noisy. In turn, at the channel end Bob receives a corrupted pulse, from which he extracts an approximate replica of the original message, $m'=(y_1,\ldots, y_n)$.
Practically, the optical transmitter is composed of three subsequent elements: the optical source, generating a carrier optical field, the modulator, that modifies the shape of the carrier field according to the encoded information (by typically varying its amplitude or phase), and a coupling device adapting the beam to the optical transmission medium, e.g. optical fibers, free space, water, \ldots. 
Similarly, the receiver is composed of the cascade of a coupling device and a detector; therefore it ultimately implements a measurement of the received pulse, whose outcome provides a classical random variable $Y$ correlated to $X$.

In general, the receiver may be designed to accomplish different objectives, that, accordingly, identify different kinds of communication protocols. Some common examples are the following:
\begin{itemize}
\item[i)] {\it quantum state discrimination}, whose task is to distinguish between a set of possible symbols with the minimum decision error probability;
\item[ii)] {\it optical communication}, in which the goal is to reliably transmit classical information over the channel at the maximum rate;
\item[iii)] {\it quantum key distribution}, where the two parties aim at sharing a common random secure key, even if the channel is attacked by a third malicious party, the eavesdropper.
\end{itemize}

In this thesis, we will widely discuss cases i) and iii), that will be addressed in Parts~II and~III, respectively.
On the contrary, here we maintain a general approach and present some basic features of both optical signaling and information theory that will be helpful throughout the rest of the work.

\subsection{Encoding of optical signals}\label{sec3:OptSig}

As previously mentioned, information is usually transmitted by electromagnetic waves that propagate throughout the physical channel. The sender encodes the input symbols into linearly polarized optical signals, e.g. laser pulses, located in temporal slots of duration $T=B^{-1}$, where $B$ is the so-called {\it slot rate}, or {\it symbol period}, characterizing the width of the spectrum in the frequency domain \cite{Banaszek2020, Essiambre2010, Cariolaro2015}.
Given that a laser source produces a narrowband radiation field, each single pulse is described by a quasi-monochromatic wavepacket:
\begin{align}
\psi(t)= u(t) e^{-i \omega_0 t} \, ,
\end{align}
where $\omega_0$ is the carrier angular frequency, such that $2\pi B \ll \omega_0$, and $u(t)$ is the complex envelope of the waveform.
The commonly adopted choices for $u(t)$ are twofold. The first one is to employ square waves, equal to:
\begin{align}
u(t)=
\left\{\begin{array}{ll} 
u_0 &\mbox{for } 0<t<T \, , \\[1ex]
0 &\mbox{elsewhere} \, ,
\end{array}
\right.
\end{align}
for some complex constant $u_0=|u_0|e^{i\phi_0} \in \mathbb{C}$. This choice will provide our benchmark scenario throughout the whole thesis. The latter is to use sinc waves, $u(t)=u_0 \sin(\pi B t)/(\pi Bt)$, overlapping in the temporal domain, in which case the slot rate $B$ coincides with the signal bandwidth in the Fourier space \cite{Banaszek2020, Essiambre2010, Cariolaro2015}.

From a classical viewpoint, the electric field associated with the wavepacket $\psi(t)$ is obtained as:
\begin{align}
E(t)= \mathscr{E}_0 \Big[u(t) e^{-i \omega_0 t}  + u^*(t) e^{i \omega_0 t} \Big] \, ,
\end{align}
$\mathscr{E}_0=\sqrt{\hbar \omega_0/(2\epsilon A_{\rm eff})}$ being the corresponding electric field per single photon, where $\hbar$, $\epsilon$ and $A_{\rm eff}$ are Planck's reduced constant, the electric permittivity of the propagation medium and the effective area of the transverse signal spatial mode, respectively \cite{GerryKnight}.
Unlike the previous Section, for a clearer understanding here we choose not to adopt natural units, to emphasize the role of Planck's constant.
In turn, the average optical power is obtained as:
\begin{align}
P= \frac{1}{T} \int_0^{T} dt \, \int_{A_{\rm eff}} d^2{\bf r} \left(\frac12 \epsilon \, |E(t)|^2 \right) = B \hbar \omega_0 \int_0^{T} dt \, |u(t)|^2 \, ,
\end{align}
that reduces to $P=B \hbar \omega_0 |u_0|^2$ in the presence of square waves.

On the contrary, at the quantum limit, each wavepacket is described by (possibly mixed) a quantum state $\rho$ with mean energy, i.e. mean number of photons, equal to $\bar{n}= P/(B \hbar \omega_0)$; such that the previously introduced time integral of the complex envelope, $\int_0^{T} dt |u(t)|^2$, corresponds to the mean pulse energy. In particular, laser pulses (with stable phase) are well described by coherent states $\rho=|\alpha\rangle \langle \alpha|$, with $|\alpha\rangle= e^{-|\alpha|^2/2} \sum_n \alpha^n/\sqrt{n!}|n\rangle$, $\alpha\in\mathbb{C}$ and $|n\rangle$ being the $n$-photon state, containing $\bar{n}=|\alpha|^2$ mean photons \cite{Banaszek2020}. In particular, for square-wave encoding we have $\bar{n}=\int_0^{T} dt |u(t)|^2= |u_0|^2 T \equiv |\alpha|^2$, establishing the connection between the wavepacket and the coherent state amplitudes, respectively.

\subsection{Elements of information theory}\label{sec3:InfoTh}

We now present the basic features of information theory, providing the framework to describe communication protocols operated in both the classical and quantum regimes.

In an optical communication protocol between a sender (Alice) and a receiver (Bob), the complex amplitude $u_0$ for a wavepacket $\psi(t)=u(t) e^{-i\omega_0 t}$ located in a given slot is randomly selected according to the outcome $x$ of an input random variable $X$, associated with probability $p_A(x)$. Thereafter, Alice injects the signal into the noisy channel, corrupting the encoded symbol, and, after propagation, Bob, performs a suitable measurement, retrieving an outcome $y$, associated to a random variable $Y$ \cite{Holevo2011, Cariolaro2015, Banaszek2020, Notarnicola2024:JASC}.

In a classical description in the absence of memory effects, the channel is characterized by a conditional probability distribution $p_{B|A}(y|x)$, such that the overall probability distribution of $Y$ reads $p_B(y)=\sum_x p_A(x) p_{B|A}(y|x)$ \cite{Cover1999}. 
As established by the second Shannon coding theorem, the amount of information about $X$ extractable from $Y$ is equal to the mutual information:
\begin{align}
I(X;Y) = H(Y) - H(Y|X) \, ,
\end{align}
where $H(Y)= {\sf H}[p_B(y)]$ and $H(Y|X)= \sum_x p_A(x) \, {\sf H}[p_{B|A}(y|x)]$ are the average and conditional Shannon entropies, respectively \cite{Cariolaro2015, Holevo2011, Banaszek2020}. Mutual information optimized over all possible input distributions yields the channel capacity:
\begin{align}
C = \max_{p_A(x)} I(X;Y) \, ,
\end{align}
expressed in bits per time slot, i.e. per channel use, and representing the maximum amount of information extractable from the correlated variables $X$ and $Y$. Accordingly, the maximum attainable transmission rate ${\sf R}$, expressed in bits per unit time, is given by ${\sf R}= B C$ \cite{Banaszek2020}.

However, in many different conditions, e.g. high-loss transmission or long-distance communications, the probed signals are so attenuated that the granulosity of electromagnetic radiation emerges, thus a quantum description in terms of photons is required.
In a quantum picture, Alice encodes symbols $x$ onto quantum states of radiation $\rho_x$, the quantum channel is modeled as a quantum CP map ${\cal E}$, while Bob's detection is described by a POVM $\{\Pi_y\}_y$, $\Pi_y \ge 0$ and $\sum_y \Pi_y = \Id$. In turn, the conditional distribution follows from the Born rule, $p_{B|A}(y|x)= \Tr[{\cal E}(\rho_x) \Pi_y]$ \cite{Paris2012}. Within this framework, for each combined choice of both the carrier quantum states and the POVM at Bob's side we define a different classical channel, mapping the input variable $X$ into its counterpart $Y$. 
Therefore, the classically-evaluated capacity  $C$ determines the maximum achievable information rate from given input ensemble $\{\rho_x\}_x$ and POVM $\{\Pi_y\}_y$. 

Given this scenario, we may proceed beyond classical limits, and determine the ultimate channel capacity by optimization over all quantum measurements and state ensembles, thus obtaining the maximum information rate compatible with quantum mechanics laws. 
To this aim, we first resort to the Holevo theorem, establishing an upper bound to the mutual information of a channel whose symbols are encoded onto quantum states \cite{Holevo1973}. That is, for any choice of state ensemble and quantum measurement, associated with random variables $X$ and $Y$, we have:
\begin{align}\label{eq:HolevoBoundth}
I(X;Y) \le \chi(A;B) \, ,
\end{align}
where $\chi(A;B)$ is the Holevo information between Alice and Bob:
\begin{align}
\chi(A;B) =  {\sf S} \left[{\cal E}\left( \sum_x p_A(x) \rho_x \right)\right] - \sum_x p_A(x) {\sf S} [{\cal E}(\rho_x)] \, ,
\end{align}
${\sf S}[\rho]=-\Tr[\rho \log_2 \rho]$ being the von Neumann entropy of state $\rho$ \cite{Holevo2011}.
We note that $\chi(A;B)$ has the same formal structure of the mutual information, being the difference between the von Neumann entropy of the average output quantum state after propagation and the average von Neumann entropy of individual output states.
Importantly, it only depends on the input state ensemble $\{\rho_x\}$ and modulation $\{p_A(x)\}$ and not on the employed quantum measurement. Moreover, the coding theorem established by Holevo \cite{Holevo1979CAP, Holevo1998} and by Schumacher–Westmoreland \cite{Schumacher1997} proves that the bound~(\ref{eq:HolevoBoundth}) is saturable by a particular POVM, being, in general, a collective measurement. 
Accordingly, the ultimate transmission rate is provided by the so-called Holevo capacity, or classical capacity, equal to \cite{Holevo1998, Giovannetti2014}:
\begin{align}\label{HolevoCAP:3}
C_{\rm H} = \max_{\{\rho_x, p_A(x)\}} \chi(A;B) \, ,
\end{align}
that can be regarded as the analog of the  previously introduced second Shannon coding theorem.

As a final remark, we note that our analysis was focused on the transmission of classical messages, achieved by the so-called ``classical–quantum" channels. Actually, this only provides the simplest scenario, operating single-letter encoding, such that a message $m=(x_1\ldots, x_n)$ is encoded onto a factorized quantum state $\rho_{x_1}\otimes \ldots \otimes \rho_{x_n}$. Further improvements can be obtained by letting the sender use as input any (possibly entangled) quantum state, or by introducing cooperation between the input and output of the quantum channel, leading to the concept of  ``quantum channel capacity", as opposed to the classical channel capacity in Eq.~(\ref{HolevoCAP:3}) \cite{Holevo2011}.

\subsubsection{The capacity of the thermal-loss channel}

As a milestone example, we now apply the previous analysis to assess the capacity of the thermal-loss channel, modeling transmission inside optical fibers in the presence of additive Gaussian excess noise, thus being also known with the term additive-white-Gaussian noise (AWGN) channel \cite{Giovannetti2004, Holevo2001, Banaszek2020}. The channel is characterized by a transmissivity $T \le 1$, quantifying signal attenuation, and a mean number of excess noise photons $\nn$ added to the signal mode in each time slot.
In particular, if Alice performs modulation of coherent states with average input energy $\bar{n}$, after transmission Bob probes a rescaled displaced-thermal state ensemble, with mean received energy
\begin{align}
\ns= T \bar{n} \, ,
\end{align}
and thermal variance on both quadratures equal to $1+2\nn$, expressed in shot-noise units (SNU).

\begin{figure}
\includegraphics[width=0.6\columnwidth]{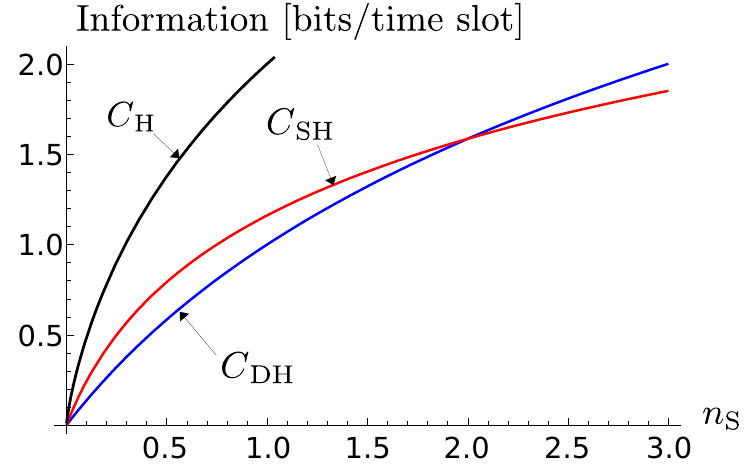}
\centering
\caption{Plot of the Shannon capacities $C_{\rm r}$, $\rm r=SH,DH$, in Eq.~(\ref{eq:ShannonCapacity}), and the Gordon-Holevo capacity $C_{\rm H}$ in~(\ref{eq:HolevoCapacity}) as a function of the mean received energy $\ns$ for zero excess noise photons $\nn=0$.}\label{fig01:sec3.3.1_Fundamentals}
\end{figure}

Within a classical description of the channel, the maximum information rate follows from the the Shannon-Hartley theorem \cite{Shannon1949}, and obtained via coherent-state encoding and measurement of either one or both canonical field quadratures $q$ and $p$, that is by single- (SH) or double-homodyne (DH) detection, respectively \cite{Olivares2021, Banaszek2020}.  The corresponding Shannon channel capacity is reached with Gaussian modulation of the coherent state amplitude $\alpha=x_A+i y_A$. For the homodyne case, we set $y_A \equiv 0$ and the uni-variate modulation $p_A(x_A)={\cal N}_{\sigma^2}(x_A)$, where
\begin{align}
{\cal N}_{\sigma^2}(x) = \frac{\exp\big[-x^2/(2\sigma^2)\big]}{\sqrt{2\pi \sigma^2}} \, ,
\end{align}
whereas in the presence of DH detection we adopt a bi-variate modulation, namely $p_A(x_A,y_A)= {\cal N}_{\sigma^2}(x_A){\cal N}_{\sigma^2}(y_A)$. In turn, the average quantum state at Bob's side contains $\ns= \sigma^2$  and $\ns= 2\sigma^2$ mean photons, respectively. Then, the Shannon capacities read:
\begin{subequations}\label{eq:ShannonCapacity}
\begin{align}
C_{\rm SH} &=\frac12 \log_2 \left(1+\frac{4\ns}{1+2 \nn} \right)\, , \\[1ex] 
C_{\rm DH} &= \log_2 \left(1+\frac{\ns}{1+\nn} \right)\, ,
\end{align}
\end{subequations}
and are reported in Fig.~\ref{fig01:sec3.3.1_Fundamentals} as functions of the mean received signal energy for zero excess noise photons.
When $\nn=0$, we have $C_{\rm SH} \ge C_{\rm DH}$ for $\ns\le 2$, whereas for higher energy the DH capacity becomes larger than the single quadrature one. In fact, the joint measurement of both the non-commuting $q$ and $p$ operators, corresponding to the DH detection, introduces an ineludible excess noise equal to the shot-noise vacuum fluctuations, thus reducing the available signal-to-noise ratio (SNR) \cite{Banaszek2020}. In turn, there is a tradeoff between this reduced SNR and the increase of accessible information due to the bi-variate signal modulation, such that if the signal energy is sufficiently low SH detection becomes preferable.
Instead, in the opposite limit of large excess noise, $\nn \gg 1$, we have:
\begin{align}
C_{\rm SH} \approx \frac12 \log_2 \left( 1 + 2\frac{\ns}{\nn} \right) <
\log_2 \left( 1 + \frac{\ns}{\nn} \right) \approx C_{\rm DH} \, ,
\end{align}
and for all SNR values DH detection is the preferable choice.

However, the Shannon capacity limit is obtained from two underlying assumptions, namely the adoption of coherent states as information carrier at the input and quadrature detection at the output. 
When both these limits are relaxed, we obtain the ultimate capacity of the channel, optimized over all measurement and quantum states, corresponding to the Gordon-Holevo capacity:
\begin{align}\label{eq:HolevoCapacity}
C_{\rm H} = g(\ns+\nn) - g(\nn) \, ,
\end{align}
where $g(x)=(x+1)\log_2(x+1) -x \log_2 x$, see Fig.~\ref{fig01:sec3.3.1_Fundamentals}. 
In particular, for pure-loss transmission, $\nn=0$, we have $C_{\rm H}=g(\ns)$ and, in the high-energy regime $\ns\gg 1$:
\begin{align}
C_{\rm H} = g(\ns) = \log_2 (1+\ns)
+ \log_2 e - \frac{\log_2 e }{2\ns}
+ O(\ns^{-2}) \, .
\end{align}
The first term coincides with the Shannon DH capacity, therefore for high signal energy the Holevo capacity introduces a constant information enhancement equal to $1~\text{nat} = \log_2 e \approx 1.44~\text{bits}$. On the contrary, when $\nn \gg 1$, we have:
\begin{align}
C_{\rm H} &= g(\ns+\nn)-g(\nn) \nonumber \\
&= C_{\rm DH}+ \left( \frac{1}{\nn} - \frac{1}{\ns + \nn}\right) \log_2 e + O(\nn^{-2}) \, ,
\end{align}
and DH detection re-approaches $C_{\rm H}$.

Remarkably, we note that also the Gordon-Holevo capacity is reached by Gaussian modulation of coherent states, as for the Shannon case. However, the optimal POVM achieving~(\ref{eq:HolevoCapacity}) is a collective measurement, operating simultaneously on multiple time slots, whose associated detection scheme is still unknown \cite{Holevo2011, Banaszek2020, Cariolaro2015}. 
Thus, the interest has been directed to either design simple collective measurements approximating the Holevo capacity in particular energy regimes \cite{Guha2011, Banaszek2017}, or to find feasible suboptimal individual measurement performed on single time slots, e.g. photon-number measurement \cite{Bowen1968, Shamai1990, Martinez2007, Cheraghchi2019, Lukanowski2021, Notarnicola2024:JASC}.

\section*{Part II: Quantum state discrimination theory}\label{part2}
\addcontentsline{toc}{section}{Part II: Quantum state discrimination theory}
\def\PK{P_{\rm K}}
\def\PD{P_{\rm D}}
\def\PHel{P_{\rm Hel}}
\def\PSQL{P_{\rm SQL}}
\def\PIK{P_{\rm IK}}
\def\PDISP{P_{\rm DF}}
\def\PHY{P_{\rm HY}}
\def\PHFF{P_{\rm HF}}
\def\opt{{\rm opt}}
\def\p{{\rm p}}
\def\nth{n_{\rm th}}
\def\OUTPUT{\Phi}
\def\sigmamax{\sigma_{\rm max}}

\section{Introduction to quantum decision theory}\label{chap:GeneralFeatures}\thispagestyle{plain}

The problem of performing discrimination of quantum states in efficient way is ubiquitous in quantum information, being at the basis of many communication and computing protocols, e.g. quantum key distribution and probabilistic computing algorithms. Furthermore, it also plays a crucial role in hypothesis testing scenarios, such as gravitational waves detection \cite{Bergou2010}.
In fact, in all quantum information schemes, information carriers are provided by quantum states of a physical system. In turn, when the set of all possible output states is known, one has to infer which particular state was exploited in order to read out the encoded information. However, this task can be easily accomplished only with a family of mutually orthogonal states, whereas, in the presence of non-orthogonal states, quantum mechanics laws forbid exact discrimination. As a consequence, any discrimination strategy is associated with a decision error probability, and the optimum receiver is the one achieving the minimum error probability compatible with the quantum mechanical limits.
Moreover, one can adopt different discrimination strategies according to the specific context under investigation, e.g. quantum quantum decision theory, unambiguous discrimination, and maximum confidence strategies \cite{Bergou2010}.

Historically, the problem of quantum state discrimination has been firstly raised in the 1970's by the seminal works of Helstrom \cite{Helstrom1976} and Holevo \cite{Holevo2011}. Following the subsequent developments of quantum information theory, there has been a revived interest in the 1990's, when Bennett et al. recognized that non-orthogonal states could be also used for quantum key distribution \cite{B92}.
In parallel, the topic has been also investigated in the context of telecommunication engineering, whose technological progresses led to the experimental demonstration of non-negligible noise of quantum origin in long-distance and deep-space communications, that ultimately introduce a bit error rate on the information read out at the receiver \cite{Cariolaro2015}.

In this Section, we provide the general framework of the theory, and then focus on the relevant scenario of binary quantum decision theory. In particular, we address discrimination of coherent states and present the most relevant examples of feasible quantum receivers, analyzing their performance also in the presence of realistic defects, e.g. non-unit quantum efficiency, dark counts, visibility reduction and phase diffusion noise.
The structure of the Section is the following. In Sec.~\ref{sec:4-Frame} we schematize the general features of quantum state discrimination, whereas Sec.~\ref{sec4:QDT} focuses itself on quantum decision theory, with particular reference to the binary case.
Thereafter, in Sec.~\ref{sec4:BinaryCoh} we address binary discrimination of coherent states, and consider binary phase-shift keying modulation, in which information is encoded on the phase of a coherent signal with fixed mean energy. Furthermore, we present the benchmark examples of quantum receivers achieving a genuine quantum advantage over conventional schemes based on quadrature detection, namely the Kennedy, Dolinar, displacement feed-forward (DFFRE) and Sasaki-Hirota receivers. Then, in Sec.~\ref{sec4:Hybrid}, we propose two new hybrid schemes, the hybrid near-optimum (HYNORE) and the hybrid feed-forward (HFFRE) receivers, and obtain a further reduction of the error probability with respect to displacement-photon counting methods. Finally, in Sec.s~\ref{sec4:IneffDiscr} and~\ref{sec4:PhNDiscr}, we analyze the performance of the proposed receivers in the presence of detection imperfections and phase noise, respectively.

\subsection{Quantum state discrimination: the general framework}\label{sec:4-Frame}

Generally speaking, the problem of quantum state discrimination is formulated as follows \cite{Cariolaro2015, Bergou2004, Barnett2009, Bergou2007, Bergou2010, Chefles2010, Bae2015, Molmer2015}. We have a physical system that can be prepared in $M\ge 2$ non-orthogonal quantum states $\{\rho_k\}_k$, $k=0,\ldots,M-1$. The set of the possible states is typically referred to with the term {\it constellation}. A quantum source randomly generates one of the $M$ states, preparing $\rho_k$ with a priori probability $0\le q_k\le 1$, $\sum_k q_k=1$.
The task is to implement a receiver to infer which was the prepared state. 
That is, we look for a POVM $\{\Pi_x\}_x$, $x \in {\cal X}$, $\Pi_x\ge 0$, and $\sum_x \Pi_x=\hat{\Id}$, $\hat{\Id}$ being the identity operator over the whole Hilbert space, associated with a decision rule, such that if the outcome $x$ falls into a certain confidence region $\Delta_j$, we infer the state generated by the source to be $\rho_j$.

The problem has no trivial solution, as perfect discrimination of non-orthogonal quantum states is not allowed by quantum mechanics laws. To better understand this point, we consider a paradigmatic example.
We address binary discrimination of pure states, when the source emits either state $|\gamma_0\rangle$ or $|\gamma_1\rangle$, with $\langle \gamma_0|\gamma_1\rangle= X\ne 0$. In this case, we should design a binary receiver, associated with a POVM $\{\Pi_0,\Pi_1\}$, $\Pi_0+\Pi_1= \hat{\Id}$.
If the receiver were able to perform perfect discrimination, outcome ``0" could not be retrieved from state ``1" and vice versa, thus we would have:
\begin{align}
\Pi_0 |\gamma_1\rangle = 0 \quad  \mbox{and} \quad \Pi_1 |\gamma_0\rangle = 0 \, .
\end{align}
However, this would also imply that:
\begin{align}
0=\langle \gamma_0 | \Pi_0 |\gamma_1\rangle +\langle \gamma_0 | \Pi_1 |\gamma_1\rangle = \langle \gamma_0 | \left(\Pi_0+ \Pi_1\right) |\gamma_1\rangle = \langle \gamma_0 |\gamma_1\rangle = X \, ,
\end{align}
leading to contradiction. The former argument can be straightforwardly extended to statistical mixtures $\rho_k= \sum_{s}\lambda_s^{(k)} |\gamma_s^{(k)}\rangle \langle \gamma_s^{(k)}|$, $k=0,1$, with $\sum_s \lambda_s^{(k)}=1$. In this case, the orthogonality condition becomes ${\cal S}_1 \perp {\cal S}_2$, ${\cal S}_k$ being the linear subspace spanned by states $\{ |\gamma_s^{(k)}\rangle \}_s$, providing the support of operator $\rho_k$ \cite{Bergou2010}.
As a consequence, in the presence of non-orthogonal quantum states it is not possible to determine with certainty which state was prepared. This is one of the fundamental result in quantum theory, being in strict connection with the no-cloning theorem, according to which no physical operation can produce an identical copy of a given unknown quantum state \cite{Wootters1982, Dieks1982,Buzek2001}.

In turn, when the outcome $x$ is obtained from the measurement $\{\Pi_x\}_x$, the receiver can take the decision $j$ even if the state $k\ne j$ was probed, resulting in a decision error.
Thus, the paradigm of perfect discrimination should be abandoned and different strategies may be adopted, according to the specific context under investigation.
Generally speaking, in literature there exists three main scenarios:
\begin{itemize}
\item {\it quantum decision theory}: it represents a conclusive discrimination strategy, being the first scenario historically addressed by Helstrom and Holevo \cite{Helstrom1976, Holevo2011}. In this case, a final decision is always performed, even in the possible presence of decision errors. To have conclusive results only, each measurement outcome must be associated with one and only one of the constellation states, resulting in a finite-valued POVM, with exactly $M$ elements. Then, the receiver is designed to minimize the overall decision error probability.

\item {\it unambiguous discrimination strategy}: guaranteeing error-free discrimination, at the cost of introducing inconclusive measurement outcomes. That is, the receiver is described by a POVM with more than $M$ elements: if we obtain an outcome associated with one of the constellation states, we perform a decision with certainty; otherwise, if an inconclusive outcome is retrieved, no decision is performed. In turn, the goal becomes to minimize the probability of obtaining inconclusive outcomes. The strategy was first introduced in \cite{Ivanovic1987} and then solved for binary pure-state discrimination in \cite{Dieks1983, Peres1988}.

\item {\it maximum confidence strategy}: this provides an intermediate solution between the two former strategies, introduced in \cite{Croke2006}. The receiver is optimized to maximize the confidence, namely the a posteriori probability $P(\rho_j|j)$ of inferring state $\rho_j$ when outcome $j$ is retrieved.
\end{itemize}

In the following, we only deal with the first scenario, being of particular relevance in the field of quantum communications.

\subsection{Quantum decision theory}\label{sec4:QDT}

\begin{figure}
\centerline{\includegraphics[width=0.8\columnwidth]{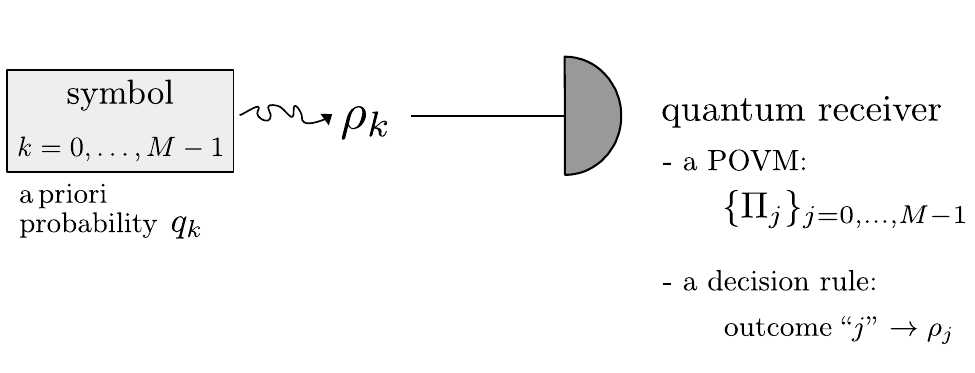}}
\centering
\caption{Schematic description of a quantum decision problem. A source encodes a classical symbol $k=0,\ldots,M-1$, generated with a priori probability $q_k\le 1$, onto non-orthogonal quantum states $\rho_k$. The task is to design a quantum receiver, namely a POVM $\{\Pi_j\}_j$, $j=0,\ldots,M-1$, associated with a decision rule, that minimizes the decision error probability.}\label{fig:01:sec4.2-QDT}
\end{figure}

The paradigm of quantum decision theory can be easily embedded in the quantum communications scheme introduced in Section~\ref{chap:QComm}, where a sender and a receiver share classical information in the following way, schematized in Fig.~\ref{fig:01:sec4.2-QDT}. The sender encodes a classical symbol $k=0,\ldots, M-1$ onto one of the $M$ constellation states $\{\rho_k\}$ and, thereafter, sends them to the receiver who implements a quantum receiver to ``decode" the value of the sent symbol.
Following the previous considerations, the quantum receiver is described by a $M$ valued POVM $\{\Pi_j\}$, $j=0,\ldots, M-1$, $\Pi_j \ge 0$, $\sum_j \Pi_j= \hat{\Id}$, such that obtaining outcome $j$ infers symbol $j$. However, due to the non-orthogonality of the constellation states, any receiver is associated with an error probability, equal to:
\begin{align}\label{eq:Perr}
P_{\rm err} = 1 - {\cal P}_{c} \, ,
\end{align}
where 
\begin{align}
{\cal P}_{c} = \sum_{k=0}^{M-1} q_k \, p(k|k) = \sum_{k=0}^{M-1} q_k \, \Tr[\rho_k \Pi_k]
\end{align}
is the (overall) correct decision probability, $p(j|k)=\Tr[\rho_k \Pi_j]$ being the probability of performing decision $j$, when symbol $k$ was sent.
In turn, the goal is to find the {\it optimum} receiver, reaching the minimum error probability, or, equivalently, the maximum correct decision probability, compatible with quantum mechanics laws \cite{Bergou2010, Helstrom1976, Holevo2011, Cariolaro2015}.

In general, minimizing Eq.~(\ref{eq:Perr}) over all possible POVMs represents a functional optimization problem, being non-easy to handle. The problem is rather simplified for the binary discrimination case, whereas treating $M$-ary state discrimination, with $M>2$, requires the employment of advanced tools from linear algebra. For these reasons, in the following we will only focus on the binary case, completely characterized the mid-1970s by Helstrom's theory, where the minimum error probability gets analytical expression \cite{Helstrom1976}. On the contrary, we leave the discussion of discrimination of multiple states for the next Section.

\subsubsection{Quantum discrimination in the binary case}
In the binary discrimination scenario, the receiver has to distinguish between two states $\rho_0$ and $\rho_1$, associated with a priori probabilities $q_0$ and $q_1$, respectively. Thus, it is described by a binary POVM $\{\Pi_0,\Pi_1\}$, where $\Pi_1=\hat{\Id}-\Pi_0$.
This allows to re-express the error probability~(\ref{eq:Perr}) as:
\begin{align}\label{eq:ReExpPErr}
P_{\rm err} &= q_0 p(1|0) + q_1 p(0|1) \nonumber \\[1ex]
&= q_0 \Tr[\rho_0 \Pi_1] + q_1 \Tr[\rho_1 \Pi_0] \nonumber \\[1ex]
&= q_0 \Tr\left[\rho_0 \left(\hat{\Id}-\Pi_0\right)\right] + q_1 \Tr[\rho_1 \Pi_0]  \nonumber \\[1ex]
&= q_0 + \Tr[ \Lambda \Pi_0] = q_1 - \Tr[ \Lambda \Pi_1] \, ,
\end{align}
where we introduced the Hermitian operator:
\begin{align}
\Lambda= q_1 \rho_1 - q_0 \rho_0 \, .
\end{align}
Furthermore, we consider the spectral decomposition $\Lambda$, namely:
\begin{align}\label{eq:SpecLambda}
\Lambda= \sum_{\lambda \ge 0} \lambda \,|\lambda\rangle \langle \lambda| + \sum_{\lambda< 0} \lambda\, |\lambda \rangle \langle \lambda| \, ,
\end{align}
where $|\lambda\rangle$ is the eigenstate associated with eigenvalue $\lambda\in \mathbb{R}$, and the contributions of the eigenspaces associated with positive and negative eigenvalues have been separated \cite{Helstrom1976, Bergou2010, Cariolaro2015}.
Given Eq.~(\ref{eq:ReExpPErr}), minimizing $P_{\rm err}$ is equivalent to find a POVM satisfying the following conditions:
\begin{align}
\min_{\Pi_0} \Big\{ \Tr[ \Lambda \Pi_0] \, \Big\} \quad \mbox{and} \quad  \max_{\Pi_1} \Big\{ \Tr[ \Lambda \Pi_1] \, \Big\}  \, .
\end{align}
Thanks to~(\ref{eq:SpecLambda}), we straightforwardly obtain the optimum POVM as:
\begin{align}
\Pi_0^{(\opt)}=  \sum_{\lambda< 0}  |\lambda \rangle \langle \lambda| \quad \mbox{and} \quad \Pi_1^{(\opt)}= \sum_{\lambda \ge 0}  |\lambda\rangle \langle \lambda| \, ,
\end{align}
coinciding wit projection over the positive and negative parts of $\Lambda$, respectively.
An intriguing particular case emerges when negative eigenvalues do not exist, as noted by Hunter \cite{Hunter2003}. In this case, we would have $\Pi_0^{(\opt)}=0$ and $\Pi_1^{(\opt)}=\hat{\Id}$, and the minimum error probability is achieved by always inferring state ``1", without performing any measurement. Analogous considerations hold in the opposite case, namely in the absence of positive eigenvalues.

The corresponding minimum error probability associated with the optimum POVM, referred to as the {\it Helstrom bound}, reads:
\begin{align}\label{eq:HB}
\PHel &= \frac12 \Big\{q_0 + \Tr[ \Lambda \Pi_0] + q_1 - \Tr[ \Lambda \Pi_1] \Big\} \nonumber \\[1ex]
&=\frac12 \Big\{1 + \Tr[ \Lambda (\Pi_0- \Pi_1)] \Big\} \nonumber \\[1ex]
&=\frac12 \left(1 +  \sum_{\lambda<0} \lambda - \sum_{\lambda\ge0} \lambda \right)
=\frac12 \left(1 -  \sum_{\lambda}  |\lambda| \right) \nonumber \\[1ex]
&= \frac12 \left[1- \Tr \big( |\Lambda | \big) \right] \, ,
\end{align}
where $|\Lambda|=(\Lambda^\dagger \Lambda)^{1/2}$. We also note that $\Tr ( |\Lambda |) = \lVert \Lambda\rVert_1$, $\lVert \cdot \rVert_1$ being the $1$-norm, thus $\PHel= (1- \lVert q_1 \rho_1 - q_0 \rho_0 \rVert_1)/2$, corresponding to the trace distance bewteen the weighted operators $q_1 \rho_1$ and $q_0 \rho_0$ \cite{Helstrom1976, Bergou2010, Cariolaro2015}.

The former expression further simplifies for pure-state discrimination of states $|\gamma_k\rangle$,  $k=0,1$, with $X=\langle \gamma_0|\gamma_1\rangle\ne0$. In this case, the eigenvalues of operator $\Lambda=q_1 |\gamma_1\rangle \langle \gamma_1|-q_0 |\gamma_0\rangle \langle \gamma_0|$ are equal to $\lambda_{\pm}=(q_1-q_0 \pm \sqrt{1-4q_0q_1 |X|^2})/2$, and the corresponding Helstrom bound becomes:
\begin{align}\label{eq:HBpureStates}
\PHel = \frac12 \left[1- \sqrt{1- 4 q_0 q_1 |\langle \gamma_0|\gamma_1\rangle|^2} \right] \, .
\end{align}
The optimum projectors become $\Pi_0^{(\opt)}=|\lambda_{-}\rangle \langle \lambda_{-}|$ and $\Pi_1^{(\opt)}=|\lambda_{+}\rangle \langle \lambda_{+}|$, where:
\begin{align}
|\lambda_{-}\rangle = \frac{1}{{\cal N}_{-}} \left( |\gamma_0\rangle + \frac{q_0 X^*}{q_1-\lambda_{-}} |\gamma_1\rangle\right) \quad \mbox{and} \quad |\lambda_{+}\rangle = \frac{1}{{\cal N}_{+}} \left( |\gamma_1\rangle + \frac{q_1 X}{q_0+\lambda_{+}} |\gamma_0\rangle\right) \, , 
\end{align}
${\cal N}_\pm$ being the normalization constant. That is, the optimum receiver is realized by a $1$-rank projective measurement over a suitable linear combination of the encoded states \cite{Helstrom1976, Bergou2010, Cariolaro2015}.

\subsection{Binary discrimination of coherent states}\label{sec4:BinaryCoh}
\begin{figure}
\centerline{\includegraphics[width=0.8\columnwidth]{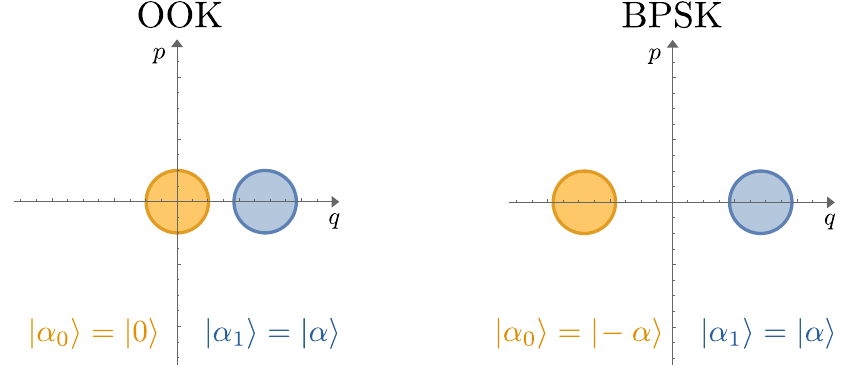}}
\centering
\caption{Phase space representation of the OOK (left) and BPSK (right) encodings. In the former case, information is encoded on the absence or presence of a given field with amplitude $\alpha>0$, i.e. $|\alpha_0\rangle=|0\rangle$ and $|\alpha_1\rangle=|\alpha\rangle$, whilst in the latter scenario symbols are encoded on the field phase, $|\alpha_0\rangle=|-\alpha\rangle$ and $|\alpha_1\rangle=|\alpha\rangle$. We note that the overlap $|\langle \alpha_0|\alpha_1\rangle|^2$ is lower for BPSK modulation, leading to a lower value of the Helstrom bound.}\label{fig:01:sec4.3-BMod}
\end{figure}

We now apply the tools of quantum decision theory, developed in the previous section, to the quantum communications systems presented in Section~\ref{chap:QComm}, where information is encoded onto coherent states, describing the radiation emitted by stable laser sources. That is, in each time slot, the sender encodes a binary symbol $k=0,1$ onto the coherent pulse $|\alpha_k\rangle$, generated with equal a priori probabilities $q_k=1/2$. Thereafter, pulses are sent to the receiver though a channel, assumed here to be noiseless. Ultimately, we implement a quantum receiver, namely a binary POVM $\{\Pi_0, \Pi_1\}$, with $\Pi_1= \Id- \Pi_0$, to infer the transmitted symbol, associated with a nonzero error probability. 
In the presence of binary modulation, the encoding stage may be deployed according to two typical strategies \cite{Cariolaro2015}: 
\begin{itemize}
\item {\it on-off keying} (OOK): it represents the simplest scheme, in which symbol ``0" is encoded onto the vacuum state $|\alpha_0\rangle= |0\rangle$, while symbol ``1" corresponds to a coherent state with given amplitude $\alpha>0$, i.e. $|\alpha_1\rangle = |\alpha\rangle$. Practically, this scheme is realized by either amplitude modulating a laser source at fixed frequency or, more simply, by switching on and off the laser itself according to the symbol to be transmitted;

\item {\it binary phase-shift keying} (BPSK): now, a coherent state of mean energy $\alpha^2$, $\alpha>0$, is generated in all time slots, and information is encoded in the field phase, namely:
\begin{align}
|\alpha_k\rangle = |e^{i (k+1) \pi} \alpha \rangle \,,  \qquad k=0,1 \, ,
\end{align}
such that $|\alpha_0\rangle= |-\alpha\rangle$ and $|\alpha_1\rangle= |\alpha\rangle$. That is, the two states has the same energy but are phase-shifted by $\pi$.
This kind of encoding is practically implemented by phase modulation of an input laser beam.
\end{itemize}

The two modulation formats are schematized in Fig.~\ref{fig:01:sec4.3-BMod}. In the following, we will draw our attention on the sole BPSK case, since, as we can see, the overlap between the two encoded pulses is lower for the BPSK case than the OOK, leading to a lower value of the Helstrom bound, see Eq.~(\ref{eq:HBpureStates}). In fact, we have $|\langle 0|\alpha\rangle|^2= \exp(-\alpha^2)$ for OOK and $|\langle -\alpha|\alpha\rangle|^2= \exp(-4\alpha^2)$ for BPSK. However, from a more practical point of view, the preference for one format over the other is not only limited to theoretical reasons, but is also due to practical characteristics, concerning the technologies of the adopted equipment, the accuracy in the phase stabilization of the encoded signals, the achievable symbol repetition rates and so on \cite{Cariolaro2015}.

In the presence of BPSK, thanks to Eq.~(\ref{eq:HBpureStates}), the minimum error probability, i.e. the Helstrom bound, becomes:
\begin{align}\label{eq:BinaryHB}
\PHel = \frac12 \left[1- \sqrt{1-e^{-4\alpha^2}} \right] \, .
\end{align}
The optimal measurement strategy achieving such a minimum is the \textit{``cat state" measurement}, defined by the two-valued POVM $\{\Pi_0, \Id-\Pi_0\}$, $\Pi_0 = |\psi_{\rm cat} \rangle \langle \psi_{\rm cat}|$, where $|\psi_{\rm cat} \rangle = c_0(\alpha) |\alpha_0\rangle + c_1(\alpha) |\alpha_1\rangle $ is an optimized ``cat state" \cite{Helstrom1976}. However, a concrete realization of such a POVM is not an easy task. 

On the contrary, the conventional binary receivers adopted in optical communication systems are based on homodyne detection, being sensitive to the field phase. The homodyne receiver works as follows. We perform measurement of the quadrature $q$ and infer symbol ``0" when the outcome $x \ge 0$ is retrieved, and symbol ``1" when a negative value $x<0$ is obtained. We remind that the homodyne distribution of states $|\alpha_k\rangle$ is equal to:
\begin{align}
p_{\rm HD}(x|k) = \frac{\exp\left[-(x-2\alpha_k)^2/2\right]}{\sqrt{2\pi}} \, ,
\end{align}
expressed in shot-noise units, being limited by shot-noise due to vacuum fluctuations. In turn, there is a nonzero probability of retrieving outcomes $x\ge0$ when state $|\alpha_0\rangle$ was sent, and vice versa. Accordingly, the error probability for the homodyne receiver reads
\begin{align}\label{eq:BinarySQL}
\PSQL &= \frac12 \left[ \int_0^\infty dx \, p_{\rm HD}(x|0) + \int_{-\infty}^{0} dx \, p_{\rm HD}(x|1)\right]  \nonumber \\[1ex]
 &=\frac{1-\erf \left(\sqrt{2}\alpha\right)}{2} \, ,
\end{align}
referred to as the {\it standard quantum limit} (SQL), or shot-noise limit. The SQL represent the best error probability achievable by semi-classical means in ideal conditions, being suboptimal with respect to the Helstrom bound, as $\PSQL> \PHel$.
Given this scenario, the task of quantum state discrimination theory is to design a feasible receiver outperforming the SQL and being as close as possible to the Helstrom bound. 
Several proposals of feasible optimum or near-optimum receivers have been advanced in literature, based on either single-shot discrimination or feedback-based strategies. In the following, we present the main ones, employing displacement operations and photon counting.

\subsubsection{The Kennedy receiver}\label{subsec4:Kennedy}
\begin{figure}
\includegraphics[width=0.6\columnwidth]{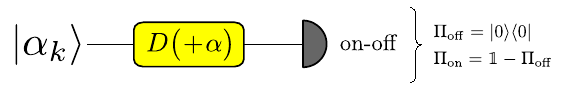}
\centering
\caption{Setup of the Kennedy receiver. The incoming signal $|\alpha_k\rangle$, $k=0,1$, undergoes a displacement operation $D(\alpha)$ followed by on-off detection. The final decision is performed according to the rule: ``off"~$\rightarrow$~``0" and ``on"~$\rightarrow$~``1".}\label{fig:sec4.3_KennedySetup}
\end{figure}

The first quantum receiver beating the SQL has been proposed in 1973 by Kennedy \cite{Kennedy1973}, whose scheme is reported in Fig.~\ref{fig:sec4.3_KennedySetup}. 

\begin{figure}
\includegraphics[width=0.6\columnwidth]{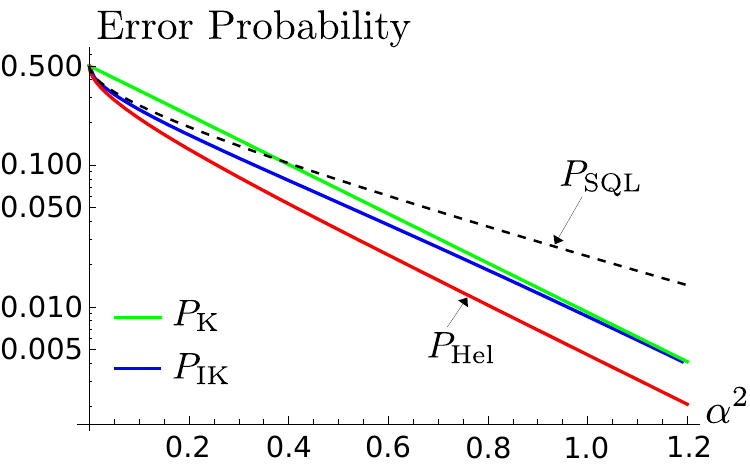}
\centering
\caption{Log plot of the error probabilities of the standard and improved Kennedy receivers $\PK$ and $\PIK$, respectively, as a function of the signal energy $\alpha^2$. $\PSQL$ and $\PHel$ refer to the SQL~(\ref{eq:BinarySQL}) and the Helstrom bound~(\ref{eq:BinaryHB}), respectively.}\label{fig:sec4.3_KennedyPlot}
\end{figure}

In the {\it Kennedy receiver}, or displacement receiver, the incoming signal $|\alpha_k\rangle$ undergoes the displacement operation \cite{Olivares2021} $D(\alpha)$, followed by on-off detection.
As discussed in Sec.~\ref{sec2:EvoQO}, the displacement may be implemented practically by letting the signals interfere with a suitable intense local oscillator at a beam splitter with large transmissivity \cite{Paris1996}.
We note that $D(\alpha)$ performs a nulling operation, leading to the mapping:
\begin{align}
|-\alpha\rangle \rightarrow |0\rangle \quad \mbox{and} \quad |\alpha\rangle \rightarrow |2\alpha\rangle \, .
\end{align}
This is the so-called ``nulling" displacement, since one of the two signals is displaced into the vacuum state.
Therefore, BPSK is turned into OOK and on-off detection provides the optimal measurement choice, with the following decision rule: ``off"~$\rightarrow$~``0" and ``on"~$\rightarrow$~``1". Thus, an error occurs when an ``off" result is retrieved from state $|\alpha_1\rangle$, leading to the error probability
\begin{align}\label{eq:Kennedy}
\PK =  q_1 \, p({\rm off}|1)= \frac12 \big| \langle 0 | 2 \alpha \rangle \big|^2 = \frac{e^{-4\alpha^2}}{2} \, ,
\end{align}
depicted in Fig.~\ref{fig:sec4.3_KennedyPlot} as a function of the signal energy and compared to both the SQL and the Helstrom limit.
As we can see, in the high-energy limit $\alpha^2\gg 1$ the Kennedy receiver is near-optimum, namely proportional to the Helstrom bound, as $\PK \approx 2 \PHel$. In fact:
\begin{align}
\PHel = \frac12 \left[ 1- \sqrt{1- e^{-4\alpha^2}} \right] \approx \frac12 \left[ 1- \left(1-\frac{e^{-4\alpha^2}}{2} \right) \right] = \frac{\PK}{2} \, ,
\end{align}
where we used the Taylor expansion $\sqrt{1-x}= 1 -x/2 +O(x^2)$.
Futhermore, the receiver also beats the SQL for $\alpha^2 > \alpha^2_{\rm K}$, with $\alpha^2_{\rm K}\approx 0.38$ \cite{Kennedy1973, Notarnicola2023:HYNORE}.

The feasibility of the Kennedy setup has been demonstrated experimentally in \cite{Lau2006, Wittmann2008, Tsujino2009, Shcherbatenko2020}. The main drawback towards a practical implementation is represented by the local oscillator laser required to realize the displacement $D(\alpha)$, which need to be finely tuned in both frequency and phase with respect to the incoming signal. To overcome this problem, to date, many practical realizations employ a single laser source from which both the signal and the local oscillator are generated. This laser emits a high-intensity coherent state, being then splitted at a beam splitter with small transmissivity, such that the transmitted (weak) pulse is sent to the phase modulator, becoming the encoded state, while the reflected (strong) pulse plays the role of the local oscillator.

\begin{figure}
\includegraphics[width=0.6\columnwidth]{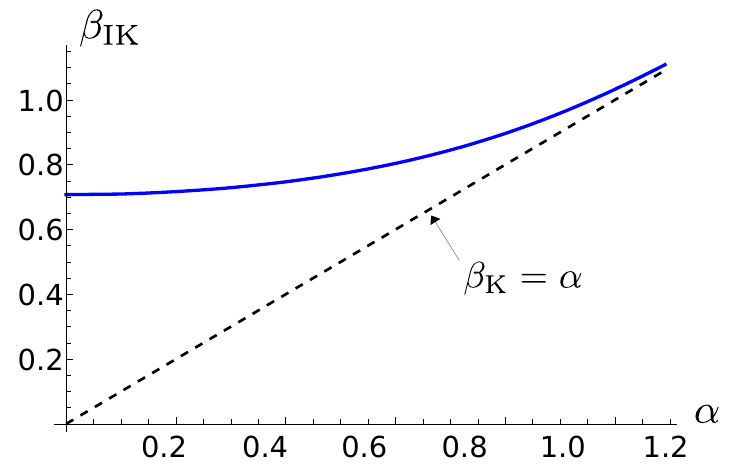}
\centering
\caption{Optimized displacement $\beta_{\rm IK}$ of the IK receiver as a function of the coherent amplitude $\alpha$ of the incoming signal. The dashed line $\beta_{\rm K}= \alpha$ represents the (fixed) displacement amplitude of the standard Kennedy receiver.}\label{fig:sec4.3_TakeokaDisp}
\end{figure}

An improved version of the Kennedy receiver has been obtained by Takeoka and Sasaki by optimizing the displacement amplitude \cite{Takeoka2008}.
In their {\it improved Kennedy} (IK) receiver, the ``nulling" displacement $D(\alpha)$ is replaced with a generic $D(\beta)$, $\beta>0$, whose value is optimized to minimize the overall error probability. In turn, the encoded states are mapped into:
\begin{align}
|-\alpha\rangle \rightarrow |\beta-\alpha\rangle \quad \mbox{and} \quad |\alpha\rangle \rightarrow |\beta+\alpha\rangle \, .
\end{align}
The final decision still follows from the outcomes of on-off detection but, differently from the standard Kennedy receiver, now, an error occurs either when a ``off" result is retrieved from state $|\alpha_1\rangle$ or when a ``on" is obtained from $|\alpha_0\rangle$. The resulting error probability reads:
\begin{align}
\PIK = \max_{\beta>0} \PIK(\beta) \, ,
\end{align}
where
\begin{align}
\PIK(\beta)= q_1 \, p({\rm off}|1) +  q_0 \,p({\rm on}|0) =  \frac12 \left[ e^{-(\beta+\alpha)^2} + 1- e^{-(\beta-\alpha)^2} \right] \, .
\end{align}
By nulling the derivative of $\PIK(\beta)$ with respect to $\beta$, i.e. $d\PIK(\beta)/d\beta=0$, we find that the optimal displacement amplitude $\beta_{\rm IK}$ shall satisfy the following transcendental equation:
\begin{align}\label{eq:ConditionBeta}
\frac{\beta_{\rm IK}-\alpha}{\beta_{\rm IK}+\alpha} = e^{-4\alpha \beta_{\rm IK}} \, ,
\end{align}
to be solved numerically. The solution of Eq.~(\ref{eq:ConditionBeta}) is plotted in Fig.~\ref{fig:sec4.3_TakeokaDisp} as a function of the coherent amplitude $\alpha$ of the encoded signal, together with $\beta_{\rm K}=\alpha$, being the displacement amplitude of the standard Kennedy receiver. As we see, the optimized displacement amplitude $\beta_{\rm IK} $ is always larger than $\alpha$, but in the high-energy limit the two lines approach each other, thus making the IK receiver coincide with the Kennedy.
In turn, as showed in Fig.~\ref{fig:sec4.3_KennedyPlot}, the IK receiver outperforms the Kennedy in the low-energy regime, whereas in the limit $\alpha^2 \gg 1$, $\PIK \approx \PK$ and the improvement due to the displacement optimization becomes negiglibile. Remarkably, the enhancement for small energies allows to beat the SQL for all $\alpha^2>0$.

Finally, it is worth to mention a further variation of the original Kennedy scheme, the so-called {\it displacement-photon-number-resolving receiver} (DPNR), originally introduced in \cite{Wittmann2010, Wittmann2010b, DiMario2018, DiMario2019}. It consists of the same setup depicted in Fig.~\ref{fig:sec4.3_KennedySetup}, where the on-off detector is replaced by a photon-number resolving (PNR) detector with finite resolution $M$, see Sec.~\ref{subsec2:PNR}.
The final decision rule is then suitably adjusted to either include a given inconclusive-result probability \cite{Wittmann2010, Wittmann2010b} or account for detection imperfections and noise \cite{DiMario2018, DiMario2019}.

\subsubsection{The Dolinar receiver}\label{subsec4:Dolinar}

The Kennedy receiver presented above represents the typical benchmark for all sub shot-noise-limited quantum receivers, due to both its theoretical simplicity and its practical feasibility with the technologies commonly adopted in optical communications, based on linear optics and on-off detection. As a consequence, its scheme has provided a building block to construct more sophisticated receivers with better performances.

A paradigmatic example is represented by the {\it Dolinar receiver} \cite{Dolinar1973}. By suitably generalizing the Kennedy scheme, in 1973 Dolinar proposed a feedback receiver employing conditional displacements and continuous-time photodetection, proving it to be optimum for all input energies.

The basic principle of the Dolinar receiver is to implement a real-time adjustment of the displacement operation of the Kennedy setup in both amplitude and phase, according to the obtained outcome of the photon counter, via closed-loop feedback control performed during the time processing of the encoded signal \cite{Dolinar1973, Assalini2011, Cariolaro2015}.
Thus, to compute the error probability we should resort to the temporal description presented in Sec.~\ref{sec3:OptSig}. The encoded pulse $|\alpha_k\rangle$, $k=0,1$, corresponds to a wavepacket $\psi_k(t)$ located in a time slot of duration $T$, equal to \cite{Assalini2011, Cariolaro2015}:
\begin{align}
\psi_k(t)=e^{i\pi (k+1)} \psi \,  e^{-i \omega t} \, , \qquad 0 < t \le T \, ,
\end{align}
$\omega$ being the carrier signal frequency and $\psi>0$, with mean photon number
\begin{align}
\bar{n}_k=\int_0^T|\psi_k(t)|^2dt=\psi^2 T \equiv \alpha^2 \, .
\end{align}
The overlap betweent the pulses is then retrieved as $|\langle \alpha_0|\alpha_1\rangle|^2= \exp[- \int_0^T dt S(t)]$, where
\begin{align}
S(t)=|\psi_0(t)-\psi_1(t)|^2  = 4 \psi^2
\end{align}
is the photon counting rate for a plane wave of complex envelope $\psi_0(t)-\psi_1(t)$ \cite{Dolinar1973}. In turn, we have $|\langle \alpha_0|\alpha_1\rangle|^2= \exp(-4\psi^2 T)= \exp(-4\alpha^2)$, as expected.
The transition from the coherent pulse representation to the continuous-time picture is obtained as follows. We perform a coarse graining in time and divide the time slot into many temporal bins of duration $\delta t \ll T$.
Thanks to the properties of coherent states, each time bin still contains a coherent state but with smaller amplitude, such that the reduced pulse in the time bin comprised between $t$ and $t+\delta t$ is equal to $|\tilde{\alpha}_k(t)\rangle=|\alpha_k \sqrt{\delta t/T}\rangle$. Accordingly, for $t'\in (t,t+\delta t]$, the field value is equal to $\psi_k(t') \approx \psi_k(t)$, with mean energy $\psi^2 \delta t$.

\begin{figure}
\includegraphics[width=0.65\columnwidth]{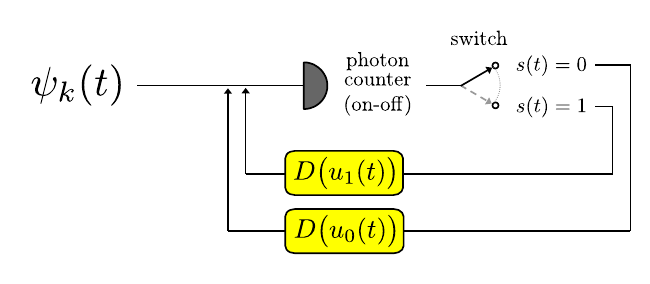}
\centering
\caption{Setup of the Dolinar receiver. The field $\psi_k(t)$ in each time bin $(t,t+\delta t]$, associated with the reduced pulse $|\tilde{\alpha}_k(t)\rangle$, 
$k=0,1$, undergoes a displacement operation $D(u_0(t))$ or $D(u_1(t))$ determined by the position of a switch $s(t)$, changing at every count of the photodetector. After time $T$, when all time bins are processed, the value $s(T)$ gives the final decision.}\label{fig:sec4.3_DolinarSetup}
\end{figure}

Given this scenario, the Dolinar receiver operates as depicted in Fig.~\ref{fig:sec4.3_DolinarSetup}. The setup consists of a photon counter performing on-off detection connected to a switch $s$, in which the field $\psi_k(t)$ in each time bin $(t, t+\delta t]$, corresponding to the reduced pulse $|\tilde{\alpha}_k(t)\rangle$, is sequentially injected \cite{Assalini2011, Notarnicola2023:FF}. The switch switches back and forth between two positions, called $s = 0$ and $s = 1$, with each click of the detector, applying alternatively two different time-varying displacement operations 
\begin{align}
&D\left(u_0(t)\right) \quad \mbox{if} \quad s(t)=0 \, , \\
&D\left(u_1(t)\right) \quad \mbox{if} \quad s(t)=1 \, ,
\end{align}
such that $D\left(u_j(t)\right) \psi_k(t)= \psi_k(t) +u_j(t)$, $j,k=0,1$.
In the above expression $s(t)$ refers to the position of the switch at time $t \le T$. The initial position of the switch is set to $s(0)=0$, meaning that the signal in the first time bin is displaced by $D(u_0(0))$. After the first photodetection, if the detector clicks the position of the switch is changed to $s(\delta t)=1$ and the second copy will be displaced by $D(u_1(\delta t))$. Otherwise, we keep still $s(\delta t)=0$ and the second copy will be displaced by $D(u_0(\delta t))$. The feedback loop continues according to this basic rule: at every click of the detector the switch changes its position. When all the time bins are processed, the final decision is obtained by reading the position of the switch, according to:
\begin{align}
&s(T)=0 \quad \rightarrow \quad \mbox{infer state } |\alpha_0\rangle \, , \\
&s(T)=1 \quad \rightarrow \quad \mbox{infer state } |\alpha_1\rangle \, . 
\end{align}

Now, the mathematical problem is to choose the functions $u_0(t)$ and $u_1(t)$ that maximize the correct detection probability at the end of the process, namely:
\begin{align}
{\cal P}_{\rm Dol}(T) = q_0 P_{00} (T) + q_1 P_{11}(T) \, ,
\end{align}
where $P_{kl}(t)$ is the probability of inferring state ``$l$" at time $t \le T$ if signal ``$k$" is sent, $k,l=0,1$ \cite{Assalini2011, Notarnicola2023:FF}. Actually, in his original proposal Dolinar inferred the optimal solution to the present problem and then verified its optimality afterwards \cite{Dolinar1973}. On the contrary, here we follow an equivalent approach leading to the same solution adopted in \cite{Assalini2011}. Furthermore, we restrict to the case of symmetric solutions, i.e. $u_0(t)=-u_1(t)=u(t)$. 

At first we assume that state $|\alpha_0\rangle= |-\alpha\rangle$ is sent. Then, $s(t)$ can be interpreted as a telegraph stochastic process \cite{Parzen1999}, being alternately driven, in each time bin $(t,t+\delta t]$, by a inhomogeneous Poisson process with rates
\begin{align}
\lambda_{+}(t)=|\psi_0(t)+u(t)|^2\qquad \text{and} \qquad \lambda_{-}(t)=|\psi_0(t)-u(t)|^2\, .
\end{align}
At time $t+\delta t$ a correct decision is performed in two cases: firstly if $s(t)=0$ and the detector detector does not click; secondly if $s(t)=1$ and the photodetector clicks. Moreover, since $\delta t \ll T$, we may safely assume that the detector effectively measures no more than one photon, thus the probability of obtaining a non-click result in the former case reads 
$p({\rm off}|0)\approx 1-\lambda_{+}(t)\delta t$, while the probability of a click in the latter scenario is equal to 
$p({\rm on}|0) \approx \lambda_{-}(t) \delta t$ \cite{Cariolaro2015}. Accordingly we have:
\begin{align}
P_{00}(t+\delta t)=P_{00}(t) \big[ 1-\lambda_{+}(t)\delta t \big]
+ \big[1-P_{00}(t)\big] \lambda_{-}(t) \delta t \, ,
\end{align}
leading to the following differential equation in the limit $\delta t \rightarrow 0$:
\begin{align}\label{eq:P00diff}
\frac{d P_{00}(t)}{dt}= \lambda_{-}(t) - \big[\lambda_{+}(t)+\lambda_{-}(t) \big] P_{00}(t) \, ,
\end{align}
to be solved with the initial condition
\begin{align}
P_{00}(0) =1 \, ,
\end{align}
as the switch is initialized in position ``0". In a similar way, we prove that $P_{11}(t)$ satisfies the same equation with initial condition $P_{11}(0)=0$, thus we conclude that also ${\cal P}_{\rm Dol}(t)$ is a solution of Eq.~(\ref{eq:P00diff}) with ${\cal P}_{\rm Dol}(0)=1/2$.

To solve it, we make the {\it ansatz} that, at any time $t\le T$ there exists some value of the displacement amplitude $u(t)$ such that the provisional correct decision probability is exactly equal to the Helstrom bound relative to binary pulses of duration $t$, namely
\begin{align}\label{eq:Ansatz}
{\cal P}_{\rm Dol} (t)=\frac{1}{2}\left[1+\sqrt{1-e^{-4\psi^2 t}}\right] \equiv \frac{1+R(t)}{2}\, .
\end{align} 
Actually, this is not a very restrictive requirement, as it has been proved that, when multiple identical copies of the encoded quantum states are available, the Helstrom bound can be reached by performing local adaptive measurements on single copies, each one being optimized according to the results measurements on the previous copies \cite{Assalini2011, Brody1996, Acin2005}. Moreover, if the number of these copies is sufficiently large, the encoded pulse is a sequence of weak coherent states and the optimum measurements are well approximated by suitable displacements and photon counting, retrieving the scenario of the Dolinar setup under investigation \cite{Assalini2011}.

Given this considerations, we plug Eq.~(\ref{eq:Ansatz}) into~(\ref{eq:P00diff}) and obtain:
\begin{align}
 \psi^2 e^{-2 i\omega t} \, \frac{1-R^2(t)}{R(t)}&=\psi^2 e^{-2 i\omega t} +u^2(t)+2\psi e^{-i\omega t} \, u(t) \nonumber \\
&\hspace{2cm}-\big[\psi^2 e^{-2 i\omega t} +u^2(t)\big]\big[1+R(t)\big] \, ,
\end{align}
which is solved by \cite{Geremia2004}
\begin{align}
u_{\rm Dol}(t)=\frac{\psi \, e^{-i\omega t}}{R(t)}=\frac{\psi \, e^{-i\omega t}}{\sqrt{1-e^{-4\psi^2 t}}} \, .
\end{align}
As a consequence, by implementing the time-varying displacement in the setup of Fig.~\ref{fig:sec4.3_DolinarSetup} with the choice $u_0(t)=u_{\rm Dol}(t)$ and $u_1(t)=-u_{\rm Dol}(t)$, at time $T$ we perform BPSK discrimination with the minimum error probability
\begin{align}
P_{\rm Dol} = 1- {\cal P}_{\rm Dol} (T) = \frac{1}{2}\left[1+\sqrt{1-e^{-4\alpha^2}}\right] \, ,
\end{align}
proving the Dolinar receiver to be optimum.
We also underline that the optimality of the Dolinar setup is guaranteed regardless the particular shape of the wavepacket $\psi_k(t)$, only provided that the corresponding optical field is described as a coherent state. In the presence of arbitary wavepackets, the optimal displacement amplitude derived by Dolinar with condition $u_0(t)=-u_1(t)=u(t)$, becomes $u_{\rm Dol}(t)= \psi_0(t)/R(t)$ \cite{Dolinar1973}.

However, even though theoretically optimum, the Dolinar scheme requires a nontrivial experimental implementation, for a twofold reason. On the one hand, to effectively process the signal in time as a sequence of shorter pulses of duration $\delta t \ll T$, the bandwidth of both the detector and electronic components must be much larger than the symbol repetition rate. Moreover, the feedback circuit needs precise control of an optical-electrical loop to achieve fast response times, to avoid delays in the local oscillator adjustment. On the other hand, continuous measurements and feedback control require detectors with high performances, i.e. high quantum efficiency, low dark count rate, high visibility and short dead time. In fact, the presence of realistic inefficiencies induces unwanted decision errors during the signal processing, which accumulate in time and affect the feedback loop, degrading the quantum advantage of the receiver and overall resulting in a bad performance.

Due to all the previous drawbacks, it is not surprising that the first attempt to implement the Dolinar receiver dates back to 2006 by Lau {\it et al.} \cite{Lau2006}, 33 years after Dolinar's paper. The authors realized both a Dolinar and a Kennedy receiver, but the obtained performance was unsatisfactory, as the measured Dolinar error probability was even larger than the Kennedy. The biggest limitation that did not allow to prove the Dolinar optimality was the limited visibility of the amplitude modulator, which was approximately $\approx 98 \%$ at the employed bandwidth, while the proper regime should have been $\approx 99.9\%$. Only a year later, in 2007 Cook {\it et al.} obtained a satisfactory practical implementation of the Dolinar receiver, close to the Helstrom bound in the low energy regime \cite{Cook2007}. 

For all these reasons, the interest has been directed to feed-forward receivers, where the signal is split into a finite number of copies, to obtain a tradeoff between minimum-error discrimination and robustness in practical contexts, as discussed in the following.

\subsubsection{The displacement feed-forward receiver}\label{subsec4:DFFRE}

\begin{figure}[t]
\includegraphics[width=0.7\columnwidth]{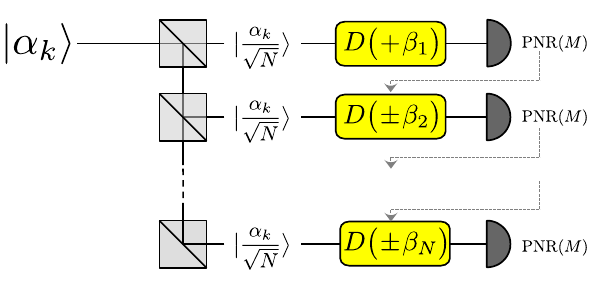}
\centering
\caption{Scheme of the displacement feed-forward receiver (DFFRE) proposed in \cite{Sych2016}. The incoming signal $|\alpha_k\rangle$, $k=0,1$, is split into $N$ copies and undergoes a sequence of conditional displacements followed by photon counting. The first copy undergoes a positive displacement, whereas the sign of the subsequent displacements is decided via Bayesian inference.}\label{fig:sec4.3_DFFREScheme}
\end{figure}

Following the previous philosophy, in 2016 Sych and Leuchs proposed a new receiver, the {\it displacement feed-forward receiver} (DFFRE), combining both the simplicity of the Kennedy scheme and the optimality of the Dolinar one \cite{Sych2016}. 
The key idea is to split the encoded state $|\alpha_k\rangle$, $k=0,1$, into a finite number of copies $N$ rather than a large number of time bins. Thereafter the displacement-photon counting scheme employed in the improved Kennedy receiver \cite{Takeoka2008} is implemented on each copy, optimizing the displacement amplitude via feed-forward Bayesian inference. The initial splitting of the signal may be implemented either by time-multiplexing \cite{Fitch2003, Schettini2007} or by spatial separation into different modes thanks to an array of splitters, providing in both cases a feasible solution overcoming the fast measurements requirement of the Dolinar proposal.

The scheme of the DFFRE is depicted in Fig.~\ref{fig:sec4.3_DFFREScheme}. As discussed, the input state $|\alpha_k\rangle$ is split into $N$ rescaled copies, that is
\begin{align}
|\alpha_k\rangle \rightarrow \bigotimes_{j=1}^{N} |\alpha_k^{(j)}\rangle \, ,
\end{align}
where $|\alpha_k^{(j)}\rangle=|\alpha_k/\sqrt{N}\rangle$. Then, each copy undergoes an optimized conditional  displacement followed by PNR($M$) detection.
We start by displacing the first copy $|\alpha_k^{(1)}\rangle$ by $D(\beta_1)$, with amplitude $\beta_1>0$ maximizing the correct decision probability, thereafter we perform PNR$(M)$ detection on the output signal $|\alpha_k/\sqrt{N}+\beta_1\rangle$. According to the maximum a posteriori probability (MAP) criterion based on Bayesian inference, the PNR$(M)$ measurement outcome $n$ is used to choose the sign of the optimized conditional displacement to be performed on the second copy. In other words, we infer the state ``0" or ``1" associated with the maximum a posteriori probability given the outcome $n$ \cite{Notarnicola2023:HYNORE, Sych2016, DiMario2018}. If ``0" is inferred we displace the second copy $|\alpha_k^{(2)}\rangle$ by $D(\beta_2)$, otherwise we apply $D(-\beta_2)$, where $\beta_2>0$ is chosen to maximize the correct decision probability, too. Then, we perform again photodetection and repeat the process until the $N$-th copy.

With ideal detectors, the previous criterion is equivalent to performing on-off detection on each displaced copy. The $j$-th copy, $j=1,\ldots, N$, undergoes the displacement operation $D(\sigma_j \beta_j)$, where $\beta_j>0$ is the optimized amplitude and $\sigma_j=\pm 1$ is the sign of the displacement. The first displacement has a fixed sign, namely, $\sigma_1=+1$. The other values of $\sigma_j$ are assigned according to the following decision rule: if we get outcome ``off" from the $(j-1)$-th measurement we set $\sigma_j= \sigma_{j-1}$, otherwise if a ``on" is retrieved we switch $\sigma_j= -\sigma_{j-1}$. Ultimately, the outcome obtained from the last copy determines the final decision. Therefore, the outcome ``off" infers state $|-\sigma_{N} \alpha\rangle$, outcome ``on" infers state $|\sigma_{N} \alpha\rangle$, $\sigma_{N}$ being the sign of the last displacement. 

\begin{figure}[t]
\includegraphics[width=0.65\columnwidth]{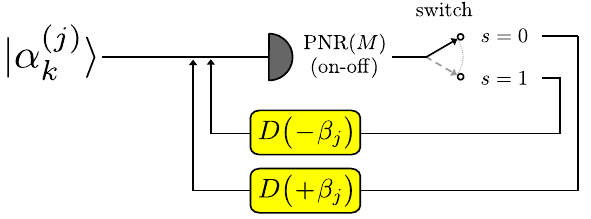}
\centering
\caption{Equivalent scheme of the DFFRE. Each copy $|\alpha_k^{(j)}\rangle$ undergoes a displacement operation whose sign is determined by the position of a switch $s$.}
\label{fig:sec4.3_DFFREEQScheme}
\end{figure}

Therefore, the DFFRE mimics the functioning of the Dolinar receiver, albeit with a discrete number of modes, being then equivalent to scheme reported in Fig.~\ref{fig:sec4.3_DFFREEQScheme}. As in Sec.~\ref{subsec4:Dolinar}, we have a PNR($M$) detector performing on-off detection connected to a switch $s$, switching at every click between $s = 0$ and $s = 1$, but, now, we process only $N$ copies $|\alpha_k^{(j)}\rangle$, $j=1,\ldots, N$ \cite{Notarnicola2023:FF}. According to the position of the switch $s^{(j)}$ after the $j$-th copy is processed, we apply one of the two displacement operations:
\begin{align}
&D(+\beta_j) \quad \mbox{if} \quad s^{(j)}=0 \, , \\
&D(-\beta_j) \quad \mbox{if} \quad s^{(j)}=1 \, ,
\end{align}
where the value $\beta_j$ is optimized to maximize the correct decision probability in each step \cite{Notarnicola2023:FF}.
If the initial position of the switch is set to $s^{(0)}=0$, by retracing the passages in Sec.~\ref{subsec4:Dolinar}, we obtain the correct decision probability after $j$ steps, namely ${\cal P}_{\rm DF}^{(j)} = q_0 P_{00}^{(j)} +q_1 P_{11}^{(j)}$, as:
\begin{align}\label{eq:PcorrDFF}
{\cal P}_{\rm DF}^{(j)} &= 
\max_{\beta_{j}} \Bigg\{ {\cal P}_{\rm DF}^{(j-1)}
\, q_{\rm off}\Big( \lambda_{-}^{(j)}(\alpha)\Big) 
+\left[1-{\cal P}_{\rm DF}^{(j-1)}\right] q_{\rm on}\Big( \lambda_{+}^{(j)}(\alpha)\Big) 
\Bigg\} \, ,
\end{align}
to be solved with the initial condition ${\cal P}_{\rm DF}^{(0)}=1/2$, where
\begin{align}\label{eq:q01}
q_{\rm off}(x) = e^{-x} \quad \mbox{and} \quad q_{\rm on}(x) = 1-e^{-x} \,,
\end{align}
are the probabilities of ``off" and ``on" results, respectively, and
\begin{align}\label{eq:lambdapm}
\lambda_{\pm}^{(j)}(\alpha)= \Big|\beta_j \pm\frac{\alpha}{\sqrt{N}}\Big|^2 \, ,
\end{align}
is the mean photon number of the resulting displaced copies.
Ultimately, we retrieve the discrimination error probability of the DFFRE after $N$ copies as:
\begin{align}\label{eq:PDFF}
\PDISP^{(N)} = 1-{\cal P}_{\rm DF}^{(N)} \,.
\end{align}

\begin{figure}[t]
\includegraphics[width=0.49\columnwidth]{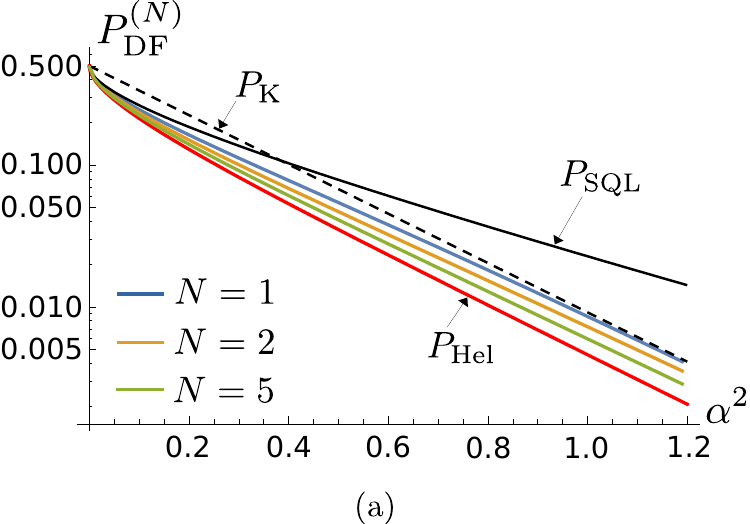}
\includegraphics[width=0.49\columnwidth]{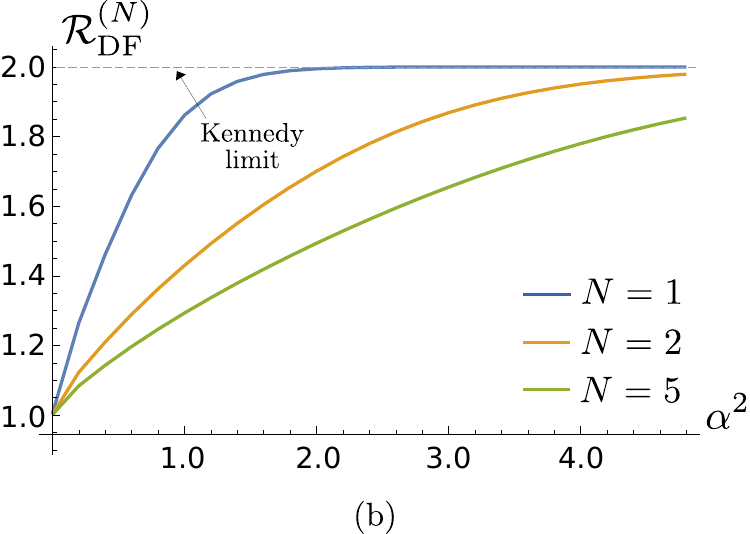}
\centering
\caption{(a) Log plot of $\PDISP^{(N)}$ as a function of the signal energy $\alpha^2$ for different number of copies $N$. $\PSQL$, $\PHel$ and $\PK$ refer to the SQL~(\ref{eq:BinarySQL}), the Helstrom bound~(\ref{eq:BinaryHB}), and the Kennedy error probability~(\ref{eq:Kennedy}), respectively.
(b) Plot of ${\cal R}_{\rm DF}^{(N)}$  as a function of $\alpha^2$ for different $N$. For $\alpha^2\gg 1$ all ratios approach the Kennedy limit.}\label{fig:sec4.3_DFFREProb}
\end{figure}

Plots of  $\PDISP^{(N)}$ are depicted in Fig.~\ref{fig:sec4.3_DFFREProb}(a) as a function of the input energy $\alpha^2$ and different values of $N$. The receivers is near-optimum and beat the SQL for all energies. For $N=1$ the DFFRE coincides with the improved Kennedy receiver \cite{Takeoka2008}, and by increasing $N$ the error probability is further reduced for $\alpha^2 \ll 1$, coming closer to the Helstrom bound~(\ref{eq:BinaryHB}), as emerges by computing the ratio
\begin{align}
{\cal R}_{\rm DF}^{(N)} = \frac{P_{\rm DF}^{(N)}}{\PHel} \, ,
\end{align}
plotted in Fig.~\ref{fig:sec4.3_DFFREProb}(b).
In the regime $\alpha^2 \ll 1$, the larger the number of copies, the smaller the ratio ${\cal R}_{\rm DF}^{(N)}$, whereas in the asymptotic limit $\alpha^2 \gg 1$ the DFFRE approaches the Kennedy receiver for any $N$.

\subsubsection{The Sasaki-Hirota receiver}\label{subsec4:Sasaki}
To conclude, we present a further proposal of optimum receiver, suggested by Sasaki and Hirota in 1996 \cite{Sasaki1996}, who proved that it is possible to reach the Helstrom bound by recasting the problem into the two-dimensional subspace spanned by the encoded states. 

At first, we perform the ``nulling" displacement $D(\alpha)$ to he encoded signals $|\alpha_k\rangle$, $k=0,1$, as in the Kennedy scheme, shifting the discrimination problem to states $|0\rangle $ and $|2\alpha\rangle$.
Now, we consider the following orthonormal basis of the subspace ${\cal S}$ spanned by $|0\rangle $ and $|2\alpha\rangle$:
\begin{align}\label{eq:BasisSH}
|\eta_0\rangle = |0\rangle \quad \mbox{and} \quad |\eta_1\rangle = \frac{1}{\sqrt{1-X^2}} \big( |2\alpha\rangle - X \, |0\rangle\big) \, ,
\end{align}
where $X=\langle 0|2\alpha\rangle$, and construct the unitary operator
\begin{align}
U(\theta)= \cos\theta \, \bigg( |\eta_0\rangle\langle \eta_0| + |\eta_1\rangle\langle \eta_1| \bigg)+
 \sin\theta \, \bigg( |\eta_0\rangle\langle \eta_1| - |\eta_1\rangle\langle \eta_0| \bigg) \, ,
\end{align}
depending on a free parameter $\theta$.
In the {\it Sasaki-Hirota receiver} the operator $U(\theta)$ is applied to the displaced signals $|0\rangle $ and $|2\alpha\rangle$, followed by a projective measurement onto the basis~(\ref{eq:BasisSH}), namely:
\begin{align}\label{eq:PVMSH}
\Pi_0=|\eta_0\rangle\langle \eta_0| \quad \mbox{and} \quad \Pi_1= |\eta_1\rangle\langle \eta_1| \, .
\end{align}
By numerical optimization of $\theta$, the authors showed that the present scheme reaches the Helstrom bound \cite{Sasaki1996, Cariolaro2015}.

We note that, since $\langle 0|\eta_1\rangle=0$, the projective measurement~(\ref{eq:PVMSH}) may be safely replaced by on-off detection, making it feasible in realistic conditions. On the contrary, the unitary $U(\theta)$ is a non-Gaussian operation, whose realization would require highly non-linear optical elements, thus making this kind of receiver not realizable with the usual practical linear optics components.

\subsection{Hybrid receivers}\label{sec4:Hybrid}

As discussed in the previous section, most sub shot-noise-limited receivers are constructed via the displacement-photon counting technique, probing the particle-like behaviour of the encoded quantum field. On the contrary, the SQL is achieved by homodyne detection, which gives information on the field phase, thus probing the wave-like properties of radiation.
Therefore, a natural question arises, that is whether or not it is possible to design hybrid receivers employing both of the field properties to obtain a better performance in terms of error probability.

To this aim, in this section we firstly propose an innovative receiver: the {\it hybrid near-optimum receiver} (HYNORE), a single-shot receiver based on the combination of the homodyne-like detection presented in Sec.~\ref{subsec2:HL} and the displacement-photon counting scheme of the Kennedy setup \cite{Notarnicola2023:HYNORE}. Later on, we extend this approach to multi-copy receivers and construct the {\it hybrid feed-forward receiver} (HFFRE), by embedding homodyne-like detection into the DFFRE scheme \cite{Notarnicola2023:FF}.
In both our proposals, we adopt homodyne-like detection (employing PNR detectors) instead of the traditional homodyne scheme (implemented by p-i-n photodiodes). The motivation behind this choice is merely practical. In fact, from an experimental point of view, a hybrid scheme involving both homodyne detection and displacement-photon counting would require to employ different types of detectors for the two components of the setup: two proportional photodiodes producing macroscopic photocurrents to implement the standard homodyne measurement, and a PNR detector for the displacement receiver. Moreover, in the context of quantum discrimination, pulsed homodyne detection would be preferable for an experimental realization at telecom wavelength, due to the reduced response time of the measurement \cite{Raymer1995, Hansen2001, Zavatta2002}. On the contrary, employing homodyne-like and low-intensity local oscillator provides a more fascinating solution since the resulting receiver is obtained with the use of sole PNR detectors.

In the following, we present in detail both the HYNORE and the HFFRE, comparing it to the Kennedy receiver and the DFFRE, respectively. The results obtained in both this section and the following ones are all original.

\subsubsection{The hybrid near-optimum receiver}\label{subsec4:HYNORE}

\begin{figure}
\includegraphics[width=0.8\columnwidth]{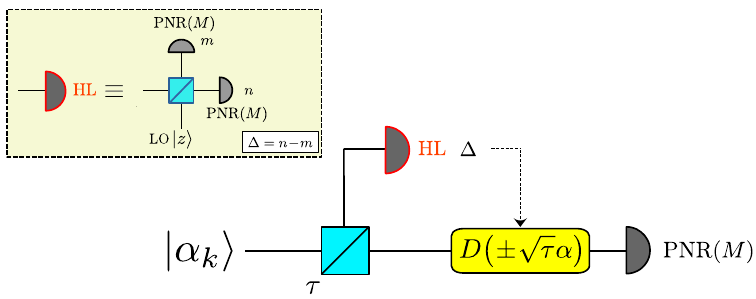}
\centering
\caption{Scheme of the HYNORE. The incoming signal is split at a beam splitter with transmissivity $\tau$, thereafter HL detection is implemented on the reflected branch. The outcome $\Delta = n - m$ is exploited to decide the displacement operation implemented on the transmitted signal.
(Inset) Setup of the weak-field homodyne, or homodyne-like, detection. The signal is mixed at a balanced beam splitter with a low-intensity local oscillator (LO), and PNR$(M)$ detection is performed on the output modes.}
\label{fig:sec4.4_HYNOREQScheme}
\end{figure}

The scheme of the HYNORE is depicted in Fig.~\ref{fig:sec4.4_HYNOREQScheme}. The idea is to exploit a displacement-PNR($M$) (DPNR) setup where the nulling displacement is not assigned a priori, but is conditioned on the outcome of a homodyne-like (HL) detection performed on a fraction of the input signal.
More in detail, we split the input coherent state $|\alpha_k\rangle$, $k=0,1$, at a beam splitter of variable transmissivity $\tau$ (this can be obtained, for instance, considering the polarization of the input states and by using a polarizing beam splitter), such that:
\begin{align}\label{eq: BS}
|\alpha_k\rangle \rightarrow |\alpha_k^{(r)}\rangle \otimes |\alpha_k^{(t)}\rangle = |-\sqrt{1-\tau}\alpha_k\rangle \otimes |\sqrt{\tau}\alpha_k\rangle \, .
\end{align}

Then, we perform HL detection on the reflected branch $|\alpha_k^{(r)}\rangle$ and, thereafter, apply a feed-forward nulling displacement operation on the transmitted part of the signal $|\alpha_k^{(t)}\rangle$ conditioned on the difference photocurrent $\Delta=n-m$ retrieved from the homodyne-like measurement:
\begin{subequations}\label{eq: Displacements}
\begin{align}
    &\Delta\ge0  \rightarrow \mbox{apply } D\left(\sqrt{\tau} \alpha\right) \, , \label{eq: Displacements:a} \\
    &\Delta<0  \rightarrow \mbox{apply } D\left(-\sqrt{\tau} \alpha\right) \, , \label{eq: Displacements:b}
\end{align}
\end{subequations}

Finally, on the resulting displaced state we perform a PNR($M$) measurement in terms of on-off detection: the photon number resolution of the detector will turn out to be useful in the presence of detection imperfections, as we will see in the following. The intuitive motivation behind the feed-forward rule of Eqs.~(\ref{eq: Displacements}) is the following. If we suppose that $|\alpha_0\rangle$ was sent, from the definition of the beam splitter operation of Eq.~(\ref{eq: BS}) it is more likely to obtain $\Delta>0$. As a consequence, we decide to perform a positive displacement sending the transmitted signal into the vacuum such that the PNR($M$) detector does not click and we refer to this event as ``off''. Of course there is still a non-zero probability to get $\Delta<0$, and in that case we decide to apply a negative displacement such that the on-off detector is more likely count some photon. This event is called ``on''. Finally, for the case $\Delta=0$, the displacement amplitude is chosen to be positive simply by convention. Analogous considerations may be obtained by considering state $|\alpha_1\rangle$. Given this scenario, the \textit{decision rule} at the end of the final measurement is chosen according to Table~\ref{tab:01-HYNOREIdealCase}.

\begin{table}[t]
\tbl{Decision strategy for the HYNORE in Fig.~\ref{fig:sec4.4_HYNOREQScheme}.\label{tab:01-HYNOREIdealCase}}
{\begin{tabular}{@{}ccc@{}} \toprule
outcomes & &  decision \\  \colrule
   $\Delta \geq 0$ $\quad$ off \, & & ``0" \\ 
    $\Delta < 0$ $\quad$ on \, & & ``0" \\ 
    $\Delta < 0$ $\quad$ off \, & & ``1" \\ 
    $\Delta \geq 0$ $\quad$ on \, & & ``1" \\ \botrule
\end{tabular}}
\end{table}

Since:
\begin{align}
    p(\Delta \geq 0; \mbox{on} | 0)= p(\Delta < 0;\mbox{off} | 1) = 0\, ,
\end{align}
the error probability for the HYNORE reads:
\begin{align}\label{eq:HYNOREpERRnopt}
    \PHY(\tau, z) &= \frac12 \left[ \ p(\Delta < 0; \mbox{off} | 0) +  p(\Delta \geq 0; \mbox{off} | 1)  \right] \nonumber \\[1ex]
    &= \frac12 \left[ \sum_{\Delta=-M}^{-1} {\cal S}_{\Delta}\big(\alpha^{(r)}_0\big) e^{-4 \tau \alpha^2}
    + \sum_{\Delta=0}^{M} {\cal S}_{\Delta}\big(\alpha^{(r)}_1\big) e^{-4 \tau \alpha^2} \right] \nonumber \\[1ex]
    &= \frac{e^{-4 \tau \alpha^2}}{2} \left[\sum_{\Delta=-M}^{-1} {\cal S}_{\Delta}(\sqrt{1-\tau}\alpha) 
    +  \sum_{\Delta=0}^{M} {\cal S}_{\Delta}(-\sqrt{1-\tau}\alpha) \right] \, ,
\end{align}
depending on the transmissivity $\tau$, ruling the splitting of the incoming state, and the amplitude of the local oscillator (LO) $|z\rangle$, $z\ge0$, of HL detection. We recall that the HL probability distribution reads:
\begin{align}\label{eq:HLdistr}
\mathcal{S}_\Delta&(\alpha^{(r)}_k) =
     \sum_{n,m=0}^{M} p_n\big(\mu_{+}(\alpha^{(r)}_k)\big) \
    p_m\big(\mu_{-}(\alpha^{(r)}_k)\big) \, \delta_{(n-m),\Delta}
\end{align}
where $\delta_{k,j}$ is the Kronecker delta, $M$ is the photon-number resolution, 
\begin{align}\label{eq:RatesMu}
\mu_{\pm}(\alpha^{(r)}_k)= \frac{|\alpha^{(r)}_k \pm z|^2}{2}\, ,
\end{align}
is the mean energy on the two output branches, respectively, and
\begin{align}\label{eq:PNRMprob}
    p_n(\mu) =
    \left\{\begin{array}{l l}
    {\displaystyle e^{-\mu} \ \frac{\mu^{n}}{n!}}  & \mbox{if}~n<M \ , \\[2ex]
    {\displaystyle 1- e^{-\mu} \sum_{j=0}^{M-1} \frac{\mu^{j}}{j!}} & \mbox{if}~n = M \ .
    \end{array}
    \right.
\end{align}

For completeness, we note that performing standard homodyne detection instead of homodyne-like, the error probability of the previous equation becomes
\begin{align}\label{eq: pERR HD}
    \PHY^{\rm (HD)}(\tau) = \frac{e^{-4 \tau \alpha^2}}{2} \ \left\{1- \erf \left[\sqrt{2(1-\tau)} \alpha \right] \right\} \, .
\end{align}
We also note that if $\tau=0$ we have the homodyne receiver, whereas if $\tau=1$ we retrieve the Kennedy one. This can be understood since when $\tau=1$ the information coming from the homodyne receiver is inconclusive, as it measures the vacuum leading to a positive or negative outcome with $50\%$ of probability. Therefore, known the outcome sign, we can apply the same inference strategy as in the Kennedy receiver.

Given this outline, in the following we compute the error probability of the HYNORE by considering two alternative scenarios involving either ideal photodetectors, namely PNR($M$) detectors with $M=\infty$, or realistic PNR detectors with finite resolution $M$.

\paragraph{HYNORE with ideal photodetectors.}

At first, let us consider PNR($\infty$) detectors, and a LO $|z\rangle$ with fixed intensity $z^2$. In this case, the HL probability distribution~(\ref{eq:HLdistr}) approaches a Skellam distribution, as discussed in Sec.~\ref{subsec2:HL}.
Under these conditions, we retrieve the HYNORE error probability by optimizing Eq.~(\ref{eq:HYNOREpERRnopt}) with respect to $\tau$, i.e. finding the transmissivity $\tau_\opt^{\rm (id)}$, that in general is a function of $\alpha^2$, minimizing the value of $\PHY(\tau,z)$ for every $\alpha^2$. Consequently, we obtain the optimized error probability of our receiver as
\begin{align}
\PHY^{\rm (id)} = \min_{\tau} \PHY(\tau,z) \, \qquad \text{for PNR($\infty$) detection} \, ,
\end{align}
depicted in Fig.~\ref{fig:sec4.4_HYNOREProb}(a). As we see, the HYNORE proves to be near-optimum, outperforming the Kennedy receiver for all $\alpha^2$.

To better enlighten this advantage, it is also relevant to introduce the ratio with the standard Kennedy receiver~(\ref{eq:Kennedy}), 
\begin{align}\label{eq: RatiohK}
    R_{h/K}^{\rm (id)}= \frac{\PHY^{\rm (id)}}{P_{K} }\,  .
\end{align}
Plots of $R_{h/K}^{\rm (id)}$ and $\tau_{\rm opt}^{\rm (id)}$ (in the inset) are displayed in 
Fig.~\ref{fig:sec4.4_HYNOREProb}(b) for different LO intensity $z^2$.
It emerges that $\tau_{\rm opt}^{\rm (id)}=0$ up to a threshold energy $N_{\rm th}(z)$ which depends on the LO amplitude $z$, while for $\alpha^2 > N_{\rm th}(z)$ it is an increasing function of the energy and reaches asymptotically 1. Note that in the limit $\tau\rightarrow 1$ some information about the signal reaches the homodyne receiver and we still have an improvement of the performance.

\begin{figure}[t]
\includegraphics[width=0.49\columnwidth]{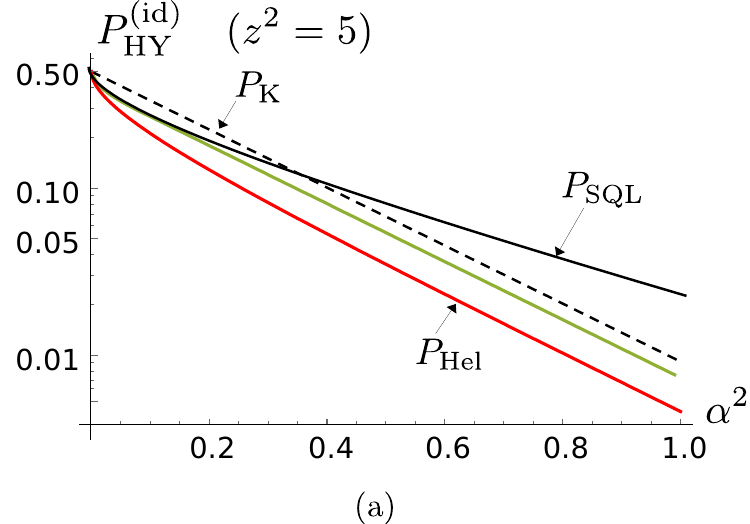} 
\includegraphics[width=0.49\columnwidth]{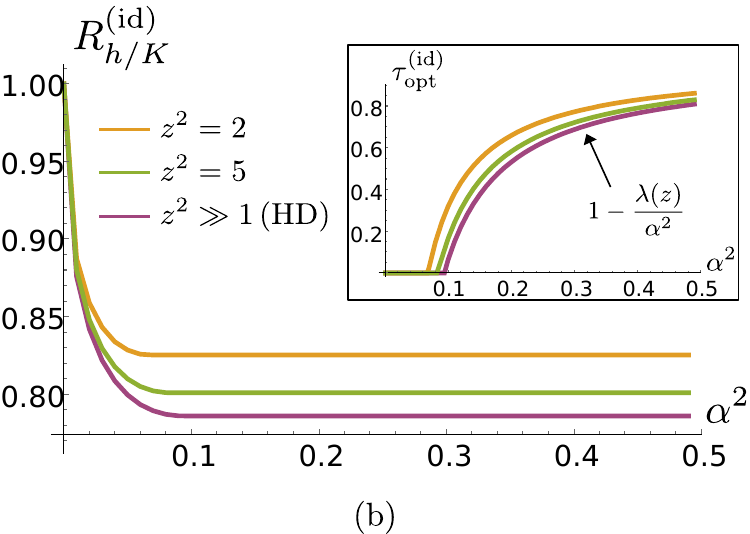}
\centering
\caption{(a) Log plot of $\PHY^{\rm (id)}$ as a function of $\alpha^2$ for LO intensity $z^2=5$ compared to the Kennedy receiver (\ref{eq:Kennedy}), the SQL (\ref{eq:BinarySQL}) and the Helstrom bound (\ref{eq:BinaryHB}). (b) Plot of the ratio $R_{h/K}^{\rm (id)}$ as a function of $\alpha^2$ for several values of the LO intensity $z^2$. In the inset, plot of the optimized transmissivity $\tau_{\rm opt}^{\rm (id)}$ as a function of $\alpha^2$. For $\alpha^2>N_{\rm th}(z)$ we have $\tau_{\rm opt}^{\rm (id)}= 1-\lambda(z)/\alpha^2$. In both the pictures we consider PNR($\infty$) detectors.}\label{fig:sec4.4_HYNOREProb}
\end{figure}

If $\alpha^2 \leq N_{\rm th}(z)$ the optimized strategy is realized with the sole homodyne-like setup, whereas for larger energies the more efficient scheme is obtained by the appropriate interplay between the homodyne-like and the DPNR parts of our receiver. The choice of the optimal $\tau$ makes the receiver near-optimum with a ratio $R_{h/K}^{\rm (id)}$ saturating to the value $R_{\infty}^{\rm (id)}<1$ for every value of the LO intensity $z^2$.

As we noticed, if we increase the value of $z^2$, the performance of the homodyne-like detection approaches the standard homodyne one and the HYNORE performs better and better. In fact, the variance of the homodyne-like quadrature probability distribution decreases as the local oscillator energy becomes quite larger with respect to the input signal one \cite{Olivares2021}. In this case, the ratio in Eq.~(\ref{eq: RatiohK}) reads
\begin{align}\label{eq: Ratio HD}
    R_{h/K}^{\rm (HD)}= \frac{\PHY^{\rm (HD)}}{P_{K}} = \frac{e^{4 (1-\tau) \alpha^2}}{2} \ \left\{1- \erf \left[\sqrt{2(1-\tau)} \alpha \right] \right\} \, .
\end{align}

The saturation of $R_{h/K}^{\rm (id)}$ for large $\alpha^2$ suggests the following ansatz on the expression of the optimized $\tau_{\rm opt}^{\rm (id)}$, namely:
\begin{align}\label{eq: Ansatz}
    \tau_{\rm opt}^{\rm (id)} = 1- \frac{\lambda(z)}{\alpha^2} \quad \mbox{for} \quad \alpha^2>N_{\rm th}(z) \, ,
\end{align}
where $\lambda(z)\in {\mathbbm R}_{+}$ and depends on the LO amplitude $z$.
As an example, for the homodyne limit $z^2\rightarrow \infty$, by computing the derivative of Eq.~(\ref{eq: Ratio HD}) with respect to $\tau$ and inserting the expression in Eq.~(\ref{eq: Ansatz}) we get the following relation that must be satisfied by $\lambda \equiv \lambda(z=\infty)$ :
\begin{align}
    \sqrt{\frac{2}{\pi \lambda}}- 4 e^{2\lambda} \ \Big[1- \erf \bigl(\sqrt{2\lambda} \bigr) \Big] = 0\, ,
\end{align}
that leads to the numerical solution $\lambda \approx 0.094$.
Then, the threshold $N_{\rm th}^{\rm (HD)}\equiv N_{\rm th}(z=\infty)$ can obtained by setting $\tau_{\rm opt}^{\rm (id)}=0$, bringing to $N_{\rm th}^{\rm (HD)}= \lambda$ and the saturation ratio reads:
\begin{align}
    R_{\infty}^{\rm (HD)} = e^{4 \lambda}  \Bigl[1- \erf \bigl(\sqrt{2\lambda} \bigr) \Bigr] \approx 0.786 \ .
\end{align}
An identical analysis can be performed for the homodyne-like case, where we may expect $\lambda(z)<\lambda$.

\paragraph{HYNORE with finite photon-number resolution.}

\begin{figure}
\includegraphics[width=0.6\columnwidth]{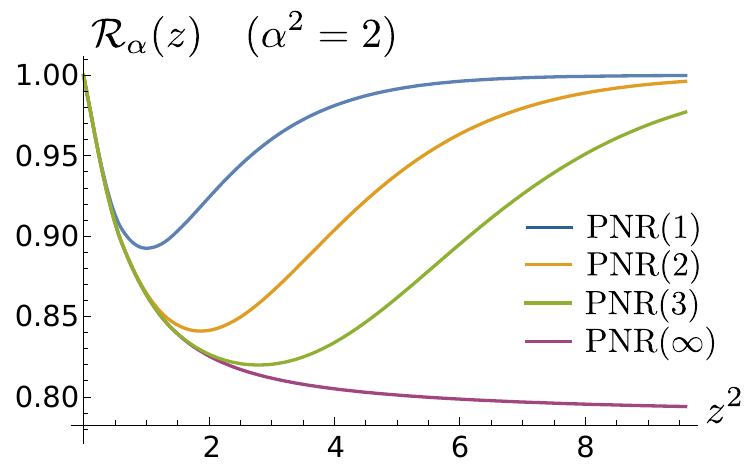}
\centering
\caption{Plot of ${\cal R}_\alpha(z)$ as a function of the LO intensity $z^2$ for different resolution $M$ and fixed signal energy $\alpha^2=2$. For $M<\infty$, ${\cal R}(z)$ exhibits a minimum at a finite LO intensity.}\label{fig:sec4.4_HYFiniteRes}
\end{figure}

We now consider the more realistic case of PNR($M$) detectors having a finite photon number resolution $M$, being only able to resolve any number of photons $n$ up to $M$. Clearly, PNR(1) is a on-off photodetector. In the absence of detection imperfections, the presence of a reduced resolution affects the sole homodyne-like setup, reducing the inferable information on the field phase, since the displacement-photon counting scheme performed on the transmitted branch is associated with on-off decision strategy regardless the value of $M$.

Differently from the ideal case, when $M<\infty$ the error probability~(\ref{eq:HYNOREpERRnopt}) is not a monotonous function of the LO intensity. This emerges by computing the following $z$-dependent ratio, for fixed input signal energy and varying LO:
\begin{align}\label{eq:Ralphaz}
{\cal R}_\alpha(z)= \frac{\min_\tau \PHY(\tau,z)}{\PK} \, ,
\end{align}
depicted in Fig.~\ref{fig:sec4.4_HYFiniteRes} for different values of $M$. As discussed above, in the ideal case, i.e. $M= \infty$, Eq.~(\ref{eq:Ralphaz}) descreases monotonically with $z^2$, proving the homodyne limit to be the best working regime. On the contrary, for $M<\infty$, ${\cal R}_\alpha(z)$ exhibits a minimum at a finite LO intensity. Indeed, if only few photons can be resolved, increasing the LO is useless since much of its energy could not be detected. 

\begin{figure}[t]
\includegraphics[width=0.49\columnwidth]{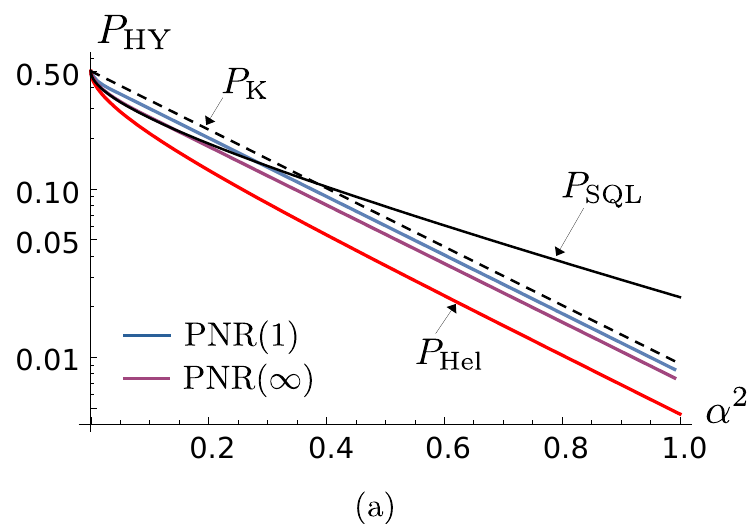}
\includegraphics[width=0.49\columnwidth]{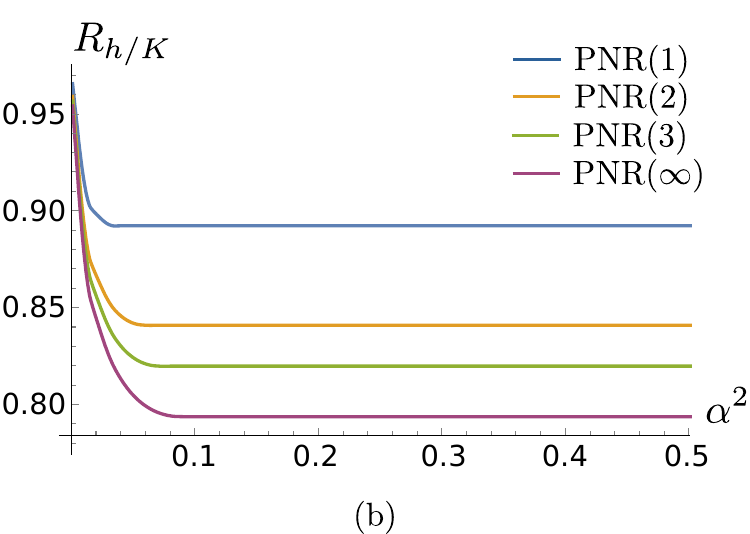}
\centering
\caption{(a) Log plot of $\PHY$ as a function of $\alpha^2$ for different resolution $M$ compared to the Kennedy receiver (\ref{eq:Kennedy}), the SQL (\ref{eq:BinarySQL}) and the Helstrom bound (\ref{eq:BinaryHB}). (b) Plot of the ratio $R_{h/K}$ as a function of $\alpha^2$ for different $M$. The case $M=\infty$ refers to the homodyne limit~(\ref{eq: pERR HD}) optimized over transmissivity.}\label{fig:sec4.4_HYNOREProb}
\end{figure}

In turn, to establish the performance of the HYNORE, we are entitled to optimize Eq.~(\ref{eq:HYNOREpERRnopt}) over both $\tau$ and $z$, leading to the optimized error probability
\begin{align}\label{eq:PHYNOREM}
\PHY= \min_{\tau,z} \PHY(\tau,z) \, \qquad \text{for PNR($M$) detection} \, ,
\end{align}
together with the ratio
\begin{align}
    R_{h/K}= \frac{\PHY}{P_{K} }\, ,
\end{align}
where, for the sake of simplicity, the dependence on the resolution $M$ has not been explicitly reported.
Plots of $\PHY$ and $R_{h/K}$ are depicted in Fig.~\ref{fig:sec4.4_RatioHKPNRM}(a) and (b), respectively, in which the case $M=\infty$ refers to the homodyne limit~(\ref{eq: pERR HD}) optimized over transmissivity. As we can see, the effect of the finite resolution is to decrease the saturation ratio $R_{\infty}$, which in any case is still less than $1$, maintaining the advantages of HYNORE with respect to the Kennedy. We also note that employing high resolution detectors is not strictly required to obtain a performance close to the ideal case. Furthermore, the optimized transmissivity $\tau_\opt$ shows analogous behavior to the one depicted in the inset of Fig.~\ref{fig:sec4.4_HYNOREProb}(b), namely $\tau_\opt= 1- \lambda(M)/\alpha^2$, $\lambda(M)$ being a increasing function of the PNR resolution $M$ such that $\lambda(M)<\lambda\approx 0.094$. In contrast, the optimized LO intensity is 
of the order of the resolution for all input energies; in fact we have $z_\opt^2 \lesssim M$.

\subsubsection{The hybrid feed-forward receiver}\label{subsec4:HFFRE}

\begin{figure}[t]
\includegraphics[width=0.7\columnwidth]{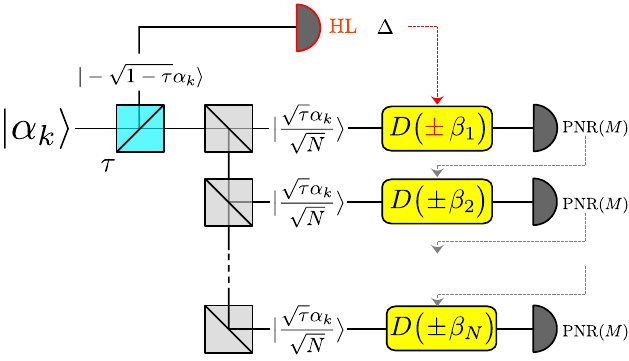}
\centering
\caption{Scheme of the HFFRE. We split the incoming signal $|\alpha_k\rangle$, $k=0,1$, at a beam splitter of variable transmissivity $\tau$. We perform HL detection on the reflected branch, whereas we implement the displacement feed-forward setup on the transmitted one. We exploit the HL outcome to decide the sign of the displacement operation on the first copy of the transmitted signal.}\label{fig:sec4.4_HFFREsetup}
\end{figure}

As discussed above, employing a hybrid receiver like the HYNORE turns out to be beneficial for quantum discrimination, obtaining a better performance than the Kennedy receiver thanks to the splitting of the incoming signal into two beams, followed by a suitable adaptive operation.
Given this consideration, a natural extension emerges. That is, the homodyne-like setup may be suitably embedded into the multi-copy approach described in Sec.~\ref{subsec4:DFFRE} to construct a new kind of receiver, referred to as the {\it hybrid feed-forward receiver}  (HFFRE) \cite{Notarnicola2023:FF}. In more detail, the HFFRE is obtained by suitably merging the setups of both the HYNORE and the DFFRE, resulting in the scheme depicted in Fig.~\ref{fig:sec4.4_HFFREsetup}.

The insight is to exploit a HL measurement to guide the choice of the first displacement operation sign in the DFFRE.
As for the HYNORE, we divide the incoming signal $|\alpha_k\rangle$, $k=0,1$, at a beam splitter with variable transmissivity $\tau$, see Eq.~(\ref{eq: BS}).
The reflected signal $|\alpha_k^{(r)}\rangle$ undergoes HL detection with difference photocurrent outcome $\Delta$. Then, we split the transmitted state $|\alpha_k^{(t)}\rangle$ into $N$ copies, $|\alpha_k^{(t)}/\sqrt{N}\rangle$, and implement the same procedure described in Sec.~\ref{subsec4:DFFRE}. The only difference with respect to the displacement feed-forward receiver lies in the displacement operation performed on the first copy. Indeed, the difference photocurrent $\Delta$ provides us with a priori information exploitable to decide the sign of the first optimized displacement operation, according to the HYNORE adaptive rule, namely:
\begin{align}
\left\{\begin{array}{ll}
\Delta \ge 0 \quad &\rightarrow \quad \mbox{apply } D(\beta_1) \, \\[1ex]
\Delta < 0 \quad &\rightarrow \quad \mbox{apply }  D(-\beta_1) \, ,
\end{array}
\right.
\end{align}
$\beta_1>0$. Displacements on the other copies are still conditioned on the outcomes of the $(j-1)$-th PNR($M$) measurement.

Thus, the probability of performing a correct decision ${\cal P}_{\rm HF}^{(j)}(\tau,z)$ after $j$ steps gets the same form of Eq.~(\ref{eq:PcorrDFF}):
\begin{align}\label{eq:PcorrHYB}
{\cal P}_{\rm HF}^{(j)}&(\tau,z) = \max_{\beta_{j}}
\Bigg\{ {\cal P}_{\rm HF}^{(j-1)}(\tau,z)  q_{\rm off}\Big( \lambda_{-}^{(j)}(\sqrt{\tau}\alpha)\Big)
\notag\\[1ex]
& \hspace{2.5cm}
+ \left[
1 -  {\cal P}_{\rm HF}^{(j-1)}(\tau,z)
\right]q_{\rm on}\Big( \lambda_{+}^{(j)}(\sqrt{\tau}\alpha)\Big) \Bigg\} \, ,
\end{align}
with the rates $\lambda_{\pm}^{(j)}$ in Eq.~(\ref{eq:lambdapm}), albeit to be solved with a different initial condition, that is:
$${\cal P}_{\rm HF}^{(0)}(\tau,z)=\frac12 \Bigg[ \sum_{\Delta=-M}^{-1} {\cal S}_\Delta\Big(\alpha^{(r)}_1\Big)
+ \sum_{\Delta=0}^{M} {\cal S}_\Delta\Big(\alpha^{(r)}_0\Big) \Bigg] \, , $$ 
corresponding to the probability of correct decision after the HL measurement. Clearly, if $\tau=1$ we retrieve the results of the DFFRE.

\begin{figure}
\includegraphics[width=0.6\columnwidth]{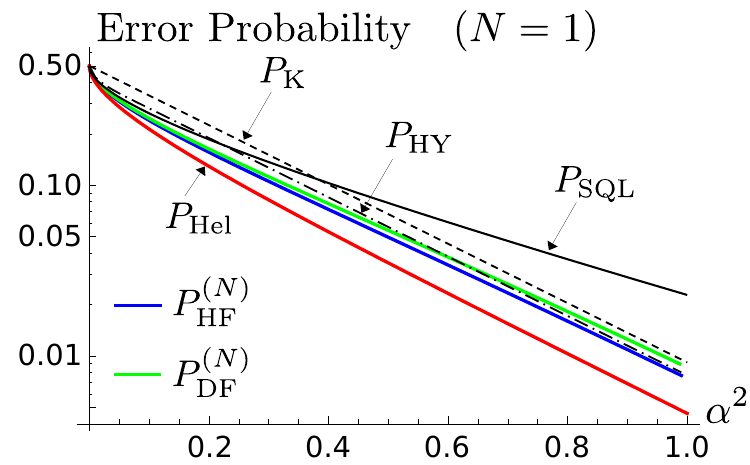}
\centering
\caption{Log plot of $\PHFF^{(N)}$ and $\PDISP^{(N)}$ as a function of the signal energy $\alpha^2$ for $N=1$. The PNR resolution is $M=2$. $\PSQL$, $\PHel$, $\PK$ and $\PHY$ refer to the SQL~(\ref{eq:BinarySQL}), the Helstrom bound~(\ref{eq:BinaryHB}), and the error probabilities of the Kennedy receiver~(\ref{eq:Kennedy}) and the HYNORE~(\ref{eq:PHYNOREM}), respectively.}\label{fig:sec4.4_RatioHKPNRM}
\end{figure}

As both $\tau$ and $z$ are free parameters, after $N$ copies the error probability reads
\begin{align}\label{eq:PHFFRE}
\PHFF^{(N)} = 1- \max_{\tau,z}{\cal P}_{\rm HF}^{(N)}(\tau,z)  \, ,
\end{align}
depicted in Fig.~\ref{fig:sec4.4_RatioHKPNRM} as a function of the input energy $\alpha^2$ and compared to the DFFRE error probability $\PDISP^{(N)}$ in Eq.~(\ref{eq:PDFF}).
The HFFRE outperforms the DFFRE, $\PHFF^{(N)} \le \PDISP^{(N)}$. Both the receivers are near-optimum and beat the SQL for all energies, but we observe different asymptotic scalings. Indeed, for $\alpha^2 \gg 1$, the DFFRE approaches the Kennedy receiver, $\PDISP^{(N)} \approx \PK$, whereas the HFFRE reaches the HYNORE, $\PHFF^{(N)} \approx \PHY$, with the $\PHY$ in Eq.~(\ref{eq:PHYNOREM}). As a consequence, exploiting information on both the phase and the photon statistics of the field proves to be a powerful tool to reduce the error probability.

\begin{figure}[t]
\includegraphics[width=0.49\columnwidth]{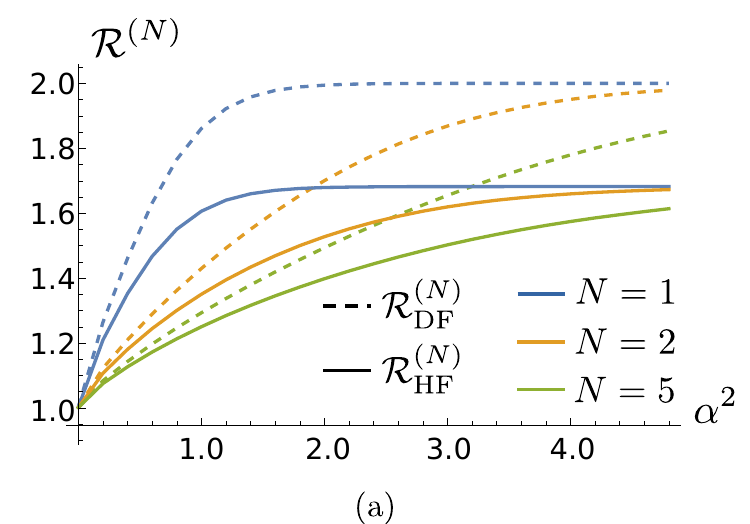} 
\includegraphics[width=0.49\columnwidth]{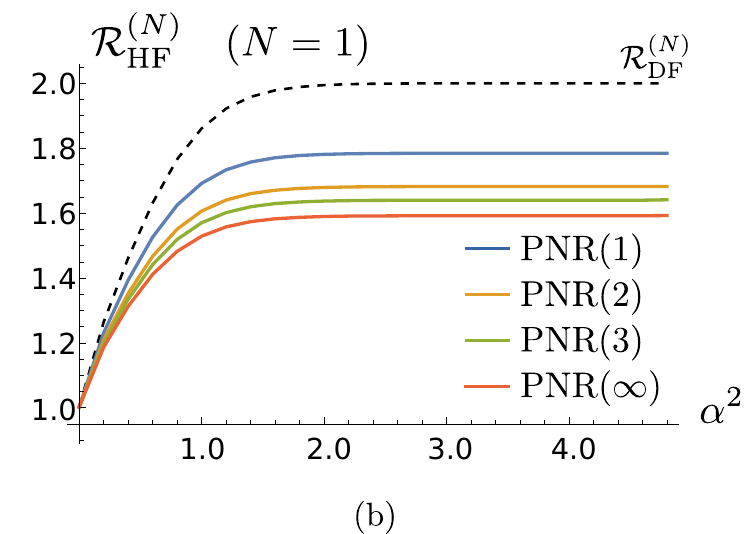}
\centering
\caption{(a) Plot of ${\cal R}_{\p}^{(N)}$, $\p={\rm DF,HF}$, as a function of $\alpha^2$ for different number of copies $N$. The PNR resolution is $M=2$. (b) Plot of ${\cal R}_{\rm HF}^{(N)}$ as a function of $\alpha^2$ for $N=1$ and different PNR resolutions $M$. The dashed line corresponds to ${\cal R}_{\rm DF}^{(N)}$ for $N=1$.}\label{fig:sec4.4_RatioHels}
\end{figure}

Furthermore, by increasing the number of copies $N$ the performance of both the feed-forward receivers improves for $\alpha^2 \ll 1$, coming closer to the Helstrom bound~(\ref{eq:BinaryHB}), as emerges by computing the ratio
\begin{align}
{\cal R}_{\p}^{(N)} = \frac{P_{\p}^{(N)}}{\PHel} \, , \quad (\p={\rm DF,HF}) \, ,
\end{align}
plotted in Fig.~\ref{fig:sec4.4_RatioHels}(a).

In the regime $\alpha^2 \ll 1$, the larger the number of copies, the smaller the ratio ${\cal R}_{\p}^{(N)}$, whereas in the asymptotic limit $\alpha^2 \gg 1$ the displacement and hybrid receiver converge to Kennedy and HYNORE, respectively, regardless the value of $N$. Moreover, the ratio for the hybrid receiver ${\cal R}_{\rm HF}^{(N)}$ may be further reduced by increasing the PNR resolution $M$, as shown in Fig.~\ref{fig:sec4.4_RatioHels}(b). In particular, the asymptotic ratio is reduced for greater values of $M$ and reaches its minimum value for PNR$(\infty)$ detectors, i.e. ideal photodetectors, in which case the HL distribution in Eq.~(\ref{eq:HLdistr}) becomes a Skellam distribution.

\subsection{Quantum receivers in the presence of detection imperfections}\label{sec4:IneffDiscr}

So far, we described the structure of quantum receivers by considering an ideal scenario involving perfect detection schemes and excluding imperfections within each element of the setups. In these conditions, we proved the hybrid receivers, namely the HYNORE and HFFRE, to outperform the receivers based on displacement strategies, i.e. the Kennedy receiver and the DFFRE, respectively.
Now, we may wonder whether or not the obtained enhancement in the error probability could be effectively realized in practical experiments. This raises the problem of the robustness of the proposed receivers in the presence of realistic conditions, e.g. limited quantum detection efficiency, dark counts and visibility reduction in the adopted displacement operations.

As one may expect, in a realistic scenario neither the hybrid nor the displacement receiver remain near-optimum, therefore they are not able to approach the Helstrom bound~(\ref{eq:BinaryHB}) anymore. Accordingly, a new goal emerges, that is to show whether or not these receivers are still able to beat the SQL~(\ref{eq:BinarySQL}) even in the presence of practical imperfections. Indeed, in this case we would get a robust quantum advantage with respect to the best receiver achievable with semi-classical means.

In the following, we compare the performance of the proposed hybrid receivers with respect to their corresponding displacement-photon counting schemes in the presence of the typical imperfections occurring in PNR detection \cite{Notarnicola2023:HYNORE, Notarnicola2023:FF}. In particular, we consider a non-unit quantum efficiency $\eta\le1$ of the PNR($M$) detectors, as well as the presence of dark counts. Moreover, since the displacement operation is realized into practice by letting the signal interfere with a suitable LO at a beam splitter \cite{Paris1996}, we also address the effects of non-unit visibility $\xi\le1$.

\subsubsection{HYNORE vs displacement receiver}\label{subsec:HYNOREvsDPNR}
At first, we start by addressing the robustness of single-copy receivers, namely the HYNORE and the displacement receiver. For a better clarity, we discuss separately the impact on the receiver performance of three experimental defects above presented, namely quantum efficiency, dark counts and reduced visibility. 

\paragraph{Reduced quantum efficiency.}

\begin{figure}
\includegraphics[width=0.6\columnwidth]{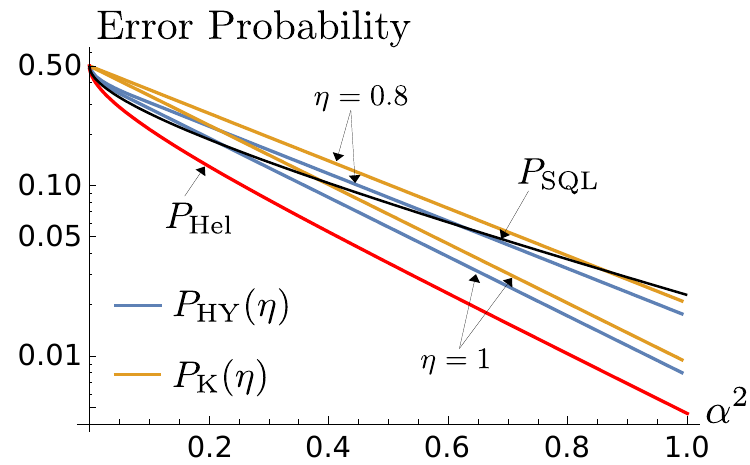}
\centering
\caption{Log plot of $\PHY(\eta)$ and $\PK(\eta)$ as a function of the signal energy $\alpha^2$ for different quantum efficiency $\eta$. The PNR resolution is $M=2$. $\PSQL$ and $\PHel$ refer to the SQL~(\ref{eq:BinarySQL}) and the Helstrom bound~(\ref{eq:BinaryHB}), respectively.}\label{fig01:sec4.5.1_QEFF}
\end{figure}

Concerning the inefficient photodetection, the introduction of a quantum efficiency $\eta$ has the effect of re-scaling all the coherent amplitudes of the measured pulses by a factor $\sqrt{\eta}$, since it corresponds to a photon loss. 

Thus, for the Kennedy receiver employing inefficient on-off detection, the error probability is changed into:
\begin{align}\label{eq: KennedyIneff}
    \PK  (\eta) = \frac{e^{-4 \eta \alpha^2}}{2} \, .
\end{align}
Instead, in the HYNORE, the efficiency affects both the HL and the PNR($M$) measurement schemes. For the HL detection the rates in Eq.~(\ref{eq:RatesMu}) are changed into $\mu_{\pm} \rightarrow \eta \mu_{\pm}$, obtaining:
\begin{align}\label{eq: pDelta with eff}
    \mathcal{S}_\Delta(\eta;\alpha^{(r)}_{k}) =
     \sum_{n,m=0}^{M} p_n\big(\eta \, \mu_{+}(\alpha^{(r)}_k)\big) \
    p_m\big(\eta \, \mu_{-}(\alpha^{(r)}_k)\big) \, \delta_{(n-m),\Delta} \, .
\end{align}

On the other hand, an inefficient on-off detection by the PNR implies the substitution $\exp(-4 \tau \alpha^2) \rightarrow \exp(-4 \eta \tau \alpha^2)$.
By performing these substitutions into Eq.~(\ref{eq:HYNOREpERRnopt}) we get the corresponding error probability:
\begin{align}
\PHY(\eta) = \min_{\tau,z} \PHY(\tau,z; \eta) \, ,
\end{align}
with
\begin{align}\label{eq: Phyb with eff}
    \PHY(\tau,z; \eta) =\frac{e^{-4 \eta \tau \alpha^2}}{2} \left[ \sum_{\Delta=-M}^{-1} \mathcal{S}_\Delta(\eta;\alpha^{(r)}_{0})
    + \sum_{\Delta=0}^{M} \mathcal{S}_\Delta(\eta;\alpha^{(r)}_{1})\right] \, .
\end{align}

\begin{figure}
\includegraphics[width=0.49\columnwidth]{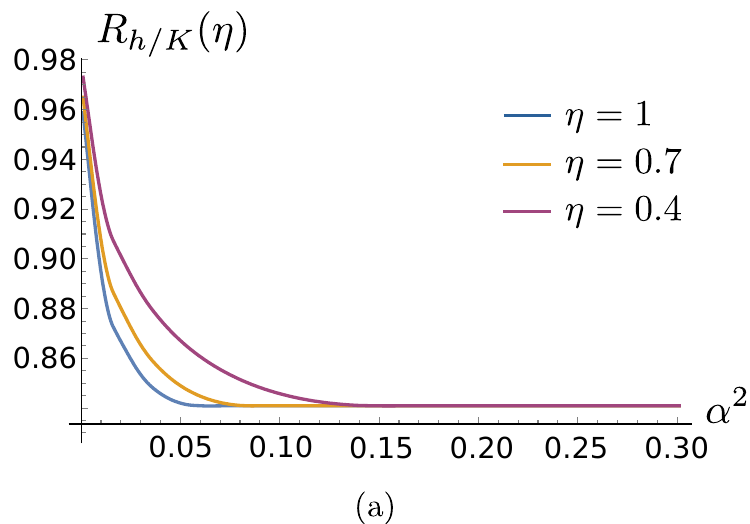}
\includegraphics[width=0.49\columnwidth]{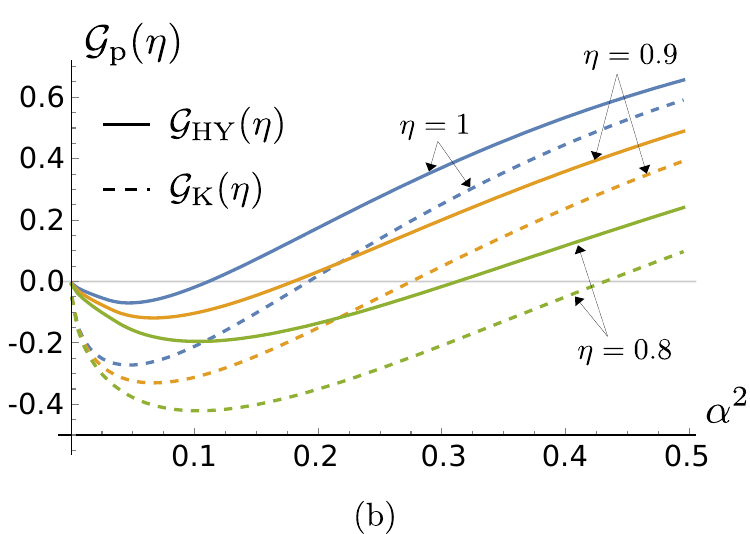}
\centering
\caption{(a) Plot of the ratio $R_{h/K}(\eta)$ as a function of $\alpha^2$ for different $\eta$. (b) Plot of the gain ${\cal G}_\p(\eta)$, $\p= {\rm K, HY}$, as a function of $\alpha^2$ for different $\eta$. The PNR resolution is $M=2$.}\label{fig02:sec4.5.1_QEFF1}
\end{figure}

The error probabilities $\PK(\eta)$ and $\PHY(\eta)$ are depicted in Fig.~\ref{fig01:sec4.5.1_QEFF} for PNR($2$) receivers. The behavior is analogous for all resolution $M$.
As expected, the performance of both detector is degraded for lower quantum efficiency, but, interestingly, for a given value of $\eta$, exploiting the HYNORE is always preferable than the Kennedy, as $\PHY(\eta)\le \PK (\eta)$.
In particular, the relative ratio
\begin{align}
    R_{h/K}(\eta) = \frac{\PHY(\eta)}{\PK (\eta)} \, ,
\end{align}
reported in Fig.~\ref{fig02:sec4.5.1_QEFF1}(a), saturates to the same $R_{\infty}$ obtained for $\eta=1$, regardless the value of quantum efficiency.
Furthermore, in the high-energy regime, both receivers beat the SQL~(\ref{eq:BinarySQL}). To better highlight this feature, we consider the gain
\begin{align}
    {\cal G}_\p(\eta) = 1-\frac{P_\p(\eta)}{\PSQL} \, , \quad \p={\rm K, HY} \, ,
\end{align}
plotted in Fig.~\ref{fig02:sec4.5.1_QEFF1}(b). Accordingly, the SQL is outpermed when ${\cal G}_\p(\eta) >0$.
We observe that there exists a threshold energy $\alpha^2_\p(\eta)$ after which the discussed receivers beat the SQL, that is ${\cal G}_\p(\eta) >0$ for $\alpha^2> \alpha^2_\p(\eta)$, and we have $\alpha^2_{\rm HY}(\eta) \le \alpha^2_{\rm K}(\eta)$. By reducing the quantum efficiency $\eta$, the gain and the threshold energy decrease and increase, respectively.

\paragraph{Dark counts.}

Dark counts are random clicks of the PNR due to environmental noise and so not directly correlated to the properties of the coherent measured pulse. They can be described in terms of Poisson counting \cite{Humer2015}, occurring at rate $\nu$ which in many realistic conditions takes values $\nu \lesssim 10^{-3}$ \cite{Izumi2012, DiMario2018QPSK, Izumi2021,Thekkadath2021OptLett,Sidhu2021}. Generally speaking, the outcome $n$ of an ideal PNR measurement on a generic coherent state $|\zeta\rangle$ in the presence of dark counts turns out to be the sum of two Poisson variables and, therefore, still follows a Poisson distribution with rate equal to $|\zeta|^2+\nu$ \footnote{The sum of two Poisson independent random variables is still a Poisson random variable. If $x \sim \mathbb{P}(\mu)$ and $y\sim \mathbb{P}(\lambda)$ are two Poisson independent random variables with rates $\mu$ and $\lambda$ respectively, the probability that $x+y$ gets the value $k$ reads $p(x+y=k) = \sum_{l=0}^k p(x=l) p(y=k-l) = e^{-\mu-\lambda} \sum_{l=0}^k \mu^l \lambda^{k-l} /(l! (k-l)!) = e^{-\mu-\lambda} (\mu+\lambda)^k/k! \sim \mathbb{P(\mu+\lambda)}$.}. In turn, in the presence of a PNR($M$) we have a probability $p_n(\mu)$ as in Eq.~(\ref{eq:PNRMprob}) but with rate $\mu=|\zeta|^2+\nu$.

The presence of dark counts has a significant effect on the performances of quantum receivers.
In particular, it becomes detrimental for the Kennedy receiver, as the receiver registers environmental clicks uncorrelated to the probed signal, inducing unwanted decision errors and undermining the ``nulling" displacement technique. In fact, in such a situation the on-off detector may click even if the vacuum is measured.

\subparagraph{Displacement-PNR receiver.}
As anticipated in Sec.~\ref{subsec4:Kennedy}, to counteract this effect, DiMario and Becerra proposed the displacement-PNR (DPNR) receiver, namely a Kennedy setup employing PNR($M$) detectors instead of on-off, and exploit the photon number resolution to choose the decision rule for discrimination in a more accurate way \cite{DiMario2018, DiMario2019}.
In fact, in place of the usual on-off strategy, in the presence of dark counts the decision rule should be changed according to the \textit{maximum a posteriori probability criterion} (MAP), discussed in App.~\ref{app:MAPcriterion}. If $|\alpha_0 \rangle$ is sent the probability of detecting $n$ photons is $p_n(\nu)$, whereas if $|\alpha_1 \rangle$ is sent the probability is $p_n(4\alpha^2+\nu)$. The error probability for the DPNR receiver is then obtained as:
\begin{align}\label{eq: DPNRM with DC}
    \PD(\nu)= 1- \frac12 \sum_{n=0}^{M} \, \max\Big[p_n(\nu), \, p_n(4\alpha^2+\nu) \Big] \, .
\end{align}
The procedure of maximizing the a posteriori probability is equivalent to defining a threshold count $\nth(\nu)\le M$ such that all measurement outcomes $n\ge \nth(\nu) $ are assigned to state ``1" and all $n< \nth(\nu) $ are assigned to state ``0". The threshold number is obtained by equating the photon number distributions of the two displaced states, namely $p_{\bar{n}}(\nu)= p_{\bar{n}}(4\alpha^2+\nu)$, $\bar{n} \in \mathbb{R}$, and considering the lowest integer greater than the obtained root $\bar{n}$, namely $\nth(\nu)= \ceil{\bar{n}}$, where $\ceil{\cdot}$ is the ceiling function. Ultimately, we have:
\begin{align}\label{eq: nTH DC}
    n_{\rm th}(\nu) = \min \left[\left\lceil\frac{4 \alpha^2}{\displaystyle \ln \left(1 + 4\alpha^2 / \nu \right)}\right\rceil, \, M \right] \, ,
\end{align}
We note that the threshold is a function of $\alpha^2$, namely $\nth(\nu)=\nth(\nu;\alpha^2)$. For the case of PNR($1$) we have $n_{\rm th}(\nu)=1$, retrieving the on-off discrimination of the standard Kennedy receiver. Furthermore, in the limit $\nu\rightarrow 0$, $n_{\rm th}(\nu)$ approaches $1$, retrieving the usual Kennedy configuration.

\begin{figure}
\includegraphics[width=0.49\columnwidth]{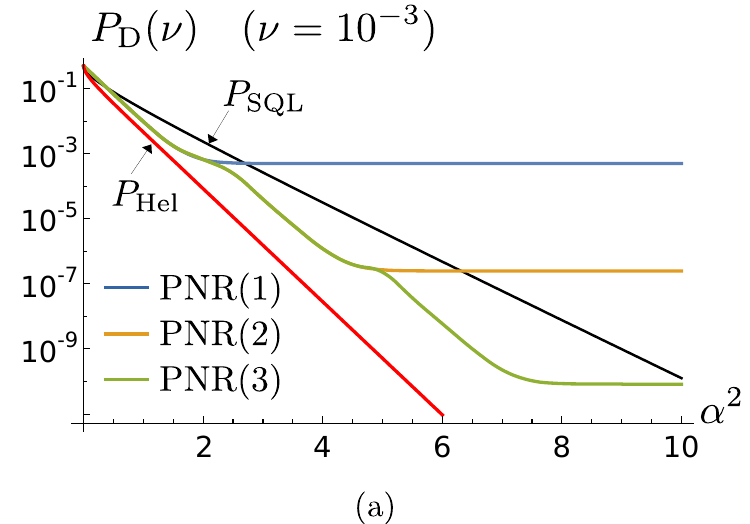} 
\includegraphics[width=0.49\columnwidth]{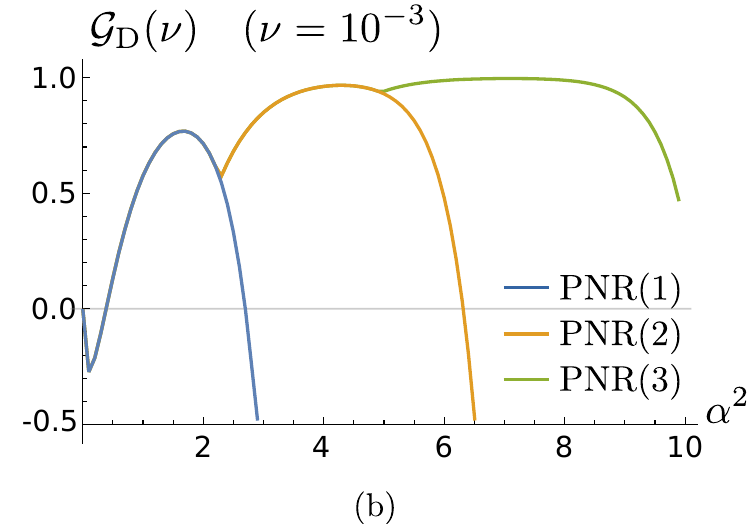}
\centering
\caption{(a) Log plot of $\PD(\nu)$ as a function of the signal energy $\alpha^2$ for different resolution $M$. $\PSQL$ and $\PHel$ refer to the SQL~(\ref{eq:BinarySQL}) and the Helstrom bound~(\ref{eq:BinaryHB}), respectively. (b) Plot of the gain ${\cal G}_{\rm D}(\nu)$ as a function of $\alpha^2$ for different $M$. In both the pictures, the dark count rate is set to $\nu=10^{-3}$.}\label{fig03:sec4.5.1_PDISP}
\end{figure}

Plots of the error probabilities for different PNR($M$) detectors are depicted in Fig.~\ref{fig03:sec4.5.1_PDISP}(a), where it emerges that dark counts have a drastic effect for large energies, making the error probability saturating. 
The step-like behaviour of the curves follows from the adopted discrimination strategy: for $\alpha^2 \ll 1$, according to~(\ref{eq: nTH DC}), the optimized discrimination threshold is equal to $\nth(\nu) = 1$, equivalent to on-off detection, whereas, for increasing $\alpha^2$, $\nth(\nu)$ jumps to higher integer values up to $\nth = M$
in the regime $\alpha^2 \gg 1$. In turn, at every change in the threshold, the corresponding error probability exhibit a cusp.
Moreover, when $\nth(\nu)=M$, the sole outcome $M$ will infer state ``1" and all other outcomes smaller than $M$ will infer state ``0". In such a situation the receiver makes the wrong decision only if a $M$ outcome were actually induced by the state $|\alpha_0\rangle$. Then, the error probability for large $\alpha^2$ should be:
\begin{align}\label{eq:saturationDC}
 \PD(\nu) \approx \frac{p_M(\nu)}{2} = \frac12 \left[1- e^{-\nu} \sum_{j=0}^{M-1} \frac{\nu^{j}}{j!}\right] \, ,
\end{align}
being independent of the pulse energy $\alpha^2$ and making $\PD(\nu)$ saturate.

Finally, we note that, in the presence of dark counts, neither the DPNR is near optimum, since it outperforms the SQL only for particular values of the signal energy. To highlight this, we consider the gain
\begin{align}
{\cal G}_{\rm D}(\nu)= 1 - \frac{\PD(\nu)}{\PSQL} \, ,
\end{align}
plotted in Fig.~\ref{fig03:sec4.5.1_PDISP}(b). As expected, the gain is not monotonic with $\alpha^2$, but it exhibits $M$ jumps before decreasing monotonously. As we can see, the DPNR receiver outperforms the SQL in the low-energy limit and only in particular intervals of $\alpha^2$.


\begin{table}[b]
\tbl{Decision strategy for the HYNORE in the presence of a nonzero dark count rate $\nu$.\label{tab:02-HYNORE-DC}}
{\begin{tabular}{@{}ccc@{}} \toprule
outcomes &  & decision \\  \colrule
    $\Delta \geq 0$ $\quad$ $n< n_{\rm th}(\nu)$ \ & & ``0" \\ 
    $\Delta < 0$ $\quad$ $n\geq n_{\rm th}(\nu)$ \ & & ``0" \\ 
    $\Delta < 0$ $\quad$ $n<n_{\rm th}(\nu)$ \ & & ``1" \\ 
    $\Delta \geq 0$ $\quad$ $n\geq n_{\rm th}(\nu)$  \ & & ``1" \\ \botrule
\end{tabular}}
\end{table}

\subparagraph{HYNORE.}
On the contrary, when considering the HYNORE setup, the presence of dark counts afflicts both PNR($M$) detection on the transmitted branch and HL detection on the reflected one. Indeed, the probability of obtaining the photocurrent difference $\Delta=-M, \ldots ,M$ now becomes:
\begin{align}\label{eq: pDelta with DC}
    \mathcal{S}_\Delta(\nu;\alpha^{(r)}_{k}) =
     \sum_{n,m=0}^{M} p_n\big(\mu_{+}(\alpha^{(r)}_k)+\nu\big) \
    p_m\big(\mu_{-}(\alpha^{(r)}_k) +\nu \big) \, \delta_{(n-m),\Delta} \, .
\end{align}
Given all the previous considerations, the decision rule for the HYNORE in presence of dark counts should be modified into that of Table \ref{tab:02-HYNORE-DC}, provided that the threshold count $\nth$ in Eq.~(\ref{eq: nTH DC}) is now computed by considering only the transmitted fraction of the energy $\tau \alpha^2$, namely $\nth(\nu)=\nth(\nu;\tau \alpha^2)$.

Ultimately, the error probability reads:
\begin{align}
\PHY(\nu) = \min_{\tau,z} \PHY(\tau,z; \nu) \, ,
\end{align}
with
\begin{align}\label{eq: Phyb with DC}
    \PHY(\tau,z; \nu) &= \frac12 \left[p(\Delta < 0; n<\nth(\nu) | 0) + p(\Delta \geq 0; n\geq \nth(\nu) | 0) \right] \nonumber \\[1ex]
&\hspace{0.5cm} + \frac12 \left[p(\Delta < 0, n\geq \nth(\nu) | 1) + p(\Delta \geq 0, n<\nth(\nu) | 1) \right] \nonumber \\[2ex]
 &= \frac12 \sum_{n=0}^{\nth(\nu)-1} p_n(4\tau \alpha^2+\nu) \left[  \sum_{\Delta=-M}^{-1} \mathcal{S}_{\Delta}(\nu;\alpha^{(r)}_0) + \sum_{\Delta=0}^{M} \mathcal{S}_{\Delta}(\nu;\alpha^{(r)}_1) \right] \nonumber\\[1ex]
     &\hspace{0.5cm} + \frac12 \sum_{n=\nth(\nu)}^{M} p_n(\nu) \left[  \sum_{\Delta=-M}^{-1} \mathcal{S}_{\Delta}(\nu;\alpha^{(r)}_1) + \sum_{\Delta=0}^{M} \mathcal{S}_{\Delta}(\nu;\alpha^{(r)}_0) \right] \, .
\end{align}

\begin{figure}
\includegraphics[width=0.49\columnwidth]{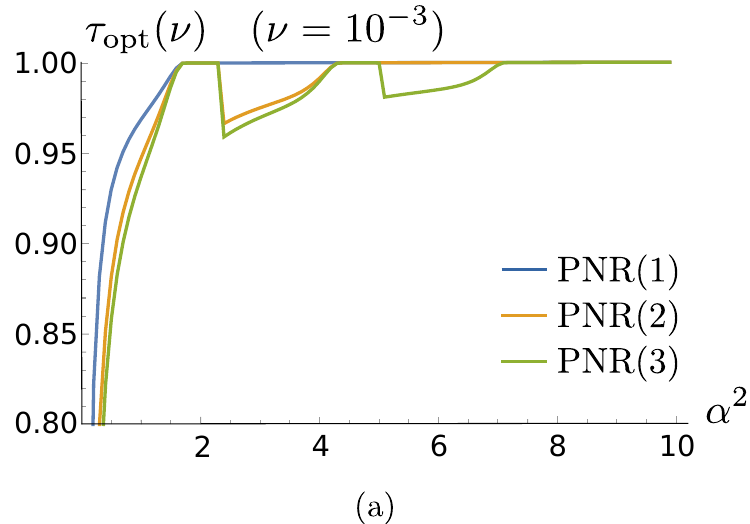} 
\includegraphics[width=0.49\columnwidth]{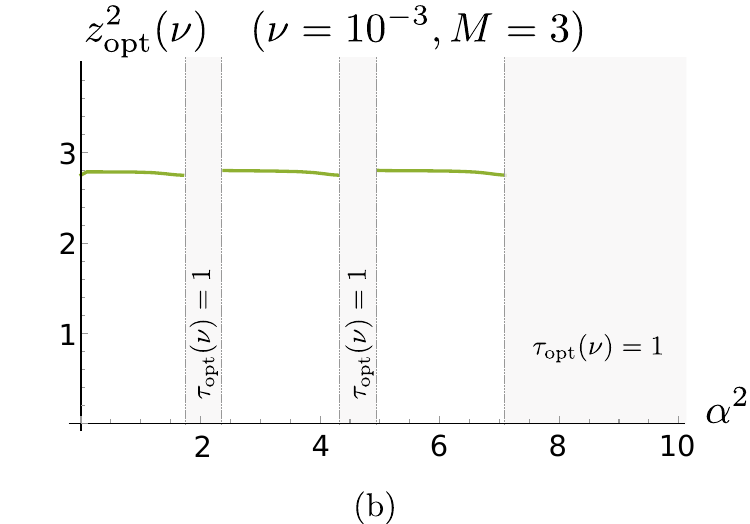}
\centering
\caption{(a) Plot of the optimized transmissivity $\tau_\opt(\nu)$ as a function of $\alpha^2$ for different PNR resolution $M$. (b) Plot of the optimized LO intensity $z^2_\opt(\nu)$ as a function of $\alpha^2$ for $M=3$. In the shaded regions we have $\tau_\opt(\nu)=1$ and the HYNORE performs as a DPNR receiver.}\label{fig04:sec4.5.1_DCoptpar}
\end{figure}

Plots of the optimized transmissivity $\tau_\opt(\nu)$ and LO intensity $z^2_\opt(\nu)$ are reported in Fig.~\ref{fig04:sec4.5.1_DCoptpar}(a) and (b), respectively.
We see that, differently from the ideal scenario described in Sec.~\ref{subsec4:HYNORE}, the optimized transmissivity $\tau_\opt(\nu)$ is not anymore a monotonous function of $\alpha^2$ asymptotically reaching $1$. On the contrary, for nonzero dark count rate, at first the value of $\tau_\opt(\nu)$ increases with $\alpha^2$ until to reach exactly the value $1$, i.e. performing as a DPNR receiver. For larger energies, according to the resolution $M$, there appears $M-1$ ``sawteeth'', that is other $M-1$ regions in which $\tau_\opt(\nu)$ decreases to a value smaller than $1$ and increases further to reach again $1$. Finally, in the high-energy limit $\alpha^2 \gg 1$, we have $\tau_\opt(\nu) \equiv 1$ and the HYNORE leads to the same performance of the DPNR, namely saturation to the value~(\ref{eq:saturationDC}).
Accordingly, when $\tau_\opt(\nu) < 1$, the optimized LO is $z^2_\opt(\nu) \lesssim M$.

\begin{figure}
\includegraphics[width=0.49\columnwidth]{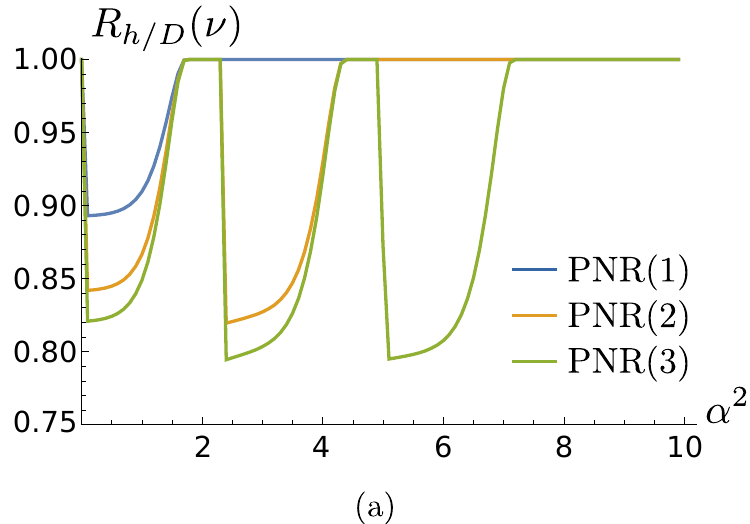} 
\includegraphics[width=0.49\columnwidth]{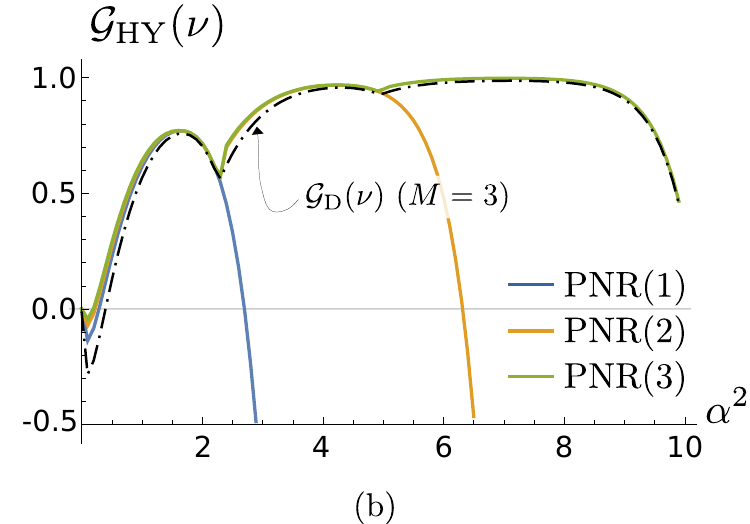}
\centering
\caption{(a) Plot of the ratio $R_{h/D}(\nu)$ as a function of $\alpha^2$ for different $M$. (b) Plot of the gain ${\cal G}_{\rm HY}(\nu)$ as a function of $\alpha^2$ for different $M$. The dot-dashed line refers to the DPNR gain ${\cal G}_{\rm D}(\nu)$ for PNR($3$) detection.}\label{fig05:sec4.5.1_RatioDC}
\end{figure}

In turn, the HYNORE outperforms the DPNR receiver only in some energy regimes. For a better visualization of the advantages brought by the hybrid receiver, we consider the relative ratio
\begin{align}
    R_{h/D}(\nu) = \frac{\PHY(\nu)}{\PD(\nu)} \, ,
\end{align}
and the gain
\begin{align}
    {\cal G}_{\rm HY}(\nu) = 1-\frac{\PHY(\nu)}{\PSQL}\, ,
\end{align}
depicted in Fig.~\ref{fig05:sec4.5.1_RatioDC}(a) and (b), respectively.
Consistently with the previous discussion, $R_{h/D}(\nu)$ is not a monotonous function of $\alpha^2$ and exhibits $M$ sawteeth when $\tau_\opt(\nu) <1$, in which case the corresponding gain is ${\cal G}_{\rm HY}(\nu) > {\cal G}_{\rm D}(\nu)$. In particular, the advantage over the DPNR is increased for larger PNR resolution $M$. This happens because of the HL part of the setup, retrieving more information on the reflected pulse when increasing $M$. Instead, when $\tau_\opt(\nu) =1$ the HYNORE performs as a DPNR and ${\cal G}_{\rm HY}(\nu) = {\cal G}_{\rm D}(\nu)$. Furthermore, the saturation of the error probabilities forbids to beat the SQL in the high-energy regime. Indeed, both the gains ${\cal G}_\p(\nu)$, $\p= {\rm D,HY}$, are positive up to a maximum energy $\alpha^2_\p(\nu)$, coinciding for both DPNR and HYNORE.

\paragraph{Visibility reduction.}

Finally, we address the effects of the interference visibility of the displacement operations employed in the realization of the receivers. This effect is consequence of the mode mismatch at the beam splitter which implements practically a displacement. We introduce the value $\xi \le1$ to quantify the overlap between the spatial areas of the signal and the auxiliary field mixed at the beam splitter. As a consequence, interference is only achieved between the field fractions being effectively superimposed, while the remaining parts do not interact with each other, resulting in an imperfect realization of mode-mixing operations \cite{Gupta2020}. A detailed model of the visibility reduction process is derived in App.~\ref{app:VisibilityModel}. 
In realistic conditions, the values of $\xi$ ranges from $0.90$ to $0.999$, according to the accuracy of the experimental setup \cite{DiMario2018, DiMario2019, Lau2006, Gupta2020}.
As discussed in \cite{DiMario2018, DiMario2019}, a reduction of the visibility affects crucially the performances of quantum receivers.

Generally speaking, we consider a coherent state $|\zeta\rangle$ which we want to displace by a quantity $\beta$ into the state $|\zeta+\beta\rangle$. For the sake of simplicity, we assume $\zeta,\beta \in \mathbb{R}$. Then we can describe the effect induced by imperfect mode matching by stating that the outcome $n$ of the subsequent PNR measurement follows a Poisson distribution with rate 
\begin{align}\label{eq:MuVis}
\mu= \zeta^2+\beta^2 + 2\xi \zeta \beta \neq (\zeta+\beta)^2 \, .
\end{align}
As for the case of dark counts, non-unit visibility is detrimental for the Kennnedy receiver, making the DPNR as the more adequate solution to implement a displacement receiver.
As in the previous subsection, in the following we first analyze the case of DPNR receiver and then address the HYNORE.

\begin{figure}
\includegraphics[width=0.6\columnwidth]{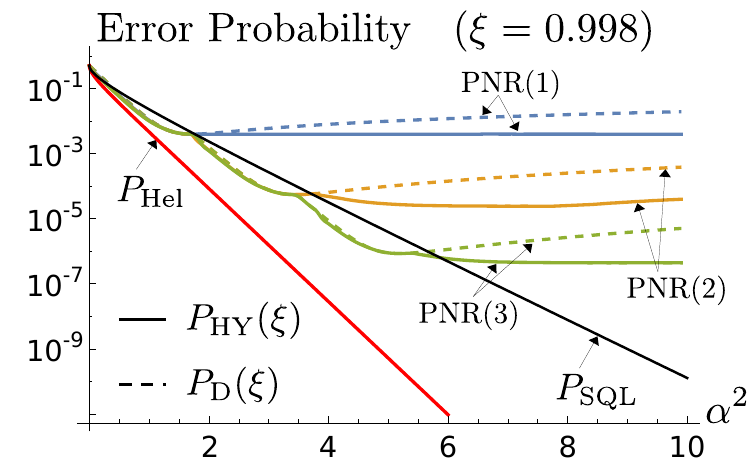}
\centering
\caption{Log plot of $\PHY(\xi)$ and $\PK(\xi)$ as a function of the signal energy $\alpha^2$ for different PNR resolution $M$ and $\xi=0.998$. $\PSQL$ and $\PHel$ refer to the SQL~(\ref{eq:BinarySQL}) and the Helstrom bound~(\ref{eq:BinaryHB}), respectively.}\label{fig06:sec4.5.1_VISRED}
\end{figure}

\subparagraph{D-PNRM receiver.} In the presence of a visibility reduction the approach is quite similar to the dark count case. Given Eq.~(\ref{eq:MuVis}), if $|\alpha_0 \rangle$ is sent the probability of detecting outcome $n$ is $p_n(2\alpha^2(1-\xi))$, whereas for $|\alpha_1 \rangle$ the probability becomes $p_n(2\alpha^2(1+\xi))$. By following the MAP criterion, the error probability then reads
\begin{align}\label{eq:PDVis}
    \PD(\xi)= 1 - \frac12 \sum_{n=0}^{M} \ \max\left[p_n(g_{-}), \, p_n(g_{+}) \right] \ ,
\end{align}
where 
\begin{align}\label{eq:gpm}
	g_{\pm} = 2\alpha^2(1\pm\xi) \, ,
\end{align}
associated to the threshold outcome $\nth(\xi)= \nth(\xi;\alpha^2)$:
\begin{align}
    \nth(\xi) = \min \left[
    \left\lceil \frac{4 \xi \alpha^2}{\displaystyle \ln \left(1+\xi\right) - \ln \left(1-\xi\right)} \right\rceil, \, M \right] \, .
\end{align}
We recall that the case of PNR($1$) is equivalent to the on-off Kennedy receiver. The consequences of a non-unit visibility on the error probability is shown in Fig.~\ref{fig06:sec4.5.1_VISRED}. As for dark counts, the visibility reduction makes the error probability non monotonic, and in particular increasing for large $\alpha^2$. As before, this is a consequence of the finite resolution $M$. In the regime of large $\alpha^2$ the threshold outcome becomes $\nth(\xi)=M$, thus the error probability is due to outcomes $M$ induced by the state $|\alpha_0\rangle$ which is not perfectly ``nulled" due to the imperfect displacement operation. Therefore we have:
\begin{align}
    \PD(\xi) \approx \frac{p_M(g_{-})}{2}= \frac12 \left[1- e^{-2\alpha^2(1-\xi)} \sum_{j=0}^{M-1} \frac{\big(2\alpha^2(1-\xi)\big)^{j}}{j!}\right] \, ,
\end{align}
which is an increasing function of $\alpha^2$.


\begin{table}[t]
\tbl{Decision strategy for the HYNORE in the presence of visibility reduction $\xi \le 1$.\label{tab:03-HYNORE-Vis}}
{\begin{tabular}{@{}ccc@{}} \toprule
outcomes &  &decision \\  \colrule
    $\Delta \geq 0$ $\quad$ $n< \nth(\xi)$ \ & & ``0" \\ 
    $\Delta < 0$ $\quad$ $n\geq \nth(\xi)$ \ & & ``0" \\ 
    $\Delta < 0$ $\quad$ $n<\nth(\xi)$ \ & & ``1" \\ 
    $\Delta \geq 0$ $\quad$ $n\geq \nth(\xi)$  \ & & ``1" \\  \botrule
\end{tabular}}
\end{table}

\subparagraph{HYNORE.} In the HYNORE, we should also include the effect of visibility reduction in the balanced beam splitter inside the HL detector. As a consequence, the probability of measuring the photocurrent $\Delta=-M,\ldots,M$ is changed into:
\begin{align}\label{eq: pDelta with Vis}
    \mathcal{S}_\Delta(\xi;\alpha^{(r)}_{k}) =
     \sum_{n,m=0}^{M} p_n\big(\mu_{+}(\alpha^{(r)}_k;\xi)\big) \,
    p_m\big(\mu_{-}(\alpha^{(r)}_k;\xi)\big) \, \delta_{(n-m),\Delta} \, ,
\end{align}
where
\begin{align}\label{eq: mu with Vis}
    \mu_{\pm}(\alpha^{(r)}_{k};\xi)&= \frac{\big(\alpha^{(r)}_{k}\big)^2 + z^2 \pm 2 \xi \, z \,  \alpha^{(r)}_{k}}{2}\, .
\end{align}

The decision rule for the HYNORE, displayed in Table~\ref{tab:03-HYNORE-Vis}, is identical to the case of dark counts, with the threshold $\nth(\xi)= \nth(\xi; \tau \alpha^2)$.
The error probability then reads:
\begin{align}
\PHY(\xi) = \min_{\tau,z} \PHY(\tau,z; \xi) \, ,
\end{align}
where
\begin{align}\label{eq: Phyb with DC}
    \PHY(\tau,z; \xi) &= \frac12 \left[p(\Delta < 0; n<\nth(\xi) | 0) + p(\Delta \geq 0; n\geq \nth(\xi) | 0) \right] \nonumber \\[1ex]
&\hspace{0.5cm} + \frac12 \left[p(\Delta < 0, n\geq \nth(\xi) | 1) + p(\Delta \geq 0, n<\nth(\xi) | 1) \right] \nonumber \\[2ex]
&=\frac12 \sum_{n=0}^{\nth(\xi)-1} p_n(\tau g_{+}) \left[  \sum_{\Delta=-M}^{-1} \mathcal{S}_{\Delta}(\xi;\alpha^{(r)}_0) + \sum_{\Delta=0}^{M} \mathcal{S}_{\Delta}(\xi;\alpha^{(r)}_1) \right] \nonumber \\
     &\hspace{1cm}   + \frac12 \sum_{n=\nth(\xi)}^{M} p_n(\tau g_{-}) \left[  \sum_{\Delta=-M}^{-1} \mathcal{S}_{\Delta}(\xi;\alpha^{(r)}_1) + \sum_{\Delta=0}^{M} \mathcal{S}_{\Delta}(\xi;\alpha^{(r)}_0) \right] \, ,
\end{align}
with the $g_{\pm}$ in Eq.~(\ref{eq:gpm}).

\begin{figure}
\includegraphics[width=0.49\columnwidth]{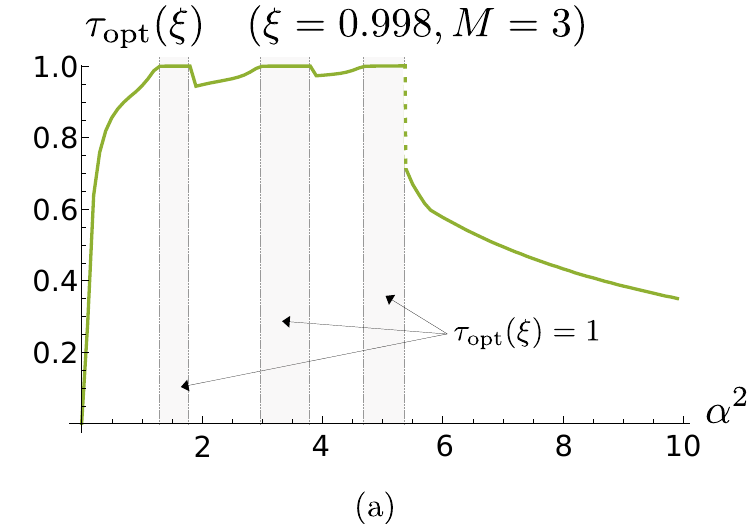} 
\includegraphics[width=0.49\columnwidth]{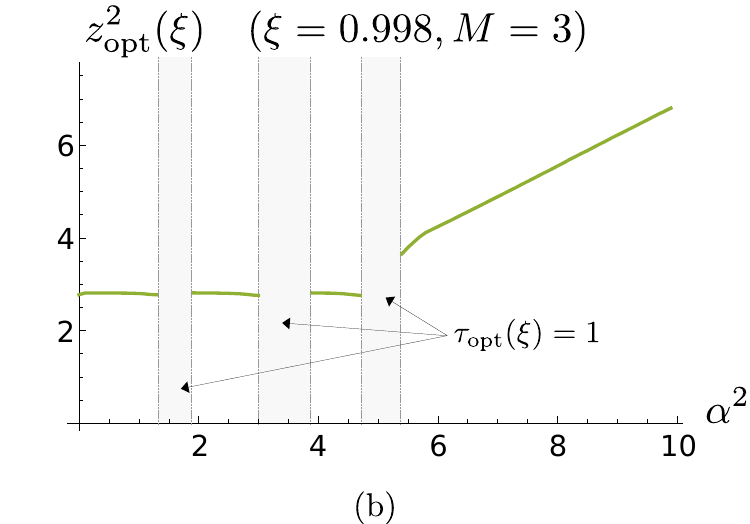}
\centering
\caption{Plot of the optimized transmissivity $\tau_\opt(\xi)$ (a) and LO intensity $z^2_\opt (\xi)$ (b) as a function of $\alpha^2$ for $\xi=0.998$. The PNR resolution is $M=3$. In the shaded regions we have $\tau_\opt(\xi)=1$ and the HYNORE performs as a DPNR receiver.}\label{fig07:sec4.5.1_VISoptpar}
\end{figure}

Plots of $\PHY(\xi)$ are reported in Fig.~\ref{fig06:sec4.5.1_VISRED}. If $\alpha^2$ is small we observe the same step-like behaviour of the dark count case, and the HYNORE beats the DPNR only for particular values of the signal energy. On the contrary, for large $\alpha^2$ the HYNORE significantly outperforms the DPNR, as $\PHY(\xi) < \PD(\xi)$. The difference between the two regimes becomes clearer by looking at the optimized transmissivity $\tau_\opt(\xi)$ and LO  $z^2_\opt(\xi)$, depicted in Fig.~\ref{fig07:sec4.5.1_VISoptpar}(a) and (b), respectively.
In the low-energy regime, similarly to Fig.~\ref{fig04:sec4.5.1_DCoptpar}, $\tau_\opt(\xi)$ is a non-monotonous function of $\alpha^2$, exhibiting $M-1$ sawteeth, whilst the optimized LO is $z^2_\opt(\xi) \lesssim M$. On the other hand, for large $\alpha^2$ the transmissivity changes discontinuously, and becomes a decreasing function of $\alpha^2$, saturating for $\alpha^2 \gg 1$ to an asymptotic value $\tau_\infty \neq 0$. Remarkably, $\tau_\infty <1$, thus by appropriately choosing the energy of the signals undergoing the HL and the DPNR measurements it is possible to regain part of the information lost by to the finite resolution of the detectors. As a result, the interplay between the two schemes allows to mitigate the negative effects introduced by the visibility reduction.
In these conditions, $z^2_\opt(\xi)$ is not constant anymore, and increases with $\alpha^2$, being a linear function for $\alpha^2$.

\begin{figure}
\includegraphics[width=0.49\columnwidth]{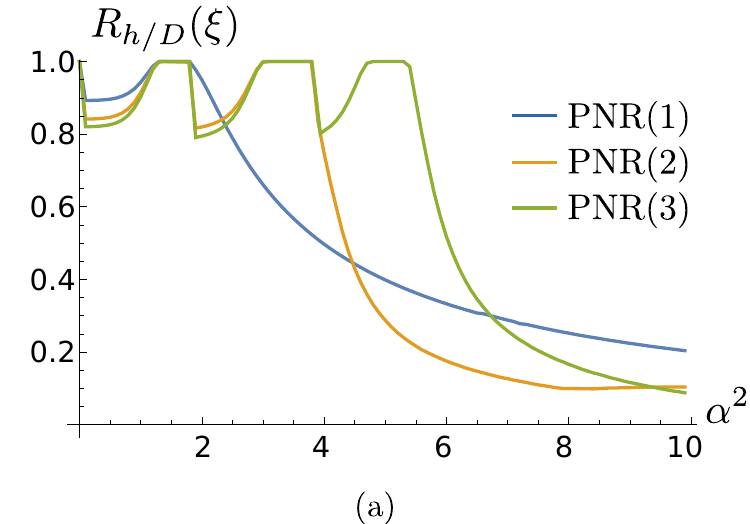} 
\includegraphics[width=0.49\columnwidth]{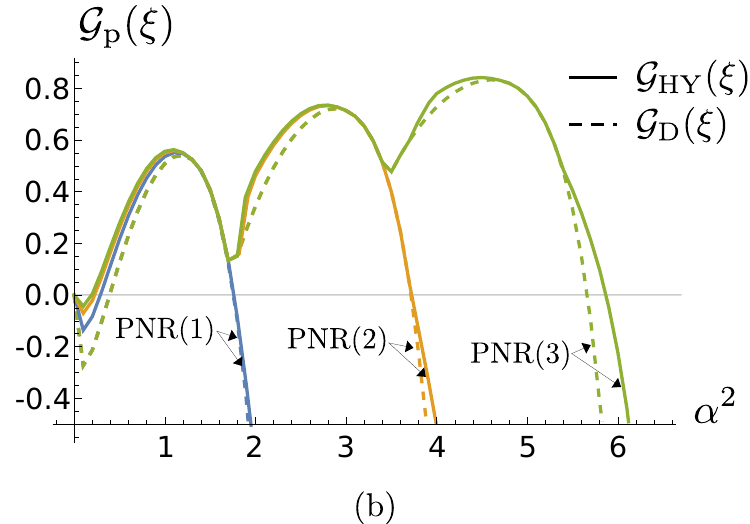}
\centering
\caption{(a) Plot of the ratio $R_{h/D}(\xi)$ as a function of $\alpha^2$ for different $M$. (b) Plot of the gain ${\cal G}_{\p}(\xi)$, $\p= {\rm D,HY}$, as a function of $\alpha^2$ for different $M$. We set the value $\xi=0.998$.}\label{fig08:sec4.5.1_RatioVIS}
\end{figure}

The existence of two different energy regimes affects also the relative ratio
\begin{align}
    R_{h/D}(\xi) = \frac{\PHY(\xi)}{\PD(\xi)} \, ,
\end{align}
shown in Fig.~\ref{fig08:sec4.5.1_RatioVIS}(a). In fact, in the low-energy regime, $R_{h/D}(\xi)$ exhibits $M$ sawteeth, whilst, after the jump in the transmissivity $\tau_\opt(\xi)$, becomes a decreasing function of $\alpha^2$.
Finally, to quantify the quantum advantage over the SQL, we consider the gain
\begin{align}
    {\cal G}_\p(\xi) = 1-\frac{P_\p(\xi)}{\PSQL} \, , \quad \p={\rm D, HY} \, ,
\end{align}
plotted in Fig.~\ref{fig08:sec4.5.1_RatioVIS}(b) for different PNR resolution. As for dark counts, both DPNR and HYNORE beat the SQL only in particular energy regimes. Even in this case, the gains are positive up to a maximum energy $\alpha^2_\p(\xi)$, which, now, is different between the two receivers, as $\alpha^2_{\rm HY}(\xi) \ge \alpha^2_{\rm D}(\xi)$.

\subsubsection{HFFRE vs DFFRE}

After widely discussing the robustness of single-copy receivers, we now extend the comparison to multi-copy receivers, namely DFFRE and HFFRE. In both cases, splitting the encoded signal into many rescaled copies makes the decision errors induced by practical imperfections accumulate during the feed-forward loop, leading to a non-trivial behaviour.
For the sake of simplicity, in the following we will perform the analysis by considering the sole HFFRE. In fact, as discussed in Sec.~\ref{subsec4:HFFRE}, the error probability associated with the DFFRE may be retrieved in an analogous way from the DFFRE scheme in Fig.~\ref{fig:sec4.4_HFFREsetup} by setting $\tau=1$.

\paragraph{Reduced quantum efficiency.}
The presence of a quantum efficiency $\eta \le 1$ requires only to rescale the coherent amplitudes of all the measured pulses by a factor $\sqrt{\eta}$, as no mixedness is introduced at the detectors.
Thereafter, in the HFFRE scheme of Fig.~\ref{fig:sec4.4_HFFREsetup} the HL probability distribution of the reflected signal $|\alpha^{(r)}_k\rangle$ becomes ${\cal S}_{\Delta} (\eta \alpha^{(r)}_k)$
with the $\mu_{\pm}$ in Eq.~(\ref{eq:RatesMu}).
The effect is the same on the transmitted branch, where the average photon numbers of the displaced copies $\lambda_{\pm}$, see Eq.~(\ref{eq:lambdapm}), are replaced by $\eta \lambda_{\pm}$. In turn, the correct decision probability ${\cal P}_{\rm HF}^{(j)}(\eta;\tau,z)$ becomes
\begin{align}
&{\cal P}_{\rm HF}^{(j)}(\eta;\tau,z) =\notag\\[1ex]
&\hspace{0.5cm}
\max_{\beta_{j}} \Bigg\{ {\cal P}_{\rm HF}^{(j-1)}(\eta;\tau,z)\,
q_{\rm off}\Big( \eta \lambda_{-}^{(j)}(\sqrt{\tau}\alpha)\Big)\notag \\[1ex]
&\hspace{1cm}
+\bigg[1 - {\cal P}_{\rm HF}^{(j-1)}(\eta;\tau,z)\bigg] q_{\rm on}\big( \eta\lambda_{+}^{(j)}(\sqrt{\tau}\alpha)\big) \Bigg\} \, ,
\end{align}
with the quantities introduced in~(\ref{eq:q01}), to be solved with the initial condition
$${\cal P}_{\rm HF}^{(0)}(\eta;\tau,z)=
\frac12 \Bigg[ \sum_{\Delta=-M}^{-1} {\cal S}_\Delta\Big(\eta\alpha^{(r)}_1\Big)
+ \sum_{\Delta=0}^{M} {\cal S}_\Delta\Big(\eta\alpha^{(r)}_0\Big) \Bigg] \, , $$ 
and the associated error probability reads
\begin{align}\label{eq:Phyb_eta}
\PHFF^{(N)}(\eta) = 1- \max_{\tau,z}{\cal P}_{\rm HF}^{(N)}(\eta;\tau,z)  \, .
\end{align}

\begin{figure}[t]
\includegraphics[width=0.6\columnwidth]{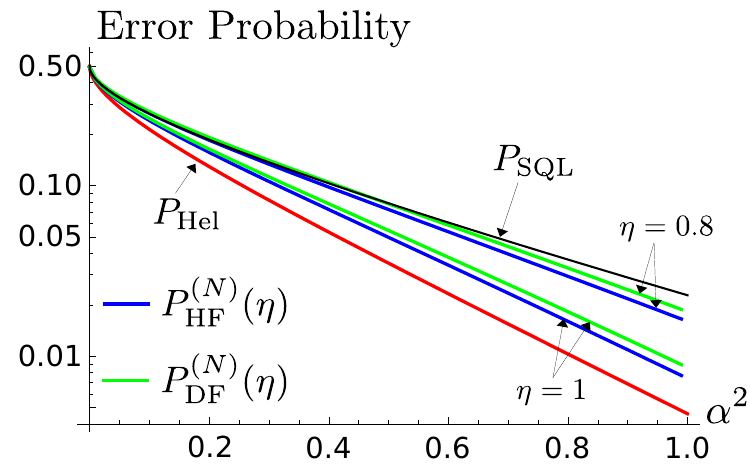}
\centering
\caption{Log plot of $\PHFF^{(N)}(\eta)$ and $\PDISP^{(N)}(\eta)$ as a function of the signal energy $\alpha^2$ for $N=1$ and different values of $\eta$. The PNR resolution is $M=2$.}\label{fig01:sec4.5.2_ErrorP_eta}
\end{figure}

\begin{figure}[t]
\includegraphics[width=0.49\columnwidth]{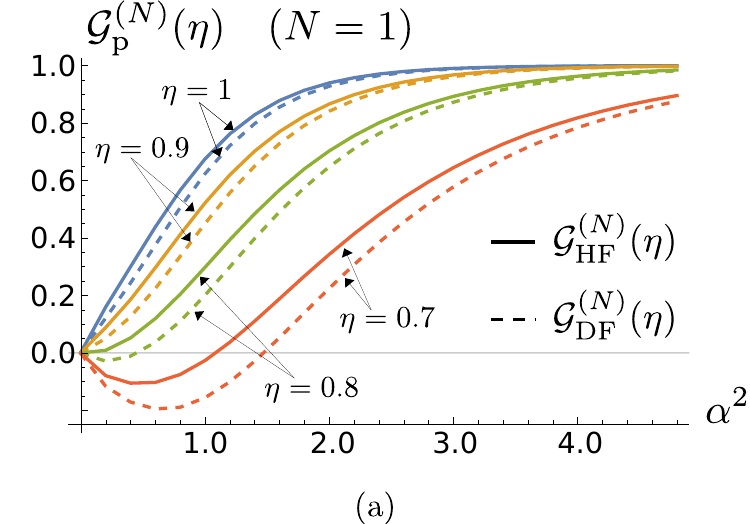} 
\includegraphics[width=0.49\columnwidth]{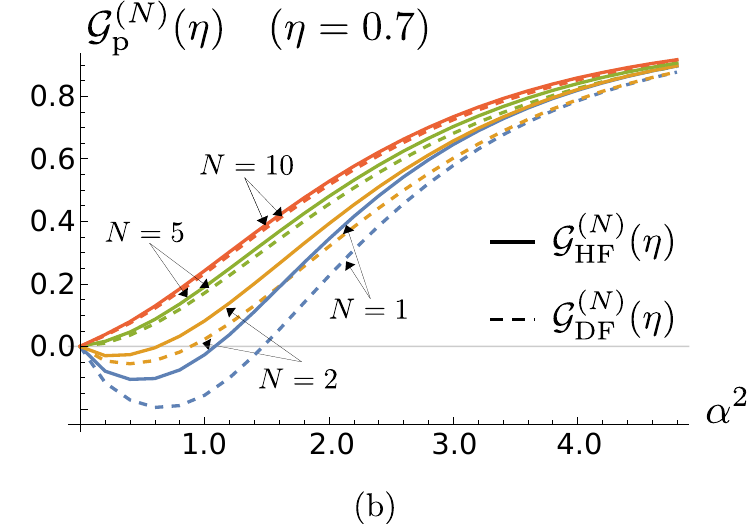}
\centering
\caption{(a) Plot of the gain ${\cal G}_{\p}^{(N)}(\eta) $, $\p={\rm DF,HF}$, as a function of the signal energy $\alpha^2$ for $N=1$ and different quantum efficiency $\eta$. (Bottom) Plot of the gain ${\cal G}_{\p}^{(N)}(\eta) $, $\p={\rm DF,HF}$, as a function of $\alpha^2$ for $\eta=0.7$ and different number of copies $N$. In both the pictures, the PNR resolution is $M=2$.}\label{fig02:sec4.5.2_RatioSQL_eta}
\end{figure}

The error probability for the DFFRE $\PDISP^{(N)}(\eta)$ may be derived from the previous equations by fixing $\tau=1$. Plots of $\PHFF^{(N)}(\eta)$ and $\PDISP^{(N)}(\eta)$ are depicted in Fig.~\ref{fig01:sec4.5.2_ErrorP_eta}, showing that the presence of a non-unit quantum efficiency increases the error probability, preventing the receivers to approach the Helstrom bound. Nevertheless, we still have $\PHFF^{(N)}(\eta) \le \PDISP^{(N)}(\eta)$ and, remarkably, in the high-energy regime both the discussed receivers beat the SQL~(\ref{eq:BinarySQL}). To better highlight this feature, we consider the gain 
\begin{align}
{\cal G}_{\p}^{(N)}(\eta) = 1-\frac{P_{\p}^{(N)}(\eta)}{\PSQL} \, , \quad (p={\rm DF, HF} \, ,
\end{align}
plotted in Fig.s~\ref{fig02:sec4.5.2_RatioSQL_eta}(a) and (b). Accordingly, the SQL is outperformed when ${\cal G}_{\p}^{(N)}(\eta)\ge 0$.

If we consider a fixed number of copies $N$, see Fig.~\ref{fig02:sec4.5.2_RatioSQL_eta}(a), there exists a threshold energy $\alpha^2_{\p}(N,\eta)$ after which the discussed receivers beat the SQL, that is ${\cal G}_{\p}^{(N)}(\eta)\ge 0$ for $\alpha^2 \ge \alpha^2_{\p}(N,\eta)$. By reducing the quantum efficiency $\eta$, the gain and the threshold energy decrease and increase, respectively. More interestingly, in the opposite scenario where we fix $\eta$ and let $N$ vary, as in Fig.~\ref{fig02:sec4.5.2_RatioSQL_eta}(b), we see that increasing the number of copies mitigates the detriments of the quantum efficiency, and makes the gain increase. In particular, for a sufficiently large $N$, $\alpha^2_{\p}(N,\eta)$ may be made arbitrarily small, maintaining ${\cal G}_{\p}^{(N)}(\eta)\ge 1$ for all energies. In all cases, the HFFRE outperforms the DFFRE, as ${\cal G}_{\rm HF}^{(N)}(\eta)\ge {\cal G}_{\rm DF}^{(N)}(\eta)$ and $\alpha^2_{\rm HF}(N,\eta) \le \alpha^2_{\rm DF}(N,\eta)$.

\paragraph{Dark counts.}
\begin{figure}
\includegraphics[width=0.49\columnwidth]{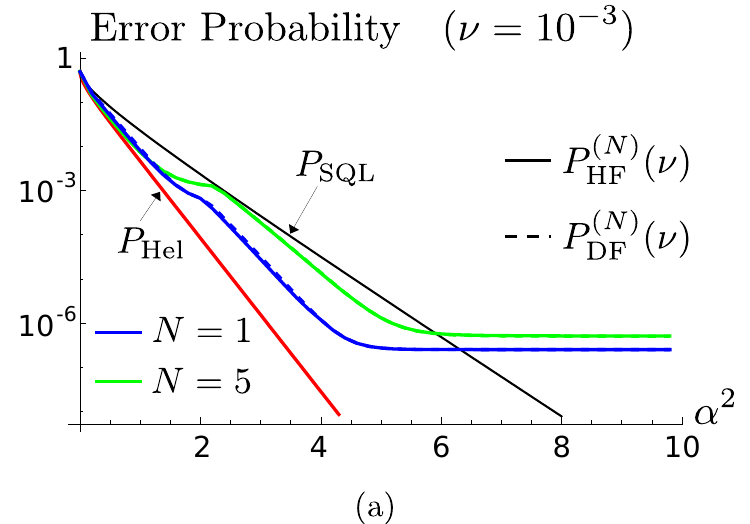}\,
\includegraphics[width=0.49\columnwidth]{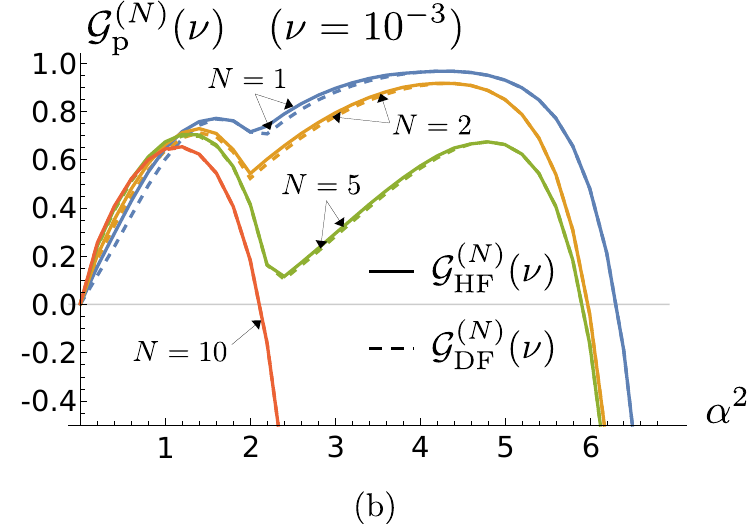}
\centering
\caption{(a) Log plot of $\PHFF^{(N)}(\nu)$ and $\PDISP^{(N)}(\nu)$ as a function of the signal energy $\alpha^2$ for different values of $N$. (b) Plot of the gain ${\cal G}_{\p}^{(N)}(\nu)$, $\p={\rm DF,HF}$, as a function of $\alpha^2$ for different $N$. In both the pictures, the PNR resolution is $M=2$ and the dark count rate is $\nu=10^{-3}$.}\label{fig03:sec4.5.2-DC}
\end{figure}

More drastic effects appear in the presence of dark counts. The HL probability distribution of the reflected signal $|\alpha^{(r)}_k\rangle$ is equal to~(\ref{eq: pDelta with DC}), as discussed in the former section.
A more detrimental effect is observed in the displacement-photon counting scheme performed on the transmitted signal. Indeed, in the presence of dark counts the MAP criterion does not coincide anymore with on-off discrimination and, in principle, one should perform a different Bayesian inference process after each detection stage. However, for the sake simplicity here we adopt a simpler decision rule. We introduce a fixed threshold outcome $1\le \nth\le M$ such that if we get outcome $n<\nth$ from the $(j-1)$-th PNR($M$) measurement we set $\sigma_j=\sigma_{j-1}$ and, then, displace the $j$-th copy by $D(\sigma_{j}\beta_j)$; otherwise if $n\ge \nth$ we choose $\sigma_j=-\sigma_{j-1}$. In the ideal scenario with zero dark count rate we have $\nth=1$. The final decision rule becomes: $n<\nth \rightarrow |-\sigma_{N} \alpha\rangle$ and $n\ge \nth\rightarrow |\sigma_{N} \alpha\rangle$.

As a consequence, the correct decision probability ${\cal P}_{\rm HF}^{(j)}(\nu;\tau,z)$ satisfies:
\begin{align}\label{eq:PcorrDC}
{\cal P}_{\rm HF}^{(j)}(\nu;\tau,z) &= \max_{\beta_{j}} \Bigg\{
{\cal P}_{\rm HF}^{(j-1)}(\nu;\tau,z)\,
\widetilde{q}_0\big( \lambda_{-}^{(j)}(\sqrt{\tau}\alpha;\nu); \nth\big)\notag\\[1ex]
&\hspace{.5cm}+
\bigg[ 1 - {\cal P}_{\rm HF}^{(j-1)}(\nu;\tau,z) \bigg]
\widetilde{q}_1\big(  \lambda_{+}^{(j)}(\sqrt{\tau}\alpha;\nu); \nth \big)
\Bigg\} \, ,
\end{align}
where
\begin{align}\label{eq:q01TH}
\widetilde{q}_0(x;\nth) &= \sum_{s=0}^{\nth-1} e^{-x} \frac{x^s}{s!} \, ,  \\[2ex]
\widetilde{q}_1(x;\nth)  &= 1-\widetilde{q}_0(x;\nth)  \, ,
\end{align}
and
\begin{align}
\lambda_{\pm}^{(j)}(\alpha;\nu) = \lambda_{\pm}^{(j)}(\alpha)+\nu \, .
\end{align}
The initial condition of Eq.~(\ref{eq:PcorrDC}) reads
$${\cal P}_{\rm HF}^{(0)}(\nu;\tau,z)=
\frac12 \Bigg[ \sum_{\Delta=-M}^{-1} {\cal S}_\Delta\Big(\nu;\alpha^{(r)}_1\Big)
+ \sum_{\Delta=0}^{M} {\cal S}_\Delta\Big(\nu;\alpha^{(r)}_0\Big) \Bigg] \, , $$ 
and the associated error probability is obtained as
\begin{align}\label{eq:Phyb_nu}
\PHFF^{(N)}(\nu) = 1- \max_{\tau,z, \nth}{\cal P}_{\rm HF}^{(N)}(\nu;\tau,z)  \, ,
\end{align}
where, differently from the other cases, we perform optimization also over the threshold discrimination outcome $\nth$. As before, with the choice $\tau=1$ we retrieve the probability $\PDISP^{(N)}(\nu)$ associated with the displacement receiver.

The plots of $\PHFF^{(N)}(\nu)$ and $\PDISP^{(N)}(\nu)$ are reported in 
Fig.~\ref{fig03:sec4.5.2-DC}(a) for different number of copies $N$ and $M=2$. 
As discussed in Sec.~\ref{subsec:HYNOREvsDPNR}, the step-like behaviour of the curves follows from the adopted discrimination strategy: for $\alpha^2 \ll 1$ the optimized discrimination threshold is equal to $\nth=1$, equivalent to on-off detection, whereas, for increasing $\alpha^2$, $\nth$ jumps to higher integer values up to $\nth=M$ in the regime $\alpha^2 \gg 1$. In turn, at every change in the threshold, the corresponding error probabilities exhibit a cusp.

Remarkably, in the presence of dark counts the performance of the receivers is not improving anymore with larger number of copies. In fact, increasing $N$ induces a reduction of the error probability only for $\alpha^2 \ll 1$. On the contrary, for large energies employing many copies becomes detrimental. Indeed, it has been shown in the previous section that dark counts induce decision errors, letting the error probability saturate for $\alpha^2 \gg 1$. Accordingly, when we split the signal into $N$ copies, the decision errors induced by dark counts accumulate, letting the error probability reach higher saturating values. 

To quantify the present effect, some analytical results may be retrieved in the limit $\alpha^2 \gg 1$.
For the DFFRE, numerical results show that, in the regime $\alpha^2 \gg 1$, the optimized displacement amplitudes are $\beta_j\approx \alpha/\sqrt{N}$ and $\nth=M$. Thus, we have $\lambda_{-}^{(j)}(\alpha;\nu)=\nu$ and $\lambda_{+}^{(j)}(\alpha;\nu)=\nu+ 4 \alpha^2/N \gg 1$. This implies that an error occurs only when the outcome $M$ is obtained from the input $|\alpha_0\rangle$, in turn the correct decision probability at the $j$-th step reads:
\begin{align}
{\cal P}_{\rm DF}^{(j)}(\nu) \approx {\cal P}_{\rm DF}^{(j-1)}(\nu) \, \widetilde{q}_0(\nu) + \left[1-{\cal P}_{\rm DF}^{(j-1)}(\nu) \right] \, ,
\end{align}
with $\widetilde{q}_0(\nu)=\widetilde{q}_0(\nu; M)$.
By iteration, we get:
\begin{align}
\PDISP^{(N)}(\nu) \approx 1- \left\{\frac{\left[\widetilde{q}_0(\nu)-1\right]^N}{2} +\frac{1-\left[\widetilde{q}_0(\nu)-1\right]^N}{1-\left[\widetilde{q}_0(\nu)-1\right]} \right\} \, ,
\end{align}
being independent of the energy $\alpha^2$ and, therefore, letting the error probability saturate.
The same result also holds for the HFFRE, since the optimized transmissivity  $\tau_\opt$ in the high-energy regime is equal to $\tau_\opt=1$.

Finally, we note that the benefits of the hybrid scheme are more relevant for $N \lesssim 5$. For larger number of copies the improvement becomes negligible: as we can see, for instance, in Fig.~\ref{fig03:sec4.5.2-DC}(a), the curves associated with the HFFRE and DFREE lines for $N=10$ are superimposed and fully indistinguishable.

The saturation of the error probability forbids to beat the SQL in the large energy regime. Indeed, the gain
\begin{align}
{\cal G}_{\p}^{(N)}(\nu) = 1-\frac{P_{\p}^{(N)}(\nu)}{\PSQL} \, , \quad \p={\rm DF,HF} \, ,
\end{align}
plotted in Fig.~\ref{fig03:sec4.5.2-DC}(b), is positive up to a maximum energy $\alpha^2_{\p}(N,\nu)$. Here the tradeoff between the number of copies and the error probability is clearer: for larger values of $N$ we increase ${\cal G}_{\p}^{(N)}(\nu)$ in the low-energy regime $\alpha^2 \ll 1$, at the expense of reducing also $\alpha^2_{\p}(N,\nu)$. If on the one hand we reduce the error probability for low energies, on the other one we inevitably reduce the range in which the receivers exhibit a quantum advantage.

\paragraph{Visibility reduction.}
\begin{figure}
\includegraphics[width=0.49\columnwidth]{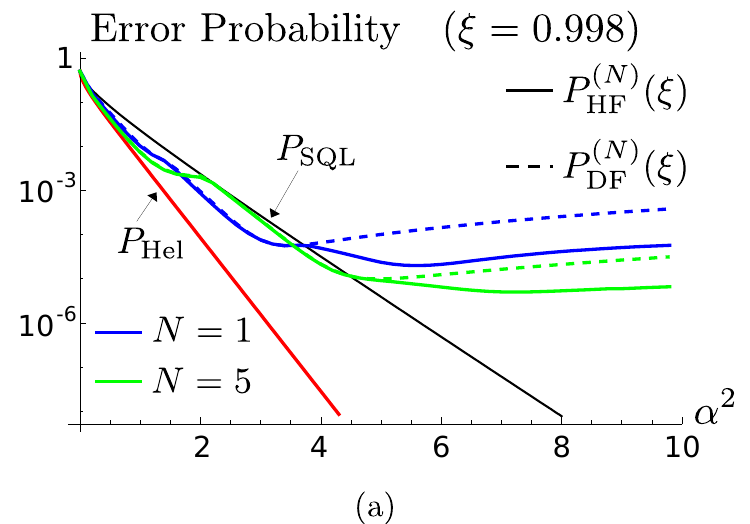} 
\includegraphics[width=0.49\columnwidth]{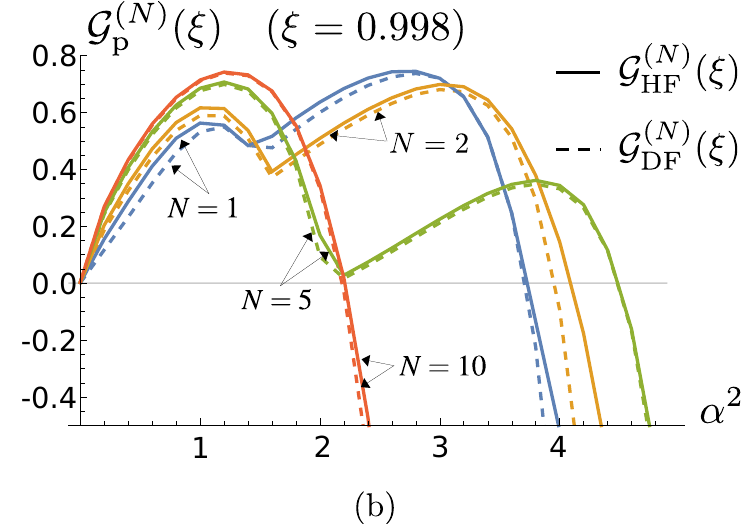}
\centering
\caption{(a) Log plot of $\PHFF^{(N)}(\xi)$ and $\PDISP^{(N)}(\xi)$ as a function of the signal energy $\alpha^2$ for different values of $N$. (b) Plot of the gain ${\cal G}_{\p}^{(N)}(\xi)$, $\p={\rm DF,HF}$, as a function of $\alpha^2$ for different $N$. In both the pictures, the PNR resolution is $M=2$ and the visibility is $\xi=0.998$.}\label{fig04:sec4.5.2-Vis}
\end{figure}

Finally, we address the impact of reduced visibility $\xi \le 1$. 
In the HFFRE we observe a visibility reduction both in the HL setup, where the signal is mixed with the LO $|z\rangle$, and in the conditional displacement operations governed by the feed-forward rule.
The HL probability distribution of the reflected signal $|\alpha^{(r)}_k\rangle$ is reported in Eq.~(\ref{eq: pDelta with Vis}), whereas for the feed-forward rule on the transmitted signal, we proceed as for the case of dark counts and introduce the threshold outcome $1\le\nth \le M$.

Accordingly, the correct decision probability ${\cal P}_{\rm HF}^{(j)}(\xi;\tau,z)$ satisfies:
\begin{align}\label{eq:PcorrVis}
{\cal P}_{\rm HF}^{(j)}(\xi;\tau,z) &=\max_{\beta_{j}} \Bigg\{ {\cal P}_{\rm HF}^{(j-1)}(\xi;\tau,z)\,
 \widetilde{q}_0\Big( \lambda_{-}^{(j)}(\sqrt{\tau}\alpha;\xi);\nth\Big)\notag\\
&\hspace{.5cm} + \bigg[ 1 -  {\cal P}_{\rm HF}^{(j-1)}(\xi;\tau,z) \bigg]
 \widetilde{q}_1\Big(  \lambda_{+}^{(j)}(\sqrt{\tau}\alpha;\xi); \nth \Big) 
 \Bigg\} \, ,
\end{align}
with the same $\widetilde{q}_{k}(x;n_{\rm th})$, $k=0,1$, introduced in Eq.~(\ref{eq:q01TH}), the rates
\begin{align}
\lambda_{\pm}^{(j)}(\alpha;\xi) = \frac{\alpha^2}{N} + \beta_j^2 \pm \frac{2 \xi \beta_j \alpha}{\sqrt{N}} \, ,
\end{align}
and the initial condition
$${\cal P}_{\rm HF}^{(0)}(\xi;\tau,z)=
\frac12 \Bigg[ \sum_{\Delta=-M}^{-1} {\cal S}_\Delta\Big(\xi;\alpha^{(r)}_1\Big)
+ \sum_{\Delta=0}^{M} {\cal S}_\Delta\Big(\xi;\alpha^{(r)}_0\Big) \Bigg] \, . $$ 
Finally, the error probability writes:
\begin{align}\label{eq:Phyb_xi}
\PHFF^{(N)}(\xi) = 1- \max_{\tau,z, \nth}{\cal P}_{\rm hyb}^{(N)}(\xi;\tau,z)  \, ,
\end{align}
whereas for $\tau=1$ we obtain the corresponding $\PDISP^{(N)}(\xi)$.

As depicted in Fig.~\ref{fig04:sec4.5.2-Vis}(a), the behaviour of  $\PHFF^{(N)}(\xi)$ and $\PDISP^{(N)}(\xi)$ is similar to the case of dark counts, with a step-like behaviour induced by the jump in the threshold $\nth$. Even in this case, increasing the number of copies $N$ reduces the error probability for low energies, $\alpha^2 \ll 1$, but, differently from the dark counts case, this reduction holds also in the high-energy regime $\alpha^2 \gg 1$. In fact, the detriments of the visibility reduction are more relevant for strong signals and, in turn, the error probability for $\alpha^2 \gg 1$ becomes an increasing function of the energy \cite{Notarnicola2023:HYNORE}. As a consequence, splitting the incoming signal into a larger number of copies $N$ reduces the energy of each displaced copy, thus partially mitigating the effects of the imperfect displacements. 

With a similar argument to the one adopted for dark counts, we can obtain the analytic expression for the error probability in the high-energy regime. For the DFFRE, we have:
\begin{align}
\PDISP^{(N)}(\xi) \approx 1- \left\{\frac{\left[\widetilde{q}_0(g)-1\right]^N}{2} +\frac{1-\left[\widetilde{q}_0(g)-1\right]^N}{1-\left[\widetilde{q}_0(g)-1\right]} \right\} \, ,
\end{align}
with $\widetilde{q}_0(g)= \widetilde{q}_0(g;M)$ and $g=2\alpha^2(1-\xi)/N$, being an increasing function of $\alpha^2$.
On the contrary, the HFFRE beats the DFFRE since the optimized transmissivity $\tau\opt$ for the HFFRE is $<1$, and combining HL and displacement results in a lower error probability.

Anyway, there still exist an intermediate region, comprised between the regimes $\alpha^2 \ll 1$ and $\alpha^2 \gg 1$, where increasing $N$ is not beneficial anymore. Moreover, we note in the high-energy regime the HFFRE outperforms significantly the DFFRE, because of the higher degree of robustness of HL with respect to visibility reduction \cite{Notarnicola2023:HYNORE}.

The existence of three different energy regimes affects also the gain with respect to the SQL,
\begin{align}
{\cal G}_{\p}^{(N)}(\xi) = 1-\frac{P_{\p}^{(N)}(\xi)}{\PSQL} \, , \quad \p={\rm DF,HF} \, ,
\end{align}
plotted in Fig.~\ref{fig04:sec4.5.2-Vis}(b). As for the case of dark counts, we have ${\cal G}_{\p}^{(N)}(\xi)  \ge 0$ up to a maximum energy $\alpha^2_{\p}(N,\xi)$, but the behaviour of $\alpha^2_{\p}(N,\xi)$ is not monotonic with the number of copies $N$. For $N\lesssim 5$, splitting the signal into more copies improves the robustness of the quantum advantage, letting $\alpha^2_{\p}(N,\xi)$ increase. On the contrary, for larger $N$ the error probabilities surpass the SQL already in the intermediate energy regime and, in turn, $\alpha^2_{\p}(N,\xi)$ decreases.

\subsection{Quantum receivers in the presence of phase noise}\label{sec4:PhNDiscr}

Throughout this Section, we performed a detailed analysis of binary quantum decision theory in the presence of coherent-state encoding, namely BPSK. Within this framework, the goal is to perform discrimination between two symbols $k=0,1$, encoded into two $\pi$ phase-shifted (pure) coherent states. As discussed, the Dolinar receiver provides the optimum POVM, even though its practical implementation remains a challenging task, whereas both displacement receivers, e.g. Kenendy and DFFRE, and hybrid receivers, namely HYNORE and HFFRE, provide feasible near-optimum schemes, beating the SQL and enhancing information transfer over attenuating (Gaussian) channels \cite{Notarnicola2023:KB}. 
Furthermore, they are robust against the typical practical inefficiencies and preserve the quantum advantage over the SQL in several regimes.

Beside this, the performance of quantum receivers in the presence of noisy non-Gaussian channels is another fundamental task towards the realistic implementation of quantum optical communications.
A paradigmatic example is provided by phase noise \cite{Banerjee2007, Ishii2011,Amiri2022,Bertaina2024}, which represents the most detrimental source of noise for phase-shift encoding, destroying the coherence and the purity of the employed coherent pulses \cite{Genoni2011,Cialdi2020}.

\begin{figure}
\centerline{\includegraphics[width=0.75\columnwidth]{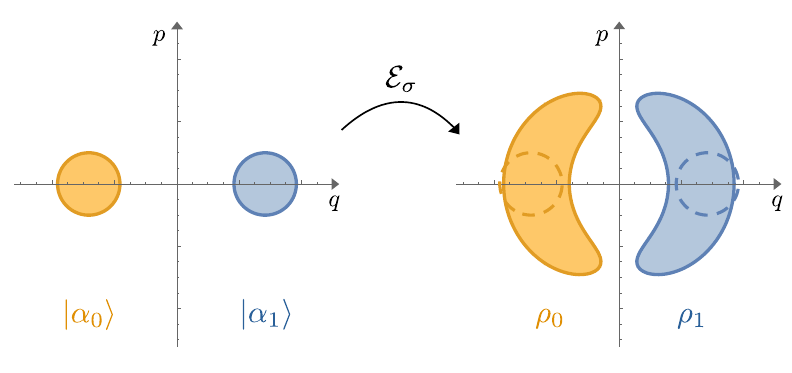}}
\centering
\caption{Phase space representation of the BPSK encoding before (left) and after (right) the application of phase diffusion. The phase diffusion CP map $\mathcal{E}_\sigma$ transforms the initial coherent signals into a Gaussian mixture of phase-shifted coherent states, reducing both purity and coherence.}\label{fig:01:sec4.6-PhS}
\end{figure}

In the presence of a phase diffusion channel, the problem of BPSK discrimination is remarkably different with respect to the scenario discussed in the previous sections. In fact, the encoded coherent states evolve according to a suitable master equation \cite{Genoni2011},
being equivalent to the completely positive (CP) map $\mathcal{E}_\sigma$, such that:
\begin{align}\label{eq:Map}
|\alpha_k\rangle \xrightarrow[]{\mathcal{E}_\sigma} \rho_k=\int_\mathbb{R}  d\phi \, g_\sigma(\phi) \, | \alpha_k e^{-i\phi} \rangle \langle \alpha_k e^{-i\phi} | \, , \qquad k=0,1 \, ,
\end{align}
where $g_\sigma(\phi)= \exp[-\phi^2/(2\sigma^2)]/\sqrt{2\pi \sigma^2}$ is a Gaussian distribution whose standard deviation $\sigma>0$ quantifies the amount of noise.
That is, the overall effect of phase diffusion is the application of a Gaussian-distributed random phase shift to the incoming signal, resulting in a overall non-Gaussian CP map. 

\begin{figure}
\includegraphics[width=0.6\columnwidth]{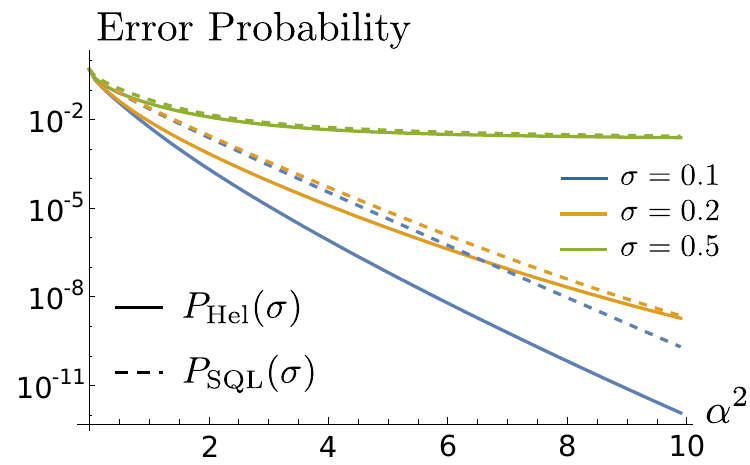}
\centering
\caption{Log plot of the Helstrom bound $\PHel(\sigma)$ and the SQL $\PSQL(\sigma)$  as a function of the signal energy $\alpha^2$ for different values of phase noise $\sigma$. In the limit of large noise, the homodyne receiver becomes near optimum and $\PSQL(\sigma) \approx \PHel(\sigma)$.}\label{fig02:sec4.6-HBandSQL}
\end{figure}

Given the previous considerations, in the presence of BPSK the effect of phase diffusion is detrimental: it reduces both the coherence and the purity of the encoded coherent states as emerges from Fig.~\ref{fig:01:sec4.6-PhS}, reporting the phase space representation of the quantum states before and after the noisy channel.
In turn, the quantum receiver has to discriminate between the two mixed phase-diffused states $\rho_0$ and $\rho_1$. The Helstrom bound becomes \cite{Helstrom1976, Bergou2010, Cariolaro2015, Olivares2013, Notarnicola2023:PhN}:
\begin{align}\label{eq:HELPN}
\PHel(\sigma)= \frac12 \left[1- \Tr \big( |\Lambda | \big) \right] \, ,
\end{align}
with
\begin{align}
\Lambda &=  \frac12 (\rho_0 -  \rho_1) \nonumber \\[1ex]
&= \frac{e^{-\alpha^2}}{2} \sum_{n,m=0}^{\infty} \frac{e^{- (n-m)^2 \sigma^2/2}}{\sqrt{n! m!}} \alpha^{n+m} \left[(-1)^{n-m} - 1 \right] \, |n\rangle \langle m| 
\, ,
\end{align}
expanded in the Fock basis $\{|n\rangle\}_n$.
Plots of $\PHel(\sigma)$ are reported in Fig.~\ref{fig02:sec4.6-HBandSQL} as a function of the mean signal energy $\alpha^2$ for different values of noise. As expected, the presence of the noise makes the error probability increase. In particular, for small $\alpha$ we may truncate $\Lambda$ at low dimension, achieving the analytic expression
\begin{align}
\PHel(\sigma) \approx \frac12 \left[1- \alpha \, e^{-\sigma^2/2} \right] \, , \qquad \mbox{for } \alpha \ll 1 \, ,
\end{align}
whereas in the limit of large noise, i.e. $\sigma \gg 1$, we have:
\begin{align}\label{eq:HELexp}
\PHel(\sigma) \approx \frac12 \left[1- f_{\rm Hel}(\alpha) e^{-\sigma^2/2} \right] \, , \qquad \mbox{for } \sigma \gg 1 \, ,
\end{align}
$f_{\rm Hel}(\alpha)$ being a decreasing function of $\alpha$ \cite{Olivares2013}.

On the contary, the SQL, obtained with homodyne detection, reads \cite{Olivares2013, Notarnicola2023:PhN}:
\begin{align}\label{eq:SQLPN}
\PSQL(\sigma) = \frac12 \left[ \int_0^\infty dx \, p_{\rm HD}^{(\sigma)} (x|0) + \int_{-\infty}^{0} dx \, p_{\rm HD}^{(\sigma)}(x|1)\right] \, ,
\end{align}
depicted in Fig.~\ref{fig02:sec4.6-HBandSQL}, where
\begin{align}
p_{\rm HD}^{(\sigma)}(x|k) = \int_{\mathbb{R}} d\phi \, g_\sigma(\phi) \, \frac{\exp\left[-(x-2\alpha_k \cos\phi)^2/2\right]}{\sqrt{2\pi}}
\end{align}
is the homodyne probability of obtaining outcome $x$ given the state $| \alpha_k \rangle$, expressed in shot-noise units.
For small values of the coherent amplitude we get the analytic expression 
\begin{align}
\PSQL(\sigma) \approx \frac12 \left[ 1- \alpha \sqrt{\frac{2}{\pi}} e^{-\sigma^2/2} \right] \, , \qquad \mbox{for } \alpha \ll 1 \, ,
\end{align}
whereas in the limit of large noise, i.e. $\sigma \gg 1$, we have:
\begin{align}\label{eq:SQLexp}
\PSQL(\sigma) \approx \frac12 \left[1- f_{\rm SQL}(\alpha) e^{-\sigma^2/2} \right] \, ,\qquad \mbox{for } \sigma \gg 1 \, ,
\end{align}
where $f_{\rm SQL}(\alpha)$ is a decreasing function of $\alpha$ such that $f_{\rm SQL}(\alpha) < f_{\rm Hel}(\alpha)$ which approaches $f_{\rm Hel}(\alpha)$ for increasing coherent amplitude $\alpha$ \cite{Olivares2013}.

As we can see from the plot, the homodyne receiver is quite robust with respect to the noise, as, for $\sigma \lesssim 0.2$ and low energy $\alpha^2$, the value of $\PSQL$ is almost constant. In contrast, when $\sigma>0$, in the high-energy limit $\alpha^2 \gg 1$ we have $\PSQL\approx \PHel$ and homodyne detection becomes near-optimum, as emerges from the asymptotic expansions~(\ref{eq:HELexp}) and~(\ref{eq:SQLexp}).
Furthermore, as $\sigma$ increases, the gap between $\PSQL$ and $\PHel$ is closed and, for $\sigma \gtrsim 0.5$, the two quantities almost coincide.

Given this scenario, a natural question arises, that is to quantify the performance of other quantum receivers, such as displacement and hybrid receivers, in the presence of phase diffusion, and assess whether or not their quantum advantage over the SQL is maintained \cite{Notarnicola2023:PhN}.
For the sake of simplicity, in the following we investigate only single-copy receivers, namely the displacement receiver and the HYNORE, leaving the analysis of multi-copy schemes as an open problem for future developments.
Moreover, we note that the presence of phase noise is detrimental for the Kennedy receiver \cite{Olivares2013}. Indeed, the Kennedy receiver is no longer near-optimum  and its performance is severely degraded for $\sigma>0$, since the displacement does not implement anymore a ``nulling" operation, as in the presence of dark counts and visibility reduction \cite{Olivares2013}. Following the same philosophy adopted in the former section, we take the DPNR receiver as a benchmark, whose performance is discussed in the following  \cite{DiMario2018, DiMario2019}.

\subsubsection{DPNR receiver in the presence of phase diffusion}\label{subsec:DPNR-PhN}

In the presence of phase noise, the displacement operation $D(\alpha)$ of the DPNR scheme maps states $\rho_k$ into $\OUTPUT_k= D(\alpha) \rho_k D(\alpha)^{\dagger}$, equal to:
\begin{align}\label{eq:OutStates}
\OUTPUT_k&= \int_\mathbb{R} d\phi \, g_\sigma(\phi) \, \left|\sqrt{\mu_k(\alpha^2,\phi)} \, e^{-i\phi/2}\right\rangle \left\langle \sqrt{\mu_k(\alpha^2,\phi)} \, e^{-i\phi/2} \right| \, ,
\end{align}
where
\begin{align}\label{eq:muk}
\mu_0(\alpha^2,\phi) = 4\alpha^2 \sin^2(\phi/2) \quad \mbox{and} \quad \mu_1(\alpha^2,\phi) = 4\alpha^2 \cos^2(\phi/2) \, .
\end{align}
Differently from the noiseless case, the nulling operation implemented by $D(\alpha)$ is not perfect and the output state $\OUTPUT_0$ still contains some photons.
Thereby, on-off detection is not the most appropriate strategy anymore and the DPNR setup is expected to outperform the Kennedy.

The PNR$(M)$ probability distribution of the displaced states $\OUTPUT_k$ reads:
\begin{align}
{\sf P}_\sigma(n|k) = \int_\mathbb{R} d\phi \, g_\sigma(\phi) \, p_n\left(\mu_k(\alpha^2,\phi) \right) \, , \quad (n=0,\ldots,M) \, ,
\end{align}
with the probability $p_n(\mu)$ and the count rates  $\mu_k(\alpha^2,\phi)$ defined in Eq.s~(\ref{eq:PNRMprob}) and~(\ref{eq:muk}), respectively.
The final decision is performed according to the maximum a posteriori probability (MAP) criterion:  given the outcome $n=0,\ldots,M$, we infer the state “0” or “1” associated with the maximum a posteriori probability \cite{DiMario2018, DiMario2019}. Again, this is equivalent to introducing a threshold $\nth= \nth(\alpha^2,\sigma)\le M$ such that all outcomes $n < \nth$ correspond to decision ``0", while outcomes $n \ge \nth$ infer state ``1" \cite{Notarnicola2023:HYNORE}. The threshold is obtained numerically by equating the photon number distributions of the two displaced phase-diffused states, namely ${\sf P}_\sigma(\bar{n}|0)={\sf P}_\sigma(\bar{n}|1)$, $\bar{n} \in {\mathbbm R}$, and considering the lowest integer greater than the obtained root $\bar{n}$, namely $\nth(\alpha^2,\sigma)=\ceil{\bar{n}}$.
In turn, the error probability of the DPNR receiver reads:
\begin{align}
\PD(\sigma)= \frac12 \left[ \sum_{n=0}^{\nth-1} {\sf P}_\sigma(n|1) + \sum_{n=\nth}^{M} {\sf P}_\sigma(n|0)\right] \, ,
\end{align}
and with PNR$(1)$ detection we retrieve the Kennedy receiver.

\begin{figure}
\includegraphics[width=0.49\columnwidth]{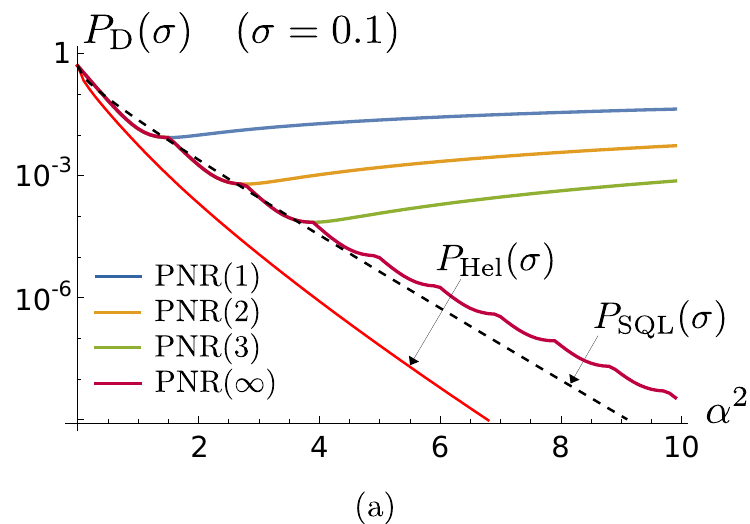} 
\includegraphics[width=0.49\columnwidth]{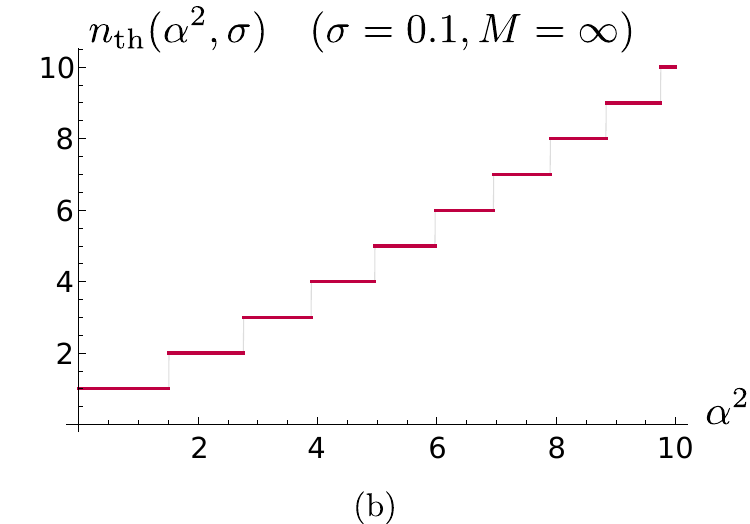}
\centering
\caption{(a) Error probability $\PD(\sigma)$ of the DPNR receiver as a function of the signal energy $\alpha^2$ for different photon number resolution $M$. The PNR$(1)$ case corresponds to the Kennedy receiver.  (b) Threshold count $\nth$ as a function of the signal energy $\alpha^2$ for PNR$(\infty$) detectors. In both the pictures we fix the noise value to $\sigma=0.1$.}\label{fig01:sec4.6.1-DPNR-vs-alpha}
\end{figure}

Plots of $\PD(\sigma)$ as a function of the signal energy $\alpha^2$ are reported in Fig.~\ref{fig01:sec4.6.1-DPNR-vs-alpha}(a) for the realistic noise value $\sigma=0.1$ \cite{Chin2021}.
The error probability is not a monotonic function of $\alpha^2$ and, as demonstrated in \cite{Olivares2013}, the Kennedy receiver is not near-optimum anymore in the presence of noise. 
The Kennedy is beaten by DPNR receivers with higher resolution $M$, whose corresponding error probabilities exhibit a step-like behaviour. This follows from the application of MAP criterion: as displayed in Fig.~\ref{fig01:sec4.6.1-DPNR-vs-alpha}(b), for $\alpha^2\ll 1$ the threshold decision count is equal to $\nth=1$, equivalent to on-off detection, whilst for larger $\alpha^2$ it jumps to higher integer values, until reaching $\nth=M$ in the high-energy limit $\alpha^2\gg 1$. 
Accordingly, the error probability has a cusp at every change in the value of $\nth$ and, once $\nth=M$, it becomes an increasing function of the energy.
In fact, in the high-energy limit a decision error occurs only when outcome $n=M$ is retrieved from state $\rho_0$, therefore the error probability is equal to \cite{Notarnicola2023:HYNORE, Notarnicola2023:FF}:
\begin{align}
\PD(\sigma) \approx \frac{{\sf P}_\sigma(M|0)}{2}= \frac12 \left[1-\sum_{j=0}^{M-1} \int_{\mathbb{R}} d\phi \, g_\sigma(\phi) \, e^{-\mu_0(\phi)}\frac{\mu_0(\phi)^j}{j!}\right] \, ,
\end{align}
being an increasing function of $\alpha^2$.
On the contrary, PNR$(\infty)$ detectors do not have a finite resolution, therefore $\nth(\sigma)$ can get arbitrary large values and the step-like behaviour is observed in every energy regime, as shown in Fig.~\ref{fig01:sec4.6.1-DPNR-vs-alpha}(a).

\begin{figure}[t]
\includegraphics[width=0.6\columnwidth]{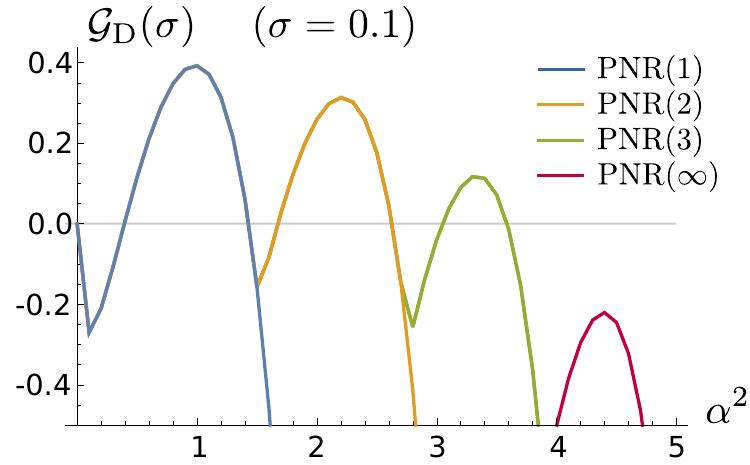}
\centering
\caption{Gain ${\cal G}_{\rm D}(\sigma)$ of the DPNR receiver with respect to the SQL as a function of the signal energy $\alpha^2$ for different photon number resolution $M$, when ${\cal G}_{\rm D}(\sigma) > 0$ we beat the SQL. The PNR$(1)$ case corresponds to the Kennedy receiver. We fix the noise value to $\sigma=0.1$.}\label{fig02:sec4.6.1-DPNR-Gain}
\end{figure}

Differently from the noiseless case, the SQL is beaten by DPNR receivers only in particular energy regimes. To highlight this, we consider the gain
\begin{align}
{\cal G}_{\rm D}(\sigma)= 1 - \frac{\PD(\sigma)}{\PSQL(\sigma)} \, ,
\end{align}
plotted in Fig.~\ref{fig02:sec4.6.1-DPNR-Gain}. In turn, we have a genuine quantum advantage over the SQL when ${\cal G}_{\rm D}(\sigma) > 0$. As expected, the gain ${\cal G}_{\rm D}(\sigma)$ is not monotonic with $\alpha^2$, but it exhibits $M$ jumps before decreasing monotonously. The DPNR receiver outperforms the SQL in the low-energy limit and only in particular intervals of $\alpha^2$. Remarkably, for a given noise $\sigma$ we obtain the maximal region of positive gain with PNR$(M)$ detectors having sufficiently small $M$, whereas increasing the resolution further is not necessary to enhance the violation of the SQL.

\begin{figure}
\includegraphics[width=0.49\columnwidth]{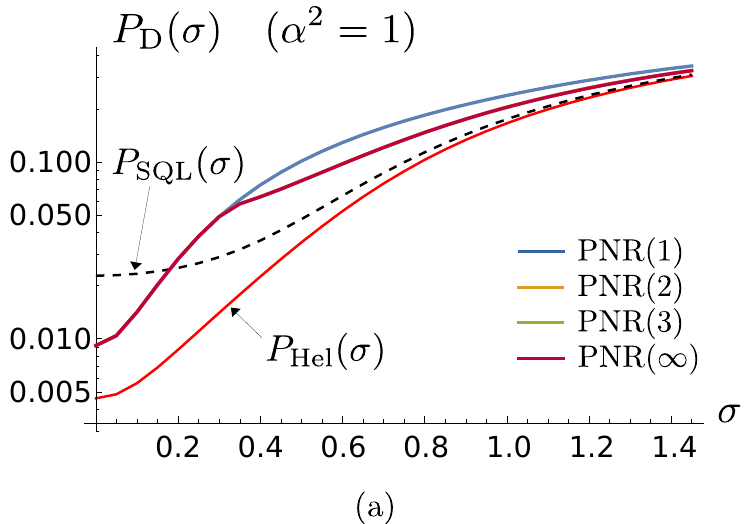} 
\includegraphics[width=0.49\columnwidth]{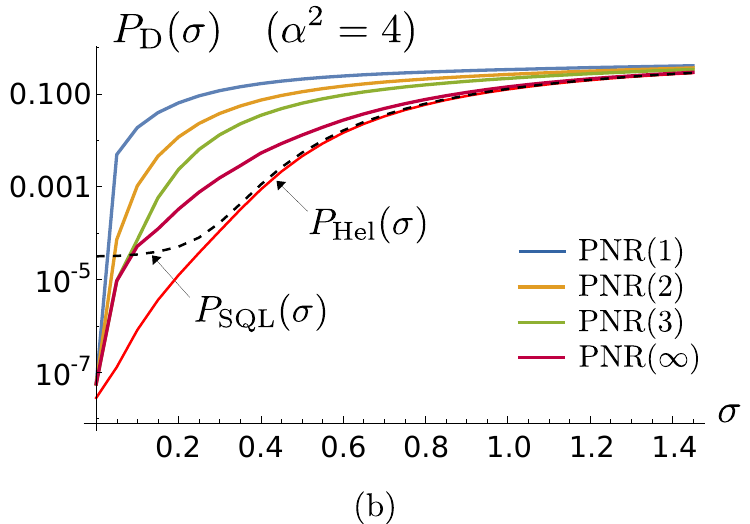}
\centering
\caption{Error probability $\PD(\sigma)$ of the DPNR receiver as a function of the noise $\sigma$ for $\alpha^2=1$ (a) and $\alpha^2=4$ (b). For $\alpha^2=1$, the curves of PNR$(M)$ detection with $M \ge 2$ are superimposed and, thus, indistinguishable. Given the energy $\alpha^2$, the DPNR receiver outperforms the SQL in the small-noise regime, whereas for large noise the SQL becomes near-optimum.}\label{fig03:sec4.6.1-DPNR-vs-sigma}
\end{figure}

In Fig.~\ref{fig03:sec4.6.1-DPNR-vs-sigma}(a) and (b) we report the error probability $\PD(\sigma)$ as a function of the noise $\sigma$ for low and high energy values $\alpha^2=1$ and $\alpha^2=4$, respectively. In both the cases DPNR receivers beat the SQL only for small noise, whilst in the large-noise limit the SQL becomes near optimum \cite{Olivares2013}. We also note that for $\alpha^2=1$ the performance of PNR$(M)$ detectors with $M\ge2$ is the same, since in this case the threshold count is equal to $\nth=2$. On the contrary, for $\alpha^2=4$ increasing the PNR resolution is beneficial to reduce the error probability.

\begin{figure}
\includegraphics[width=0.6\columnwidth]{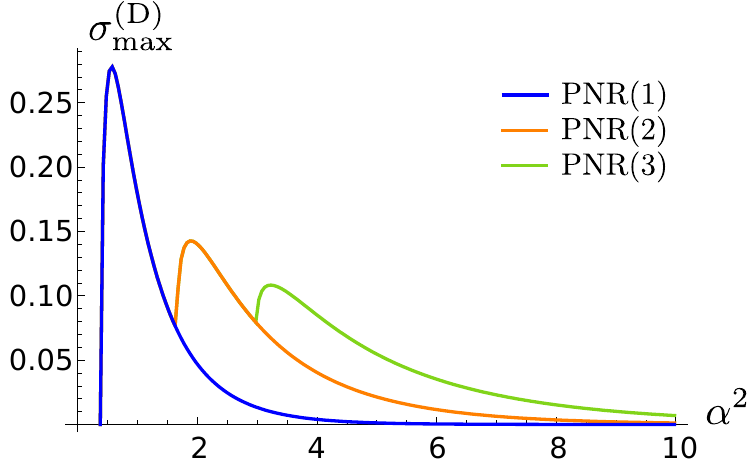}
\centering
\caption{Maximum tolerable phase noise $\sigmamax^{(\rm D)}$ as a function of the signal energy $\alpha^2$ for different photon number resolution $M$. The DPNR receiver beats the SQL in the undergraph region, namely $\sigma < \sigmamax^{(\rm D)}$.}\label{fig04:sec4.6.1-DPNR-sigmaMAX}
\end{figure}

Given the previous considerations, we introduce as a figure of merit the maximum tolerable phase noise $\sigmamax^{(\rm D)}$, namely the maximum level of noise for which ${\cal G}_{\rm D}(\sigma)\ge 0$ for a given signal energy $\alpha^2$, depicted in Fig.~\ref{fig04:sec4.6.1-DPNR-sigmaMAX}. Thus, the DPNR receiver outperforms the SQL if $\sigma < \sigmamax^{(\rm D)}$, corresponding to the undergraph region of $\sigmamax^{(\rm D)}$. We have $\sigmamax^{(\rm D)}=0$ for $\alpha^2<\alpha^2_{\rm K}\approx 0.38$, since in that regime the DPNR does not beat the SQL neither in the noiseless case (in which it performs as a Kennedy); then the plot exhibits $M$ peaks and, thereafter, it decreases towards $0$.

\subsubsection{HYNORE in the presence of phase diffusion}\label{subsec:HYNORE-PhN}
Now, we address the role of the HYNORE for BPSK discrimination of phase-diffused coherent states. As discussed above, quantum receivers based on either quadrature measurements or displacement and photon counting show different degrees of robustness against phase noise, therefore a hybrid scheme like the HYNORE, based on the combination of both of them, provides a good candidate to better mitigate the impact of the noise.

To evaluate the performance of the HYNORE we proceed as follows. After the beam splitter with transmissivity $\tau$, the dephased signal $\rho_{k}$ is split into the separable bipartite state
\begin{align}
    \Xi_{k} = \int_\mathbb{R} d\phi\, g_\sigma(\phi)  \left|\alpha_k^{(r)} e^{-i \phi} \right\rangle \left\langle \alpha_k^{(r)} e^{-i \phi}\right| \otimes \left|\alpha_k^{(t)} e^{-i \phi} \right\rangle \left\langle \alpha_k^{(t)} e^{-i \phi} \right| \, ,
\end{align}
with $\alpha_k^{(r)}$ and $\alpha_k^{(t)}$ introduced in Eq.~(\ref{eq: BS}).
Then, we perform HL detection on the first branch obtaining outcome $\Delta$ and displace the conditional state on the second branch accordingly, obtaining the (not normalized) state $\OUTPUT_k(\Delta)= \Tr_r[U \, \Xi_{k} \, U^{\dag}]$, where the reflected beam has been traced out, and
\begin{align}
U=\mathbb{P}_\Delta\otimes D\big\{\Theta(\Delta) \sqrt{\tau}\alpha+[1-\Theta(\Delta)] (-\sqrt{\tau}\alpha)\big\}\, ,
\end{align}
$\mathbb{P}_\Delta$ being the projection operator over the eigenspace associated with the outcome $\Delta$ and $\Theta(\Delta)$ is the Heaviside Theta function, returning $1$ for $\Delta\ge 0$ and $0$ elsewhere.
In turn, we have:
\begin{align}
\OUTPUT_k(\Delta) =& \int_\mathbb{R}  d\phi\, g_\sigma(\phi) \, {\cal S}(\Delta|\alpha_k^{(r)} e^{-i \phi}) \left| \alpha_k(\tau,\phi) \, e^{-i\phi/2}\right\rangle \left\langle  \alpha_k(\tau,\phi) \, e^{-i\phi/2}\right| \, ,
\end{align}
with:
\begin{align}
\alpha_k(\tau,\phi)=\Theta(\Delta) \sqrt{\mu_k(\tau\alpha^2,\phi)}+ [1-\Theta(\Delta)] \sqrt{\mu_{k \oplus 1}(\tau\alpha^2,\phi)} \, ,
\end{align}
${\cal S}(\Delta|\alpha_k^{(r)} e^{-i \phi})$ being the HL probability of Eq.~(\ref{eq:HLdistr}) and ``$\oplus$'' denoting the mod~$2$ sum.
Finally, we implement PNR($M$) detection on states $\OUTPUT_k(\Delta)$. The resulting joint probability of outcomes $-M\le \Delta\le M$ and $n=0,\ldots,M$ reads:
\begin{align}
{\sf P}_\sigma(\Delta,n|k)&= \int_\mathbb{R}  d\phi\, g_\sigma(\phi) \, {\cal S}(\Delta|\alpha_k^{(r)} e^{-i \phi}) \nonumber \\
& \hspace{1.0cm} \times p_n \bigg\{ \Theta(\Delta)\, \mu_k(\tau\alpha^2,\phi) + [1-\Theta(\Delta)] \, \mu_{k \oplus 1}(\tau\alpha^2,\phi) \bigg\} \, .
\end{align}

We perform discrimination according to the MAP criterion, i.e. by considering the threshold count depicted in Fig.~\ref{fig01:sec4.6.1-DPNR-vs-alpha}(b), with the remark that the energy value to be considered is now the transmitted fraction $\tau \alpha^2$, namely $\nth=\nth(\tau\alpha^2,\sigma)$. Accordingly, the decision rule is modified as in Table~\ref{tab:04-HYNORE-PhN}.


\begin{table}[t]
\tbl{Decision strategy for the HYNORE in the presence of phase diffusion.\label{tab:04-HYNORE-PhN}}
{\begin{tabular}{@{}ccc@{}} \toprule
outcomes &  & decision \\  \colrule
    $\Delta \geq 0$ $\quad$ $n< \nth$ \ & & ``0" \\ 
    $\Delta < 0$ $\quad$ $n\geq \nth$ \ &  & ``0" \\ 
    $\Delta < 0$ $\quad$ $n<\nth$ \ & & ``1" \\ 
    $\Delta \geq 0$ $\quad$ $n\geq \nth$  \ &  &``1" \\  \botrule
\end{tabular}}
\end{table}

The error probability is obtained as
\begin{align}\label{eq: HY-PhN}
\PHY(\sigma) = \min_{\tau,z} \PHY(\tau,z;\sigma) \, ,
\end{align}
where
\begin{align}
    \PHY(\tau,z;\sigma)  &= \frac12 \big[{\sf P}_\sigma(\Delta < 0, n<\nth | 0) + {\sf P}_\sigma(\Delta \ge 0, n\ge \nth| 0) \nonumber \\
&\hspace{1.cm} +{\sf P}_\sigma(\Delta < 0, n\ge\nth| 1) + {\sf P}_\sigma(\Delta \ge 0, n<\nth | 1) \big] \nonumber \\[2ex]
    &= \frac12 \int_\mathbb{R} d\phi \,  g_\sigma(\phi) \ \Biggl\{ \sum_{n=0}^{\nth-1} p_n\left(\mu_1(\tau \alpha^2,\phi) \right) \nonumber \\[1ex]
&\hspace{1.cm} \times \Biggl[  \sum_{\Delta=-M}^{-1} \mathcal{S}(\Delta|\alpha_0^{(r)} e^{-i \phi}) + \sum_{\Delta=0}^{M} \mathcal{S}(\Delta|\alpha_1^{(r)} e^{-i \phi}) \Biggr]\nonumber \\[1ex]
    & \hspace{1.5cm}  + \sum_{n=\nth}^{M} p_n\left(\mu_0(\tau \alpha^2,\phi) \right) \nonumber \\[1ex]
&\hspace{2.cm} \times \Biggl[  \sum_{\Delta=-M}^{-1} \mathcal{S}(\Delta|\alpha_1^{(r)} e^{-i \phi}) + \sum_{\Delta=0}^{M} \mathcal{S}(\Delta|\alpha_0^{(r)} e^{-i \phi}) \Biggr] \Biggr\} \ .
\end{align}

\begin{figure}
\includegraphics[width=0.49\columnwidth]{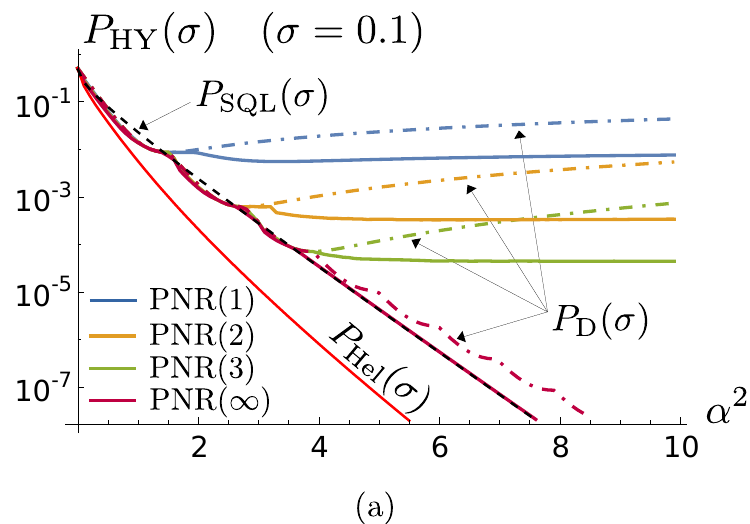} 
\includegraphics[width=0.49\columnwidth]{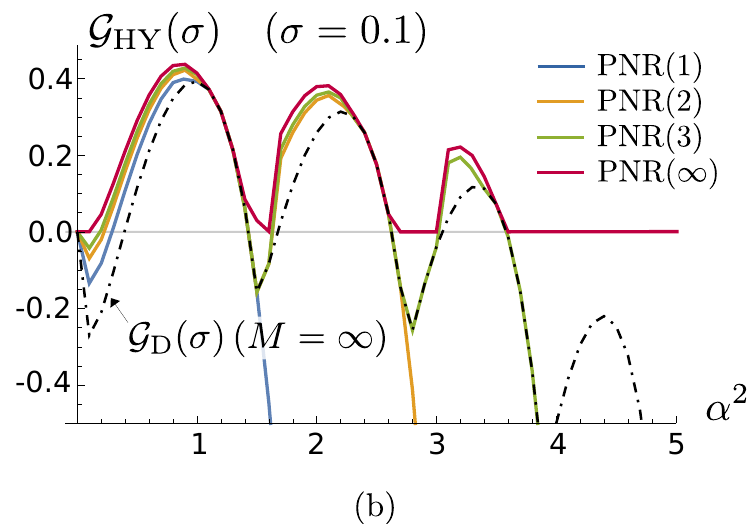}
\centering
\caption{(a) Error probability $\PHY(\sigma)$ of the HYNORE as a function of the signal energy $\alpha^2$ for different photon number resolution $M$. The dot-dashed lines are the error probabilities of the DPNR receiver. (b) Gain ${\cal G}{\rm HY}(\sigma)$ of the HYNORE with respect to the SQL as a function of the signal energy $\alpha^2$. In both the pictures we fix the noise value to $\sigma=0.1$.}\label{fig05:sec4.6.1-HYNORE-vs-alpha}
\end{figure}

Plots of $\PHY(\sigma)$ are reported in Fig.~\ref{fig05:sec4.6.1-HYNORE-vs-alpha}(a). Like the DPNR receiver, the error probability $\PHY(\sigma)$ exhibits a step-like behaviour induced by the change in the threshold $\nth$. The HYNORE outperforms the DPNR, $\PHY(\sigma) \le \PD(\sigma)$, especially in the high-energy limit $\alpha^2\gg 1$, where the error probability is reduced of a factor $\approx 5,15,20$ for $M=1,2,3$, respectively, showing higher robustness in mitigating the phase noise.
Moreover, we observe a quantum advantage also in the low-energy regime, as emerges by computing the gain
\begin{align}
{\cal G}_{\rm HY}(\sigma)= 1 - \frac{\PHY(\sigma)}{\PSQL (\sigma)} \, ,
\end{align}
depicted in Fig.~\ref{fig05:sec4.6.1-HYNORE-vs-alpha}(b). We have ${\cal G}_{\rm HY}(\sigma) \ge {\cal G}_{\rm D}(\sigma)$ and, differently from the DPNR case, improving the resolution $M$ makes the gain increase, since the HL scheme performs better and better, coming closer to the homodyne limit. As one may expect, the best performance is obtained with PNR$(\infty)$ detectors, where the HL performs as standard homodyne detection and ${\cal G}_{\rm HY}(\sigma) \ge 0$ for all energies.

\begin{figure}
\includegraphics[width=0.49\columnwidth]{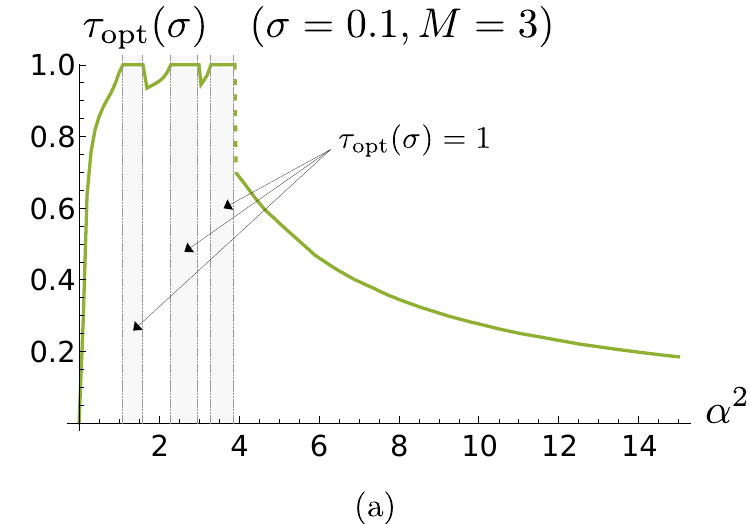} 
\includegraphics[width=0.49\columnwidth]{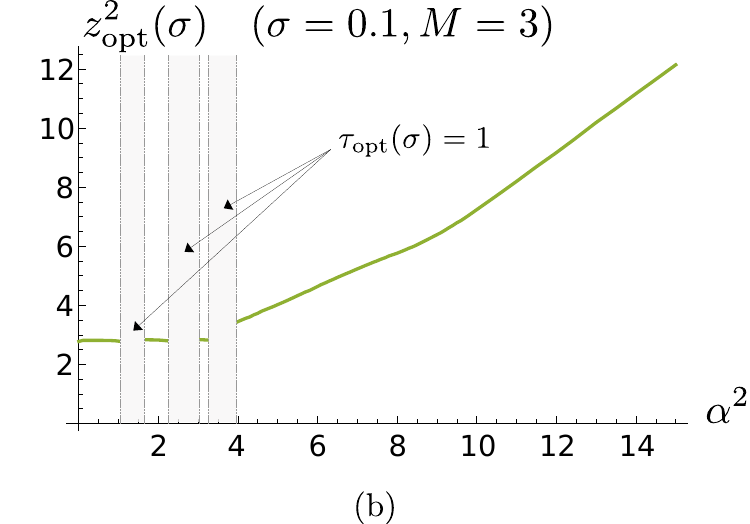}
\centering
\caption{Optimized transmissivity $\tau_\opt(\sigma)$ (a) and LO $z^2_\opt(\sigma)$ (b) as a function of the signal energy $\alpha^2$ for PNR$(3)$ detectors. Both the quantities have been obtained by numerical optimization. In the shaded regions we have $\tau_\opt(\sigma)=1$ and the HYNORE performs as a DPNR receiver.  We fix the noise value to $\sigma=0.1$.}\label{fig06:sec4.6.1-OptPar}
\end{figure}

The physical meaning of the present results is clearer when considering the optimized transmissivity $\tau_\opt(\sigma)$ and LO amplitude $z^2_\opt(\sigma)$ obtained after the minimization in Eq.~(\ref{eq: HY-PhN}), reported in Fig.~\ref{fig06:sec4.6.1-OptPar}(a) and (b), respectively, for the case of PNR$(3)$ detectors. Analogous results can be retrieved for other values of the resolution $M$. The results are similar those obtained in Sec.~\ref{subsec:HYNOREvsDPNR} in the presence of reduced visibility $\xi \le 1$. In fact, in the low-energy limit, the transmissivity $\tau_\opt(\sigma)$ increases with $\alpha^2$ up to reach $1$ (corresponding to DPNR). Thereafter, we observe $M-1$ ``sawteeth", namely regions where $\tau_\opt(\sigma)<1$ before increasing to again reach $1$. Accordingly, when $\tau_\opt(\sigma)<1$ the LO is $z^2_\opt (\sigma)\approx M$ and the HYNORE outperforms the DPNR, whilst when $\tau_\opt(\sigma)=1$ all signal is sent to the transmitted DPNR setup.
On the contrary, in the high-energy limit the transmissivity jumps discontinuously and becomes a decreasing function of $\alpha^2$, saturating for $\alpha^2\gg 1$ to an asymptotic value $\tau_\infty \ne 0$. Remarkably, $\tau_\infty<1$, therefore the optimal strategy is obtained with a proper combination of both the HL and the DPNR schemes. In this regime, $z^2_\opt(\sigma)$ increases with $\alpha^2$, being a linear function for $\alpha^2 \gg 1$.

\begin{figure}
\includegraphics[width=0.49\columnwidth]{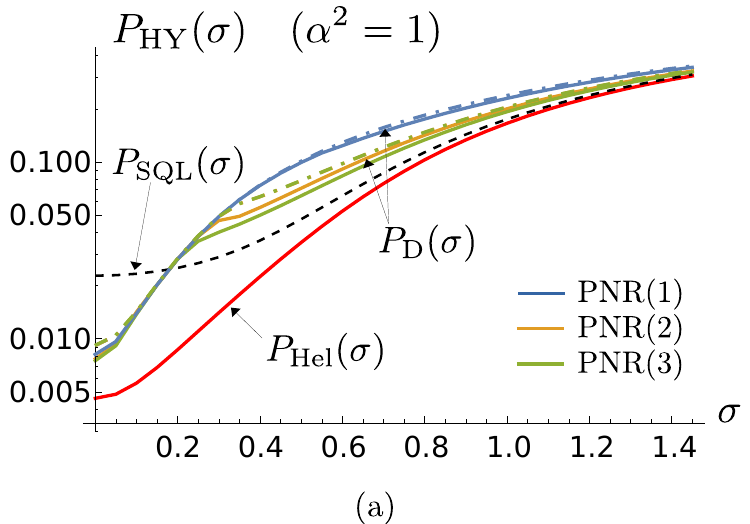} 
\includegraphics[width=0.49\columnwidth]{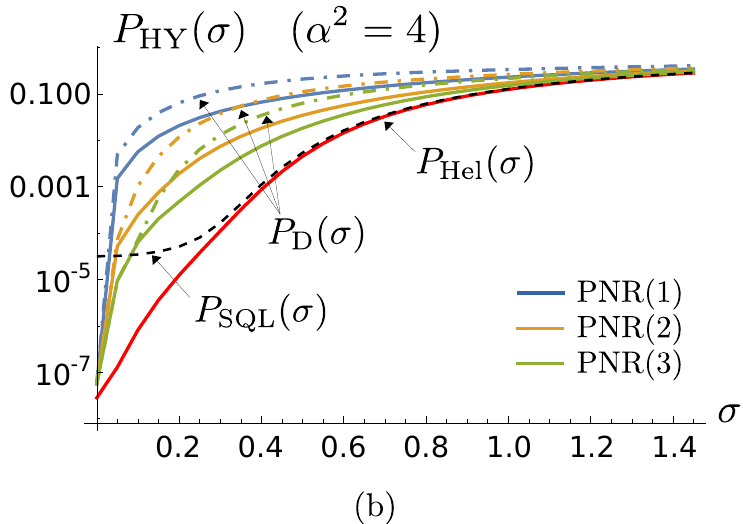}
\centering
\caption{Error probability $\PHY(\sigma)$ of the DPNR receiver as a function of the noise $\sigma$ for $\alpha^2=1$ (a) and $\alpha^2=4$ (b), compared to the DPNR receiver.}\label{fig07:sec4.6.1-HY-vs-sigma}
\end{figure}

\begin{figure}
\includegraphics[width=0.6\columnwidth]{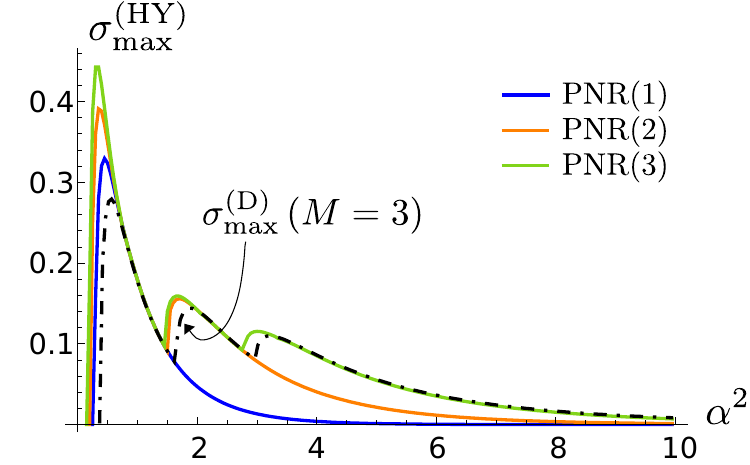}
\centering
\caption{Maximum tolerable phase noise $\sigmamax^{(\rm HY)}$ as a function of the signal energy $\alpha^2$ for different photon number resolution $M$. The HYNORE receiver beats the SQL in the undergraph region, namely $\sigma < \sigmamax^{(\rm HY)}$. The dot-dashed line refers to $\sigmamax^{(\rm D)}$ for $M=3$.}\label{fig08:sec4.6.1-HY-sigmaMAX}
\end{figure}

Finally, in Fig.~\ref{fig07:sec4.6.1-HY-vs-sigma} we plot $\PHY(\sigma)$ (solid lines) as a function of the noise $\sigma$ for $\alpha^2=1$ (left panel) and $\alpha^2=4$ (right panel), respectively, comparing it to the DPNR (dot-dashed lines). We see that $\PHY(\sigma)  \le \PD(\sigma)$ in both the small- and large-noise limits and the enhancement is more relevant for large $\alpha^2$, consistently with the previous analysis. As a consequence, the HYNORE increases the maximum tolerable phase noise $\sigmamax^{(\rm HY)}$, as depicted in Fig.~\ref{fig08:sec4.6.1-HY-sigmaMAX}. In fact, we have $\sigmamax^{(\rm HY)}\ge \sigmamax^{(\rm D)}$ for all energies, and $\sigmamax^{(\rm HY)}=0$ for $\alpha^2<\alpha^2_{\rm HY}(M)$, enlarging the region of quantum advantage with respect to the DPNR. Moreover, increasing the resolution $M$ lets the height of the peaks increase, improving further the robustness of the receiver. 

\def\PGM{{\rm PGM}}
\def\HET{{\rm DH}}
\def\Pmin{P_{\rm min}}
\def\PBon{P_{\rm Bon}}
\def\PQD{P_{\rm QD}}
\def\PQDF{P_{\rm QDF}}

\section{Discrimination in multilevel quantum communications systems}\label{chap:MaryDisc}

In this Section, we proceed beyond binary decision theory and tackle the problem of discrimination of a constellation of $M\ge 2$ non-orthogonal quantum states. This represents a relevant issue in digital communication systems operated at the quantum limit, that leads to the appearance of a nonzero bit error rate due to decision errors induced by the nonzero overlap between the encoded states. 
In particular, the decision problem can be recast into a convex optimization problem, and, thanks to advanced tools of linear algebra, there are well established necessary and sufficient conditions to be fulfilled by the optimum receiver. To date, unlike the binary case, the general expression of the optimum POVM is not known, and the optimum receiver is explicitly constructed only in the presence of pure states with geometrically uniform symmetry (GUS) \cite{Yuen1970, Yuen1970paper, Holevo1973, Yuen1975, Kennedy1973, Cariolaro2015}.
To overcome these limitations, suboptimal methods have been established, e.g. the pretty good measurement (PGM) method, yielding the optimal decision under certain conditions \cite{Ban1997, Hausladen1994, Hausladen1996, Sasaki1998, Sasaki1998b, Kato1999, Eldar2001, Eldar2002, Eldar2004}.

The Section is organized as follows. In Sec.~\ref{sec:MixedMarydiscr}, we present a comprehensive review of the results of $M$-ary quantum decision theory, together with the characterization of the optimum receiver guaranteed by Yuen's theorem. On the other hand, Sec.~\ref{sec:PureMarydiscr} focuses in detail on pure-state discrimination, providing a further characterization of quantum receivers. Thereafter, in Sec.~\ref{sec:GUS} we introduce the concept of GUS and perform explicit construction of the optimum POVM achieving the minimum decision error probability.
Instead, in Sec.~\ref{sec:PGM}, we analyze the PGM method, that, in general, determines a suboptimal receiver (not achieving minimum error probability), whilst becoming optimal in the presence of pure-state discrimination and GUS.
Finally, as a paradigmatic case study, in Sec.~\ref{sec:QPSKdiscr} we apply the obtained results to quadrature phase-shift keying (QPSK) discrimination of coherent states, presenting some relevant examples of quantum receivers, namely the Bondurant, quaternary displacement (QDRE) and quaternary displacement feed-forward (QDFFRE) receivers.

\subsection{Discrimination of $M$-ary constellations}\label{sec:MixedMarydiscr}

In the previous Section we widely investigated the problem of binary discrimination, firstly developed by Helstrom's theory, and, thereafter, we considered a relevant application for quantum communications, namely coherent state discrimination, providing a few examples of feasible quantum receivers.
Now, we proceed beyond the binary case and present the general theory for $M$-ary state discrimination, mainly developed by Yuen, Holevo and Kennedy \cite{Yuen1970, Yuen1970paper, Holevo1973, Yuen1975, Kennedy1973, Cariolaro2015}.
Differently from Helstrom's theory, in this case the solution to the decision problem, that is the optimum POVM achieving minimum error probability, is highly nontrivial and requires advanced tools from linear algebra.

To begin with, we start by reviewing the scenario under investigation. We have a source that encodes a set of classical symbols $k=0,\ldots, M-1$, onto a {\it constellation} of non-orthogonal quantum states ${\cal C}=\{\rho_k\}_k$, $\rho_k$ being a positive semidefinite operator acting on a Hilbert space ${\cal H}$, with $\Tr[\rho_k]=1$, and generated with a priori probability $0\le q_k\le 1$, $\sum_k q_k=1$. The number $M$ of the constellation states is typically referred to as {\it modulation order}.
The task is to implement a quantum receiver to infer which was the state emitted by the source. The receiver is described by a $M$-valued POVM ${\bf \Pi}=\{\Pi_j\}_j$, $j=0,\ldots, M-1$, with $\Pi_j\ge0$ and $\sum_j\Pi_j=\hat{\Id}$, associated with a decision rule, such that, when outcome $j$ is retrieved, we infer the probed state to be $\rho_j$. 
Due the non orthogonality of the constellation states, any receiver is associated with an error probability $P_{\rm err}=1-{\cal P}_c$, ${\cal P}_c$ being the correct decision probability, equal to:
\begin{align}
{\cal P}_c = \sum_{k=0}^{M-1} q_k \, p(k|k)\,,
\end{align}
where $p(j|k)=\Tr[\rho_k \Pi_j]$ is the probability of inferring symbol $j$ when state $\rho_k$ was sent.
Therefore, the goal is to identify the optimum receiver, achieving the minimum error probability $P_{\rm err}^{\rm (min)}$ or, equivalently, the maximum correct decision probability ${\cal P}_c^{\rm(max)}$ compatible with quantum mechanics laws \cite{Cariolaro2015}. 

The decision problem presented above can be recast into the framework of convex {\it semidefinite programming} (SDP), namely optimization of a linear cost function with (linear) constraints over the closed convex cone of positive semidefinite operators \cite{Vandenberghe1996, Csiszar2011, Cariolaro2015}.
In fact, for a given constellation, we should determine the POVM that maximizes the functional:
\begin{align}\label{eq:Functional}
{\cal J}({\bf \Pi}) = \sum_{j=0}^{M-1} \Tr[\widetilde{\rho}_j \Pi_j] \, , \quad \Pi_j \in {\cal B} \, ,
\end{align}
where $\widetilde{\rho}_j=q_j \rho_j$ are the weighted density operators and ${\cal B}$ is the set of Hermitian operators on $\cal H$. That is, we should solve the following SDP problem:
\begin{align}\label{eq:SDPproblem}
&\max_{\Pi_j \in {\cal B}} {\cal J}({\bf \Pi}) \, , \nonumber \\[1ex]
&\mbox{subject to:} \nonumber \\
&\Pi_j\ge0 \, , \quad  j=0,\ldots, M-1 \,, \quad \mbox{and} \quad \sum_j\Pi_j=\hat{\Id} \, .
\end{align}
The first constraint in~(\ref{eq:SDPproblem}) defines the cone set of Hermitian positive semidefinite operators, being the proper domain of functional ${\cal J}$, while the second one provides a linear constraint on the variables $\Pi_j$.

Within this framework, a first characterization of the optimum POVM has been carried out independently in the early 1970s by both Yuen \cite{Yuen1970, Yuen1970paper} and Holevo \cite{Holevo1973}. In particular, in 1970 Yuen determined a class of necessary and sufficient conditions to be satisfied in the presence of equiprobable symbols, even though his proof contained redundant constraints and did not hold anymore in the presence of infinite dimensional Hilbert spaces. On the contrary, in 1972, Holevo derived the sufficient condition for optimality with different approach, but he did not establish that it was also necessary. Instead, he found a necessary condition, different than the sufficient one, being valid for any functional of the POVM ${\bf \Pi}$, even non-linear.
A unified description was finally achieved by Yuen {\it et al.} in 1975 \cite{Yuen1975}, after private communications with Holevo. 
This ultimate proof removes the redundancies of the previous version and properly extends its validity to non-uniform a priori probability and infinite dimensional spaces.

Yuen's proof exploits the Lagrange duality theorem in convex programming. The principle behind it is to transform constrained maximization into a dual minimization problem, where the constraints are included as Lagrange multipliers \cite{Vandenberghe1996, Csiszar2011, Luenberger1997}.
The construction of the dual problem requires technical concepts and properties from topology theory, therefore we decided not to report it here, as it goes beyond the purpose of this dissertation. Its derivation is explicitly reported in \cite{Yuen1975} for both finite and infinite dimensional Hilbert spaces.
Ultimately, Yuen {\it et al.} derived the following equivalence:
\begin{align}\label{eq:construction}
\max_{\substack{\Pi_j \in {\cal B} \\[.5ex] \Pi_j \ge 0 \\[.2ex] \sum_j \Pi_j=\hat{\Id}}} \sum_j\Tr[\widetilde{\rho}_j \Pi_j] = \min_{\substack{\Lambda \in {\cal T} \\[.3ex] \Lambda -\widetilde{\rho}_j\ge 0}} \Tr[\Lambda] \, ,
\end{align}
where $\Lambda \in {\cal T}$, ${\cal T} \subset {\cal B}$ being the subset of the trace-class Hermitian operators on ${\cal H}$, namely ${\cal T}= \{T \in {\cal B} : \Tr[T] <\infty\}$. The right-hand side problem,
\begin{align}\label{eq:dualproblem}
&\min_{\Lambda \in {\cal T}} \Tr[\Lambda] \, , \nonumber \\[1ex]
&\mbox{subject to:} \nonumber \\
&\Lambda -\widetilde{\rho}_j \ge 0 \, , \quad j=0,\ldots, M-1 \, ,
\end{align}
is the dual problem of~(\ref{eq:SDPproblem}).
Thanks to topological arguments, they established the existence of a solution to problems~(\ref{eq:SDPproblem})-(\ref{eq:dualproblem}), therefore there exists a POVM $\bf \Pi_\opt$ and a corresponding trace-class operator $\Lambda_\opt$ such that:
\begin{align}
{\cal J}({\bf \Pi}_\opt) = \Tr[\Lambda_\opt] = {\cal P}_c^{\rm (max)} \, ,
\end{align}
retrieving the maximum correct decision probability.

Furthermore, they formulated the theorem, from now on referred to as ``Yuen's theorem", providing a complete characterization of the optimum POVM, determining necessary and sufficient conditions to be fulfilled by $\bf \Pi_\opt$. To prove it, we need the following lemma.

\begin{lemma}\label{LemmaYuen}
Let $X$ and $Y$ be two positive semidefinite operators of an arbitrary Hilbert space. Then
\begin{align}
\Tr[ X Y] \ge 0 \, ,
\end{align}
and $\Tr[XY] = 0$ if and only if $X Y = Y X= 0$.
\end{lemma}

\begin{proof}
Since both $X$ and $Y$ are positive, they admit a unique positive semidefinite square root $X^{1/2}$ and $Y^{1/2}$ such that $X=\left(X^{1/2}\right)^2$ and $Y=\left(Y^{1/2}\right)^2$, respectively. Then:
\begin{align}\label{eq:prooflemma}
\Tr[X Y] &=\Tr \left[\left(X^{1/2}\right)^2  \left(Y^{1/2}\right)^2\right]  \nonumber \\
&= \Tr \left[ \left( X^{1/2} Y^{1/2}\right)^\dagger \left(X^{1/2} Y^{1/2}\right) \right] \ge 0 \, .
\end{align}
From~(\ref{eq:prooflemma}) and the ciclicity of the trace, it follows that $\Tr[X Y ]=0$ if and only if $X^{1/2} Y^{1/2}=Y^{1/2}X^{1/2}=0$.
Then, if $X^{1/2} Y^{1/2}=0$, we have $X Y=0$. On the other hand, $X Y=0$ also implies that $(XY)^\dagger=YX=0$, thus $X$ and $Y$ commute, i.e. $[X,Y]=0$. In turn, all their powers commute with one another and, in particular, $[X^{1/2},Y]=[X^{1/2},Y^{1/2}]=0$. Therefore, $XY= (X^{1/2} Y^{1/2})^\dagger (X^{1/2} Y^{1/2})=0$, implying $X^{1/2} Y^{1/2}=0$.
\end{proof}

We are now ready to enunciate Yuen's theorem.
 \begin{theorem}[{\bf Yuen {\it et al.}, 1975}] \label{YuenThm}
In a $M$-ary system characterized by the weighted density operators $\widetilde{\rho}_j=q_j\rho_j$, $j=0,\ldots,M-1$, the POVM ${\bf \Pi}=\{\Pi_j\}_j$ is optimal if and only if the following two conditions hold:
\begin{subequations}\label{eq:SuffCond}
\begin{align}
&\sum_j \widetilde{\rho_j} \Pi_j = \sum_j  \Pi_j \widetilde{\rho_j} \, , \label{eq:SuffCond1}\\[1ex]
&\sum_s \widetilde{\rho_s} \Pi_s - \widetilde{\rho}_j \ge 0 \, , \quad j=0,\ldots,M-1 \label{eq:SuffCond2} \, .
\end{align}
\end{subequations}
\end{theorem}

\begin{proof}
$``\Rightarrow"$: we start by proving the necessity of conditions~(\ref{eq:SuffCond}). Let ${\bf \Pi}^{(0)}=\{\Pi_j^{(0)}\}_j$ and $\Lambda^{(0)}$ solve the SDP problems~(\ref{eq:SDPproblem}) and~(\ref{eq:dualproblem}), respectively. Then, thanks to~(\ref{eq:construction}), we have:
\begin{align}
\sum_j \Tr\left[\Pi_j^{(0)}\left(\Lambda^{(0)} - \widetilde{\rho}_j\right)\right] =0 \, .
\end{align}
Lemma~\ref{LemmaYuen} then yields:
\begin{align}\label{eq:PiLambda=LambdaPi}
\Pi_j^{(0)}\left(\Lambda^{(0)} - \widetilde{\rho}_j\right) =\left(\Lambda^{(0)} - \widetilde{\rho}_j\right)\Pi_j^{(0)}=0 \, , \quad j=0,\ldots, M-1 \, .
\end{align}
We perform summation over index $j$ and obtain:
\begin{align}\label{eq:Loperator}
\Lambda^{(0)} =  \sum_j \widetilde{\rho}_j \Pi_j^{(0)} =  \sum_j \Pi_j^{(0)} \widetilde{\rho}_j \, ,
\end{align}
thus proving~(\ref{eq:SuffCond1}). Moreover, Eq.~(\ref{eq:Loperator}), together with the dual problem constraint $\Lambda_\opt \ge \rho_j$ for all $j$, leads to~(\ref{eq:SuffCond2}).

$``\Leftarrow"$: to show sufficiency we first derive a general property. Let ${\bf \Pi}=\{\Pi_j\}_j$ and $\Lambda$ be arbitrary operators that only satisfy the constraints of the SDP problem. In particular, we have $\Lambda-\widetilde{\rho}_s \ge 0$ for all $s=0,\ldots,M-1$. Then, by Lemma~\ref{LemmaYuen}, we have $\Tr[\Pi_s (\Lambda-\widetilde{\rho}_s)] \ge 0$, therefore:
\begin{align}\label{eq:LEq}
\Tr \left[ \sum_s \widetilde{\rho}_s \Pi_s \right] \le \Tr[\Lambda] \, .
\end{align}
Now, let ${\bf \Pi}^{(0)}=\{\Pi_j^{(0)}\}_j$ be a POVM that also satisfies~(\ref{eq:SuffCond}). We claim that ${\bf \Pi}^{(0)}$ is optimal. In fact, we define the operator $\Lambda^{(0)}= \sum_s \widetilde{\rho}_s \Pi_s^{(0)}$, which satisfies $\Lambda^{(0)} -\widetilde{\rho}_j \ge 0$ thanks to~(\ref{eq:SuffCond2}) and saturates the equality in Eq.~(\ref{eq:LEq}), thus maximizing $\Tr \left[ \sum_s \widetilde{\rho}_s \Pi_s \right] $.
\end{proof}

Yuen's theorem is a cornerstone result of quantum decision theory, even though it does not provide a method to construct the optimal measurement. Moreover, we note that the optimum POVM ${\bf \Pi}_\opt$ is not unique in general. As an example, we consider the case where the constellation states $\{\rho_k\}_k$ commute with one another, namely $[\rho_k,\rho_j]=0$ \cite{Helstrom1967, Liu1970}. In this case, there exists a common set of eigenstates $\{|\mu_{l_j}\rangle\}_j$, $l_j \in \{0,\ldots,M-1\}$ for all $j$, that diagonalizes all states. The optimum POVM $\{\Pi_j\}_j$ is constructed from the projectors $\mathbb{P}_{l_j}=|\mu_{l_j}\rangle \langle \mu_{l_j}|$, according to the maximum a posteriori probability (MAP) criterion: given projector $\mathbb{P}_{l_n}$, we infer the state $\rho_n$ with the highest a posteriori probability $\Tr[\widetilde{\rho}_n \mathbb{P}_{l_n}] > \Tr[\widetilde{\rho}_k \mathbb{P}_{l_n}]$, for all $k\ne n$. In this case, we set $\Pi_n= \mathbb{P}_{l_n}$. On the contrary, if there exists $l'_n$ such that, for some $\rho_m$ and $\rho_n$, we have:
\begin{align}
\Tr[\widetilde{\rho}_m \mathbb{P}_{l'_{n}} ] = \Tr[\widetilde{\rho}_n \mathbb{P}_{l'_{n}} ] > \Tr[\widetilde{\rho}_k \mathbb{P}_{l'_{n}}] \, \quad \forall k \neq m \neq n \, ,
\end{align}
then we may choose either $\Pi_n= \mathbb{P}_{l'_{n}}$ or $\Pi_m= \mathbb{P}_{l'_{n}}$ with the same overall correct decision probability \cite{Helstrom1967, Liu1970}.

Furthermore, from Yuen's theorem we derive the following result.
\begin{corollary}\label{YuenCorollary}
If the POVM ${\bf \Pi}=\{\Pi_j\}_j$ is optimum, then
\begin{align}
\Pi_j \left( \sum_s \widetilde{\rho_s} \Pi_s - \widetilde{\rho}_j \right) = \left( \sum_s \widetilde{\rho_s} \Pi_s - \widetilde{\rho}_j \right) \Pi_j = 0 \, ,
\end{align}
for all $j=0,\ldots, M-1$. 
\end{corollary}
\begin{proof}
The result follows directly from Eq.s~(\ref{eq:PiLambda=LambdaPi}) and~(\ref{eq:Loperator}).
\end{proof}
The corollary provides us with a recipe to construct the optimal measurement, according to the following outline:
\begin{itemize}
\item at first, we solve the dual problem~(\ref{eq:dualproblem}), retrieving the optimum operator $\Lambda_\opt$ and the corresponding maximum correct decision probability ${\cal P}_c^{\rm(max)}= \Tr[\Lambda_\opt]$;
\item then, from Theorem~\ref{YuenThm} we know that the optimum POVM ${\bf \Pi}_\opt= \{\Pi_j^{(\opt)}\}_j$ is related to $\Lambda_\opt$ via the equality $\Lambda_\opt= \sum_s \widetilde{\rho}_s \Pi_s^{(\opt)}$. Therefore, thanks to Corollary~\ref{YuenCorollary}, we obtain the optimal measurement operators as solutions of the system of equations:
\begin{align}
(\Lambda_\opt -\widetilde{\rho}_j) \Pi_j^{(\opt)}=0 \, ,
\end{align}
for all $j=0,\ldots, M-1$, with the further request $\sum_j \Pi_j^{(\opt)}=\hat{\Id}$. In the matrix notation, we have:
\begin{align}
\begin{pmatrix} 
\Lambda_\opt -\widetilde{\rho}_0 & 0 & \cdots & 0 \\
0 & \Lambda_\opt -\widetilde{\rho}_1  &  \cdots & 0 \\
 & & \ddots & \\
0 & 0 & \cdots & \Lambda_\opt -\widetilde{\rho}_{M-1} \\
\hat{\Id} & \hat{\Id} & \cdots & \hat{\Id}
\end{pmatrix}
\begin{pmatrix} \Pi_0^{(\opt)} \\ \Pi_1^{(\opt)} \\ \vdots \\ \Pi_{M-1}^{(\opt)} \end{pmatrix}
=
\begin{pmatrix} 0 \\ 0 \\ \vdots \\ 0 \\ \hat{\Id} \end{pmatrix} \, ,
\end{align}
where we deal with a rectangular $(M+1) \times M$ matrix to include the identity resolution constraint \cite{Eldar2003}.
\end{itemize}
However, generally speaking, analytic solution to this problem cannot be reached. Nevertheless, the dual problem~(\ref{eq:dualproblem}) can be numerically solved efficiently thanks to SDP algorithms, providing the maximum correct decision probability in several cases of practical interest \cite{Vandenberghe1996, Csiszar2011, Cariolaro2015}.

Finally, to conclude the presentation of the general quantum decision theory, we report a further property satisfied by the optimum POVM, derived by Eldar {\it et al.} in 2003, that holds in the presence of finite-dimensional Hilbert spaces ${\cal H}$ \cite{Eldar2003}.

\begin{proposition}\label{PropositionEldar}
In the presence of a finite dimensional Hilbert space $\cal H$, ${\rm dim}({\cal H})=d<\infty$, the optimal POVM elements $\Pi_j$ have rank not higher than that of the associated constellation state, namely:
\begin{align}
{\rm rank}(\Pi_j) \le {\rm rank}(\rho_j) \, , \quad j=0,\ldots, M-1 \, .
\end{align}
\end{proposition}

The proof of the proposition invokes the {\it rank-nullity theorem}, a result from linear algebra providing a relation between the rank and nullity of a linear operator $A$ acting on a finite dimensional vector space $V$. The rank of $A$ is equal to the dimension of its image, i.e. ${\rm rank}(A)={\rm dim}({\rm Im}(A))$, while the nullity is the dimension of its kernel, i.e. the null space ${\rm ker}(A)=\{v\in V: Av =0 \}$. The theorem states that ${\rm rank} (A) + {\rm dim}({\rm ker}(A)) = {\rm dim}(V)$ \cite{Friedberg2014}.

\begin{proof}
We first note that it is not restrictive to assume that the constellation states span the whole Hilbert space $\cal H$. Otherwise, we can rephrase the problem on the subspace ${\cal S} \subset {\cal H}$ spanned by the eigenstates of $\{\rho_k\}_k$, having finite dimension too, as ${\rm dim}({\cal S}) \le {\rm dim}({\cal H}) < \infty$.
Now, let ${\bf \Pi}=\{\Pi_j\}_j$ be an optimum POVM, and $\Lambda=\sum_s \widetilde{\rho}_s \Pi_s$ its associated solution to the dual problem. Then, by Corollary~\ref{YuenCorollary}, it follows that, for all $j$, the image of $\Pi_j$ lies in the kernel of $\Lambda-\widetilde{\rho}_j$, as all vectors in the form $\Pi_j|h\rangle$, for some $|h\rangle \in {\cal H}$, satisfy $(\Lambda-\widetilde{\rho}_j)\Pi_j|h\rangle=0$. In turn, ${\rm Im}(\Pi_j) \subset {\rm ker}(\Lambda-\widetilde{\rho}_j)$ and, consequently, 
\begin{align}\label{eq:rankPiLambda}
{\rm rank}(\Pi_j) \le d - {\rm rank}(\Lambda-\widetilde{\rho}_j) \, .
\end{align}
We also note that $\Lambda$ is a full-rank operator. In fact, we have $\Lambda-\widetilde{\rho}_j \ge 0$ for all $j$ thanks to the constraints of the dual problem and, since and the eigenvectors of the constellation states span ${\cal H}$, for any $|h\rangle \in {\cal H}$, there exists an index $k$ such that $\langle h |\rho_k|h\rangle >0$. These two conditions imply $\langle h |\Lambda|h\rangle >0$ for all $|h\rangle \in {\cal H}$, therefore the kernel of $\Lambda$ only contains the null vector, ${\rm ker}(\Lambda)=\{0\}$.
From the subadditivity of the rank, i.e. $ {\rm rank} (A+B) \le {\rm rank} (A)+ {\rm rank} (B)$ for any two linear operators $A$ and $B$ \cite{Friedberg2014}, we have $d= {\rm rank}(\Lambda) \le {\rm rank}(\Lambda-\widetilde{\rho}_j) + {\rm rank}(\widetilde{\rho}_j)$ which, together with Eq.~(\ref{eq:rankPiLambda}), leads to:
\begin{align}
{\rm rank}(\Pi_j) \le {\rm rank}(\widetilde{\rho}_j) = {\rm rank}(\rho_j) \, ,
\end{align}
proving the desired result.
\end{proof}

As a final remark, we stress that Proposition~(\ref{PropositionEldar}) is only valid if the dimension of the Hilbert space $\cal H$ is finite, or, at least, if the subspace spanned by the eigenstates of the constellation states is finite-dimensional. Otherwise, in the presence of infinite-dimensional vector spaces the rank-nullity theorem does not hold anymore, and the argument of the proof vanishes.

\subsection{Pure-state discrimination}\label{sec:PureMarydiscr}

The problem of the optimal decision is considerably simplified if the constellation ${\cal C}=\{\rho_k\}_k$ is composed of linearly independent pure states, namely:
\begin{align}
\rho_k = |\gamma_k\rangle \langle \gamma_k| \, , \qquad k=0,\ldots, M-1 \, .
\end{align}
To quantify the overlap between the encoded states, it is convenient to introduce the {\it Gram matrix} $G$, that is the $M\times M$ matrix of the inner products:
\begin{align}\label{eq:Grammatrix}
G= \big(\langle \gamma_l | \gamma_k\rangle \big)_{l,k} \,, \qquad l,k=0,\ldots,M-1 \, ,
\end{align}
which is $G \ne \Id_M$ for non-orthogonal states, $\Id_M$ being the $M\times M$ identity matrix.
It is not restrictive to reduce the problem to the $M$ dimensional subspace spanned by the encoded states, ${\cal S} = {\rm span} \{ |\gamma_k\rangle : k=0,\ldots, M-1 \}$ and to find a $M$-valued POVM ${\bf \Pi}=\{\Pi_j\}_j$, such that
\begin{align}\label{eq:ResolutionoverS}
\Pi_j \ge 0 \quad \mbox{and} \quad \sum_j \Pi_j= \mathbb{P}_{\cal S} \, ,
\end{align}
$ \mathbb{P}_{\cal S}$ being the projection operator onto subspace $\cal S$ \cite{Cariolaro2011, Cariolaro2015}. In fact, if we decompose each POVM element as $\Pi_j= \Pi'_j + \Pi''_j$, with $\Pi'_j$ and $\Pi_j''$ having support equal to $\cal S$ and ${\cal S}^\perp$, respectively, we would have $\Pi''_j |\gamma_k\rangle =0$ for all $k$, thus the two POVM sets $\{\Pi_j\}_j$ and $\{\Pi_j'\}_j$ would be associated with the same correct decision probability.
Moreover, due to the linearly independence of constellation states, subspace ${\cal S}$ has dimension equal to $M$, therefore Proposition~\ref{PropositionEldar} holds. In more detail, the following theorem, firstly proved by Kennedy in 1973, provides characterization of the optimum POVM \cite{Kennedy1973Theorem}.
\begin{theorem}
[{\bf Kennedy, 1973}] In a $M$-ary system specified by $M$ pure states $\{|\gamma_k \rangle\}_k$, the optimum POVM ${\bf \Pi}=\{\Pi_j\}_j$ is a $1$-rank projective measurement, namely:
\begin{align}\label{eq:Kennedythm}
\Pi_j= |\mu_j\rangle \langle \mu_j| \, ,  \qquad j=0,\ldots, M-1 \, ,
\end{align}
for some measurement vectors $\{|\mu_j\rangle\}_j$ satisfying $\langle \mu_j|\mu_k\rangle = \delta_{jk}$.
\end{theorem}

\begin{proof}
Thanks to Proposition~\ref{PropositionEldar}, the optimal measurement is composed of $1$-rank operators, hence Eq.~(\ref{eq:Kennedythm}) follows. It only remains to prove the orthonormality relation among the measurement vectors. Following the same arguments reported above, it is not restrictive to assume that $|\mu_j \rangle \in {\cal S} $ for all $j$.
Then, the POVM elements resolve the identity over $\cal S$, $\sum_j |\mu_j\rangle \langle \mu_j| = \mathbb{P}_{\cal S}$, and
\begin{align}
|\mu_k\rangle = \mathbb{P}_{\cal S} |\mu_k\rangle =  \sum_j |\mu_j\rangle \langle \mu_j|\mu_k \rangle \, ,
\end{align}
which yields $\langle \mu_j|\mu_k\rangle = \delta_{jk}$ .
\end{proof}

In particular, the measurement vectors $\{|\mu_j\rangle\}_j$ provide an orthonormal basis of subspace $\cal S$.

\subsubsection{Characterization of $M$-ary pure-state discrimination receivers}\label{sec:M=GammaA}

Given the Kennedy theorem, the decision problem in the presence of pure states may be recast into a geometric optimization task. Indeed, we introduce the state and measurement (row) matrices \cite{Notarnicola2023:KB}:
\begin{align}
\Gamma = \bigg(|\gamma_0\rangle, \ldots, |\gamma_{M-1}\rangle\bigg) \quad \mbox{and} \quad \mathbb{M} = \bigg(|\mu_0\rangle, \ldots, |\mu_{M-1}\rangle\bigg) \, ,
\end{align}
respectively. With this notation, the Gram matrix~(\ref{eq:Grammatrix}) is retrieved as $G= \Gamma^\dagger \, \Gamma$, whereas, due to the orthonormality of the measurement vectors $\mathbb{M}^\dagger \, \mathbb{M}= \Id_M$, $\Id_M$ being the $M \times M$ identity matrix. 
Since $\{|\mu_j\rangle\}_j \subset {\cal S}$, the measurement vectors are expressible as a linear combination of the state vectors, $|\mu_j\rangle = \sum_k a_{kj} |\gamma_k\rangle$, $a_{kj} \in \mathbb{C}$, or equivalently,
\begin{align}\label{eq:Mmatrix}
\mathbb{M}= \Gamma \, A \, ,
\end{align}
$A$ being a $M\times M$ matrix with coefficients $(a_{kj})_{k,j}$.

In turn, we provide characterization of any quantum receiver by its corresponding matrix $A$, subject to the constraint:
\begin{align}\label{eq:ConstA}
A A\dag = G^{-1} \, ,
\end{align}
guaranteeing the identity resolution of the resulting POVM \cite{Notarnicola2023:KB}.

Eq.~(\ref{eq:ConstA}) can be derived as follows. For all vectors $|\psi\rangle \in {\cal S}$, we have $\mathbb{P}_{{\cal S}}|\psi\rangle = |\psi\rangle$, where $|\psi\rangle = \sum_s b_s |\gamma_s\rangle$, $b_s \in \mathbb{C}$. Thanks to Eq.s~(\ref{eq:ResolutionoverS}) and~(\ref{eq:Mmatrix}), the following equations hold:
\begin{align}
\Bigg[\sum_j \Big( \sum_{k,l} a_{kj} a_{lj}^*  |\gamma_k\rangle \langle \gamma_l | \Big) \Bigg] \sum_s b_s |\gamma_s\rangle = \sum_t b_t |\gamma_t\rangle \, , \\[1ex]
\sum_k \Big(\sum_{j,l,s} a_{kj} a_{jl}\dag G_{ls} b_s \Big) |\gamma_k\rangle = \sum_t b_t |\gamma_t\rangle \, ,
\end{align}
where $G_{ls}= \langle \gamma_l|\gamma_s\rangle$. In the matrix notation we have $A A\dag G  \, {\bf b} = {\bf b}$ to be satisfied for all ${\bf b} = (b_0,\ldots, b_{M-1})$. This implies:
\begin{align}
A A\dag G = \Id_M \, ,
\end{align}
and, ultimately, $A A\dag = G^{-1}$.

\subsubsection{Consequences of Yuen's theorem}\label{sec:Gamma=MB}

A further characterization of the receiver, being somehow specular to~(\ref{eq:Mmatrix}), can be obtained by expanding the state vectors on the orthonormal basis composed of the measurement vectors, $|\gamma_k\rangle = \sum_j b_{jk} |\mu_j\rangle$ or, equivalently
\begin{align}\label{eq:Bmatrix}
\Gamma= \mathbb{M} \, B \, ,
\end{align}
where:
\begin{align}
B_{kj} = \langle \mu_j|\gamma_k\rangle \, , \qquad k,j=0,\ldots, M-1 \, ,
\end{align}
are the inner products between the elements of the two vector systems, describing the geometry between the state and measurement vectors.
Since $\mathbb{M}^\dagger \, \mathbb{M}= \Id_M$, the relation between matrices $A$ and $B$ is:
\begin{align}
B= \mathbb{M}^\dagger \, \Gamma = A^\dagger \, G \, ,
\end{align}
implying $B^\dagger \, B = \Gamma^\dagger \, \Gamma = G$.

If matrix $A$ provides a characterization of the quantum receiver, as $\mathbb{M}=\Gamma \, A$, the properties of matrix $B$ determine its optimality for the decision problem. In fact, the maximum correct decision probability can be re-expressed as a function of the sole matrix $B$, since the conditional probability of obtaining outcome $j$ if the $k$-th state is probed is given by
\begin{align}\label{eq:CondP}
p(j|k)= \Tr [ \rho_k \, \Pi_j ]= |\langle \mu_j |\gamma_k\rangle|^2 = |B_{kj}|^2 \, .
\end{align}
In turn, we have:
\begin{align}\label{eq:CondP}
{\cal P}_c = \sum_{k=0}^{M-1} q_k \, |B_{kk}|^2 \, .
\end{align}
Yuen's theorem can be specified to the pure-state discrimination scenario, finding necessary and sufficient conditions that should be fulfilled by $B$. Accordingly, Theorem~\ref{YuenThm} leads to the following corollary \cite{Cariolaro2015}.

\begin{corollary}\label{CorollaryYuen+Kennedy}
In a $M$-ary system specified by $M$ pure states $\Gamma=\{|\gamma_k \rangle\}_k$, generated with a priori probabilities $\{q_k\}_k$, the optimum measurement vectors $\mathbb{M}=\{|\mu_j \rangle \}_j$ must verify the following conditions:
\begin{subequations}\label{eq:YuenPureState}
\begin{align}\label{eq:YuenPureState1}
q_j B_{jj}^{*} B_{kj} - q_k B_{kk} B_{jk}^{*} = 0 \, , \quad  k,j=0,\ldots, M-1 \, ,  
\end{align}
\begin{align}\label{eq:YuenPureState2}
\left(\sum_{s=0}^{M-1} q_s B_{ss} |\mu_s\rangle \langle \gamma_s| \right) - q_j |\gamma_j\rangle \langle \gamma_j| \ge 0 \, , \quad j=0,\ldots, M-1 \, .
\end{align}
\end{subequations}
\end{corollary}

\begin{proof}
Eq.~(\ref{eq:YuenPureState1}) follows from condition~(\ref{eq:SuffCond1}) of Yuen's theorem, namely $L= \sum_j q_j \rho_j \Pi_j= \sum_j q_j \Pi_j \rho_j= L^\dagger$. In the pure-state discrimination scenario, we have:
\begin{align}
L &= \sum_j q_j \langle \gamma_j| \mu_j \rangle \, |\gamma_j\rangle \langle \mu_j|  \nonumber \\
&= \sum_j q_j B_{jj}^* \left(\sum_k B_{kj} |\mu_k\rangle \right) \langle \mu_j|  \nonumber \\[1ex]
&= \sum_{jk} q_j B_{jj}^* B_{kj} |\mu_k\rangle \langle \mu_j| \, .
\end{align}
Imposing $L=L^\dagger$ leads to:
\begin{align}
\sum_{jk} \left(q_j B_{jj}^* B_{kj} - q_k B_{kk} B_{jk}^* \right) |\mu_k\rangle \langle \mu_j| = 0 \, ,
\end{align}
from which we obtain~(\ref{eq:YuenPureState1}).
With analogous argument, we retrieve Eq.~(\ref{eq:YuenPureState2}) from condition~(\ref{eq:SuffCond2}).
\end{proof}

In principle, one may claim to construct the optimal matrix $B$ by solving the nonlinear system determined by Eq.s~(\ref{eq:YuenPureState1}) together with condition $B^\dagger \, B= G$, and, then, retrieve the maximum correct decision probability. 
In particular, the system is composed of $M (M+1)/2$ equations, and its physical solutions will be only those verifying~(\ref{eq:YuenPureState2}). However, it has been showed that this procedure is rather cumbersome, as the search of an exact solution is nontrivial \cite{Cariolaro2015}. On the contrary, numerical algorithms based on the SDP approach remain the preferable choice due to their computational efficiency, even in the presence of pure states.

\subsection{The geometrically uniform symmetry}\label{sec:GUS}

Generally speaking, the search for the optimum POVM, both in the presence of mixed- and pure-state discrimination, turns out to be simpler if the constellation ${\cal C}=\{\rho_k\}_k$ exhibits some degree of symmetry. In fact, the symmetries of the encoded states may be exploited to further characterize the optimal measurement, thus reducing the complexity of the SDP problem.
In particular, the most relevant example is provided by the {\it geometrically uniform symmetry} (GUS), being also verified in several practical optical communications systems, e.g. phase-shift keying \cite{Cariolaro2015, Forney1991, Ban1997, Eldar2001, Eldar2004, Assalini2010}.

We start by considering the most general scenario in which $\rho_k$ are mixed states.
The constellation ${\cal C}=\{\rho_k\}_k$, $k=0,\ldots, M-1$, satisfies the GUS if there exists a unitary {\it symmetry operator} $S$, with $S^\dagger S= \hat{\Id}$, such that:
\begin{align}
\rho_k = S^k \, \rho_0 \left(S^\dagger\right)^k \qquad \mbox{and} \qquad S^M = \hat{\Id} \, ,
\end{align}
$\hat{\Id}$ being the identity operator over the Hilbert space $\cal H$.
That is, in the presence of GUS, all states $\rho_k \in {\cal C}$ can be retrieved from a single ``reference" state $\rho_0$ by sequential application of the symmetry operator $S$.
From now on, for consistency with the symmetry of $\cal C$, we will assume {\it equal a priori probabilities}
\begin{align}
q_k=\frac{1}{M} \, , \quad k=0,\ldots, M-1\, ,
\end{align}
such that also the weighted operators $\{\widetilde{\rho}_k\}_k$ have the GUS, i.e. $\widetilde{\rho}_k = S^k \, \widetilde{\rho}_0 (S^\dagger)^k$ \cite{Cariolaro2015, Ban1997, Eldar2001, Eldar2004}.
Even though this choice may appear more restrictive, it is still of great interest since uniform sampling represents the standard scenario occurring in practical quantum communications formats \cite{Cariolaro2015, Notarnicola2023:KB, Notarnicola2024:SEC}.
In this condition, the GUS can be also embedded into the optimal measurement, according to the following proposition \cite{Assalini2010, Cariolaro2015}.

\begin{proposition}\label{PropositionGUS}
If the weighted constellation states $\{\widetilde{\rho}_k\}_k$ verify the GUS for some operator $S$, then it is not restrictive to assume that also the optimum POVM ${\bf \Pi}=\{\Pi_j\}_j$ verifies the GUS for the same symmetry operator, that is:
\begin{align}
\Pi_j = S^j \, \Pi_0  \left(S^\dagger\right)^j \, , \quad j=0,\ldots, M-1 \, .
\end{align}
\end{proposition}

\begin{proof}
Let ${\bf \Pi}=\{\Pi_j\}_j$ be an optimum POVM satisfying Theorem~\ref{YuenThm}, and maximizing the functional ${\cal J}({\bf \Pi})$ defined in~(\ref{eq:Functional}). We note that, in general, this measurement does not satisfy the GUS. To prove the proposition, starting from ${\bf \Pi}$, we construct another POVM ${\bf \Upsilon}=\{\Upsilon_j\}_j$, that now satisfies the GUS for operator $S$, and prove it to be optimum too.
To this aim, we set:
\begin{subequations}
\begin{align}
\Upsilon_0= \frac{1}{M} \sum_{n=0}^{M-1} \left(S^\dagger\right)^n \Pi_n \, S^n \, ,
\end{align}
\begin{align}
\Upsilon_j = S^j \, \Upsilon_0 \left(S^\dagger\right)^j\, , \,  j=1,\ldots, M-1 \, ,
\end{align}
\end{subequations}
where $q_n$ is the a priori probabilities associated with state $\rho_n$.
The collection ${\bf \Upsilon}$ is a POVM, as $\Upsilon_j \ge 0$ by construction and
\begin{align}
\sum_j \Upsilon_j &= \frac{1}{M} \sum_{j, n} S^{j-n} \, \Pi_n  \big(S^\dagger\big)^{j-n} 
= \frac{1}{M} \sum_{m,n} S^{m} \, \Pi_n \big(S^\dagger\big)^{m}  \nonumber \\[1ex]
&=\frac{1}{M} \sum_m  S^{m} \left( \sum_n \Pi_n \right) \big(S^\dagger\big)^{m} = \hat{\Id} \, ,
\end{align}
where we performed the change of variables $m=j-n$ and exploit the periodicity of the symmetry operator $S$.

Now, we compute:
\begin{align}
{\cal J}({\bf \Upsilon}) &= \sum_j \Tr[\widetilde{\rho}_j \Upsilon_j] = \sum_j \Tr\left[S^j \widetilde{\rho}_0 \Upsilon_0 \big(S^\dagger\big)^{j} \right] \nonumber \\
&= \sum_j \Tr[\widetilde{\rho}_0 \Upsilon_0] = M \Tr[\widetilde{\rho}_0 \Upsilon_0] \nonumber \\
&= \sum_n \Tr\left[\widetilde{\rho}_0 \left(S^\dagger\right)^n \Pi_n \, S^n \right] = \sum_n \Tr\left[S^n \widetilde{\rho}_0 \left(S^\dagger\right)^n \Pi_n \right] = {\cal J}({\bf \Pi}) \, .
\end{align}
Thus, the two POVMs ${\bf \Pi}$ and ${\bf \Upsilon}$ has the same correct decision probability. We conclude that ${\bf \Upsilon}$ is an optimum measurement that satisfies the GUS. 
\end{proof}

From the previous proof, we note that, in the presence of a POVM ${\bf \Pi}=\{\Pi_j\}_j$ satisfying the GUS, the calculation of the correct decision probability reduces to:
\begin{align}\label{eq:PcGUS}
{\cal P}_c = M \Tr[\widetilde{\rho}_0 \Pi_0] = \Tr[\rho_0 \Pi_0]   \, .
\end{align}
Furthermore, the SDP problem~(\ref{eq:construction}) is also simplified as:
\begin{align}
\max_{\Pi_0 \ge 0} \Tr[\rho_0 \Pi_0] = \min_{\substack{\Lambda -\widetilde{\rho}_0\ge 0 \\[.3ex] [\Lambda, S]=0}} \Tr[\Lambda] \, ,
\end{align}
where the dual problem is now subject to the two constraints $\Lambda -\widetilde{\rho}_0\ge 0$ and $[\Lambda, S]= \Lambda S - S \Lambda =0$ \cite{Cariolaro2015, Assalini2010}.
Therefore, in the presence of GUS (and equiprobable symbols), the optimum POVM is completely specified by operators $S$, $\rho_0$ and $\Pi_0$, inducing a simplification in the numerical SDP algorithms providing solutions to the dual problem \cite{Cariolaro2015}.

\subsubsection{Pure-state discrimination}
We now address the pure-state discrimination case, where the presence of GUS provides analytic solution to the decision problem, leading to exact expressions for both the maximum correct decision probability and the optimum POVM \cite{Notarnicola2023:KB}.
 
In the presence of a pure-state constellation, specified by the state matrix $\Gamma=\big(|\gamma_k\rangle\big)_k$, $k=0,\ldots,M-1$, and equiprobable symbols $q_k=1/M$, the GUS is satisfied if there exists a unitary symmetry operator $S$, $S^\dagger S =\hat{\Id}$, such that \cite{Notarnicola2023:KB, Cariolaro2015, Ban1997, Eldar2001, Eldar2004}
\begin{align}
|\gamma_k\rangle= S^k \, |\gamma_0\rangle \qquad \mbox{and} \qquad S^M=\hat{\Id} \, .
\end{align}
In this case, the Gram matrix $G=\Gamma^\dagger \, \Gamma$ is a circulant matrix, having the form \cite{Notarnicola2023:KB, Cariolaro2015, Davis1970}
\begin{align}
G= 
\left(
\begin{array}{cccc} 
G_{00} & G_{M-1 \,0} & \ldots  & G_{10} \\
G_{10} & G_{00} &   \ldots & G_{20} \\
\vdots & \vdots & \ddots & \vdots \\
G_{M-1 \, 0} & G_{M-2 \,0} & \ldots & G_{00} \\
\end{array}
\right) \, .
\end{align}
That is, its elements only depend on the difference modulo $M$ between the indices, i.e. $G_{jk}= G_{(j-k) \, {\rm mod} M,0}$: indeed, $G_{jk}= \langle \gamma_j|\gamma_k\rangle = \langle \gamma_0| S^{k-j} |\gamma_0\rangle = \langle \gamma_{(j-k) \, {\rm mod} M}|\gamma_0\rangle$.

Furthermore, the following property holds.
\begin{proposition}\label{PropSandTcommute}
In the presence of GUS, the symmetry operator $S$ commutes with the Gram operator
\begin{align}\label{eq:GramOperator}
T \equiv \Gamma \, \Gamma^\dagger =\sum_{k=0}^{M-1} |\gamma_k\rangle \langle \gamma_k | \, ,
\end{align}
that is $[S,T]= ST-TS=0$.
\end{proposition}
\begin{proof}
By the definition of the Gram operator and the unitarity of $S$, for which $S^\dagger= S^{-1}$, we have $T= \sum_k S^k |\gamma_0\rangle \langle \gamma_0 | S^{-k}$, therefore:
\begin{align}
TS&= \sum_k S^k |\gamma_0\rangle \langle \gamma_0 | S^{-k+1}= S \sum_k S^{k-1} |\gamma_0\rangle \langle \gamma_0 | S^{-(k-1)} \nonumber \\[1ex]
& = S \sum_j S^{j} |\gamma_0\rangle \langle \gamma_0 | S^{-j} = ST \, .
\end{align}
\end{proof}

\subsubsection{Properties of circulant matrices}\label{sec:Circulant}
As will become clearer, the circulant structure is a fundamental tool to construct the optimal solution to the decision problem.
In particular, circulant matrices satisfy the following properties, that will be helpful throughout the text.
\begin{itemize}
\item Any circulant matrix $C=(C_{jk})_{jk}$, $j,k=0,\ldots, M-1$, $C_{jk} = C_{(j-k)\, {\rm mod} M, 0}$, is diagonalizable by the unitary matrix $\mathbb{U}=  \mathbb{F}^{-1}$, $\mathbb{F}=(F_{jk})_{jk}$ being the discrete Fourier transform matrix, with: \cite{Davis1970}
\begin{align}\label{eq:inverseDFT}
F_{jk}= \frac{e^{-i 2\pi j k/M}}{\sqrt{M}} \, ,  \quad j,k=0,\ldots,M-1 \, .
\end{align}
\item The spectral decomposition of $C$ then reads: 
\begin{align}
C= \mathbb{U} \, \Lambda_C \, \mathbb{U}\dag \, ,
\end{align}
where $\Lambda_C= {\rm diag}(\lambda_0,\ldots, \lambda_{M-1})$, is the diagonal matrix composed of the eigenvalues $\{\lambda_j\}_j$ of $C$, given by the discrete Fourier transform of the circulant vector ${\bf r}=(r_{p})_p \equiv (C_{p \, {\rm mod} M, 0})_p$, $p=0,\ldots, M-1$, namely \cite{Cariolaro2015, Davis1970}:
\begin{align}
\lambda_p = \sum_{q=0}^{M-1} {\mathbb{U}}_{pq} r_q \, .
\end{align}
\item Since all circulant matrices are diagonalized by the same unitary $\mathbb{U}$, we conclude that circulant matrices form a commutative algebra. That is, for any pair of circulant matrices $C_1$ and $C_2$, also $C_1C_2$ and $C_2 C_1$ are circulant, and $[C_1,C_2]=0$ \cite{Davis1970}. In particular, if $C$ is circulant, then $C\dag$ is also circulant and $[C,C\dag]=0$.
\end{itemize}

\subsubsection{Construction of the optimum POVM for pure-state constellations with GUS}\label{sec:OptwithGUS}
Thanks to the circulant property, we are now able to provide an exact derivation of the optimum POVM in the presence of a pure-state constellation satisfying the GUS for operator $S$ \cite{Notarnicola2023:KB}. To our knowledge, this method is original and, remarkably, it provides a simpler strategy to retrieve the optimum receiver than the traditional derivation, based on the symmetry of suboptimal receivers, that will be described in Sec.~\ref{sec:PGMderivation}.

By invoking Proposition~\ref{PropositionGUS}, we assume that also the optimal measurement vectors $\mathbb{M}=\big(|\mu_j\rangle\big)_j$, $j=0,\ldots,M-1$, exhibit the GUS for the same operator $S$, thus the optimum POVM $\{\Pi_j\}_j$ is identified by a single ``reference" measurement vector:
\begin{align}
|\mu_0\rangle = \sum_{k=0}^{M-1} a_{k0} \, |\gamma_k\rangle \, ,
\end{align}
$a_{k0} \in \mathbb{C}$, while all the others will be retrieved as $|\mu_j\rangle= S^j |\mu_0\rangle$, $j=0,\ldots, M-1$. Consequently, the matrix $A$ in~(\ref{eq:Mmatrix}) is a circulant matrix, as $a_{kj}= a_{(k-j) \, {\rm mod} M, 0}$.
Its eigendecomposition is given by 
\begin{align}
A= \mathbb{U} \, \Lambda_A \, \mathbb{U}\dag \, ,
\end{align}
where $\Lambda_A= {\rm diag}(\lambda_0,\ldots, \lambda_{M-1})$, is the diagonal matrix composed of the eigenvalues $\{\lambda_j\}_j$ of $A$. Furthermore, $A\dag$ is also circulant and commutes with $A$, thereby, $A\dag= \mathbb{U} \Lambda_A\dag \mathbb{U}\dag$ and Eq.~(\ref{eq:ConstA}) becomes:
\begin{align}
\mathbb{U} \, \left|\Lambda_A\right|^2 \, \mathbb{U}\dag = G^{-1} \, ,
\end{align}
where $|\Lambda_A|^2= {\rm diag}(|\lambda_0|^2,\ldots, |\lambda_{M-1}|^2)$. We conclude that $A$ and $G^{-1}$ are simultaneously diagonalizable and $|\lambda_j|^2= g_j^{-1}$, $\{g_j\}_j$ being the eigenvalues of the Gram matrix~(\ref{eq:Grammatrix}) listed in increasing order, that is, $g_0\ge g_1\ge \ldots\ge g_{M-1}$. In conclusion, the matrix $A$ may be re-expressed in the following form:
\begin{align}\label{eq:Aphi}
A\equiv A_{\boldsymbol{\phi}}=\mathbb{U} \, \Lambda_A^{(\boldsymbol{\phi})} \, \mathbb{U}\dag \,,
\end{align}
where:
\begin{align}
\Lambda_A^{(\boldsymbol{\phi})}= {\rm diag}\Bigg(\bigg\{\lambda_j^{(\boldsymbol{\phi})} \bigg\}_{j=0,\ldots, M-1}\Bigg) \, ,
\end{align}
and
\begin{align}
\lambda_j^{(\boldsymbol{\phi})} = e^{i \phi_j} g_j^{-1/2} \, ,
\end{align}
in which the relative phases $\boldsymbol{\phi}=(\phi_0,\ldots,\phi_{M-1})$ provide the only free parameters. Furthermore, the matrix $A_{\boldsymbol{\phi}}$ is defined up to an overall phase due to~(\ref{eq:Mmatrix}), therefore we may fix $\phi_0=0$, ending up with $M-1$ phases whose value can be arbitrarily chosen.
We conclude that, in the presence of GUS, every pure-state discrimination receiver is ultimately identified by the set of phases $\boldsymbol{\phi}$, which may be properly chosen to maximize the correct decision probability ${\cal P}_c$, thus transforming a convex functional optimization task into optimization of a real function with $M-1$ real variables \cite{Notarnicola2023:KB}.

Thanks to Eq.~(\ref{eq:PcGUS}), the correct decision probability, when both the state and measurment vectors verify GUS and $q_k=1/M$, reduces to:
\begin{align}
{\cal P}_c = |\langle \mu_0|\gamma_0\rangle |^2 = |B_{00}|^2 \, ,
\end{align}
with the matrix $B=A^\dagger G$ introduced in~(\ref{eq:Bmatrix}).
Since both $A^\dagger$ and $G$ are circulant, we conclude that also $B$ is circulant, and equal to:
 \begin{align}\label{eq:Bphi}
B\equiv B_{\boldsymbol{\phi}}=\mathbb{U} \, \Lambda_B^{(\boldsymbol{\phi})} \, \mathbb{U}\dag \,,
\end{align}
where:
\begin{align}
\Lambda_B^{(\boldsymbol{\phi})}= {\rm diag}\Bigg(\bigg\{e^{-i \phi_j} g_j^{1/2} \bigg\}_{j=0,\ldots, M-1}\Bigg) \, ,
\end{align}
with the quantities introduced in Eq.~(\ref{eq:Aphi}).
In turn, for every tuple of relative phases $\boldsymbol{\phi}=(\phi_0,\ldots,\phi_{M-1})$, the corresponding correct decision probability becomes:
\begin{align}
{\cal P}_c (\boldsymbol{\phi}) = \left|\left(B_{\boldsymbol{\phi}}\right)_{00}\right|^2 = \left| \frac{1}{M}\sum_{j=0}^{M-1} \left(e^{-i \phi_j} g_j^{1/2}\right) \right|^2\, .
\end{align}
That is, the correct decision probability is determined by the square modulus of the sum of the $M$ complex numbers $e^{-i \phi_j} g_j^{1/2}$. From geometric considerations, we conclude that its maximum value is achieved when the relative phases $\phi_j$ are all equal with one another. In turn, the optimum receiver is defined by the choice:
\begin{align}\label{eq:phioptDISC}
\boldsymbol{\phi}_\opt= {\bf 0}= (0,\ldots, 0) \, ,
\end{align}
with the corresponding maximum correct decision probability:
\begin{align}\label{eq:PcMAXwithGUS}
{\cal P}_c^{\rm (max)} = {\cal P}_c(\boldsymbol{\phi}_\opt) = \left| \frac{1}{M} \sum_{j=0}^{M-1} g_j^{1/2} \right|^2 = \left|\left(G^{1/2}\right)_{00}\right|^2\, ,
\end{align}
where $G^{1/2}$ is the square root of the Gram matrix.
Finally, the optimal matrix $A$ becomes 
\begin{align}\label{eq:AoptDISC}
A_\opt= G^{-1/2} \, ,
\end{align}
that defines the optimal measurement vectors via $\mathbb{M}_\opt = \Gamma \, A_\opt$ \cite{Notarnicola2023:KB}.

In conclusion, the presence of GUS makes the problem of designing the optimum receiver for pure-state discrimination solvable. The optimum POVM, as well as the maximum correct decision probability, is completely characterized by the Gram matrix $G$, i.e. by the geometry of the constellation.
However, from a wider viewpoint, we remark that Eq.~(\ref{eq:Aphi}) provides characterization of any $M$-valued projective measurement to be performed on a pure-state constellation, and its validity goes further beyond the framework of quantum discrimination. As a consequence, the phases $\boldsymbol{\phi}$ can be determined to optimize any desired figure of merit according to the context under investigation, not only the correct decision probability. As will be discussed in Sec.~\ref{sec: Konrad}, this will be the starting point to address the role of the class of $M$-valued quantum receivers for continuous-variable quantum key distribution.

\subsection{The pretty good measurement method}\label{sec:PGM}

In the previous sections, we described the fundamental results of quantum decision theory, providing characterization of the optimum POVM for $M$-ary discrimination. The optimum receiver has to satisfy Yuen's theorem, see Theorem~\ref{YuenThm}, which, however, does not yield a method to construct the associated POVM. In general, optimization can be carried out numerically by solving the dual problem~(\ref{eq:dualproblem}), and a closed-form analytic expression for the optimum POVM is known only in few cases, e.g. binary discrimination or $M$-ary pure-state constellations with equiprobable symbols satisfying the GUS.

In turn, one may claim for simpler methods that, although being suboptimal, reduce the complexity of the problem. That is we look for a suboptimal POVM that does not achieve the maximum correct decision probability anymore, but, in turn, that can be constructed in a simpler way with respect to the optimum receiver satisfying Theorem~\ref{YuenThm}. 
Within suboptimal methods, the most important is represented by the {\it pretty good measurement} (PGM) method, being also referred to as {\it square root measurement} (SRM) or {\it least square measurement} (LSM) method \cite{Ban1997, Hausladen1994, Hausladen1996, Sasaki1998, Sasaki1998b, Kato1999, Eldar2001, Eldar2002, Eldar2004}.

The method was firstly proposed in 1994 by Hausladen {\it et al.} \cite{Hausladen1994, Hausladen1996}, who considered the problem of distinguishing between an arbitrary set of pure states which, in general, can be linearly dependent. The authors propose a quantum receiver composed of $1$-rank operators that performs ``pretty good” discrimination, even though not being, in general, optimum.
Later on, this measurement has been also called SRM, since the pretty good measurement is obtained by computing the square root of the Gram operator \cite{Sasaki1998, Sasaki1998b, Kato1999}.
Instead, the LSM method was subsequently developed by Eldar and Forney in 2001 \cite{Eldar2001, Eldar2002}, with the goal of designing a $1$-rank POVM, projecting on measurement vectors that minimize the sum of the squared norms of the differences between each corresponding state and measurement vector.
In particular, Eldar and Forney investigated the connection between their method and the other types of measurements, proving the LSM technique to be equivalent to the SRM. Thereafter, the method has been extended to mixed state discrimination \cite{Eldar2004}.

The PGM is a cornerstone example of suboptimal POVM. In fact, its construction is simple, as it is directly derived from the given collection of states, it is ``pretty good” when the states to be distinguished are equiprobable and almost orthogonal, namely it is asymptotically optimal \cite{Hausladen1994, Hausladen1996, Holevo1979}. Remarkably, it becomes optimal in the presence of pure-state constellations satisfying the GUS.
For these reasons, in more recent times, it has been systematically applied to the performance evaluation of typical quantum communications systems \cite{Cariolaro2010, Cariolaro2010b}, and also experimentally implemented in the framework of cavity quantum electrodynamics \cite{Sasaki1998, Sasaki1998b}.

In the following, we present the detailed derivation of the PGM following the approach of Eldar and Forney, which provides a simple geometric interpretation. 
We start by reviewing a fundamental theorem of linear algebra, namely the singular value decomposition (SVD) of a matrix \cite{Cariolaro2015, Eldar2001, Friedberg2014}, which is widely invoked in the construction of the measurement.

\begin{theorem}[{\bf Singular value decomposition (SVD)}]\label{SVDThm}
 Let $\Phi$ be an arbitrary $n \times m$ complex matrix of rank $r \le \min\{n,m\}$.
Then, there exist a unitary $n \times n$ matrix $U$, a diagonal $n \times m$ matrix $\Sigma$, and a unitary $m\times m$ matrix $V$ such that:
\begin{align}
\Phi = U \, \Sigma \, V^\dagger = \sum_{i=0}^{r-1} \sigma_i \,  |u_i \rangle \langle v_i| \, ,
\end{align}
where:
\begin{enumerate}
\item[i)] $\Sigma$ is a diagonal $n \times m$ matrix whose first $r$ diagonal elements are $\sigma_i>0$, and whose remaining $m-r$ diagonal elements are $0$. The numbers $\sigma_i$, referred to as the {\it singular values} of $\Phi$, are equal to $\sigma_i= \sqrt{\lambda_i}$, where $\{\lambda_i\}_j$, $i=0,\ldots, r-1$, are the eigenvalues of the $r$-rank positive semidefinite matrix $S=\Phi^\dagger \Phi$;
\item[ii)] $V$ is a $m \times m$ unitary matrix whose first $r$ columns are the orthonormal eigenvectors $|v_i \rangle$, spanning a subspace ${\cal V} \subseteq \mathbb{C}^m$, and whose remaining $m-r$ columns span the orthogonal complement ${\cal V}^\perp \subseteq \mathbb{C}^m$. In particular, $V$ diagonalizes matrix $S=\Phi^\dagger \Phi$, as $\Phi^\dagger \Phi = V (\Sigma^\dagger \Sigma) V^\dagger = \sum_{i=0}^{r-1} \sigma_i^2 \,  |v_i \rangle \langle v_i|= \sum_{i=0}^{r-1} \lambda_i \,  |v_i \rangle \langle v_i|$;
\item[iii)] $U$ is a $n \times n$ unitary matrix whose first $r$ columns are the orthonormal eigenvectors $|u_i \rangle$, spanning a subspace ${\cal U} \subseteq \mathbb{C} ^n$, and whose remaining $n-r$ columns span the orthogonal complement ${\cal U}^\perp \subseteq \mathbb{C}^n$. In particular, $U$ diagonalizes matrix $T=\Phi \Phi^\dagger$, as $\Phi\Phi^\dagger = U (\Sigma \Sigma ^\dagger) U^\dagger = \sum_{i=0}^{r-1} \sigma_i^2 \,  |u_i \rangle \langle u_i|= \sum_{i=0}^{r-1}  \lambda_i\,  |u_i \rangle \langle u_i|$.
\end{enumerate}
\end{theorem}

In particular, we note that the eigenvectors of $U$ and $V$ provide a complete orthonormal system of $\mathbb{C}^n$ and $\mathbb{C}^m$, respectively, and further satisfy:
\begin{align}
\sum_{i=0}^{r-1}  |u_i\rangle \langle u_i|= \mathbb{P}_{\cal U} \qquad \mbox{and} \qquad \sum_{i=0}^{r-1}  |v_i\rangle \langle v_i|= \mathbb{P}_{\cal V} \, ,
\end{align}
$\mathbb{P}_{\cal U}$ and $\mathbb{P}_{\cal V}$ being the projectors onto subspaces $\cal U$ and $\cal V$, respectively.

However, in practical applications, computing the full SVD, including a full unitary decomposition of the null-space of $\Phi$, is rather useless. Instead, if the matrix $\Phi$ has no full rank, i.e. $r<\min\{n,m\}$, it suffices to adopt the {\it reduced SVD}, namely:
\begin{align}
\Phi = \sum_{i=0}^{r-1} \sigma_i \,  |u_i \rangle \langle v_i| = U_r \, \Sigma_r \, V^\dagger_r \, ,
\end{align}
where the matrices $U_r$ and $V_r$ only contain the first $r$ eigenvectors of $U$ and $V$, respectively, and $\Sigma_r$ is a $r \times r$ diagonal matrix that only contains the nonzero singular values $\sigma_i>0$. In turn, $U_r$ and $V_r$ become $n \times r$ and $r \times m$ matrices, respectively \cite{Cariolaro2015, Friedberg2014}.

Given these results, we are now ready to introduce the PGM method. Following Eldar and Forney's approach, we present the theory for pure-state discrimination and equiprobable symbols, as in the original proposal \cite{Eldar2001, Eldar2002}. The extension to mixed states will be briefly mentioned therafter.

%
%

\subsubsection{Derivation of the pretty good measurement}\label{sec:PGMderivation}
\begin{figure}
\includegraphics[width=0.5\columnwidth]{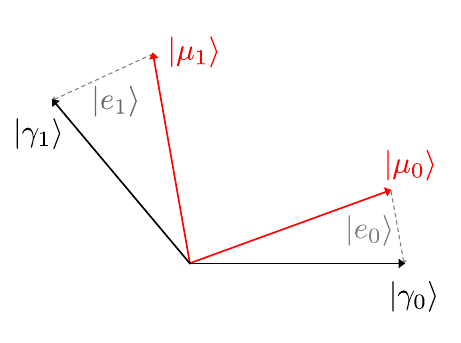}
\centering
\caption{Two-dimensional example of the PGM method. Given the constellation states $\{|\gamma_k\rangle\}_k$, we seek for a $1$-rank POVM $\{\Pi_j\}_j$, with $\Pi_j=|\mu_j\rangle \langle \mu_j|$, such that the error vectors $|e_j\rangle= |\gamma_j\rangle - |\mu_j\rangle$ minimize the squared error $E$ in Eq.~(\ref{eq:SQERR}), subject to the normalization constraint $\sum_j |\mu_j\rangle \langle \mu_j|= \mathbb{P}_{\cal S}$.}\label{fig:sec5.4_PGM}
\end{figure}

To begin with, we consider the problem of $M$-ary pure state discrimination of the constellation ${\cal C}= \{|\gamma_k\rangle\langle \gamma_k|\}_k$, $k=0,\ldots, M-1$, described by the state matrix:
\begin{align}
\Gamma = \bigg( |\gamma_0\rangle, \ldots, |\gamma_{M-1}\rangle \bigg) \, ,
\end{align}
see Sec.~\ref{sec:M=GammaA}. Even though in practical scenarios the constellation states are linearly independent, here we assume the possible presence of linearly dependent vectors, and let $\Gamma$ have rank $r \le M$, such that the subspace ${\cal S}={\rm span}\{|\gamma_k\rangle: k=0,\ldots, M-1\}$ has dimension $r\le M$.  Furthermore, we assume equiprobable symbols, namely equal a priori probabilities:
\begin{align}
q_k= \frac{1}{M} \, , \quad k=0,\ldots, M-1 \, .
\end{align}
The method can be extended to non-uniform generation by substituting the constellation states with the weighted states $\sqrt{q_k} |\gamma_k\rangle$ \cite{Eldar2001}.
The goal of the PGM scheme is to define a suitable $M$-valued $1$-rank POVM $\{\Pi_j\}_j$, $j=0,\ldots, M-1$, with $\Pi_j=|\mu_j\rangle \langle \mu_j|$, associated with the measurement matrix:
\begin{align}
\mathbb{M}= \bigg( |\mu_0\rangle, \ldots, |\mu_{M-1}\rangle \bigg) \, .
\end{align}
We may safely assume the measurement vectors to be $|\mu_j\rangle \in {\cal S}$, thus we have $\mathbb{M}=\Gamma\, A$ for some $M\times M$ matrix $A$, see Sec.~\ref{sec:M=GammaA}, and:
\begin{align}\label{eq:normSMR}
\mathbb{M} \mathbb{M}^\dagger = \sum_j |\mu_j\rangle \langle \mu_j|= \mathbb{P}_{\cal S} \, .
\end{align}
Instead, in general, we do not require the vectors $\{|\mu_j\rangle\}_j$ to be orthogonal or normalized \cite{Eldar2001, Eldar2002, Cariolaro2015, Notarnicola2023:KB}.

To construct such a measurement, we seek for measurement vectors $\{|\mu_j\rangle\}_j$ ``close" to the states $\{|\gamma_k\rangle\}_k$, i.e. making the difference vectors $|e_j\rangle =|\gamma_j\rangle - |\mu_j\rangle$ as ``small" as possible, as schematized in Fig.~\ref{fig:sec5.4_PGM}. More precisely, we look for a measurement vector set $\{|\mu_j\rangle\}_j$ that minimize the squared error:
\begin{align}\label{eq:SQERR}
E= \sum_{j=1}^{M} \langle e_j|e_j\rangle = \sum_{j=1}^{M} \bigg(\langle \gamma_j| - \langle \mu_j| \bigg) \bigg(|\gamma_j\rangle - |\mu_j\rangle \bigg) \, ,
\end{align}
subject to the normalization constraint~(\ref{eq:normSMR}) \cite{Eldar2001, Eldar2002, Cariolaro2015}. We note that, if the state vectors were orthonormal, the minimum of $E$, compatible with~(\ref{eq:normSMR}), would be trivially reached by the choice $|\mu_j\rangle=|\gamma_j\rangle$ for all $j$, with $E_{\rm min} = 0$.
On the contrary, the problem is nontrivial in the presence of non-orthogonal states. To determine the solution, we first re-express Eq.~(\ref{eq:SQERR}) in terms of the state and measurement matrices as:
\begin{align}\label{eq:SQERR1}
E= \Tr \left[ \left(\Gamma - \mathbb{M} \right)^\dagger \left(\Gamma - \mathbb{M} \right)\right]= \Tr \left[ \left(\Gamma - \mathbb{M} \right)  \left(\Gamma - \mathbb{M} \right)^\dagger\right] \, .
\end{align}
We now employ the reduced SVD of the state matrix $\Gamma$, i.e. $\Gamma= U_r \, \Sigma_r \, V^\dagger_r$. 
Since the columns of $\Gamma$ are composed by states $|\gamma_k\rangle \in {\cal S}$, we conclude that $\Gamma$ is a square $M\times M$ matrix
with rank $r\le M$. In turn, $U_r$ is a $M \times r$ unitary matrix that diagonalizes the Gram operator $T=\Gamma \, \Gamma^\dagger = \sum_k |\gamma_k\rangle\langle \gamma_k|$, whose eigenvectors provide a complete orthonormal system of the subspace $\cal S$, i.e. $\sum_{i=0}^{r-1} |u_i\rangle \langle u_i|= \mathbb{P}_{\cal S}$.
Given this considerations, we perform the trace in~(\ref{eq:SQERR1}) over the $U_r$-eigenbasis, obtaining:
 \begin{align}
E= \sum_{i=0}^{r-1} \langle u_i| \left(\Gamma - \mathbb{M} \right)  \left(\Gamma - \mathbb{M} \right)^\dagger |u_i\rangle = \sum_{i=0}^{r-1} \langle d_i| d_i\rangle \, ,
\end{align}
where
 \begin{align}
| d_i\rangle = \left(\Gamma - \mathbb{M} \right)^\dagger |u_i\rangle = \Gamma^\dagger |u_i\rangle - |a_i\rangle \, , \quad i=0,\ldots, r-1 \, ,
\end{align}
with $|a_i\rangle = \mathbb{M}^\dagger |u_i\rangle$.
Thanks to the reduced SVD of $\Gamma^\dagger$, i.e. $\Gamma^\dagger_r = V_r \, \Sigma_r \, U^\dagger_r = \sum_{i=0}^{r-1} \sigma_i |v_i\rangle \langle u_i|$, we have $\Gamma^\dagger |u_i\rangle = \sigma_ i |v_i\rangle$ for $i=0\ldots, r-1$. Moreover, vectors $\{|a_i\rangle\}_i$, $i=0,\ldots, r-1$, provide an orthonormal set of $\cal S$. In fact, we have $\langle a_i|a_j\rangle= \langle u_i| \mathbb{M} \mathbb{M}^\dagger |u_j\rangle = \langle u_i|\mathbb{P}_{\cal S} |u_j\rangle= \delta_{ij}$.

Then, the problem of designing a POVM minimizing~(\ref{eq:SQERR}) reduces to finding a set of $r$ orthonormal vectors $\{|a_i\rangle\}_i$ that minimizes
 \begin{align}
E= \sum_{i=0}^{r-1} \langle d_j| d_j\rangle = \sum_{i=0}^{r-1} \left[ \left(1+\sigma_i^2\right) - 2 \sigma_i {\rm Re}\left(\langle a_i|v_i\rangle \right) \right] \, .
\end{align}
The minimum is achieved when ${\rm Re}(\langle a_i|v_i\rangle)$ gets its largest value for all $i=0,\ldots, r-1$. Since also vectors $\{|v_i\rangle\}_i$ are orthonormal, we conclude that the minimizing vectors should be $|a_i\rangle = |v_i\rangle$,  $i=0,\ldots, r-1$ \cite{Eldar2001, Eldar2004, Cariolaro2015}.

In turn, the PGM is defined by the relation $\mathbb{M}_\PGM |u_i\rangle = |v_i\rangle$ or, equivalently:
\begin{align}\label{eq:PGM1}
\mathbb{M}_\PGM = U_r V_r^\dagger = \sum_{i=0}^{r-1} |u_i\rangle \langle v_i| \, ,
\end{align}
associated with the minimum square error $E_{\rm min}= \sum_{i=0}^{r-1} (1-\sigma_i)^2$.
Thanks to the reduced SVD of $\Gamma$, Eq.~(\ref{eq:PGM1}) can be expressed in equivalent way as a function of both the Gram matrix $G= \Gamma^\dagger \, \Gamma$ and the Gram operator $T= \Gamma \ \Gamma^\dagger$ as:
\begin{align}\label{eq:PGM}
\mathbb{M}_\PGM = U_r V_r^\dagger =  T^{-1/2} \Gamma = \Gamma \, G^{-1/2} \, ,
\end{align}
where $T^{-1/2}$ and $G^{-1/2}$ should be intended as square-root Moore-Penrose pseudo-inverses of $T$ and $G$, respectively, when matrix $\Gamma$ is not full-rank, i.e. $r<M$.
In particular, from Eq.~(\ref{eq:PGM}) we obtain the PGM measurement vectors as:
\begin{align}
|\mu_j\rangle_\PGM = T^{-1/2} |\gamma_j\rangle \, , \quad j=0,\ldots, M-1 \, ,
\end{align}
and $\Pi_j^{(\PGM)}= T^{-1/2} |\gamma_j\rangle \langle \gamma_j| T^{-1/2}$.
Equivalently, given the relation $\mathbb{M}=\Gamma \, A$ derived in Sec.~\ref{sec:M=GammaA}, the coefficient matrix $A$ for the PGM reads:
\begin{align}
A_\PGM= G^{-1/2} \, .
\end{align}
We also note, that the PGM measurement vectors are orthogonal only in the presence of linearly independent constellation states, when matrix $\Gamma$ has full rank $r=M$. Indeed, in that case we have 
\begin{align}
\mathbb{M}_\PGM^\dagger \, \mathbb{M}_\PGM = \bigg({}_\PGM\langle \mu_j|\mu_k \rangle_\PGM\bigg)_{jk}=G^{-1/2} \, G \, G^{-1/2} =  \Id_M \, .
\end{align}
On the contrary, if $r<M$, we should consider Moore-Pensore pseudo-inverses and the previous argument does not hold anymore. Then, the $M$ measurement vectors cannot be mutually orthonormal since they span the $r$- dimensional subspace $\cal S$.

Finally, we now evaluate the correct decision probability associated with the PGM with the methods of Sec.~\ref{sec:Gamma=MB}. We obtain the matrix $B=\mathbb{M}^\dagger \, \Gamma = A^\dagger \, G$ as:
\begin{align}
B_\PGM= G^{1/2} \, ,
\end{align}
such that the probability of obtaining outcome $j$ when state $|\gamma_k\rangle$ was sent reads $p_\PGM (j|k) = |{}_\PGM\langle \mu_j|\gamma_k\rangle|^2 = |B_{kj}|^2 = |(G^{1/2})_{kj}|^2$. The correct decision probability then becomes:
\begin{align}
{\cal P}_c^{(\PGM)}= \frac{1}{M} \sum_{k=0}^{M-1} \left|\left(G^{1/2}\right)_{kk} \right|^2 \, ,
\end{align}
where we recall that we considered equiprobable symbols.

\subsubsection{On the optimality of the PGM}\label{sec:OptimalityPGM}
Generally speaking, the PGM provides a suboptimal POVM for the decision problem, leading to a lower bound of the maximum achievable correct decision probability. This raises the problem to understand how ``close" it is with the respect to the optimum receiver. 
As a first fundamental result, the PGM is asymptotically optimal, namely it becomes optimal when the Gram matrix tends to the identity $G \to \Id_M$. As firstly, proved by Holevo in 1979 with independent methods, the minimum error probability $P_{\rm min}= 1 - {\cal P}_c^{(\rm max)}$ is related to the PGM error probability $P_\PGM= 1 -{\cal P}_c^{(\PGM)}$ by: \cite{Holevo1979}
\begin{align}
P_{\rm min} \le P_\PGM \le \frac{2}{M} \Tr \left[\Id_M - G^{1/2}\right] \, ,
\end{align}
thus in the limit $G \to \Id_M$ we have $P_\PGM \approx P_{\rm min} $.

More importantly, Eldar and Forney proved that the PGM becomes optimal when the constellation states satisfy the GUS with equiprobable symbols \cite{Eldar2001, Cariolaro2015}.
Indeed, we note that the PGM matrix $A_\PGM=G^{-1/2}$ coincides with the optimum one derived in Sec.~\ref{sec:OptwithGUS} for pure-state constellations with GUS. Equivalently, we prove that matrix $B_\PGM= G^{1/2}$ satisfies the necessary and sufficient conditions for optimality outlined in Corollary~\ref{CorollaryYuen+Kennedy}.
That is, we have to prove that $B_\PGM=(B_{kj})_kj$ verifies:
\begin{subequations}\label{eq:YuenPureStatePGM}
\begin{align}\label{eq:YuenPureState1PGM}
B_{jj}^{*} B_{kj} - B_{kk} B_{jk}^{*} = 0 \, , \quad  k,j=0,\ldots, M-1 \, ,  
\end{align}
\begin{align}\label{eq:YuenPureState2PGM}
\left(\sum_{s=0}^{M-1} B_{ss} |\mu_s\rangle_\PGM \langle \gamma_s| \right) - |\gamma_j\rangle \langle \gamma_j| \ge 0 \, , \quad j=0,\ldots, M-1 \, ,
\end{align}
\end{subequations}
where we considered equiprobable symbols, i.e. $q_k=1/M$.
We remark that, in the presence of GUS, $G$ is circulant, thus $B_\PGM= G^{1/2}$ is also circulant and its element only depends on the difference between indices, i.e. $B_{jk} = \langle \mu_j|\gamma_k\rangle = w(k-j)$, where $w$ is a complex-valued function satisfying $w^*(j)=w(-j)$. Then, we have:
 \begin{align}\label{eq:YuenPureState1PGM}
B_{jj}^{*} B_{kj} - B_{kk} B_{jk}^{*} = w(0)^* w(j-k) - w(0) w(k-j)^*= 0 \, ,  
\end{align}
and condition~(\ref{eq:YuenPureState1PGM}) is satisfied.
To prove~(\ref{eq:YuenPureState2PGM}), we consider the operator:
\begin{align}
L=\sum_{s} B_{ss} |\mu_s\rangle_\PGM  \langle \gamma_s| = w(0) \sum_{s} |\mu_s\rangle_\PGM \langle \gamma_s| \, .
\end{align}
We now construct the reduced SVD of $\Gamma$ in the presence of GUS.
First of all, we note that, in the presence of GUS, $\Gamma$ is full-rank and the Gram matrix $G$ is diagonalized by the unitary $\mathbb{F}^{-1}$, $\mathbb{F}$ being the discrete Fourier transform matrix, see Sec.~\ref{sec:Circulant}. Therefore, the right unitary $V_r$ in the SVD corresponds to $\mathbb{F}^{-1}$, its column vectors being equal to $|F_i\rangle = (e^{2\pi i j k}{M}/\sqrt{M})_j$, $i,j=0,\ldots, M-1$. Furthermore, the matrix of the singular values $\Sigma={\rm diag}(\{\sigma_j\}_j)$ has nonzero diagonal values, and rank equal to $M$. Then, it is easy to check that the choice $\mathbb{Y}= \Gamma (\mathbb{F}^{-1}) \Sigma^{-1}$ yield the left unitary matrix $U_r$ of the SVD. In turn, we have:
\begin{align}
\Gamma = \mathbb{Y} \, \Sigma \left(\mathbb{F}^{-1}\right)^\dagger = \sum_{i=0}^{M-1} \sigma_i \, |u_i\rangle \langle F_i| \, ,
\end{align}
where the eigenvectors of $\mathbb{Y}$ are equal to $|u_i\rangle = \Gamma\, |F_i\rangle/\sigma_i$, $i=0,\ldots, M-1$.
The PGM is obtained as:
\begin{align}
\mathbb{M}_\PGM= \mathbb{Y} \left(\mathbb{F}^{-1}\right)^\dagger= \sum_{i=0}^{M-1} |u_i\rangle \langle F_i| \, .
\end{align}
As a consequence, the following relations hold:
Then,
\begin{align}
|\gamma_s\rangle &= \mathbb{Y} \Sigma |F_s\rangle \quad \mbox{and}\quad  |\mu_s\rangle_\PGM = \mathbb{Y} |F_s\rangle \, ,
\end{align}
and Eq.~(\ref{eq:YuenPureState2PGM}) reduces to:
\begin{align}
\mathbb{Y} \Big( w(0)\Sigma - \Sigma |F_j\rangle \langle F_j | \Sigma \Big) \mathbb{Y}^\dagger \ge 0 \,, \quad  j=0,\ldots, M-1 \, .
\end{align}
It is, therefore, sufficient to show that operator $T_j=w(0)\Sigma - \Sigma |F_j\rangle \langle F_j | \Sigma $ is positive semidefinite for all $j$. To this aim, we recall that $w(0)= {}_\PGM\langle \mu_j|\gamma_j\rangle = \langle F_j|\Sigma |F_j\rangle$ and we invoke the Cauchy-Schwartz inequality \cite{Friedberg2014}. Thereby, for any $|h\rangle \in {\cal S}$, we have:
 \begin{align}
\langle h| T_j |h \rangle &= \langle F_j|\Sigma |F_j\rangle \langle h| \Sigma |h \rangle - |\langle h| \Sigma |F_j\rangle|^2 \nonumber \\[1ex]
&\ge \langle F_j|\Sigma |F_j\rangle \langle h| \Sigma |h \rangle - \langle h| \Sigma |h \rangle \langle F_j|\Sigma |F_j\rangle\nonumber \\[1ex]
&=0 \, .
\end{align}
Thus, also condition~(\ref{eq:YuenPureState2PGM}) is verified, proving the PGM as the optimum receiver in the presence of GUS.

\subsubsection{Extension to mixed states}
The PGM has been originally formulated for pure-state discrimination, in which case the construction of the POVM follows the geometric intuition depicted in Fig.~\ref{fig:sec5.4_PGM}.
However, in 2004 Eldar and Forney derived an extension of the method for mixed-state discrimination of constellations ${\cal C}=\{\rho_k\}_k$, $k=0,\ldots, M-1$. \cite{Cariolaro2015, Eldar2004}.
The principle behind the generalization is the so-called {\it factor decomposition} of the density operators $\rho_k$ \cite{Cariolaro2015}.

The factor decomposition of a general positive semidefinite matrix $\rho$ is equal to:
\begin{align}
\rho=\gamma \gamma^\dagger \, ,
\end{align} 
achieved for some matrix $\gamma$, which does not need to be diagonal. The existence of a factor follows from the positivity of $\rho$. In particular, given the spectral decomposition of $\rho$, i.e. $\rho=\sum_{j=0}^{r-1} \rho_j |\phi_j\rangle \langle \phi_j|= \Phi D \Phi^\dagger$, with $\rho_j\ge 0$, $r={\rm rank}(\rho)$, $D={\rm diag}(\{\rho_j\}_j)$ and $\Phi=(|\phi_0\rangle, \ldots, |\phi_{r-1}\rangle)$, we immediately derive $\gamma= \Phi \sqrt{D}$ as a factor of $\rho$ \cite{Cariolaro2015}.

Once obtained the factors $\gamma_k$ for each state $\rho_k$, we define the state matrix $\Gamma$ as:
\begin{align}
\Gamma= \bigg( \gamma_0 , \ldots, \gamma_{M-1} \bigg) \, ,
\end{align}
where the number of columns is now equal to $H\ge M$, with the associated Gram matrix and operator:
\begin{align}
G= \Gamma^\dagger \Gamma=\left(\gamma^\dagger_j \gamma_k\right)_{jk} \quad \mbox{and} \quad T= \Gamma \, \Gamma^\dagger = \sum_{j=0}^{M-1} \gamma_j \gamma_j^\dagger \, ,
\end{align}
where $G_{jk}=\gamma^\dagger_j \gamma_k$ now is the $(j,k)$-block of matrix $G$, $j,k=0,\ldots, M-1$ \, .

We now look for a POVM $\{\Pi_j\}_j$ in the form $\Pi_j=\mu_j \mu_j^\dagger$, where now the $\{\mu_j\}_j$ represent the measurement factors. Then, the construction in Sec.~\ref{sec:PGMderivation} still holds and the PGM is defined, accordingly, by:
\begin{align}
\mathbb{M}_\PGM= (\mu_0, \ldots, \mu_{M-1}) = T^{-1/2} \, \Gamma = G^{-1/2} \, \Gamma \, .
\end{align}
We have $B_\PGM=\mathbb{M}_\PGM^\dagger \Gamma= G^{1/2}$, with the corresponding correct decision probability:
\begin{align}
{\cal P}_c^{(\PGM)}= \frac{1}{M} \sum_{k=0}^{M-1} \Tr\left[\left(G^{1/2}_{kk}\right)^\dagger G^{1/2}_{kk} \right] \, ,
\end{align}
$G_{kk}^{1/2}$ being the $(k,k)$-block of $G^{1/2}$.

As regards the optimality, we underline that, differently from the case of pure states, in the presence of mixed-state constellation with GUS, the PGM, in general, is not optimal anymore.
To achieve optimality, a further condition is required, namely:
\begin{align}\label{eq:Gusmixed}
\left(B_\PGM\right)_{00}= G^{1/2}_{00} = \alpha \, \Id_{H} \, ,
\end{align}
where $\alpha$ is an arbitrary proportionality constant and $\Id_H$ is the $H \times H$ identity matrix \cite{Eldar2004}. Satisfying condition~(\ref{eq:Gusmixed}) is nontrivial. As an example, in the commonly exploited quantum communication systems, based on phase-shift keying and pulse-position modulation, Eq.~(\ref{eq:Gusmixed}) is not verified, thus the PGM is not optimum even in the presence of GUS \cite{Cariolaro2015}.

\subsection{Quantum receivers for quadrature phase-shift keying discrimination}\label{sec:QPSKdiscr}
\begin{figure}
\centerline{\includegraphics[width=0.5\columnwidth]{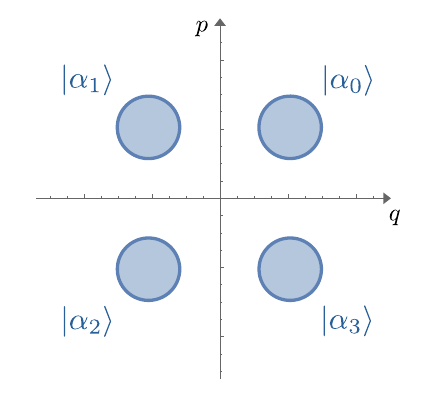}}
\centering
\caption{Phase space representation of the QPSK encoding, where information is encoded in the coherent states $|\alpha_k\rangle=|\alpha \, e^{\pi (2k+1)/M}\rangle$, $k=0,\ldots, M-1$ and $M=4$. The constellation satisfies the GUS for the symmetry operator $S_\theta= \exp(-i \theta a^\dagger a)$, $\theta=2\pi/M$, namely the phase-shift operator.}\label{fig:01:sec5.5-QPSK}
\end{figure}

As in the previous Section, we now consider the application of the quantum decision theory developed in the former section to a realistic quantum communication scenario, involving $M$-ary coherent state encoding, see Sec.~\ref{chap:QComm}.
Differently from Sec.~\ref{sec4:BinaryCoh}, in the presence of multilevel communication systems, there exist different formats for coherent state modulation, e.g. phase-shift keying, quadrature amplitude modulation, amplitude phase-shift keying, whose corresponding constellations exhibit different kinds of symmetry.
Here, we focus our attention on a paradigmatic example, namely phase-shift keying (PSK), that provides the simplest scheme for practical implementation, as it only requires phase modulation of a carrier laser beam with given intensity \cite{Cariolaro2015}.
In more detail, a PSK($M$) constellation is composed of the $M$ coherent states:
\begin{align}
|\alpha_k\rangle= |\alpha \, e^{i \pi (2k+1)/M} \rangle \, , \qquad k=0,\ldots, M-1 \, , 
\end{align}
where $\alpha\ge0$, generated with equal a priori probabilities $q_k=1/M$. That is, the constellation is composed of $M$ coherent states with the same energy $\alpha^2$ and phase-shifted by $\theta=2\pi/M$; therefore, it satisfies the GUS for the phase-shift symmetry operator $S_\theta= \exp(- i \theta \hat{n})$, $\hat{n}$ being the photon-number quantum operator \cite{Cariolaro2015, Notarnicola2023:KB}. 
In the following, we focus on the special case $M=4$, reported in Fig.~\ref{fig:01:sec5.5-QPSK}, also referred to as {\it quadrature phase-shift keying} (QPSK). The QPSK constellation is a cornerstone example in $M$-ary quantum communications, being investigated in several frameworks, ranging from quantum decision theory \cite{Cariolaro2015, Bondurant1993, Becerra2011, Muller2012, Izumi2012, Izumi2013, Becerra2013, Muller2015, DiMario2018, Izumi2020, Sidhu2021}, information transmission over a quantum channel \cite{Blahut1988, Griffin2002, Kramer2003, Ly-Gagnon2006, Gursoy2007, Bhadani2020}, and continuous variable quantum key distribution \cite{Notarnicola2023:KB, Leverrier2009, Liao2018, Ghorai2019, Lin2019, Lin2020, Papanastasiou2021, Lupo2022, Zhao2024}.

The QPSK scheme satisfies the GUS, therefore the PGM provides the optimal solution to the decision problem. That is, the optimum receiver, leading to the maximum correct decision probability, is obtained by the $1$-rank projective POVM $\{\Pi_j\}_j$, $\Pi_j=|\mu_j\rangle \langle \mu_j|$, $j=0,\ldots, M-1$, satisfying $\mathbb{M}= \Gamma \, A$ with the optimal coefficient matrix $A=G^{-1/2}$, see Sec.~\ref{sec:OptwithGUS}, where the Gram matrix $G=(G_{jk})_{jk}$, $j,k=0,\ldots, 3$, is defined as:
\begin{align}
G_{jk}= \langle \alpha_j|\alpha_k\rangle &= \exp\left\{-\alpha^2 \left[1- \cos\left(\frac{\pi (k-j)}{2} \right)\right] \right\} \times \nonumber \\[2ex]
& \hspace{2cm} \exp\left\{ - i \, \alpha^2 \sin\left(\frac{\pi (k-j)}{2} \right) \right\} \, .
\end{align}

Thereafter, we obtain the minimum error probability allowed by quantum mechanics as:
\begin{align}\label{eq:QuaternaryHB}
\Pmin = 1 - \left|\left(G^{1/2}\right)_{00}\right|^2 = 1- \frac{1}{16} \left( \sum_{j=0}^{3} g_j^{1/2} \right)^2 \, ,
\end{align}
where $\{g_j\}_j$ are the eigenvalues of the Gram matrix. Since $G$ is circulant, it is diagonalized by the inverse discrete Fourier transform matrix $\mathbb{F}^{-1}$ in Eq.~\ref{eq:inverseDFT}, therefore $g_j = (\mathbb{F} \, G \, \mathbb{F}^{-1})_{jj}$. Straightforward calculation leads to:
\begin{align}
g_0&=2 e^{-\alpha^2}(\cosh\alpha^2 + \cos\alpha^2)\, , \nonumber \\
g_1&=2 e^{-\alpha^2}(\sinh\alpha^2 + \sin\alpha^2)\, , \nonumber \\
g_2&=2 e^{-\alpha^2}(\cosh\alpha^2 - \cos\alpha^2)\, , \nonumber \\
g_3&=2 e^{-\alpha^2}(\sinh\alpha^2 - \sin\alpha^2)\, .
\end{align}
In literature, the error probability~(\ref{eq:QuaternaryHB}) is sometimes referred to as the {\it QPSK Helstrom bound} (with partial abuse of language), since derivation of the optimum QPSK receiver was also carried out by Helstrom in~\cite{Helstrom1976}.
We also note that the optimum POVM, equal to the PGM, corresponds to projection over a suitable linear superpositions of the constellation states $\{|\alpha_k\rangle\}_k$, i.e. $|\mu_j\rangle = \sum_k (G^{-1/2})_{kj} |\alpha_k\rangle$, as for binary coherent state discrimination.
However, differently from the binary case where the PGM is implemented via the Dolinar receiver \cite{Dolinar1973, Assalini2011}, in the presence of QSPK designing a feasible optimum receiver is an open problem, and, from a practical point of view, there is no clear idea on its experimental implementation \cite{Notarnicola2023:KB}.

On the other hand, the standard quantum limit (SQL), namely the error probability associated with conventional QPSK receivers, is achieved by double homodyne detection. That is, we perform joint measurement of quadratures $q$ and $p$ on the incoming signal, retrieving a pair of real outcomes ${\bf x}=(x,y)\in\mathbb{R}^2$, and adopt the following decision rule: if both $x,y \ge 0$ we infer state ``0", if $x<0$ and $y \ge 0$ we infer ``1", if both $x,y<0$ we infer ``2" and, finally, if $x\ge 0$ and $y<0$, we infer ``3".
The double homodyne probability distribution of state $|\alpha_k\rangle$ reads:
\begin{align}
p_\HET ({\bf x}|k) 
= \frac{1}{4\pi} 
& \exp\left\{- \frac{\left[x-2 \, {\rm Re}(\alpha_k)\right]^2+\left[y-2 \,{\rm Im}(\alpha_k)\right]^2}{4} \right\} \, ,
\end{align}
expressed in shot-noise units, such that the probability $p_\HET(j|k)$ of performing the decision $j$ when state $k$ is sent is equal to $p_\HET(j|k)= \int_{Q_j} d{\bf x} \, p_\HET ({\bf x}|k) $, where $Q_j$ is the $(j+1)$-th quadrant of the $(x,y)$ plane, corresponding to the confidence region derived from the previous decision rule. In particular, for all $k=0,\ldots, 3$, we have:
\begin{align}
p_\HET(k|k)= p_\HET(0|0) = \int_{0}^{\infty} dx \int_{0}^{\infty} dy \, p_\HET ({\bf x}|k) = \left[\frac{1 + \erf \left(\alpha/\sqrt{2}\right)}{2} \right]^2 \, ,
\end{align}
and, accordingly, we retrieve the SQL as:
\begin{align}\label{eq:QuaternarySQL}
\PSQL = 1-\frac14 \sum_{k=0}^3 p_\HET(k|k) = 1- \frac14 \left[1 + \erf \left(\frac{\alpha}{\sqrt{2}}\right) \right]^2 \, ,
\end{align}
such that $\PSQL > \Pmin$. In particular, in the high-energy limit $\alpha^2 \gg 1$, we have $\PSQL \approx \sqrt{2/(\pi \alpha^2)} e^{-\alpha^2/2}$ and $\Pmin \approx e^{-2\alpha^2}/2$: thereby, the two error probabilities show different dependence on the input energy, and the ratio $\PSQL/\Pmin \to \infty$ when $\alpha^2 \to \infty$.

As a consequence, we claim to design quantum receivers that beat the SQL and provide a genuine quantum advantage, even though they are not able to reach the minimum error probability.
Following the philosophy adopted for BPSK discrimination, the first idea in this direction is to provide suitable generalization of the Dolinar receiver, and investigate its optimality. This approach has been carried out by Bondurant in 1993, who designed a QPSK feedback receiver based on conditional nulling displacements \cite{Bondurant1993}. Unfortunately, the Bondurant receiver is not optimum, and, to date, it is not known whether or not the PGM can be implemented by optical feedback and linear optics. Nevertheless, the Bondurant receiver outperforms the SQL in the high-energy limit, thus providing anyway a cornerstone example in the field of $M$-ary quantum communication systems. Its functioning is explained in detail here below.

\subsubsection{The Bondurant receiver}

Starting from the Dolinar receiver, that provides optimal binary discrimination by conditional time-dependent displacements and feedback control, see Sec.~\ref{subsec4:Dolinar}, it is possible to design feedback receivers for $M$-ary discrimination by suitable extension of the scheme in Fig.~\ref{fig:sec4.3_DolinarSetup} \cite{Jabir2020}.
The first attempt in this direction has been made in 1993 by Bondurant, who proposed a feedback receiver for QPSK discrimination \cite{Bondurant1993}. In particular, he designed two kinds of receivers, referred to as {\it type I} and {\it type II Bondurant receiver}, respectively, and obtain a near-optimum performance, beating the SQL in the high-energy regime.

As in Sec.~\ref{subsec4:Dolinar}, since we deal with coherent states, the analysis can be conducted by considering the time-dependent wavepackets $\psi_k(t)$, associated with the encoded states $|\alpha_k\rangle$, $k=0,\ldots, 3$, equal to:
\begin{align}
\psi_k(t)=e^{i\pi (2k+1)/4} \, \psi \,  e^{-i \omega t} \, , \qquad 0 < t \le T \, ,
\end{align}
with $\psi>0$, and $\omega$ and $T$ being the carrier signal frequency and the time slot duration, respectively. Accordingly, the mean energy of each pulse reads $\bar{n}_k=\psi^2 T= \alpha^2$. 

\begin{figure}
\includegraphics[width=0.65\columnwidth]{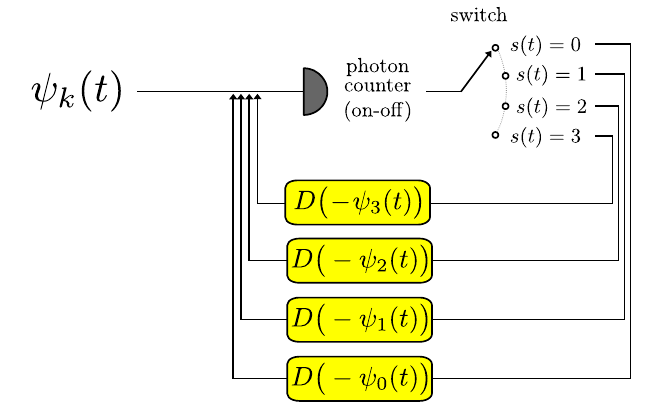}
\centering
\caption{Setup of the Bondurant receiver. The field $\psi_k(t)$ undergoes a nulling displacement operation $D(-\psi_j(t))$, $j,k=0,\ldots,3$, the value of $j$ being determined by the position of a switch $s(t)$, with initial condition is $s(0)=0$. In the type I receiver, $s(t)$ increases in sequential fashion at every count of the photodetector, whereas in the type II receiver $s(t)$ is suitably determined from the the time arrival distribution of the clicks. After time $T$, the value $s(T)$ gives the final decision.}\label{fig01:sec5.5.1_BondurantSetup}
\end{figure}

Similarly to the Dolinar scheme, the Bondurant receiver implements a feedback loop, applying a time-dependent ``nulling" displacement $D(-\psi_j(t))$ to the incoming signal $\psi_k(t)$, $j,k=0,\ldots,3$, where the nulled symbol $j$ is decided according to the outcome of a photodetector, performing continuous-time measurement
The receiver is realized by a photon counter performing on-off detection connected to a switch $s$, that, now, can assume the four positions, namely $s=0,\ldots, 3$, as depicted in Fig.~\ref{fig01:sec5.5.1_BondurantSetup}.
The position $s(t)$ at time $t\le T$ determines the amplitude of a nulling displacement operation $D(-\psi_{j}(t))$, with $j=s(t)$, to be performed on the field $\psi_k(t)$. To determine the value of $s(t)$ at each time, different switching rules are considered for the type I and type II scheme.

In particular, the type I Bondurant receiver tries to null out the incoming field {\it in sequential order}.
The initial position of the switch is set to $s(0)=0$, and the signal at time $t=0$ is displaced by $D(-\psi_0(0))$. Then, at every click of the detector, the switch increases its position by $1$, proceeding in sequential fashion, namely $0\to 1\to 2\to 3$. As an example, if the detector register clicks at times $t_1$, $t_2$ and $t_3$, the displacement is changed to $D(-\psi_1(t))$ for $t_1 < t \le t_2$, $D(-\psi_2(t))$ for $t_2 < t \le t_3$ and $D(-\psi_3(t))$ for $t_3 < t \le T$.
After the whole signal is processed, the switch position $s(T)$ infers the probed state to be $|\alpha_{s(T)}\rangle$.

\begin{figure}
\includegraphics[width=0.6\columnwidth]{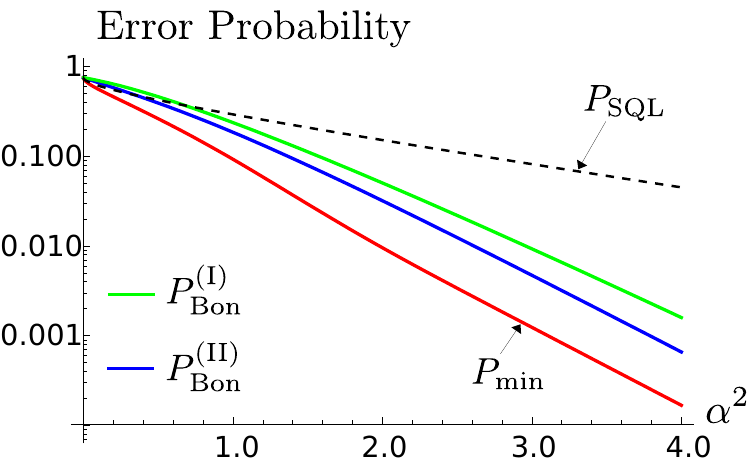}
\centering
\caption{Log plot of $\PBon^{(\rm p)}$, $\rm p=I,II$, as a function of the signal energy $\alpha^2$. $\PSQL$ and $\Pmin$ refer to the SQL~(\ref{eq:QuaternarySQL}) and the minimum error probability~(\ref{eq:QuaternaryHB}) achieved by the PGM, respectively. Type I and type II receivers beat the SQL for $\alpha^2 \ge \alpha^2_{\rm p}$, $\rm p=I,II$, whereas, in the limit $\alpha^2 \gg 1$, we have $\PBon^{(\rm I)}\approx \alpha^2 e^{-2\alpha^2}$ and $\PBon^{(\rm II)}\approx e^{-2\alpha^2}$, proving type II receiver to be near-optimum.}\label{fig02:sec5.5.1_BondurantPlot}
\end{figure}

The calculation of the error probability follows from the properties of Poisson stochastic processes \cite{Parzen1999}.
In particular, if state $k$ is probed and symbol $j$ is nulled, the displaced field $D(-\psi_j(t)) \psi_k(t)= \psi_k(t) - \psi_j(t)$, is associated with Poisson photon counting statistics, and the outcome of the photon counter is described as a stationary Poisson stochastic process with rate:
\begin{align}
\lambda_{jk}= |\psi_k(t)-\psi_j(t)|^2 = 2 \psi^2 \left\{1- \cos\left[\frac{\pi (j-k)}{2}\right]\right\} \, ,
\end{align}
such that the probability of registering a click in a time bin of duration $\delta t \ll T$ is equal to $\lambda_{jk}\delta t$ \cite{Parzen1999}.
In turn, the probability $\pi_{jk}^{(0)}(t_1,t_2)$ that the detector does not click in the time interval $(t_1,t_2]$ reads:
\begin{align}
\pi_{jk}^{(0)}(t_1,t_2) = \lim_{\delta t \to 0} \Big(1- \lambda_{jk} \delta t \Big)^{\frac{t_2-t_1}{\delta t}} = e^{-\lambda_{jk}(t_2-t_1)} \, .
\end{align}
Given this results, the receiver performs a decision error when not enough photocounts are registered, i.e. when the detector at time $T$ experiences $n_{\rm c}<k$ clicks if state $k$ is probed, with the consequence that the ``nulling" displacement $D(-\psi_k(t))$ is not implemented at all.
Accordingly, the error probability $\PBon^{\rm (I)}(k)$ given state $k$ is equal to
\begin{align}
\PBon^{\rm (I)}(k)= \sum_{n_{\rm c}=0}^{k-1} p(n_{\rm c}|k)\, ,
\end{align}
where $p(n_{\rm c}|k)$ is the probability of retrieving $n_{\rm c}$ clicks if state $k$ was sent.
If $k=0$ no errors are made, since state ``0" is always nulled thanks to the initial condition $s(0)=0$. On the contrary, if $k=1$ an error occurs if no clicks are registered over the whole interval $(0,T]$, as in this case the receiver always implements $D(-\psi_0(t))$ and, ultimately, performs the decision ``0". The error probability $\PBon^{\rm (I)}(1)$ reads:
\begin{align}
\PBon^{\rm (I)}(1)= p(n_{\rm c}=0|1)= \pi_{01}^{(0)}(0,T) = e^{-2\psi^2 T}= e^{-2\alpha^2} \, .
\end{align}
If state $k=2$ is sent, we have a decision error if either zero or one click is registered. The probability in the former case is equal to $p(n_{\rm c}=0|2)= \pi_{02}^{(0)}(0,T)=\exp(-4\alpha^2)$, whereas in the latter one we have a single click at time $t_1\le T$, thus:
\begin{align}
p(n_{\rm c}=1|2) &= \int_0^T dt_1 \, \pi_{02}^{(0)}(0,t_1) \, \lambda_{02} \, \pi_{12}^{(0)}(t_1,T) \nonumber \\
&= \int_0^T dt_1 \, e^{-4\psi^2 t_1} \left(4\psi^2 \right) e^{-2\psi^2 (T-t_1)} \nonumber \\
&= 2 e^{-2\alpha^2} \left(1-e^{-2\alpha^2} \right) \, .
\end{align}
Summing up the two contributions, we get:
\begin{align}\label{eq:pBon2}
\PBon^{\rm (I)}(2)= 2 e^{-2\alpha^2} -e^{-4\alpha^2} \, .
\end{align}
Finally, in the case $k=3$ errors are obtained if the photon counter clicks at most twice. We have:
\begin{subequations}\label{eq:p013}
\begin{align}
p(n_{\rm c}=0|3) &= e^{-2\alpha^2} \, , \\
p(n_{\rm c}=1|3) &= e^{-2\alpha^2} -e^{-4\alpha^2} \, ,
\end{align}
\end{subequations}
while the probability of getting two clicks at times $t_1\le T$ and $t_2>t_1$ reads:
\begin{align}\label{eq:p23}
p(n_{\rm c}=2|3) &= \int_0^T dt_1 \int_{t_1}^T dt_2 \, \pi_{03}^{(0)}(0,t_1) \, \lambda_{03} \, \pi_{13}^{(0)}(t_1,t_2) \, \lambda_{13} \, \pi_{23}^{(0)}(t_2,T) \nonumber \\
&= \int_0^T dt_1 \int_{t_1}^T dt_2 \, e^{-2\psi^2 t_1} \left(2\psi^2 \right) e^{-4\psi^2 (t_2-t_1)} \left(4\psi^2 \right)e^{-2\psi^2 (T-t_2)} \nonumber \\
&= 4 \alpha^2 e^{-2\alpha^2} -2 e^{-2\alpha^2} \left(1-e^{-2\alpha^2} \right) \, .
\end{align}
Ultimately, $\PBon^{\rm (I)}(3)$ reads:
\begin{align}
\PBon^{\rm (I)}(3)=  4 \alpha^2 e^{-2\alpha^2} + e^{-4\alpha^2} \, ,
\end{align}
and the overall error probability of the type I Bondurant receiver is obtained as:
\begin{align}
\PBon^{\rm (I)}=\frac14 \sum_{k=0}^{3} \PBon^{\rm (I)}(k)= e^{-2\alpha^2} \left(\alpha^2 + \frac34 \right)  \, .
\end{align}
The plot of $\PBon^{\rm (I)}$ is reported in Fig.~\ref{fig02:sec5.5.1_BondurantPlot}, together with the SQL~(\ref{eq:QuaternarySQL}) and the minimum error probability~(\ref{eq:QuaternaryHB}). As we can see, the receiver is not optimum, as in the high-energy limit $\alpha^2 \gg 1$ we have $\PBon^{\rm (I)} \approx \alpha^2 e^{-2\alpha^2}$, whilst $\Pmin \approx e^{-2\alpha^2} $. Nevertheless, it still beats the SQL for $\alpha^2 \ge \alpha^2_{\rm I} \approx 0.68$, providing a quantum advantage for a wide range of signal energies.

The performance of the type I receiver can be enhanced by suitably improving the the ordering of the pulse nulling. In fact, we note that the most detrimental effect on $\PBon^{\rm (I)}$ is determined by the probability $p(n_{\rm c}=2|3)$ in Eq.~(\ref{eq:p23}), containing the term proportional to $\alpha^2 e^{-2\alpha^2}$.
This contribution can be significantly reduced by extracting some information on the signal from the time arrival distribution of the first two clicks. In fact, if state $k=3$ is sent, the counting rate switches from $\lambda_{03}=2\psi^2$ to $\lambda_{13}=4\psi^2$ when the first click is registered, therefore we expect the second click to occur more quickly then the first one. That is, if the photon counter clicked at times $t_1$ and $t_2$, it would be more likely that $t_1 > (t_2-t_1)$. On the contrary, we would have $t_1 \le (t_2-t_1)$ if state $k=2$ were sent, as, now, the count rate would be reduced after the first click, since $\lambda_{02}=4\psi^2$ and $\lambda_{12}=2\psi^2$.
Following this considerations, we construct the type II receiver from the same scheme in Fig.~\ref{fig01:sec5.5.1_BondurantSetup}, albeit with the following improved switching rule:
\begin{itemize}
\item the switch is initialized in position $0$, $s(0)=0$. If no clicks are registered, i.e. $n_{\rm c}=0$, at time $T$ we infer state ``0";
\item if a click is registered at time $t_1 \le T$, the switch moves to position $s(t)=1$ for $t> t_1$. If no more clicks are registered, at time $T$ we infer state ``1";
\item if a second click is obtained at $t_2>t_1$, then:
\begin{itemize}
\item[a)] if $t_1 \le (t_2-t_1)$, i.e. $t_2 \ge 2 t_1$, we set $s(t)=2$ and perform the displacement $D(-\psi_2(t))$ for $t>t_2$. If no more counts are registered until time $T$, we infer state ``2"; otherwise, if another count occurs, we infer state ``3".
\item[b)] if $t_1 > (t_2-t_1)$, i.e. $t_2 < 2 t_1$, we set $s(t)=3$ and perform the displacement $D(-\psi_3(t))$ for $t>t_1$. If no more counts are registered until time $T$, we infer state ``3"; otherwise, if another count occurs, we infer state ``2".
\end{itemize}
\end{itemize}
In turn, the error probabilities for given input state $\PBon^{\rm (II)}(k)$ only differ for $k=2,3$, while $\PBon^{\rm (II)}(0)=\PBon^{\rm (I)}(0)=0$ and $\PBon^{\rm (II)}(1)=\PBon^{\rm (I)}(1)=\exp(-2\alpha^2)$. On the contrary, if state $k=2$ was sent, errors occur when either the number of total clicks is $n_{\rm c} <2$, see Eq.~(\ref{eq:pBon2}), or $n_{\rm c} =2$ clicks are obtained at times $t_1$ and $t_2$, with $t_2< \min\{2 t_1,T\}$. In this latter case, we have:
\begin{align}
p(n_{\rm c}=2,& t_2 <\min\{2 t_1,T\} |2) \nonumber \\[1ex]
&=\int_0^T dt_1 \int_{t_1}^{\min\{2 t_1,T\}} dt_2 \, \pi_{02}^{(0)}(0,t_1) \, \lambda_{02} \, \pi_{12}^{(0)}(t_1,t_2) \, \lambda_{12} \, \pi_{32}^{(0)}(t_2,T) \nonumber \\
&= \int_0^T dt_1 \int_{t_1}^{\min\{2 t_1,T\}} dt_2 \, e^{-4\psi^2 t_1}\, \left(4\psi^2 \right) \, e^{-2\psi^2 (t_2-t_1)} \,\left(2\psi^2 \right)  \, e^{-2\psi^2 (T-t_2)} \nonumber \\
&= 2 e^{-4\alpha^2} \left(1-e^{\alpha^2}\right)^2 \, ,
\end{align}
therefore:
\begin{align}\label{eq:pBon2}
\PBon^{\rm (II)}(2)= e^{-4\alpha^2} \left(1-2e^{\alpha^2}\right)^2 \, .
\end{align}
Finally, if state $k=3$ we have a decision error when either $n_{\rm c} <2$, see Eq.~(\ref{eq:p013}), or when $n_{\rm c} =2$ clicks are obtained at times $t_1$ and $t_2$, with $t_2 \ge \min\{2 t_1,T\}$, with associated probability:
\begin{align}
p(n_{\rm c}=2,& t_2 \ge\min\{2 t_1,T\} |3) \nonumber \\[1ex]
&=\int_0^T dt_1 \int_{\min\{2 t_1,T\}}^{T} dt_2 \, \pi_{03}^{(0)}(0,t_1) \, \lambda_{03} \, \pi_{13}^{(0)}(t_1,t_2) \, \lambda_{13} \, \pi_{23}^{(0)}(t_2,T) \nonumber \\
&= \int_0^T dt_1 \int_{\min\{2 t_1,T\}}^T dt_2 \, e^{-2\psi^2 t_1}\, \left(2\psi^2 \right) \, e^{-4\psi^2 (t_2-t_1)} \,\left(4\psi^2 \right)  \, e^{-2\psi^2 (T-t_2)} \nonumber \\
&= 2 e^{-4\alpha^2} \left(1-e^{\alpha^2}\right)^2 \, ,
\end{align}
such that
\begin{align}\label{eq:pBon3}
\PBon^{\rm (II)}(3)=\PBon^{\rm (II)}(2)= e^{-4\alpha^2} \left(1-2e^{\alpha^2}\right)^2 \, .
\end{align}
Ultimately, we retrieve the error probability of the type II Bondurant receiver as:
\begin{align}
\PBon^{\rm (II)}=\frac14 \sum_{k=0}^{3} \PBon^{\rm (II)}(k)= \frac34 e^{-4\alpha^2} - 2 e^{-3\alpha^2} + 2e^{-2\alpha^2}  \, ,
\end{align}
reported in Fig.~\ref{fig02:sec5.5.1_BondurantPlot}. We have $\PBon^{\rm (II)} \ge \PBon^{\rm (I)}$ for all energies and beats the SQL for $\alpha^2 \ge \alpha^2_{\rm II} \approx 0.35 < \alpha^2_{\rm I}$. Remarkably, in the high-energy regime we have:
\begin{align}
\PBon^{\rm (II)}\approx 2e^{-2\alpha^2} = 4 \Pmin   \qquad \mbox{for} \,\, \alpha^2 \gg 1 \, ,
\end{align}
proving the type II receiver to be near-optimum.

The two Bondurant receivers represent a benchmark for all the QPSK receivers proposed thereafter. In particular, in 2015 M\"uller {\it et al.} designed an improved version by optimizing the displacement amplitude \cite{Muller2015}.
That is, they replaced the ``nulling" displacements $D(-\psi_j(t))$ with optimized displacements $D(u_j(t))$, with
\begin{align}
u_j(t)=e^{i\pi (2j+1)/4} \, u \,  e^{-i \omega t} \, , \qquad j=0,\ldots, 3\, ,
\end{align}
where the wavepacket amplitude $u$ is optimized to minimize the resulting error probability. Differently to exact-nulling schemes, now the total number of detection events is, in general, unbounded, as the clicks registered by the photon counter at time $T$ may be $n_{\rm c} > 3$. Therefore, the authors considered two different switching rules: either cyclic probing, where $s(t)$ is changed in cyclic order $0\to 1\to 2\to 3\to 0 \to \ldots$, or Bayesian probing, based on the maximum a posteriori probability (MAP) criterion. In both the cases, they obtain a quantum advantage for all input energy, beating the SQL for all $\alpha^2>0$ and outperforming also the type I Bondurant receiver \cite{Muller2015}.

However, despite its theoretical relevance, due to the technical difficulties in implementing the feedback loop, the first experimental realization of the Bondurant scheme has been obtained only in 2020 by Jabir {\it et al.} \cite{Jabir2020}.
The authors implement M\"uller's improved scheme with the cyclic probing rule, where the real-time displacement adjustment is implemented by a field-programmable gate array (FPGA), stimulated by the electric pulses produced by the photon detector.
The experimental data prove a quantum advantage over the SQL only in the low energy regime, due to the presence of a reduced visibility $\xi\approx 0.997$ that prevents an efficient decision strategy for large energies, making the error probability increase when $\alpha^2 \gg 1$.

\subsubsection{The quaternary displacement receiver}

\begin{figure}
\includegraphics[width=0.75\columnwidth]{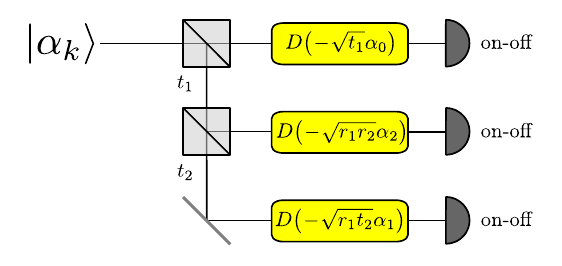}
\centering
\caption{Scheme of the QDRE proposed in \cite{Izumi2012}. 
The incoming signal $|\alpha_k\rangle$ is split into $3$ branches thanks to a pair of beam splitters with transmissivity $t_{1(2)}$ (and corresponding reflectivity $r_{1(2)}=1-t_{1(2)}$. On the $3$ signals, we implement the ``nullling" displacements $D(-\sqrt{t_1}\alpha_0)$, $D(-\sqrt{r_1 r_2}\alpha_2)$ and $D(-\sqrt{r_1 t_2}\alpha_1)$, respectively, followed by on-off detection. We perform the final decision according to the outcomes of the $3$ detection schemes: if a ``off" is obtained on the first branch, we infer state ``0"; if ``on" and ``off" are retrieved from the first and second branch, respectively, we infer ``2"; if two ``on" and a ``off" are registered on the first, second and third branch, respectively, we infer ``1"; otherwise, we perform decision ``3".
}\label{fig01:sec5.5.2_QDSetup}
\end{figure}

As discussed above, the Bondurant receiver represents a challenging solution from a practical point of view, as its implementation requires continuous photo-detection and fast electrical feedback.
As a consequence, simpler feasible receivers have been proposed thereafter, splitting the coherent signal into a finite number of copies, and employing the displacement-photon counting technique. These receivers have been designed in different forms, either employing feed-forward strategies \cite{Becerra2011, Becerra2013, Izumi2012, Izumi2013, Muller2015, DiMario2018} or not \cite{Muller2012, Izumi2012, Izumi2013, Zhao2024}.

In particular, displacement receivers without feed-forward can be constructed as a generalization of the Kennedy receiver, described in Sec.~\ref{subsec4:Kennedy}, where, now, the input signal should be divided in $3$ copies, as at most $3$ nulling displacements are necessary to perform a conclusive decision. The first proposal in this direction has been raised by Izumi {\it et al.} in 2012 \cite{Izumi2012}, whose scheme is reported in Fig.~\ref{fig01:sec5.5.2_QDSetup}. In the following, we will refer to this scheme as the {\it quaternary displacement receiver} (QDRE).

In the QDRE, the incoming signal $|\alpha_k\rangle$, $k=0,\ldots, 3$, is split in three branches thanks to a pair of beam splitters with transmissivity $t_{1(2)}$ (and corresponding reflectivity $r_{1(2)}=1-t_{1(2)}$). Following Kennedy's philosophy, the signal in the first branch, $|\sqrt{t_1} \alpha_k\rangle$, undergoes the displacement $D(-\sqrt{t_1}\alpha_0)$, nulling symbol ``0", followed by on-off detection. If the measurement outcome is ``off", we directly infer state $|\alpha_0\rangle$, and the corresponding probability $p(0|k)$ reads:
\begin{align}
p(0|k)= e^{- t_1 |\alpha_k - \alpha_0|^2}= e^{-2 t_1 \alpha^2 \left[1-\cos(k \pi/2)\right]} \, .
\end{align}
Otherwise, if we obtain ``on" from the first detection, we discard the hypothesis ``0" and consider the subsequent branch.
The signal $|\sqrt{r_1 r_2} \alpha_k\rangle$ on the second branch is then displaced by $D(-\sqrt{r_1 r_2}\alpha_2)$. We choose to null symbol ``2" instead of ``1" because $|\alpha_2 - \alpha_0|^2 \ge |\alpha_1 - \alpha_0|^2$, therefore, if a ``on" is retrieved on the first branch, it would be more likely that state ``2" was sent. As before, we perform again on-off detection: if the result is “off,” then we infer state $|\alpha_2\rangle$, otherwise symbol ``2" is discarded and a final decision between states ``1" and ``3" is performed according to the result on the third branch.
The probability $p(2|k)$ of performing decision ``2" when state $k$ is sent is then equal to:
\begin{align}
p(2|k)&= \left( 1- e^{- t_1 |\alpha_k - \alpha_0|^2}\right) e^{- r_1 r_2 |\alpha_k - \alpha_2|^2} \nonumber \\[1ex]
&=\left(1-e^{-2 t_1 \alpha^2 \left[1-\cos(k \pi/2)\right]} \right) e^{-2 r_1 r_2 \alpha^2 \left[1-\cos((k-2) \pi/2)\right]}  \, .
\end{align}
Finally, if the result is ``on" also on the second branch, we displace the signal on the third one $|\sqrt{r_1 t_2} \alpha_k\rangle$ by $D(-\sqrt{r_1 t_2}\alpha_1)$: if the result is “off,” we infer state $|\alpha_1\rangle$, otherwise we infer state $|\alpha_3\rangle$. The corresponding probabilities then read:
\begin{align}
p(1|k)&= \left( 1- e^{- t_1 |\alpha_k - \alpha_0|^2}\right) \left(1-e^{- r_1 r_2 |\alpha_k - \alpha_2|^2}\right) e^{- r_1 t_2 |\alpha_k - \alpha_1|^2} \nonumber \\[1.5ex]
&=\left(1-e^{-2 t_1 \alpha^2 \left[1-\cos(k \pi/2)\right]} \right) \left(1-e^{-2 r_1 r_2 \alpha^2 \left[1-\cos((k-2) \pi/2)\right]} \right) \times \nonumber \\[1.5ex]
&\hspace{5cm} e^{-2 r_1 t_2 \alpha^2 \left[1-\cos((k-1) \pi/2)\right]}   \, ,
\end{align}
and
\begin{align}
p(3|k)&= \left( 1- e^{- t_1 |\alpha_k - \alpha_0|^2}\right) \left(1-e^{- r_1 r_2 |\alpha_k - \alpha_2|^2}\right) \left(1-e^{- r_1 t_2 |\alpha_k - \alpha_1|^2} \right) \nonumber \\[1.5ex]
&=\left(1-e^{-2 t_1 \alpha^2 \left[1-\cos(k \pi/2)\right]} \right) \left(1-e^{-2 r_1 r_2 \alpha^2 \left[1-\cos((k-2) \pi/2)\right]} \right) \times \nonumber \\[1.5ex]
&\hspace{5cm} \left(1-e^{-2 r_1 t_2 \alpha^2 \left[1-\cos((k-1) \pi/2)\right]} \right)   \, .
\end{align}
Summing up all the contributions, we obtain the error probability as:
\begin{align}\label{eq:Perrt12}
\PQD(t_1,t_2) = 1 - \frac14 \sum_{k=0}^{3} p(k|k) \, ,
\end{align}
that depends on the values of the transmissivities $t_{1(2)}$.

\begin{figure}
\includegraphics[width=0.6\columnwidth]{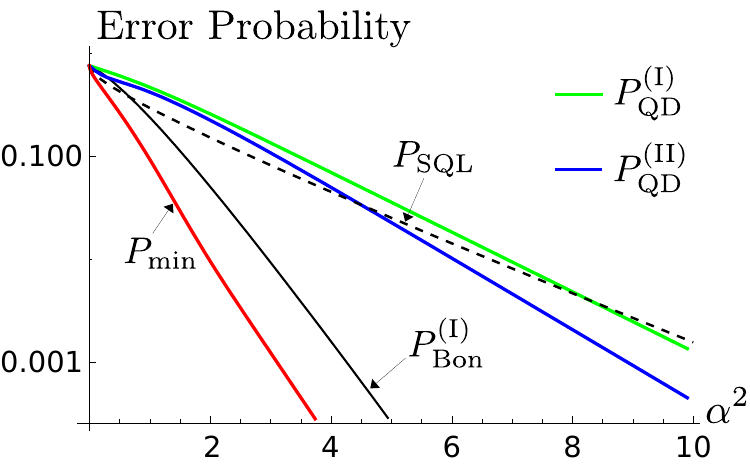}
\centering
\caption{Log plot of $\PQD^{(\rm p)}$, $\rm p=I,II$, as a function of the signal energy $\alpha^2$. $\PSQL$, $\Pmin$, and $\PBon^{\rm (I)}$ refer to the SQL~(\ref{eq:QuaternarySQL}), the minimum error probability~(\ref{eq:QuaternaryHB}) achieved by the PGM, and the error probability of type I Bondurant receiver~(\ref{eq:pBon2}).}\label{fig02:sec5.5.2_QDREPlot}
\end{figure}

We identify two scenarios. In the former, referred to as case I, we consider coherent-state splitting in equal copies, corresponding to $t_1=1/3$ and $t_2=1/2$; in the latter, called case II, we optimize the values $t_{1(2)}$ for each $\alpha^2$ to minimize Eq.~(\ref{eq:Perrt12}). That is,
\begin{align}
\PQD^{\rm (I)} &= \PQD (t_1=1/3,t_2=1/2) \nonumber \\[1ex]
&= \frac14 e^{-8 \alpha^2/3} \left(1- e^{2 \alpha^2/3} + 4 e^{2 \alpha^2}\right) \, ,
\end{align}
and
\begin{align}
 \PQD^{\rm (II)} = \min_{t_1,t_2} \PQD (t_1,t_2) \, .
\end{align}
Plots of $\PQD^{\rm (p)}$, $\rm p=I,II$, are reported in Fig.~\ref{fig02:sec5.5.2_QDREPlot} as a function of $\alpha^2$.
As we can see, in both the cases the QDRE beats the SQL for sufficiently high $\alpha^2$, and $\PQD^{\rm (II)} \le \PQD^{\rm (I)}$, proving optimization of the splitted signal fractions as crucial factor to maximize the receiver performance.
In particular, the numerically optimized transmissivities in the limit $\alpha^2 \gg 1$ are $t_1\approx 2/5$ and $t_2 \approx 1/3$, therefore:
\begin{align}
 \PQD^{\rm (II)} \approx \frac54 e^{-4 \alpha^2/5} \qquad \mbox{for} \, \, \alpha^2 \gg 1 \, ,
\end{align}
whereas $ \PQD^{\rm (I)} \approx \exp(-2 \alpha^2/3) >  \PQD^{\rm (II)} $.

\subsubsection{The quaternary displacement feed-forward receiver}\label{sec:QDFFRE}

\begin{figure}
\includegraphics[width=0.75\columnwidth]{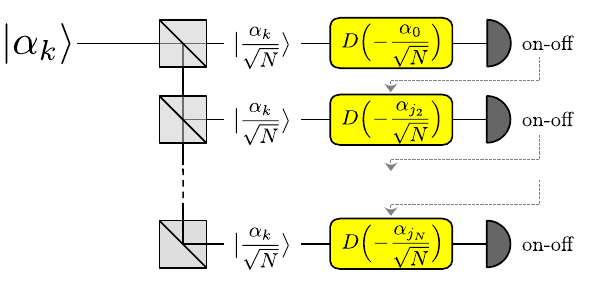}
\centering
\caption{Scheme of the QDFFRE proposed in \cite{Izumi2012}. The incoming signal $|\alpha_k\rangle$ is split into $N$ copies. Each copy $m=1,\ldots, N$ undergoes a conditional displacement $D(-\alpha_{j_{m}}/\sqrt{N})$ followed by on-off detection. For the first copy we have $j_1=0$. For the others, the outcome of the $(m-1)$-th detection sets out the displacement amplitude $j_{m}$ to be implemented on the following copy.}\label{fig01:sec5.5.3_QDFSetup}
\end{figure}

The QDRE performance may be significantly improved by considering displacement feed-forward receivers. This class of receivers provide a particularly attractive solution, leading to error probabilities closer to the Bonduarant limit, albeit with few adaptive steps, simpler setup and less-demanding requirements in terms of detection efficiency, dark count rate and visibility reduction \cite{Notarnicola2023:KB, Izumi2012}.
As a paradigmatic example, here, we focus on the proposal of the {\it quaternary displacement feed-forward receiver} (QDFFRE) presented in \cite{Izumi2012} and depicted in Fig.~\ref{fig01:sec5.5.3_QDFSetup}. 

The QDFFRE is based on the slicing property of coherent states: indeed, thanks to an array of splitters, the incoming signal $|\alpha_k\rangle$ is split into $N \ge 3$ identical copies with reduced amplitude $|\alpha_k/\sqrt{N}\rangle$. Then, each $m$-th copy, $m=1,\ldots, N$, undergoes a conditional displacement operation followed by an on-off detection which returns a click-no click result. The first copy is displaced by $D(-\alpha_{j_1}/\sqrt{N})$, with $j_1=0$, being mapped into the coherent state $|(\alpha_k-\alpha_{j_1})/\sqrt{N}\rangle$. In turn, if $k=0$ the incoming signal is displaced into the vacuum and the subsequent on-off detector will not click, whereas if $k\neq 0$ the detector is more likely to click with a probability $1-p_k$, where
\begin{align}\label{eq:pk}
p_0&=1 \,,  \nonumber \\
p_1&= p_3= e^{-2\alpha^2/N} \,,  \nonumber \\
p_2&= e^{-4\alpha^2/N} \, .
\end{align}
According to the result of the first detection, we decide what would be the value of the amplitude of the displacement $D(-\alpha_{j_2}/\sqrt{N})$ applied to the second copy: if an ``off" result is registered, that is the detector does not click, we set $j_2=j_1=0$; otherwise we discard hypothesis ``$k=0$", set $j_2=j_1+1$ and probe the final hypothesis from the remaining set $k=1,2,3$. We proceed iteratively in this way until the last copy is processed, following the feed-forward rule: if the $(m-1)$-th detection gives outcome ``off" we displace the $m$-th copy by $D(-\alpha_{j_{m}}/\sqrt{N})$ with $j_{m}=j_{m-1}$, if an ``on" is obtained we set $j_{m}=j_{m-1}+1$, discard all states $j\le j_{m-1}$ and restrict the decision to the states $j_m,\ldots, 3$. The outcome of the last detection determines the final decision. If an ``off" is retrieved, we decide the state $j=j_m$ has been sent, otherwise we perform a random decision among the remaining states.

The conditional probabilities $p^{(N)}(j|k)$ of inferring the state $j=0,\ldots,3$ after $N$ copies if state $k=0,\ldots,3$ was sent read:
\begin{subequations}\label{eq:CondQDFFRE}
\begin{align}
p^{(N)}(0|k) &= p_k^N \, , \\[2ex]
p^{(N)}(1|k) &= \sum_{t=0}^{N-2} p_k^t \, (1-p_k) \, p_{(k-1) \bmod 4}^{N-1-t} 
\, + \, \frac{p_k^{N-1}(1-p_k)}{3} \, ,\\[2ex]
p^{(N)}(2|k) &= \sum_{t=0}^{N-3} \, \sum_{s=0}^{N-3-t} p_k^t \, (1-p_k) \, p_{(k-1) \bmod 4}^s \, (1-p_{(k-1) \bmod 4}) \times \, \nonumber \\[1ex]
&\, p_{(k-2) \bmod 4}^{N-2-t-s} \, + \, \sum_{t=0}^{N-2} p_k^t \, (1-p_k) \, \frac{p_{(k-1) \bmod 4}^{N-2-t}(1-p_{(k-1) \bmod 4})}{2} \nonumber \\[1ex]
&\, + \, \frac{p_k^{N-1}(1-p_k)}{3} \, ,\\[2ex]
p^{(N)}(3|k)&= \sum_{t=0}^{N-3} \, \sum_{s=0}^{N-3-t}  \,\, \sum_{u=0}^{N-3-t-s}  p_k^t \, (1-p_k) \, p_{(k-1) \bmod 4}^s \times \, \nonumber \\[1ex]
&\, (1-p_{(k-1) \bmod 4}) \, p_{(k-2) \bmod 4}^{u} \, (1-p_{(k-2) \bmod 4}) \, p_{(k-3) \bmod 4}^{N-3-t-s-u} \nonumber \\[1ex]
& \, + \, \sum_{t=0}^{N-2} p_k^t \, (1-p_k) \, \frac{p_{(k-1) \bmod 4}^{N-2-t}(1-p_{(k-1) \bmod 4})}{2} \nonumber \\[1ex]
&\, + \, \frac{p_k^{N-1}(1-p_k)}{3} \, .
\end{align}
\end{subequations}

\begin{figure}
\includegraphics[width=0.6\columnwidth]{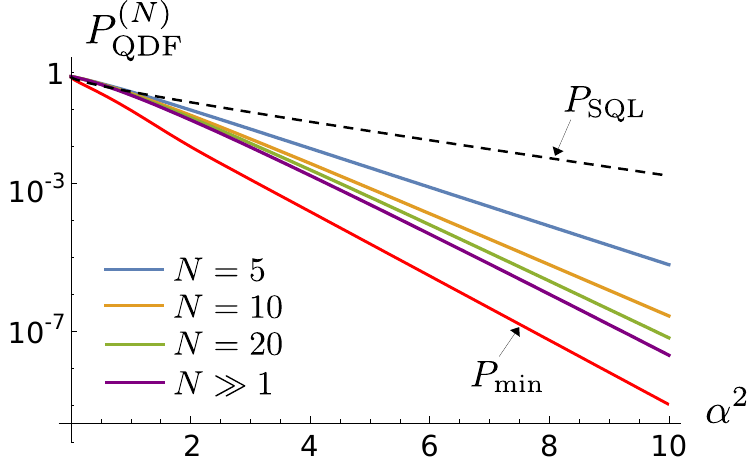}
\centering
\caption{Log plot of the decision error probability $\PQDF^{(N)}$ as a function of the signal energy $\alpha^2$ for different $N$, compared to both the SQL~(\ref{eq:QuaternarySQL}) and the minimum error probability~(\ref{eq:QuaternaryHB}) achieved by the PGM. The QDFFRE beats the SQL only in the regime $\alpha^2 \gg 1$ and, for large $N$, scales as $\PQDF^{(N)} \approx \alpha^2 e^{-2\alpha^2}$, approaching the type I Bondurant receiver, whilst the minimum error probability is $\Pmin \approx e^{-2\alpha^2}$.}\label{fig02:sec5.5.3_QDFPlot}
\end{figure}

Then, the associated decision error probability reads:
\begin{align}
\PQDF^{(N)}=1-\frac{1}{4}\sum_{k=0}^{3} p^{(N)}(k|k)\, , 
\end{align}
depicted in Fig.~\ref{fig02:sec5.5.3_QDFPlot} as a function of $\alpha^2$ for different $N$.
As emerges from the plot, the present displacement receiver outperforms the SQL achieved with double homodyne detection only in the high-energy regime $\alpha^2 \gg 1$.
We also note that, in the limit $N\gg1$, we have $\PQDF^{(N)} \approx e^{-2\alpha^2}(\alpha^2+3/4)$, and the QDFFRE approaches the type I Bondurant receiver \cite{Bondurant1993}.

The gap between $\PQDF^{(N)}$ and $\Pmin$ can be further reduced by optimizing the receiver setup in Fig.~\ref{fig01:sec5.5.3_QDFSetup}.
In particular, Izumi {\it et al.} obtained a better performance in the low-energy regime by changing the order of the nulling displacements from the sequenatial fashion $0\to 1\to 2\to 3$ to $0\to 2\to 1\to 3$, following the scheme of the QDRE, such that, if the first ``on" is retrieved from the $(m-1)$-th copy, we displace the $m$-th one by $D(-\alpha_{2}/\sqrt{N})$ instead of $D(-\alpha_{1}/\sqrt{N})$. This choice reduces the error probability for $\alpha^2 < 1$, but it worsen the receiver performance in the asymptotic limit $\alpha^2 \gg 1$ \cite{Izumi2012}.
Further improvements in the low-energy regime may be obtained by replacing on-off detectors by photon-number resolving detectors to adopt the maximum a posteriori probability criterion \cite{Izumi2013, Becerra2011}, and by optimizing the displacement amplitude, following M\"uller's approach \cite{Muller2015, Becerra2013}.
With these methods, we obtain an improved version of the QDFFRE, being able to beat the SQL for all energies and maximizing the performance of the displacement-photon counting setup.

\section*{Part III: Continuous variable quantum key distribution}\label{part3}
\addcontentsline{toc}{section}{Part III: Continuous variable quantum key distribution}

\def\rrangle{\rangle\!\rangle}
\def\llangle{\langle\!\langle}
\def\bmsigma{\boldsymbol\sigma}
\def\Re{{\mathrm{Re}}}
\def\Im{{\mathrm{Im}}}
\def\DW{{\rm DW}}
\def\GG{{\rm GG}}
\def\bmsigma{\boldsymbol\sigma}
\def\sigmamA{\boldsymbol\sigma^{\rm(m)}_{A}}
\def\sigmamB{\boldsymbol\sigma^{\rm(m)}_{B}}
\def\sigmamAB{\boldsymbol\sigma^{\rm(m)}_{A(B)}}
\def\sigmaz{\boldsymbol\sigma_z}
\def\G{{ \rm G}}
\def\a{ {\mathrm{I}} }
\def\b{ {\mathrm{II}} }
\def\p{ {\mathrm{p}} }
\def\opt{ {\mathrm{opt}} }

\section{Quantum key distribution: the general framework}\label{chap:CVQKD}

The third Part of the thesis is devoted to the field of quantum key distribution (QKD), that has received increasing attention in the latest years, as it guarantees unconditionally secure communication between two distant parties connected by an untrusted quantum channel.
In fact, while the current cryptographic systems, e.g. public-key cryptography, only offer {\it conditional} security based on assumptions on the computational complexity of specific tasks, QKD makes sender and receiver share a random secure key with {\it unconditional} security, regardless the action of any third malicious party, e.g. an eavesdropper \cite{Gisin2002, Scarani2009, Diamanti2016, PirandolaREV}.
This powerful property directly follows from the quantum mechanics laws, that impose ineludible limits on any possible eavesdropping strategy, and, accordingly, lead to general security proofs only based on few basic assumption. 
Nevertheless, in practical conditions dealing with unconditional security may be excessive, as a realistic eavesdropper may encounter further limitations due to both the available state-of-art technologies and the particular composition of the adopted experimental setup, therefore, in time, different security layers have been established, according to the level of trust of the equipment, e.g. the trusted-device scenario and the wiretap channel assumption.

In this Section, we discuss the fundamental aspects of QKD, with particular attention to continuous variable QKD (CVQKD), in which the key is distilled after the exchange of coherent states of radiation. Then, we provide a detailed analysis of the unconditional security framework. At first, we present the GG02 protocol, providing the milestone CVQKD scheme proposed by Grosshans and Grangiér in 2002, employing Gaussian modulation of coherent states and Gaussian detection at the receiver \cite{Grosshans2002, Grosshans2003-1, Grosshans2003-2, Grosshans2005}. Then, we prove the ``optimality of Gaussian attacks" theorem, independently established in 2006 by both Navascués {\it et al.} \cite{Navascues2006} and García-Patrón {\it et al.} \cite{GarciaPatron2006, LeverrierThesis}, that provides a sufficient condition to assess unconditional security of protocols employing non-Gaussian modulation and Gaussian detection, and study discrete modulation CVQKD with both the phase-shift keying (PSK) and quadrature amplitude modulation (QAM) formats.

The structure of the Section is the following. At first, in Sec.~\ref{sec: BasicQKD} we give a general overview on QKD, together with an historical outline, and highlight the fundamental features of all protocols, introducing the key generation rate (KGR) as the main figure of merit. Thereafter, in Sec.~\ref{sec: BasicCVQKD}, we widely discuss CVQKD, and outline the different security frameworks to conduct the security analysis.
Given these premises, Sec.~\ref{sec: GG02} presents the GG02 protocol, for which the KGR can be exactly computed.
Subsequently, in Sec.~\ref{sec:OptGaussUnc} we prove the ``optimality of Gaussian attacks" theorem, and exploit it in Sec.~\ref{sec:DM_CVQKD} to address unconditional security of the PSK and QAM protocol, employing discrete modulation.

\subsection{Basic notions on quantum key distribution}\label{sec: BasicQKD}

The cryptographic systems currently employed for secure communications are mostly based on public-key cryptography, which allows two distant parties to exchange confidential messages without pre-sharing a secret
key. It provides a convenient solution for practical applications but, unfortunately, only offers conditional security, as it usually relies on some assumptions on both the computational resources of a possible eavesdropper
and the complexity of an underlying mathematical problem, e.g. the difficulty of factoring large integers in the Rivest-Shamir-Adleman (RSA) algorithm \cite{RSA}.
On the other hand, the one-time pad \cite{Vernam1926}, together with analogous cryptographic techniques, offer unconditional security (guaranteed by information theory), albeit with much less practical implementation,
as they require the involved parties to share in advance a secret key with the same length as the confidential message, to keep it secret, and to use it only once \cite{Shannon1949}. 
In practice, the one-time pad technique shifts the security problem from the transmission of the confidential message to the distribution of a secure key. However, since distributing long keys is practically not convenient and may pose a significant security risk, public-key cryptography is more widely used than the one-time pad.

Nevertheless, the rapidly emerging progress in the field of quantum computation represents a potential threat for conventional classical cryptosystems. In fact, the quantum factoring algorithm developed by Shor allows to perform probabilistic factorization of non-trivial integers in bounded-error polynomial time \cite{Shor1997, NielsenChuang}.
A further weakening for public-key protocols may also arise from future advances in number theory, where an efficient factorization algorithm for classical Turing machines may be developed \cite{PrimesinP}.
In contrast, a possible solution to the problem of secure key distillation is offered by quantum key distribution (QKD), allowing to share a secret key through the exchange of quantum states \cite{Gisin2002, Scarani2009, Diamanti2016, PirandolaREV}. The very laws of quantum mechanics, like the uncertainty principle, the no-cloning theorem, or the monogamy of entanglement, guarantee the unconditional security of QKD protocols, making the two communicating parties detect the intrusion of any malicious eavesdropper \cite{Wootters1982}. 

Historically, the first QKD protocols have been designed for discrete variable (DV) systems, i.e. qubits. Preliminary ideas in this direction dates back to the early 1970s, when Wiesner speculated about the design of bank notes robust to counterfeiting \cite{Wiesner1983, Brassard2005}, until 1984, when Bennett and Brassard introduced the first seminal DVQKD protocol, referred to as BB84 \cite{BB84}, which nowadays is the main one being commercially distributed. 
In time, several DVQKD schemes have been proposed, e.g. employing entangled photons \cite{E91}, non-orthogonal quantum states \cite{B92} and decoy states \cite{Wang2005, Lo2005, WangPRA}. However, from a practical point of view, the large-scale implementation of DV systems is highly nontrivial for a twofold reason. The main obstacle is represented by the generation of single photons, whose polarization provides the proper degree of freedom to encode the secure bits. Furthermore, the technology required by DVQKD is not compatible with the that of classical telecommunication systems, which, instead, are based on exchange of laser pulses and homodyne or double homodyne (DH) measurements of optical signals \cite{Lodewyck2005, Lodewyck2007, Fossier2009, Banaszek2020}.

For these reasons, from the late 1990s, proposals of QKD protocols on continuous variable (CV) platforms were developed \cite{Ralph1999, Ralph2000, Hillery2000, Reid2000, Cerf2001, Gottesman2001}. Ultimately, in 2002 Grosshans and Grangier proposed the first genuine CVQKD protocol based on Gaussian modulation of coherent states and single quadrature detection at the receiver's side, referred to as GG02 \cite{Grosshans2002, Grosshans2003-1, Grosshans2003-2, Grosshans2005}. Later, a no-switching scheme where the single quadrature measurement is replaced by DH detection has also been proposed by Weedbrook {\em et al.} \cite{Weedbrook2004}.
More recently, also squezeed-state protocols have been developed, obtaining further advantages in the case of long-distance transmission \cite{Usenko2011, Usenko2015, Usenko2018, Usenko2020, Derkach2020}.

\subsubsection{The general structure of a QKD protocol}\label{subsec:StructQKD}

Regardless the particular platform employed to transmit quantum states between the two parties, either DV or CV, all QKD protocols share the same fundamental structure, being divided into two main parts, that is a quantum communication stage followed by classical post-processing \cite{PirandolaREV}.

The first one represents the quantum part of the protocol. Here, the sender, Alice, encodes the outcomes of a classical random variable $\alpha$, generated with a priori probability $p_A(\alpha)$, onto an ensemble of quantum states, being not necessarily orthogonal with one another. These states are, then, sent to the receiver, Bob, throughout a quantum channel.
More precisely, with the term ``quantum channel" we refer to the physical support connecting Alice and Bob, in which the encoded signals propagate. In practical contexts, it describes different physical systems according to the adopted platform, e.g. optical fibers, free-space settings, fading channels or underwater communications \cite{Lodewyck2005, PirandolaFree, Papanastasiou2018, Tang2022}. Beside, the channel is typically noisy, and introduces distortions of the input signals; therefore, it is considered as {\em untrusted}, assuming these distortions to be produced by a third malicious party, the eavesdropper (Eve), who is interested in extracting the secure key generated by Alice and Bob.
After the channel, Bob probes his received signals by performing a measurement, described by a suitable POVM, retrieving a random outcomes $\beta$, associated with probability $p_B(\beta)$, being partially correlated to Alice's ones.
In turn, after repeated iterations of the present scheme, Alice and Bob share a set of raw data described by the two correlated classical variables $\alpha$ and $\beta$.

For what concerns the post-processing part, we can divide it into three steps.
\begin{itemize}
\item {\em Channel evaluation}: Alice and Bob use part of their raw data to assess the characteristics of the channel, by performing estimation of the channel parameters, such as its transmissivity and added noise.
\item {\em Reconciliation}: starting from the results of channel estimation, Alice and Bob partially share a subset of their data to perform error correction, which allows them to detect and eliminate errors induced by the signal transmission, and ultimately, agree on a common raw key. We note that, at this stage, this raw common bit string can be partially known by the eavesdropper which can intercept the flow of information.
\item{\em Privacy amplification}: the raw key undergoes a stage of privacy amplification, implemented via numerical codes based on hash functions,e.g. low density parity check (LDPC) codes \cite{Bennett1995,Bloch2006}, which allows the trusted parties to reduce the eavesdropped information to a negligible amount, at the cost of reducing the length of the common bit string. The result of this stage provides the secure key, which is typically much shorter than the raw one.
\end{itemize}
In some particular protocols, like BB84, a further step, called {\em sifting}, is introduced: in this case, the two parties perform classical communication to agree on a subset of their raw data, while discarding the rest, according to the
measurement bases that they chose independently in each repetition of the protocol \cite{PirandolaREV, Gisin2002, Scarani2009}. 

Remarkably, we underline that the reconciliation step requires a public exchange of information, performed on a classical authenticated channel. Therefore the secure key can be distilled only thanks to proper interplay between quantum and classical communication stages; whereas the sole quantum communication is insufficient to the task.
Moreover, reconciliation can be performed in two alternative ways, according to the party that publicly reveals part of his data.
We have {\em direct reconciliation} (DR) if this party is Alice, and Bob post-processes its outcomes accordingly to infer Alice's encodings. This procedure is typically realized via forward classical communication Alice $\to$ Bob. In the opposite scenario, referred to as {\em reverse reconciliation} (RR), the situation is reversed; now, we have backward classical communication Bob $\to$ Alice, and Alice post-processes her data to infer Bob's variable \cite{PirandolaREV}.

\subsubsection{Eavesdropping strategies}\label{subsec:EveQKD}

The figure of merit to assess the security of the protocol is the key generation rate (KGR), also referred to as secret key rate, expressed in bits per time slot, i.e. per channel use, and defined as the difference between the amount of information shared by Alice and Bob and the information lost throughout channel propagation, assumed to be intercepted by Eve.
Given this premise, security can be investigated in two different conditions. The former, called {\em asymptotic security}, requires the two parties to perform $N\gg 1$ repetitions of the protocol, thus possessing an infinite dataset of variables $(\alpha,\beta)$. In particular, this implies that the channel parameters can be estimated with no uncertainty, according to the Cramér-Rao theorem \cite{Paris2009}. Clearly, the asymptotic security provides a simpler, although less realistic, scenario. On the other hand, when the number of repetitions $N$ is not large enough to reach the asymptotic regime, we deal with {\em finite size effects}, which introduce ineludible inefficiencies in all the post-processing stages, thus weakening security of the scheme.

Let us start with the asymptotic case. Now, the possible attacks that Eve may launch can be divided into three main class, namely individual, collective, and coherent, according to the amount of resources in her hands.
In {\em individual attacks}, Eve performs an independent and identically distributed (i.i.d.) attack on each single intercepted signal. That is, at every repetition of the protocol, she prepares a fresh ancillary state which interacts with the transmitted signal and is individually measured thereafter. The individual measurements can be either performed on-the-fly or delayed at the end of the protocol, letting Eve optimize them according to the public information shared on the classical authenticated channel.
Therefore, in this case the three parties, Alice, Bob and Eve, end up with three classical correlated random variables $\alpha$, $\beta$ and $\gamma$, respectively.
The asymptotic KGR is then obtained as the difference between the mutual information shared by the various parties, namely:
\begin{subequations}
\begin{align}
K_{\rm ind}&= \beta I(A;B)-I(A;E) \qquad \mbox{ for DR,} \\[1ex]
K_{\rm ind}&= \beta I(A;B)-I(B;E) \qquad \mbox{ for RR,} 
\end{align}
\end{subequations}
where $\beta \le 1$ is the reconciliation efficiency, quantifying the procedural errors of both error correction and privacy amplification, $I(A;B)$ is Alice and Bob’s mutual information associated with the variables $\alpha$ and $\beta$, whilst $I(A;E) [I(B;E)]$ is the mutual information shared by Alice (Bob) and Eve, corresponding to the variables $\alpha$ ($\beta$) and $\gamma$.
Clearly, the protocol is successful iff the KGR is larger than $0$, meaning that the information shared between Alice and Eve is larger than that intercepted by Eve.

On the contrary, we have {\em collective attacks} when Eve still launches an i.i.d. attach using a fresh ancilla per channel use, but now, she stores all the interpreted states into a quantum memory, being collectively measured only at the end of the protocol. If so, we should assume Eve to achieve the maximum amount of information allowed by quantum mechanics laws, corresponding to the Holevo information (which, indeed, is achieved by collective measurement).
Thus, the KGR becomes:
\begin{subequations}
\begin{align}
K_{\rm coll}&= \beta I(A;B)-\chi(A;E) \qquad \mbox{ for DR,} \\[1ex]
K_{\rm coll}&= \beta I(A;B)-\chi(B;E) \qquad \mbox{ for RR,}
\end{align}
\end{subequations}
where $\chi(A;E)= {\sf S}[\rho_E] - \int d\alpha \, p_A(\alpha) {\sf S}[\rho_{E|\alpha}]$ and $\chi(B;E)= {\sf S}[\rho_E] - \int d\beta \, p_B(\beta) {\sf S}[\rho_{E|\beta}]$ are the Holevo information between Alice and Eve, and Bob and Eve, respectively. In the former expression, $\rho_E$ represents the average state in Eve's hands, $\rho_{E|\alpha(\beta)}$ is Eve's state conditioned to the outcome $\alpha(\beta)$, associated with probability $p_A(\alpha) [p_B(\beta)]$, and ${\sf S}[\varrho]=-\Tr[\varrho \log_2 \varrho]$ is the von Neumann entropy of state $\varrho$.

Finally, in the case of {\em coherent attacks}, the i.i.d. hypothesis is relaxed. That is, Eve prepares a global non-factorized ancillary state on a set of correlated modes, which jointly interacts with all the encoded signals via collective unitary operation. The output state is then stored in a quantum memory, and collectively measured at the end of the protocol.
Coherent attacks provide the most powerful eavesdropping strategy; however, in the asymptotic scenario, Renner proved that they can be reported back to collective attacks thanks to quantum de Finetti reduction \cite{renner1,renner2,renner-cirac}.

Beyond asymptotic security, finite-size effects arise in the presence of a finite number $N$ of protocol repetitions. In this case, the KGR should be appropriately modified to take into account the inefficiencies introduced in each post-processing step, namely parameter estimation, error correction and privacy amplification. On the one hand, channel evaluation is not exact with a finite statistical sample, and the estimated parameters are associated with a confidence interval of finite width; on the other hand, error correction and privacy amplification are practically implemented by probabilistic routines, thus being associated with a nonzero failure probability that scales with the length of the processed dataset.
In turn, in the finite-size setting perfect security cannot be achieved, but we are limited to $\varepsilon$-security: that is, we introduce a (possibly small) $\varepsilon$ parameter, quantifying the error probability of each protocol step, meaning that the protocol is successful with probability $\ge 1-\varepsilon$.
The security proofs conducted in this framework falls under the {\it composable security} paradigm, currently established for many DV and CV schemes \cite{renner-scarani,Sheridan,renner-comp,furrer-comp, Leverrier2017, Pirandola2021, PirandolaFree, lupo-comp, Papanastasiou2021, Lupo2022}.

\subsection{Continuous variable QKD}\label{sec: BasicCVQKD}

In this thesis, we restrict ourselves to the analysis of CVQKD schemes, which can be divided into two main categories, according to the quantum states of radiation employed by Alice to encode her random variable, being either coherent or squeezed signal states \cite{PirandolaREV, Laudenbach2018}.
In more detail, we only consider protocols employing coherent-state modulation, focusing on asymptotic security under collective attacks.
Instead, a detailed discussion on the fundamental issues of the finite-size setting is reported in \cite{Pirandola2021, PirandolaFree, LeverrierThesis}.
Moreover, we focus on RR strategies, that provide a more powerful solution for long-distance communications, whilst DR is intrinsically bounded to a $3$ dB-limit of channel losses, as more than $50\%$ of the signal must arrive to the receiver in order to have the information shared between sender and receiver larger than the one shared between sender and eavesdropper \cite{PirandolaREV,Laudenbach2018}.

Here, we present the basic features of coherent-state protocols. First of all, we underline the security analysis can be carried out under two equivalent frameworks. The former, referred to as the {\em prepare and measure} (PM) protocol, represents the most intuitive and feasible scheme. Here, Alice samples a random variable drawn from a suitable probability distribution, either discrete or continuous, and encodes its value onto optical signals. The signals are then sent  to Bob, who implements a quantum measurement to infer the value of the encoded variable. The latter is the {\em entanglement-based} (EB) protocol, being a theoretical scheme, equivalent to the PM, in which Alice's preparation is modeled as a quantum measurement over one branch of a two-mode entangled state, that projects the remaining mode onto the signal state transmitted throughout the channel  \cite{PirandolaREV, Laudenbach2018}. Despite this more elaborated structure, the EB scheme turns out to provide a simpler theoretical analysis, as will become clearer in the following.

\subsubsection{Prepare and measure protocol}\label{subsec:PM}

\begin{figure}
\includegraphics[width=0.8\columnwidth]{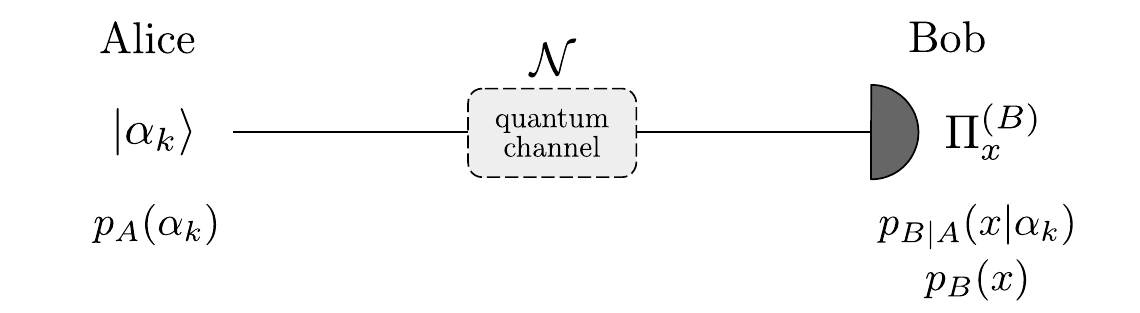}
\centering
\caption{Scheme of the PM version of CVQKD. Alice randomly generates a coherent state $|\alpha_k\rangle$ with probability $p_A(\alpha_k)$ and sends it to Bob throughout an untrusted noisy quantum channel, described by the quantum CP map $\cal N$. Bob performs the POVM $\{\Pi_x^{(B)}\}_x$ on the received signals, obtaining a set of outcomes $x$ correlated to Alice's ones.}
\label{fig01:sec7.2.1_PMproto}
\end{figure}

The prepare and measure (PM) version of CVQKD, corresponding to the practical implementation scheme of the protocol, is depicted in Fig.~\ref{fig01:sec7.2.1_PMproto}. In the PM protocol, Alice prepares a coherent state drawn from a constellation $\{|\alpha_k\rangle\}_k$, $k \in {\cal K}$, sampling the $k$-th state with a priori probability $p_A(\alpha_k)$, such that $\sum_k p_A(\alpha_k)=1$. Here, we adopt a general description, where the random variable of Alice's source may be either continuous or discrete \cite{Laudenbach2018, Denys2021}. 
When the constellation contains an infinite number of states, the set ${\cal K}$ has infinite cardinality and we deal with {\it continuous modulation} \cite{Grosshans2002, Grosshans2003-1, Grosshans2003-2, Grosshans2005, Weedbrook2004}, whereas in the presence of a finite number of the constellation states, ${\cal K}$ contains a finite number of elements $k=0,\ldots,M-1$, and we have {\it discrete modulation} \cite{Leverrier2009, Leverrier2010:8state, Leverrier2011,Becir2012, Hirano2017, Qu2017, Ghorai2019, Lin2019, Liao2020, Kanit2022, Denys2021,Roumestan2021, Roumestan2022, Notarnicola2024:SEC, Djordjevic2019, Almeida2021, Pereira2022}. 
The average state generated at Alice's side then reads:
\begin{align}\label{eq:rhoPM}
\rho = \sum_{k  \in {\cal K}} p_A(\alpha_k) |\alpha_k\rangle \langle \alpha_k| \, .
\end{align}
We note that $\rho$ is a density operator acting on the subspace spanned by the constellation states, namely $\rho \in {\cal L}({\cal S})$, ${\cal S}= {\rm span} \{|\alpha_k\rangle : k\in {\cal K}\}$, which in the presence of continuous modulation typically coincides with the whole Hilbert space. On the contrary, in discrete modulation schemes, $\rho$ is a convex mixture of $M$ linearly independent vectors, thus ${\rm rank}(\rho)=M$.
In turn, the mean photon number employed at the modulation stage, referred to as the {\it modulation energy} reads:
\begin{align}\label{eq:ModEnergy}
\bar{n} = \sum_k p_A(\alpha_k) |\alpha_k|^2 \, .
\end{align}

After modulation, the encoded pulses are injected into the untrusted noisy channel, described by a quantum CP map ${\cal N}$ until to reach Bob, who performs a suitable quantum measurement on the received signals, described by the POVM $\{ \Pi_x^{(B)}\}_x$, $\Pi_x^{(B)}\ge 0$, $\sum_x \Pi_x^{(B)}= \hat{\Id}$, and retrieves the outcome $x \in {\cal X}$. Typically, one consider Gaussian measurements, either homodyne or DH. 
The outcomes $x$ are distributed according to the conditional probability distribution $p_{B|A}(x|\alpha_k)$ when Alice sent the $k$-th state, from which we retrieve the overall probability distribution $p_B(x)= \sum_k p_A(\alpha_k) p_{B|A}(x|\alpha_k)$.

\subsubsection{Entanglement based protocol}\label{subsec:EB}

\begin{figure}
\includegraphics[width=0.8\columnwidth]{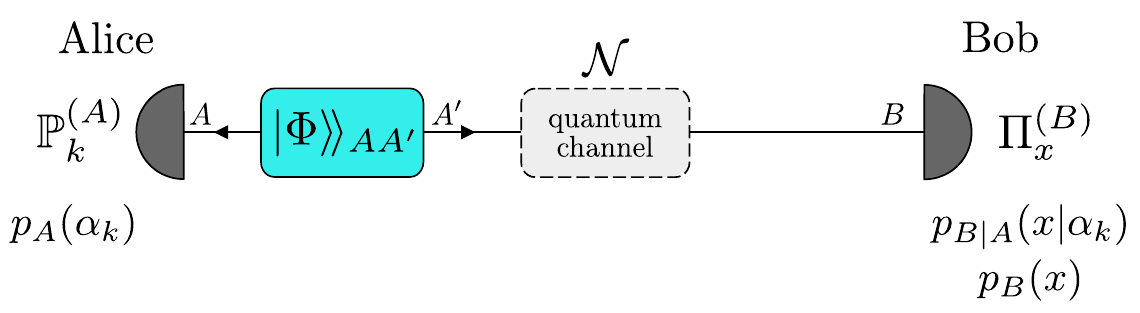}
\centering
\caption{Scheme of the EB version of CVQKD. Now, Alice holds the two-mode entangled state $|\Phi\rrangle_{A A'}$ on two modes $A$ and $A'$; when she performs the projective measurement $\mathbb{P}_k^{(A)}$ on mode $A$, mode $A'$ is projected into the coherent state $|\alpha_k\rangle_{A'}$ with probability $p_A(\alpha_k)$. The prepared state is then injected into the noisy untrusted channel, associated with the CP map ${\cal N}:A'\to B$, to Bob, who implements his POVM $\{\Pi_x^{(B)}\}_x$.}
\label{fig01:sec7.2.2_EBproto}
\end{figure}

An equivalent version of the former PM protocol is the entanglement based (EB) scheme. While the PM protocol represents the actual practical implementation of CVQKD, the corresponding EB version provides a simpler theoretical analysis. In fact, the two schemes are indistinguishable from the perspective both of Bob and Eve, therefore they share the same security and are equivalent between each other \cite{PirandolaREV, Laudenbach2018, Denys2021}. 

The key idea to construct the EB scheme is to model Alice's state preparation as the result of a quantum measurement performed onto an ancillary physical system.
As depicted in Fig.~\ref{fig01:sec7.2.2_EBproto}, in the EB protocol Alice holds a bipartite entangled state $|\Phi\rrangle_{A A'}$ on two modes $A$ and $A'$, expressed in the form:
\begin{align}\label{eq:PurStateFull}
|\Phi\rrangle_{A A'} = \sum_k  \sqrt{p_A(\alpha_k)} \,  |\psi_k\rangle_{A} \, |\alpha_k\rangle_{A'}\, ,
\end{align}
where ${ }_A\langle \psi_j |\psi_k\rangle_A = \delta_{jk}$, with $j,k \in {\cal K}$. Thereby, when Alice performs the projective measurement $\mathbb{P}_k^{(A)}= |\psi_k\rangle_A\langle \psi_k|$ on mode $A$, mode $A'$ is projected into the coherent state $|\alpha_k\rangle_{A'}$ with probability $p_A(\alpha_k)$.
The prepared signal on $A'$ is then injected into the untrusted channel and probed by Bob, as in the PM protocol \cite{Laudenbach2018, Denys2021}.
Moreover, 
we note that the average state on mode $A'$ being sent to Bob is equal to the state $\rho$ reported in Eq.~(\ref{eq:rhoPM}):
\begin{align}\label{eq:Pur}
\Tr_A\Big[|\Phi\rrangle_{A A'} \llangle \Phi|\Big] = \sum_k p_A(\alpha_k) |\alpha_k\rangle_{A'}\langle \alpha_k| = \rho_{A'} \, ,
\end{align}
where the pedix $A'$ underlines the optical mode on which the state is generated. This guarantees the equivalence of the EB protocol with the PM. 
However, we remind that the choice of the purification is highly not unique, therefore there exist infinitely many choices of both $|\Phi\rrangle_{A A'}$ and $\mathbb{P}_k^{(A)}$, all of them being equivalent from the perspective of security analysis.


Given this outline, we now perform explicit construction of the entangled state in Alice's hands, satisfying both Eq.s~(\ref{eq:Pur}) and~(\ref{eq:Req}) \cite{Leverrier2009, LeverrierThesis, Denys2021}. 
We start from the eigenvalue decomposition of the state $\rho$ in~(\ref{eq:rhoPM}), namely $\rho=\sum_j \lambda_j |\phi_j\rangle \langle \phi_j|$, $\lambda_j\ge 0$, and consider the particular purification obtained by the Schmidt decomposition:
\begin{align}\label{eq:PurState}
|\Phi\rrangle_{A A'} &\equiv \sum_j \sqrt{\lambda_j} \, |\phi^*_j\rangle_A \, |\phi_j\rangle_{A'}  \nonumber \\
&= \sum_j \sqrt{\lambda_j} \, |\phi^*_j\rangle_A \left( \sum_{n=0}^{\infty} \langle n |\phi_j\rangle |n\rangle_{A'} \right)  \nonumber \\
&= \sum_{n=0}^{\infty} \left(\sum_j \sqrt{\lambda_j} \, |\phi^*_j\rangle_A \langle \phi^*_j| \right) |n\rangle_{A} |n\rangle_{A'} \nonumber \\[1ex]
&= \left[ (\rho^{*})^{1/2} \otimes \Id \right] \, |{\rm EPR}\rrangle_{AA'} \, ,
\end{align}
where we introduced the Fock basis $\{|n\rangle\}_n$, such that $\langle n|\phi_j\rangle=\langle \phi_j^*|n\rangle$, and the EPR state $|{\rm EPR}\rrangle= \sum_{n=0}^{\infty} |n\rangle |n\rangle$, corresponding to a (un-normalizable) two-mode squeezed vacuum state with infinite amount of squeezing. 
The $*$ in Eq.~(\ref{eq:PurState}), denoting complex conjugation, only represents a technical detail of small practical relevance. In fact, the main constellation schemes employed in optical communications, e.g. phase-shift keying and quadrature-amplitude modulation, exhibit symmetry with respect to complex conjugation, and one typically has $|\phi^*_j\rangle= |\phi_j\rangle$ and $\rho^* =\rho$.

The state~(\ref{eq:PurState}) is pure and normalized, and satisfies the physical request~(\ref{eq:Pur}).
Moreover, it provides a symmetric configuration between the two subsystems $A$ and $A'$, as performing partial trace over mode $A'$ yields:
\begin{align}\label{eq:Req}
\Tr_{A'}\Big[|\Phi\rrangle_{A A'} \llangle \Phi|\Big] = \sum_k p_A(\alpha_k) |\alpha_k^*\rangle_{A}\langle \alpha_k^*| = \rho_{A}^* \, .
\end{align}
In turn, the overall state on mode $A$ has the same mean energy as that on $A'$, i.e. $\bar{n}_A=\bar{n}_{A'}=\bar{n}$, see Eq.~(\ref{eq:ModEnergy}), and, beside complex conjugation, it describes the same statistical ensemble being injected into the channel towards Bob. 


We now prove that this purification can be brought back to the expression~(\ref{eq:PurStateFull}). To this aim, we define the projective measurement $\mathbb{P}_k^{(A)}$ associated with $|\Phi\rrangle_{A A'}$ by introducing the measurement vectors \cite{Denys2021}:
\begin{align}\label{eq:PurMeas}
|\psi_k\rangle \equiv \sqrt{p_A(\alpha_k)} \, (\rho^*)^{-1/2} \, |\alpha_k^*\rangle \, ,
\end{align}
where, in general, $(\rho^*)^{-1/2}$ represents the square-root Moore-Penrose pseudo-inverse of $\rho^*$, accounting for the case in which $\rho$ gets finite rank, thus being not invertible. In particular, this implies that $\rho^* (\rho^*)^{-1} =\mathbb{P}_{{\cal S}^*}$, $\mathbb{P}_{{\cal S}^*}$ being the projector onto the subspace ${\cal S}^*$ spanned by the conjugated constellation states $\{|\alpha_k^*\rangle\}_k$.

Eq.~(\ref{eq:PurMeas}) implies the equivalence between states~(\ref{eq:PurState}) and~(\ref{eq:PurStateFull}), as their overlap ${\cal O}_{AA'}$ is equal to:
\begin{align}
{\cal O}_{AA'} &= \Bigg\{\sum_k  \sqrt{p_A(\alpha_k)} \langle \psi_k|\langle \alpha_k | \Bigg\} \Bigg\{ [ (\rho^{*})^{1/2} \otimes \Id ] \sum_{n=0}^{\infty} |n\rangle |n\rangle \Bigg\} \nonumber \\[1ex]
&= \sum_{kn} p_A(\alpha_k) \langle \alpha_k^* | \langle \alpha_k | \left[ (\rho^{*})^{-1/2} \otimes \Id \right] \left[ (\rho^{*})^{1/2} \otimes \Id \right] |n\rangle |n\rangle\nonumber \\
&= \sum_{kn} p_A(\alpha_k) \langle \alpha_k^* | \langle \alpha_k| (\mathbb{P}_{{\cal S}^*}\otimes \Id ) |n\rangle |n\rangle\nonumber \\
&= \sum_n \langle n| \Bigg( \sum_k p_A(\alpha_k)  | \alpha_k\rangle \langle \alpha_k| \Bigg) |n\rangle  = \Tr[\rho] = 1\, ,
\end{align}
where we used the properties $\langle \alpha_k^* |n \rangle= \langle n |\alpha_k \rangle$ and $ \mathbb{P}_{{\cal S}^*}|\alpha_k^* \rangle = |\alpha_k^* \rangle$.

Finally, in order to complete the EB construction, we need to prove that the measurement vectors $\mathbb{M}=\{|\psi_k\rangle : k\in {\cal K} \}$ form indeed an orthonormal system in the subspace ${\cal S}^*$, such that the associated measurement $\mathbb{P}_k^{(A)}$ is projective. The completeness relation follows directly from~(\ref{eq:PurMeas}):
\begin{align}\label{eq:Identity}
\sum_k|\psi_k\rangle\langle \psi_k| &=\sum_k p_A(\alpha_k) (\rho^*)^{-1/2} |\alpha_k^*\rangle \langle \alpha_k^*|  (\rho^*)^{-1/2} \nonumber \\
&=  (\rho^*)^{-1/2} \rho  \, (\rho^*)^{-1/2} = \mathbb{P}_{{\cal S}^*}\, ,
\end{align}
On the contrary, the obtain the orthogonality condition $\langle \psi_j|\psi_k\rangle= \delta_{jk}$, we proceed as follows. At first, we compute the reduced (not normalized) state $|\chi_j\rangle_{A'}$ obtained after projecting the state $|\Phi\rrangle_{AA'}$ in~(\ref{eq:PurState}) onto $|\psi_j\rangle_A$, that yields the coherent state $|\alpha_j\rangle_{A'}$:
 \begin{align}\label{eq:CohRed}
|\chi_j\rangle_{A'} &= { }_A\langle \psi_j |\Phi \rrangle_{AA'} \nonumber \\
&= \sqrt{p_A(\alpha_j)} \,  \sum_n {}_A\langle \alpha_j^*| \left[(\rho^*)^{-1/2} \otimes \Id \right] \left[(\rho^*)^{1/2} \otimes \Id \right] |n\rangle_A |n\rangle_{A'} \nonumber \\
&= \sqrt{p_A(\alpha_j)}  \,\sum_n {}_A\langle \alpha_j^*| \, \mathbb{P}_{{\cal S}^*} |n\rangle_A |n\rangle_{A'} \nonumber \\
&=\sqrt{p_A(\alpha_j)}  \,\sum_n \langle n|\alpha_j\rangle |n\rangle_{A'} = \sqrt{p_A(\alpha_j)} |\alpha_j\rangle_{A'} \, .
\end{align}
On the other hand, from~(\ref{eq:PurStateFull}) we get $|\chi_j\rangle_{A'}= \sum_k \sqrt{p_A(\alpha_k)} \langle \psi_j|\psi_k \rangle |\alpha_k\rangle_{A'}$, being equivalent to Eq.~(\ref{eq:CohRed}) iff $\langle \psi_j|\psi_k\rangle= \delta_{jk}$, thus proving the measurement vectors to form a complete orthonormal set.


Bearing this in mind, the overall state $\rho_{AB}$ shared by Alice and Bob after propagation through the noisy channel can be written as:
\begin{align}\label{eq:rhoABN}
\rho_{AB}=  \left(\hat{\Id}_A \otimes {\cal N} \right) \,  \Big[|\Phi\rrangle_{A A'} \llangle \Phi| \Big]\, ,
\end{align}
where, now, the CP map $\cal N$ acts on mode $A'$, being transformed into $B$.
In turn, the conditional output statistics at Bob's side is obtained as $p_{B|A}(x|\alpha_k)= \Tr [\rho_{AB} \, \mathbb{P}_k^{(A)} \otimes \Pi_x^{(B)}]$.

\subsubsection{Addressing physical layer security}

\begin{figure}
\includegraphics[width=0.7\columnwidth]{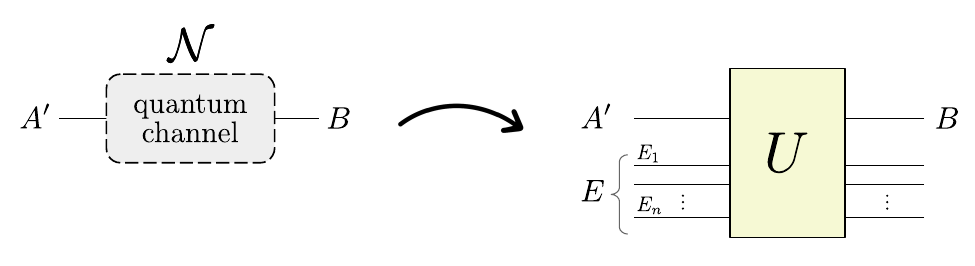}
\centering
\caption{Unitary dilation of the channel noisy map $\cal N$, obtained according to Kraus theorem. It is described by a set of additional modes $E=(E_1,\ldots, E_n)$, $n>1$, coupled to the signal mode $A'$ via a suitable joint unitary operation $U$.}
\label{fig01:sec7.2.3_Dilation}
\end{figure}

As discussed along the previous lines, regardless the adopted version of the protocol, in CVQKD Alice and Bob have only access to the input ensemble $\{ p_A(\alpha_k), |\alpha_k\rangle \}_k$ and the output statistics $p_{B|A}(x|\alpha_k)$, $k \in \mathcal{K}$, and $p_B(x)$. However, these quantities are not sufficient to perform full characterization of the untrusted channel $\cal N$, which remains only partially known, being, in general, described in terms of few relevant parameters to be estimated during the channel evaluation stage. 
As an example, in fiber-optic practical realizations (which will provide the main case study in this thesis) the quantum channel is typically described by means of a thermal-loss channel with transmissivity 
\begin{align}
T=10^{-\kappa d/10} \, ,
\end{align}
where $d$ is the transmission distance in kilometers and $\kappa=0.2$ dB/km is typical loss rate for fibers at telecom wavelength (1550 nm) \cite{Lodewyck2005,Lodewyck2007, Banaszek2020, Agrawal2002, Jouguet2013}; and with the excess noise $\epsilon \ge 0$, introduced by realistic
defects in the experimental apparatus, being of the order of $10^{-3}\div10^{-2}$ shot-noise units \cite{Lodewyck2005,Lodewyck2007}.
In particular, the excess noise is introduced both at the modulation stage, thanks to imperfect generation of the signals; during propagation, by non-idealities of the fiber support; and in the detection apparatus, e.g. arising from electronic noise, phase-mismatch between signal and local oscillator in the presence of homodyne and DH measurement, ...
Equivalently, this effect can be also modeled in terms of equivalent number of thermal photons; that is by assuming a single-mode thermal bath interfering with the encoded signal mode, such that the mean photon number at the receiver’s side is equal to
\begin{align}
\bar{n}_{\rm rec}= T  \bar{n} + \bar{n}_{\rm b} \, ,
\end{align}
where $\bar{n}$ is the mean energy at transmitter, reported in~(\ref{eq:ModEnergy}), whereas $\bar{n}_{\rm b}= T \epsilon /2$ is the number of background photons added to the signal mode, resulting in excess noise equal to $\epsilon$. We precise that the adoption of the excess noise as a figure of merit is typical in fiber-optic communications, whilst the description in terms of equivalent background photons is more common for free-space channels \cite{Pirandola2021, PirandolaFree}.

Given these considerations, an eavesdropper may exploit Alice and Bob's incomplete knowledge to his favor, and manipulate the channel in suitable way to extract the largest amount of information without being detected by the two trusted parties.
In fact, the noisy map $\cal N$ is, in general, not unique, namely, there exists different CP maps that lead to the same local statistics at the sender's and receiver's side. According to Kraus theorem, each of these maps is associated with a different unitary dilation, described by a set of additional modes $E=(E_1,\ldots, E_n)$, for some $n>1$, coupled to $A'$ via a suitable joint unitary operation $U$, as schematized in Fig.~\ref{fig01:sec7.2.3_Dilation}. 
Then, Eve may optimize her strategy and choose the unitary dilation that provides her with the maximum amount of information compatible with both the fundamental limits imposed by quantum mechanics and the local statistics probed by Alice and Bob, thus being completely undetected by them.

Accordingly, to address security of the protocol, we should first answer to the question: {\em How much is the channel untrusted?} Indeed, different assumptions on the setup would lead to different security levels, corresponding to more or less constraints on Eve's action. In light of this, we identify three main different security frameworks:

\begin{itemize}
\item {\em Unconditional security}: the channel is completely untrusted, thus Eve performs arbitrary channel manipulation, i.e. her attack implements the most-informative unitary dilation among those that preserve the local statistics at Alice and Bob's sides, without any further constraint.

\item {\em Trusted-device scenario}: there is some level of trust in the equipment, e.g. trusted detection losses and noise. This implies that only few additional modes $(E_1,\ldots, E_m)$ in $\bf E$, with $m<n$, are at Eve's disposal, whilst the remaining $n-m$ are assumed to be under Alice and Bob's control, thus reducing the set of possible eavesdropping strategies.

\item {\em Quantum wiretap channel}: in this case, Alice and Bob hold characterization of the quantum channel, obtained thanks to either reasonable assumptions or prior information, therefore they have access to a specific noise map $\cal N$. This also establishes the channel connecting Alice to Eve
in terms of the unitary dilation of $\cal N$, therefore arbitrary channel manipulation by Eve is not allowed \cite{Pan2019, Pan2020}.

\end{itemize}

The two latest scenarios provide examples of restricted eavesdropping, being also referred in literature with the terms {\em practical security} or {\em conditional security}, as opposed to the unconditional approach. Here below we will firstly focus on the unconditional security analysis, whilst the features of restricted eavesdropping will be addressed in Section~\ref{chap:RESTREAV}.

\subsubsection{Unconditional security}\label{subsec:UncGeneral}

In the unconditional security framework, Eve is unrestricted; therefore, the most powerful attack that Eve may launch is the so-called {\em purification attack}, where she is assumed to ``purify" the state $\rho_{AB}$ in Eq.~(\ref{eq:rhoABN}). That is, she has full access to the unitary dilation of the noise map $\cal N$ depicted in Fig.~\ref{fig01:sec7.2.3_Dilation}, and controls all the additional modes $E$. Therefore, the tripartite system $ABE$ is closed and isolated, and the state of system is a pure state $|\Psi\rangle_{ABE}$ such that $\rho_{AB}=\Tr_E[|\Psi\rangle_{ABE}\langle \Psi|]$ \cite{Laudenbach2018}.
This allows to explicitly evaluate the Holevo information $\chi(B;E)=S(E)-S(E|B)$, namely the maximum amount of information extractable by Eve, where $S(E)={\sf S}[\rho_E]$ is the von Neumann entropy of Eve's overall state $\rho_E$, and $S(E|B)= \sum_x p_B(x) {\sf S} [\rho_{E|x}]$, where $\rho_{E|x}$ is the conditional Eve’s state related to Bob’s measurement outcome $x$, obtained with probability $p_B(x)$.

In fact, we consider the Schmidt decomposition of the (pure) state of $ABE$:
\begin{align}
|\Psi\rangle_{ABE}= \sum_s \sqrt{\lambda_s} \, |\varphi_s\rangle_{AB} \otimes |\varsigma_s\rangle_E \, , 
\end{align}
$\lambda_s \ge 0$, and obtain the quantum states of the reduced subsystems $AB$ and $E$ as:
\begin{align}
\rho_{AB}&= \Tr_{E} \Big[|\Psi\rangle_{ABE}\langle \Psi|\Big] = \sum_s \lambda_s |\varphi_s\rangle_{AB} \langle \varphi_s | \, , \nonumber \\[1ex]
\rho_{E}&= \Tr_{AB} \Big[|\Psi\rangle_{ABE}\langle \Psi|\Big] = \sum_s \lambda_s |\varsigma_s\rangle_{E} \langle \varsigma_s | \, ,
\end{align}
respectively, being diagonal in the Schmidt bases \cite{Laudenbach2018, NielsenChuang}. As a consequence, they have the same von Neumann entropy, equal to:
\begin{align}
S(E)=S(AB)= -\sum_s \lambda_s \log_2 \lambda_s \, , 
\end{align}
where $S(AB)={\sf S}[\rho_{AB}]$.

Analogously, when Bob performs a $1$-rank measurement $\Pi_x= |\pi_x\rangle_B \langle \pi_x|$, retrieving outcome $x$ with probability $p_B(x)$, the joint conditional state of modes $AE$ is pure and equal to $|\Xi\rrangle_{AE|x}= {}_B\langle \pi_x| \Psi\rangle_{ABE}/\sqrt{p_B(x)}$, therefore the two reduced conditional states on modes $A$ and $E$, obtaining after partial trace, are isentropic: we have ${\sf S} [\rho_{A|x}]={\sf S} [\rho_{E|x}]$.
In turn, the average conditional entropy $S(E|B)= \sum_x p_B(x){\sf S} [\rho_{E|x}]$ and $S(A|B)= \sum_x p_B(x){\sf S} [\rho_{A|x}]$ are equal, i.e. $S(E|B)=S(A|B)$. We note that the choice of a $1$-rank measurement is not too restrictive, as in practical realizations one often deals with homodyne or DH detection.
We conclude that, for a given channel map $\cal N$,
\begin{align}\label{chibe}
\chi(B;E)&=S(E)-S(E|B) \nonumber \\[1ex]
&= S(AB) - S(A|B) \, ,
\end{align}
completely characterizing Eve's information in terms of the state $\rho_{AB}$ shared by Alice and Bob and reported in~(\ref{eq:rhoABN}).
However, in the unconditional security approach, we remind that Alice and Bob do not perform characterization state $\rho_{AB}$, but have only limited information on it, arising from the statistics $p_A(\alpha_k)$, $p_{B|A}(x|\alpha_k)$, and $p_B(x)$, respectively. Accordingly, there exists different states $\rho_{AB}$, or, equivalently, different channel CP maps $\cal N$, leading to the same statistics, and the KGR is obtained by performing optimization over all the possible CP maps that preserve these local statistics at Alice's and Bob's side, namely:
\begin{align}\label{eq:DWbound}
K_{\DW} = \beta I(A;B) - \sup_{{\cal N}: A'\to B} \chi(B;E) \, ,
\end{align}
referred to as the {\em Devetak-Winter bound} (DW) \cite{DevetakWinter}, where $\beta \le 1$ is the reconciliation efficiency.

\subsection{The GG02 protocol}\label{sec: GG02}

\begin{figure}
\includegraphics[width=0.7\columnwidth]{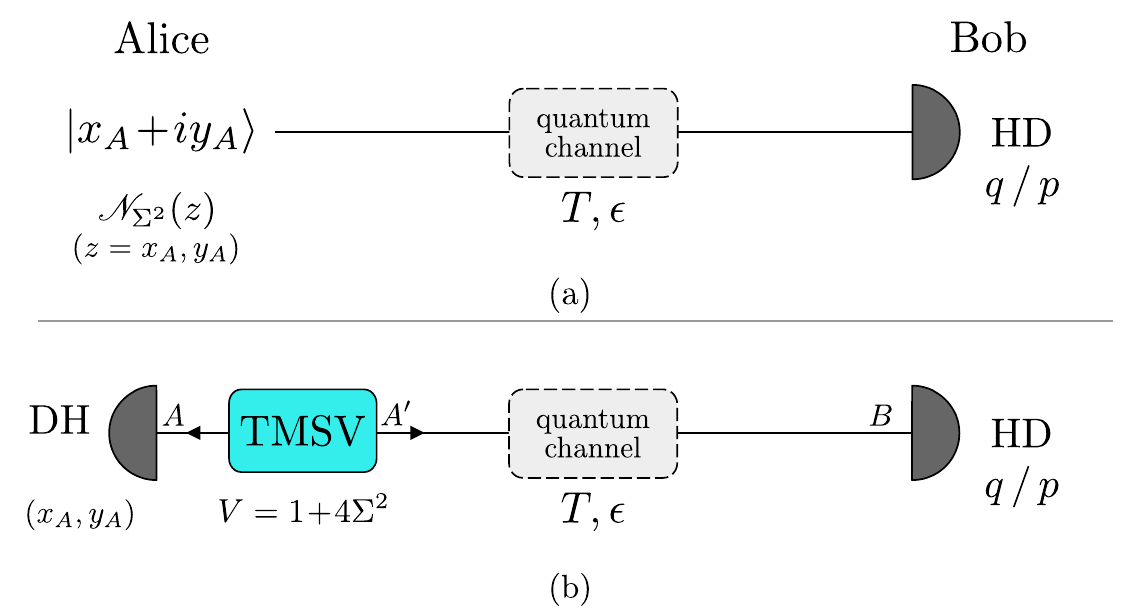}
\centering
\caption{Scheme of the GG02 protocol both in the PM (a) and EB version (b). In the PM protocol, Alice generates a random coherent state $|x_A+iy_A\rangle$, where the amplitudes $z=x_A,y_A$ are drawn from a normal distribution $\mathscr{N}_{\Sigma^2}(z)$ with variance $\Sigma^2$. Instead, in the EB scheme the signal generation is obtained by performing DH detection on the first branch of a TMSV with modulation variance $V=1+4\Sigma^2$; when she retrieves the outcome $(x_A,y_A)$, the second arm is projected onto a coherent state with amplitude proportional to $x_A+iy_A$.}
\label{fig01:sec7.3.1_GG02}
\end{figure}

We now introduce the GG02 protocol, being the first seminal CVQKD protocol, proposed by Grosshans and Grangier in 2002 and proving the benchmark for all the subsequent progress in the field \cite{Grosshans2002, Grosshans2003-1, Grosshans2003-2, Grosshans2005}.

The scheme of the protocol in its PM version is reported in Fig.~\ref{fig01:sec7.3.1_GG02}(a).
Here, Alice implements Gaussian modulation of coherent states; that is, in each repetition of the protocol, she generates a coherent state $|x_A+ i y_A\rangle$, where the variables $z=x_A,y_A$ are (independently) sampled from the normal distribution:
\begin{align}\label{eq:pA_GG}
\mathscr{N}_{\Sigma^2}(z)= \frac{e^{- z^2/(2 \Sigma^2)}}{\sqrt{2 \pi \Sigma^2}} \, , \qquad z=x_A,y_A \, ,
\end{align}
with zero mean and variance $\Sigma^2 \ge 0$, such that the probability of preparing a coherent state with amplitude $x_A+i y_A$ reads $p_A(x_A,y_A)= \mathscr{N}_{\Sigma^2}(x_A)\mathscr{N}_{\Sigma^2}(y_A)$.
The overall statistical mixture generated by Alice then reads:
\begin{align}\label{eq:rhoGG}
\rho &= \int_{\mathbb{R}^2} dx_A dy_A \, p_A(x_A,y_A) \, |x_A+ i y_A\rangle \langle x_A+ i y_A| \nonumber \\[1ex]
&= \frac{1}{1+2 \Sigma^2} \sum_{n=0}^{\infty} \left( \frac{2 \Sigma^2}{1+2 \Sigma^2} \right)^n |n \rangle \langle n| \nonumber \\[1ex]
&= \nu^{\rm th} (2\Sigma^2) \, ,
\end{align}
corresponding to a thermal state with $\bar{n}= 2\Sigma^2$ mean photons.
The encoded signals are then injected into the untrusted quantum channel, described as a thermal-loss channel with transmissivity $T \le 1$ and excess noise $\epsilon\ge 0$, until to reach Bob, who performs Gaussian detection on his received pulses, either homodyne detection of a random quadrature chosen between $q$ and $p$, as in the original proposal \cite{Grosshans2002}, or DH detection, as in the no-switching scheme proposed in \cite{Weedbrook2004}.
In the following we will follow the original proposal and, thanks to the symmetry of the modulation scheme with respect to both quadratures, we safely assume that Bob always homodynes quadrature $q$. The DH protocol leads to analogous results. 
Given this considerations, the conditional probability that Bob obtains outcome $x_B$, measuring $q$, when Alice sent the state $|x_A+ i y_A\rangle$ reads:
\begin{align}\label{eq:pB|A_GG}
p_{B|A}(x_B|x_A) = \frac{\exp\left[- (x_B- 2 \sqrt{T} x_A)^2/(2(1+ T \epsilon)) \right]}{\sqrt{2\pi (1+ T \epsilon)}} \, , 
\end{align}
expressed in shot-noise units (SNU), which will be always considered thereafter. As we can see, $p_{B|A}(x_B|x_A)$ is independent of $y_A$.
Accordingly, the overall distribution probed by Bob reads:
\begin{align}\label{eq:pB_GG}
p_B(x_B) &= \int_{\mathbb{R}} dx_A \, \mathscr{N}_{\Sigma^2}(x_A) \, p_{B|A}(x_B|x_A) \nonumber \\[1ex]
&= \frac{\exp\left[- x_B^2/(2 \Sigma^2_B) \right]}{\sqrt{2\pi\Sigma^2_B}} \, 
\end{align}
where $\Sigma^2_B=1+ T (4\Sigma^2+ \epsilon)= 1+ T (2\bar{n}+ \epsilon)$.
We underline that all the probed statistics, corresponding to Alice's modulation and Bob's detection, are Gaussian: an important property that will be useful for the security analysis.
Given the probability distributions~(\ref{eq:pA_GG}),~(\ref{eq:pB|A_GG}), and~(\ref{eq:pB_GG}), we evaluate the mutual information by recalling that the Shannon entropy of a Gaussian distribution $\mathcal{G}(\mu,\sigma^2)$ with mean $\mu$ and variance $\sigma^2$ is equal to ${\sf H}[\mathcal{G}(\mu,\sigma^2)]= \log_2(2\pi e \sigma^2)/2$. In turn, we get:
\begin{align}\label{eq:IGG}
I_\GG (A;B) &= H(B) - H(B|A) \nonumber \\
&= \frac12 \log_2 \left( 1+ \frac{2 T \bar{n}}{1+ T \epsilon} \right) \, ,
\end{align}
which coincides with the Shannon capacity for the signal-to-noise ratio (SNR):
\begin{align}
{\rm SNR}= \frac{2 T \bar{n}}{1+ T \epsilon} \, ,
\end{align}
whose numerator represents the mean signal power, proportional to the mean number of photons at the receiver's side, while the denominator provides the added noise on quadratures in SNU, equal to $1+ T \epsilon$.

Equivalently, we design the EB version of the protocol, by considering a purification of state~(\ref{eq:rhoGG}), provided by the two-mode squeezed vacuum state (TMSV):
\begin{align}
|{\rm TMSV} \rangle\!\rangle_{A A'} =
\sqrt{1-\lambda^2}\sum_{n=0}^{\infty} \lambda^n |n\rangle |n \rangle \, ,
\end{align}
where $\lambda= \sqrt{ (V-1)/(V+1) }$, $V=1+2\bar{n}= 1 +4 \Sigma^2$ being the modulation variance \cite{Grosshans2003-2, Laudenbach2018}. Then, Alice performs DH measurement on mode $A$, described as a $1$-rank projection onto the coherent state $|\alpha\rangle_A$, where $\alpha=x_A+iy_A$; when she obtains the outcomes $(x_A,y_A)$, the branch $A'$ is projected onto a coherent state, as
\begin{align}
{}_A\langle \alpha |{\rm TMSV} \rrangle_{A A'} \propto |\lambda \, \alpha \rangle_{A'} \, .
\end{align}
This guarantees the equivalence with the PM scheme, provided that Alice rescales her outcomes by a constant factor $\lambda$. The scheme of the EB protocol is displayed in Fig.~\ref{fig01:sec7.3.1_GG02}(b).

\subsubsection{Unconditional security}

To assess unconditional security, we underline that in the GG02 protocol all the statistics probed by Alice and Bob are Gaussian. As a consequence, the quantum channel has also to be Gaussian; otherwise some non-Gaussianity introduced throughout signal propagation would be registered at the receiver's side.
This imposes a constraint on the possible eavesdropping strategies by Eve, being limited to implement a Gaussian purification attack; making it straightforward to compute the DW~(\ref{eq:DWbound}), as no optimization over all channel maps $\cal N$ yielding Alice's and Bob's statistics is required.

Then, approaching the problem in the EB description, we start by considering the covariance matrix (CM) of Alice's TMSV state, namely:
\begin{align}
\bmsigma_{AA'} =
\begin{pmatrix} V \, \Id_2 & Z \, \sigmaz \\ Z \, \sigmaz & V \, \Id_2 \end{pmatrix} 
 \, ,
\end{align}
where $Z=\sqrt{V^2-1}$, $\Id_2$ is a $2\times 2$ identity matrix and $\sigmaz$ is the Pauli $z$-matrix \cite{Ferraro2005,Serafini2017}. 
Thereafter, mode $A'$ is injected into the thermal-loss channel, with associated transmissivity $T \le1$ and thermal noise $\bar{n}_T= T \epsilon/(2(1-T))$, described via a Gaussian completely positive (CP) map associated with the matrices \cite{Serafini2017}:
\begin{align}\label{eq:TLxy}
X_{\rm TL} = \sqrt{T} \, \Id_2 \quad \mbox{and} \quad Y_{\rm TL} = (1-T)(1+2\bar{n}_T) \Id_2 \,.
\end{align}  
Ultimately, the state $\rho_{AB}$ shared by Alice and Bob is a Gaussian state with zero first moments and CM $\bmsigma_{AB}=(\Id_2 \oplus X_{\rm TL} )  \bmsigma_{AA'} (\Id_2\oplus X_{\rm TL})^{\mathsf{T}}+({\bf 0} \oplus Y_{\rm TL} )$, ${\bf 0}$ being the null $2\times 2$ matrix. Straightforward calculations lead to:
\begin{align}\label{eq: Gamma_GG02}
\bmsigma_{AB} =
\begin{pmatrix}  \bmsigma_A & \bmsigma_{Z} \\[1ex] \bmsigma_Z^{\sf T} & \bmsigma_{B} \end{pmatrix} 
=
\begin{pmatrix} V \, \Id_2 & \sqrt{T} Z \, \sigmaz \\ \sqrt{T} Z \, \sigmaz & T(V+\chi) \, \Id_2 \end{pmatrix} 
 \, ,
\end{align}
 where
\begin{align}
\chi= \frac{1-T}{T}+\epsilon 
\end{align}
provides the total added noise on quadratures, due to both the vacuum and the thermal excess noise contributions.

We also note that Eq.~(\ref{eq:IGG}) can be re-derived as follows. Alice and Bob performs Gaussian detection on their local modes, namely DH and homodyne of $q$, being associated with the CMs:
\begin{align}
\sigmamA = \Id_2 \quad \mbox{and} \quad \sigmamB = \lim_{z\rightarrow 0} 
\begin{pmatrix}  z & 0 \\ 0 & z^{-1}\end{pmatrix} \, ,
\end{align}
respectively.
In turn, the mutual information between Alice and Bob is obtained directly from~(\ref{eq: Gamma_GG02}) as:
\begin{align}\label{eq: IAB GG02}
I_\GG(A;B) &= \frac12 \log_2 \Bigg\{\frac{\det\big[\bmsigma_A+\sigmamA\big] \det\big[\bmsigma_B+\sigmamB \big]}{\det\big[\bmsigma_{AB}+(\sigmamA  \oplus \sigmamB )\big]} \Bigg\} \\[1ex]
&= \frac12 \log_2 \left[ 1+ \frac{T (V-1)}{1+ T \epsilon} \right]  \, .
\end{align}

Furthermore, the Holevo information $\chi(B;E)=S(AB)-S(A|B)$ can be also retrieved from the CM~(\ref{eq: Gamma_GG02}), by exploiting the tools of the Gaussian formalism. In fact, the von Neumann entropy of the (Gaussian) state $\rho_{AB}$ depends only on its CM and reads:
\begin{align}
S(AB)=h\left(\frac{{\rm d}_1-1}{2}\right)+ h\left(\frac{{\rm d}_2-1}{2}\right) \, ,
\end{align}
where
\begin{align}\label{eq:hfunc}
h(x)= (x+1) \log_2( x+1) - x \log_2 x\, ,
\end{align}
and
\begin{align}
    {\rm d}_{1(2)}= \sqrt{\frac{\Delta \pm \sqrt{\Delta^2-4 I_4}}{2}} \, ,
\end{align}
being the symplectic eigenvalues of $\bmsigma_{AB}$, with $I_{1(2)}= \det(\bmsigma_{A(B)})$, $I_3= \det(\bmsigma_Z)$, $I_4= \det(\bmsigma_{AB})$ and $\Delta= I_1+I_2+2I_3$. Moreover, the conditional state of Alice $\rho_{A|x_B}$ given Bob's outcome $x_B$ is still a Gaussian state with CM:
\begin{align}
    \bmsigma_{A|B} &= \bmsigma_A - \bmsigma_Z \bigg[ \bmsigma_B + \sigmamB \bigg]^{-1} \bmsigma_Z^\mathsf{T} \\[1ex]
&= \begin{pmatrix}  V - \frac{V^2-1}{V+\chi}& 0 \\[1ex] 0 & V\end{pmatrix}
 \, ,
\end{align}
being independent of $x_B$, thus $S(A|B)=h(({\rm d}_3-1)/2)$, where $d_3 = \sqrt{\det(\bmsigma_{A|B})}$. Accordingly, Eve's Holevo information writes:
\begin{align}\label{eq: chiBE GG02}
\chi_\GG(B;E) = h\left(\frac{{\rm d}_1-1}{2}\right)+ h\left(\frac{{\rm d}_2-1}{2}\right) -h\left(\frac{{\rm d}_3-1}{2}\right) \, .
\end{align}

\begin{figure}
\includegraphics[width=0.8\columnwidth]{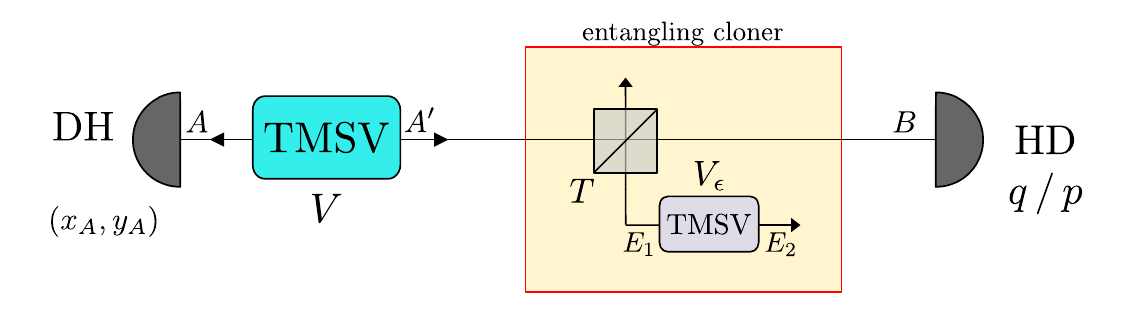}
\centering
\caption{Schematic description of the entangling cloner attack, performed in the GG02 protocol. Eve replaces the quantum channel by a lossless channel where she inserts a beam splitter with transmissivity $T$. To mimic the presence of the excess noise, she generates a TMSV with variance $V_\epsilon= 1+2 \bar{n}_T$
on two modes $\boldsymbol{E}=(E_1,E_2)$ and lets Alice's signal mode $A'$ interfere with $E_1$ at the beam splitter. Then, she keeps the reflected beam for herself while sending the transmitted one to Bob. This scheme allows her to be undetected from Alice and Bob, as performing partial trace over modes $\boldsymbol{E}$ yields a thermal-loss channel CP map.}
\label{fig02:sec7.3.1_EntCloner}
\end{figure}

As demonstrated in~\cite{Laudenbach2018}, the optimal purification eavesdropping strategy is practically implemented by the so-called {\it entangling cloner attack}, introduced by Weedbrook {\it et al.} in \cite{Weedbrook2012} and displayed in Fig.~\ref{fig02:sec7.3.1_EntCloner}. It represents an active eavesdropping strategy, allowing Eve to mimic the effect of the channel losses and excess noise.
In more detail, Eve replaces the quantum channel by a lossless channel where she inserts a beam splitter with transmissivity $T$.
She prepares a TMSV with variance $V_\epsilon= 1+2 \bar{n}_T= 1+ T \epsilon/(1-T)$ on two modes $\boldsymbol{E}=(E_1,E_2)$ and lets Alice's signal mode $A'$ interfere with $E_1$ at the beam splitter. Then, she keeps the reflected beam for herself while sending the transmitted one to Bob.
In this way, when Alice generates a coherent state $|x_A+iy_A\rangle$, Bob will receive a displaced thermal state, with displacement amplitude $\sqrt{T}(x_A+iy_A)$ and mean number of thermal photons equal to $T\epsilon/2$, leading to the same results of a thermal-loss channel with parameters $(T,\epsilon)$ {\it in the absence of eavesdropper}. Thereby, the entangling cloner attack allows Eve to hide herself behind the channel losses and noise, being completely undetected by Alice and Bob.

\begin{figure}
\includegraphics[width=0.49\columnwidth]{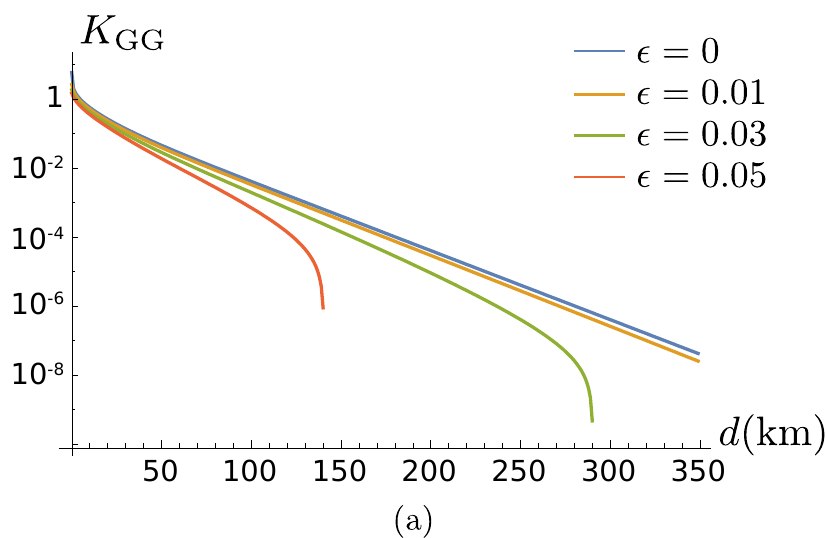} \,
\includegraphics[width=0.49\columnwidth]{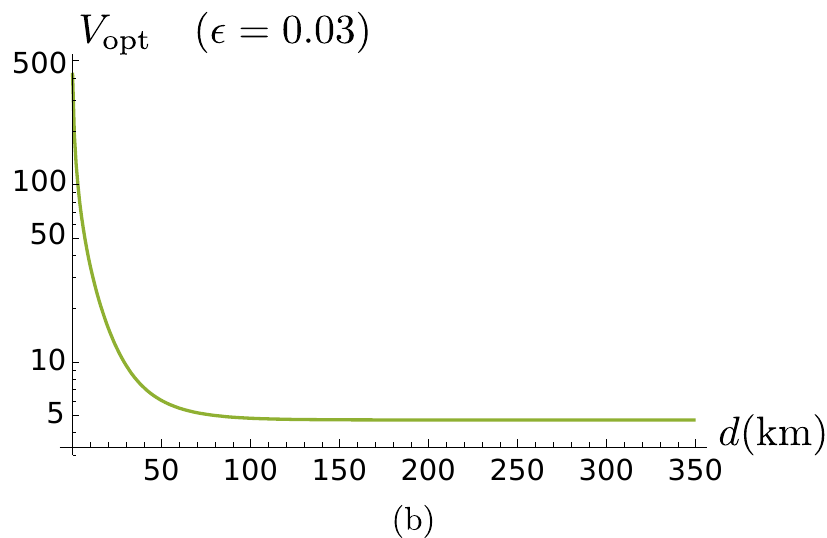}
\centering
\caption{(a) Log plot of the KGR $K_\GG$ as a function of the transmission distance $d$ in km for different values of the excess noise. For $\epsilon>0$ there exists a maximum transmission distance $d_{\rm max}$ after which the KGR drops to $0$. (b) Log plot of the optimized modulation variance $V_{\rm opt}$ for the GG02 protocol as a function of $d$ for $\epsilon=0.03$. In both the pictures we set the reconciliation efficiency $\beta=0.95$ and the loss rate $\kappa=0.2$ dB/km.}\label{fig03:sec7.3.1_KGRGG}
\end{figure}

Given this considerations, we obtain the KGR associated with the GG02 scheme as 
\begin{align}
    K_{\GG}= \max_{V} \Big\{\beta I_\GG(A;B) -\chi_\GG(B;E) \Big\}\, ,
\end{align}
where we perform optimization over the modulation variance $V$ for fixed realistic values of reconciliation efficiency $\beta = 0.95$, loss rate $\kappa=0.2$ dB/km, and channel excess noise $\epsilon$ \cite{Bloch2006, Leverrier2009, Denys2021}.

Plots of the resulting KGR as a function of the transmission distance $d$ in km is reported in Fig.~\ref{fig03:sec7.3.1_KGRGG}(a). 
As demonstrated in \cite{Grosshans2002, Grosshans2005, Laudenbach2018}, in the absence of excess noise, $\epsilon=0$, and for unit reconciliation efficiency $\beta=1$, the KGR is $K_\GG >0$ for all $d\ge 0$, allowing to share secret keys at arbitrary large distances. Otherwise, as we can see from the plot, when $\epsilon>0$, $K_\GG$ is positive up to a maximum transmission distance $d_{\rm max}$, i.e. $K_\GG >0$ for $d\le d_{\rm max}$, after which the KGR drops to $0$ and no secure communication is possible. 
The resulting maximum transmission distance $d_{\rm max}$ is then a function of both the excess noise $\epsilon$ and the reconciliation efficiency $\beta$, i.e.  $d_{\rm max}= d_{\rm max}(\epsilon,\beta)$. In particular, for $\beta=0.95$ and $\epsilon=0.03$ ($\epsilon=0.05$), we have $d_{\rm max}\approx 290$ km ($d_{\rm max}\approx 140$ km).

For completeness, in Fig.~\ref{fig03:sec7.3.1_KGRGG}(b), we also show the optimized input modulation $V_\opt$, being a decreasing function of $d$, that, differently from the KGR, is only weakly dependent on the excess noise value $\epsilon$.

\begin{figure}
\includegraphics[width=0.6\columnwidth]{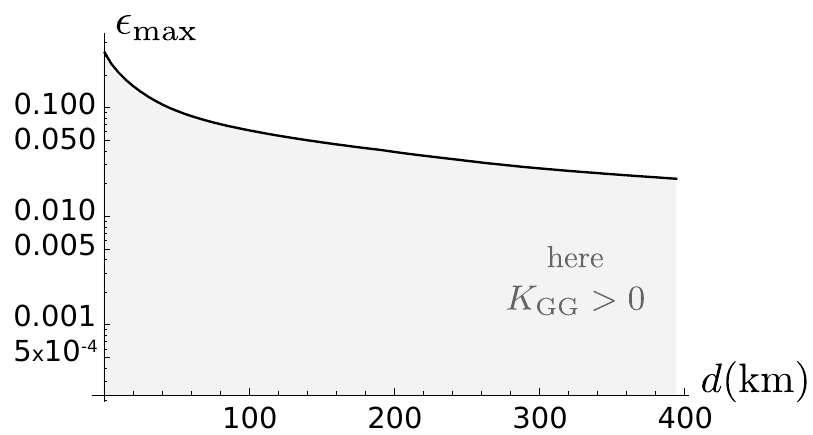}
\centering
\caption{Log plot of the maximum tolerable excess noise $\epsilon_{\rm max}$ for the GG02 protocol as a function of the distance $d$ in ${\rm km}$. The shaded area, corresponding with the undergraph of $\epsilon_{\rm max}$, represents the region where $K_\GG >0$. We set the reconciliation efficiency $\beta=0.95$.}\label{fig04:sec7.3.1_epsilonmaxGG}
\end{figure}

Finally, another relevant figure of merit to evaluate the performance of the protocol is the maximum tolerable excess noise $\epsilon_{\rm max}$, being a decreasing function of the transmission distance $d$, as depicted in Fig.~\ref{fig04:sec7.3.1_epsilonmaxGG}. It represents the maximum value of $\epsilon$ for which the KGR is positive. That is, at the distance $d$ the KGR is positive as long as $\epsilon < \epsilon_{\rm max}$.

\subsection{The optimality of Gaussian attacks}\label{sec:OptGaussUnc}

The GG02 protocol provides the cornerstone example of CVQKD, representing the benchmark for all other proposals of possible protocols. Moreover, its security analysis can be easily carried out, since the protocol involves only Gaussian resources: a particular feature that, indirectly, constrains Eve to implement Gaussian attacks, i.e. the entangling cloner. However, in more general schemes, non-Gaussian elements may be introduced, e.g. non-Gaussian modulation of coherent states at Alice's side, or non-Gaussian channels, in which case the statistics probed by Alice and Bob are not sufficient to characterize the quantum state $\rho_{AB}$ shared by them in the EB protocol, making it hard to compute the DW. In fact, in these conditions, the DW should be evaluated by optimization over all possible channel quantum CP maps compatible with the probed statistics.

A convenient solution is to look for simpler suitable bounds, upper bounding Eve's Holevo information and, accordingly, providing a lower bound on the DW, and a sufficient condition to establish unconditional security.
Following this outline, a fundamental result is the so-called ``{\it optimality of Gaussian attacks}" theorem, being independently proved in 2006 with different methods by both Navascués {\it et al.} \cite{Navascues2006} and García-Patrón and Cerf \cite{GarciaPatron2006, LeverrierThesis}.
In particular, provided Bob’s measurement to be Gaussian, the theorem establish an upper bound on Eve's information by the Holevo
information of the Gaussian state having the same first and second momenta of the quantum state $\rho_{AB}$ in~(\ref{eq:rhoABN}).
The resort to Gaussian formalism leads to a simple lower bound on the DW, being useful to assess security in protocols adopting either non-Gaussian modulation by Alice or propagation through non-Gaussian channels.

Here, we prove the theorem following Navascués approach \cite{Navascues2006}, which exploits the extremality properties of Gaussian states and operations. On the contrary, the equivalent proof developed by García-Patrón  and Cerf in \cite{GarciaPatron2006, LeverrierThesis} involves theorems from functional analysis, being more technical, and, therefore, it is hard to provide a physical interpretation.

To begin with, in the following subsection we derive some fundamental results that will be exploited throughout the proof.

\subsubsection{Theoretical framework}\label{subsec:GOLemmas}
Let us now consider an arbitrary quantum state $\varrho$ of a physical system $A$, being a quantum operator over a Hilbert space ${\cal H}$, and let $F: {\cal B}({\cal H}) \rightarrow \mathbb{C}$ be a functional over the set of Hermitian operators ${\cal B}({\cal H})$.
We now introduce the state $\varrho_\G$ as the Gaussian state having the same first moments and CM as $\varrho$. If $F$ is the value of some quantity computed for state $\varrho$ and $F_\G$ the corresponding value related to state $\varrho_\G$, we define the difference:
\begin{equation}\label{delta:gauss}
\Delta F \equiv F_\G - F\,.
\end{equation}

To prove the optimality of Gaussian attacks it is useful to state the following lemmas.
The first lemma involves the average conditional entropy.
\begin{lemma}\label{Lemma1:GOpt}
Let $\{\Pi_x\}_x$ be a positive-operator valued measurement (POVM) performed on a system 
$A$ and associated with a classical register $X$, such that $\Pi_x=M_x^\dagger M_x$ and $\sum_x \Pi_x = \hat{\Id}$. Let $p(x)=\Tr[\varrho \Pi_x]$ and $\varrho_{|x}= M_x \varrho M_x^\dagger / p(x)$ be the probability of retrieving outcome $x$ and the corresponding conditional state of $A$, respectively.
Then, the average conditional entropy $S(A|X)= \sum_x p(x) S(\varrho_{|x})$ is equal to:
\begin{align}\label{eq:provelemma1gopt}
S(A|X) = S(\overline{AX}) - H(X) \, ,
\end{align}
where $S(\overline{AX})$ is the von Neumann entropy of the state
\begin{align}
R_{AX}= \sum_x p(x) \varrho_{|x} \otimes |x\rangle \langle x| \, ,
\end{align}
$|x\rangle$ being the (classical) state of register $X$ associated with outcome $x$, and $H(X)=-\sum_x p(x) \times $ $ \log_2 p(x)$ is the Shannon entropy of the distribution $p(x)$.
\end{lemma}

\begin{proof}
To prove Eq.~(\ref{eq:provelemma1gopt}), we should compute the von Neumann entropy of state $R_{AX}$, namely $S(\overline{AX}) = -\Tr \left[R_{AX} \log_2 R_{AX} \right]$. To begin with, we evaluate the operator 
\begin{align}
{\cal R}=\log_2 R_{AX}= \log_2 \left(\sum_x \widetilde{\varrho}_{|x} \otimes |x\rangle \langle x| \right) \, .
\end{align}
with $\widetilde{\varrho}_{|x}= p(x) \varrho_{|x}$.
For given $x$, we consider the spectral decomposition $\widetilde{\varrho}_{|x}$, equal to $\widetilde{\varrho}_{|x}= \sum_j \lambda_j(x) |\phi_j(x) \rangle \langle \phi_j(x) |$, with $\lambda_j(x) \ge 0$ and $\{ |\phi_j(x) \rangle \}_j$ being a complete orthonormal system.
Then, state $R_{AX}$ becomes:
\begin{align}\label{eq: Rax}
R_{AX}= \sum_x \sum_j \lambda_j(x) |\phi_j(x) \rangle \langle \phi_j(x) | \otimes |x\rangle \langle x|\, .
\end{align}
We note that Eq.~(\ref{eq: Rax}) corresponds to the spectral decomposition of $R_{AX}$, provided a suitable reordering of the indices $j$ and $x$. Therefore, we straightforwardly obtain the spectral decomposition of ${\cal R}$ as:
\begin{align}
{\cal R}&= \sum_x \underbrace{\sum_j \log_2 \left(\lambda_j(x)\right) |\phi_j(x) \rangle \langle \phi_j(x) |}_{\equiv \log \widetilde{\varrho}_{|x}} \otimes |x\rangle \langle x| \nonumber \\[1ex]
&= \sum_x \log_2 \widetilde{\varrho}_{|x} \otimes |x\rangle \langle x| \, .
\end{align}
As a consequence, the von Neumann entropy $S(\overline{AX})$ is equal to:
\begin{align}
S(\overline{AX}) &= -\Tr \left[R_{AX} \log_2 R_{AX} \right] \nonumber \\[1ex]
& = - \Tr_{AX} \left\{ \left(\sum_y p(y) \varrho_{|y} \otimes |y\rangle \langle y| \right) \left[\sum_x \log_2 \left(p(x) \varrho_{|x} \right) \otimes |x\rangle \langle x| \right]\right\} \, ,
\end{align}
where the trace is performed over both systems $A$ and $X$.
At first, we perform partial trace over $X$, obtaining:
\begin{align}
S(\overline{AX}) &= - \sum_x \sum_y p(y) \Tr_{A} \Big\{ \varrho_{|y} \log_2 \left(p(x) \varrho_{|x} \right) |\langle y|x\rangle|^2 \Big\} \nonumber \\[1ex]
&= - \sum_x p(x) \Tr_{A} \left\{ \varrho_{|x} \log_2 \left(p(x) \varrho_{|x} \right) \right\} \nonumber \\
&= - \sum_x p(x) \Tr_{A} \{ \varrho_{|x} \left( \log_2 p(x) + \log_2\varrho_{|x} \right) \} \nonumber \\
&= - \sum_x p(x) \log_2 p(x) \underbrace{\Tr_{A} \{ \varrho_{|x} \}}_{=1} - \sum_x p(x) \underbrace{\Tr_{A} \{ \varrho_{|x} \log_2\varrho_{|x} \}}_{=S[\varrho_{|x}]}   \nonumber \\
&= H(X) + S(A|X) \, ,
\end{align}
$H(X)$ being the Shannon entropy of $p(x)$. In turn, we have $S(A|X)= S(\overline{AX}) - H(X)$. 
Moreover, in the second line we used the property $\langle x|y\rangle  = \delta_{xy}$, being valid since the register states are classical and, thus, distinguishable.
\end{proof}

The second lemma introduces the relation between $\Delta S(A)$, namely, the difference between the von Neumann entropies of $\varrho_\G$ and $\varrho$, see Eq.~(\ref{delta:gauss}), and their relative entropy.
\begin{lemma}\label{Lemma2:GOpt}
For any quantum state $\varrho$ of a system $A$, we have
\begin{align}
\Delta S(A) = S(\varrho \parallel \varrho_\G) \quad \mbox{and} \quad S(\varrho \parallel \varrho_\G) \ge 0 \, ,
\end{align}
where $S(\varrho \parallel  \varrho_\G) = \Tr[\varrho \log_2 \varrho]-\Tr[\varrho \log_2 \varrho_\G]$ is the relative von Neumann entropy.
\end{lemma}

\begin{proof}
We re-express the quantity $\Delta S(A)= S_\G(A) - S(A)= -\Tr[\varrho_\G \log \varrho_\G] + \Tr[\varrho \log \varrho] $ as $\Delta S(A) = S(\varrho \parallel \varrho_\G) + \Delta$, where
\begin{align}
\Delta= \Tr[\left(\varrho-\varrho_\G\right) \log_2 \varrho_\G]  \, .
\end{align}
We remind that any $n$-mode Gaussian state can be written as a Gibbs state in the quadrature vector operators $\hat{{\bf r}}=(q_1,p_1,q_2,p_2,\ldots, q_n,p_n)^\mathsf{T}$. That is, $\varrho_\G= \exp(-\zeta \hat{\mathbb{H}})/{\cal Z}$, with $\zeta>0$, ${\cal Z}= \Tr[\exp(-\zeta \hat{\mathbb{H}})]$, and $\hat{\mathbb{H}}= \hat{{\bf r}}^\mathsf{T} {\bf d} +\hat{{\bf r}}^\mathsf{T} \mathbb{H} \hat{{\bf r}}/2$, for some displacement vector ${\bf d} \in  \mathbb{R}^{2n}$ and $2n \times 2n$ symmetric matrix $\mathbb{H} \in {\rm Sym}(2n)$ \cite{Serafini2017}.
In turn, we have $\log_2 \varrho_\G= - \log_2 {\cal Z} -\zeta \hat{\mathbb{H}}$, being a polynomial operator of the second order in the canonical variables $\{q_k,p_k\}_k$. This, together with the fact that, by construction, $\varrho$ and $\varrho_\G$ share the same CM, leads to $\Delta= \Tr[\left(\varrho-\varrho_\G\right) \log_2 \varrho_\G]=0$. Ultimately, we have $\Delta S(A) = S(\varrho \parallel \varrho_\G)$.
\end{proof}
We remark that, since the relative entropy is a non-negative quantity, Lemma~\ref{Lemma2:GOpt} implies that the state of maximal entropy for fixed first moments and CM is Gaussian.
With analogous methods the lemma can be also proved for probability distributions $p(x)$, $x \in {\cal X}$. That is:
\begin{align}
\Delta H(X) = H(p \parallel p_\G)\ge 0 \, , 
\end{align}
where $p_\G(x)$ is the Gaussian distribution having the same FM and CM as $p(x)$, and $H(p \parallel p_\G)= \sum_x p(x) \log_2\left(p(x)/p_\G(x)\right)$ is the Shannon relative entropy \cite{Jaynes1957}.

Finally, the last lemma follows from the monotonicity of quantum relative entropy, firstly proved by Lindblad \cite{Lindblad1974, Breuer2002}.
\begin{lemma}\label{Lemma3:GOpt}
For any Gaussian CPTP map ${\cal E}_\G$ acting the system $A$, one has:
\begin{align}
\Delta S(A) \ge \Delta S\big({\cal E}_\G(A)\big)\,,
\end{align}
and $\Delta S(A) = \Delta S\big({\cal E}_\G(A)\big)$ iff the state of $A$ is Gaussian.
\end{lemma}

\begin{proof}
Lemma~\ref{Lemma3:GOpt} follows from the monotonicity of the quantum relative entropy: that is, for any two states $\varrho_1$ and $\varrho_2$ and any quantum CP map ${\cal E}$, we have:
\begin{align}\label{eq:mon}
S(\varrho_1 \parallel \varrho_2) \ge S\left({\cal E}(\varrho_1) \parallel {\cal E}(\varrho_2)\right)\,.
\end{align}
If ${\cal E}= {\cal E}_\G$ is a Gaussian map and $\varrho$ is an arbitrary quantum state of system $A$, the choices $\varrho_1=\varrho$ and $\varrho_2=\varrho_\G$, together with Lemma~\ref{Lemma2:GOpt}, imply that:
\begin{align}
\Delta S(A) \ge S\left({\cal E}_\G(\varrho) \parallel {\cal E}_\G(\varrho_\G)\right) \, . 
\end{align}

To conclude the proof, we should demonstrate that the state $\Xi'={\cal E}_\G(\varrho_\G)$ is indeed the Gaussian state associated with $\Xi={\cal E}_\G(\varrho)$, that is $\Xi_\G=\Xi'$. 
To this aim, we approach the problem in the Heisenberg picture and consider the input-output relations associated with ${\cal E}_\G$. 
At first, we remind that any open Gaussian dynamics can be retrieved from a Gaussian interaction with a Gaussian environment, having null first moments and CM $\bmsigma_E$, and being thereafter traced out \cite{Serafini2017}. In turn, if $\hat{{\bf r}}_{\rm in (out)}$ are the input (output) quadrature operators of the channel, we have:
\begin{align}
\hat{{\bf r}}_{\rm out} = \mathsf{X} \, \hat{{\bf r}}_{\rm in} + M  \hat{{\bf r}}_{E} \, \quad \mbox{and} \, \quad {\sf Y}=M \bmsigma_E M^{\sf T}
\end{align} 
where $\hat{{\bf r}}_{E}$ is the quadrature vector operator of the environmental modes, and $\sf X$ and $\sf Y$ are the matrices describing the evolution of first and second momenta under ${\cal E}_\G$, see Sec.~\ref{subsec2:GaussianDyn} \cite{Serafini2017}.
Given these considerations, we evaluate the first moments and CM of state $\Xi$ as:
Then, the first moments and CM of state $\Xi$ read:
\begin{subequations}
\begin{align} 
    \mathbf{R}_{\Xi}&= \langle \hat{{\bf r}}_{\rm out} \rangle =  \mathsf{X} \, \mathbf{R}_{\varrho} \, , \\
    \boldsymbol\sigma_{\Xi} &= \frac12  \langle \{\hat{{\bf r}}_{\rm out},\hat{{\bf r}}_{\rm out}^{\sf T} \} \rangle - \mathbf{R}_{\Xi}\mathbf{R}_{\Xi}^\mathsf{T}
= {\sf X} \bmsigma_{\varrho} {\sf X}^{\sf T} + {\sf Y} \, ,
\end{align}
\end{subequations}
where $\mathbf{R}_{\varrho}$ and $\bmsigma_{\varrho}$ are the first moments and CM of the input state $\varrho$. Analogous results can be obtained for state $\Xi'$, provided the substitution $\varrho\to \varrho_\G$. However, by construction $\mathbf{R}_{\varrho_\G}=\mathbf{R}_{\varrho}$ and $\bmsigma_{\varrho_\G}=\bmsigma_{\varrho}$, thus, we conclude that $\Xi_\G=\Xi'$.
\end{proof}

\subsubsection{Proving Gaussian optimality}

\begin{figure}
\includegraphics[width=0.8\columnwidth]{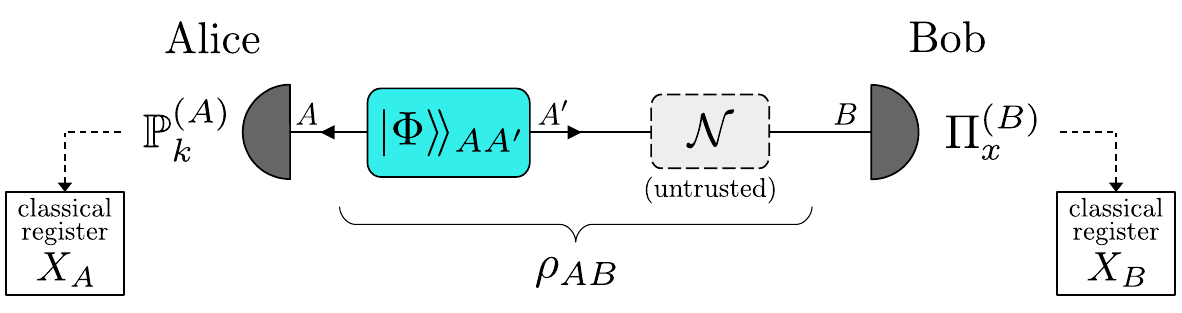}
\centering
\caption{Extended scheme of the EB version of a general CVQKD protocol, adopted for the proof of the optimality of Gaussian attacks. Differently than Fig.~\ref{fig01:sec7.2.2_EBproto}, we now expand the protocol description and include the two classical registers $X_{A(B)}$ in which Alice and Bob store the results of the quantum measurement performed on the first and second branch of the state $\rho_{AB} =(\hat{\Id}_A \otimes {\cal N}) [|\Phi\rrangle_{A A'} \llangle \Phi|]$, respectively.}
\label{fig01:sec7.4.1_OptG}
\end{figure}

The preliminary results derived above allow us to rephrase the scheme of a general CVQKD protocol in more precise fashion, 
as schematized in Fig.~\ref{fig01:sec7.4.1_OptG}.
We start from the EB protocol depicted in Fig.~\ref{fig01:sec7.2.2_EBproto}, and widely discussed in Sec.s~\ref{subsec:EB} and~\ref{subsec:UncGeneral}, in which Alice holds an entangled pure state $|\Phi\rrangle_{AA'}$ and sends branch $A'$ to Bob, such that the two parties eventually share the state $\rho_{AB}$ in Eq.~(\ref{eq:rhoABN}). Then, following Lemma~\ref{Lemma1:GOpt}, we construct an extended scheme of the previous EB picture, in which the classical outcomes of Alice's and Bob's measurements are stored in a classical register $X_{A(B)}$, respectively. This description provides explicit modeling of the quantum-to-classical decoding induced by the performed quantum measurements but, as a matter of fact, it only represents a technicality that does not change the amount of information shared by the parties. Within this approach, we rewrite the mutual information shared by Alice and Bob as
$I(A;B)=H(X_A)+H(X_B)-H(X_AX_B)$, where $H(\cdot)$ is the Shannon entropy associated with the classical variables $X_{A(B)}$. Similarly, the Holevo information shared by Bob and Eve becomes $\chi(B;E)=S(E)-S(E|X_B)$, where $S(E)$ is the von Neumann entropy of the overall state in Eve's hands, and $S(E|X_B)$ is the average conditional entropy of Eve related to Bob’s measurement outcomes, being stored in register $X_B$. 

We are now ready prove the Gaussian optimality theorem.

\begin{theorem}[{\bf Optimality of Gaussian attacks}]\label{Theorem1:GOpt}
Provided Bob's measurement to be Gaussian, for any state $\rho_{AB}$, the Holevo information $\chi(B;E)$ is upper bounded by:
\begin{align}\label{eq:OptGAtt}
\chi(B;E) \le \chi_\G (B;E) \, ,
\end{align}
where $\chi_\G(B;E)$ is the Holevo information computed for the Gaussian state having the same CM as $\rho_{AB}$.
\end{theorem}

\begin{proof}
At first, following the purification argument adopted in Eq.~(\ref{chibe}), we re-express Eve's information as $\chi(B;E)=S(E)-S(E|X_B)= S(AB)-S(AB|X_B)$.
To prove the theorem, we focus on the quantity $\Delta  \chi(B;E)$, see~(\ref{delta:gauss}). Then, we have:
\begin{align}
\Delta  \chi(B;E) & =\chi_\G(B;E) -\chi(B;E) \nonumber\\
&= \Delta S(AB) - \Delta S(AB|X_B) \nonumber\\
&= \Delta S(AB) - \Delta S\left(\overline{ABX_B}\right) + \Delta H(X_B)
\end{align}
where we used Lemma~\ref{Lemma1:GOpt} to get the last equality. Since the map $AB \rightarrow \overline{ABX_B}$ is a Gaussian CP map, as Bob's detection is Gaussian, Lemma~\ref{Lemma3:GOpt} implies:
\begin{align}
\Delta S(AB) - \Delta S\left(\overline{ABX_B}\right) + \Delta H(X_B) \ge \Delta H(X_B) \, ,
\end{align}
and, finally, exploiting Lemma~\ref{Lemma2:GOpt} we find $\Delta H(X_B) \ge 0$, or, summarizing the previous relations, $\chi_\G(B;E) \ge \chi(B;E)$, that proves the theorem.
\end{proof}

As a consequence, Eq.~(\ref{eq:OptGAtt}) provides a lower bound on the DW:
\begin{align}\label{eq:LBoundDW}
K= \beta I(A;B) - \chi_\G(B;E) \ge K_{\rm DW} \, ,
\end{align} 
where $I(A;B)$ is the (exact) mutual information between Alice and Bob, whereas $\chi_\G(B;E)$ is the Holevo information retrieved in a virtual EB protocol where Alice and Bob share a Gaussian state with the CM $\bmsigma_{AB}$ of state $\rho_{AB}$. 
The evaluation $\bmsigma_{AB}$ is based on the amount of knowledge that Alice and Bob have on the channel. In particular, we identify two main scenarios, corresponding to the assumption of a linear or nonlinear channel.

\paragraph{Linear channel.} The channel ${\cal N}: A' \to B$ is linear if described by the following input-output relations of the quadrature operators in Heisenberg picture,
\begin{align}\label{eq:inoutLinear}
q_B= \sqrt{T} q_{A'} + q_{\rm b} \qquad \mbox{and} \qquad p_B= \sqrt{T} p_{A'} + p_{\rm b} \, ,
\end{align}
where $q_{\rm b}$ and $p_{\rm b}$ are the quadrature operators of some additional background noise modes, uncorrelated with the signal mode, such that $\langle q_{\rm b}\rangle=\langle p_{\rm b}\rangle=0$, and $\langle q_{\rm b}^2\rangle= \langle p_{\rm b}^2\rangle= 1-T + T \epsilon$.
In this case, $\bmsigma_{AB}$ can be expressed as a function of the CM $\bmsigma_{AA'}$ of the state $|\Phi\rrangle_{AA'}$ generated by Alice, see Eq.~(\ref{eq:PurStateFull}), that reads:
\begin{align}
\bmsigma_{AA'} = \begin{pmatrix} V \, \Id_2 &  Z \,\bmsigma_z \\ Z \,\bmsigma_z & V\, \Id_2 \end{pmatrix} \, ,
\end{align} 
where: 
\begin{align}
V=1+2\bar{n} \qquad \mbox{and} \qquad Z= \frac12 \llangle \Phi | (q_A q_{A'} -p_A p_{A'}) |\Phi\rrangle \, ,
\end{align} 
in which $\bar{n}= \sum_k p_A(\alpha_k) |\alpha_k|^2$ is the mean number of photons per symbol, and $\bmsigma_z$ is the Pauli $z$-matrix.
The correlation term $Z$ can be also expressed as a function of the average quantum state generated by Alice, equal to $\rho=\sum_k p_A(\alpha_k) |\alpha_k\rangle \langle \alpha_k|$, see Eq.~(\ref{eq:rhoPM}), as:
\begin{align}\label{eq:ZDenys}
Z= 2 \Tr\left[\rho^{1/2} a  \, \rho^{1/2}a^\dagger \right] \, ,
\end{align} 
where $a$ and $a^\dagger$ are the creation and annihilation operators, respectively \cite{Denys2021}. 
Then, thanks to~(\ref{eq:inoutLinear}), we obtain $\bmsigma_{AB}$ as:
\begin{align}\label{eq:sigmaABlinear}
\bmsigma_{AB} = \begin{pmatrix} V \, \Id_2 & \sqrt{T} \, Z \,\bmsigma_z \\[1ex]  \sqrt{T} \,Z \,\bmsigma_z & T\left(V+\chi\right)\, \Id_2 \end{pmatrix} \, , \qquad \mbox{for a linear channel} \, .
\end{align}

\paragraph{Nonlinear channel.} The linear channel provides a simple model to describe beam propagation in optical media, e.g. fibers, being often adopted in classical communications. However, from the CVQKD perspective, it may represent a too restrictive condition. In fact, if the channel from Alice to Bob is linear, we implicitly assume to know all the statistical moments of Bob's output state, whereas the quantities evaluated in a realistic CVQKD protocol are only the first and second moments of Bob's conditional state when Alice sends state $k$, and the second moment of Bob’s overall state, as discussed in Sec.~\ref{sec: BasicCVQKD}.
For these reasons, it is also worth to investigate the more general case of an arbitrary ${\cal N}:A' \to B$, being associated with a nonlinear input-output relation.
This scenario has been addressed in detail by Denys {\it et al.} in~\cite{Denys2021}. Now, the CM $\bmsigma_{AB}$ gets the different expression: 
\begin{align}
\bmsigma_{AB} = \begin{pmatrix} V \, \Id_2 & \widetilde{Z} \,\bmsigma_z \\[1ex]  \widetilde{Z} \,\bmsigma_z & T\left(V+\chi\right)\, \Id_2 \end{pmatrix} \,, \qquad \mbox{for a nonlinear channel} \, .
\end{align}
where the off-diagonal block terms $\widetilde{Z}$ satisfy:
\begin{align}\label{eq:Zbound}
\widetilde{Z} \ge 2 \sqrt{T} \Tr\left[\rho^{1/2} a  \, \rho^{1/2}a^\dagger \right] - \sqrt{2 T \epsilon w} \, ,
\end{align}
in which we introduce the quantity:
\begin{align}
w= \sum_k p_A(\alpha_k) \Big[ \langle \alpha_k|a_\rho^\dagger a_\rho |\alpha_k\rangle - |\langle \alpha_k| a_\rho |\alpha_k\rangle|^2 \Big] \, ,
\end{align}
with $a_\rho= \rho^{1/2} a \, \rho^{-1/2}$.
To assess unconditional security, one can either perform numerical calculation of $\widetilde{Z}$ \cite{Ghorai2019, Lupo2022}, or take the right hand side of~(\ref{eq:Zbound}) as a lower bound to the CM $\bmsigma_{AB}$ and evaluate the corresponding KGR \cite{Denys2021}.

\subsubsection{Final remarks and comments}

Remarkably, we underline that referring to Theorem~\ref{Theorem1:GOpt} as the ``optimality of Gaussian attacks" theorem is rather misleading. 
The theorem merely states that we may safely assess security of a CVQKD protocol by considering the Gaussian state associated with the unknown state $\rho_{AB}$ actually shared by Alice and Bob. Therefore, it does not provide identification of the most powerful attack at Eve's disposal, which, in general, is non Gaussian.
In fact, the bound~(\ref{eq:OptGAtt}) can be saturated iff $\rho_{AB}$ itself is a Gaussian state, in which case we retrieve the usual GG02 protocol, where the optimal eavesdropping coincides with a Gaussian attack, i.e. the entangling cloner \cite{Laudenbach2018}. On the contrary, in all other protocols, the bound is not attainable, and the optimal attack is no longer Gaussian.

Finally, we note that Eq.~(\ref{eq:LBoundDW}) requires to evaluate the exact mutual information $I(A;B)$ shared by the two parties, whose numerical calculation should be preferably approached in the PM picture. However, with analogous techniques, a simpler Gaussian bound on $I(A;B)$ can be obtained in the presence of Gaussian modulation at Alice's side. In this case, the following theorem holds.
\begin{theorem}\label{Theorem2:GOpt}
If Bob's measurement is Gaussian and Alice employs Gaussian modulation, for any state $\rho_{AB}$, the mutual information $I(A;B)$ is lower bounded by:
\begin{align}\label{eq:OptGAtt}
I(A;B) \ge I_\G(A;B) \, ,
\end{align}
where $I_\G(A;B)$ is the mutual information computed for the Gaussian state having the same CM as $\rho_{AB}$, see Eq.~(\ref{eq: IAB GG02}) .
\end{theorem}

\begin{proof}
Starting from the expression $I(A;B)=H(X_A)+H(X_B)-H(X_AX_B)$, we evaluate the quantity $\Delta  I(A;B)$, see~(\ref{delta:gauss}):
\begin{align}
\Delta I(A;B) & =I_\G(A;B)-I(A;B) \nonumber\\
&= \Delta H(X_A) + \Delta H(X_B)- \Delta H(X_AX_B) \nonumber \\
&\le \Delta H(X_A) \, , 
\end{align}
where we used Lemma~\ref{Lemma3:GOpt} as the map $X_AX_B \rightarrow X_B$, corresponding to averaging over the (Gaussian) random variable $X_A$, is Gaussian. Then, since Alice implements Gaussian modulation, we have $\Delta H(X_A) \equiv 0$, concluding the proof.
\end{proof}
Theorem~\ref{Theorem2:GOpt} implies that the KGR of a Gaussian-modulated protocol with Gaussian detection is lower bounded by that of an all-Gaussian protocol where Alice and Bob share the Gaussian state with CM $\bmsigma_{AB}$, namely:
\begin{align}
K'= \beta I_\G(A;B) - \chi_\G(B;E) \ge K \ge K_{\rm DW} \, ,
\end{align} 
which provides a lower bound to~(\ref{eq:LBoundDW}), that avoids the exact numerical computation of the mutual information $I(A;B)$. In particular, the present bound is applicable to CVQKD schemes over non-Gaussian channels.


\subsection{Discrete modulation protocols}\label{sec:DM_CVQKD}

Despite its simplicity, the GG02 scheme discussed in the previous sections raises important issues about its practical implementation with the state-of-the-art technologies in optical communications, especially regarding the continuous modulation of the coherent pulses.
In fact, although justified on a theoretical level, Gaussian modulation involves
several practical difficulties, and, so far, its application is limited to short-distance communications \cite{Jouguet2012, WeedbrookREV, DjordjevicDGM}. 
The main drawback lies in the error correction stage needed for the reconciliation process, 
being implemented via suitable codes originally designed for discrete variables, that perform worse in the presence of Gaussian modulation \cite{Leverrier2009, LeverrierThesis, DjordjevicDGM}.
In fact, as discussed in Sec.~\ref{sec: GG02}, to perform GG02 in the long-distance regime, the protocol should operate at low values of modulation energy, corresponding to few mean photons per time slot, see Fig.~\ref{fig03:sec7.3.1_KGRGG}(b); thus, accordingly, we should consider low values of SNR. 
On the other hand, reconciliation of Gaussian variables, even with the most advanced error correction codes, e.g. turbo codes \cite{BB84} or LDPC codes \cite{Bencheikh2001}, can be efficiently performed only in the high SNR regime, whereas the common techniques, e.g. the slice reconciliation method \cite{VanAssche2004}, even if assisted by LDPC, leads to low reconciliation efficiency if the SNR is too low \cite{Bloch2006, Leverrier2009, LeverrierThesis}.

To date, the problem of obtaining good reconciliation efficiencies at low SNR is still open; therefore a possible solution is to design CVQKD protocols employing discrete modulation formats of appropriate order. In this way, the dataset in Alice's and Bob's hands would be the same of a classical communication scheme based on discrete signaling and additive white Gaussian noise channel, for which there exists more efficient codes working in the low SNR regime, e.g. multi-edge type LDPC codes \cite{Richardson2002}.

For these reasons, CVQKD employing discrete modulation has been recently addressed in literature, with the intent of adopting a feasible modulation technique being as close as possible to the Gaussian modulation limit \cite{Leverrier2009, Leverrier2010, Leverrier2011, Becir2012, Hirano2017, Qu2017, Ghorai2019, Lin2019, Liao2020, Ghalaii_DM, Papanastasiou2021, Denys2021, Roumestan2021, Roumestan2022, Kanitschar2022, Djordjevic2019, Almeida2021, Pereira2022}.
To this aim, in the following we address two paradigmatic formats, namely, phase-shift keying (PSK), a well known stratgey in literature, and quadrature amplitude modulation (QAM), here investigated for the first time in the context of unconditionally secure CVQKD.

\subsubsection{Phase-shift keying (PSK)}\label{subsec:PSKproto}

\begin{figure}
\centerline{\includegraphics[width=0.9\columnwidth]{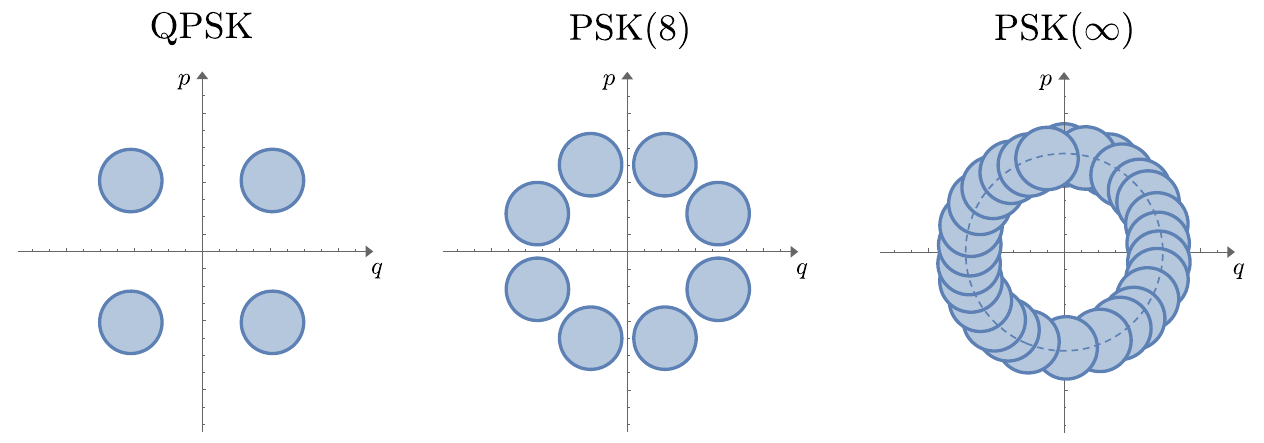}}
\centering
\caption{Phase space representation of several PSK($M$) constellations, with different $M>1$. The case PSK($4$) is also referred to as quadrature phase-shift keying (QPSK). We also note that, in the limit $M\gg 1$, we approach the PSK($\infty$), or continuous phase-shift keying, modulation, in which the constellation is composed of infinitely many coherent states with the same amplitude $\alpha>0$, and different phase, getting continuous values in the range $\phi \in [0,2\pi)$.}\label{fig:01:sec7.5.1-PSKM}
\end{figure}

The first discrete modulation protocol was proposed in 2009 by Leverrier and Grangier \cite{Leverrier2009}, involving the {\it phase-shift keying} (PSK) format already presented in Sec.~\ref{sec:QPSKdiscr}.
The choice of PSK modulation is mainly due for twofold reason. Firstly, a well known result in classical communication theory is that PSK constellations, of proper order $M\ge 2$, approximate the Shannon capacity, achieved by Gaussian modulation, in the low-energy regime \cite{Blahut1987, Cover1999}; thus making it worth of interest to assess their relevance also for CVQKD. Secondly, from a practical point of view, PSK$(M)$ protocols represent the simplest hybrid schemes, combining both the physical implementation of GG02, and the practicality of discrete modulation protocols, for which the error correction stage, occurring during reconciliation, is much simpler to perform. 

Given these considerations, in the presence of PSK($M$) modulation, Alice randomly prepares a coherent state drawn from the constellation of the $M$ states $\{|\alpha_k\rangle\}_k$, with:
\begin{align}
|\alpha_k\rangle= |\alpha \, e^{i \pi (2k+1)/M} \rangle \, , \qquad k=0,\ldots, M-1 \, , 
\end{align}
where $\alpha\ge0$, generated with equal a priori probabilities $p_A(\alpha_k)=1/M$ \cite{Cariolaro2015, Notarnicola2023:KB}.
We remind that the case $M=4$ is also referred to as quadrature phase-shift keying (QPSK).
The phase space representation of PSK($M$) constellations is reported in Fig.~\ref{fig:01:sec7.5.1-PSKM} for the typical values $M=4,8$ commonly employed in optical communications.

Accordingly, the overall quantum state generated at Alice's side is equal to:
\begin{align}\label{eq:rhoPSK}
\rho= \frac{1}{M} \sum_{k=0}^{M-1} |\alpha_k\rangle\langle \alpha_k| = \sum_{k=0}^{M-1} \lambda_k |\phi_k\rangle \langle \phi_k| \, ,
\end{align}
whose right hand side provides its spectral decomposition, associated with eigenvalues:
\begin{align}
\lambda_k &= e^{-\alpha^2} \sum_{n=0}^\infty \frac{\alpha^{2(nM +k)}}{(nM+k)!} =\frac{e^{-\alpha^2}}{M} \sum_{j=0}^{M-1} e^{-i\frac{2\pi}{M} jk}  \exp(\alpha^2 e^{i \frac{2\pi}{M} j}) \, ,
\end{align}
and eigenstates:
\begin{align}
|\phi_k\rangle = \frac{e^{-\alpha^2/2}}{\sqrt{ \lambda_k}} \sum_{n=0}^\infty \frac{\alpha^{nM +k}}{\sqrt{(nM+k)!}} \,  |nM+k\rangle \, ,
\end{align}
expanded in the Fock basis.
As an example, for the QPSK case we have:
\begin{align}\label{lambdaqpsk}
\lambda_{0(2)}&= \frac{e^{-\alpha^2}}{2} \Big[\cosh (\alpha^2) \pm \cos(\alpha^2) \Big] \, , \nonumber \\[1.5ex]
\lambda_{1(3)}&= \frac{e^{-\alpha^2}}{2} \Big[ \sinh (\alpha^2) \pm \sin(\alpha^2) \Big]\, .
\end{align}

After signal modulation, Alice injects the pulses into the unstrusted quantum channel, associated with transmissivity $T \le 1$ and excess noise $\epsilon \ge 0$; finally, Bob collects the output signals and performs Gaussian detection. Here, we consider the case of homodyne detection, and, without loss of generality, assume that quadrature $q$ is measured.
In turn, when Alice sends state $|\alpha_k\rangle$, Bob's conditional probability of obtaining the outcome $x_B$ is equal to:
\begin{align}\label{eq:pB|A_PSK}
p_{B|A}(x_B|\alpha_k) = \frac{\exp\left\{- \left[x_B- 2 \sqrt{T} \alpha \cos\left(\frac{\pi (2k+1)}{M}\right)\right]^2 \big/\left(2(1+ T \epsilon)\right) \right\}}{\sqrt{2\pi (1+ T \epsilon)}} \, , 
\end{align}
whose corresponding Shannon entropy gets the analytic expression
\begin{align}\label{eq:pB_PSK}
{\sf H}\left[p_{B|A}(x_B|\alpha_k) \right] &= \int_{\mathbb{R}} dx_B \, p_{B|A}(x_B|\alpha_k) \log_2 p_{B|A}(x_B|\alpha_k) \nonumber \\[1ex] 
&= \frac12 \log_2 \Big[2\pi e (1+T\epsilon) \Big] \, ,
\end{align}
being independent of the signal amplitude.
Instead, Bob's overall homodyne distribution reads:
\begin{align}\label{eq:pB_PSK}
p_B(x_B) &= \frac{1}{M}\sum_{k=0}^{M-1} p_{B|A}(x_B|\alpha_k) \, .
\end{align}
In turn, we obtain the mutual information shared by the two parties as:
\begin{align}\label{eq:IAB_PSK}
I_{AB}(\alpha^2) = H_B -  \frac12  \log_2 \Big[2\pi e (1+T\epsilon) \Big] \, ,
\end{align}
where $H_B$ is the Shannon entropy of $p_B(x_B)$, to be evaluated numerically.

We now prove unconditional security of the PSK($M$) protocol, by invoking the optimality of Gaussian attacks presented in the previous section. That is, we consider the EB description, where Alice and Bob share the non-Gaussian state $\rho_{AB}$ in~(\ref{eq:rhoABN}), and
provide an upper bound to the actual Holevo infomation shared between Bob and Eve by the Holevo information obtained in the associated EB Gaussian protocol, where Alice and Bob share the Gaussian state with the same CM as $\rho_{AB}$, equal to $\bmsigma_{AB}$. 
Moreover, here we adopt the linear channel assumption, as in the original proposal \cite{Leverrier2009}, and, thanks to Eq.~(\ref{eq:sigmaABlinear}), we get:
\begin{align}\label{eq:sigmaABPSK}
\bmsigma_{AB} =\begin{pmatrix}  \bmsigma_A & \bmsigma_{Z} \\[1ex] \bmsigma_Z^{\sf T} & \bmsigma_{B} \end{pmatrix} 
=
\begin{pmatrix} V \, \Id_2 & \sqrt{T} \, Z_M \,\bmsigma_z \\[1ex]  \sqrt{T} \,Z_M \,\bmsigma_z & T\left(V+\chi\right)\, \Id_2 \end{pmatrix} \, , 
\end{align}
$V=1+2\alpha^2$ being the modulation variance, $\chi=(1-T)/T+\epsilon$, and with the correlation $Z_M=2\Tr[\rho^{1/2} a \, \rho^{1/2} a^\dagger]$, see Eq.~(\ref{eq:ZDenys}). To explicitly compute it, we exploit Eq.~(\ref{eq:rhoPSK}) and note that:
\begin{align}
\langle \phi_j |\alpha_k\rangle = \sqrt{\lambda_j} e^{i\frac{2\pi}{M} jk}  \quad \text{and} \quad a \, |\phi_k\rangle = 
\alpha \, \lambda_{(k-1) \bmod M}^{1/2}\lambda_k^{-1/2} |\phi_{(k-1) \bmod M}\rangle \, .
\end{align}
Straightforward calculations lead to \cite{Leverrier2009, LeverrierThesis, Denys2021}:
\begin{align}
Z_M=2\alpha^2 \sum_{k=0}^{M-1}\frac{\lambda_k^{3/2}}{\lambda_{k+1}^{1/2}} \, .
\end{align}
Then, we obtain the Holevo information as:
\begin{align}\label{eq: chiBE PSK}
\chi_{BE}(\alpha^2) = h\left(\frac{{\rm d}_1-1}{2}\right)+ h\left(\frac{{\rm d}_2-1}{2}\right) -h\left(\frac{{\rm d}_3-1}{2}\right) \, ,
\end{align}
with the $h$ function in Eq.~(\ref{eq:hfunc}), ${\rm d}_{1(2)}$ being the symplectic eigenvalues of $\bmsigma_{AB}$, and $d_3 = \sqrt{\det(\bmsigma_{A|B})}$, where:
\begin{align}
    \bmsigma_{A|B}&= \bmsigma_A - \bmsigma_Z \bigg[ \bmsigma_B + \sigmamB \bigg]^{-1} \bmsigma_Z^{\mathsf{T}} \, ,
\end{align}
in which 
\begin{align}
\sigmamB = \lim_{z\rightarrow 0} 
\begin{pmatrix}  z & 0 \\ 0 & z^{-1}\end{pmatrix}
\end{align}
is the $2\times 2$ CM associated with homodyne detection.
Ultimately, the KGR reads:
\begin{align}\label{eq:KGRalphaM}
K(\alpha^2) = \beta I_{AB}(\alpha^2) - \chi_{BE}(\alpha^2) \, ,
\end{align}
$\beta \le 1$ being the reconciliation efficiency, where we highlighted the dependence on the modulation energy $\alpha^2$.
Since we are interested in determining the highest achievable key rate as a function of the transmission
distance $d$, we perform optimization over $\alpha^2$ and obtain:
\begin{align}\label{eq:KGRUNCQPSK}
K = \max_{\alpha^2} K(\alpha^2) \, ,
\end{align}
together with the distance-dependent optimized mean energy $\alpha^2_\opt$.

\begin{figure}
\includegraphics[width=0.49\columnwidth]{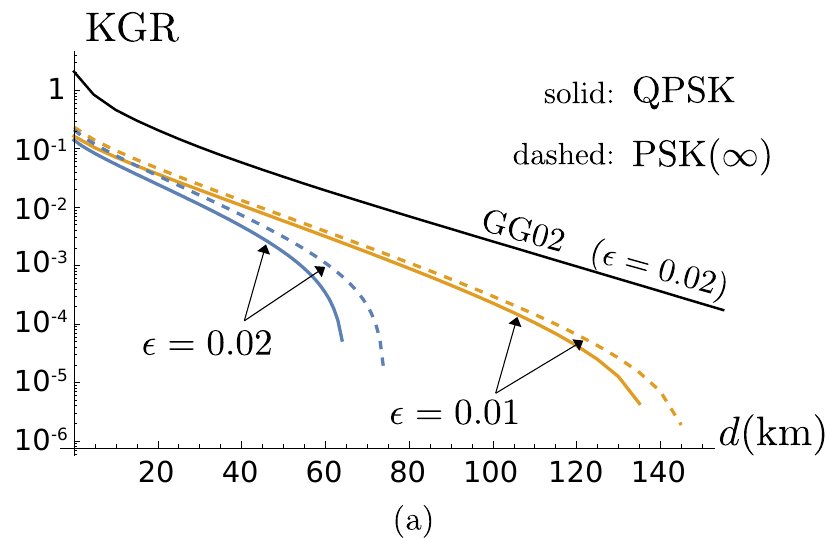} 
\includegraphics[width=0.49\columnwidth]{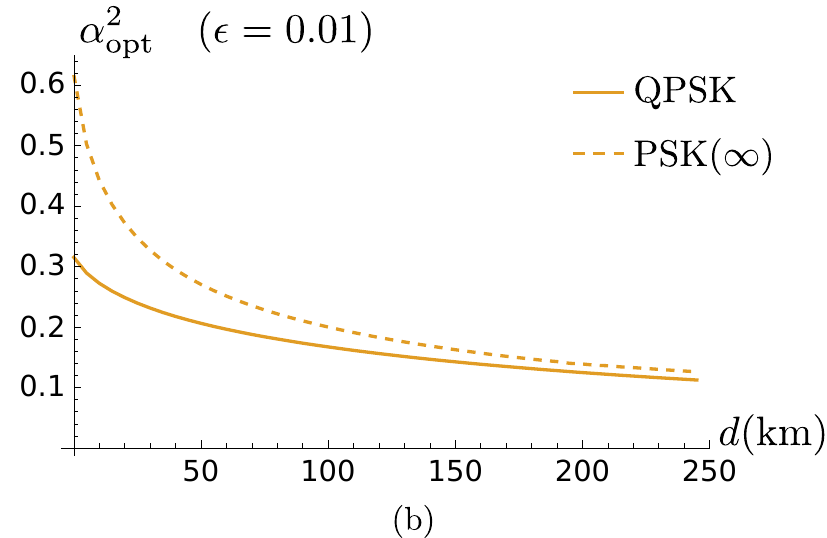}
\centering
\caption{(a) Log plot of the optimized KGR $K$ as a function of the transmission distance $d$ in km for QPSK (solid lines) and PSK($\infty$) (dashed lines) and different values of the excess noise. The black line corresponds to the KGR $K_\GG$ of the GG02 protocol for $\epsilon=0.02$. As we can see, $K < K_\GG$, proving PSK($M$) modulation to be strongly suboptimal with respect to Gaussian modulation. (b) Plot of the optimized modulation energy $\alpha^2_\opt$ as a function of $d$ for QPSK (solid line) and PSK($\infty$) (dashed line) and $\epsilon=0.01$. In both the pictures we set the reconciliation efficiency $\beta=0.95$ and the loss rate $\kappa=0.2$ dB/km.}\label{fig02:sec7.5.1_KGRPSK}
\end{figure}

In Fig.~\ref{fig02:sec7.5.1_KGRPSK}(a) we report plots of the optimized KGR $K$ as a function of the transmission distance $d$ for the QPSK case, i.e. $M=4$, and different channel excess noise~$\epsilon$. Similarly to the GG02 protocol, the KGR is positive up to a maximum transmission distance $d_{\rm max}$ decreasing with the excess noise $\epsilon$. However, as we can see, for a given excess noise, we have $K < K_\GG$, and both the values of KGR and maximum distance of the PSK($M$) protocol are much lower than the corresponding ones achieved by the GG02.
This is a ineludible consequence both of the type of constellation employed and the values of the resulting optimized mean energy $\alpha^2_\opt$, reported in Fig.~\ref{fig02:sec7.5.1_KGRPSK}(b). In fact, at given energy $\alpha^2$, the mutual information $I_{AB}(\alpha^2)$ is close to that shared in GG02, equal to~(\ref{eq:IGG}) and coinciding with the Shannon capacity, only in the range $\alpha^2\lesssim 10^{-1}$. In this regime, the correlation term $Z_M$ of the CM~(\ref{eq:sigmaABPSK}) is $Z_M \lesssim Z_\GG$, where $Z_\GG=\sqrt{V^2-1}$ is the off diagonal CM term of the TSMV state employed in GG02, and, accordingly, also $\chi_{BE}(\alpha^2)$ is close to the Holevo information achieved by the Gaussian modulation protocol.
On the contrary, in high-energy regime, the overlap between the PSK encoded states vanishes and $I_{AB}(\alpha^2)$ saturates to the maximum possible entropy of the constellation, equal to $\log_2(M)$.
In turn, for high $\alpha^2$, the encoded symbols are more ``distinguishable", allowing Eve to retrieve more information, and making the KGR~(\ref{eq:KGRalphaM}) deviate from the GG02 limit and drop below $0$.
The tradeoff between these two energy regimes leads to optimized values of the modulation energy comprised between $0.2 \le \alpha^2_\opt \le 0.6$, being at least one order of magnitude lower than those achieved by GG02. This, ultimately, leads to lower values of optimized KGR and makes PSK modulation strongly suboptimal with respect to Gaussian modulation.

Further improvements may be obtained by increasing the modulation order $M$, e.g. considering the PSK($8$) protocol proposed by Becir {\it et al.} \cite{Becir2012}, for which both the optimized energy and key rate are larger. In light of this, the best performance of PSK modulation is achieved in the limit $M\gg 1$, where we approach the PSK($\infty$), or continuous phase-shift keying, modulation. Now, the constellation is composed of an infinite number of coherent pulses in the form $|\alpha_\phi\rangle= |\alpha e^{i \phi} \rangle $, with the same amplitude $\alpha>0$ and different phase, getting continuous values in the range $\phi \in [0,2\pi)$, and being chosen with a priori probability $p_A(\alpha_\phi)=1/2\pi$, see Fig.~\ref{fig:01:sec7.5.1-PSKM} \cite{Kunz2019}. Then, the average state at Alice's side is the phase-averaged  (PHAV) state \cite{Allevi2012}:
\begin{align}
\rho_{\rm PHAV}= \frac{1}{2\pi} \int_0^{2\pi} d\phi \,  |\alpha e^{i \phi} \rangle\langle \alpha e^{i \phi} | = e^{-\alpha^2} \sum_{n=0}^{\infty} \frac{\alpha^{2n}}{n!} |n\rangle \langle n | \, ,
\end{align}
corresponding to a Poisson-distributed ensemble of Fock states $\{|n\rangle\}_n$, while the correlation term $Z_M$ in the CM~(\ref{eq:sigmaABPSK}) becomes:
\begin{align}
Z_\infty= 2 e^{-\alpha^2} \sum_{n=0}^{\infty} \frac{\sqrt{n+1}}{n!} \alpha^{2n+1} \quad \mbox{for PSK($\infty$) modulation,}
\end{align}
where the (convergent) series has to be evaluated numerically. Accordingly, we follow the same procedure above outlined, and compute the optimized KGR and modulation energy, plotted in Fig.~\ref{fig02:sec7.5.1_KGRPSK}(a) and~(b), respectively. As we can see, PSK($\infty$) induces an enhancement both in the KGR and the maximum transmission distance with respect to QPSK, being more accentuated for higher values of excess noise $\epsilon$. However, as we can see, the gap with respect to GG02 is not closed, thus the sole phase modulation, even of infinitely many coherent states, is not sufficient to approach the peformance Gaussian modulation protocol.
Furthermore, we note that numerical calculations prove that PSK($8$) is sufficient to well approximate the continuous phase-shift keying limit.

\begin{figure}
\includegraphics[width=0.55\columnwidth]{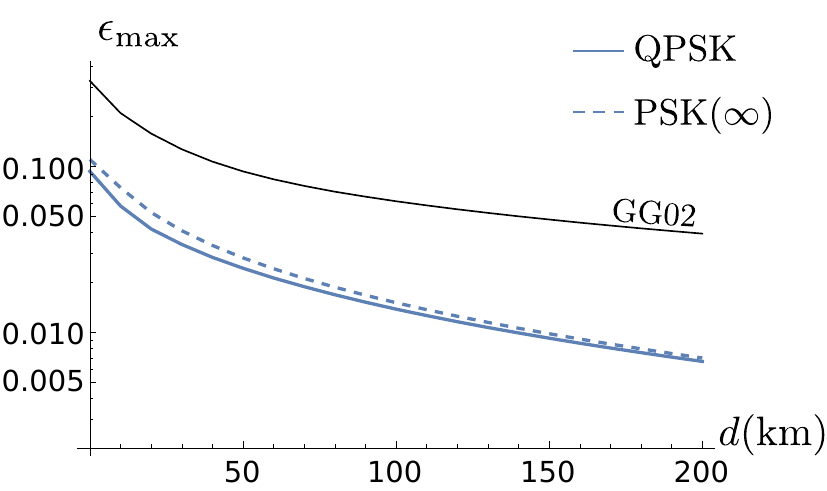}
\centering
\caption{Log plot of the maximum tolerable excess noise $\epsilon_{\rm max}$ for the QPSK (solid line) and PSK($\infty$) (dashed line) protocols as a function of the distance $d$ in ${\rm km}$. The black line corresponds to the maximum tolerable noise $\epsilon_{\rm max}^{(\GG)}$ of GG02. As we see, PSK($M$) protocol are much less tolerant to the channel excess noise than the Gaussian modulation protocol. We set the reconciliation efficiency $\beta=0.95$.}\label{fig03:sec7.5.1_emaxPSK}
\end{figure}

Finally, we compute the maximum tolerable excess noise $\epsilon_{\rm max}$ for both the QPSK and PSK($\infty$) protocols, depicted in Fig.~\ref{fig03:sec7.5.1_emaxPSK}, and compared to the maximum tolerable noise $\epsilon_{\rm max}^{(\GG)}$ of GG02. As expected, we have $\epsilon_{\rm max}< \epsilon_{\rm max}^{(\GG)}$, consistently with the previous discussion.

\subsubsection{Quadrature amplitude modulation (QAM)}\label{subsec:QAMproto}

Despite their practicality for the error correction procedure, in the previous subsection we showed that, for a given reconciliation efficiency $\beta \le 1$, PSK($M$) protocols are strongly suboptimal with respect to GG02, in terms of both KGR and maximum tolerable excess noise.
Therefore, we may look for other suitable discrete modulation formats, that, in principle, could close the gap with the Gaussian modulation protocol, and still provide efficient reconciliation.

Recently, quadrature amplitude modulation (QAM) of a regular grid of signals has been proposed as a promising solution \cite{Denys2021, Roumestan2021, Roumestan2022}. In fact, differently from PSK, QAM constellations may employ a non-uniform discrete probability distribution of the symbols that approximates better the Gaussian one, thus obtaining a higher KGR closer to GG02. To implement this non-uniform sampling, probabilistic amplitude shaping (PAS) is a practical coded modulation scheme that combines QAM, probabilistic constellation shaping, and forward error correction (FEC) to closely approach optimal channel capacity \cite{bocherer2015bandwidth,Buchali:JLT2016,fehenberger2016probabilistic}.
PAS uses a distribution matcher to map uniformly distributed information
bits on QAM symbols with the desired target distribution \cite{CCDM:schulte2016,HIDM:civelli2020,ESS:gultekin2020}.
In particular, a Maxwell--Boltzmann target distribution is considered,
which maximizes the source entropy for a given discrete constellation
and mean energy per symbol \cite{kschischang1993optimal} (in practice,
lower-energy symbols are used more often than higher-energy symbols,
reducing the energy required to achieve a certain information rate).

\begin{figure}
\centerline{\includegraphics[width=0.45\columnwidth]{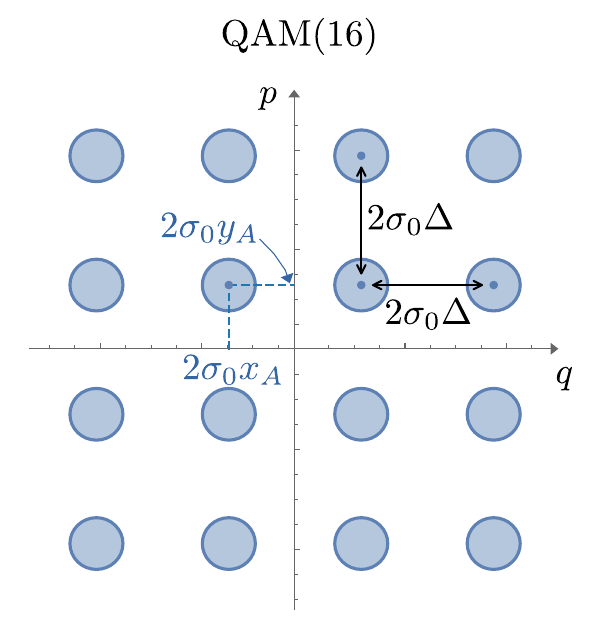}}
\centering
\caption{Phase space representation of the QAM($M^2$) constellation, with $M=4$, composed of a regular grid of $M \times M$ coherent states, centered in $(2\sigma_{0}x_{A},2\sigma_{0}y_{A})$ and with pace $2\sigma_{0}\Delta$, $\sigma_0^2$ being the shot noise variance.}\label{fig:01:sec7.5.2-QAM}
\end{figure}

In more detail, the QAM format adopted by Alice works as follows \cite{proakis01,Cariolaro2015}.
Alice generates a coherent state $|\alpha_{x_A,y_A}\rangle=|x_A+ i y_A\rangle$, where each couple $(x_A, y_A)$ is drawn from the finite set
$\mathcal{A}=\Lambda\times\Lambda$,
where
\begin{align}
\Lambda=\left\{n\Delta\ :\ n=-\frac{M-1}{2},\ldots ,\frac{M-1}{2}\right\}
\end{align} 
contains $M= 2^{k}$ points, for some $k\in\mathbb{N}$.
The points in $\Lambda$ are placed at distance $\Delta \ge 0$ between one another, 
$\Delta \in\mathbb{R}$ being a parameter that determines the mean energy per symbol, which is hence referred to indifferently as {\it scaling factor} or {\it symbol spacing}  \cite{proakis01}.
In turn, the resulting constellation consists in a square lattice of $M \times M$ coherent states, centered in $(2\sigma_{0}x_{A},2\sigma_{0}y_{A})$ and with pace $2\sigma_{0}\Delta$, $\sigma_0^2$ being the shot noise variance, depicted in Fig.~\ref{fig:01:sec7.5.2-QAM} for the relevant case of $M^2=16$ symbols.
As before, we will consider shot noise units (SNU), fixing $\sigma_0^2=1$ in the rest of the analysis.
Unlike PSK, now the pulses encoded by Alice are associated with different mean energies $|\alpha_{x_A,y_A}|^2$, therefore different probability distributions may be investigated to sample the alphabet $\mathcal{A}$. In particular, here we discuss two alternative possibilities.
The former, referred to as case $\a$, is the uniform distribution:
\begin{align}
\mathcal{P}(z)= \frac{1}{M} \, , \quad z=x_{A},y_{A} \, , \qquad \mbox{(case $\a$)} \, ,
\end{align}
commonly exploited in classical communications \cite{proakis01} and quantum state-discrimination schemes \cite{Cariolaro2015}. The latter, case $\b$, is the Maxwell--Boltzmann (MB) distribution: \cite{Cover1999,kschischang1993optimal}
\begin{align}\label{eq:MBdistr}
\mathcal{M}_{\xi}(z)=\frac{e^{-\xi z^{2}}}{Z} \, , \quad z=x_{A},y_{A} \, , \qquad \mbox{(case $\b$)} \, ,
\end{align}
$Z=\sum_{z}e^{-\xi z^{2}}$ being the normalization constant, that depends on the free parameter $\xi$, referred to as the {\it inverse temperature}.
The MB represents the maximum-entropy distribution for a discrete random variable with given variance (mean energy) \cite{Cover1999} and, for a sufficiently large QAM constellation, it has been shown to closely approach the Shannon capacity of the AWGN channel \cite{bocherer2015bandwidth}.
For this reason, it provides a good candidate to also enhance discrete modulation CVQKD, closing the gap between the PSK($M$) and the GG02 protocols. We also note that in the limit $\xi\to\infty$, only the lowest-energy level of the MB have non-zero probability, so that the resulting constellation tends to a simple QAM($4$), i.e. with $M=2$, being equivalent to QPSK. On the other hand, for $\xi=0$, all the levels of the MB distribution have the same probability, and we retrieve the uniform modulation adopted in case $\a$.

We start the security analysis by considering case $\a$. At first, we determine the value of the symbol spacing $\Delta^{(\a)}$ from the mean energy $\bar{n}$ of the constellation. The overall state generated by Alice is:
\begin{align}\label{eq:rhoQAMa}
\rho^{(\a)}= \frac{1}{M^2} \sum_{x_A\in \Lambda} \sum_{y_A\in \Lambda} |x_A+ i y_A\rangle \langle x_A+ i y_A| \, ,
\end{align}
its mean energy being equal to $\bar{n}= 2/M \sum_{x_A} x_A^2= 2 \Delta^2/M \sum_{n=-(M-1)/2}^{(M+1)/2} n^2$, which can be inverted to obtain:
\begin{align}
\Delta^{(\a)}= \sqrt{\frac{6 \bar{n}}{M^2-1}} \, .
\end{align}
Accordingly, the mean energy $\bar{n}$ provides a free parameter in Alice's hands to properly adjust the size of the QAM constellation. Thereafter, once the signals have been generated, Alice injects them into the quantum channel of parameters $(T,\epsilon)$. Bob receives them and performs homodyne detection, associated with the mutual information:
\begin{align}
I_{AB}^{(\a)}(\bar{n}) =H_{B}^{(\a)}-   \frac12  \log_2 \Big[2\pi e (1+T\epsilon) \Big] \, ,
\end{align}
where $H_B^{(\a)}$ is the Shannon entropy of Bob's overall distribution:
\begin{align}\label{eq: pBa}
p_{B}^{(\a)}(x_{B})&= \frac{1}{M} \sum_{x_A}  p_{B|A}(x_{B} | x_{A}) \, ,
\end{align}
with
\begin{align}
p_{B|A}(x_B|x_A) = \frac{\exp\left[- (x_B- 2 \sqrt{T} x_A)^2 /(2(1+ T \epsilon)) \right]}{\sqrt{2\pi (1+ T \epsilon)}} \, . 
\end{align}
Instead, the Holevo information $\chi_{BE}^{(\a)}(\bar{n})$ is computed thanks to the optimality of Gaussian attacks. We exploit Eq.~(\ref{eq: chiBE PSK}) and the CM~(\ref{eq:sigmaABPSK}), where, now, the correlation term $Z_M$ should be changed into:
\begin{align}
Z^{(\a)}= 2\Tr\left\{ \big[\rho^{(\a)}\big]^{1/2} a \, \big[\rho^{(\a)}\big]^{1/2} a^\dagger \right\} \, ,
\end{align}
to be evaluated numerically from the quantum state~(\ref{eq:rhoQAMa}).
The resulting KGR then reads:
\begin{align}
K^{(\a)} = \max_{\bar{n}} \left[ \beta I_{AB}^{(\a)}(\bar{n}) - \chi_{BE}^{(\a)}(\bar{n}) \right] \, ,
\end{align}
together with the optimized energy $\bar{n}_\opt^{(\a)}$.

On the contrary, in case $\b$, the modulation stage is associated with two free parameters: the constellation energy $\bar{n}$ and the inverse temperature $\xi$. Now, the statistical operator at Alice's side becomes:
\begin{align}\label{eq:rhoQAMb}
\rho^{(\b)}= \sum_{x_A,y_A} \mathcal{M}_{\xi}(x_{A})\mathcal{M}_{\xi}(y_{A}) \, |x_A+ i y_A\rangle \langle x_A+ i y_A| \, ,
\end{align}
therefore we have:
\begin{align}
{\bar n}= 2 \sum_{x_A} \mathcal{M}_{\xi}(x_{A}) \, x_A^2 = 2 \Delta^2 \sum_{n=-\frac{M-1}{2}}^{\frac{M+1}{2}} \mathcal{M}_{\xi}(n \Delta) \,n^2 \, ,
\end{align}
to be inverted numerically, and whose corresponding solution $\Delta^{(\b)}(\xi)$ exhibits an implicit dependence also on $\xi$, making the spacing dependent on both the energy and the inverse temperature.
The mutual information shared by Alice and Bob becomes:
\begin{align}
I_{AB}^{(\b)}(\bar{n},\xi) =H_{B}^{(\b)}-   \frac12  \log_2 \Big[2\pi e (1+T\epsilon) \Big] \, ,
\end{align}
where $H_B^{(\b)}$ is the Shannon entropy of the distribution:
\begin{align}\label{eq: pBb}
p_{B}^{(\b)}(x_{B})&=\sum_{x_A} {\cal M}_\xi(x_A) \, p_{B|A}(x_{B} | x_{A}) \, ,
\end{align}
whereas the Holevo information $\chi_{BE}^{(\b)}(\bar{n},\xi)$ is computed by Eq.~(\ref{eq: chiBE PSK}) and the CM~(\ref{eq:sigmaABPSK}), with the correlation term:
\begin{align}
Z^{(\b)}= 2\Tr\left\{ \big[\rho^{(\b)}\big]^{1/2} a \, \big[\rho^{(\b)}\big]^{1/2} a^\dagger \right\} \, .
\end{align}
The obtained KGR is equal to:
\begin{align}
K^{(\b)} = \max_{\bar{n},\xi} \left[ \beta I_{AB}^{(\b)}(\bar{n},\xi) - \chi_{BE}^{(\b)}(\bar{n},\xi) \right] \, ,
\end{align}
with the optimized energy $\bar{n}_\opt^{(\b)}$ and inverse temperature $\xi_\opt$.

\begin{figure}
\includegraphics[width=0.49\columnwidth]{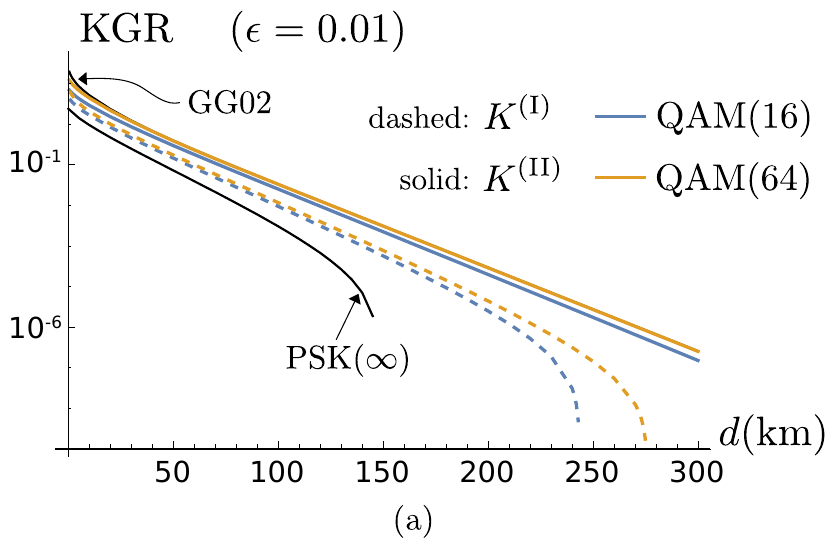}
\includegraphics[width=0.49\columnwidth]{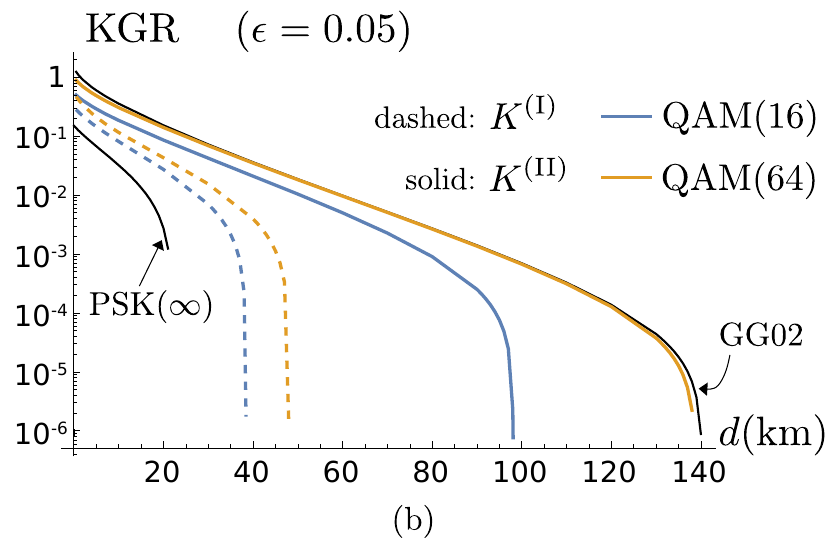}
\centering
\caption{Log plot of the optimized KGRs $K^{(\p)}$, $\p=\a,\b$, as a function of the transmission distance $d$ in km for $\epsilon=0.01$ (a) and $\epsilon=0.05$ (b). As we can see, QAM modulation outperforms the results obtained for PSK($M$) protocols for all $M$, both in the presence of uniform and MB sampling. Moreover, when QAM is further assisted by PAS, namely in case $\b$, by increasing the number of symbols $M$, we progressively close the existing gap between the PSK and the Gaussian modulation. In both the pictures we set the reconciliation efficiency $\beta=0.95$ and the loss rate $\kappa=0.2$ dB/km.}\label{fig02:sec7.5.2_KGRQAM}
\end{figure}

Plots of $K^{(\p)}$, $\p=\a,\b$, are reported in Fig.~\ref{fig02:sec7.5.2_KGRQAM} as a function of the transmission distance $d$ with excess noise $\epsilon=0.01$ (a) and $\epsilon=0.05$ (b), for the relevant constellations QAM($16$) and QAM($32$), corresponding to $M=4,8$, respectively.
As we can see, QAM modulation provides a preferable choice over PSK, beating the PSK($\infty$) protocol for both cases $\p=\a,\b$, and leading to higher KGR and larger maximum transmission distance.
The MB sampling outperforms the uniform one, as $K^{(\b)}\ge K^{(\a)}$, and, remarkably, its corresponding KGR is also able to approach the GG02 scheme for a large enough number of symbols $M$.
In particular, the QAM($32$) constellation is able to well approximate the results of GG02 for excess noises $\epsilon \le 0.05$. In turn, PAS proves itself as a crucial factor to combine both the practical necessity of discrete modulation formats and the possibility to obtain high values of key rate. 

\begin{figure}
\includegraphics[width=0.49\columnwidth]{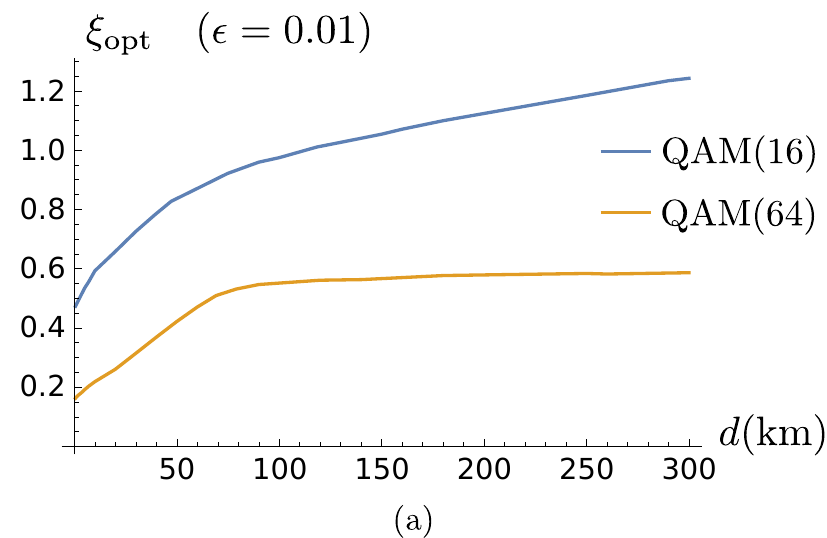} 
\includegraphics[width=0.49\columnwidth]{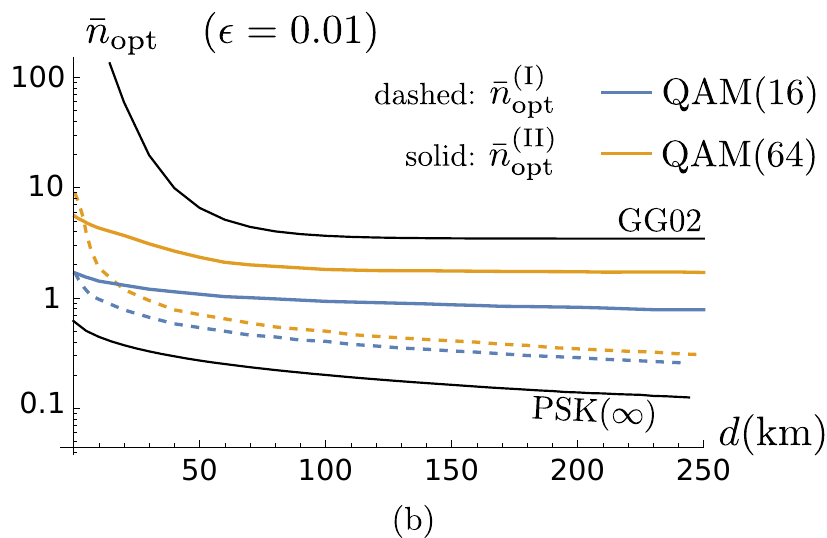}
\centering
\caption{Plot of the optimized parameters $\xi_\opt$ (a) and $\bar{n}_\opt$ (b) for cases $\p=\a,\b$, as a function of the transmission distance $d$ in km. In both the pictures we set the values $\epsilon=0.01$, $\beta=0.95$, and $\kappa=0.2$ dB/km.}\label{fig03:sec7.5.2_OptParQAM}
\end{figure}

The enhancement brought by PAS is due to a nontrivial optimization of the MB distribution, as emerges by considering the optimized inverse temperature $\xi_\opt$, reported in Fig.~\ref{fig03:sec7.5.2_OptParQAM}(a).
In fact, $\xi_\opt$ an increasing function of the transmission distance such that $0<\xi_\opt<\infty$; thus, we do not retrieve neither the QPSK nor the uniform modulation limit, and the best-working performance of the protocol is achieved by exploiting the full cardinality of the constellation, in which the lower-energy symbols are more likely than the higher-energy ones. Moreover, $\xi_\opt$ saturates for large $d$ and is a decreasing function of $M$.
For the sake of completeness, in Fig.~\ref{fig03:sec7.5.2_OptParQAM}(b) we report the optimized energy $\bar{n}_\opt^{(\p)}$, $\p=\a,\b$, which is a decreasing function of the distance and saturates as $d$ increases. As one may expect, increasing the size of the constellation increases also $\bar{n}_\opt^{(\p)}$, until to reach the value of the GG02 scheme. Moreover, for large $d$ we have $\bar{n}_\opt^{(\b)} \ge \bar{n}_\opt^{(\a)}$, and the MB reaches its maximum KGR for higher energies.

\begin{figure}
\includegraphics[width=0.55\columnwidth]{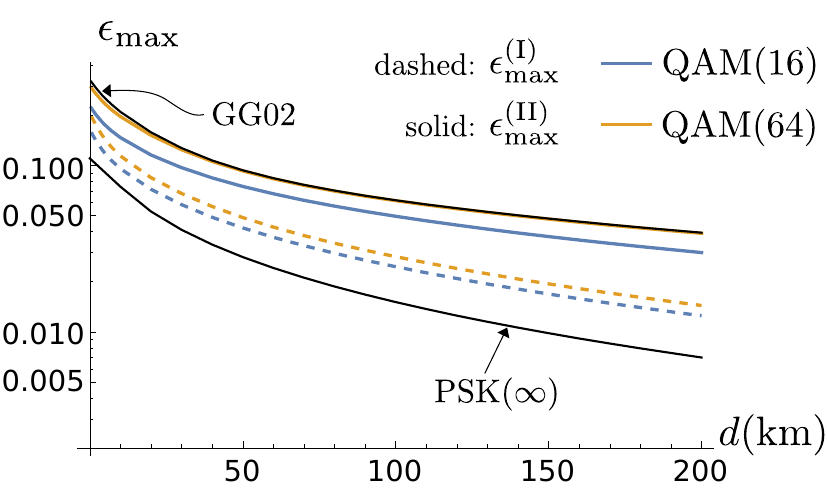}
\centering
\caption{Log plot of the maximum tolerable excess noise $\epsilon_{\rm max}$ for the QAM protocols in cases $\p=\a,\b$, together with the maximum tolerable noises of the PSK($\infty$) and GG02 protocol. Consistently with the previous results, QAM modulation closes the gap with respect to GG02, especially when assisted by PAS. We set the reconciliation efficiency $\beta=0.95$.}\label{fig04:sec7.5.2_emaxQAM}
\end{figure}

Finally, we compute the maximum tolerable excess noise $\epsilon_{\rm max}^{(\p)}$, $\p=\a,\b$, plotted in Fig.~\ref{fig04:sec7.5.2_emaxQAM}, and compared to the maximum tolerable noise of both the PSK($\infty)$ and the GG02 protocols. Consistently with the former analysis, we have $\epsilon_{\rm max}^{(\b)} \ge \epsilon_{\rm max}^{(\a)}$, and, in both the cases, the maximum tolerable noise is larger than that of PSK($\infty$). Remarkably, by increasing the modulation order $M$, the QAM constellation assisted by PAS closes the gap with respect to GG02.

\def\ch{{\rm ch}}
\def\D{{\rm d}}
\def\E{{\bf E}}

\section{CVQKD in the presence of restricted eavesdropping}\label{chap:RESTREAV}

In this Section, we study the two other main security frameworks presented in the previous one, namely the trusted-device scenario and the wiretap channel assumption. They both provide examples of restricted eavesdropping, in which the eavesdropper, Eve, cannot perform arbitrary channel manipulation, unlike the unconditional security case.

In more detail, in the trusted-device scenario we assume a composite quantum channel, given by the composition of two subsequent noisy maps, of which only one is actually controlled by Eve. This scenario models the realistic case of noisy detection at the receiver's side, where we may safely assume that detection noise and losses are simply lost to the environment and cannot be intercepted by Eve, who, instead, controls both the losses and noise acquired during signal transmission. 
On the contrary, the wiretap channel provides an example of specific eavesdropping. Indeed, the term wiretap channel refers to a quantum channel from Alice to Bob being completely characterized in terms of its unitary dilation; accordingly, if we consider the ancillary environmental modes to be controlled by Eve, the wiretap channel not only described signal propagation from both Alice to Bob, but also determines the quantum map Alice $\to$ Eve, ultimately identifying a particular eavesdropping strategy.

The Section is organized as follows. In Sec.~\ref{sec: Trusted}, we study the trusted-device scenario: we extend the validity of the optimality of Gaussian attacks to this framework, and compute the corresponding KGR for the QPSK protocol, as a paradigmatic example. Subsequently, in Sec.~\ref{sec: Wiretap} we introduce the wiretap channel description and perform the associated CVQKD security analysis. Ultimately, we compute the KGR for the QPSK protocol, comparing the obtained results under both the unconditional security and wiretap channel assumptions. 

\subsection{The trusted-device scenario}\label{sec: Trusted}

As widely discussed in the previous Section, the traditional security proofs for CVQKD have been established under unconditional security. This implicitly assumes the presence of an omnipotent eavesdropper having full access to the quantum channel connecting Alice and Bob, namely being able to both collect all the lost photons and control the noise acquired during the propagation \cite{Laudenbach2018}. Within this framework, security has been firstly guaranteed by the optimality of Gaussian attacks, providing a lower bound to DW, while, more recently, an almost exact calculation of DW have been achieved by a sophisticated approach based on SDP \cite{Lin2019, Lin2020, Ghorai2019}. 

In realistic conditions, however, Bob will also experience further noise due to non-idealities in his measurement device, which may not be directly accessible to Eve. 
Therefore, it is justified to address physical layer security under different frameworks, considering different degrees of trust for each setup component. In particular, we retrieve unconditional security if all components are completely untrusted, whereas when some of the setup elements (e.g. detection losses and noise) 
are trusted, we deal with the {\it trusted-device scenario} \cite{Pirandola2021, Laudenbach2019, Notarnicola2023:MJ}. 
This latter scenario includes also loss-compensation strategies, e.g. exploiting phase-insensitive amplifiers to perform signal restoration before detection \cite{Caves1982}, being noisy operations that are useless under the unconditional security approach, but whose application, if trusted, may be beneficial \cite{Fossier2009, Notarnicola2023:MJ}.

To date, security in the trusted scenario has been carried out only for Gaussian protocols, where the optimal eavesdropping strategy is again realized by entangling cloner attack \cite{Laudenbach2019, Notarnicola2023:MJ}, leaving the problem open for more general schemes, e.g. involving discrete modulation.
The first obstacle in this direction is to identify the optimal attack that Eve may launch.
In fact, while in the existing unconditional security proofs Eve's information is computed by a purification ansatz, see Sec.~\ref{subsec:UncGeneral}, this argument does not hold anymore 
in the case of some lack of information on Eve's side, that, now, cannot purify the whole state shared among the parties.
This makes the security analysis nontrivial, as the framework presented in both~\cite{Navascues2006,LeverrierThesis, GarciaPatron2006} and~\cite{Lin2019,Lin2020, Ghorai2019} cannot be applied straightforwardly.

To this aim, in the following we extend for the first time the validity of Gaussian attacks optimality for general CVQKD protocols in the presence of trusted devices. In particular, we prove that, provided Bob's measurement to be Gaussian, the information extracted by Eve is upper bounded by the Holevo information of the Gaussian state having the same first and second momenta of the quantum state shared between Alice, Bob and the trusted parties. The results of this section are original.

\subsubsection{Extending the optimality of Gaussian attacks}\label{subsec: OptGtrusted}

\begin{figure}
\includegraphics[width=0.9\columnwidth]{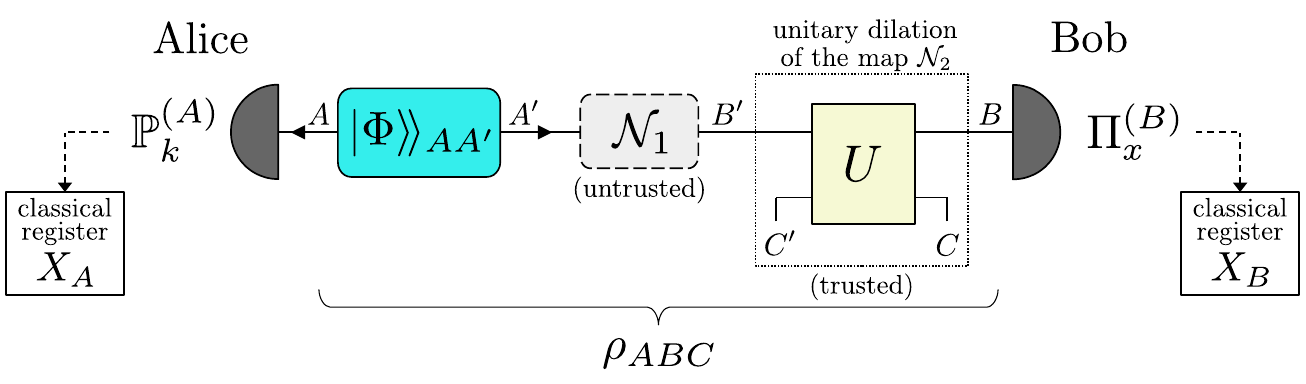}
\centering
\caption{Extended scheme of the EB version of a CVQKD protocol in the presence of trusted devices. Alice prepares the state $|\Phi\rrangle_{A A'}$ and injects mode $A'$ into the quantum channel. The noisy evolution is composed of the two CP maps ${\cal N}_1$ (untrusted) and ${\cal N}_2$ (trusted).
Thereafter, Bob performs a Gaussian measurement $\{\Pi_x^{(B)}\}_x$, storing the obtained outcomes in a classical register $X_B$.}
\label{fig01:sec8.1.1_OptGTrust}
\end{figure}

To investigate security, we adopt the EB framework depicted in Fig.~\ref{fig01:sec8.1.1_OptGTrust}. We model the noisy evolution as a composition of two distinct CP maps ${\cal N}_1$ and ${\cal N}_2$, describing the quantum channel under Eve's control and the further losses and noise, respectively. As a matter of fact, ${\cal N}_1$ is untrusted, whilst ${\cal N}_2$ could be assumed to be trusted, and here we assume it is.

Overall, the signal prepared by Alice on mode $A'$ experiences the subsequent evolutions ${\cal N}_1$ and ${\cal N}_2$ and reaches Bob, who implements a 1-rank Gaussian measurement $\{\Pi_{x}^{(B)}\}_{x}$ and stores the outcomes $x$ in a classical register $X_B$. We denote by $B'$ and $B$ the output modes retrieved after the maps ${\cal N}_1$ and ${\cal N}_2$, respectively (see Fig.~\ref{fig01:sec8.1.1_OptGTrust}).
Since ${\cal N}_2$ is trusted, we have access to its unitary dilation guaranteed by Kraus theorem, that is a set of $n$ trusted ancillary modes $C'=(C'_1,\ldots, C'_n)$ and a unitary operation $U$ coupling modes $B'$ and $C'$ \cite{Paris2012}. The output ancillary modes after $U$ are referred to as $C$. Ultimately, Alice and Bob share the state $\rho_{ABC}$ of modes $ABC$.


Given this scenario, the optimal eavesdropping strategy is the ``purification attack" as proved in \cite{Laudenbach2018, Laudenbach2019}, where the eavesdropper is assumed to ``purify" the state $\rho_{ABC}$. Then, Eve has access to the unitary dilation of ${\cal N}_1$: the state of the global (closed and isolated) system of modes $ABCE$ can be written as a pure state 
$|\Psi\rangle_{ABCE}$ such that $\rho_{ABC}= \Tr_E\left[|\Psi\rangle_{ABCE}\langle \Psi|\right]$.
In turn, thanks to the property of von Neumann entropy \cite{Laudenbach2018, Laudenbach2019, Notarnicola2023:NLA}, the Holevo information between Bob and Eve becomes $\chi(B;E) = S(ABC) - S(ABC|X_B)$, depending only on $\rho_{ABC}$.

Starting from these considerations, the extension of the optimality of Gaussian attacks becomes straightforward. In fact, we now prove that, for any state $\rho_{ABC}$:
\begin{align}\label{eq:OptGausstrust}
\chi(B;E) \le \chi_\G(B;E) \, ,
\end{align}
where $\chi_\G(B;E)$ is the Holevo information computed for the $(n+2)$-mode Gaussian state having the same CM as $\rho_{ABC}$.
To this aim, we exploit the results derived in framework presented in Sec.~\ref{subsec:GOLemmas} and evaluate the quantity $\Delta  \chi(B;E)$, see Eq.~(\ref{delta:gauss}), whose positivity confirms Gaussian optimality. Indeed:
\begin{align}
\Delta  \chi(B;E) & =\chi_\G(B;E) -\chi(B;E) \nonumber \\
&= \Delta S(ABC) - \Delta S(ABC|X_B) \nonumber \\
&= \Delta S(ABC) - \Delta S\left(\overline{ABCX_B}\right) + \Delta H(X_B)
\end{align}
where we used Lemma~\ref{Lemma1:GOpt} to get the last equality. Since the map $ABC \rightarrow \overline{ABCX_B}$ is Gaussian, as Bob's detection is Gaussian, Lemma~\ref{Lemma3:GOpt} implies:
\begin{align}
\Delta S(ABC) - \Delta S\left(\overline{ABCX_B}\right) + \Delta H(X_B) \ge \Delta H(X_B) \, ,
\end{align}
and, finally, exploiting Lemma~\ref{Lemma2:GOpt} we get $\Delta H(X_B) \ge 0$.

As for the unconditional security framework, Eq.~(\ref{eq:OptGausstrust}) bounds the maximum amount of information that Eve may extract during the protocol, leading to a lower bound of the KGR achievable by Alice and Bob. Remarkably, this bound can be saturated iff $\rho_{ABC}$ itself is a Gaussian state, in which case the optimal eavesdropping coincides with an entangling cloner attack \cite{Laudenbach2018, Laudenbach2019}. Moreover, if also the map ${\cal N}_2$ were untrusted, we would retrieve the standard proof of Gaussian optimality under unconditional security by tracing out modes $C$ and computing the CM of $\rho_{AB}=\Tr_{C}[\rho_{ABC}]$.

The calculation of the CM $\bmsigma_{ABC}$ associated with $\rho_{ABC}$ is further simplified under some realistic assumptions. 
First of all, if the map ${\cal N}_2$ is Gaussian, it suffices to compute the CMs $\bmsigma_{AB'}$ and $\bmsigma_{C'}$ of the states of $AB'$ and $C'$, respectively, and:
\begin{align}\label{eq:sigmaABC}
\bmsigma_{ABC} = \left(\Id_2 \oplus S \right) \bmsigma_{AB'}\oplus \bmsigma_{C'} \left(\Id_2 \oplus S \right)^\mathsf{T} \, ,
\end{align}
where $\oplus$ denotes direct sum, $\Id_2$ is the $2\times 2$ identity matrix and $S$ is the symplectic matrix associated with the unitary operator $U$ in Fig.~\ref{fig01:sec8.1.1_OptGTrust} \cite{Ferraro2005, Serafini2017}.
Furthermore, if the channel ${\cal N}_1$ is linear and characterized by a transmissivity $0\le T_\ch \le 1$ and excess noise at the transmitter $\epsilon_\ch\ge 0$, such that the total added noise is $\chi_\ch=(1-T_\ch)/T_\ch + \epsilon_\ch$, the CM $\bmsigma_{AB'}$ reads:
\begin{align}\label{eq:sigmaABprime}
\bmsigma_{AB'} = \begin{pmatrix} V \, \Id_2 & \sqrt{T_\ch} \, Z \,\bmsigma_z \\ \sqrt{T_\ch} \,Z \,\bmsigma_z & T_\ch\left(V+\chi_\ch\right)\, \Id_2 \end{pmatrix} \, ,
\end{align}
where $V=1+2\bar{n}$, $\bar{n}= \sum_k p_A(\alpha_k) |\alpha_k|^2$ being the mean number of photons per symbol and $Z=\llangle \Phi | (q_A q_{A'} -p_A p_{A'}) |\Phi\rrangle/2= 2 \Tr[\rho^{1/2} a  \, \rho^{1/2}a^\dagger]$, in which $\rho=\sum_k p_A(\alpha_k) |\alpha_k\rangle \langle \alpha_k|$ is overall quantum state at Alice's side, see Eq.~(\ref{eq:ZDenys}).
In the more general case of an arbitrary ${\cal N}_1$, we should consider the bound in Eq.~(\ref{eq:Zbound}) for the off-diagonal block terms of $\bmsigma_{AB'}$, as discussed in Sec.~\ref{sec:OptGaussUnc}.

The present extension of the Gaussian optimality theorem provides a cornerstone to prove security under restricted eavesdropping in wider scenarios.
First of all, it guarantees a sufficient condition to assess security in non-Gaussian protocols with Gaussian measurements, e.g. involving discrete modulation or non-Gaussian channels \cite{Notarnicola2023:NLA, Ghalaii, Ghalaii_DM}.
Thereafter, it provides a scalable approach that may be extended to composite quantum channels, such as multi-span links \cite{Notarnicola2023:MJ, Jarzyna2019, Lukanowski2023}, free-space settings with multiple noise sources \cite{Pirandola2021, Usenko2012, Ruppert2019,Derkach2020}, or trusted preparation noise \cite{Usenko2016, Laudenbach2019}. In fact, if the channel connecting Alice and Bob is modeled by a sequence of $M\ge 2$ CP maps ${\cal N}_j$, $j=1,\ldots,M$, of which only $m<M$ are trusted, we bound Eve's Holevo information from the CM of the state of the quantum system composed of Alice, Bob and the ancillary modes associated with the $m$ maps. Moreover, by increasing $m$, the modes under Eve's control are progressively reduced and the KGR increases accordingly.
As a final remark, we note that the philosophy adopted here can be also embedded into the SDP approach of \cite{Lin2019,Lin2020}, where, now, the quantum state shared between Alice, Bob and the trusted parties should be considered to perform the convex optimization algorithm.

\subsubsection{A case study: the QPSK protocol}\label{subsec: DMTrusted}

We now apply the optimality of Gaussian attacks to a paradigmatic example: the quadrature phase-shift keying (QPSK) protocol originally proposed in \cite{Leverrier2009}.
As discussed in Sec.~\ref{subsec:PSKproto}, in the QPSK protocol Alice samples one of the four coherent states $|\alpha_k\rangle = |\alpha e^{i \pi (2k+1)/4}\rangle$, generated with equal a priori probability $p_A(\alpha_k)=1/4$. The signal is then injected into the untrusted channel ${\cal N}_1$, assumed to be a linear channel with parameters $(T_\ch,\epsilon_\ch)$, until to reach Bob, who implements 
homodyne detection of either quadrature $q_B$ or $p_B$. 

We perform the security analysis by considering trusted noisy detection at Bob's side, including non-unit quantum efficiency $\eta \le 1$ and nonzero electronic noise.
In this case, the map ${\cal N}_2$ describes detection noise, modeled as a thermal-loss (Gaussian) channel, where detection losses are described as a beam splitter with transmissivity $\eta$, and the electronic noise is a thermal noise arising from a two-mode squeezed state (TMSV) on two ancillary modes $C'=(C'_1,C'_2)$, whose first branch $C_1'$ is injected into the auxiliary port of the previous beam splitter \cite{Lodewyck2007, Fossier2009, Usenko2016, Pirandola2021}. The TMSV has $\bar{n}_\D= \eta T_\ch \epsilon_\D/[2(1-\eta)]$ mean photons, $\epsilon_\D\ge0$ being the detection excess noise.
In turn, state $\rho_{C'}$ is a Gaussian state with CM:
\begin{align}\label{eq:sigmaCprime}
\bmsigma_{C'} = \begin{pmatrix} V_\D \, \Id_2 & Z_\D \,\bmsigma_z \\ Z_\D \,\bmsigma_z & V_\D\, \Id_2 \end{pmatrix} \, ,
\end{align}
where $V_\D=1+2\bar{n}_\D$ and $Z_\D=\sqrt{V_\D^2-1}$ \cite{Lodewyck2007}.
The global channel, given by the subsequent applications of ${\cal N}_1$ and ${\cal N}_2$, is then associated with a total transmissivity $T_{\rm tot}= \eta T_\ch$ and excess noise $\epsilon_{\rm tot}=\epsilon_\ch + \epsilon_\D$ \cite{Pirandola2021, Laudenbach2019}.

To address physical layer security, we identify three scenarios associated with different trust levels at the {\it detection}:
\begin{itemize}
\item  trusted losses and noise (${\rm tL; tN}$);
\item trusted losses and untrusted noise (${\rm tL; uN}$);
\item untrusted losses and noise (${\rm uL; uN}$).
\end{itemize}
The latter scenario falls under the unconditional security framework. In all the cases the lower bound on the KGR is computed as 
\begin{align}
K^{\rm (sL;sN)} = \max_{\alpha^2} \left[\beta I_{AB}(\alpha^2) - \chi_{BE}^{\rm(sL;sN)} (\alpha^2) \right]
\, , \qquad \rm s=t,u\, ,
\end{align}
where $\beta\le 1$ is the reconciliation efficiency, $I_{AB}(\alpha^2)$ is the exact mutual information shared between Alice and Bob in Eq.~(\ref{eq:IAB_PSK}), while $\chi_{BE}^{\rm(sL;sN)} (\alpha^2)$ is the Gaussian bound over Eve's Holevo infomation, to be computed in different ways according to the scenario under investigation.
For case (${\rm tL; tN}$), both modes $C'$ are trusted, therefore $\chi_{BE}^{\rm(sL;sN)} (\alpha^2)$ is retrieved from the $8 \times 8$ CM $\bmsigma_{ABC}$ in Eq.~(\ref{eq:sigmaABC}), with the symplectic matrix $S_\eta$ of the detection-losses beam splitter, namely:
\begin{align}
S_\eta = \begin{pmatrix} \sqrt{\eta} \, \Id_2 & \sqrt{1-\eta} \, \Id_2\\-\sqrt{1-\eta} \, \Id_2 & \sqrt{\eta}\, \Id_2 \end{pmatrix} \, ,
\end{align}
and the CMs $\bmsigma_{C'}$ in~(\ref{eq:sigmaCprime}) and $\bmsigma_{AB'}$ in~(\ref{eq:sigmaABprime}), with the choice of parameters $V=1+2\alpha^2$ and
\begin{align}
Z= 2 \alpha^2 \sum_{k=0}^{3} \frac{\lambda_{k}^{3/2}}{\lambda_{k+1}^{1/2}}  \, ,
\end{align}
with $\lambda_{0,2}= e^{-\alpha^2}[\cosh(\alpha^2) \pm \cos(\alpha^2)]/2$, and $\lambda_{1,3}= e^{-\alpha^2}[\sinh(\alpha^2) \pm \sin(\alpha^2)]/2$ \cite{Leverrier2009, LeverrierThesis}. The von Neumann entropy of the Gaussian state associated with $\bmsigma_{ABC}$ is then retrieved by Eq.~(\ref{eq:EntropyGS}).
On the contrary, for case (${\rm tL; uN}$) detection noise is untrusted, therefore the mode $C_2$ shall be assumed under Eve's control and the Holevo information $\chi_{BE}^{\rm(sL;sN)} (\alpha^2) $ is computed from the $6\times 6$ CM $\bmsigma_{ABC_1}$ of state $\rho_{ABC_1}= \Tr_{C_2}[\rho_{ABC}]$, obtained by selecting the sub-blocks of $\bmsigma_{ABC}$ associated with modes $ABC_1$. Similarly, for case (${\rm uL; uN}$) both modes $C$ are untrusted, therefore $\chi_{BE}^{\rm(sL;sN)} (\alpha^2) $ is obtained with the unconditional security approach, by considering the state $\rho_{AB}= \Tr_{C}[\rho_{ABC}]$ and its associated $4\times 4$ CM $\bmsigma_{AB}$.

\begin{figure}
\includegraphics[width=0.49\columnwidth]{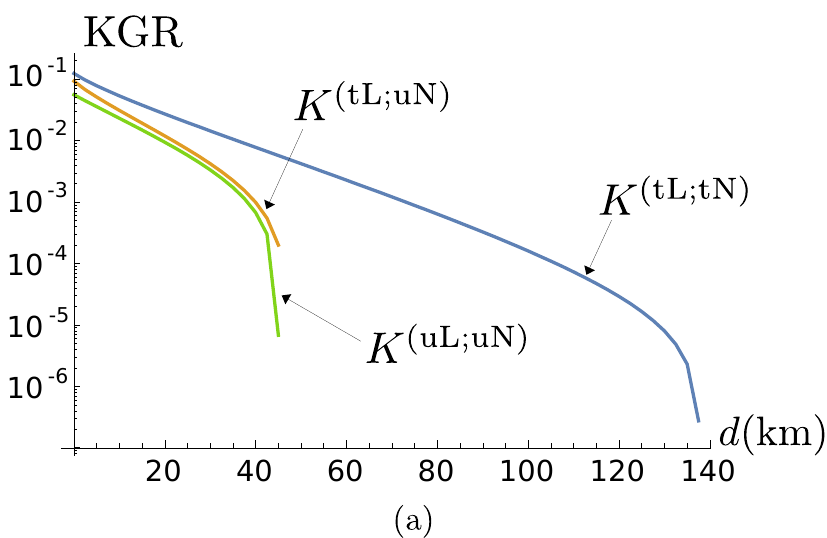}
\includegraphics[width=0.49\columnwidth]{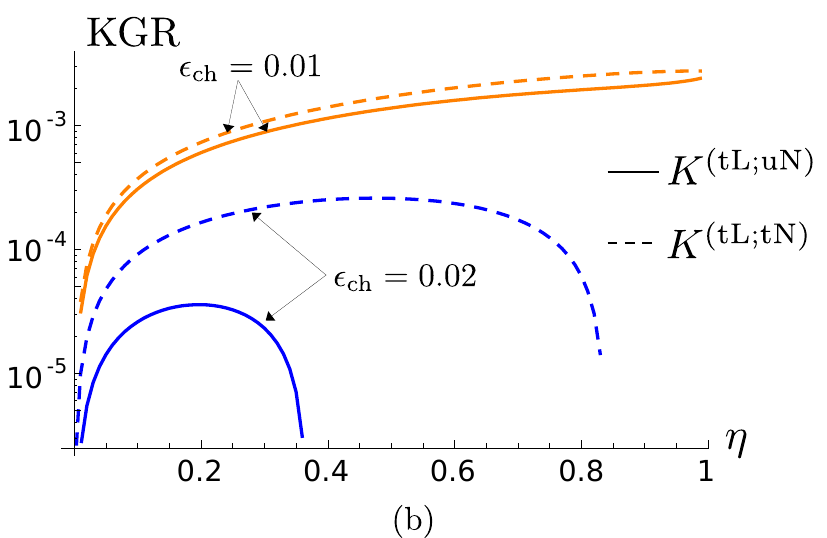}
\centering
\caption{(a) Log plot of the optimized KGRs of the QPSK protocol $K^{({\rm sL; sN})}$, $\rm s=t,u$, as a function of the transmission distance $d$ in km, for the realistic parameters $\eta=0.7$, $\epsilon_\D=0.01$, and $\epsilon_\ch=0.01$ \cite{Pirandola2021}.
(b) Log plots of the optimized KGRs $K^{({\rm tL; uN})}$ and $K^{({\rm tL; tN})}$, as a function of the quantum efficiency $\eta$ for fixed transmission distance $d=60$ km and detection noise $\epsilon_\D=0.001$, and different channel excess noise $\epsilon_\ch$. For large $\epsilon_\ch$, there is an increase in the KGR in the trusted-device scenario. In both the pictures we set the reconciliation efficiency $\beta=0.95$ and the loss rate $\kappa=0.2$ dB/km.}\label{fig01:sec8.1.2_KGRtrusted}
\end{figure}

The three resulting KGRs are reported in Fig.~\ref{fig01:sec8.1.2_KGRtrusted}(a) as a function of the transmission distance $d$,
for the realistic values of quantum efficiency $\eta=0.7$ and detection noise $\epsilon_\D=0.01$ \cite{Pirandola2021}, and channel excess noise $\epsilon_\ch=0.01$. In all cases, the modulation variance $V$ has been optimized to maximize the key rate value. As we see, $K^{({\rm uL; uN})} \le K^{({\rm tL; uN})} \le K^{({\rm tL; tN})}$ and in the trusted-device scenario both the KGR and the maximum transmission distance are increased. In particular, a crucial role is played by the presence of trusted detection excess noise.
ndeed, the detection noise accounts for both electronic noise of realistic measurement devices and phase mismatch introduced by an imperfect LO in the homodyne setup \cite{Lodewyck2007, Pirandola2021}, being typically larger than the channel excess noise, $\epsilon_\D \gtrsim \epsilon_\ch$. Therefore, 
a trusted $\epsilon_\D$ guarantees a huge increase in the transmission distances reached by the protocol.

Furthermore, in the presence of trusted detector we observe the ``fighting noise with noise" effect. That is, as displayed in Fig.~\ref{fig01:sec8.1.2_KGRtrusted}(b), for cases (${\rm tL; uN}$) and (${\rm tL; tN}$) and large channel excess noise $\epsilon_\ch$, trusted losses and noise {\it increase} the key rate with respect to the lossless detection scheme \cite{Renner2005, GarciaPatronThesis, Laudenbach2019}.
It happens when the values of $\eta$ and $\epsilon_\D$ decrease Eve's information more than the mutual information $I_{AB}(\alpha^2)$, resulting in a higher KGR. While this effect is fragile against detection noise for case (${\rm tL; uN}$), it shows more robustness for case (${\rm tL; tN}$) even for large $\epsilon_\D$.

\subsection{The wiretap channel}\label{sec: Wiretap}

\begin{figure}
\includegraphics[width=0.7\columnwidth]{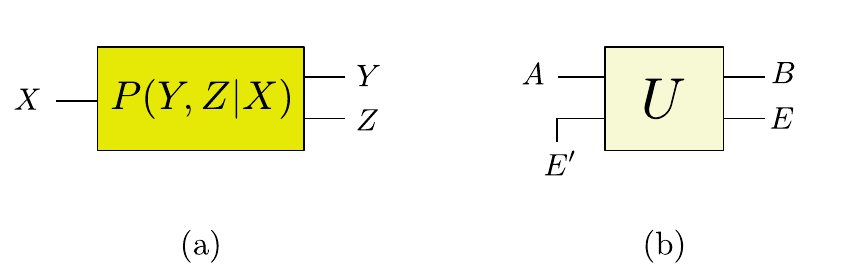}
\centering
\caption{Schematic representation of a classical (a) and quantum (b) wiretap channel.}\label{fig01:sec8.2.1_Wiretap}
\end{figure}

Finally, we discuss the last proposed security framework and address secret-key distillation in the presence of a wiretap channel. 
Generally speaking, the concept of wiretap channel has been first introduced by Wyner in 1975 in the context of classical communications \cite{Wyner1975}. He addressed the problem of reliable transmission of classical information over a memory-less channel being wiretapped at the receiver's side, in which the wiretapper views the channel output by a second (separately parameterized) memory-less channel. In turn, in classical information theory, 
the wiretap channel is completely characterized by the conditional probability distribution $P(Y,Z|X)$ between the sender, described by a classical random variable $X$, the legitimate receiver $Y$, {\it and} the wiretapper $Z$, see Fig.~\ref{fig01:sec8.2.1_Wiretap}(a).

Moving to the quantum realm, the wiretap scenario involves a quantum channel  ${\cal N}: A \to B$ between the sender, holding a quantum system $A$, and the receiver $B$, {\it in the presence} of a third malicious party, namely the eavesdropper Eve, $E$, that wiretaps the channel output by a second quantum channel $A' \to E$.
Accordingly, the quantum wiretap channel, shown in Fig.~\ref{fig01:sec8.2.1_Wiretap}(b), is formally described by the unitary dilation $U$ of the CP map $\cal N$ guaranteed by Kraus theorem. The transformation $U: AE' \to BE$ couples Alice's system $A$ to an ancillary quantum system $E'$, leading, at the output, to a joint correlated system $BE$.
We assume that Eve controls the input system $E'$, being prepared in a state statistically independent of $A$, such that, at the output, subsystem $B$ is sent to Bob, while she keeps subsystem $E$ for herself.
Furthermore, from the perspective of CVQKD, the wiretap channel description implies that the action of the eavesdropper is known, as the unitary $U$ is specified, and, unlike the unconditional security case, no arbitrary channel manipulation is allowed. This captures the realistic idea that Eve is not omnipotent and cannot perform any quantum operation, but is partially limited to implement realistic attacks compatible with the state-of-the-art technologies \cite{Pan2019,Pan2020}. 

\begin{figure}
\includegraphics[width=0.9\columnwidth]{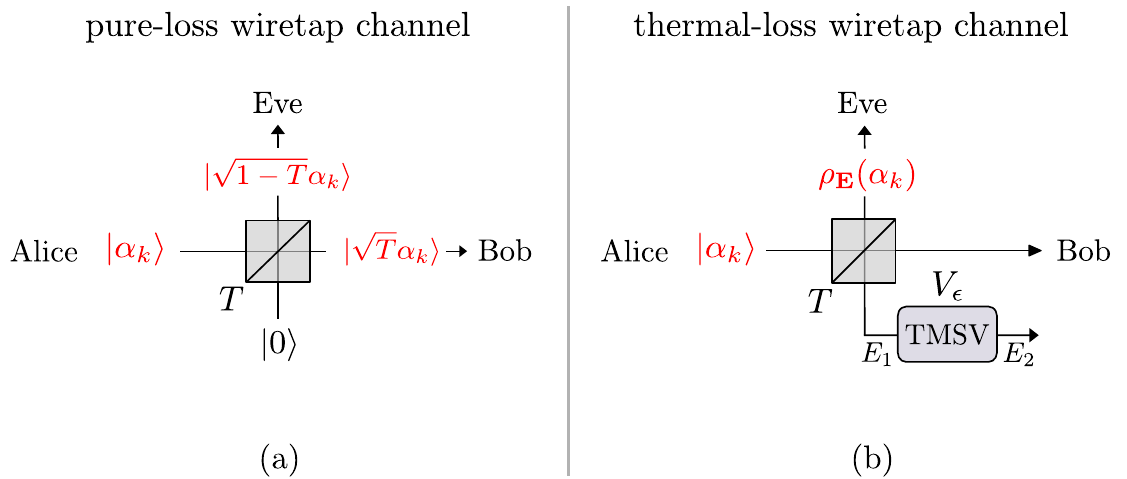}
\centering
\caption{Scheme of the PM CVQKD protocol in the presence of a pure-loss (a) and thermal-loss (b) wiretap channel. In the former case, Eve only collects the fraction of the signals lost during the
propagation through the channel, whereas in the latter, she performs an entangling-cloner attack: that is she injects one arm of a TMSV into the channel beam splitter, retrieving the final output state.}\label{fig02:sec8.2.1_PTWiretap}
\end{figure}

The wiretap channels associated with the usual CVQKD protocols over either pure-loss and thermal-loss channels, are depicted in Fig.~\ref{fig02:sec8.2.1_PTWiretap}(a) and (b), respectively, where, now, the prepare and measure (PM) approach provides the natural framework to adopt, as the unitary dilation of the channel map is known \cite{Laudenbach2019, Pan2019, Pan2020}. In both the cases, transmission losses are described by a beam splitter dynamics of suitable transmissivity $T=10^{-\kappa d/10}$, where $\kappa$ is the photon loss rate of the link. In the pure-loss case, the auxiliary port of the beam splitter is left in the vacuum, and Eve performs passive eavesdropping: that is she only collects the reflected fraction of the encoded signals. On the contrary, in the latter case, she also generates the channel excess noise acquired by the pulses, thus implementing active eavesdropping. We model this effect via the entangling cloner scheme, see Fig.~\ref{fig02:sec7.3.1_EntCloner}. In more detail, Eve locally prepares a TMSV with variance $V_\epsilon=1+T \epsilon/(1-T)$ on two modes ${\bf E}=(E_1,E_2)$, and makes branch $E_1$ interfere with the pulse sent by Alice; ultimately, she collects the output reflected state. 
Physically, this implies that Eve is limited to implement a Gaussian attack, in which she both collects all the lost photons and hides behind the channel excess noise.

We conclude that, in general, performing CVQKD over a quantum wiretap channel yields a higher KGR than the unconditional security scenario, as we are considering a particular Gaussian unitary dilation of the noisy map ${\cal N}: A\to B$. On the contrary, the GG02 protocol provides a special case where the two approaches lead to same KGR, given that the optimal attack is known, and equal to the entangling-cloner.

\subsubsection{The QPSK protocol over a wiretap channel}\label{subsec: Secondini}

\begin{figure}
\includegraphics[width=0.6\columnwidth]{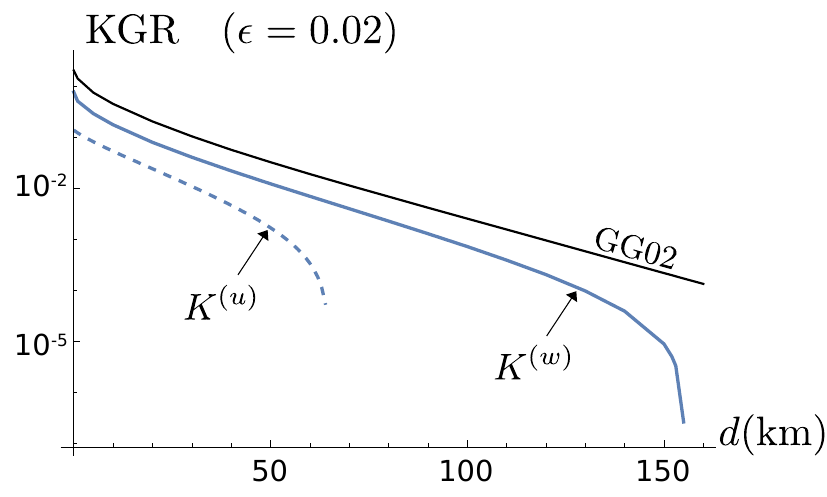}
\centering
\caption{Log plot of the optimized KGR $K^{(w)}$ of the QPSK protocol over a wiretap channel, as a function of the transmission distance $d$ in km, compared to both the KGR $K^{(u)}$ of the QPSK protocol under unconditional security and the GG02 protocol. We set the reconciliation efficiency $\beta=0.95$, the loss rate $\kappa=0.2$ dB/km and the excess noise $\epsilon=0.02$.}\label{fig03:sec8.2.1_KGR}
\end{figure}

As a paradigmatic example, we now compute the KGR for the QPSK protocol, introduced in Sec.~\ref{subsec:PSKproto}, in the presence of a thermal-loss wiretap channel.
According to the previous discussions, when performing the entangling-cloner attack, Eve is undetected by Alice and Bob, who then share the same mutual information $I_{AB}(\alpha^2)$ in Eq.~(\ref{eq:IAB_PSK}).
The difference with respect to the unconditional security approach lies in the amount of information achievable by Eve. In fact, since the wiretap channel represents a particular unitary dilation, we expect Eve to have access to the Holevo information $\chi_{BE}^{(w)} (\alpha^2)$, being lower than that in Eq.~(\ref{eq: chiBE PSK}).

However, in the present case, the computation of $\chi_{BE}^{(w)} (\alpha^2)$ is not straightforward and requires the advanced tools of Gaussian formalism summarized in Sec.~\ref{subsec:Gaussianstates}.
In particular, we adopt the PM description depicted in Fig.~\ref{fig02:sec8.2.1_PTWiretap}(b), noting that, if Alice samples the coherent state $|\alpha_k\rangle=|\alpha e^{i \pi (2k+1)/4}\rangle$, $k=0,\ldots, 3$, both Eve's overall and conditional states $\rho_{\bf E }(\alpha_k)$ and $\rho_{{\bf E} |x_B}(\alpha_k)$, respectively, are Gaussian states.
In more detail, in a thermal-loss wiretap channel, Eve generates a TMSV state with variance $V_\epsilon=1+T \epsilon/(1-T)$, with zero first moment (FM) vector, ${\bf r}_{\bf E }^{(0)} =0$, and CM
\begin{align}
\boldsymbol{\sigma}_{\E }^{(0)} =
\begin{pmatrix}
V_\epsilon \, \Id_2 & Z_\epsilon \, \boldsymbol{\sigma}_z \\ 
Z_\epsilon\,  \boldsymbol{\sigma}_z  & V_\epsilon \, \Id_2
\end{pmatrix} \, ,
\end{align}
with $Z_\epsilon=\sqrt{V_\epsilon^2-1}$ and $\boldsymbol{\sigma}_z$ being the Pauli $z$-matrix.
Moreover, if Alice samples a state with amplitude $\alpha_k$, she gets a single-mode Gaussian state with FM ${\bf r}_A^{(0)}= 2\alpha \,[\cos((2k+1)\pi/4),\sin((2k+1)\pi/4)]$ and CM $\boldsymbol{\sigma}_A^{(0)}= \Id_2$.
Thereafter, Alice's pulse interferes at the channel beam splitter with Eve's mode $E_1$, resulting in a tripartite Gaussian state $\rho_{A\E }(\alpha_k)$ characterized by FM and CM 
equal to
\begin{align}\label{eq.TripartiteAE}
{\bf r}_{A\E } = S \, \left({\bf r}_A^{(0)} \oplus {\bf r}_{\E }^{(0)}\right) \qquad \mbox{and} \qquad
\boldsymbol{\sigma}_{A\E } = S \, \left(\boldsymbol{\sigma}_A^{(0)} \oplus \boldsymbol{\sigma}_{\E }^{(0)}\right) \, S^\mathsf{T} \, ,
\end{align}
with $S= S_{\rm BS} \oplus \Id_2$ and
\begin{align}
S_{\rm BS} =
\begin{pmatrix}
\sqrt{T} \, \Id_{2} & \sqrt{1-T} \, \Id_{2} \\
-\sqrt{1-T} \, \Id_{2} & \sqrt{T} \, \Id_{2}
\end{pmatrix}
\end{align}
being the symplectic matrix associated with the beam splitter operation. From~(\ref{eq.TripartiteAE}) we straightforwardly derive both Eve's overall state $\rho_{\E }(\alpha_k)$ and conditional state after Bob's homodyne measurement $\rho_{\E |x_B}(\alpha_k)$ \cite{Ferraro2005, Serafini2017}.
%
In turn, the overall quantum states at Eve's sides read:
\begin{subequations}\label{eq:StatesExN}
\begin{align}
\rho_{\E} &= \frac14 \sum_{k=0}^{3} \rho_{\E}(\alpha_k) \, ,\\
\rho_{\E |x_{B}} &= \frac{1}{4 p_{B}(x_{B})}\sum_{k=0}^{3} p_{B|A}(x_B |\alpha_k) \, \rho_{\E |x_B}(\alpha_k) \, , 
\end{align}
\end{subequations}
with the probability distributions in~(\ref{eq:pB|A_PSK}) and~(\ref{eq:pB_PSK}).
Ultimately, we obtain the Holevo information as:
\begin{eqnarray}
\chi_{BE}^{(w)}(\alpha^2) =S\big[\rho_{\E }\big]-\int dx_{B} \, p_B (x_{B}) \, S\big[\rho_{\E |x_{B}} \big]  \, ,
\end{eqnarray}
that can be computed numerically by suitably expanding states~(\ref{eq:StatesExN}) onto the Fock basis \cite{Quesada2019}.
The resulting KGR then reads:
\begin{eqnarray}
K^{(w)}= \max_{\alpha^2} \Big[ \beta I_{AB}(\alpha^2)-\chi_{BE}^{(w)}(\alpha^2) \Big] \, ,
\end{eqnarray}
optimized over the input modulation energy.

Plot of $K^{(w)}$ is reported in Fig~\ref{fig03:sec8.2.1_KGR} as a function of the transmission distance $d$ in km for $\epsilon=0.02$, compared to the KGR $K^{(u)}$ of the QPSK protocol under unconditional security in Eq.~(\ref{eq:KGRUNCQPSK}). As expected, we have $K^{(w)} \ge K^{(u)}$, and the wiretap channel assumption leads to higher values of key rate and maximum transmission distance. Incidentally, we also note that the gap between $K^{(w)}$ and $K^{(u)}$ provides a measure of the non-Gaussianity of the protocol, as the entangling cloner scheme embedded in the wiretap model provides the optimal eavesdropping strategy for a Gaussian protocol. In turn, we expect the separation between the two key rates to be progressively reduced when considering modulation formats of increasing order than better approximate Gaussian modulation, e.g. QAM.

\def\Re{{\mathrm{Re}}}
\def\Im{{\mathrm{Im}}}
\def\Id{{\mathbbm 1}}
\def\dag{^{\dagger}}
\def\p{{\rm p}}
\def\q{{\rm q}}
\def\caseI{{\rm I}}
\def\caseII{{\rm II}}
\def\a{{\rm a}}
\def\b{{\rm b}}
\def\no{{\rm n}}
\def\unc{u}
\def\comp{c}

\def\Ttot{T_\no}
\def\Ntot{\chi_\no}
\def\bmsigma{\boldsymbol\sigma}
\def\sigmamA{\boldsymbol\sigma_{\rm DH}}
\def\sigmamBq{\boldsymbol\sigma_{\rm a}}
\def\sigmamBp{\boldsymbol\sigma_{\rm b}}
\def\sigmamAB{\boldsymbol\sigma^{\rm(m)}_{A(B)}}
\def\sigmaz{\boldsymbol\sigma_z}

\def\V{a}
\def\W{b}
\def\Z{z}

\def\NLA{\mathcal{T}}
\def\id{ {\mathrm{id}} }

\section{Optical amplification for long-distance CVQKD}\label{chap:Amplifiers}

The previous Section showed that one of the crucial limitations of CVQKD is provided by the channel losses and noises, that both reduce the amount of extractable secure bits, i.e. the KGR, and introduce a maximum transmission distance, after which the KGR drops to $0$. Accordingly, a challenging strategy to enhance CVQKD in the long-distance regime is to adopt optical amplification techniques to perform signal restoration after transmission, or, equivalently, loss mitigation.
However, optical amplifiers are limited in the quantum regime, and deterministic noiseless amplification is untenable without the introduction of excess noise of quantum origin. In turn, optical amplification has to be described in terms of CP maps, and amplifiers are commonly divided into two main classes: conventional amplifiers and probabilistic noiseless linear amplifiers (NLAs). With the first tem, we refer to either phase-insensitive amplifiers (PIAs), performing noisy signal amplification by two-mode squeezing, with the ineludible introduction of thermal noise on both the field quadratures, and phase-sensitive amplifiers (PSAs), namely single-mode squeezing operations that noiselessly amplify a single field quadrature at the expense of the conjugate one. On the other hand, NLAs are probabilistic heralded schemes that perform noiseless amplification by coupling the signal mode to an ancillary system, being measured thereafter. Amplification is successful with a given probability, provided that a particular outcome is retrieved from the measurement of the ancillary system.
Given this scenario, in this Section we address the role of optical amplification in CVQKD protocols, assessing the potentiality and limitations of each type of amplifiers.

The Section is organized as follows. In Sec.s~\ref{sec9-LD} and~\ref{sec2:Amplification} we introduce the problems of long-distance CVQKD and optical amplification at the quantum limit, respectively. Thereafter, we study the application of both conventional amplifiers and NLAs within the context of CVQKD. In particular, in Sec.~\ref{sec9-PIAPSAQKD}, we discuss the role of PIAs and PSAs arranged in a multi-span configuration, addressing security under both the unconditional security framework and the trusted-device scenario, where only a single span of the link is untrusted. Then, in Sec.~\ref{sec:CVQKD-NLA} we analyze NLA-assisted CVQKD under unconditional security, by considering both the application of an ideal NLA and feasible physical NLAs, that is quantum scissors (QS) and single-photon catalysis (SPC).

\subsection{The problem of long-distance CVQKD}\label{sec9-LD}

As widely discussed in the previous Section, CVQKD makes a sender and a receiver share a common secure key in the presence of an untrusted channel up to a maximum transmission distance $d_{\rm max}$, at which the key generation rate (KGR) vanishes. The value of $d_{\rm max}$ depends on several factors, such as  the channel characteristics, e.g. the loss rate and the amount of thermal excess noise \cite{Lodewyck2005}, the non-unit reconciliation efficiency \cite{Bloch2006}, the presence of finite-size effects \cite{Leverrier2010} and the degree of trust of the channel, conditioning the amount of information at the eavesdropper's disposal \cite{PirandolaFree, Pirandola2021}. In turn, all these limitations crucially affect the KGR and prevent long-distance communication in practical realizations.
As an example, we note that the first implementation of CVQKD at telecom wavelength has been achieved in 2007 by Lodewyck {\em et al.} \cite{Lodewyck2007}, reaching the maximum distance of $\approx 25$~km. Since then, in 2013 and 2016 the maximum transmission distance has been increased to $80$~km \cite{Jouguet2013} and $100$~km \cite{Huang2016}, respectively. More recently, thanks to the exploitation of ultra-low-losses fibers, CVQKD has been established up to distances ranging from $100$ to $200$~km \cite{Zhang2020, Pi2023, Bian2023, Hajomer2023}.
In contrast, the performance of realistic DVQKD, implemented by the so-called decoy states \cite{Wang2003, Lo2005, Wang2005}, is much more robust, reaching distances from $15$ to $100$~km in the early 2000s \cite{Zhao2006, Peng2007, Rosenberg2007, Schmitt2007}, until the record transmission distance of $421$~km achieved by Zbinden {\em et al.} with ultra-low-loss fibers \cite{Record2018}.

A challenging task to face those CVQKD issues is to embed strategies in the original protocols allowing to increase as much as possible the maximum transmission distance. 
In principle, we may follow two different approaches, aiming at either reducing the amount of channel losses or mitigating the impact of excess noise.
In this Section, we focus on the first scenario and consider the role of the optical amplifiers, introduced in
the following section,
to perform signal restoration after transmission, with the intent of increasing the effective transmissivity of the channel.
In more detail, in the following we take the GG02 protocol, presented in Sec.~\ref{sec: GG02}, as a benchmark, and design an improved scheme assisted by two different classes of amplifiers:
\begin{itemize}
\item at first, we adopt conventional optical amplifiers, namely phase-insensitive (PIAs) and phase-sensitive amplifiers (PSAs), in a multi-span configuration, where the untrusted channel is composed of many regenerative stations interspersed with lossy links. In this case, we investigate security under both the unconditional and the trusted-device frameworks;
\item thereafter, we discuss a more intriguing solution provided by heralded noiseless linear amplifiers (NLAs), being probabilistic operations that amplify signals without additional noise, provided that a particular outcome is retrieved from the measurement of some ancillary modes. We consider noiseless amplification at the receiver's side and, for the sake of simplicity, we only address unconditional security, comparing both ideal and physical feasible examples of NLAs.
\end{itemize}

\subsection{Amplification of optical signals}\label{sec2:Amplification}

The problem of optical amplification is one of the relevant issues in quantum communications.
In fact, since the development of masers in the 1950's, it has been realized that active optical media, e.g. laser systems without cavity, can be effectively used to amplify the power of a laser beam without converting it into an electrical signal \cite{Bachor2019}. For this reason, optical amplifiers can be exploited as optical repeaters for signal restoration in long distance fiber-optic communication networks, as well as front-end devices for optical receivers, in which a weak optical signal is amplified before detection.
The range of possible applications makes it fundamental to assess the ultimate quantum-mechanical limitations of these devices, which has been firstly analyzed by Caves in~\cite{Caves1982}.

Ideally, at the quantum limit, we would like an optical amplifier to be described by a quantum operation that, when a coherent state $|\alpha\rangle$ is considered at the input, produces as output another coherent state with amplitude $g\alpha$, for a given real (amplitude) gain $g\ge 1$. Accordingly, $G=g^2$ represents the amplifier power gain \cite{NLARalphLund}.
This suggests the following input-output relation between the input radiation mode $a$, $[a, a^\dagger]=1$, and the output mode $a_\OUT$:
\begin{align}\label{eq:AmpIdinout}
a_\OUT=   g \, a \, ,
\end{align}
that, however, violates the canonical commutation rule:
\begin{align}
[a_\OUT,a_\OUT^\dagger]= g^2 \,[a, a^\dagger]= G \ne 1 \, .
\end{align}
We conclude that the present transformation does not represent a physical operation, and a unitary description of optical amplification is untenable.
In turn, we should recast the problem within the framework of quantum CP maps, in which case two different approaches are possible, involving either deterministic maps or heralded non-deterministic (non-unitary) transformations.

\subsubsection{Phase-insensitive and phase-sensitive amplifiers}\label{subsec2:PIA-PSA}

\begin{figure}
\includegraphics[width=0.8\columnwidth]{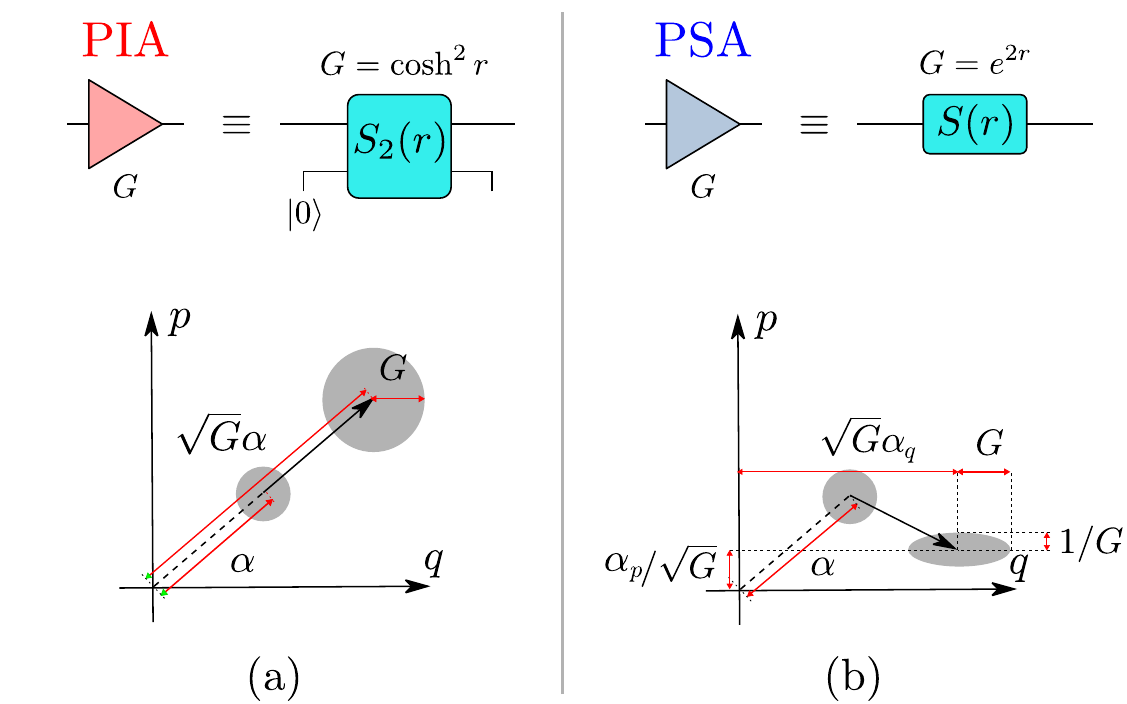}
\centering
\caption{Schemes of the phase-insensitive amplifier (PIA) (a) and phase-sensitive amplifier (PSA) (b). In both the cases the amplification power gain $G$ is related to the squeezing parameter $r$. The PIA applies the same amplification to both quadratures and introduces additional noise while PSA amplifies one of the quadratures and deamplifies the second rescaling the variances accordingly as seen on the example for a coherent state with amplitude $\alpha=\alpha_q+i\alpha_p$.}\label{fig01:sec9.1.1_Amplifiers}
\end{figure}

From the perspective of deterministic CP maps, the usual conclusion is that an additional noise operator should be added to the input-output relation~(\ref{eq:AmpIdinout}) in order to preserve the commutation rule.
That is, we introduce an ancillary optical mode $b$ prepared in the vacuum state $|0\rangle$, such that $[b,b^\dagger]=1$ and $[a,b]=0$, and consider the following transformation:
\begin{align}\label{eq:PIAHeis}
a_\OUT= g \, a + \sqrt{g^2-1}\,  b^\dagger \, ,
\end{align}
that, now, satisfies $[a_\OUT,a_\OUT^\dagger]=1$. Then, we conclude that linear amplification is an intrinsically noisy process, amplifying not only the signal power but also its noise level. Indeed, if mode $a$ is excited in the coherent state $|\alpha\rangle$, the expectation values and variances on the two output quadratures become 
\begin{subequations}
\begin{align}
\langle q_\OUT\rangle= 2 \sigma_0 g \, \Re(\alpha) \qquad \mbox{and} \qquad \langle p_\OUT\rangle= 2 \sigma_0 g \, \Im(\alpha) \, ,
\end{align}
\begin{align}
\Delta^2 q= \Delta^2 p = (2g^2-1) \sigma_0^2 = \sigma_0^2 +N_g\, ,
\end{align}
\end{subequations}
respectively, thus introducing an excess noise equal to $N_g=2(g^2-1) \sigma_0^2$ on both quadratures, that turns the pure state $|\alpha\rangle$ into a mixed state.

From a practical point of view, the transformation~(\ref{eq:PIAHeis}) may be realized by the phase-insensitive amplifier (PIA) depicted in Fig.~\ref{fig01:sec9.1.1_Amplifiers}(a).
It is implemented by coupling the signal mode $a$ together with an ancillary mode $b$ excited in the vacuum $|0\rangle$ and performing a two-mode squeezing operation, namely
\begin{align}
S_2(r) = \exp\Big[ r \, (a\dag b\dag - a b) \Big] \, ,
\end{align}
$r\ge0$ being the squeezing parameter \cite{Bachor2019, Serafini2017}. The original input mode is then transformed into $a_\OUT = \sqrt{G} \, a + \sqrt{G- 1} \,  b^\dagger$, with $G=\cosh^2 r$, thus retrieving the result of Eq.~(\ref{eq:PIAHeis}).
Thereafter, we trace over mode $b$, ending up with an amplified signal but at the expense of introducing a further ineludible noise equal to $G-1$ on both quadratures variances.

Given these considerations, PIAs are of particular interest for systems working at high powers, whereas their application at the quantum level is more limited due to the introduced excess noise \cite{Caves1982}.
The issue of noise may be circumvented by employing phase-sensitive amplifiers (PSAs), see Figure~\ref{fig01:sec9.1.1_Amplifiers}(b), implemented via a unitary
single-mode squeezing operation
\begin{align}
S(r) = \exp\Bigg\{ \frac{r}{2} \, \Big[(a\dag)^2- a^2 \Big] \Bigg\} \, ,
\end{align}
$r\ge0$ \cite{Bachor2019, Serafini2017}. The PSA amplifies the quadrature $q$ by a factor $\sqrt{G}=\exp(r)\ge 1$ at the expense of squeezing, i.e. de-amplifying, quadrature $p$ by $1/\sqrt{G} \le 1$. Consequently, the quadrature variances are also amplified and de-amplified by $G$ and $1/G$, respectively. Crucially, the input commutation relations between the quadratures are preserved without introducing any further noise.

\subsubsection{Noiseless linear amplifiers}\label{subsec2:NLA}

Even if conventional amplifiers, namely PIAs and PSAs, can efficiently perform restoration of classical signals, their application at the quantum limit is more limited. Indeed, both PIAs and PSAs introduce partial distortions of the original signal, by introducing either thermal noise on both quadratures or phase-sensitive effects, respectively.
Therefore, we may look for a more sophisticated solution, that produces noiseless linear amplification, albeit in probabilistic fashion. In fact, in principle, the input-output relation~(\ref{eq:AmpIdinout}) can be realized by a non-deterministic (non-unitary) transformation $\NLA$, such that $\NLA |\alpha\rangle= \gamma |g\alpha\rangle$, for some constant $\gamma \in \mathbb{C}$. 
In particular, Eq.~(\ref{eq:AmpIdinout}) is retrieved with the choice $\NLA=g^{\hat{n}}$, $\hat{n}=a^\dagger a$ being the photon-number operator of the input radiation mode, where $\NLA$ is a non-unitary unbounded operator.
In turn, we have:
\begin{align}
\NLA |\alpha\rangle = e^{-|\alpha|^2/2} \sum_{n=0}^{\infty} \frac{\alpha^n}{\sqrt{n!}} \, g^n |n \rangle = \gamma |g \alpha\rangle \, ,
\end{align}
for the constant $\gamma= \exp[(g^2-1)|\alpha|^2/2]$.
Given this consideration, we define an ideal noiseless linear amplifier (NLA) as the device implementing the CP map:
\begin{align}\label{eq:Tid}
{\cal E}_\id (\rho) = P_\id \, \frac{\NLA \rho \, \NLA^\dagger}{\Tr[\NLA \rho \, \NLA^\dagger]} + (1-P_\id) |0\rangle \langle 0| \, .
\end{align}
That is, we assume that an heralding signal identifies which state is produced in every run of the device, such that
noiseless linear amplification of the input signal is achieved with probability $P_\id\le 1$, whereas, in the opposite case, the output state is left into the vacuum. We also note that the success probability $P_\id \le 1$ is not equal to the trace of the post-selected state $\NLA \, \rho \, \NLA^\dagger$, since $\NLA$ is unbounded.
The map ${\cal E}_\id$ describes a physical operation, provided that the distinguishability of the amplified states is not increased on average \cite{NLARalphLund, Xiang2010}; that is, the map must not decrease the fidelity $\mathcal{F}$ between any two input quantum states \cite{NielsenChuang, Blandino2012}. In turn, for all states $\rho$ we should have:
\begin{align}
{\cal F} \left( \rho, |0\rangle \langle 0 | \right) \le {\cal F} \left( {\cal E}_\id(\rho), |0\rangle \langle 0 | \right)  \, , 
\end{align}
where we exploited the property ${\cal E}_\id(|0\rangle \langle 0 |)=|0\rangle \langle 0 |$.
In the presence of a coherent input $\rho=|\alpha\rangle\langle \alpha|$, this imposes a constraint on the success probability, namely \cite{Xiang2010}:
\begin{align}
P_\id \le \frac{1- e^{-|\alpha|^2}}{1- e^{-g^2|\alpha|^2}}  \, .
\end{align}
In summary, provided $P_\id$ to satisfy this bound, we have an heralded probabilistic device that increases the amplitude of a coherent state retaining the initial amount of noise.

Nevertheless, to date, the unitary dilation of ${\cal E}_\id$, that should provide its exact experimental implementation, is still unknown.  Therefore, the task is to design feasible physical NLAs, that approximate the map~(\ref{eq:Tid}), at least in the limit of weak amplitudes \cite{NLARalphLund, NLAFiu1, NLAXi, NLAMc, NLASPC, NLARalph, NLAFiu2, NLAJoshua1, NLAFiu3, NLAJoshua2}. Here, we present two relevant examples, the quantum scissors (QS) and the single-photon catalysis (SPC).

\begin{figure}
\includegraphics[width=0.7\columnwidth]{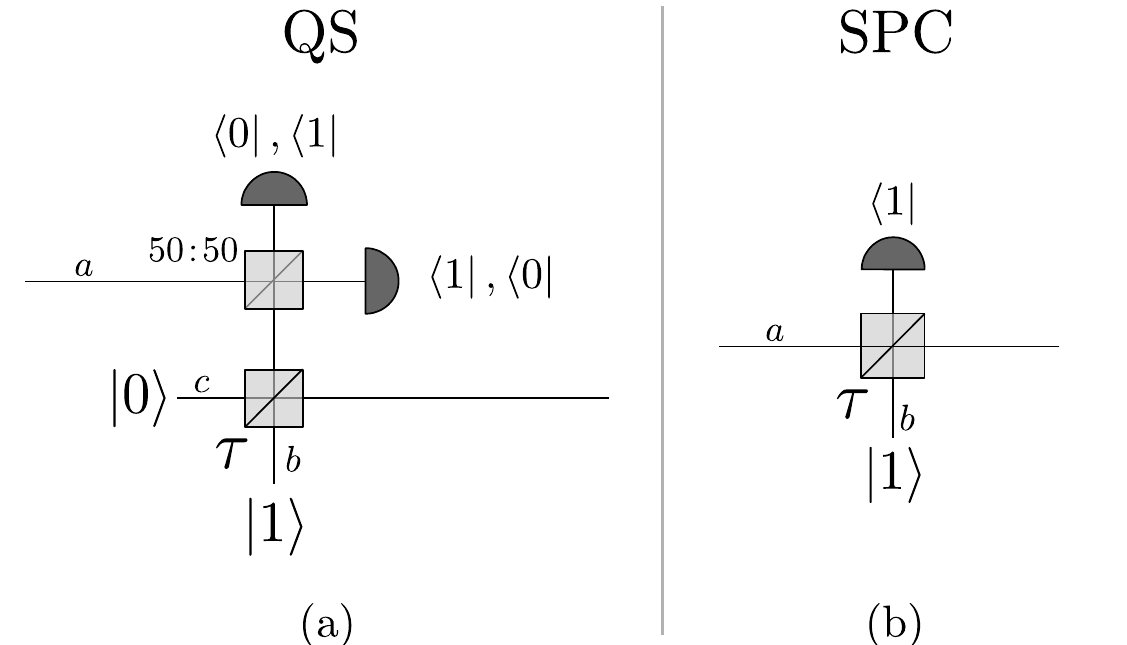}
\centering
\caption{Schematic representation of the quantum scissors (QS) (a) and the single-photon catalyis (SPC) (b).
In the QS scheme, a single photon is mixed with the vacuum at a beam splitter with transmissivity $\tau$. One of the output branches then impinges at a balanced beam splitter with the incoming signal, after which double conditional photo-detection is performed. Noiseless amplification is achieved when one of the two detectors reveals a single photon. Instead, in the SPC process a single photon interferes directly with the incoming signal at a beam splitter with transmissivity $\tau$ and then a single photon is retrieved at the end. }\label{fig02:sec9.1.1_NLA}
\end{figure}

\paragraph{Quantum scissors (QS).} The first implementation of a physical NLA has been proposed by Ralph and Lund in~\cite{NLARalphLund} via the QS scheme schematized in Fig.~\ref{fig02:sec9.1.1_NLA}(a). 

In the QS scheme, the incoming signal mode $a$, impinges at a balanced beam splitter with the first arm of the two-mode single-photon entangled state:
\begin{align}
|\varphi\rrangle_{bc}= \sqrt{\tau} \, |10\rrangle_{bc} +  \sqrt{1-\tau} \, |01\rrangle_{bc} \, ,
\end{align}
obtained by mixing a single photon with the vacuum at a beam splitter with transmissivity $\tau$, as depicted in Fig.~\ref{fig02:sec9.1.1_NLA}(a). 
If mode $a$ is excited in the coherent state $|\alpha\rangle$, the tripartite state at the output reads:
\begin{align}
|\Psi\rangle_{abc}&= U_{ab} \, D_a(\alpha) \left[\sqrt{\tau} b^\dagger + \sqrt{1-\tau} c^\dagger \right] |000\rangle  \nonumber \\[1ex]
&= \sqrt{\frac{\tau}{2}} D_a\left(\frac{\alpha}{\sqrt{2}}\right) D_b\left(-\frac{\alpha}{\sqrt{2}}\right) \left( |0 1\rangle_{ab} + |10\rangle_{ab} \right) |0\rangle_c \nonumber \\[1ex]
&\hspace{3.cm} +  \sqrt{1-\tau} \, \left| \frac{\alpha}{\sqrt{2}}\right\rangle_a \left| -\frac{\alpha}{\sqrt{2}}\right\rangle_b  |1\rangle_c \, ,
\end{align}
where $D_{a(b)}(\cdot)$ is the displacement operator on mode $a(b)$ and $U_{ab} $ is the beam splitter operator acting on modes $a$ and $b$.

Thereafter, we perform conditional photodetection on both the output modes $a$ and $b$: when one of the two detectors retrieves a single photon, the output mode $c$ is projected onto the (not normalized) quantum state:
\begin{align}
|\psi\rangle = e^{-|\alpha|^2/2} \sqrt{\frac{\tau}{2}} \left[ |0\rangle_c  \pm \sqrt{\frac{1-\tau}{\tau}} \alpha \, |1\rangle_c \right] \, ,
\end{align}
where the ``$+$" and ``$-$"  solutions are obtained by projection onto $\Pi_{10}=|10\rrangle_{ab}\llangle 10|$ and $\Pi_{01}=|01\rrangle_{ab}\llangle 01|$, respectively.
In particular, for weak amplitude coherent states, i.e. $|\alpha|^2 \ll 1$, we have $|\psi \rangle \approx |g\alpha\rangle$ for the gain:
\begin{align}
g= \sqrt{\frac{1-\tau}{\tau}} \, ,
\end{align}
being larger than $1$ for $\tau \le 1/2$. Hence, the action of the QS is to truncate the coherent state expansion up to the first order and simultaneously amplify the coherent amplitude.
Finally, the success probability is is given by twice the norm of $|\psi\rangle$, as NLA is successful when both the pairs of outcomes $(0,1)$ and $(1,0)$ are retrieved, leading to:
\begin{align}
P_{\rm QS}= 2 e^{-|\alpha|^2} \frac{\tau}{2} \Big[1+  g^2 |\alpha|^2 \Big] \approx \tau e^{(g^2-1) |\alpha|^2} \, .
\end{align}

Experimental demonstrations of the QS scheme have also been achieved by means of linear optics, parametric-down-conversion-based single-photon source, and single-photon detection \cite{Ozdemir2001, Ferreyrol2010, Ferreyrol2011}.

\paragraph{Single photon catalysis (SPC).}  It has been recently proved in~\cite{NLASPC} that also SPC provides a candidate to implement noiseless amplification. As depicted in Fig.~\ref{fig02:sec9.1.1_NLA}(b), in the SPC scheme a single ancillary mode $b$, excited in the single-photon state $|1\rangle$, impinges with the incoming signal at a beam splitter with transmissivity $\tau$. Thereafter, photodetection is implemented on the reflected branch, conditioning on $\Pi_1=|1\rangle \langle 1|$.

In turn, the reduced conditional dynamics for the signal mode $a$ is described by the Kraus operator:
\begin{align}
M_1= {}_b\langle 1| U_{ab} |1\rangle_b =\left[1-(1-\tau) a a^\dagger\right]  \,  \frac{e^{\ln(\sqrt{\tau}) a^\dagger a }}{\sqrt{\tau}}  \, ,
\end{align}
$U_{ab}=\exp[ \arccos(\sqrt{\tau}) \ (a^\dagger b - a b^\dagger ) ] $ being the beam splitter operator acting on modes $a$ and $b$.
When a weak coherent state $|\alpha\rangle \approx |0\rangle + \alpha |1\rangle$, $|\alpha|^2 \ll 1$, is considered at the input, we obtain:
\begin{align}
|\psi\rangle = M_1 |\alpha\rangle \approx \sqrt{\tau} \left[ |0\rangle - \frac{1-2\tau}{\sqrt{\tau}} \alpha \, |1\rangle \right] \approx \sqrt{\tau} \, | e^{i\pi} g\alpha\rangle \, ,
\end{align}
that approximates an amplified coherent state with a $\pi$ phase shift and gain:
\begin{align}
g= \frac{1-2\tau}{\sqrt{\tau}} \, ,
\end{align}
such that $g\ge 1$ for $\tau \le 1/4$.

Moreover, the SPC setup has been experimentally realized in \cite{Lvovsky2002, Bartley2012} with the intent of generating nonclassical states of light, whilst its application for noiseless amplification has not been practically tested to date.
However, we note that, differently from QS, in the SPC process a single photon interferes directly with the incoming signal; therefore, SPC provides a simpler scheme and may represent a feasible alternative to QS for experimental NLA realizations.
%
%

\subsection{CVQKD with phase-insensitive and phase-sensitive amplifiers}\label{sec9-PIAPSAQKD}

Starting from the discussions of the previous section, we now address for the first time the application of optical amplifiers as a possible resource to mitigate the losses impact in CVQKD schemes. The results obtained in the following sections are original.
Here, we start from the case of PIAs and PSAs, arranged in multi-span configuration.

\subsubsection{Multi-span links}\label{sec:MULTISPANth}

\begin{figure}
\includegraphics[width=0.8\columnwidth]{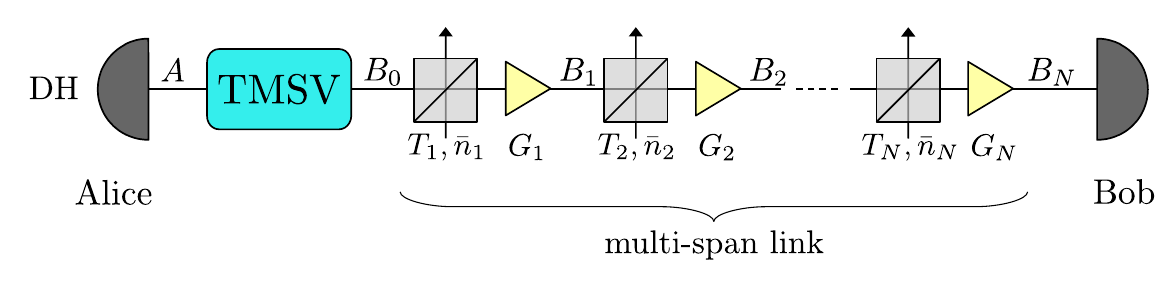}
\centering
\caption{Scheme of CVQKD in the presence of a multi-span link. A two mode squeezed vacuum state (TMSV) is distributed between Alice and Bob. Alice performs a double homodyne (DH) measurement on her mode, whereas the mode sent to Bob travels through a thermal-loss channel modeled by a series of $M$ beam splitters with transmissivities $T_j$ and added mean thermal number of photons $\bar{n}_j$. To counteract losses, the signal mode $B_j$ after each span is amplified by either PIA or PSA with power gain $G_j$. Finally, Bob performs a measurement which we assume to be either case $\caseI$: a random homodyne detection of quadratures $q/p$, or homodyne detection of either $q$, $\caseII\a$, or $p$, $\caseII\b$.}\label{fig01:sec9.2.1_MultiSpan}
\end{figure}

To compensate transmission losses and restore the signal, a first natural option is to employ conventional optical amplifiers \cite{Bachor2019, Caves1982, Notarnicola2022} and consider a multi-span link, that is, a periodic array of amplifiers connected by many independent thermal-loss channels. 
To date, this configuration have been investigated with the intent of increasing channel capacity for information transmission \cite{Yariv1990, Antonelli2014, Jarzyna2019, Lukanowski2023}, showing that both PIAs \cite{Jarzyna2019} and PSAs \cite{Lukanowski2023} induce an exponential enhancement of the ultimate capacity being more appreciable for short-distance communication.
In contrast, in the context of CVQKD the role of optical amplifiers has been investigated to compensate for detection imperfections \cite{Fossier2009}, raising the question on their possible application to the channel losses mitigation task.
In fact, multi-span links provide a simple and versatile solution for practical implementations, unlike many other solutions.
As an example, an alternative choice would be instead to insert classical-like repeaters after each span and establish keys separately between neighboring nodes. At the quantum limit, a classical repeater would be described as an intercept-resend system, performing double homodyne detection on the received signal and preparing a coherent pulse with amplitude equal to the obtained outcome.
Due to the probabilistic nature of quantum measurements, the overall effect of such intercept-resend strategy would be the introduction of an excess thermal noise, similar to the case of PIA. 
In principle, establishing keys between subsequent nodes would significantly reduce the impact of channel losses at the cost of introducing additional excesss noise due to probabilistic nature of the quadrature measurement. However, from a practical point of view, this configuration would be unfeasible in real networks, as classical-repeater nodes are expensive and, thus, cannot be placed at every few kilometers. A further solution may be the adoption of quantum repeaters \cite{Furrer2018, Pirandola2019Commun, Dias2020}, which represent an intriguing strategy from a theoretical point of view, but rather an unpractical one with current state-of-the-art technology. In fact, quantum repeaters for CV systems employ probabilistic noiseless linear amplifiers as a fundamental building block, and, thus, require the presence of a quantum memory, not yet available with the current optical communication technologies and far from a direct large-scale implementation \cite{Bersin2024, Wallucks2020}.
Thus, employing either classical or quantum repeaters can lead to a much higher key rate at the cost of making infrastructure complicated and expensive. On the contrary, optical amplifiers like PIA and PSA are much cheaper and manageable than repeaters and provide a feasible tool to enhance communication between the nodes.

Given these considerations, the scheme of CVQKD over multi-span links is depicted in Fig.~\ref{fig01:sec9.2.1_MultiSpan}. In particular, we start from the GG02 scheme in its entanglement-based version: that is, Alice has a two-mode squeezed vacuum state (TMSV) with variance $V>1$, namely
\begin{align}
|{\rm TMSV} \rangle\!\rangle =
\sqrt{1-\lambda^2}\sum_{n=0}^{\infty} \lambda^n |n\rangle |n \rangle \, ,
\end{align}
where $\lambda= \sqrt{ (V-1)/(V+1) }$ and $|n\rangle$ being the Fock state with $n$ photons \cite{Olivares2021}.
She injects the second branch into the quantum channel while performing double homodyne (DH) detection on the remaining mode, such that the conditional state sent to Bob is a coherent state. Ultimately, Bob performs a homodyne measurement on the received pulses, which in the former version of GG02 consists of a random homodyne detection of either $q$ or $p$ quadratures \cite{Grosshans2002, Grosshans2005}.

Unlike in the standard GG02 protocol, now, the quantum channel consists of a multi-span link with $M$ spans alternated by optical amplifiers. 
Each span $j=1,\ldots,M$ is modeled as an independent thermal-loss channel with transmissivity $T_j\le 1$ and excess noise $\epsilon_j\ge 0$. More precisely, the optical mode entering the $j$-th link is mixed at a beam splitter with transmissivity $T_j$ with a thermal state having $\bar{n}_j= T_j \epsilon_j/[2(1-T_j)]$ mean number of photons \cite{Olivares2021}.
Thereafter, the radiation undergoes optical amplification, either phase-insensitive or phase-sensitive, before being injected into the $(j+1)$-th span. 
For simplicity, here we assume both identical and equally spaced amplifiers, such that all spans have the same transmissivity $T_j=T$, added thermal noise $\bar{n}_j=\bar{n}_T$ and amplification power gain $G_j=G$. Note, however, that this choice may not be the optimal arrangement \cite{Jarzyna2019}.
Then, if the total transmission distance is $d$, two neighboring amplifiers are spaced by $d/M$ and we have
\begin{align}
T= 10^{-\kappa d/(10 M)} \, ,
\end{align}
$\kappa= 0.2$ dB/km being the typical loss rate of standard optical fibers \cite{Lodewyck2005,Lodewyck2007, Banaszek2020}. Moreover, we assume the added thermal photons in each span to be equal to
 \begin{align}\label{eq:thnoise}
\bar{n}_T= \frac{T^M \epsilon}{2(1-T^M)} \, .
\end{align}
With these choices, in the absence of optical amplification, that is $G=1$, we retrieve the standard GG02 scenario, that is a single-span thermal-loss channel with total transmissivity $\Ttot= T^M$ and added noise $\Ntot=(1-\Ttot)/\Ttot+\epsilon$, $\epsilon\ge 0$ being the total excess noise \cite{Laudenbach2018, Grosshans2005}.

Starting from the scheme in Fig.~\ref{fig01:sec9.2.1_MultiSpan}, we address three different cases, differing from one another by both the employed amplifier and the measurement implemented by Bob:
\begin{itemize}
\item Case $\caseI$: PIA link and random homodyne detection of quadratures $q/p$,
\item Case $\caseII\a$: PSA link and homodyne detection of quadrature $q$, namely the anti-squeezed quadrature,
\item Case $\caseII\b$: PSA link and homodyne detection of quadrature $p$, namely the squeezed quadrature.
\end{itemize}
We note that the presence of a PSA link makes the channel phase-sensitive, thus differentiating the behavior of quadratures $q$ and $p$. Therefore, in the presence of PSAs Bob may perform homodyne detection of a single quadrature for those experimental runs dedicated to key extraction, while homodyning both $q$ and $p$ for the channel evaluation stage, in order to fully characterize the quantum channel \cite{Laudenbach2018}.
We also remark the important difference between PIA and PSA: the former is a noisy operation requiring the introduction of an additional light mode lost to the environment which, in principle, can be intercepted by a malicious party, whereas the latter amplification scenario assumes unitary evolution which does not leak any information, thus being always trusted.

In the following, we compute the KGR for all three cases under both unconditional security and the trusted device scenarios, in which we assume trusted amplifiers and only a single untrusted span $0\le k\le M$.
To perform the analysis, we adopt the notation introduced in Fig.~\ref{fig01:sec9.2.1_MultiSpan}. At first Alice has two optical modes $A$ and $B_0$ excited in a TMSV state. Then, the mode $B_0$ is injected into the sequence of $M$ spans. We denote by $B_j$ the optical mode coming out from the $j$-th span and subsequently amplified by the $j$-th amplifier. Finally, we refer to the last output mode as $B=B_M$.
We start by computing the mutual information shared between Alice and Bob, addressing the cases $\caseI$ and $\caseII\p$, $\p=\a,\b$, separately. The whole analysis is carried out following the Gaussian formalism.

In fact, a thermal-loss channel with transmissivity $T \le1$ and thermal noise $\bar{n}_T$ is described via a Gaussian completely positive (CP) map associated with the matrices \cite{Serafini2017}:
\begin{align}\label{eq:TLxy}
X_{\rm TL} = \sqrt{T} \, \Id_2 \quad \mbox{and} \quad Y_{\rm TL} = (1-T)(1+2\bar{n}_T) \Id_2 \,,
\end{align}  
$\Id_2$ being the $2\times 2$ identity matrix.

As regards optical amplification, PIA are descripted by the Gaussian CP map \cite{Serafini2017}:
\begin{align}\label{eq:PIAxy}
X_{\rm PIA} = \sqrt{G} \, \Id_2 \quad \mbox{and} \quad Y_{\rm PIA} = (G-1) \Id_2 \,,
\end{align}  
$G\ge 1$ being the amplification power gain, whilst PSA are unitary maps, thus completely described by the symplectic matrix \cite{Ferraro2005}:
\begin{align}\label{eq:PSAsymp}
S_{\rm PSA}= 
\begin{pmatrix}
G^{1/2} & 0 \\ 
0 & G^{-1/2} 
\end{pmatrix} \, .
\end{align}

Thus, for case $\caseI$, namely in the presence of a PIA link, each span is given by the composition of the two Gaussian CP maps described by Eqs.~(\ref{eq:TLxy}) and~(\ref{eq:PIAxy}), resulting in a overall Gaussian CP map defined by the matrices:
\begin{align}
X^{(\caseI)} &= X_{\rm PIA} X_{\rm TL}= \sqrt{G T} \, \Id_2 \, ,\nonumber \\
Y^{(\caseI)} &= X_{\rm PIA}Y_{\rm TL} X_{\rm PIA}^{\mathsf{T}} + Y_{\rm PIA} \nonumber \\
&= \left[G(1-T)(1+2\bar{n}_T)+ (G-1) \right] \, \Id_2\, .
\end{align} 
Otherwise, for case $\caseII$, namely PSA link, each span is the composition of the CP map~(\ref{eq:TLxy}) and the symplectic evolution~(\ref{eq:PSAsymp}), resulting in the overall Gaussian CP map associated with $X^{(\caseII)} = S_{\rm PSA} X_{\rm TL}$ and $Y^{(\caseII)} = S_{\rm PSA}Y_{\rm TL} S_{\rm PSA}^{\mathsf{T}}$, namely: 
\begin{align}\label{eq:PSAxy}
X^{(\caseII)} = 
\begin{pmatrix}
\sqrt{GT} & 0 \\ 
0 & \sqrt{G^{-1} T} 
\end{pmatrix}\, ,
\end{align} 
and
\begin{align}
Y^{(\caseII)} = 
\begin{pmatrix}
G (1-T)(1+2\bar{n}_T) & 0 \\ 
0 & G^{-1} (1-T)(1+2\bar{n}_T)
\end{pmatrix}\, .
\end{align} 

\paragraph{Case $\caseI$ : PIA link}\label{sec:MI-PIA} 
The initial state before injection into the channel is a TMSV in modes $A$ and $B_0$, completely characterized by its covariance matrix (CM)
\begin{align}
\bmsigma_{AB_0} =
\begin{pmatrix} V \, \Id_2 & Z \, \sigmaz \\ Z \, \sigmaz & V \, \Id_2 \end{pmatrix} 
 \, ,
\end{align}
where $Z=\sqrt{V^2-1}$, and $\sigmaz$ is the Pauli $z$-matrix. 

The mode $B_0$ is injected into the noisy channel, which may be modeled via a sequence of Gaussian CP maps. 
More specifically, each node is described by a Gaussian CP map~(\ref{eq:PIAxy}), such that the bipartite state on modes $AB_j$ after the $j$-th span is a Gaussian state with associated CM $\bmsigma_{AB_j}^{(\caseI)}=(\Id_2 \oplus X^{(\caseI)})  \bmsigma_{AB_{j-1}}^{(\caseI)} (\Id_2\oplus X^{(\caseI)})^{\mathsf{T}}+({\bf 0} \oplus Y^{(\caseI)})$, ${\bf 0}$ being the null $2\times 2$ matrix.

Accordingly, after $M$ nodes applying PIA the state shared between Alice and Bob is still Gaussian with CM 
\begin{align}\label{eq:CM_Nnodes_PIA}
\bmsigma_{AB}^{(\caseI)}= 
\begin{pmatrix}  \bmsigma_A^{(\caseI)} & \bmsigma_{Z}^{(\caseI)} \\[1ex] \bmsigma_Z^{(\caseI) \mathsf T} & \bmsigma_{B}^{(\caseI)} \end{pmatrix} 
=
\begin{pmatrix} 
\V^{(M)} \, \Id_2 &  \Z^{(M)} \sigmaz \\
\Z^{(M)} \sigmaz & \W^{(M)} \, \Id_2  \\
 \end{pmatrix}   \, ,
\end{align}
where
\begin{subequations}
\begin{align}
\V^{(M)}&= V \, , \\[1ex]
\W^{(M)} &= T^{(M)} \bigg[ V+ \chi^{(M)} \bigg]  \, , \\[1ex]
\Z^{(M)}&= \sqrt{T^{(M)}} \, Z  \, , 
\end{align}
\end{subequations}
and
\begin{subequations}
\begin{align}
T^{(M)}&= (G T)^M \, , \\[1ex]
\chi^{(M)} &= \frac{1}{(G T)^{M-1}} \frac{1- (GT)^M}{1-GT} \bigg[\chi +\chi_G\bigg]  \, ,
\end{align}
\end{subequations}
$\chi= (1-T) (1+2\bar{n}_T)/T$ being the added noise introduced after the passage though a single span due to the channel thermal noise, while $\chi_G=(G-1)/(G T)$ is the added noise due to the PIA.
Consequently, compared to the scenario in the absence of amplifiers, the PIA link is equivalent to a thermal-loss channel with increased transmissivity $T^{(M)} \ge \Ttot$, but also increased added noise $\chi^{(M)}\ge \Ntot$.

After transmission, Alice performs DH detection on her mode, associated with the CM $\sigmamA= \Id_2$, while Bob implements a homodyne detection of either quadrature $q$ or $p$, referred to as sub-cases $\a$ and $\b$, and described by the CMs
\begin{align}
\sigmamBq = \lim_{z\rightarrow 0} 
\begin{pmatrix}  z & 0 \\ 0 & z^{-1}\end{pmatrix}
\quad \mbox{and} \quad
\sigmamBp = \lim_{z\rightarrow \infty} 
\begin{pmatrix}  z & 0 \\ 0 & z^{-1}\end{pmatrix} \, ,
\end{align}
respectively. Due to the symmetry of~(\ref{eq:CM_Nnodes_PIA}), the resulting statistics for both quadrtures are identical, therefore, we can safely assume that Bob always measures the quadrature $q$. In turn, the mutual information between Alice and Bob may be obtained directly from~(\ref{eq:CM_Nnodes_PIA}) as:
\begin{align}\label{eq: IAB-PIA}
I_{AB}^{(\caseI)} (V,G) = \frac12 \log_2 \Bigg\{\frac{\det\big[\bmsigma_A^{(\caseI)}+\sigmamA\big] \det\big[\bmsigma_B^{(\caseI)}+\sigmamBq \big]}{\det\big[\bmsigma_{AB}^{(\caseI)}+(\sigmamA  \oplus \sigmamBq )\big]} \Bigg\} \, ,
\end{align}
where we highlighted the dependence on the free parameters $V$ and $G$.

\paragraph{Case $\caseII$ : PSA link}\label{sec:MI-PSA} 

For cases $\caseII\p$, $\p=\a,\b$,
we follow analogous procedure as in the previous subsection. Now, each node is modeled by a Gaussian CP map~(\ref{eq:PSAxy}). The shared state on modes $AB_j$ has CM $\bmsigma_{AB_j}^{(\caseII)}=(\Id_2 \oplus X^{(\caseII)})  \bmsigma_{AB_{j-1}}^{(\caseII)} (\Id_2 \oplus X^{(\caseII)})^{\mathsf{T}}+({\bf 0} \oplus Y^{(\caseII)})$, thus ultimately
we obtain the CM of the state shared between Alice and Bob as:
\begin{align}\label{eq:CM_Nnodes_PSA}
\bmsigma_{AB}^{(\caseII)}= 
\begin{pmatrix} \bmsigma_A^{(\caseII)} & \bmsigma_Z^{(\caseII)} \\[1ex] \bmsigma_Z^{(\caseII) \mathsf T} & \bmsigma_{B}^{(\caseII)} \end{pmatrix} 
=
\begin{pmatrix} 
\V^{(M)} &   0 & \Z_1^{(M)} & 0 \\
 0 & \V^{(M)} & 0 & -\Z_2^{(M)}  \\
\Z_1^{(M)} & 0 & \W_1^{(M)} & 0  \\
 0& -\Z_2^{(M)} & 0& \W_2^{(M)} \\
 \end{pmatrix} \, ,
\end{align}
where
\begin{subequations}
\begin{align}\label{eq:WZpsa}
\W_{1(2)}^{(M)} = T_{1(2)}^{(M)} \bigg[ V+ \chi_{1(2)}^{(M)} \bigg]  \, , \\[1ex]
\Z_{1(2)}^{(M)}= \sqrt{T_{1(2)}^{(M)}} \, Z  \, ,
\end{align}
\end{subequations}
and
\begin{subequations}
\begin{align}
T_1^{(M)}&= (G T)^M \, , \qquad T_2^{(M)}=(G^{-1} T)^M \,,  \\[1.5ex]
\chi_1^{(M)} &= \frac{1}{(G T)^{M-1}} \frac{1- (GT)^M}{1-GT} \chi \, , \\[1.5ex]
\chi_2^{(M)} &= \frac{1}{(G^{-1} T)^{M-1}} \frac{1- (G^{-1}T)^M}{1-G^{-1}T} \chi \, ,
\end{align}
\end{subequations}
with $\chi= (1-T) (1+2\bar{n}_T)/T$.

Unlike case $\caseI$, the PSA link is a phase-sensitive channel.
Indeed, in the presence of PSA, quadrature $q$ exhibits an increased transmissivity $T_1^{(M)} \ge \Ttot$ and reduced added noise $\chi_1^{(M)} \le \Ntot$, whereas quadrature $p$ shows a reduced transmissivity $T_2^{(M)} \le \Ttot$ with increased added noise $\chi_2^{(M)} \ge \Ntot$.
As we discuss in the following, under appropriate conditions this allows Bob to hide behind the increased noise to reduce the amount of information intercepted by an eventual eavesdropper. The mutual information for the two sub-cases $\p=\a,\b$ then reads:
\begin{align}\label{eq: IAB-PSA}
I_{AB}^{(\caseII\p)} (V,G) = \frac12 \log_2 \Bigg\{\frac{\det\big[\bmsigma_A^{(\caseII)}+\sigmamA\big] \det\big[\bmsigma_B^{(\caseII)}+\bmsigma_\p \big]}{\det\big[\bmsigma_{AB}^{(\caseII)}+(\sigmamA  \oplus \bmsigma_\p )\big]} \Bigg\} \, .
\end{align}

In the next subsections, we will perform a security analysis of the above protocols by considering both the cases of unconditional security, where the entire channel is untrusted, and conditional security, assuming that only a single span is untrusted and may be intercepted by Eve. In both scenarios, we take as a benchmark the security of the associated protocol in the absence of optical amplifiers, referred to as the ``no-amplifier protocol", in which we assume Bob to perform a random homodyne measurement of either quadrature $q$ or $p$ as in GG02.
The results of the standard no-amplifier protocol can be retrieved from both cases $\caseI$ and $\caseII$ by fixing $G=1$.

\subsubsection{Unconditional security}\label{sec:MultiSpanUncSec}

At first, we analyze the performance of the discussed protocol under the unconditional security approach, where the whole transmission line is supposed to be attacked by Eve. In this framework, all elements of the multi-span link are assumed to be untrusted and the most powerful attack is the so-called purification attack \cite{Laudenbach2018, Grosshans2005}. That is, Eve intercepts all the lost photons, and collects modes associated with the channel noise and purifies the final state shared between Alice and Bob, such that the tripartite system $ABE$ is pure \cite{Laudenbach2018}.
Under these conditions employing PIAs is useless because Eve would have access also to their purification, and extract more information with respect to the no-amplifier protocol. In contrast, case $\caseII$ is still worth of interest due to the unitarity of phase-sensitive amplification.

Considering reverse reconciliation \cite{Laudenbach2018, Grosshans2005}, for the cases $\caseII\p$, $\p=\a,\b$, the KGR is given by
\begin{align}
K^{(\caseII\p)}_\unc (V,G)=\beta I_{AB}^{(\caseII\p)}(V,G)- \chi_{BE}^{(\caseII\p)}(V,G) \, ,
\end{align}
where $\beta\le 1$ is the reconciliation efficiency and $\chi_{BE}^{(\caseII\p)}(V,G)= S_E - S_{E|B}^{(\p)}$ is the Holevo information between Bob and Eve, $S_E$ and $S_{E|B}^{(\p)}$ being the Von Neumann entropies of Eve's overall state and Eve's conditional state after Bob's measurement, respectively.
Due to the purification attack and the fact that Bob's measurement is represented by a $1$-rank operator, we have $S_E=S_{AB}$ and $S_{E|B}^{(\p)}=S_{A|B}^{(\p)}$, where $S_{AB}$ and $S_{A|B}^{(\p)}$ are the Von Neumann entropies of Alice and Bob's bipartite state and Alice's conditional state, respectively.
These two latter quantities can be retrieved from the CM~(\ref{eq:CM_Nnodes_PSA}), leading to:
\begin{align}\label{eq: chiBE-UNC}
\chi_{BE}^{(\caseII\p)}(V,G) = h\left(\frac{{\rm d}_1-1}{2}\right)+ h\left(\frac{{\rm d}_2-1}{2}\right) -h\left(\frac{{\rm d}_3^{\, (\p)}-1}{2}\right) \, ,
\end{align}
where 
\begin{align}\label{eq:hfunc}
h(x)= (x+1) \log_2( x+1) - x \log_2 x\, ,
\end{align}
${\rm d}_{1}$ and ${\rm d}_2$ are the symplectic eigenvalues of~(\ref{eq:CM_Nnodes_PSA}) and ${\rm d}_3^{\, (\p)}= \sqrt{\det\big[ \bmsigma_{A|B}^{(\caseII\p)} \big]}$, with
\begin{align}
\bmsigma^{(\caseII\p)}_{A|B}= \bmsigma_A^{(\caseII)} - \bmsigma_Z^{(\caseII)} \Big[\bmsigma_B ^{(\caseII)}+ \bmsigma_\p \Big]^{-1} \bmsigma_Z^{(\caseII) \mathsf T} \, .
\end{align}
In particular, we have:
\begin{align}\label{eq:d3}
{\rm d}_3^{\, (\a(\b))}= V \sqrt{1-\frac{Z^2}{V\bigg[V+ \chi_{1(2)}^{(M)}\bigg]}} \, .
\end{align}

\begin{figure}
\includegraphics[width=0.49\columnwidth]{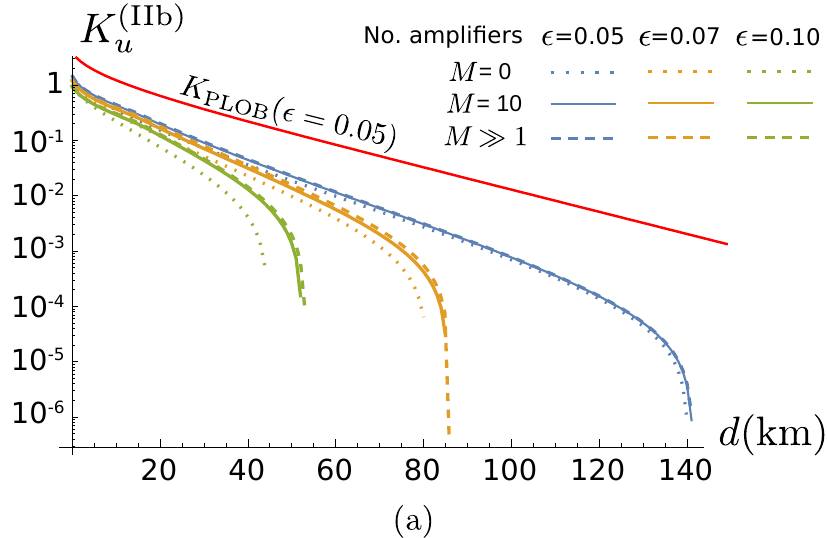} 
\includegraphics[width=0.49\columnwidth]{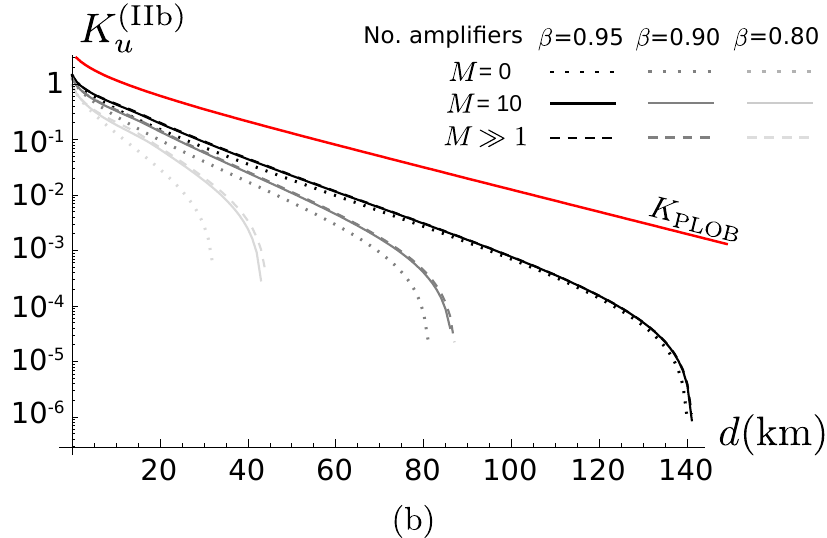}
\centering
\caption{(a) Log plot of $K_u^{(\caseII\b)}$ as a function of the transmission link length $d$ for different level of external noise and number of amplifiers $M$, with fixed reconciliation efficiency $\beta=0.95$. (b) Log plot of $K_u^{(\caseII\b)}$ as a function of $d$ for different values of reconciliation efficiency and number of amplifiers $M$, with fixed channel excess noise $\epsilon=0.05$. The enhancement introduced by PSAs is accentuated for lower $\beta$. The case $M=0$ refers to the no-amplifier protocol. The pink shaded area represents KGR greater than the PLOB  bound, computed for $\epsilon=0.05$.}\label{fig01:sec9.2.2_KGR-unc}
\end{figure}

Finally, we perform optimization over the free parameters - modulation variance $V$ and power gain $G$, obtaining
\begin{align}\label{eq:KGRUnc}
K^{(\caseII\p)}_\unc = \max_{V,G} \, K^{(\caseII\p)}_\unc (V,G) \, , \qquad (\p=\a,\b) \, ,
\end{align}
subject to the set of constraints $\W_{1}^{(j)}\le V$, see Eq.~(\ref{eq:WZpsa}), i.e.
\begin{align}\label{eq:ConditionG}
T_{1}^{(j)} \bigg[ V+ \chi_{1}^{(j)} \bigg] \le V \, , \quad (j=1,\ldots,M) \, ,
\end{align}
assuring that throughout the channel the squeezing operation does not amplify the variances of the quadratures, proportional to the total optical power, over their input values \cite{Jarzyna2019, Lukanowski2023}.
This conditions arises from a physical requirement that realistic optical fibers cannot support propagation of pulses with arbitrarily high energy without damaging the optical infrastructure or the emergence of unwanted nonlinear effects. Therefore, it is reasonable to impose a condition on the gain of the PSA, such that energy of the amplified signal after each span is not larger than the input one.

Furthermore, since we assume all amplifiers to be characterized by the same gain , it suffices to verify condition~(\ref{eq:ConditionG}) for $j=1$, satisfied if $\W_{1}^{(1)}\le V$, namely
\begin{align}\label{eq:ConditionG-PSA}
G \le G_{\rm max}^{(\caseII)} \equiv \frac{V}{1+T(V+\epsilon-1)} \, .
\end{align}

\begin{figure}
\includegraphics[width=0.55\columnwidth]{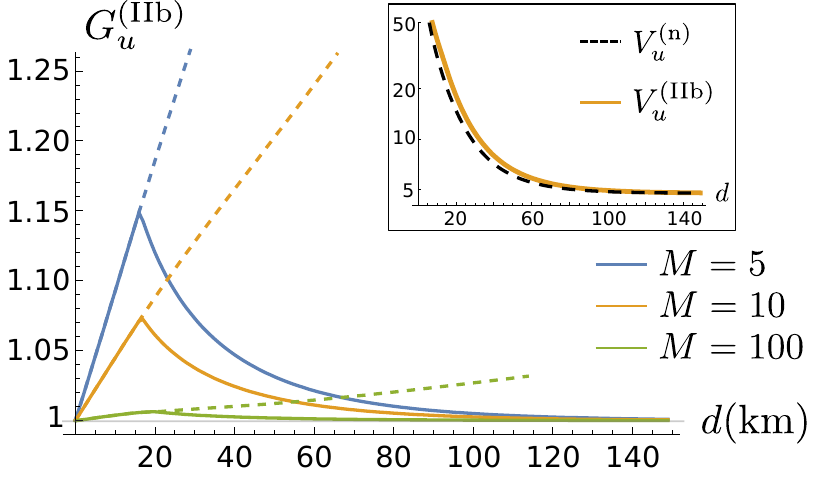}
\centering
\caption{Plot of the optimized amplifier power gain $G^{(\caseII\b)}_\unc$ as a function of link length $d$ for different number of amplifiers $M$. The dashed lines represent the maximum attainable gain $G^{(\caseII)}_{\rm max}$ computed with the optimized modulation $V^{(\caseII\b)}_\unc$, presented in the inset. We set $\epsilon=0.05$ and $\beta=0.95$.}\label{fig02:sec9.2.2_OptParUnc}
\end{figure}

In this security paradigm, the no-amplifier protocol is described by a single-span quantum channel with transmissivity $T_\no$ and added noise $\chi_\no$, which coincides with the GG02 protocol. The benchmark key rate $K^{(\no)}_\unc$ is obtained by optimizing the unamplified KGR over the modulation variance $V$:
\begin{align}
K^{(\no)}_\unc = \max_{V} \, K^{(\caseII\p)}_\unc(V,G=1) \, .
\end{align}
The obtained numerical results suggest that the optimized gain for case $\caseII\a$ is equal to $G^{(\caseII\a)}_\unc \equiv 1$ for all $d$, therefore $K^{(\caseII\a)}_\unc \equiv K^{(\no)}_\unc$ and measuring the anti-squeezed quadrature $q$ does not increase the key rate of the discussed protocol. 
On the contrary, the case $\caseII\b$ improves the security for large values of excess noise $\epsilon$, as depicted in Fig.~\ref{fig01:sec9.2.2_KGR-unc}(a). In this case, PSA links offer a higher KGR and, remarkably, increase the achievable maximum transmission distance, although the enhancement is relevant only for large excess noise, namely $\epsilon \gtrsim 0.05$ \cite{Roumestan2021, Roumestan2022}.
Furthermore, as shown in Fig.~\ref{fig01:sec9.2.2_KGR-unc}(b), at fixed excess noise $\epsilon$, the KGR increase induced by PSAs becomes larger for lower values of the reconciliation efficiency $\beta$. For the sake of completeness, in Fig.~\ref{fig01:sec9.2.2_KGR-unc} we also show the Pirandola–Laurenza–Ottaviani–Banchi (PLOB) bound \cite{PLOB}:
\begin{align}\label{eq:PLOB}
K_{\rm PLOB}= - \log_2 \big[ (1-\Ttot) \Ttot^{\bar{n}_T} \big] - h(\bar{n}_T)\, ,
\end{align}
which represents the maximum KGR achievable with the considered repeaterless thermal-loss channel.

The optimized gain $G^{(\caseII\b)}_\unc$ obtained from the maximization procedure is plotted in Fig.~\ref{fig02:sec9.2.2_OptParUnc}. For small link lengths $d$, constraint~(\ref{eq:ConditionG-PSA}) leads to $G^{(\caseII\b)}_\unc= G^{(\caseII)} _{\rm max}$ and the gain increases with link length, whereas for larger $d$ it becomes a decreasing function approaching $1$ asymptotically. Moreover, $G^{(\caseII\b)}_\unc$ decreases with the number of spans $M$, as expected. Finally, the optimized modulation $V^{(\caseII\b)}_\unc$ is a decreasing function of the link length such that $V^{(\caseII\b)}_\unc \ge V^{(\no)}_\unc$, where $V^{(\no)}_\unc$ is the optimized modulation of the no-amplifier protocol.

\begin{figure}
\includegraphics[width=0.49\columnwidth]{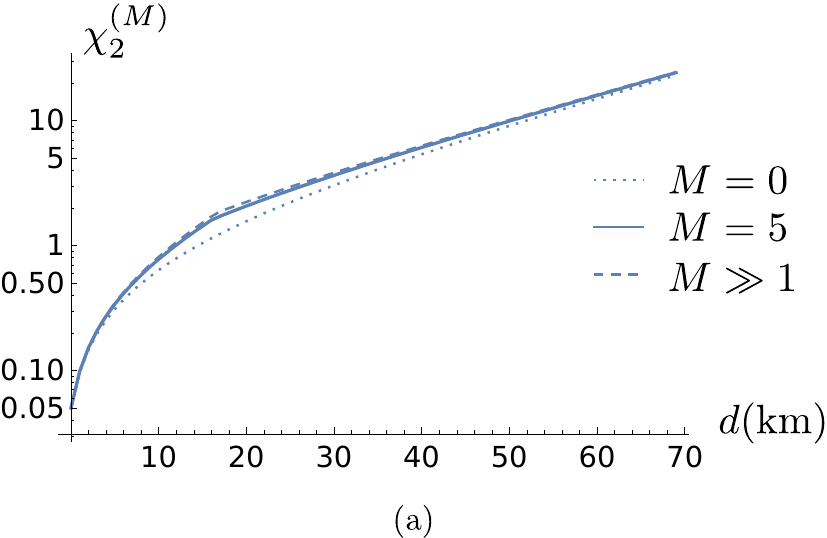} 
\includegraphics[width=0.49\columnwidth]{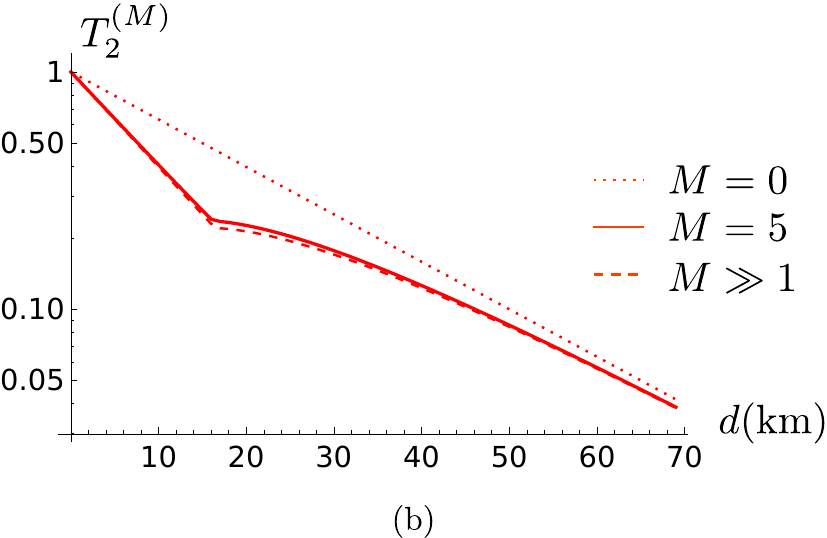}
\centering
\caption{Plot of the added noise $\chi_2^{(M)}$ (a) and the effective link transmission $T_2^{(M)}$ (b) as a function of link length $d$ for different number of amplifiers $M$ for $\epsilon=0.05$ and $\beta=0.95$. The case $M=0$ refers to the no-amplifier protocol.}\label{fig03:sec9.2.2_EffPar}
\end{figure}

The physical explanation of the previous results is the following.  When measuring the squeezed quadrature $p$, Bob observes a higher added noise with respect to the standard protocol, that is, $\chi_2^{(M)}\ge \Ntot$, and a reduced effective transmissivity $T_2^{(M)}\le \Ttot$, as depicted in Fig.~\ref{fig03:sec9.2.2_EffPar}.
In turn, the mutual information between Alice and Bob is reduced, but at the same time also Eve's Holevo information is reduced since the conditional entropy $S_{E|B}^{(\b)}$ becomes larger, according to~(\ref{eq:d3}). The tradeoff between the two types of information leads to the the existence of an optimized gain for which the Holevo information is reduced more than the mutual information, eventually resulting in a higher KGR obtained by ``hiding" behind the noise.

In light of this, the advantage introduced by PSAs shall increase with the number of spans $M$. In particular, we may obtain the maximum increase in KGR in the continuous-amplification limit, $M \gg 1$. Since $T^M=\Ttot$ is fixed, in this limit, up to a leading order in $M$, we have that $T\approx 1$, $1-T \approx - \ln T = -(\ln \Ttot)/M$ and $G^{M}=G_\infty$. Consequently, the effective transmissivities and added noises read
\begin{align}
T_1^{(\infty)}&= G_\infty \Ttot \,,  \qquad T_2^{(\infty)}= G^{-1}_\infty \Ttot \, ,
\end{align}
while the effective added noise $\chi_1^{(M)}$ on quadrature $q$ becomes:
\begin{align}
\chi_1^{(\infty)} &= \lim_{M\to \infty} \left\{ \frac{1}{G_\infty \Ttot} \frac{1- G_\infty \Ttot}{1-(G_\infty \Ttot)^{1/M}} \frac{1-T}{T} (1+2\bar{n}_T)\right\} \nonumber \\[1ex]
&=  \frac{1}{G_\infty \Ttot} \left[- \frac{1- G_\infty \Ttot}{\ln(G_\infty \Ttot)/M}  \right] \left[- \frac{\ln \Ttot}{M}  \right] (1+2\bar{n}_T) \nonumber \\[1ex]
&= \frac{1-G_\infty\Ttot}{G_\infty \Ttot} \frac{\ln \Ttot}{\ln\left(G_\infty \Ttot\right)}(1+2\bar{n}_T)\, ,  
\end{align}
where we also adopted the Taylor expansion $1-x^{1/M} = - \ln x/M +O(1/M^2)$, holding for $M\gg 1$. With analogous method, we obatin:
\begin{align}
\chi_2^{(\infty)} &= \frac{1-\Ttot/G_\infty}{\Ttot/G_\infty} \frac{\ln \Ttot}{\ln\left( \Ttot/G_\infty\right)}(1+2\bar{n}_T) \, ,
\end{align}
and we obtain the KGR by~(\ref{eq:KGRUnc}). These channel parameters, calculated for the resulting optimized gain $G_\infty$, are plotted in Fig.~\ref{fig03:sec9.2.2_EffPar}. Note also that even a few spans allow one to approach the continuous amplification limit.

\begin{figure}
\includegraphics[width=0.55\columnwidth]{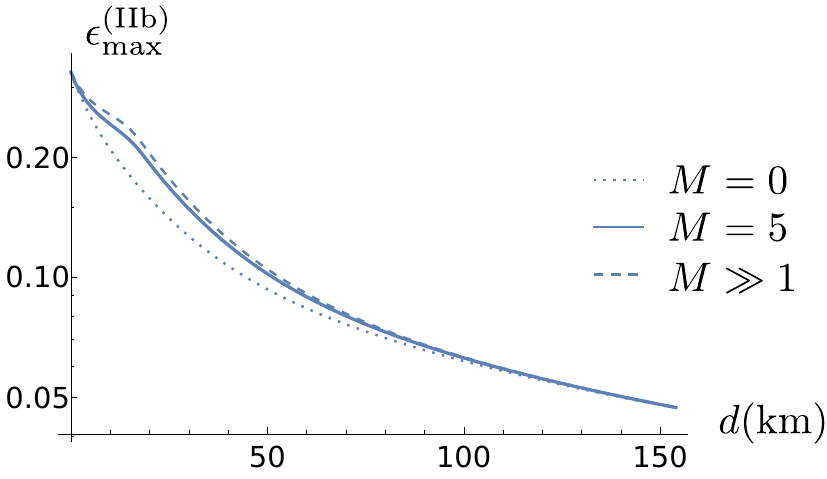}
\centering
\caption{Plot of the maximum tolerable noise $\epsilon_{\rm max}^{(\caseII\b)}$ as a function of the link length $d$ for different number of amplifiers $M$ and $\beta=0.95$. The case $M=0$ refers to the no-amplifier protocol.}\label{fig04:sec9.2.2_EpsMAx}
\end{figure}

Finally, we calculate the maximum tolerable excess noise $\epsilon_{\rm max}^{(\caseII\b)}$ as a function of the transmission distance, reported in Fig.~\ref{fig04:sec9.2.2_EpsMAx}. It represents the maximum acceptable amount of noise to maintain a positive KGR. Consistently with the previous results, the exploitation of PSAs increases the maximum tolerable excess noise with respect to the no-amplifier scheme in the metropolitan-distance regime, as $\epsilon_{\rm max}^{(\caseII\b)} \ge \epsilon_{\rm max}^{(\no)}$. As expected, the advantage introduced increases with the number of nodes.

\subsubsection{Trusted-device scenario}

\begin{figure}
\includegraphics[width=0.8\columnwidth]{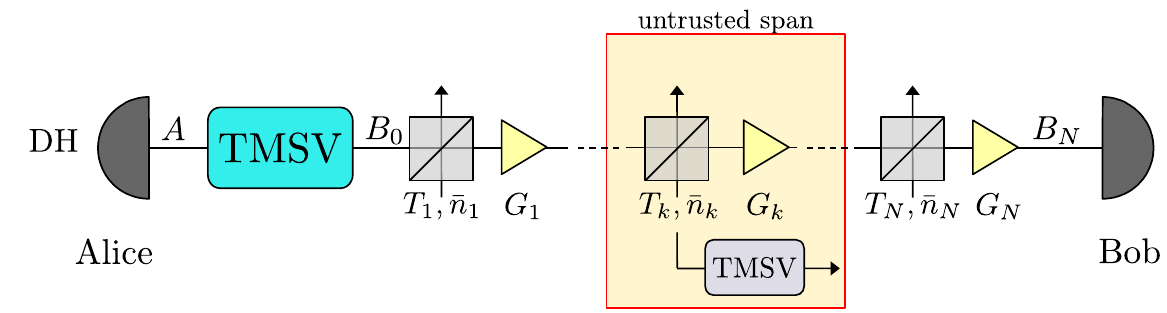}
\centering
\caption{Scheme of the CVQKD protocol under restricted eavesdropping. All the amplifiers are trusted and Eve is allowed to attack only the $k$-th span, $k=1,\ldots,M$, via active eavesdropping, that is by injecting one arm of a TMSV state into the span, hiding herself behind the introduced excess noise. }\label{fig01:sec9.2.3_Cond}
\end{figure}

We now discuss the second instance under investigation, namely the restricted eavesdropping case. In this scenario we assume Eve to attack only a single span of the link, whilst all the remaining ones as well as the employed amplifiers are considered to be trusted, thus letting our analysis to belong to the conditional security framework. In turn, only a fraction $1/M$ of the whole fiber link is untrusted.
The scheme for the eavesdropping strategy under investigation is depicted in Fig.~\ref{fig01:sec9.2.3_Cond}. Across the whole channel, only the $k$-th link, $k=1,\ldots,M$, is untrusted and may be attacked via entangling cloner attack by Eve \cite{Laudenbach2018, Pan2020}, performing active eavesdropping. That is, Eve hides herself behind the thermal noise $\bar{n}_k=\bar{n}_T$, equal to~(\ref{eq:thnoise}), by generating a TMSV state with variance $V_\epsilon= 1+2 \bar{n}_T$ on two modes $\boldsymbol{E}=(E_1,E_2)$ and injecting mode $E_1$ into the second input port of the beam splitter modeling the $k$-th span, retrieving the reflected output state. In this way she gets undetected by Alice and Bob, as performing partial trace over modes $\boldsymbol{E}$ introduces an additive thermal noise with exactly $\bar{n}_T$ mean number of photons.
In order to perform the security analysis under the above paradigm we shall compute the quantum state in Eve's possession after the entangling cloner attack. We proceed as follows, starting with the case~$\caseI$.

Since all nodes $j=1,\ldots,k-1$ are trusted, the quantum state shared by Alice and Bob injected into the $k$-th span is in the form~(\ref{eq:CM_Nnodes_PIA}), namely: 
\begin{align}
\bmsigma_{AB_{k-1}}^{(\caseI)}= 
\begin{pmatrix}
\V^{(k-1)} \, \Id_2 &  \Z^{(k-1)} \, \sigmaz \\
\Z^{(k-1)} \,\sigmaz & \W^{(k-1)} \, \Id_2  \\
\end{pmatrix} \, .
\end{align}
Instead, the CM of Eve's initial TMSV state reads:
\begin{align}
\bmsigma_{\boldsymbol{E}}= 
\begin{pmatrix}
V_\epsilon \, \Id_2 &  Z_\epsilon \, \sigmaz \\
Z_\epsilon \,\sigmaz & V_\epsilon \, \Id_2  \\
\end{pmatrix} \, ,
\end{align}
with $Z_\epsilon=\sqrt{V_\epsilon^2-1}$.
After the interference at the beam splitter, the joint quantum state of Alice, Bob and Eve is described by the CM:
\begin{align}
\bmsigma_{AB_{k} \boldsymbol{E}}^{(\caseI)} = S \, \bigg( \bmsigma_{AB_{k-1}}^{(\caseI)} \oplus \bmsigma_{\boldsymbol{E}} \bigg) \, S^{\mathsf {T}} \, ,
\end{align} 
where
\begin{align}
S=\Id_2 \oplus S_{\rm BS} \oplus \Id_2 \, ,
\end{align}
and
\begin{align}
S_{\rm BS} =
\begin{pmatrix}
\sqrt{T} \, \Id_2 &  \sqrt{1-T} \, \Id_2 \\
- \sqrt{1-T} \, \Id_2 & \sqrt{T} \, \Id_2  \\
\end{pmatrix} \, 
\end{align}
is the symplectic matrix associated with the beam splitter operation \cite{Ferraro2005, Serafini2017}.

Thereafter, we let the transmitted signal pass through the remaining $M-k$ spans, applying the techniques described in~\ref{sec:MULTISPANth}. Ultimately, the tripartite joint state after the channel is associated with the CM:
 \begin{align}
\bmsigma_{AB \boldsymbol{E}}^{(\caseI)} &= 
\begin{pmatrix}
\bmsigma_{AB}^{(\caseI)} &  \bmsigma_{C}^{(\caseI)}\\[1ex]
 \bmsigma_{C}^{(\caseI) \mathsf{T}} &\bmsigma_{\boldsymbol{E}}^{(\caseI)}  \\
\end{pmatrix} \, ,
\end{align}
with the $\bmsigma_{AB}^{(\caseI)}$ in Equation~(\ref{eq:CM_Nnodes_PIA}) and
 \begin{align}
\bmsigma_{\boldsymbol{E}}^{(\caseI)} &= 
\begin{pmatrix}
\Big[ (1-T)\W^{(k-1)} + T V_\epsilon \Big]  \, \Id_2 &  \sqrt{T} Z_\epsilon \, \sigmaz \\
\sqrt{T} Z_\epsilon \, \sigmaz & V_\epsilon \, \Id_2  \\
\end{pmatrix} \, ,	\label{eq:sigmaE-PIA}
\\[2ex]
\bmsigma_{C}^{(\caseI)} &= 
\begin{pmatrix}
\bmsigma_{A\boldsymbol{E}}^{(\caseI)} \\[0.5ex]  \hdashline \\[-2ex]
\bmsigma_{B\boldsymbol{E}}^{(\caseI)}
\end{pmatrix}
= 
\begin{pmatrix}
c^{(1)} \,  \sigmaz &  \boldsymbol{0}  \\[0.5ex]   \hdashline \\[-2ex]
c^{(2)} \, \Id_2 & c^{(3)} \,  \sigmaz  \\
\end{pmatrix} \, ,
\end{align}
being the CM of Eve's overall state and the correlation matrix between Alice and Bob and Eve, respectively, with
\begin{align}
c^{(1)} &= - \sqrt{1-T}\, \Z^{(k-1)}  \,, \\[1ex]
c^{(2)} &= \sqrt{(G T)^{M-k+1} (1-T)} \, \bigg[ V_\epsilon-\W^{(k-1)}\bigg] \,, \\[1ex]
c^{(3)} &=  \sqrt{(G T)^{M-k} G (1-T)} \,Z_\epsilon \, .
\end{align}
Subsequently, after Bob's measurement Eve is left with the conditional state associated with:
\begin{align}\label{eq:sigmaEcond-PIA}
\bmsigma^{(\caseI)}_{\boldsymbol{E}|B}= \bmsigma_{\boldsymbol{E}}^{(\caseI)} - \bmsigma_{B \boldsymbol{E}}^{(\caseI) \mathsf{T}} \, \Big[\bmsigma_B ^{(\caseI)}+ \bmsigma_\a \Big]^{-1} \bmsigma_{B \boldsymbol{E}}^{(\caseI)} \, .
\end{align}

Similarly as in the unconditional security case, the KGR resulting from the present conditional security analysis is given by the difference between the appropriately rescaled Alice and Bob's mutual information $I_{AB}^{(\caseI)}(V,G)$ and the Holevo information between Eve and Bob $\chi_{BE}^{(\caseI)}(V,G)$:
\begin{align}
K^{(\caseI)}_\comp (V,G)=\beta I_{AB}^{(\caseI)}(V,G)- \chi_{BE}^{(\caseI)}(V,G) \, ,
\end{align}
where $\beta$ denotes the reconciliation efficiency. The Holevo information can be written as
\begin{align}\label{eq:chiBE-comp}
\chi_{BE}^{(\caseI)}(V,G)= S_E^{(\caseI)} - S_{E|B}^{(\caseI)} = h({\rm d}_1^{\,(\caseI)})+ h({\rm d}_2^{\,(\caseI)}) -h({\rm d}_3^{(\caseI)})  -h({\rm d}_4^{(\caseI)}) \, ,
\end{align}
where $h(x)$ is the function in~(\ref{eq:hfunc}) and ${\rm d}_{1(2)}^{\,(\caseI)}$ and ${\rm d}_{3(4)}^{\,(\caseI)}$ are symplectic eigenvalues of the CMs~(\ref{eq:sigmaE-PIA}) and~(\ref{eq:sigmaEcond-PIA}), respectively. The resulting optimized KGR is equal to:
\begin{align}
K^{(\caseI)}_\comp = \max_{V,G} \, K^{(\caseI)}_\comp (V,G) \, ,
\end{align}
subject to the constraints of maximum power in the link $T^{(j)} [ V+ \chi^{(j)} ] \le V$ for all $j=1,\ldots, M$, or, equivalently,
\begin{align}\label{eq:ConditionG-PIA}
G \le G_{\rm max}^{(\caseI)} \equiv \frac{1+V}{2+T(V+\epsilon-1)} \, .
\end{align}

The same procedure may be followed to derive the key rate of case $\caseII$, identifying the corresponding CMs $\bmsigma_{\boldsymbol{E}}^{(\caseII)}$ and $\bmsigma^{(\caseII\p)}_{\boldsymbol{E}|B}$, $\p=\a,\b$, the latter depending on the particular quadrature measured by Bob. 
Now, the joint state of the three parties is associated with the CM:
 \begin{align}
\bmsigma_{AB \boldsymbol{E}}^{(\caseII)} &= 
\begin{pmatrix}
\bmsigma_{AB}^{(\caseII)} &  \bmsigma_{C}^{(\caseII)}\\[1ex]
 \bmsigma_{C}^{(\caseII) \mathsf{T}} &\bmsigma_{\boldsymbol{E}}^{(\caseII)}  \\
\end{pmatrix} \, ,
\end{align}
with the $\bmsigma_{AB}^{(\caseII)}$ in Equation~(\ref{eq:CM_Nnodes_PSA}) and
 \begin{align}
\bmsigma_{\boldsymbol{E}}^{(\caseII)} &= 
\begin{pmatrix}
e_1 & 0 &  \sqrt{T} Z_\epsilon & 0 \\
0 & e_2 &  0 & -\sqrt{T} Z_\epsilon \\
\sqrt{T} Z_\epsilon & 0 & V_\epsilon & 0  \\
0 & -\sqrt{T} Z_\epsilon & 0 & V_\epsilon
\end{pmatrix} \, ,	\label{eq:sigmaE-PSA}
\\[2ex]
\bmsigma_{C}^{(\caseII)} &= 
\begin{pmatrix}
\bmsigma_{A\boldsymbol{E}}^{(\caseII)} \\[1ex] 	\hdashline \\[-2ex]
\bmsigma_{B\boldsymbol{E}}^{(\caseII)}
\end{pmatrix}
= 
\begin{pmatrix}
c^{(1)}_1 & 0 &0 & 0 \\
0 & -c^{(1)}_2 & 0 & 0 \\[1ex] 	\hdashline \\[-2ex]
c^{(2)}_1 & 0 & c^{(3)}_1 & 0  \\
0 & c^{(2)}_2 & 0 & -c^{(3)}_2
\end{pmatrix} \, ,
\end{align}
with
\begin{subequations}
\begin{align}
e_{1(2)} &= \Big[ (1-T)\W_{1(2)}^{(k-1)} + T V_\epsilon \Big] \, , \\[1ex]
c^{(1)}_{1(2)} &= - \sqrt{1-T}\, \Z^{(k-1)}_{1(2)}  \,,\\[1ex]
c^{(2)}_{1} &= \sqrt{(G T)^{M-k+1} (1-T)} \, \bigg[ V_\epsilon-\W^{(k-1)}_{1}\bigg] \,, \\[1ex]
c^{(2)}_{2} &= \sqrt{(G^{-1} T)^{M-k+1} (1-T)} \, \bigg[ V_\epsilon-\W^{(k-1)}_{2}\bigg] \,, \\[1ex]
c^{(3)}_{1} &=  \sqrt{(G T)^{M-k} G (1-T)} \,Z_\epsilon \,, \\[1ex]
c^{(3)}_{2} &=  \sqrt{(G^{-1} T)^{M-k} G^{-1} (1-T)} \,Z_\epsilon \, .
\end{align}
\end{subequations}
Finally, Eve's conditional CM reads:
\begin{align}\label{eq:sigmaEcond-PSA}
\bmsigma^{(\caseII\p)}_{\boldsymbol{E}|B}= \bmsigma_{\boldsymbol{E}}^{(\caseII)} - \bmsigma_{B \boldsymbol{E}}^{(\caseII) \mathsf{T}} \, \Big[\bmsigma_B ^{(\caseII)}+ \bmsigma_\p \Big]^{-1} \bmsigma_{B \boldsymbol{E}}^{(\caseII)} \, , \qquad \p=\a,\b \, .
\end{align}
The corresponding KGR can be written as:
\begin{align}
K^{(\caseII\p)}_\comp (V,G)=\beta I_{AB}^{(\caseII\p)}(V,G)- \chi_{BE}^{(\caseII\p)}(V,G) \, , \qquad  \p=\a,\b \, ,
\end{align}
with the mutual information $I_{AB}^{(\caseII\p)}(V,G)$ given in~(\ref{eq: IAB-PSA}) and the Holevo information equal to
\begin{align}\label{eq:Holevo_BE_PSA}
\chi_{BE}^{(\caseII\p)}(V,G)&= S_E^{(\caseII)} - S_{E|B}^{(\caseII\p)} \nonumber \\
&= h({\rm d}_1^{\,(\caseII)})+ h({\rm d}_2^{\,(\caseII)}) -h({\rm d}_3^{\,(\caseII\p)})  -h({\rm d}_4^{\,(\caseII\p)}) \, ,
\end{align}
${\rm d}_{1(2)}^{\,(\caseII)}$ and ${\rm d}_{3(4)}^{\,(\caseII\p)}$ being the symplectic eigenvalues of $\bmsigma_{\boldsymbol{E}}^{(\caseII)}$ and $\bmsigma^{(\caseII\p)}_{\boldsymbol{E}|B}$, respectively. Finally, one obtains
\begin{align}
K^{(\caseII\p)}_\comp = \max_{V,G} \, K^{(\caseII\p)}_\comp (V,G) \, ,
\end{align}
subject to the constraint~(\ref{eq:ConditionG-PSA}).
Differently from Section~\ref{sec:MultiSpanUncSec}, in this scenario the no-amplifier protocol is equivalent to the case of a wiretap channel under restricted eavesdropping,
in which Eve has access only to a portion $1/M$ of the fiber link \cite{Pan2020}. That is, we may model the channel as an asymmetric three-span channel composed of three beam splitters with effective transmissivities $T_{l}= T^{k-1}$, $T_{k}= T$ and $T_{r}= T^{M-k}$, and thermal noise $\bar{n}_l=\bar{n}_k=\bar{n}_r=\bar{n}_T$, respectively, in which only the central span is attacked by Eve via entangling-cloner attack. The benchmark key rate $K^{(\no)}_\comp$ is then equal to:
\begin{align}
K^{(\no)}_\comp = \max_{V} \, K^{(\caseI)}_\comp (V,G=1) \, .
\end{align}

In the following, we show the obtained results, by comparing directly cases $\caseI$ and $\caseII\a$, in which the amplified quadrature is probed by Bob and, thereafter, by discussing case $\caseII\b$, where Bob detects the de-amplified quadrature.
%
\subsubsection{Cases $\caseI$ and $\caseII\a$ : measuring the amplified quadrature}\label{sec:AmpQuad}

\begin{figure}
\includegraphics[width=0.49\columnwidth]{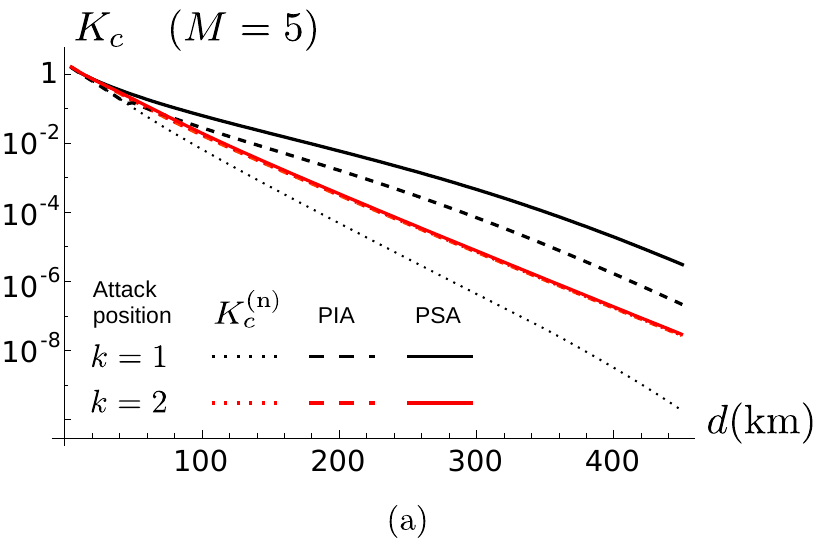} 
\includegraphics[width=0.49\columnwidth]{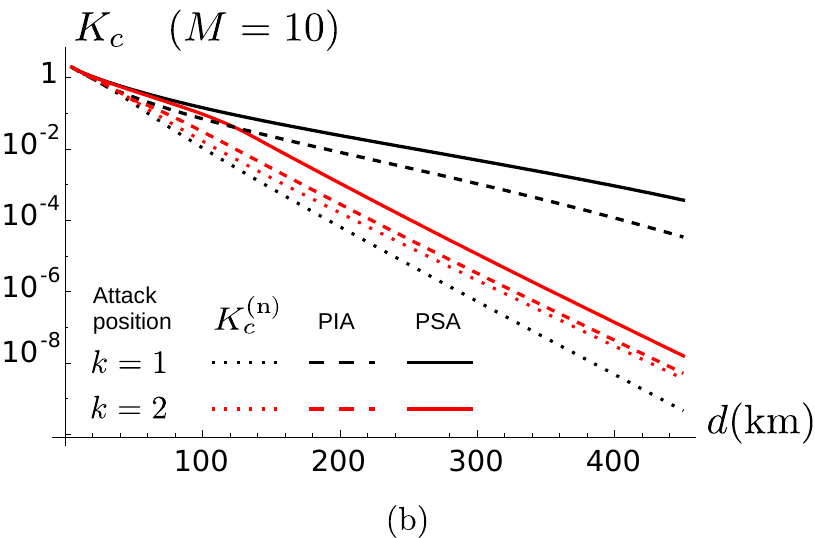} \\[1ex]
\includegraphics[width=0.49\columnwidth]{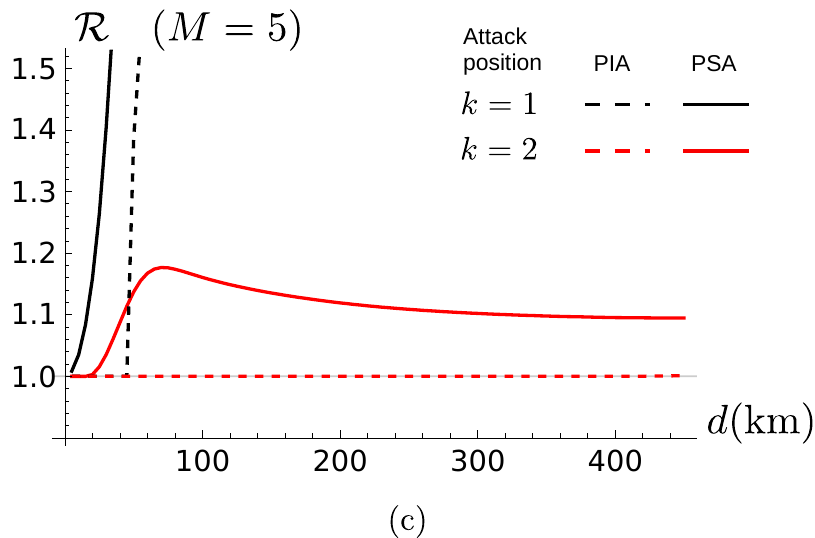} 
\includegraphics[width=0.49\columnwidth]{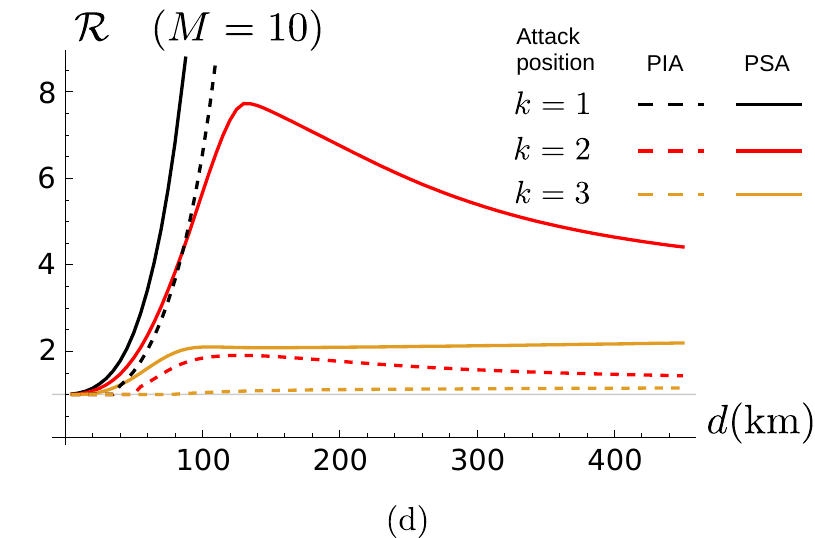} \\
\centering
\caption{Plot of the optimized KGR $K_\comp$ and key ratio $\cal R$ for cases $\caseI$ and $\caseII\a$ as a function of the transmission link length $d$ for different locations of the eavesdropper for $M=5$ (a) and (c) and $M=10$ (b) and (d), respectively. We set $\epsilon=0.05$ and $\beta=0.95$.}\label{fig02:sec9.2.3_KGR-comp}
\end{figure}

For both the discussed cases $\caseI$ and $\caseII$, plots of the KGR $K_\comp^{(\q)}$, $\q=\caseI,\caseII\a$, are presented in Figure~\ref{fig02:sec9.2.3_KGR-comp} for links with $M=5$ (a) or $M=10$ (b) amplifiers and different positions $k=1,\ldots, N$ of the untrusted span, and compared to $K_\comp^{(\no)}$ for no-amplifier protocol. We underline that the results for $M=5$ and $M=10$ can be only qualitatively compared, as we keep the assumption that only one span is untrusted and, in turn, by increasing $M$ Eve becomes more and more restricted.

In general, one can observe that, when Bob measures the amplified quadrature, both PIAs and PSAs improve the KGR with respect to the no-amplifier protocol only if Eve attacks one of the first spans of the fiber-link. The case $k=1$, where the first span is the untrusted one, represents the best-case scenario, where the key rate is increased by several orders of magnitude. Indeed, in this scenario the signal intercepted by Eve has not been amplified yet. Thus, Eve's overall state, described by the CM $\bmsigma_{\boldsymbol{E}}^{(\q)}$, is independent of the gain $G$ and the only effect of amplification is the reduction of the conditional entropy $S_{E|B}^{(\q)}$ appearing in the Holevo information Eqs.~(\ref{eq:chiBE-comp}) and~(\ref{eq:Holevo_BE_PSA}). On the other hand, for $k \ge 2$, amplifying Bob's received signal also increases Eve's overall entropy $S_{E}^{(\q)}$. In turn, the benefits of optical amplification are more and more reduced with increasing $k$. To better quantify this effect, we compute the ratio:
\begin{align}
{\cal R}^{(\q)} = \frac{K_\comp^{(\q)}}{K_\comp^{(\no)}} \,, \qquad \q=\caseI,\caseII\a \, ,
\end{align}
which is presented in Figure~\ref{fig02:sec9.2.3_KGR-comp}(c-d). 
All ratios are initially equal to $1$ up to a threshold distance, that is, ${\cal R}^{(\q)}=1$ if $d \le d_{\rm min}^{(\q)}$, thereafter for $k\ge 2$ they reach a maximum and then decrease towards an asymptotic value. Moreover, the key ratio ${\cal R}^{(\q)}$ decreases with increasing $k$ and there exists a threshold value $k_{\rm th}$ such that for $k \ge k_{\rm th}^{(\q)}$ we have ${\cal R}^{(\q)} \equiv 1$. Therefore, if Eve attacks a span located further, $k \ge k_{\rm th}^{(\q)}$, employing signal amplification is no longer beneficial. For link parameters values $\kappa=0.2\,\textrm{dB/km}$, $\epsilon=0.05$ and $\beta=0.95$ one obtains $k_{\rm th}^{(\caseI)}=2$ and $k_{\rm th}^{(\caseII\a)}=3$ for $M=5$, while for $M=10$ one gets $k_{\rm th}^{(\caseI)}=5$ and $k_{\rm th}^{(\caseII\a)}=8$.
Importantly, note that the performance of PIA links is always lower than PSA ones, as ${\cal R}^{(\caseI)} \le {\cal R}^{(\caseII\a)}$, $d_{\rm min}^{(\caseI)} \le d_{\rm min}^{(\caseII\a)}$ and $k_{\rm th}^{(\caseI)} \le k_{\rm th}^{(\caseII\a)}$. This is a direct consequence of the additional noise introduced by the phase-insensitive amplification process.

\begin{figure}
\includegraphics[width=0.49\columnwidth]{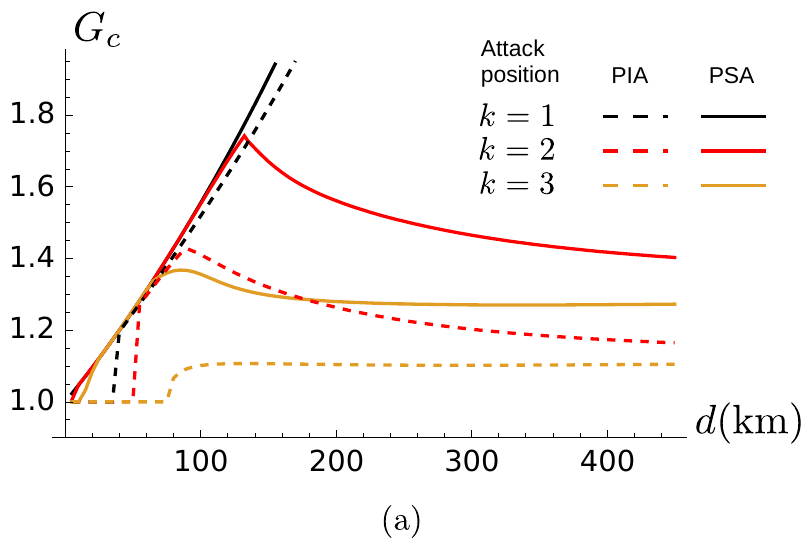}
\includegraphics[width=0.49\columnwidth]{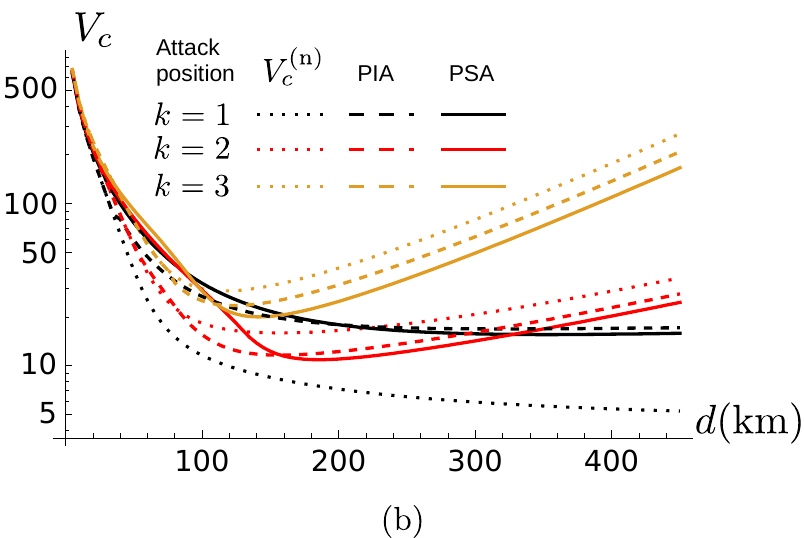}
\centering
\caption{Optimal amplifier power gain (a) and modulation (b) for cases $\caseI$ and $\caseII\a$ as a function of the link length $d$ for different locations of the untrusted span $k$ for $M=10$, $\epsilon=0.05$ and $\beta=0.95$.}\label{fig03:sec9.2.3_OptParComp}
\end{figure}

The optimized gain $G^{(\q)}_\comp$ and modulation $V^{(\q)}_\comp$, $\q=\caseI,\caseII\a$, are depicted in Figure~\ref{fig03:sec9.2.3_OptParComp} (a) and (b), respectively.
Consistent with the results from the previous paragraph, it is optimal to not amplify the signal, i.e. $G^{(\q)}_\comp=1$, for short distances $d\le d_{\rm min}^{(\q)}$. For longer link lengths the optimal gain initially increases with $d$, following constraints~(\ref{eq:ConditionG-PSA}) and~(\ref{eq:ConditionG-PIA}), and then ultimately decreases towards an asymptotic value. The optimal gain $G^{(\q)}_\comp$ also decreases with $k$, similarly to the key ratio.
On the other hand, the behavior of optimal modulation $V^{(\q)}_\comp$ is quite peculiar. For $k=1$ it is a monotonous decreasing function of the transmission distance $d$, as obtained in Section~\ref{sec:MultiSpanUncSec}. The presence of optical amplifiers increases the modulation value with respect to the no-amplifier protocol, as $V^{(\q)}_\comp \ge V^{(\no)}_\comp$. On the contrary, when $k\ge2$ the situation is completely different and in the long-distance regime the optimized modulation turns out to be an increasing function of $d$. In fact, if Eve attacks one of the last spans of the communication link she intercepts a weak pulse, therefore it is possible to safely increase the input modulation variance without preventing secure communication between Alice and Bob.

When the amplified quadrature is measured, the effective transmissivity probed by Bob, namely $T^{(M)}$ and $T^{(M)}_1$ for cases $\caseI$ and $\caseII\a$ respectively, is larger with respect to the no-amplifier protocol, $T^{(M)},\,T^{(M)}_1\ge T_{\no}$. This leads to an increase of both mutual information between Alice and Bob, and, at the same time, Holevo information on Eve's side. This is because for $k\ge2$ she also receives an amplified signal. In turn, when performing optimization over the free parameters, there emerges a tradeoff between these two types of information, resulting in the key rates shown in Figure~\ref{fig02:sec9.2.3_KGR-comp}.
In particular, for short-distance communication, $d \le d_{\rm min}^{(\q)}$, one obtains that optical amplification is useless, $G_c^{(\q)}=1$.
The difference between cases $\caseI$ and $\caseII\a$ is due to the different impact of the added noise. In fact, for case $\caseII\a$ the added noise is rescaled with respect to the no-amplifier protocol, $\chi_1^{(M)}\le \chi_{\no}$, whilst for case $\caseI$ the noise is increased because of the additive contribution $\chi_G$ due to phase-insensitive amplification, $\chi^{(M)}\ge \chi_{\no}$. In the latter case the (incoherent) added contribution $\chi_G$ detriments the mutual information between Alice and Bob, being less than its counterpart of case $\caseII\a$. Ultimately, this leads to a reduced performance of PIA links with respect to PSA  ones. 

\subsubsection{Case $\caseII\b$ : measuring the de-amplified quadrature}\label{sec:deAmpQuad}

\begin{figure}
\includegraphics[width=0.49\columnwidth]{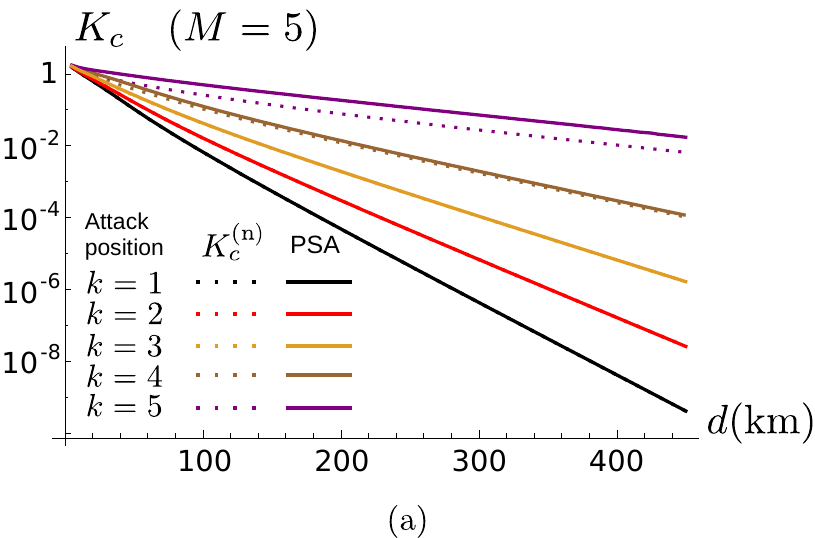} 
\includegraphics[width=0.49\columnwidth]{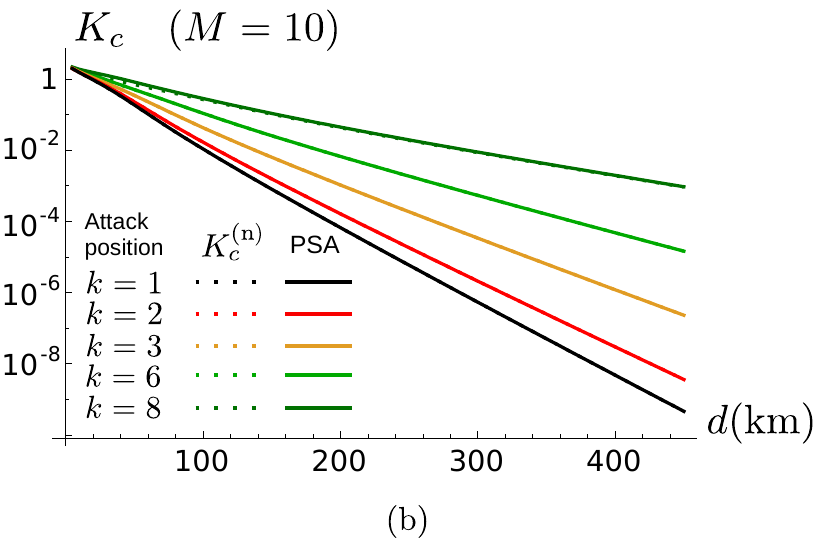} \\[1ex]
\includegraphics[width=0.49\columnwidth]{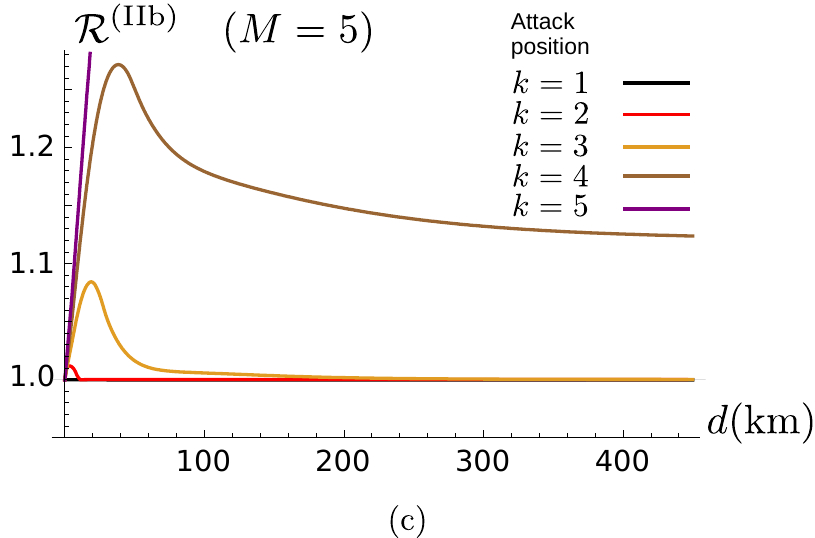} 
\includegraphics[width=0.49\columnwidth]{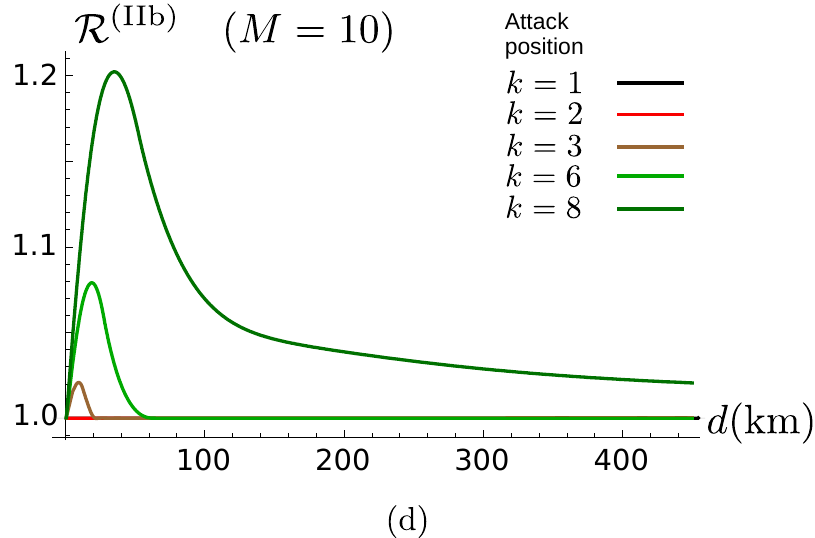} \\
\centering
\caption{Plot of the optimized KGR and key ratio $\cal R^{(\caseII\b)}$ for case $\caseII\b$ as a function of the transmission link length $d$ for different locations of the eavesdropper for $M=5$ (a) and (c) and $M=10$ (b) and (d), respectively. We set $\epsilon=0.05$ and $\beta=0.95$.}\label{fig02:sec9.2.3_KGR-comp-deamp}
\end{figure}

The KGR $K_\comp^{(\caseII\b)}$ for the Bob's measurement of the de-amplified quadrature, $\caseII\b$,  is depicted in Figure~\ref{fig02:sec9.2.3_KGR-comp-deamp} for links with $M=5$ (a) or $M=10$ (b) amplifiers and different positions $k=1,\ldots, M$ of the untrusted span, together with the key ratio
\begin{align}
{\cal R}^{(\caseII\b)} = \frac{K_\comp^{(\caseII\b)}}{K_\comp^{(\no)}} \,.
\end{align}
The scenario is reversed with respect to the previous cases. Indeed, when Bob probes the squeezed (i.e. de-amplified) quadrature, PSA links improve the resulting KGR if Eve attacks one of the last spans of the channel. The best-case scenario is provided by $k=M$, in which the KGR increases by more than an order of magnitude. Consequently, and in contrast to the results from Section~\ref{sec:AmpQuad}, one observes enhancement in the key ratio ${\cal R}^{(\caseII\b)}$ with increasing $k$.
In this scenario the PSA becomes useless if Eve attacks the first span for all $M$, namely ${\cal R}^{(\caseII\b)} \equiv 1$, since in this case she intercepts the pulse before all amplifiers and therefore, de-amplifying the signal only reduces the mutual information between Alice and Bob, maintaining a higher Holevo information at Eve's side. On the other hand, for $k \ge 2$, de-amplifying Bob's signal also reduces Eve's extracted information, thus leading to ${\cal R}^{(\caseII\b)} \ge 1$. 
In particular, there exists a threshold attack location $k_{\rm th}^{(\caseII\b)}$ such that for $k \le k_{\rm th}^{(\caseII\b)}$ one has ${\cal R}^{(\caseII\b)} \equiv 1$, being equal to $k_{\rm th}^{(\caseII\b)}= 1$ for $M=5$ and $k_{\rm th}^{(\caseII\b)}=2$  for $M=10$. 
For eavesdropping performed on a span located further within the link $k \ge k_{\rm th}^{(\caseII\b)}$ all key ratios exhibit a maximum and then decrease towards an asymptotic value, equal to $1$ for locations closer to the threshold value or greater than $1$ for those placed further, implying an improvement of security in the long-distance regime brought by the PSA link. 

We note that the absence of PSA advantage for $k=1$ does not stand in contradiction with its existence in the unconditional security framework discussed in Section~\ref{sec:MultiSpanUncSec}, where Eve is assumed to collect the reflected pulses from all spans. This is because de-amplification reduces the accessible information contained in the signals lost after the second span, eventually resulting in a enhancement of the KGR.

\begin{figure}
\includegraphics[width=0.49\columnwidth]{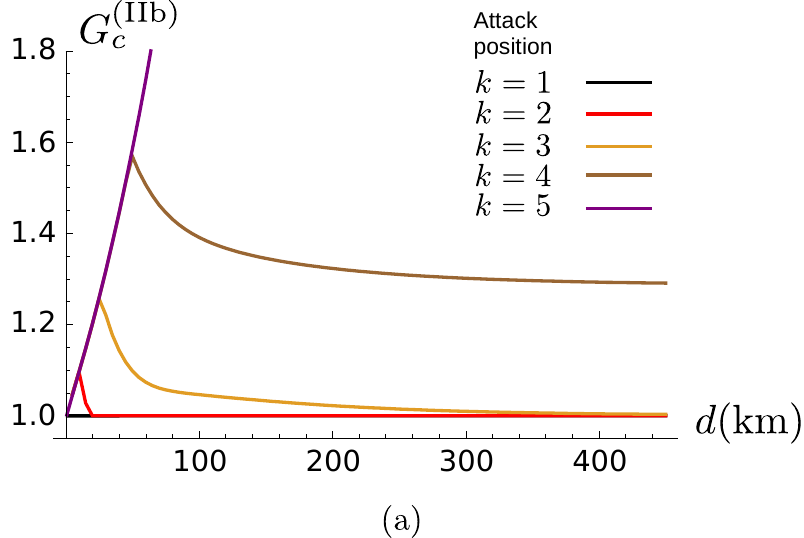} 
\includegraphics[width=0.49\columnwidth]{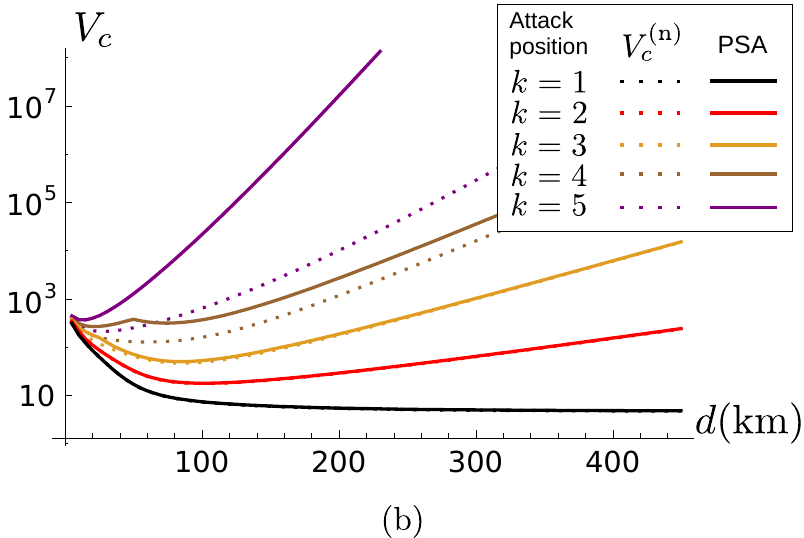}
\centering
\caption{Optimal amplifier power gain (a) and modulation (b) for case $\caseII\b$ as a function of the link length $d$ for different locations of the untrusted span $k$ for $M=10$, $\epsilon=0.05$ and $\beta=0.95$.}\label{fig05:sec9.2.3_OptParComp-de}
\end{figure}

In Figure~\ref{fig05:sec9.2.3_OptParComp-de}(a) and (b), one can see the optimized gain $G^{(\caseII\b)}_\comp$ and modulation $V^{(\caseII\b)}_\comp$, respectively.
We see that amplification is not beneficial, $G^{(\caseII\b)}_\comp \equiv 1$ , for eavesdropping performed on initial spans $k\le k_{\rm th}^{(\caseII\b)}$, whereas for attacks on latter spans the optimal gain increases with the link length following constraint~(\ref{eq:ConditionG-PSA}), until finally decreasing towards an asymptotic value. In accordance with the previous results, one needs to employ the stronger optimal amplification the further the eavesdropped span is located.
The optimized modulation increases with respect to the no-amplifier protocol $V^{(\caseII\b)}_\comp \ge V^{(\no)}_\comp$. Similarly to the results obtained in Section~\ref{sec:AmpQuad}, it is a decreasing function of the link length if the attack is performed on the first span, whilst it becomes non-monotonous for $k\ge 2$, increasing in the long-distance regime.

The physical meaning of these results is analogous to these obtained in Section~\ref{sec:MultiSpanUncSec}. Indeed, the case~$\caseII\b$ is associated with a reduced transmissivity with respect to the no-amplifier protocol, $T^{(M)}_2\le T_{\no}$, and amplified added noise $\chi^{(M)}_2\ge \chi_{\no}$. Therefore, for $k\ge 2$ by employing PSAs Bob accepts to reduce the extracted mutual information, in order to increase the conditional entropy $S_{E|B}^{(\caseII\b)}$, resulting in a lower Holevo information between Eve and himself. The tradeoff between these two quantities is such that for $k \ge k_{\rm th}^{(\caseII\b)}$ one has $G^{(\caseII\b)}_\comp \ge 1$ and PSA links increase the obtained KGR.

\subsection{CVQKD with noiseless linear amplifiers}\label{sec:CVQKD-NLA}

\def\a{ {\mathrm{QS}} }
\def\b{ {\mathrm{SPC}} }
\def\p{ {\mathrm{p}} }
\def\sigmamA{\boldsymbol\sigma^{\rm(m)}_{A}}
\def\sigmamB{\boldsymbol\sigma^{\rm(m)}_{B}}
\def\sigmamAB{\boldsymbol\sigma^{\rm(m)}_{A(B)}}
\def\sigmaz{\boldsymbol\sigma_z}
\def\aA{a_A}
\def\aB{a_B}
\def\aC{a_{B_1}}
\def\aD{a_{B_2}}
\def\bA{b_A}
\def\bB{b_B}
\def\bC{b_{B_1}}
\def\bD{b_{B_2}}
\def\alphaA{\alpha_A}
\def\alphaB{\alpha_B}
\def\alphaC{\alpha_{B_1}}
\def\alphaD{\alpha_{B_2}}
\def\betaA{\beta_A}
\def\betaB{\beta_B}
\def\betaC{\beta_{B_1}}
\def\betaD{\beta_{B_2}}\def\alphaA{\alpha_A}
\def\alphaB{\alpha_B}
\def\alphaC{\alpha_{B_1}}
\def\alphaD{\alpha_{B_2}}
\def\betaA{\beta_A}
\def\betaB{\beta_B}
\def\betaC{\beta_{B_1}}
\def\betaD{\beta_{B_2}}
\def\bmalpha{\boldsymbol\alpha}
\def\bmbeta{\boldsymbol\beta}
\newcommand{\NOTll}{\hskip 0.4mm \not \hskip -0.4mm \ll}

As outlined in the previous section, conventional amplifiers based on parametric down conversion have a relatively limited impact of CVQKD in the presence of unconditional security, whereas they prove themselves more powerful under restricted eavesdropping, when only few parts of the fiber links are untrusted.
An intriguing solution to overcome this fundamental limitation is provided by heralded noiseless linear amplification at the receiver's side. Indeed, in 2012 Blandino {\em et al.} showed that an ideal probabilistic NLA with amplitude gain $g$ leads to an increase in the maximum transmission distance proportional to $\log g$ \cite{Blandino2012}. Nevertheless, any realistic physical NLA can only approximate the ideal amplifier for low-amplitude optical signals \cite{NLARalphLund, NLAFiu1, NLAXi, NLAMc, NLASPC, NLARalph, NLAFiu2, NLAJoshua1, NLAFiu3, NLAJoshua2}. To avoid this drawback, measurement-based NLAs, performing virtual amplification based on classical data post-selection, have also been proposed \cite{MB1, MB2, MB3}. However, the low success probabilities of these operations  \cite{Bernu2014, Zhao2017} make physical NLAs still worth of investigation. Recently, CVQKD employing quantum scissors (QS) \cite{NLARalphLund} has been addressed, allowing to achieve long-distance secure communication for sufficiently low channel excess noise \cite{Ghalaii,Ghalaii_DM}. To the same goal, also single-photon catalysis (SPC) has been investigated \cite{NLASPC, QKDNonGauss}.

In the following, we firstly present the unconditional security analysis for the GG02 protocol assisted by the ideal NLA proposed by Blandino {\em et al.}. Thereafter, we investigate security in the presence of feasible physical NLAs, realized via either the quantum scissors (QS) or the single-photon catalysis (SPC) scheme, in which we consider the simplified realistic scenario where photo-detection is replaced by on-off detection, see Sec.~\ref{subsec2:NLA}. 
Moreover, we distinguish two alternative cases. In the former, we fix the NLA gain $g$ and show that also physical NLAs increase the maximum transmission distance by the same amount $\ln g$ as the ideal amplifier. In the latter, we assume $g$ to be a free parameter and optimize its value, obtaining that both physical and ideal NLAs achieve arbitrary long-distance CVQKD.
For the physical amplifiers, we also discuss the robustness in the presence of a quantum detection efficiency $\eta\le1$, showing that the detection efficiency only rescales the KGR without preventing long-distance communication.

\subsubsection{Ideal NLA}\label{subsec:idealNLAass}
\begin{figure}
\includegraphics[width=0.8\columnwidth]{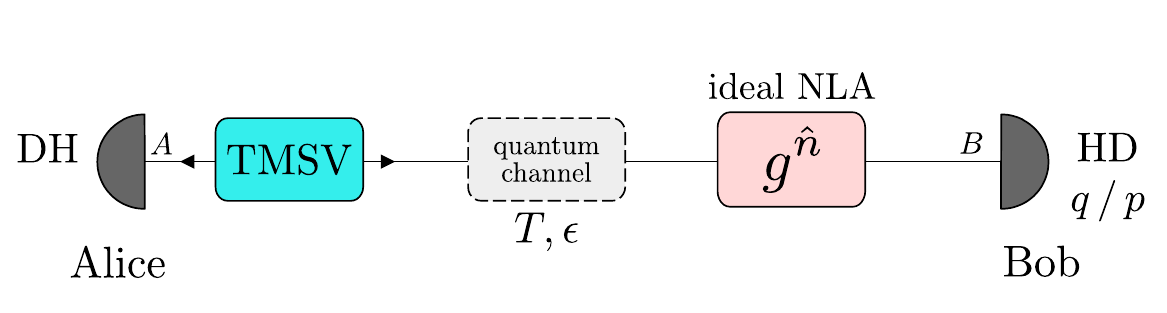}
\centering
\caption{Scheme of the CVQKD protocol assisted by the ideal NLA proposed in \cite{Blandino2012}. Alice performs double homodyne (DH) detection on the first branch of a TMSV state with variance $V>1$ to perform coherent state generation. The second one is then injected into the thermal-loss channel to Bob, who implements the ideal NLA operation $g^{\hat{n}}$ to amplify his signal, before performing homodyne (HD) detection.}
\label{fig01:sec9.3.1_IdealNLA}
\end{figure}

To begin with, we consider a GG02 scheme in which Bob employs an ideal NLA to amplify his received signal, as depicted in Fig.~\ref{fig01:sec9.3.1_IdealNLA}. 
That is, Alice prepares a two-mode squeezed vacuum (TMSV) state with modulation variance $V>1$ and injects one mode into a thermal-loss channel with transmissivity $T\le 1$ and excess noise $\epsilon \ge 0$, with corresponding added noise $\chi=(1-T)/T+\epsilon$. Thereafter, Bob implements an ideal NLA with gain $g>1$ on his received pulse, before performing a Gaussian measurement, here assumed to be homodyne (HD) detection. 
As discussed in Sec.~\ref{subsec2:NLA}, the ideal NLA is a non-deterministic operation described by the self-adjoint operator $g^{\hat{n}}$, $\hat{n}$ being the photon-number operator of the optical mode undergoing amplification, therefore it preserves Gaussianity \cite{NLARalphLund, Blandino2012}.
Therefore the protocol in Fig.~\ref{fig01:sec9.3.1_IdealNLA} is equivalent to a GG02 scheme where Alice and Bob share a Gaussian state with covariance matrix (CM):
\begin{align}\label{eq: GammaAB|p}
    \bmsigma_{AB}^{(\id)} = \begin{pmatrix} V_{\id}\, \Id_2 & \sqrt{T_\id}Z_\id \, \sigmaz \\[1ex] \sqrt{T_\id}Z_\id\, \sigmaz & T_\id (V_\id + \chi_\id) \, \Id_2 \end{pmatrix} \, ,
\end{align}
with $\chi_\id=(1-T_\id)/T_\id + \epsilon$, $Z_\id=\sqrt{V_\id^2-1}$, and the effective channel parameters, explicitly derived in App.~\ref{app:Blandino}:
\begin{subequations}\label{eq: IdealPar}
\begin{align}
V_{\id}&= V+ \frac{T (g^2-1) Z^2}{2-T (g^2-1)(V-1+\epsilon)} \, ,\\[1ex]
T_{\id} &=\frac{g^2 T}{1+ T (g^2-1)[1+T \epsilon (g^2-1)(2-\epsilon)/4-\epsilon]} \, ,\\[1ex]
\epsilon_{\id} &= \epsilon + (g^2-1) \frac{T \epsilon(2-\epsilon)}{2}  \,,
\end{align}
\end{subequations}
provided that:
\begin{align}\label{eq: Conditiong}
g \le \sqrt{1+\frac{2}{T(V+\epsilon-1)} } \,.
\end{align}
Without the last condition on the gain an unphysical un-normalizable state is obtained \cite{Blandino2012, NLARalphLund}. Equivalently, for a fixed gain Eq.~(\ref{eq: Conditiong}) corresponds to a threshold of the transmissivity, namely:
\begin{align}
T \le T_{\rm th} \equiv \frac{2}{(g^2-1)(V+\epsilon-1)} \, ,
\end{align}
preventing the use of the NLA protocol for distances $d \le~d_{\rm th}^{(\id)} = (-10 \log_{10}T_{\rm th})/\kappa$. For $d > d_{\rm th}^{(\id)}$, employing the ideal NLA is equivalent to considering an effective channel of increased transmissivity $T_{\id} \ge T$.
The resulting KGR then reads:
\begin{align}
    \widetilde{K}_{\id}(V,g)= P_{\id}(V,g) \bigg[\beta I_{AB}^{(\id)}(V,g)- \chi_{BE}^{(\id)}(V,g) \bigg] \, ,
\end{align}
where $P_{\id}(V,g)$ is the success probability of the NLA, $\beta\le 1$ is the reconciliation efficiency, whereas
 $I_{AB}^{(\id)}(V,g)$ and $\chi_{BE}^{(\id)}(V,g)$ are computed from Eq.s~(\ref{eq: IAB GG02}) and~(\ref{eq: chiBE GG02}), respectively, with the modified parameters~(\ref{eq: IdealPar}). As demonstrated in App.~\ref{app:Blandino}, the success NLA probability is bounded by $P_{\id}(V,g)\le 1/g^2$, therefore from now on we consider as a benchmark the KGR:
\begin{align}\label{eq: KGR ideal}
    K_{\id}(V,g)= \frac{1}{g^2} \bigg[\beta I_{AB}^{(\id)}(V,g)- \chi_{BE}^{(\id)}(V,g) \bigg] \, .
\end{align}

The KGR~(\ref{eq: KGR ideal}) depends on the two free parameters $V$ and $g$ that can be optimized. As discussed in the following, the choice of the gain $g$ will be a crucial task. Hence, we will discuss two separate cases. In the former case we assume a fixed $g$ and optimize only the modulation variance, obtaining the KGR:
\begin{align}\label{eq: KGR ideal Opt1}
    K_{\id}(g)= \max_V K_{\id}(V,g) \, ,
\end{align}
and the corresponding distance-dependent modulation $V_{\opt}^{(\id)}(g)$. In the latter case the optimization involves also the gain, obtaining:
\begin{align}\label{eq: KGR ideal Opt2}
    K_{\id}= \max_{V,g} K_{\id}(V,g) \, ,
\end{align}
and the associated parameters $V_{\opt}^{(\id)}$ and $g_{\opt}^{(\id)}$.
For the sake of clarity, we will review the obtained results in Sec.~\ref{eq:UncSecNLA} together with the physical NLA-assisted strategies under investigation.

\subsubsection{Physical NLAs: QS and SPC}

\begin{figure}
\includegraphics[width=0.8\columnwidth]{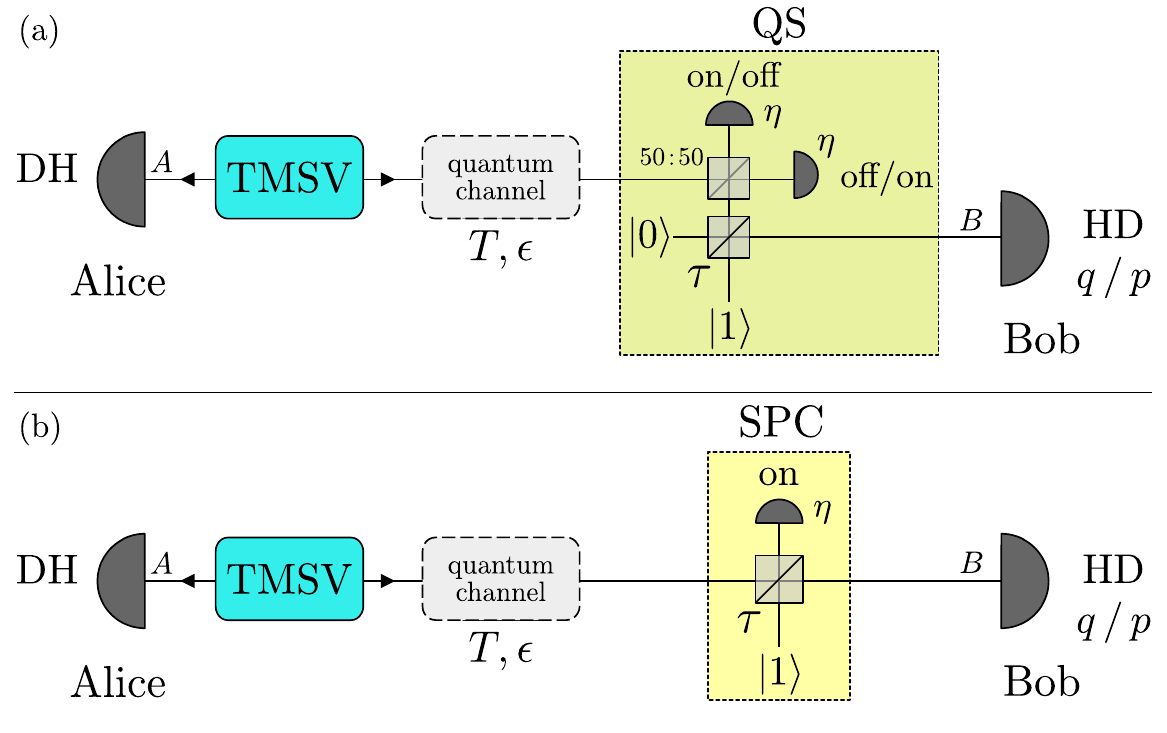}
\centering
\caption{Scheme of the CVQKD protocol assisted by the two physical NLAs discussed in the paper. (a) Strategy based on quantum scissors (QS); (b) strategy based on single-photon catalysis (SPC).}
\label{fig02:sec9.3.1_PhysicalNLA}
\end{figure}

Now, we consider the more realistic scenario in which Bob employs a physical NLA, realized via either QS or SPC and employing on-off detection rather than photon counting.
\par
In the QS scheme proposed in \cite{Ghalaii}, see Fig.~\ref{fig02:sec9.3.1_PhysicalNLA}(a), Bob prepares two ancillary modes in the Fock states $|1\rangle$ and $|0\rangle$, respectively. He mixes them at a beam splitter with transmissivity $\tau$ and lets the reflected signal interfere at a balanced beam splitter with the pulse received by Alice. Then, he performs conditional on-off detection on both the output branches (see App.~\ref{app:CovMatNLA} for details), corresponding to the positive-operator-valued measurement (POVM) $\{\Pi_{\rm off}, \Pi_{\rm on}=\Id-\Pi_{\rm off}\}$, where:
\begin{equation}
\Pi_{\rm off} = \sum_{k=0}^{\infty} (1-\eta)^{k} |k\rangle \langle k | \, ,
\end{equation}
and $\eta\leq 1$ is the detection quantum efficiency. If one of the two detectors gives the outcome ``on", Bob performs homodyne detection on the post-selected output state. The value of $\tau$ fixes the gain associated with the NLA, that for low-amplitude coherent signals reads $g= \sqrt{(1-\tau)/\tau}$ \cite{NLARalphLund}. 
Thus, to achieve the gain $g$ we set the transmissivity equal to:
\begin{align}\label{eq: gQS}
\tau_{\a}(g)= \frac{1}{1+g^2} \, .
\end{align}
\par
On the contrary, in the SPC scheme, reported in Fig.~\ref{fig02:sec9.3.1_PhysicalNLA}(b), Bob has a single ancillary mode excited in $|1\rangle$ impinging at a beam splitter with transmissivity $\tau$ with the pulse received by Alice. He performs on-off detection on the reflected branch, conditioning on outcome ``on", and homodynes the post-selected state. The associated gain is $g= (1-2\tau)/\sqrt{\tau}$ \cite{NLASPC}, which can be inverted to find the transmissivity as a function of the gain,
\begin{align}\label{eq: gSPC}
\tau_{\b}(g)= \frac18 \bigg( 4 + g^2 - g \sqrt{8 + g^2}\bigg) \, .
\end{align}
\par
In both the cases, after the NLA Alice and Bob share a non-Gaussian state $\rho_{AB}^{(\p)}$, $\p= \a,\b$. 
However, since Bob's measurement is Gaussian, the security analysis of the NLA-assisted protocol can be based on the optimality of Gaussian attacks discussed in Sec.~\ref{sec:OptGaussUnc}, which, in this scenario, maximize the amount of information extractable by Eve.
Moreover, since also Alice's DH detection is Gaussian, we further consider the Gaussian lower bound on the mutual information, and, in turn, obtain a lower bound of the exact KGR as:
\begin{align}\label{eq:K NotOpt}
K_{\p}(V,g) = P_{\p}(V,g) \bigg[ \beta I_{AB}^{(\p)}(V,g) -\chi_{BE}^{(\p)}(V,g) \bigg] \, ,
\end{align}
where $P_{\p}(V,g)$ is the success probability associated with the $\p$-th NLA and $I_{AB}^{(\p)}(V,g) $ and $\chi_{BE}^{(\p)}(V,g)$ are the mutual information and the Holevo information, respectively, both computed for a Gaussian state having the same CM of $\rho_{AB}^{(\p)}$. 
The condition $K_{\p}(V,g) \ge 0$ provides a sufficient condition to guarantee secure communication. Nevertheless, our results are in good agreement with other exact numerical approaches \cite{Ghalaii}, proving the bound~(\ref{eq:K NotOpt}) to be tight, especially in the long-distance regime $\kappa d \gg 1$.

Thus, in our approach it suffices to compute the CM $\bmsigma_{AB}^{(\p)}$ associated with $\rho_{AB}^{(\p)}$ to perform the security analysis. Straightforward calculations lead to:
\begin{align}\label{eq: GammaAB|p}
    \bmsigma_{AB}^{(\p)} = \begin{pmatrix} V_{\p}(V,g)\, \Id_2 & Z_{\p}(V,g) \, \sigmaz \\[1ex] Z_{\p}(V,g) \, \sigmaz & W_{\p}(V,g) \, \Id_2 \end{pmatrix} \, ,
\end{align}
as derived in App.~\ref{app:CovMatNLA}.
The expressions of $P_{\p}(V,g)$, $V_{\p}(V,g)$, $W_{\p}(V,g)$ and $Z_{\p}(V,g)$ are clumsy and thus only reported in App.~\ref{app:CovMatNLA}. We compute the mutual information and the Holevo information following the procedure described in Sec.~\ref{sec: GG02} by substituting $\bmsigma_{AB} \rightarrow \bmsigma_{AB}^{(\p)}$ and optimize Eq.~(\ref{eq:K NotOpt}) over the free parameters, obtaining the KGRs
\begin{align}\label{eq: K Opt}
K_{\rm p}(g)= \max_{V} \, K_{\rm p}(V,g)  \, , \qquad \p=\a,\b \,,
\end{align}
for a fixed $g$, together with the corresponding modulation $V_{\rm opt}^{(\p)}(g)$, and
\begin{align}\label{eq: K Opt}
K_{\rm p}= \max_{V,g} \, K_{\rm p}(V,g)  \, , \qquad \p=\a,\b \,,
\end{align}
if $g$ can be optimized too, with the associated optimized parameters $V_{\rm opt}^{(\p)}$ and $g_{\rm opt}^{(\p)}$.
\par
We note that in the SPC scheme there always exists a local maximum for $\tau=1$, in which case the SPC performs as the identity operator, allowing to retrieve the results of the original GG02 protocol. However, for a more fair comparison with the QS, in the optimization procedure we have neglected this point and restricted maximization over the interval $0\le\tau \le 1/2$ for which the corresponding gain is $g\geq 0$, as shown in App.~\ref{app:CovMatNLA}.

\subsubsection{Unconditional security for the NLA-assisted protocols}\label{eq:UncSecNLA}

\begin{figure}
\includegraphics[width=0.49\columnwidth]{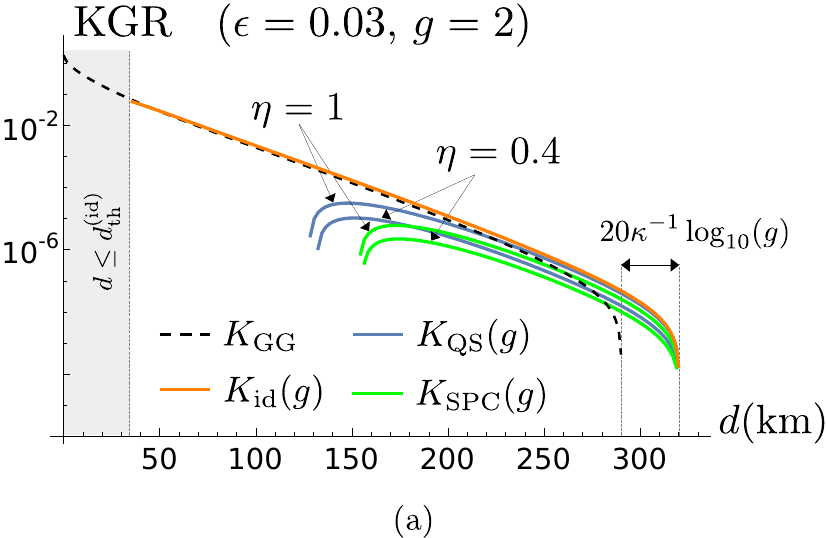} 
\includegraphics[width=0.49\columnwidth]{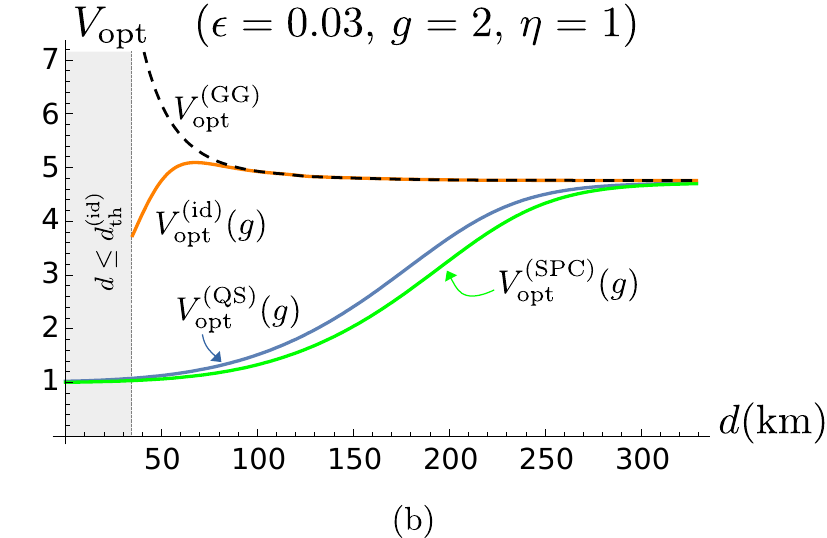}
\centering
\caption{(a) Log plot of the KGRs $K_{\p}(g)$ for different values of the quantum efficiency $\eta$ and $K_{\id}(g)$ as functions of the distance $d$ in ${\rm km}$. The dashed line is the KGR of the original protocol. (Bottom) Plot of the optimized (input) modulations $V_{\rm opt}^{(\p)}(g)$ and $V_{\rm opt}^{(\id)}(g)$ as a function of the distance $d$ in ${\rm km}$ for $\epsilon=0.03$. In both the plots, the shaded region represents the regime $d\le d_{\rm th}^{(\id)}$, where ideal NLAs generate an unphysical un-normalizable state (see the text for details). We set $\beta=0.95$, $\epsilon=0.03$ and $g=2$.}\label{fig01:sec9.3.3_KGR-Nopt}
\end{figure}

We now compare the KGRs of all the schemes under investigation, for the two cases of fixed or optimized gain.
 
\subsubsection{KGR with fixed gain $g$}\label{sec: gFixed}
For a fixed $g$, the optimized KGRs are depicted in Fig.~\ref{fig01:sec9.3.3_KGR-Nopt}(a) for $\epsilon>0$. 
As emerges from the plot, NLAs are fundamental in the long-distance regime, as for large $d$ all the NLA-assisted protocols beat the KGR $K_{\rm GG}$ of the original protocol. 
The ideal NLA increases the maximum transmission distance by the amount $(20 \log_{10} g)/\kappa$, since for $T\ll 1$ the effective transmissivity in Eq.~(\ref{eq: IdealPar}) is $T_\id \approx g^2 T$ \cite{Blandino2012}. 
Remarkably, also the physical NLA-assisted protocols achieve the same maximum transmission distance.
Moreover, the presence of inefficient conditional detection reduces the value of the KGRs, still maintaining the same increase in distance even 
for the realistic values of practical CVQKD systems where $0.4\le \eta\le 0.6$ \cite{Lodewyck2005, Lodewyck2007}.
\par
In fact, by expanding the CM~(\ref{eq: GammaAB|p}) in the long-distance regime where $T\ll 1$ up to the first order in $T$, we have:
\begin{subequations}\label{eq: apprCM-Nopt}
\begin{align}
V_{\p}(V,g) &=  V + O(T) \, ,  \\[1.ex]
W_{\p} (V,g) &=  g^2 T (V + \chi) + O(T^2)\, , \\[1.ex]
Z_{\p}(V,g) &=  \sqrt{g^2 T} \, Z + O(T^{3/2}) \, , \qquad \p=\a,\b\, ,
\end{align} 
\end{subequations}
corresponding to the CM of a GG02 scheme with transmissivity $g^2 T$, consistently with the ideal case.
The success probabilities read :
\begin{equation}
P_{\p}(V,g) \approx P_{\p}(g)=\eta \tau_{\p}(g) \, ,
\end{equation}
and, being $P_{\b}(g) \le P_{\a} (g)$, we have $K_{\b}(g) \le K_{\a} (g)$. In turn, a quantum efficiency $\eta\le 1$ only reduces the success probability and rescales the KGR, without preventing long-distance secure communication.
For completeness, we report the (input) optimized modulations in Fig.~\ref{fig01:sec9.3.3_KGR-Nopt}(b). Despite the different behaviour at small distances, for large $d$ all the protocols converge to the same asymptotic value, not depending on $\epsilon$. Numerical calculations have also shown that $V_{\rm opt}^{(\p)}(g)$ does not depend on the quantum efficiency.

\begin{figure}
\includegraphics[width=0.49\columnwidth]{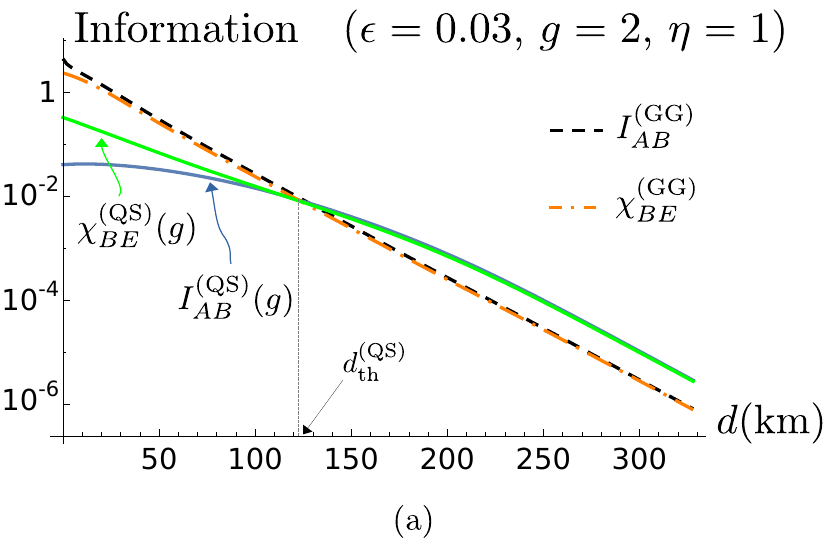} 
\includegraphics[width=0.49\columnwidth]{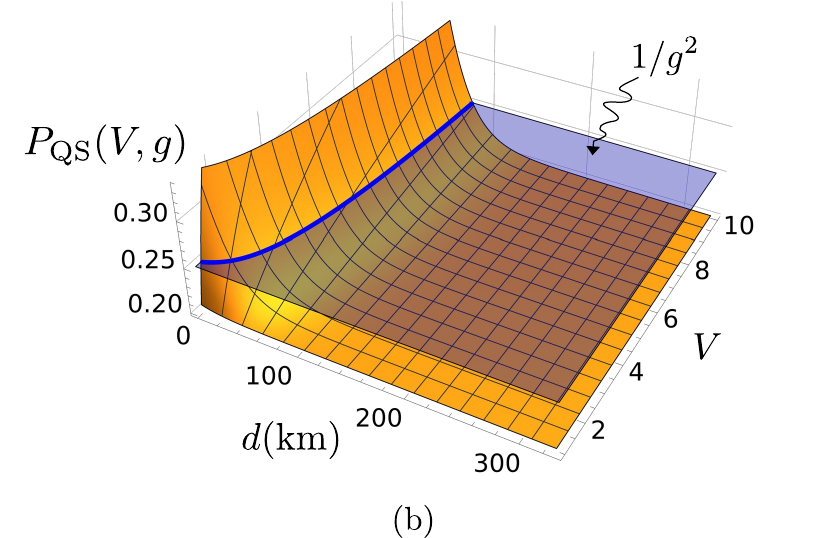}
\centering
\caption{(a) Log plot of $I_{AB}^{(\a)}(g)$ and $\chi_{BE}^{(\a)}(g)$ (solid lines) and $I_{AB}^{(\rm GG)}$ (dashed line) and $\chi_{BE}^{\rm (GG)}$ (dash-dotted line) as a function of the distance $d$ in ${\rm km}$. (b) Plot of the success probability $P_\a(V,g)$ as a function of the distance $d$ and the modulation variance $V$. The horizontal plane refers to the value $1/g^2$: when $P_\a(V,g) > 1/g^2$, the $\a$ do not perform noiseless amplification. In both the pictures we set $\beta=0.95$, $\epsilon=0.03$, $g=2$ and $\eta=1$.}\label{fig02:sec9.3.3_ShortD}
\end{figure}

We note that in the short-distance regime, where $T \approx 1$ or, equivalently, $\kappa d \ll 1$, both the physical NLAs are useless, since we obtain negative KGR up to a threshold distance $d_{\rm th}^{(\p)}$, $\p=\a,\b$. In this regime, the CM~(\ref{eq: GammaAB|p}) cannot be recast in the GG02 form of Eq.~(\ref{eq: Gamma_GG02}), and, as displayed in Fig.~\ref{fig02:sec9.3.3_ShortD}(a) for the $\a$ case, both the mutual information $I_{AB}^{(\p)} (g)= I_{AB}^{(\p)} (V_{\rm opt}^{(\p)}(g),g)$ and the Holevo information $\chi_{BE}^{(\p)} (g)= \chi_{BE}^{(\p)} (V_{\rm opt}^{(\p)}(g),g)$ are lower than their GG02 counterparts $I_{AB}^{\rm (GG)}$ and $\chi_{BE}^{\rm (GG)}$, respectively. Moreover, for $\epsilon>0$ we have $I_{AB}^{(\p)}(g) \le \chi_{BE}^{(\p)}(g)$, leading to a negative KGR which inhibits secure communication.
This effect may be understood by considering the success probability $P_\p(V,g)$ of the proposed physical NLAs, plotted in Fig.~\ref{fig02:sec9.3.3_ShortD}(b) for the $\a$ case.  Analogous considerations hold for $\b$.
When $P_\p(V,g)>1/g^2$ the $\p$ scheme does not implement a true NLA \cite{Blandino2012,Ghalaii}, and the amplification process introduces a unavoidable noise on the quadrature variances, becoming a further resource for Eve's attack.  
Accordingly, for $\kappa d \ll 1$ the optimization procedure leads to low modulation variances  $V_{\rm opt}^{(\p)}(g) \approx 1$, resulting in a lower mutual information with respect to the GG02 scheme and in a negative KGR.
On the other hand, for $\kappa d \gg 1$, $V_{\rm opt}^{(\p)} \approx V_{\rm opt}^{\rm (GG)}$ and both $I_{AB}^{(\p)}$ and $\chi_{BE}^{(\p)}$ outperform the GG02 protocol. In turn, between the short- and long-distance regimes, we identify the threshold distance such that $K_\p(g)\ge0$ for $d\le d_{\rm th}^{(\p)}$.

\begin{figure}
\includegraphics[width=0.6\columnwidth]{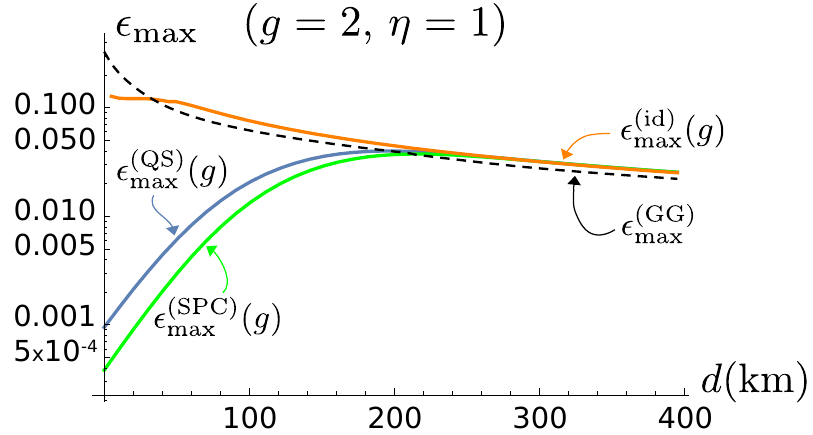}
\centering
\caption{Log plot of the maximum tolerable excess noise $\epsilon_{\rm max}^{(\id)}(g)$ and $\epsilon_{\rm max}^{(\p)}(g)$, $\p=\a,\b$, as a function of the distance $d$ in ${\rm km}$. The black dashed line corresponds to the $\epsilon_{\rm max}^{(\rm GG)}$ of the original protocol. We set $\beta=0.95$ and $\eta=1$.}\label{fig03:sec9.3.3_epsilonNOpt}
\end{figure}

Finally, in Fig.~\ref{fig03:sec9.3.3_epsilonNOpt} we plot the maximum tolerable excess noise (MTEN) $\epsilon_{\rm max}$ as a function of the distance $d$: it represents the maximum value of $\epsilon$ still leading to a positive KGR.
For the original protocol, $\epsilon_{\rm max}^{(\rm GG)}$ is a decreasing function of $d$. The behaviour is rather different for the NLA-assisted protocols. In the presence of ideal NLA the MTEN $\epsilon_{\rm max}^{(\id)}(g)$ for $d \lesssim 40$~km is lower than the original protocol due to the limitation imposed by ~(\ref{eq: Conditiong}). However, for larger distances we have $\epsilon_{\rm max}^{(\id)}(g) > \epsilon_{\rm max}^{(\rm GG)}$.
On the contrary, the MTEN associated with the physical NLAs, namely $\epsilon_{\rm max}^{(\p)}(g)$, is not a monotonous function of $d$: it is an increasing function of $d$ approaching $\epsilon_{\rm max}^{(\id)}$. A quantum efficiency $\eta\le1$ does not affect the value of $\epsilon_{\rm max}^{(\p)}$, consistently with the previous discussions.
As a consequence, for fixed $g$, in the long-distance regime the physical NLAs guarantee the same performance of the ideal NLA.

\subsubsection{KGR with optimized gain $g$}

\begin{figure}
\includegraphics[width=0.49\columnwidth]{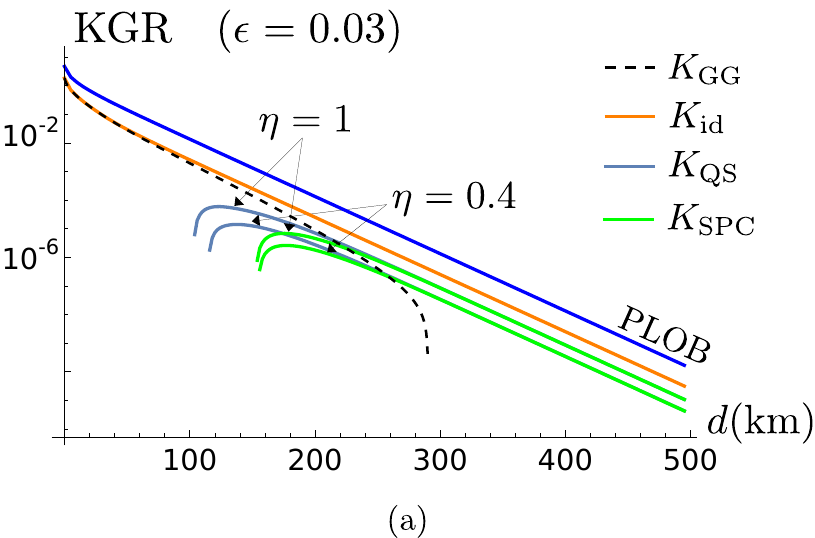} 
\includegraphics[width=0.49\columnwidth]{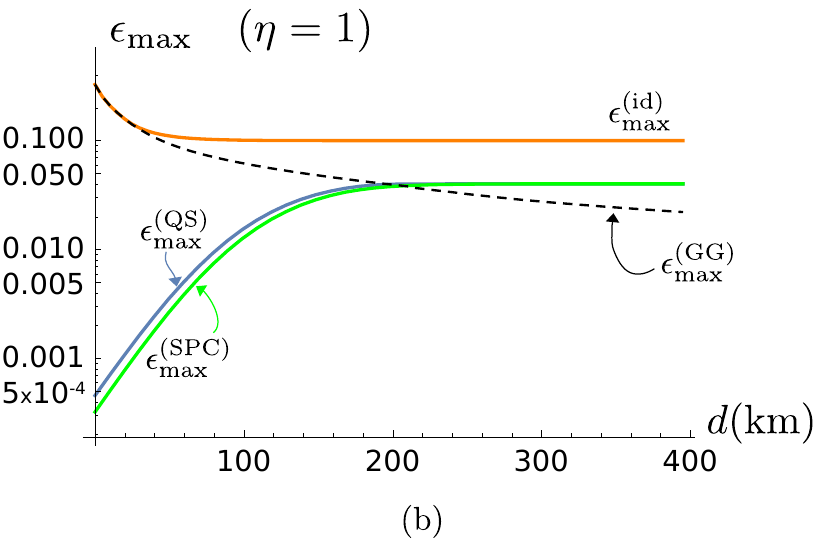}
\centering
\caption{(a) Log plot of the KGRs $K_{\p}$, $\p=\a,\b$, and $K_{\id}$ as a function of the distance $d$ in ${\rm km}$, for different values of the quantum efficiency $\eta$ and $\epsilon=0.03$ and with optimized gain $g$. The dashed line is the KGR of the original protocol and the upper line is the PLOB bound~(\ref{eq:PLOB}). (b) Log plot of the maximum tolerable excess noises $\epsilon_{\rm max}^{(\id)}$ and $\epsilon_{\rm max}^{(\p)}$, $\p=\a,\b$, as a function of the distance $d$ in ${\rm km}$, for $\eta=1$. $\epsilon_{\rm max}^{(\rm GG)}$ corresponds to the maximum tolerable excess noise of the original protocol. In both the pictures we set $\beta=0.95$.}\label{fig04:sec9.3.3_KGR-Opt}
\end{figure}

The situation is rather different if we can also optimize the gain $g$ associated with the NLAs, as reported in Fig.~\ref{fig04:sec9.3.3_KGR-Opt}(a).
Firstly, in the short-distance regime the physical NLAs still exhibits a threshold distance to obtain a positive KGR, differently from the ideal amplifier. 
Secondly, all the NLA-assisted protocols allow to reach arbitrary large distances, but the ideal amplifier outperforms the physical ones. As before, a quantum efficiency still rescales the KGR. However, differently from Sec.~\ref{sec: gFixed}, in the long distance regime $\kappa d \gg 1$, $K_{\a}$ and $K_{\b}$ are almost identical, proving SPC as a feasible alternative to QS. 
We also remark that in the long-distance regime both $K_{\id}$ and $K_{\p}$, $\p=\a,\b$, are proportional to the PLOB bound:
\begin{align}
K_{\rm PLOB}= - \log_2 \big[ (1-T) T^{\bar{n}_T} \big] - h(\bar{n}_T)\, ,
\end{align}
with $\bar{n}_T= T \epsilon/(2(1-T))$ and the $h$ function in Eq.~(\ref{eq:hfunc}), thus resulting in nearly optimal strategies.

\begin{figure}
\includegraphics[width=0.49\columnwidth]{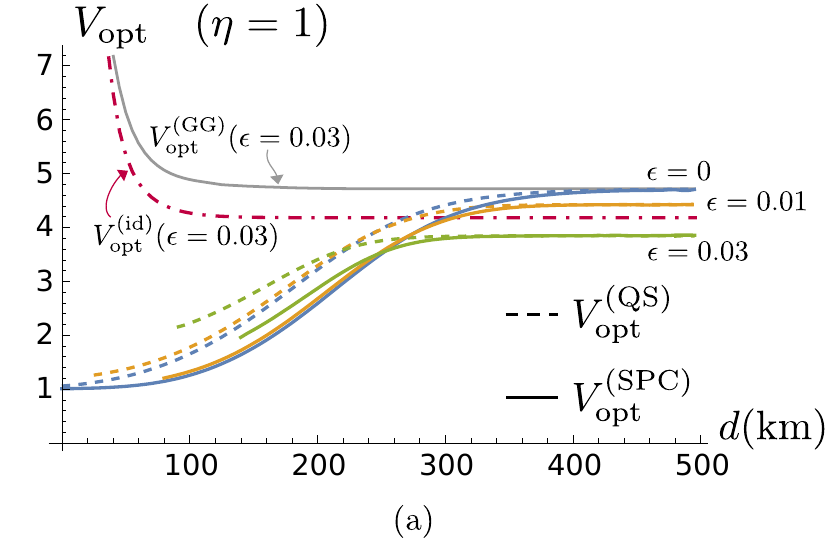} 
\includegraphics[width=0.49\columnwidth]{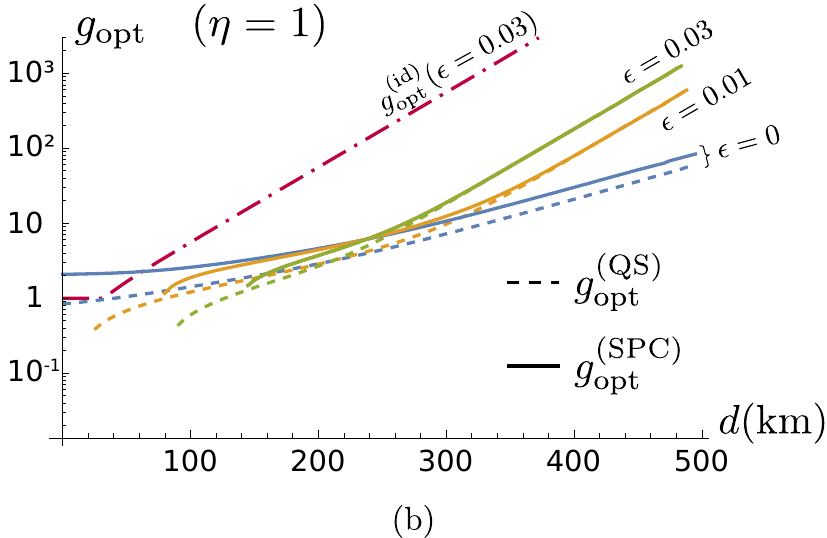}
\centering
\caption{(a) Plot of $V_{\rm opt}^{(\p)}$, $\p=\a,\b$, as a function of the distance $d$ in ${\rm km}$, for different values of excess noise $\epsilon$. The upper gray and the dash-dotted lines represent the optimized modulation for the original and the ideal NLA-assisted protocols, respectively, for $\epsilon=0.03$. (b) Log plot of $g_{\rm opt}^{(\p)}$, $\p=\a,\b$, as a function of the distance $d$ in ${\rm km}$, for different values of excess noise $\epsilon$. The plots have been performed only for the distances such that $K_{\p}>0$, $\p=\a,\b$. We set $\beta=0.95$ and $\eta=1$.}\label{fig05:sec9.3.3_OptValues}
\end{figure}

Furthermore, in Fig.~\ref{fig05:sec9.3.3_OptValues}(a) and (b) we report the optimized parameters $V_{\rm opt}^{(\p)}$ and $g_{\rm opt}^{(\p)}$, respectively. 
The modulation $V_{\rm opt}^{(\p)}$ has a different behavior with respect to Sec.~\ref{sec: gFixed}, being an $\epsilon$-dependent growing function of $d$. On the contrary, the modulations of the original and the ideal NLA-assisted protocols are decreasing functions of $d$ converging to an asymptotic value not depending on $\epsilon$, as for the case of fixed $g$.
Instead, the optimized gains $g_{\rm opt}^{(\id)}$ and $g_{\rm opt}^{(\p)}$ grow exponentially with $d$ in the long-distance regime. However, if $\epsilon=0$ this exponential scaling is not reached yet for the physical NLAs within the considered range of distances $d\le 500 \, \rm{km}$.

Finally, in Fig.~\ref{fig04:sec9.3.3_KGR-Opt}(b) we plot the MTENs as a function of $d$.
Differently from Sec.~\ref{sec: gFixed}, the MTEN associated with the physical NLAs, namely $\epsilon_{\rm max}^{(\p)}$, do not achieve the performance of the ideal one, $\epsilon_{\rm max}^{(\id)}$. Actually, both these MTENs outperform the original protocol and saturate to a value  $\epsilon_\infty$ as $\kappa d \gg1$. However, the saturation value of the physical NLAs, namely $\epsilon_\infty^{(\p)}\approx 0.04$, is lower than the ideal NLA one, that is $\epsilon_\infty^{(\id)} \approx 0.1$, see Fig.~\ref{fig04:sec9.3.3_KGR-Opt}(b). The numerical results also show that a quantum efficiency $\eta\le1$ does not affect the value of $\epsilon_\infty^{(\p)}$, consistently with the previous findings. 

\begin{figure}
\includegraphics[width=0.49\columnwidth]{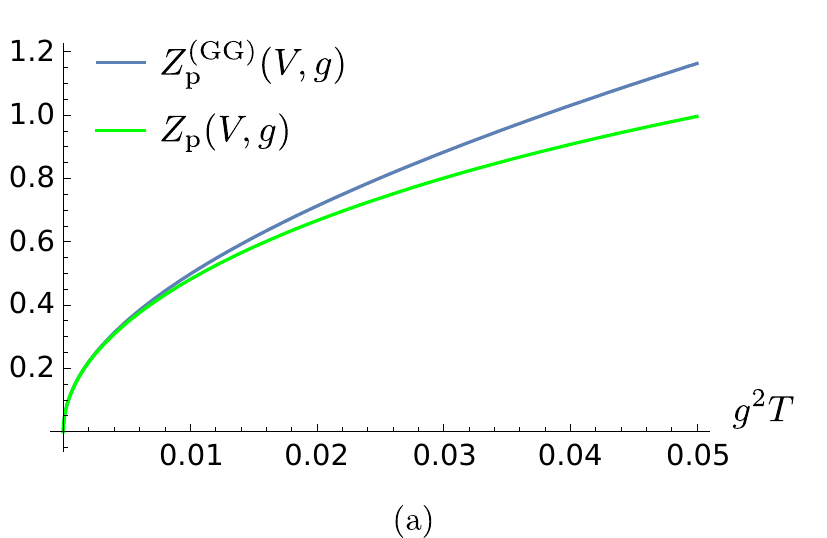} 
\includegraphics[width=0.49\columnwidth]{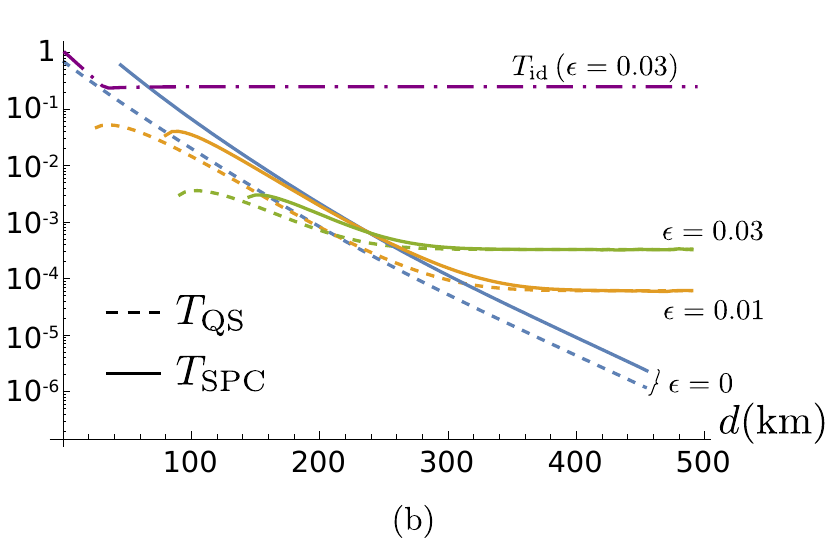}
\centering
\caption{(a) Plot of $Z_{\p}(V,g)$ and $Z_{\p}^{\rm (GG)}(V,g)$, $\p=\a,\b$, as a function of $g^2 T$ for $\epsilon=0.03$ and $V=4$. (b) Log plot of the effective transmissivity $T_{\p}$, $\p=\a,\b$, as a function of the distance $d$ in ${\rm km}$, for different values of excess noise $\epsilon$.  The plot have been performed only for the distances such that $K_{\p}>0$. In both the pictures we set $\beta=0.95$ and $\eta=1$.}\label{fig06:sec9.3.3_Teff}
\end{figure}

The difference between ideal and physical NLAs emerges by expanding the CM~(\ref{eq: GammaAB|p}) in the long-distance regime $T\ll 1$ up to the first order, keeping all the contributions of $O(g^2 T)$, due to the fact that $g_{\rm opt}^{(\p)} \gg 1$, and neglecting the other terms:
\begin{subequations}\label{eq: apprCM}
\begin{align}
V_{\p}(V,g) & \approx V + \delta V_{\p} \, ,  \\[1.ex]
W_{\p} (V,g)& \approx T_{\p} \big[V_{\p} (V,g) + \chi_{\p} \big] \, ,  \\[1.ex]
Z_{\p}(V,g) & \approx \frac{T_{\p}}{\sqrt{g^2 T}} \, Z \, , \qquad \p=\a,\b \, ,
\end{align}
\end{subequations}
where $\delta V_{\p} =T_{\p} Z^2/2$. $T_{\p}$ represents the effective transmissivity
\begin{align}
T_{\p}= \frac{g^2 \, T}{1+g^2 T (V+\epsilon-1)/2} \, , 
\end{align} 
while $\chi_{\p}= (1- T_{\p})/T_{\p} + \epsilon_{\p}$, with the effective excess noise 
\begin{align}
\epsilon_{\p}= \epsilon- \delta V_{\p} \, .
\end{align}
Employing a physical NLA is then equivalent to considering an effective channel of higher transmissivity $T_{\p}\ge T$ and lower excess noise $\epsilon_{\p}\le \epsilon$.
Nevertheless, the correspondence with a GG02 protocol does not occur anymore, as the correlation term $Z_{\p}(V,g)$ does not coincide with the one expected for a GG02 scheme, namely,
\begin{align}
Z_{\p}^{\rm (GG)}(V,g) = \sqrt{T_{\p} \, \big[V_{\p}(V,g)^2-1 \big]} \, ,
\end{align}
but rather:
\begin{align}
Z_{\p}(V,g)\le Z_{\p}^{\rm (GG)}(V,g) \, ,
\end{align}
as depicted in Fig.~\ref{fig06:sec9.3.3_Teff}(a). We have $Z_{\p}(V,g)\approx Z_{\p}^{\rm (GG)}(V,g)$ only if $g^2 T \ll 1$. As a consequence, the analogy with the ideal-NLA assisted protocol in Eq.~(\ref{eq: IdealPar}) is broken.

Now, the optimization procedure described above leads to exponential gains $g_{\rm opt}^{(\id)}$ and $g_{\rm opt}^{(\p)}$ for the ideal and physical NLAs, respectively, such that the product $g^2 T$ is kept constant for $\kappa d \gg 1$. Consequently, the effective transmissivities $T_{\id}$ and $T_{\p}$ saturate, as shown in Fig.~\ref{fig06:sec9.3.3_Teff}(b).
In turn, also the mutual information and the Holevo information saturate and the corresponding KGRs~(\ref{eq: KGR ideal Opt2}) and~(\ref{eq: K Opt}) turn out to be proportional only to the success probability of the NLAs, namely:
\begin{align}
K_{\id} \propto \frac{1}{\big(g_{\rm opt}^{(\id)}\big)^2} \propto T\, ,
\end{align}
and 
\begin{align}\label{eq: K asymp}
K_{\p} \propto P_{\p}
\approx \frac{\eta T }{2 T_{\p}} \bigg[ 1+ T_{\p}(V_{\p}+ \chi_{\p})\bigg]\, ,
\end{align}
with $P_{\p}=P_{\p}\big(V_{\rm opt}^{(\p)}, g_{\rm opt}^{(\p)}\big)$ and $V_{\p}= V_{\p}\big(V_{\rm opt}^{(\p)}, g_{\rm opt}^{(\p)}\big) $, decreasing linearly with $T$ and thus guaranteeing $K_{\p}>0$ for $\kappa d \gg 1$. The same linear scaling is achieved by the PLOB bound if $T \ll 1$:
\begin{align}
K_{\rm PLOB} \approx T \left\{\frac{2-\epsilon[1- \ln(\epsilon/2)]}{2 \ln 2} \right\}\, , 
\end{align} 
which proves both all the NLA-assisted protocols to be nearly optimal.
Furthermore, as in Sec.~\ref{sec: gFixed} a quantum efficiency $\eta\le 1$ only rescales the KGR and does not introduce any maximum transmission distance.

Moreover, the saturation value of $T_{\p}$ determines the difference between ideal and physical NLAs. Indeed, if $\epsilon_{\p}$ is small we have $T_{\p}\ll 1$ and the physical NLA-assisted protocols approximate a GG02 protocol with the effective channel parameters $T_{\p}$ and $\epsilon_{\p}$. By increasing the excess noise further, we have $T_{\p}\NOTll 1$ and $Z_{\p}(V,g)\le Z_{\p}^{\rm (GG)}(V,g) $, the state shared between Alice and Bob is less correlated and the protocol deviates more and more from GG02. This implies the reduced asymptotic maximum tolerable excess noise with respect to the ideal case.

\section{Enhancing CVQKD by non-Gaussian measurements}\label{chap:nonGauss}

\def\opt{{\rm KOR}}
\def\disp{{\rm QDF}}
\def\p{{\rm p}}
\def\Re{{\rm Re \, }}
\def\Im{{\rm Im \,}}

In this last Section, we proceed beyond the Gaussian CVQKD previously discussed, and make a first step towards the design of fully non-Gaussian protocols.
As a matter of fact, to date, all the proposed CVQKD schemes always assume Gaussian detection at the receiver, for both practical and theoretical reasons, whereas it has recently been proved that the secret key capacity provided by the PLOB bound, introduced in the previous Section, is achieved by a non-Gaussian measurement \cite{Lami2023}. Therefore, non-Gaussian CVQKD represents a new challenging topic, with unexplored potentialities. 
On the other hand, addressing security of protocols that do not exploit Gaussian detection is a nontrivial task, as, in this case, the Gaussian optimality theorem is no longer applicable, and more complex analyses have to be carried out, e.g. based on the semidefinite programming approach recently proposed by Lin {\it et al.} \cite{Lin2019, Lin2020}. 
In turn, all these considerations make non-Gaussian CVQKD both an attractive and unclear field of research.

In this Section we propose an example of CVQKD employing non-Gaussian detection. By drawing inspiration on the results of quantum state discrimination theory presented in Sec.~\ref{chap:MaryDisc}, we design an optimized state-discrimination receiver for the QPSK protocol, referred to as the key-rate optimized receiver (KOR) \cite{Notarnicola2023:KB}. For the sake of simplicity, we analyze security under a pure-loss wiretap channel, and compare the resulting KGR with that obtained from conventional double-homodyne detection, showing an enhancement in the metropolitan-network distance regime. We also consider the performance of a feasible scheme, namely the quaternary displacement feed-forward receiver (QDFFRE), obtaining an increase in the KGR with respect to Gaussian detection up to a maximum transmission distance.

The structure of the Section is the following. In Sec.~\ref{sec:10-nGQKD} we discuss the issues and the state of the art in non-Gaussian CVQKD. Thereafter, in Sec.~\ref{sec: Konrad} we perform explicit construction of the KOR for the QPSK protocol, discussing its security under the wiretap channel assumption, also comparing its performance with the QDFFRE. The results of the whole Section are original.

\subsection{Towards non-Gaussian CVQKD}\label{sec:10-nGQKD}

As we outlined throughout the previous Sections, several CVQKD protocols have been proposed in literature, employing either Gaussian modulation of coherent states \cite{Grosshans2002, Grosshans2003-1, Grosshans2003-2, Weedbrook2004, Grosshans2005} or discrete modulation formats \cite{Leverrier2009, Sych2010, Leverrier2011, Becir2012, Ghorai2019, Lin2019, Liao2020, Denys2021, Upadhyaya2021, Liu2021, Lupo2022, Notarnicola2022}.
All these schemes share a common feature: they assume Gaussian detection at Bob's side, either homodyne or double homodyne (DH).
This is mainly due for a twofold reason. The former, more practical, is that quadrature detection provides a simple and feasible solution for experimental implementations being large-scale applicable, as it is commonly adopted in the state-of-the-art communication systems at telecom wavelength \cite{Lodewyck2005, Lodewyck2007, Fossier2009}. The latter, more theoretical, is that Gaussian detection guarantees unconditional security proofs, thanks to the resort to the optimality of Gaussian attacks \cite{Navascues2006, LeverrierThesis, GarciaPatron2006}, whilst, only recently, a tight lower bound to the Devetak-Winter (DW) bound, not invoking Gaussian optimality, has been obtained \cite{Lin2019}.
However, the fundamental limitations of Gaussian CVQKD have been recently established in \cite{Lami2023}, proving a gap between the key generation rate (KGR) achievable with Gaussian operations and the secret-key channel capacity provided by the PLOB bound.
In turn, non-Gaussianity becomes necessary to enhance quantum secure communications and close the gap with respect to the PLOB bound.

These results suggest non-Gaussian measurements as a potential resource for CVQKD, also considering that in many other frameworks, they often outperform Gaussian ones. For instance, in the transmission of classical information, the classical capacity of a lossy bosonic channel is enhanced, in particular energy regimes, by resorting to photon-number resolving (PNR) detection \cite{Martinez2007, Cheraghchi2019, Lukanowski2021} and weak-field homodyne measurement \cite{Notarnicola2024:JASC}. Furthermore, in quantum state discrimination, the quantum receivers associated with the standard quantum limit (SQL), based on Gaussian detection, do not achieve the minimum decision error probability, as widely discussed in Sec.s~\ref{chap:GeneralFeatures} and~\ref{chap:MaryDisc}.
Nevertheless, addressing unconditional security of a CVQKD scheme employing non-Gaussian measurements is an open problem since the Gaussian optimality theorem does not hold anymore, and the DW bound can only be directly evaluated with the advanced methods presented in \cite{Lin2019}. Therefore, to date, the main results in the field of non-Gaussian key distribution have been limited to restricted eavesdropping settings.
In particular, Cattaneo {\em et al.} showed that weak-field homodyne detection enhances a Gaussian modulation protocol, in the presence of a quantum pure-loss wiretap channel under both individual and collective attacks, with higher increase in the KGR for high channel transmissivity, corresponding to short-distance communications \cite{Cattaneo2017}.

Following the same philosophy, we now extend the analysis in this direction and address a complementary problem, namely to design an optimized state-discrimination receiver for discrete modulation  CVQKD, by exploiting the characterization outlined in Sec.s~\ref{chap:MaryDisc}. In fact, in the present literature, only the receivers achieving the SQL have been investigated for CVQKD \cite{Liao2018}, leaving the open problem of designing a genuine quantum receiver maximizing the KGR of a $M$-ary protocol. This also raises the question about the compatibility between optimum discrimination and maximum secret-key distillation, that is whether or not the POVM minimizing the decision error probability coincides with that maximizing the KGR.

In light of this, in the following we consider a quadrature phase-shift keying (QPSK) protocol employing a quantum discrimination receiver in place of Gaussian detection and show that, in particular conditions, it is possible to theoretically design a suitable measurement outperforming the conventional quadrature detection schemes. Moreover, as in \cite{Cattaneo2017}, we investigate security under a quantum wiretap channel, where Eve may only collect the lost fraction of the encoded signals without performing arbitrary channel manipulation. 

\subsection{The QPSK protocol with state-discrimination receivers}\label{sec: Konrad}

\begin{figure}
\includegraphics[width=0.7\columnwidth]{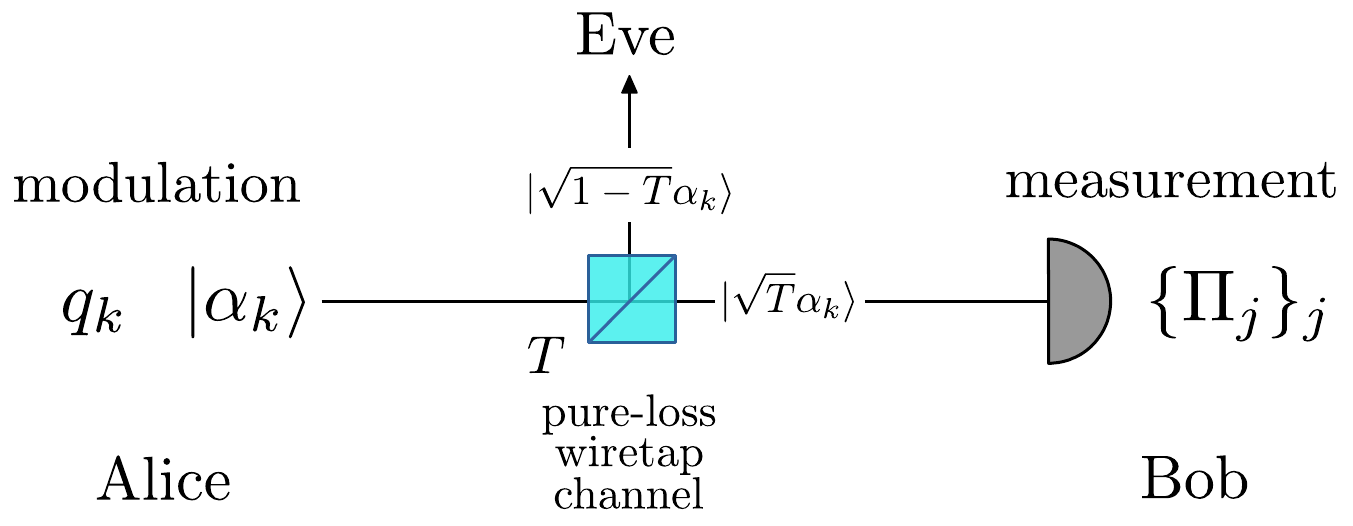}
\centering
\caption{Scheme of the QPSK protocol employing state-discrimination receivers. Alice generates one of the coherent states $|\alpha_k\rangle$, $k=0,\ldots, M-1$, with uniform probability $q_k=1/M$ and sends it via the quantum wiretap channel to Bob, who performs the finite-valued POVM $\{\Pi_j\}_j$, $j=0,\ldots, M-1$.}
\label{fig01:sec10.2.0_Protocol}
\end{figure}

Here, we investigate the potentiality of state-discrimination receivers for secure quantum communications and propose a new quantum receiver, the {\it key-rate optimized receiver} (KOR), for CVQKD protocols employing discrete modulation. In particular, we consider a QPSK protocol in which Bob implements the KOR rather than a Gaussian measurement \cite{Notarnicola2023:KB}. We compute the KGR under the wiretap channel assumption and,  for the sake of simplicity, we consider a pure-loss channel, in order to deal with discrimination of pure states at Bob's side. This latter assumption depicts a simplified scenario, providing a cornerstone fostering more advanced developments. However, it is still worth of investigation for a twofold reason. At first, in the recent literature there has been a revived interest in passive eavesdropping strategies \cite{Banaszek2021, Derkach2021, Kundu2023, Medlock2021, Ghalaii2023, Jarzyna2023}. Secondly, passive eavesdropping can be used as the first stepping stone to identify scenarios where a potential advantage of quantum receivers may be substantial, even in the possible presence of nonzero excess noise.
Starting from the general structure of quantum receivers derived in Sec.s~\ref{sec:M=GammaA} and~\ref{sec:OptwithGUS}, we design the KOR to maximize the 
KGR of the addressed protocol and compare it with the pretty good measurement (PGM), being the POVM minimizing the decision error probability for discrimination, showing that both these non-Gaussian measurements improve the KGR in the metropolitan-network distance regime with respect to the conventional Gaussian measurement scheme. Finally, we also investigate the performance of some feasible quantum receivers, by considering the quaternary displacement feed-forward receiver (QDFFRE) introduced in Sec.~\ref{sec:QDFFRE}.

Given this outline, the protocol under investigation is reported in Fig.~\ref{fig01:sec10.2.0_Protocol}. The sender, Alice, employs the QPSK modulation, that is she generates one of the $M=4$ coherent states:
\begin{align}
|\alpha_k\rangle= |\alpha \, e^{i \pi (2k+1)/M} \rangle \, , \qquad k=0,\ldots, M-1 \, , 
\end{align}
where $\alpha\ge0$, sampled with equal a priori probabilities $q_k=1/M$. We remind that the QPSK constellation satisfies the geometrically uniform symmetry (GUS) for the phase-shift symmetry operator $S_\theta= \exp(- i \, \theta \, \hat{n})$, with $\theta=2\pi/M$ and $\hat{n}$ being the photon-number quantum operator \cite{Cariolaro2015}.
After the modulation stage, Alice injects the signals into an untrusted pure-loss wiretap channel with transmissivity:
\begin{align}
T= 10^{-\kappa d/10} \,,
\end{align}
$d$ being the transmission distance (expressed in km) and $\kappa=0.2$ dB/km is the loss rate of common fibers at telecom wavelength \cite{Lodewyck2005,Lodewyck2007, Banaszek2020, Agrawal2002, Jouguet2013}. Then, the transmitted fraction $|\alpha_k^{(t)}\rangle = |\sqrt{T} \alpha_k\rangle$ reaches the receiver, Bob, whereas the eavesdropper, Eve, receives the reflected part $|\alpha_k^{(r)}\rangle = |\sqrt{1-T} \alpha_k\rangle$.
Ultimately, Bob, probes the rescaled constellation $\{|\alpha_k^{(t)}\rangle \}_k$ via a state-discrimination receiver, namely, a finite-valued POVM $\{\Pi_j\}_j$, $j=0,\ldots, M-1$, defined in the $M$-dimensional subspace $\cal S$ spanned by the transmitted pulses and satisfying the properties described in Sec.~\ref{sec:PureMarydiscr}. 
Actually, we note that, to form a truly identity-resolving set, the POVM elements $\{\Pi_j\}_j$, should be complemented with an $(M+1)$-th inconclusive element $\Pi_M= \hat{\Id}- \mathbb{P}_{\cal S}$, $\mathbb{P}_{\cal S}$ being the projection operator onto subspace $\cal S$. However, for the case under investigation, $\Pi_M$ is irrelevant, and we will neglect it in the following. This would not be true anymore in the presence of a channel excess noise; thus, registering a zero probability for this additional outcome may provide a useful way to check the reasonableness of the pure-loss hypothesis in a realistic implementation of the proposed protocol.

Provided these two assumptions, in the following we construct the optimized POVM that describes the KOR, and show it to bring advantages in some particular regimes.

\subsubsection{Construction of the key-rate optimized receiver}\label{sec:POVM}
In our protocol Bob should employ an optimized POVM to perform discrimination among the transmitted pulses, described by the state vector $\Gamma = (|\alpha_0^{(t)}\rangle, \ldots, |\alpha_{M-1}^{(t)}\rangle)$ and the Gram matrix:
\begin{align}\label{eq:GramT}
G=\Big(\big\langle \alpha^{(t)}_l \big| \alpha^{(t)}_k \big\rangle \Big)_{l,k=0,\ldots,M-1} \, ,
\end{align}
see Sec.~\ref{sec:M=GammaA}, in which the overlap $G_{lk}=\langle \alpha^{(t)}_l \big| \alpha^{(t)}_k \big\rangle$ reads \cite{Serafini2017}:
\begin{align}
G_{lk}&= \exp\left\{-\frac12 \left|\alpha^{(t)}_k -\alpha^{(t)}_l \right|^2 + \frac12 \left[ \alpha^{(t)}_k  \left(\alpha^{(t)}_l\right)^*-\left(\alpha^{(t)}_k\right)^* \alpha^{(t)}_l \right] \right\} \nonumber \\[1.5ex]
&= \exp \left(-T \alpha^2 \left\{1-\cos\left[\frac{2\pi}{M}(k-l)\right] \right\} + i \, T \alpha^2 \sin\left[\frac{2\pi}{M}(k-l)\right] \right)\, .
\end{align}

The constellation of transmitted pulses maintains the GUS for the phase-shift operator $S_\theta$, thus the set of measurement vectors $\mathbb{M}=~\{|\mu_j\rangle\}_j$, $j=0,\ldots,M-1$, with $\mathbb{M}= \Gamma A$, also satisfies the GUS, and the corresponding matrix $A$ is in the form~(\ref{eq:Aphi}), according to the results of Sec.~\ref{sec:OptwithGUS}. That is, $A \equiv A_{\boldsymbol{\phi}}=\mathbb{U} \, \Lambda_A^{(\boldsymbol{\phi})} \, \mathbb{U}\dag$, $\mathbb{U}$ being the inverse discrete Fourier transform matrix and
\begin{align}
\Lambda_A^{(\boldsymbol{\phi})}= {\rm diag}\Bigg(\bigg\{\lambda_j^{(\boldsymbol{\phi})} \bigg\}_{j=0,\ldots, M-1}\Bigg) \, ,
\end{align}
where:
\begin{align}
\lambda_j^{(\boldsymbol{\phi})} = e^{i \phi_j} g_j^{-1/2} \, ,
\end{align}
$\{g_j\}_j$ being the eigenvalues of the Gram matrix~(\ref{eq:GramT}) and $\boldsymbol{\phi}$ being the array of phases fully characterizing the receiver.
Given this structure, the KOR is obtained by optimizing the phase array $\boldsymbol{\phi}$ to maximize the KGR, and is described by the optimized ``reference" measurement vector:
\begin{align}\label{eq:mu0phi}
|\mu_0^{(\boldsymbol{\phi})}\rangle &= \sum_{k=0}^{M-1} \left(A_{\boldsymbol{\phi}}\right)_{k0} \, \left|\alpha_k^{(t)}\right\rangle \\
&= e^{-T \alpha^2/2} \sum_{n=0}^{\infty} \frac{\left(\sqrt{T}\alpha_0\right)^n}{\sqrt{n!}} \,  \lambda_{(n-1) \bmod M}^{(\boldsymbol{\phi})} \, |n\rangle\, ,
\end{align}
where $a \bmod b$ is the modulo operation, returning the reminder of the division $a/b$, $a,b\in \mathbb{Z}$, and $\{|n\rangle\}_n$ is the photon-number basis. In turn, the other measurement vectors are obtained as $|\mu_j^{(\boldsymbol{\phi})}\rangle= (S_\theta)^j |\mu_0^{(\boldsymbol{\phi})}\rangle$, $j=0,\ldots, M-1$.

Given the previous considerations, we compute the KGR of the discussed protocol, considering a reverse reconciliation scenario \cite{Grosshans2002, Grosshans2003-1, Grosshans2003-2}. 
Moreover, for the sake of simplicity, we perform the asymptotic key-rate calculation, where the channel parameters are known with no uncertainty. Under this paradigm, for a generic state-discrimination receiver described by the phase vector $\boldsymbol{\phi}$, the KGR reads:
\begin{align}\label{eq:KGRdisc}
K(\boldsymbol{\phi},\alpha^2)= \beta I_{AB}(\boldsymbol{\phi},\alpha^2)- \chi_{BE}(\boldsymbol{\phi},\alpha^2)\, ,
\end{align}
where $I_{AB}$ and $\chi_{BE}$ are the mutual information between Alice and Bob and the Holevo information \cite{Holevo1998} between Bob and Eve, respectively, and $\beta\le 1$ is the reconciliation efficiency \cite{Leverrier2009, Leverrier2011}.

The mutual information reads:
\begin{align}\label{eq:IABdisc}
I_{AB}(\boldsymbol{\phi},\alpha^2)= H\left[p_B^{(\boldsymbol{\phi})}(j)\right] - \frac{1}{M} \sum_{k=0}^{M-1} H\left[p_{B|\alpha_k}^{(\boldsymbol{\phi})}(j)\right] \, ,
\end{align}
where
\begin{align}\label{eq:pB|A}
p_{B|\alpha_k}^{(\boldsymbol{\phi})}(j) 
= \left\langle \sqrt{T} \alpha_k \left|  \Pi_j \right| \sqrt{T} \alpha_k \right\rangle 
= \left|\left(A_{\boldsymbol{\phi}}\dag G\right)_{kj}\right|^2 \, ,
\end{align}
and
\begin{align}\label{eq:pB}
p_B^{(\boldsymbol{\phi})}(j) =\frac{1}{M} \sum_{k=0}^{M-1} p_{B|\alpha_k}^{(\boldsymbol{\phi})}(j) \, ,
\end{align}
are the conditional and overall probabilities of Bob's detection associated with outcome $j=0,\ldots,M-1$, respectively, and $H[p(x)]= -\sum_x p(x) \log_2 p(x)$ is the Shannon entropy of the probability distribution $p(x)$.

To compute the Holevo information shared between Bob and Eve, that is the maximum amount of information accessible to Eve, we approach the problem in the prepare-\&-measure picture \cite{Laudenbach2018, Grosshans2002, Grosshans2003-1} and obtain:
\begin{align}\label{eq:chiBEdisc}
\chi_{BE}(\boldsymbol{\phi},\alpha^2)= S\left[\rho_E \right] - \sum_{j=0}^{M-1}  p_B^{(\boldsymbol{\phi})}(j) \, S\left[\rho_{E|j}^{(\boldsymbol{\phi})}\right] \, ,
\end{align}
where $\rho_{E|j}^{(\boldsymbol{\phi})}$ and $\rho_E$ are the conditional and overall Eve’s state, respectively, $p_B^{(\boldsymbol{\phi})}(j)$ is Bob’s probability distribution~(\ref{eq:pB}) and $S[\rho]= - \Tr[\rho \log_2 \rho]$ represents the von Neumann entropy associated with state $\rho$.
These two states may be retrieved from the joint state of Bob and Eve after the channel, that is:
\begin{align}
\rho_{BE}&=U_{\rm BS}(T) \, \rho_A \otimes |0\rangle \langle 0 | \, U_{\rm BS}(T)\dag \nonumber \\[1ex]
&= \frac{1}{M} \sum_{k=0}^{M-1} \left|\alpha_k^{(t)} \right\rangle\left \langle \alpha_k^{(t)} \right| \otimes  \left|\alpha_k^{(r)} \right\rangle \left\langle \alpha_k^{(r)} \right| \, ,
\end{align}
where $\rho_A= \sum_k q_k|\alpha_k\rangle \langle \alpha_k|$ is Alice's overall state, $|0\rangle$ is the vacuum state and $U_{\rm BS}(T)$ is unitary operator associated with a beam splitter with transmissivity $T$ \cite{Olivares2021}, as displayed in Fig.~\ref{fig01:sec10.2.0_Protocol}. In turn, we have:
\begin{align}\label{eq:rhoEdisc}
\rho_E= \Tr_B\left[\rho_{BE}\right] = \frac{1}{M} \sum_{k=0}^{M-1} \left|\sqrt{1-T} \alpha_k \right\rangle \left\langle \sqrt{1-T} \alpha_k \right| \, ,
\end{align}
and
\begin{align}\label{eq:rhoEdisccondB}
\rho_{E|j}^{(\boldsymbol{\phi})}&= \frac{1}{p_B^{(\boldsymbol{\phi})}(j)} \Tr_B\left[\rho_{BE} \, \Pi_j \otimes \hat{\Id}_E\right] \nonumber \\[1ex]
&= \frac{1}{M p_B^{(\boldsymbol{\phi})}(j)} \sum_{k=0}^{M-1} \, p_{B|\alpha_k}^{(\boldsymbol{\phi})}(j) \, \left|\sqrt{1-T} \alpha_k \right\rangle \left\langle \sqrt{1-T} \alpha_k \right| \, ,
\end{align}
$\Tr_B$ being the partial trace over Bob's mode and $\hat{\Id}_E$ being the identity operator over Eve's mode.
Finally, the von Neumann entropy of states~(\ref{eq:rhoEdisc}) and~(\ref{eq:rhoEdisccondB}) may be computed as follows.
Both of them are expressed in the form:
\begin{align}\label{eq:state}
\varrho = \sum_{k=0}^{M-1} c_k \, |\alpha_k^{(r)}\rangle \langle \alpha_k^{(r)}| \, ,
\end{align}
for some coefficients $c_k \in\mathbb{C}$. To compute the associated entropy we need to diagonalize~(\ref{eq:state}). If $|\psi\rangle$ is the eigenvector of $\varrho$ associated with eigenvalue $\omega$, we have $|\psi\rangle = \sum_m b_{m} |\alpha_m^{(r)}\rangle$ and the following chain of equations holds:
\begin{align}
&\varrho |\psi\rangle = \omega |\psi\rangle  \nonumber \\[1ex]
&\Bigg( \sum_k c_k |\alpha_k^{(r)}\rangle \langle \alpha_k^{(r)}| \Bigg) \sum_m b_{m} |\alpha_m^{(r)}\rangle =  \omega  \sum_s b_{s} |\alpha_s^{(r)}\rangle  \nonumber \\[1ex]
&\sum_k  c_k \Bigg( \sum_m  \mathbb{G}_{km} b_{m}   \Bigg) |\alpha_k^{(r)}\rangle =  \omega  \sum_s b_{s} |\alpha_s^{(r)}\rangle \, ,
\end{align}
where $\mathbb{G}_{km}= \langle \alpha_k^{(r)}|\alpha_m^{(r)}\rangle$.

As a consequence, we obtain the set of equations:
\begin{align}
\omega \, b_{k} = c_k \Bigg( \sum_{m=0}^{M-1}  \mathbb{G}_{km} b_{m}  \Bigg) \, , \quad k=0,\ldots, M-1 \, ,
\end{align}
or, equivalently,
\begin{align}
\Bigg(\frac{\omega}{c_k}-1 \Bigg) b_{k} -  \sum_{m\neq k} \mathbb{G}_{km} b_{m} = 0  \, .
\end{align}
This defines the homogeneous linear system $M {\bf b}= 0$, where ${\bf b}=(b_0,\ldots, b_{M-1})$ and
\begin{align}
M= 
\begin{pmatrix} 
\frac{\omega}{c_0}-1 & -\mathbb{G}_{01} & -\mathbb{G}_{02} & -\mathbb{G}_{03} \\
-\mathbb{G}_{10} &\frac{\omega}{c_1}-1 &  -\mathbb{G}_{12} & -\mathbb{G}_{13} \\
-\mathbb{G}_{20} & -\mathbb{G}_{21} & \frac{\omega}{c_2}-1 & -\mathbb{G}_{23} \\
-\mathbb{G}_{30} & -\mathbb{G}_{31} & -\mathbb{G}_{32} & \frac{\omega}{c_3}-1 \\
\end{pmatrix} \, .
\end{align}
The equation $M {\bf b}= 0$ always admits a trivial solution ${\bf b}= 0$, therefore to obtain a nonzero eigenvector we shall impose the condition $\det M = 0$. This provides us with the four eigenvalues $\{ \omega_j \}_j$ and the corresponding von Neumann entropy $S[\varrho] =-\sum_j \omega_j \log_2 \omega_j$.
For state $\rho_E$ in~(\ref{eq:rhoEdisc}), for which $c_k= M^{-1}$, the equation $\det M = 0$ may be solved analytically, leading to:
\begin{align}
\omega_{0(1)}&= \frac{e^{-(1-T) \alpha^2}}{2} \Bigg\{\cosh \Big[(1-T) \alpha^2\Big] \pm \cos\Big[(1-T) \alpha^2\Big]\Bigg\} \, , \nonumber \\[1.5ex]
\omega_{2(3)}&= \frac{e^{-(1-T) \alpha^2}}{2} \Bigg\{ \sinh \Big[(1-T) \alpha^2\Big] \pm \Big|\sin\Big[(1-T) \alpha^2\Big]\Big|\Bigg\} \, .
\end{align}

In our analysis, we are interested in the maximum achievable KGR as a function of the transmission distance $d$, therefore, in the end we will perform optimization over the free parameters, namely the phases $\boldsymbol{\phi}$ and the constellation energy $\alpha^2$. The final, optimized, KGR is therefore equal to
\begin{align}\label{eq:Kopt}
K_\opt= \max_{\boldsymbol{\phi},\, \alpha^2} \, K(\boldsymbol{\phi},\, \alpha^2) \, ,
\end{align}
with the optimized phases and modulation energy denoted by $\boldsymbol{\phi}_\opt= (0, \phi_1^{(\opt)},\ldots, \phi_{M-1}^{(\opt)})$ and $\alpha^2_\opt$, reminding that, thanks to the GUS, we may set $\phi_0^{(\opt)}=0$.
As a consequence, we define the KOR via Eq.~(\ref{eq:Aphi}), as the quantum receiver associated with the optimized phase vector $\boldsymbol{\phi}_\opt$. Furthermore, we compare the performance of the KOR with that associated with the PGM, defined via Eq.~(\ref{eq:Aphi}) with the choice $\boldsymbol{\phi}_\PGM={\bf 0}$, for which the optimized KGR reads:
\begin{align}\label{eq:KPGM}
K_\PGM= \max_{\alpha^2} \,  K(\boldsymbol{\phi}={\bf 0},\, \alpha^2) \, ,
\end{align}
with the optimized energy $\alpha^2_\PGM$. The results obtained for the above two receivers are discussed in the following section, where we compare both KGRs with the key rate of the DH protocol in order to highlight the advantages brought by the two non-Gaussian measurements.

The DH protocol is analogous to the one discussed above and employs DH detection at Bob's side, that is, a measurement of both field quadratures $q$ and $p$, retrieving a pair of real outcomes ${\bf x}=(x_B,y_B)\in\mathbb{R}^2$. 
We underline that, while both the KOR and the PGM are described in terms of a finite-valued POVM with $M$ possible outcomes, in the presence of heterodyne detection we have a continuous-variable measurement.
Therefore, in this case Bob's conditional probability reads:
\begin{align}
p_{B|\alpha_k}^{(\HET)}({\bf x}) 
= \frac{1}{4\pi \sigma_0^2} 
& \exp\left\{-\left[x_B-2\sigma_0 \sqrt{T}\,  \Re(\alpha_k)\right]^2/(4\sigma_0^2) \right\} \times \nonumber \\[.5ex]
& \exp\left\{-\left[y_B-2\sigma_0 \sqrt{T} \, \Im(\alpha_k)\right]^2/(4\sigma_0^2)\right\} \, ,
\end{align}
$\sigma_0^2$ being the shot-noise variance \cite{Olivares2021}, which from now on will be taken equal to $1$, performing calculations in shot-noise units (SNU).
Similarly to~(\ref{eq:IABdisc}), the obtained mutual information is given by:
\begin{align}
I_{AB}^{(\HET)} (\alpha^2)= H\left[p_B^{(\HET)} ({\bf x})\right] - \frac{1}{M} \sum_{k=0}^{M-1} H\left[p_{B|\alpha_k}^{(\HET)} ({\bf x})\right] \, ,
\end{align}
with $p_B^{(\HET)} ({\bf x})= M^{-1} \sum_k p_{B|\alpha_k}^{(\HET)}({\bf x})$. Instead, the Holevo information becomes:
\begin{align}\label{eq:holevo_het}
\chi_{BE}(\alpha^2)= S\left[\rho_E \right] - \int_{\mathbb{R}^2} d^2{\bf x} \,\, p_{B|\alpha_k}^{(\HET)}({\bf x}) \,\, S\left[\rho_{E|{\bf x}}\right] \, ,
\end{align}
with $\rho_E$ given in~(\ref{eq:rhoEdisc}) and
\begin{align}
\rho_{E|{\bf x}}
=& \frac{1}{M p_B^{(\HET)}({\bf x})} \,\, \sum_{k=0}^{M-1} p_{B|\alpha_k}^{(\HET)}({\bf x}) \,\, \left|\sqrt{1-T} \alpha_k \right\rangle \left\langle \sqrt{1-T} \alpha_k \right| \, .
\end{align}
The integration in (\ref{eq:holevo_het}) can been performed numerically by exploiting the Simpson's rule \cite{Press2007}.
Finally, the resulting 
KGR is obtained as
\begin{align}\label{eq:Khet}
K_\HET= \max_{\alpha^2} \, \left[\beta I_{AB}^{(\HET)}(\alpha^2) -\chi_{BE}^{(\HET)}(\alpha^2) \right]\, ,
\end{align}
with the optimized modulation energy $\alpha^2_\HET$.

As a final remark, towards a realistic implementation of the present protocol, we underline that both the KOR and the PGM may not represent appropriate POVMs for the channel evaluation stage. Indeed, assuming that the channel properties do not change, Alice and Bob must estimate the channel parameters, which in the present case is limited to the sole transmissivity $T$. However, unlike homodyne and DH detection, in principle the designed POVM $\{\Pi_j\}_j$ does not guarantee full channel characterization. This problem may be circumvented, at least for the asymptotic key rate calculation, 
by performing Gaussian detection on a small fraction of the exchanged pulses and reserving it for the channel estimation stage, whilst exploiting the non-Gaussian receiver only for the key extraction.
On the contrary, in the presence of a finite-size scenario, Alice and Bob estimate the channel transmissivity $T$ with a finite uncertainty $\Delta T$, thus leaving more space for Eve's intervention. Therefore, they employ a conservative strategy and compute the KGR by considering a lower value of the transmissivity, namely $T-\Delta T$. However, the main effect of this lower effective transmissivity is to reduce the range of distances for which the state-discrimination receivers outperforms the heterodyne protocol. Furthermore, the dataset for the key extraction is also finite, resulting in a lower KGR with respect to the asymptotic case.

\begin{figure}
\includegraphics[width=0.49\columnwidth]{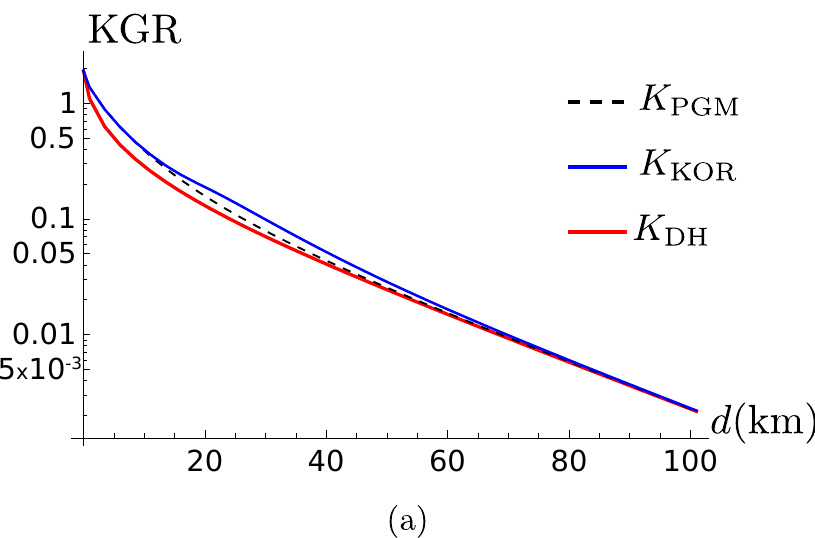} 
\includegraphics[width=0.49\columnwidth]{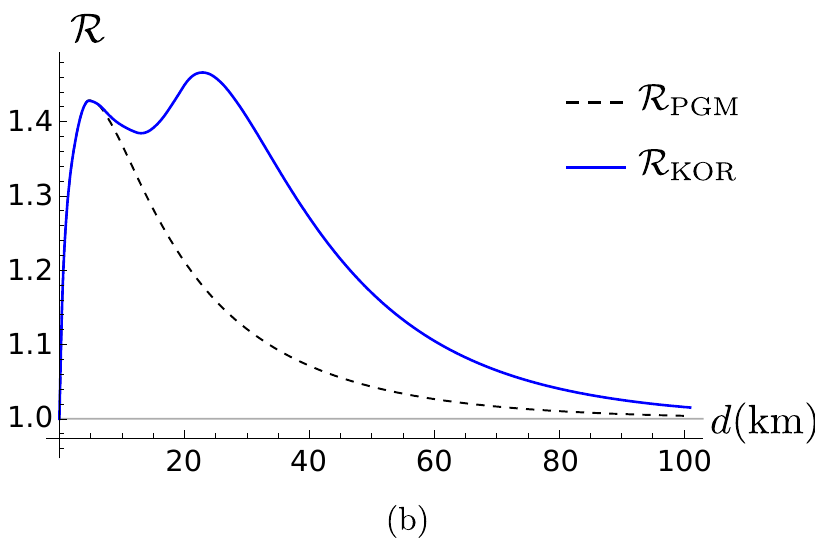}
\centering
\caption{(a) Log plot of $K_{\p}$, $\p=\PGM,\opt$, compared to $K_{\HET}$, as a function of the transmission distance $d$ in km. (b) Plot of the ratio ${\cal R}_{\p}$, $\p=\PGM,\opt$, as a function of the transmission distance $d$. State-discrimination receivers improve the KGR with respect to the DH protocol in the regime $d \le 100$~km. In both pictures we set $\beta=0.95$.}\label{fig01:sec10.2.1_KGR}
\end{figure}

\paragraph{Results.} We now compare the results derived previously. In Fig.~\ref{fig01:sec10.2.1_KGR}(a) we plot the 
KGRs~(\ref{eq:Kopt}),~(\ref{eq:KPGM}) and~(\ref{eq:Khet}) as a function of the transmission distance $d$, expressed in km. The reconciliation efficiency is fixed to $\beta=0.95$ \cite{Lodewyck2005, Lodewyck2007, Denys2021}. We see that both PGM and KOR beat the DH protocol, that is $K_{\p} \ge K_{\HET}$, $\p=\PGM,\opt$. The improvement in the 
KGR is more relevant for metropolitan-network distances, in particular for $d\le 100$~km, whereas for larger ones both $K_{\opt}$ and $K_{\PGM}$ approach $K_{\HET}$ and achieve the same asymptotic scaling. To quantify this improvement we compute the ratio
\begin{align}
{\cal R}_{\p} = \frac{K_{\p}}{K_{\HET}} \, , \qquad \p=\PGM,\opt \, ,
\end{align}
reported in Fig.~\ref{fig01:sec10.2.1_KGR}(b). Both the ratios exhibit peaks for $d\le 40$~km and then decrease towards~$1$ in the long-distance regime, but the behaviour is rather different between the two cases. In fact, ${\cal R}_{\PGM}$ achieves a single maximum at $\approx 5$~km, increasing the 
KGR with respect to $K_{\HET}$ by more than $42 \%$, and then decays monotonously to~$1$. On the contrary, $K_{\opt}$ is not a monotonic function of the transmission distance and, in turn, the associated ratio exhibits two separated peaks. The KOR coincides with the PGM up to its first maximum, that is ${\cal R}_{\opt}={\cal R}_{\PGM}$ for $d \lesssim 7$~km, while for larger $d$ we have ${\cal R}_{\opt}\ge{\cal R}_{\PGM}$. Thereafter, ${\cal R}_{\opt}$ reaches a local minimum and then achieves a second maximum at $\approx 23$~km, with $\approx 47 \%$ increase in the 
KGR. Ultimately, the curve decreases to $1$, approaching the DH protocol together with ${\cal R}_{\PGM}$.

\begin{figure}
\includegraphics[width=0.49\columnwidth]{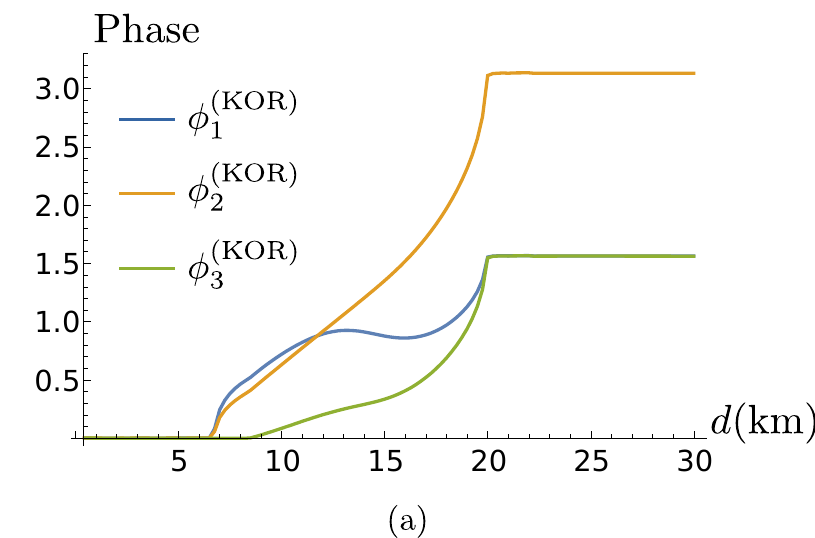} 
\includegraphics[width=0.49\columnwidth]{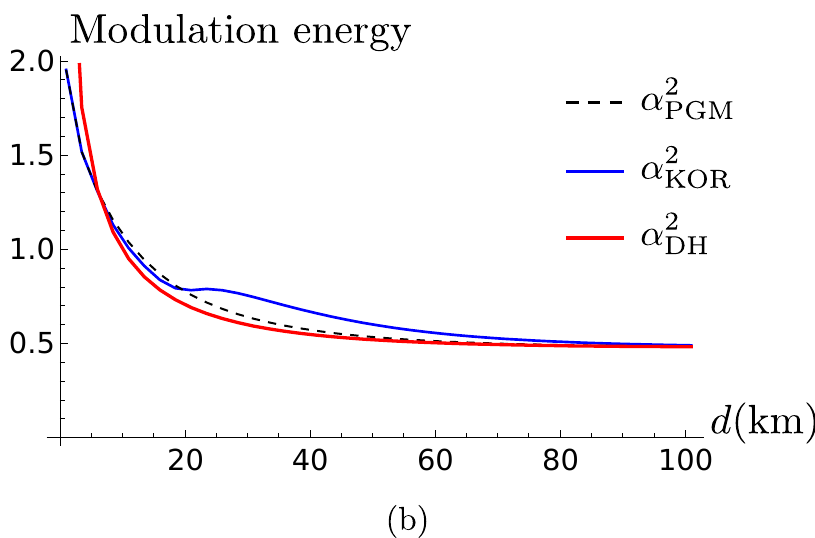}
\centering
\caption{(a) Plot of the optimized phases $\phi_j^{(\opt)}$, $j=1,\ldots,3$, as a function of the transmission distance $d$ in km. We recall that $\phi_0^{(\opt)}=0$. (b) Plot of the optimized modulation energies $\alpha^2_{\p}$, $\p=\PGM,\opt$, and $\alpha^2_{\HET}$, as a function of the transmission distance $d$.  In both pictures we set $\beta=0.95$.}\label{fig02:sec10.2.1_OptPar}
\end{figure}

The behavior of $K_{\opt}$ is a consequence of the resulting optimized phases $\phi_j^{(\opt)}$, depicted in Fig.~\ref{fig02:sec10.2.1_OptPar}(a). We recall that $\phi_0^{(\opt)}=0$ by definition. For $d \lesssim 7$~km we have $\boldsymbol{\phi}_{\opt}={\bf 0}$ and the optimized receiver is identical to the PGM, whereas for larger distances the optimized phases are nonzero and ${\cal R}_{\opt}\ge{\cal R}_{\PGM}$. Interestingly, for $d \gtrsim 20$~km the optimized phase tuple becomes distance-independent and reads $\boldsymbol{\phi}_{\opt}=(0,\pi/2,\pi,\pi/2)$. This choice allows to reach the second maximum in Fig.~\ref{fig01:sec10.2.1_KGR}(b), after which the KOR approaches the DH protocol.
For completeness, Fig.~\ref{fig02:sec10.2.1_OptPar}(b) reports also the optimized energies $\alpha^2_{\p}$, $\p=\PGM,\opt$, and $\alpha^2_{\HET}$. All curves converge to~$0.5$ average number of photons in the long-distance regime but, differently from the other cases, $\alpha^2_{\opt}$ shows the same non-monotonic trend of $K_{\opt}$.

\begin{figure}
\includegraphics[width=0.6\columnwidth]{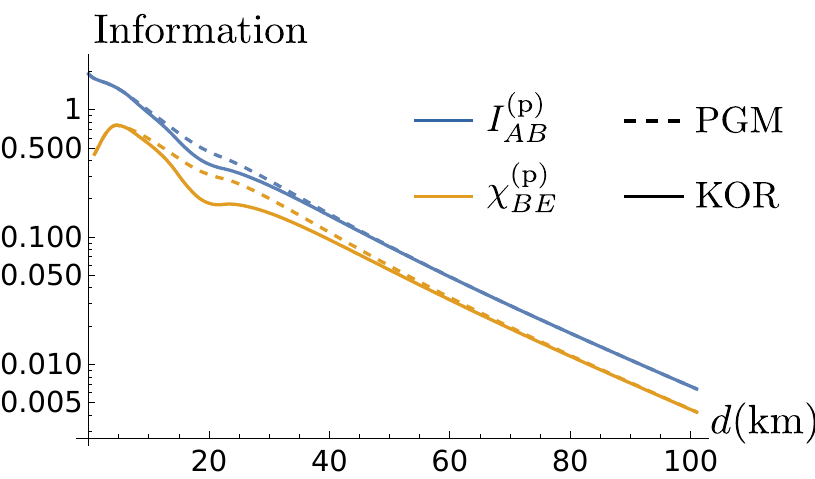}
\centering
\caption{Log plot of $I_{AB}^{(\p)}$ and $\chi_{BE}^{(\p)}$, $\p=\PGM,\opt$, as a function of the transmission distance $d$ in km. Both the quantities are computed with the optimized parameters $\alpha^2_{\p}$ and $\boldsymbol{\phi}_{\opt}$ (for the KOR). We set $\beta=0.95$.}\label{fig03:sec10.2.1_Info}
\end{figure}

The previous results prove non-Gaussian receivers as a potential tool for improving the key rate of the QPSK protocol, at least in the present restricted eavesdropping scenario. Remarkably, they also highlight that the discrete-valued POVM minimizing the error probability, namely the PGM, does not coincide with the discrete-valued POVM maximizing the KGR, namely the KOR. 
The reason becomes evident when comparing separately the mutual and the Holevo information appearing in the KGR~(\ref{eq:KGRdisc}). In Fig.~\ref{fig03:sec10.2.1_Info} we plot the quantities $I_{AB}^{(\p)}$ and $\chi_{BE}^{(\p)}$, $\p=\PGM,\opt$, computed with the same optimized energy and phases previously obtained and depicted in Fig.~\ref{fig02:sec10.2.1_OptPar}
As we can see, in the metropolitan-network distance regime the optimized receiver is associated with a reduced mutual information with respect to the PGM but, at the same time, reducing the mutual information induces also a reduction of the Holevo information extractable by Eve, thus resulting in a higher KGR.
As a consequence, differently from the state-discrimination scenario, in CVQKD there emerges a tradeoff between the goal of increasing the information accessible to Bob and the necessity of making the encoded symbols less ``distinguishable" to weaken Eve's attack.

In light of this, we may interpret the physical meaning of the optimized phases as follows.
For small transmission distances $\kappa d \ll 1$, Eve's intercepted signals are too weak to give her sufficient knowledge on which symbol was sent and the two different goals of reducing the error probability and maximizing the KGR are compatible, therefore the KOR coincides with the PGM. On the contrary, for larger $d$ the compatibility does not hold anymore, and Bob has to sacrifice part of his potential information and to reduce the mutual information shared with Alice to the detriment of the eavesdropper.

\begin{figure}
\begin{center}
\includegraphics[width=0.42\columnwidth]{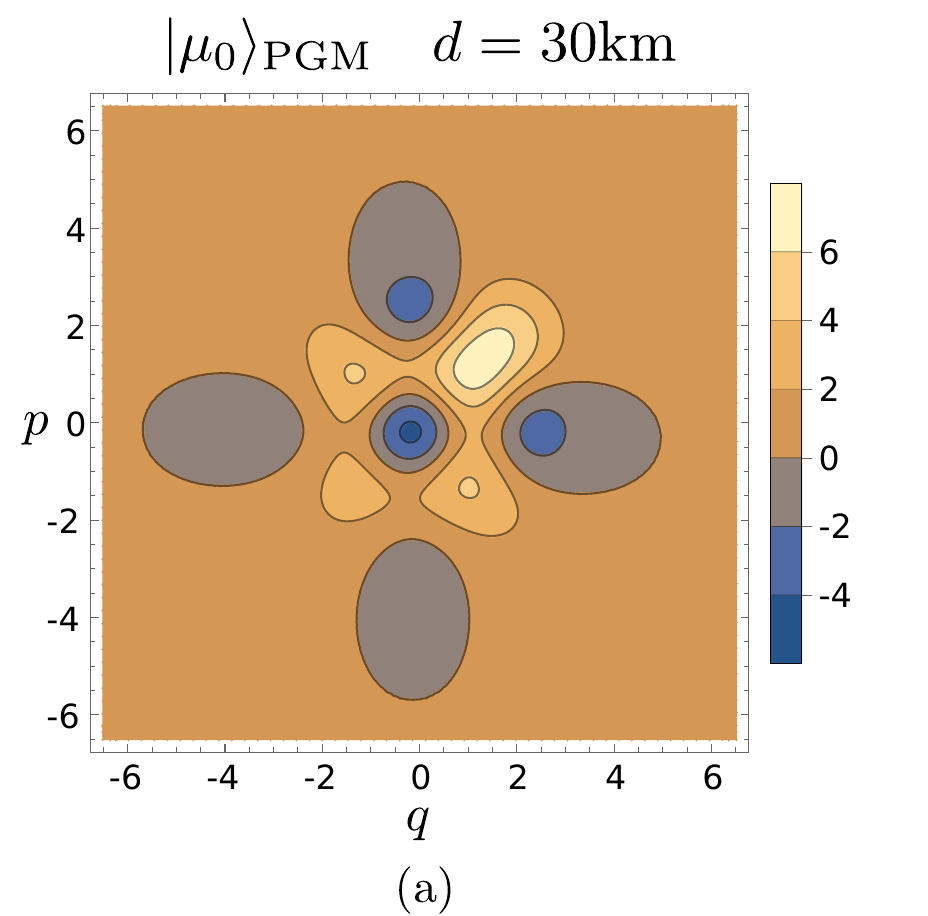} \quad
\includegraphics[width=0.42\columnwidth]{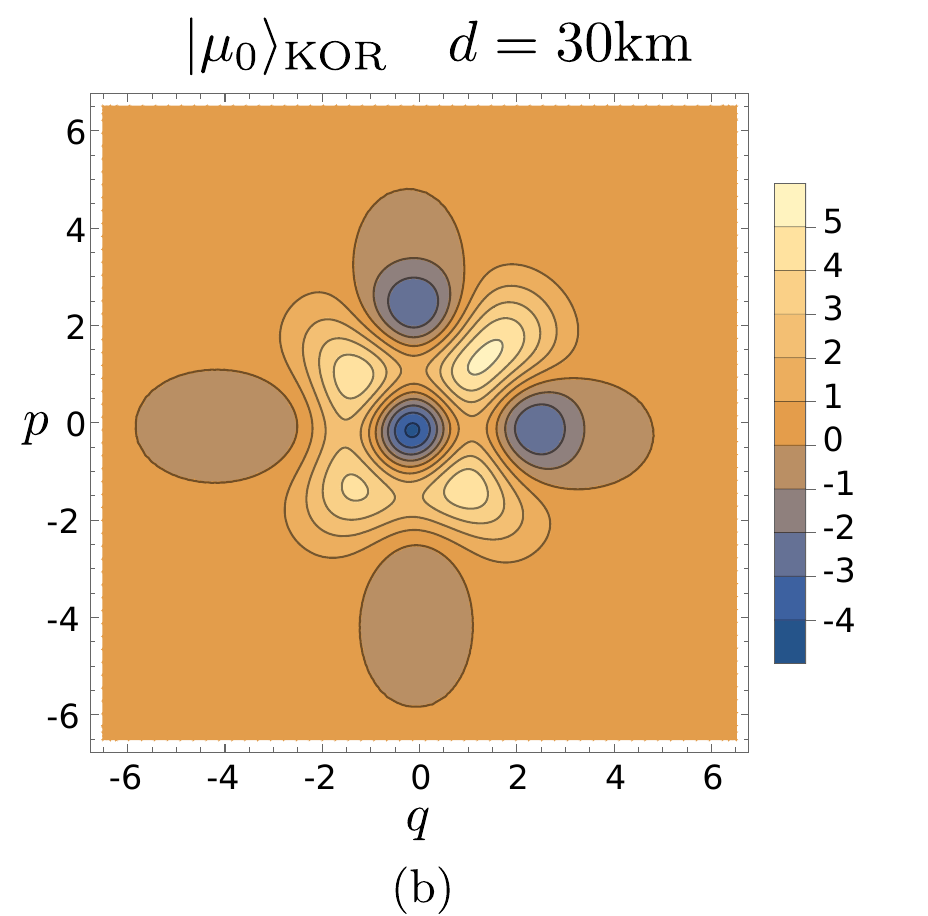} \\[2ex]
\includegraphics[width=0.42\columnwidth]{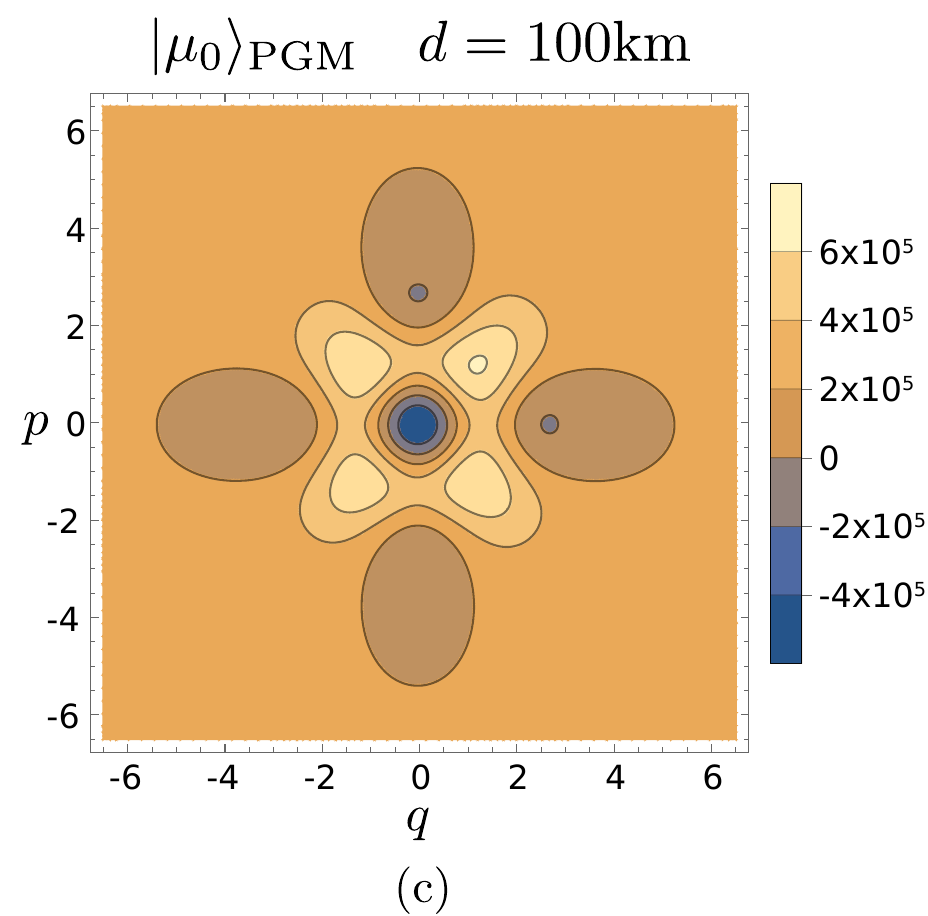} \quad
\includegraphics[width=0.42\columnwidth]{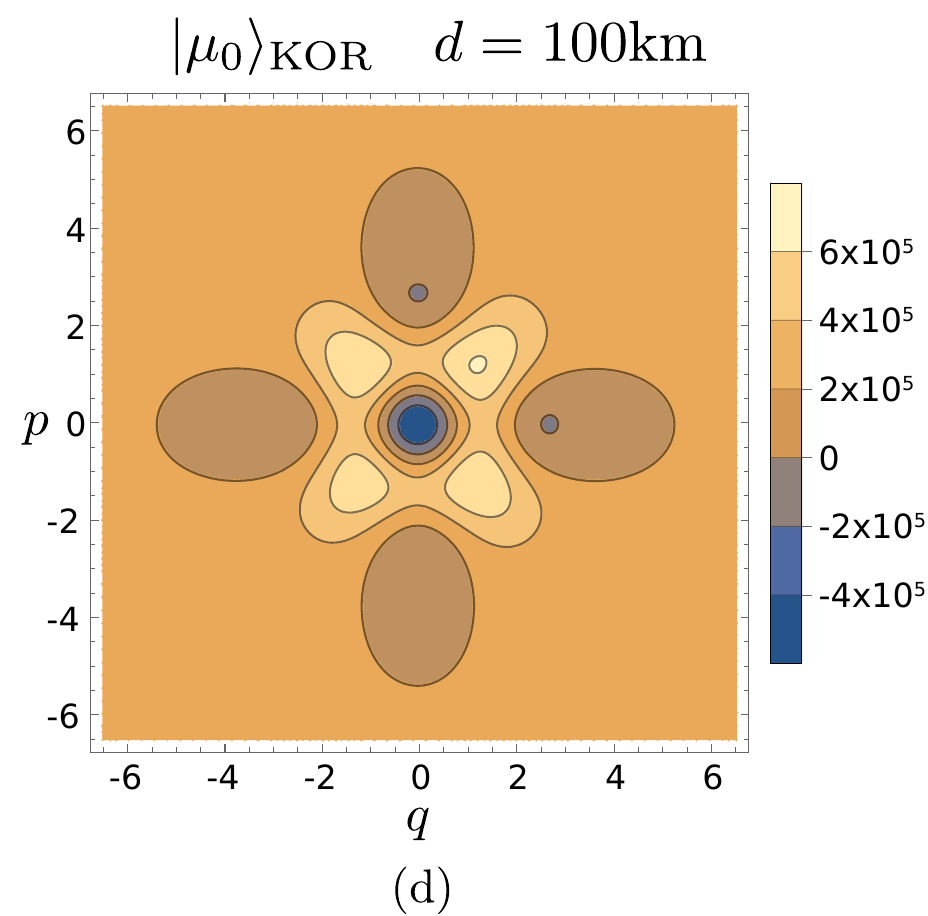}
\end{center}
\caption{Contour plot of the Wigner functions $W^{(\p)}(q,p)$ of the reference measurement vectors $|\mu_0\rangle_{\p}$, $\p=\PGM,\opt$, for either $d=30$ km (a-b) or $d=100$ km (c-d). We set $\alpha^2=1$ and $\boldsymbol{\phi}={\bf 0}$ and $\boldsymbol{\phi}=(0,\pi/2,\pi,\pi/2)$ for the PGM and the optimized receiver, respectively.}\label{fig04:sec10.2.1_Wigner}
\end{figure}

The discussed tradeoff may be qualitatively appreciated by comparing the phase-space representations of the PGM and the KOR effects. More in detail, we consider the two reference measurement vectors $|\mu_0\rangle_{\p}$, $\p=\PGM,\opt$, computed from~(\ref{eq:mu0phi}) with the phases $\boldsymbol{\phi}={\bf 0}$ and $\boldsymbol{\phi}=(0,\pi/2,\pi,\pi/2)$, respectively, and compute the associated Wigner function:
\begin{align}
W^{(\p)}(q,p) = \frac{1}{2\pi} \sum_{n=0}^{\infty} \, (-1)^n \, \langle n | D\dag(\zeta) \, \rho_\p \, D(\zeta) | n\rangle \, , \quad \p=\PGM,\opt \, ,
\end{align}
where $\zeta= (q+ i p)/2$ expressed in SNU, $\rho_\p= |\mu_0\rangle_{\p} {\, }_{\p}\langle \mu_0|$ and $D(\zeta)$ is the displacement operator \cite{Olivares2021, Serafini2017}. The contour plots of $W^{(\p)}(q,p)$ are depicted in Fig.~\ref{fig04:sec10.2.1_Wigner} for $\alpha^2=1$ and two different transmission distances $d=30$ km and $d=100$ km.
If $d=30$ km, that is for metropolitan-network distances, there is a qualitative difference between the two compared  cases, see Fig.s~\ref{fig04:sec10.2.1_Wigner}(a) and~\ref{fig04:sec10.2.1_Wigner}(b). Both Wigner functions exhibit four peaks, corresponding to the four transmitted states $|\alpha_k^{(t)}\rangle$. However, $W^{(\PGM)}(q,p)$ is well concentrated around state $|\alpha_0^{(t)}\rangle$, while $W^{(\opt)}(q,p)$ is more delocalized over the four states and the peaks are less distinguishable. This implies a reduced distinguishability of the states and, in turn, a reduced mutual information $I_{AB}^{(\opt)}$.
On the contrary, when the distance is larger, e.g. $d=100$ km, the transmitted states are weak coherent states with a greater overlap between one another. As a consequence, $W^{(\p)}(q,p)$ for respective receivers are equally delocalized over the four peaks and the differences between PGM and KOR become negligible; see Fig.s~\ref{fig04:sec10.2.1_Wigner}(c) and~\ref{fig04:sec10.2.1_Wigner}(d). In turn, the associated KGRs converge to the same value, corresponding also to the rate of the DH protocol, as depicted in Fig.~\ref{fig01:sec10.2.1_KGR}.
Furthermore, in all cases we observe a Wigner-negativity, proving both $|\mu_0\rangle_{\p}$ to be non-classical (as well as non-Gaussian) states at all distances \cite{Olivares2021,Serafini2017,Genoni2013}.

\subsubsection{Employing feasible receivers}\label{sec:IzumiQKD}

Even though both PGM and KOR discussed in the previous sections have shown interesting potentialities for CVQKD, from a practical point of view there is no clear idea on their experimental implementation. 
In fact, as discussed in Sec.~\ref{sec:QPSKdiscr}, in the presence of QSPK designing a feasible optimum receiver is an open problem. In contrast, different suboptimal receivers have been proposed, ranging from feedback ones, like the Bondurant receiver, to displacement-photon counting schemes, either without feed-forward, e.g. the quaternary displacement receiver (QDRE), or employing it, like the quaternary displacement feed-forward receiver (QDFFRE) \cite{Kennedy1973, Izumi2012, Becerra2013, DiMario2018,  DiMario2018QPSK, Izumi2020, Notarnicola2023:FF}.
Therefore, it is worth of interest to investigate also the performance of these receivers for CVQKD. Here, in particular, we focus on the QDFFRE proposed by Izumi {\em et al.} in \cite{Izumi2012} and presented in Sec.~\ref{sec:QDFFRE}.

In particular, when Bob adopts the QDFFRE in the protocol of Fig.~\ref{fig01:sec10.2.0_Protocol}, he probes the conditional probabilities reported in Eq.~(\ref{eq:CondQDFFRE}), namely:
\begin{subequations}
\begin{align}
p_{B|\alpha_k}^{(N)}(0) =& p_0^N \, , \\[2ex]
p_{B|\alpha_k}^{(N)}(1) =& \sum_{t=0}^{N-2} p_k^t \, (1-p_k) \, p_{(k-1) \bmod M}^{N-1-t} 
\, + \, \frac{p_k^{N-1}(1-p_k)}{3} \, ,\\[2ex]
p_{B|\alpha_k}^{(N)}(2) =& \sum_{t=0}^{N-3} \, \sum_{s=0}^{N-3-t} p_k^t \, (1-p_k) \, p_{(k-1) \bmod M}^s \, (1-p_{(k-1) \bmod M}) \times \, \nonumber \\[1ex]
&\, p_{(k-2) \bmod M}^{N-2-t-s} \, + \, \sum_{t=0}^{N-2} p_k^t \, (1-p_k) \, \frac{p_{(k-1) \bmod M}^{N-2-t}(1-p_{(k-1) \bmod M})}{2} \nonumber \\[1ex]
&\, + \, \frac{p_k^{N-1}(1-p_k)}{3} \, ,
\end{align}
\begin{align}
p_{B|\alpha_k}^{(N)}(3) =& \sum_{t=0}^{N-3} \, \sum_{s=0}^{N-3-t}  \,\, \sum_{u=0}^{N-3-t-s}  p_k^t \, (1-p_k) \, p_{(k-1) \bmod M}^s \times \, \nonumber \\[1ex]
&\, (1-p_{(k-1) \bmod M}) \, p_{(k-2) \bmod M}^{u} \, (1-p_{(k-2) \bmod M}) \, p_{(k-3) \bmod M}^{N-3-t-s-u} \nonumber \\[1ex]
& \, + \, \sum_{t=0}^{N-2} p_k^t \, (1-p_k) \, \frac{p_{(k-1) \bmod M}^{N-2-t}(1-p_{(k-1) \bmod M})}{2} \nonumber \\[1ex]
&\, + \, \frac{p_k^{N-1}(1-p_k)}{3} \, ,
\end{align}
\end{subequations}
where, now, $p_0=1$, $p_1= p_3= \exp(-2 T \alpha^2/N)$, and $p_2= \exp(-4 T \alpha^2/N)$, as Bob receives only the transmitted fraction of Alice’s signals. Instead, the overall Bob's probability reads $p_B^{(N)}(j) =M^{-1} \sum_{k=0}^{M-1} p_{B|\alpha_k}^{(N)}(j)$, $j=0,\ldots, M-1$.

\begin{figure}
\includegraphics[width=0.49\columnwidth]{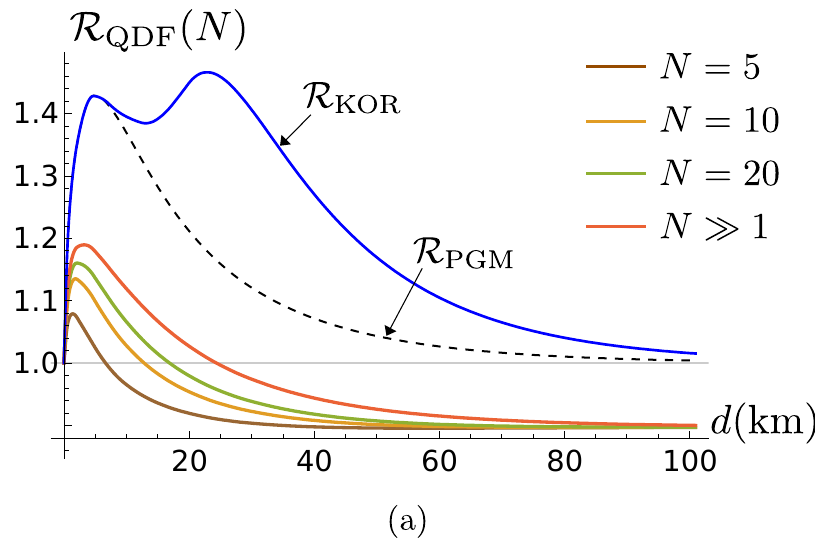} 
\includegraphics[width=0.49\columnwidth]{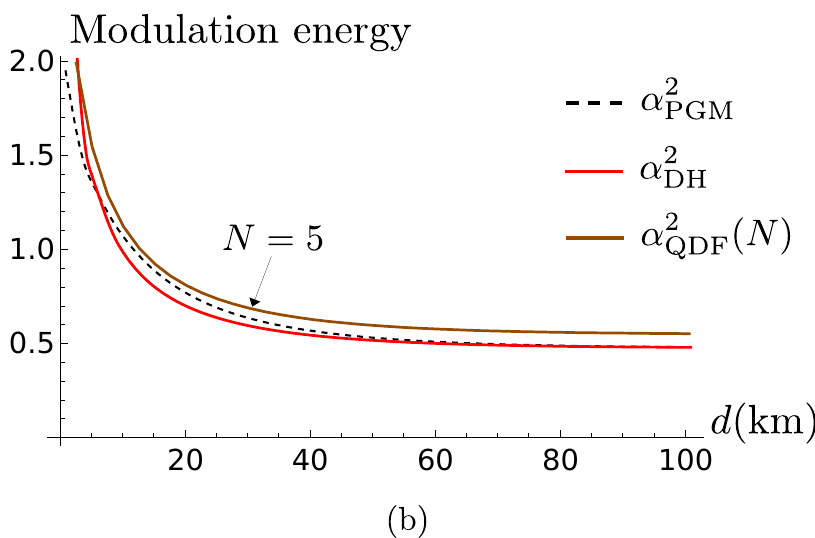}
\centering
\caption{(a) Plot of the ratio ${\cal R}_\disp(N)$ as a function of the transmission distance $d$ in km. Differently from both PGM and KOR, the QDFFRE improves the KGR with respect to the DH protocol only up to a maximum distance $d_{\rm max}(N)$, increasing with the number of copies $N$.
(b) Plot of the optimized modulation energies $\alpha^2_{\disp}(N)$, $\alpha^2_{\PGM}$, and $\alpha^2_{\HET}$, as a function of the transmission distance $d$.  In both pictures we set $\beta=0.95$. }\label{fig05:sec10.2.1_KGRQDF}
\end{figure}

To compute the KGR $K_\disp(N;\alpha^2)$ associated with the QDFFRE, we exploit Eq.s~(\ref{eq:KGRdisc}),~(\ref{eq:IABdisc}) and~(\ref{eq:chiBEdisc}), provided the substitutions $p_{B|\alpha_k}^{(\boldsymbol{\phi})}\rightarrow p_{B|\alpha_k}^{(N)}$ and $p_{B}^{(\boldsymbol{\phi})}\rightarrow p_{B}^{(N)}$, and optimize over the modulation energy, obtaining:
\begin{align}
K_\disp(N) = \max_{\alpha^2} \,  K_\disp(N; \alpha^2) \, .
\end{align}
Moreover, we also compute the ratio with respect to the DH protocol, namely,
\begin{align}
{\cal R}_\disp(N) = \frac{K_\disp(N)}{K_\HET} \, ,
\end{align}
reported in Fig.~\ref{fig05:sec10.2.1_KGRQDF}(a) for different number of copies $N$.
Unlike the PGM and the KOR, the QDFFRE outperforms the DH protocol only up to a maximum transmission distance $d_{\rm max}(N)$ whose value increases with $N$. Afterwards, we have $K_\disp(N) \le K_{\HET}$ and, in turn, ${\cal R}_\disp(N)$ saturates to an asymptotic value $\le 1$. The best performance is achieved in the limit of infinite copies, $N\gg 1$, where the receiver approximates the type-I Bondurant receiver, see Sec.~\ref{sec:QDFFRE}, obtaining a maximum increase in the KGR of about $\lesssim 20 \%$ and $d_{\rm max}(N) \lesssim 25$ km.

This behaviour is a direct consequence of the optimized modulation $\alpha^2_{\disp}(N)$, reported in Fig.~\ref{fig05:sec10.2.1_KGRQDF}(b). In fact, $\alpha^2_{\disp}(N)$ is a decreasing function of $d$, which in the long-distance regime, $\kappa d \gg 1$, reaches an asymptotic value $\gtrsim 0.5$. Numerical calculations also show this asymptote to be independent of the number of copies $N$. 
In these conditions, Bob receives a signal with $T \alpha^2_{\disp}(N) \ll 0.5$ mean photons, for which the QDFFRE does not beat the SQL, achieved by DH discrimination (see, for instance, Fig.~\ref{fig02:sec5.5.3_QDFPlot}). In turn, even the KGR of the CVQKD protocol is lower than the corresponding heterodyne protocol.
On the other hand, for $\kappa d \ll 1$, the optimized modulation is of few photons, the QDFFRE outperforms the SQL and we observe an increase also in the KGR.

In conclusion, despite its feasibility, the present displacement feed-forward scheme is not optimal for CVQKD, just as it is not optimal for coherent state discrimination. Nevertheless, it still provides an improvement of the resulting key rate in the short-distance regime, being a candidate for experimental realizations of the present protocol.

\section{Conclusions and future perspectives}\label{chap:Concl}

In this PhD thesis we have addressed some relevant aspects of quantum communications theory in continuous variable systems, with particular reference to quantum optical platforms. The field has gained much interest in time, for a twofold reason. On the one hand, quantum effects determine the ultimate limits of the transmission of classical information over optical communication links, being of particular relevance for protocols operated at low signal powers, e.g. near-space or deep-space communications. On the other hand, quantum features, like superposition, entanglement, teleportation, and Heisenberg's uncertainty relations, can be effectively exploited as a resource to design novel protocols and algorithms, and convey information in more efficient fashion. It is the case of quantum computation and quantum key distribution (QKD), where the superposition principle and the ultimate quantum noise due to commutation rules are crucial to efficiently solve hard computation problems and to distill random secure keys with unconditional security, respectively, thus overcoming the limitations of the corresponding classical schemes.

In particular, throughout the thesis, we focused ourselves on two main topics, namely quantum state discrimination and continuous variable (CV) QKD, providing a comprehensive theoretical analysis, together with an eye to possible practical implementations being compatible with the state-of-art technologies in optical telecommunications. We both addressed the general aspects of the theory and proposed new receivers and protocols that can be feasibly experimentally demonstrated, discussing also their robustness with respect to the relevant realistic inefficiencies occurring in practice.

To begin with, we dealt with the problem of quantum state discrimination. In Sec.~\ref{chap:GeneralFeatures}, we introduced the fundamental aspects of the theory, whose task is to perform conclusive decision among a set of non-orthogonal quantum states. The very laws of quantum mechanics forbid exact discrimination, making every quantum receiver associated with a nonzero decision error probability, therefore the task is to identify the optimum receiver achieving the lowest possible error probability. Subsequently, we focused on the binary discrimination scenario, for which Helstrom's theory provides a full characterization of the optimum receiver. Then, we addressed binary discrimination of coherent states of radiation, that represents one of the fundamental problems in optical communication schemes, and consider a binary phase-shhift keying (BPSK) encoding. 
Remarkably, we proposed new hybrid receivers, based on the combination of weak-field homodyne detection and conditional displacement-photon counting, that outperform conventional detection schemes, based on quadrature detection, providing a genuine quantum advantage, and closing the gap with the Helstrom bound.

In Sec.~\ref{chap:MaryDisc}, we widened our analysis to multilevel systems, studying discrimination of a constellation of $M$ quantum states, with $M\ge 2$. In this case, the decision task can be recast into a convex optimization problem, and advanced linear algebra tools lead to characterization of the optimum receiver.
Nevertheless, an explicit construction of this optimum quantum measurement is obtained only in the particular case of pure-state discrimination and geometrically uniform symmetry (GUS).
Ultimately, we addressed $M$-ary discrimination of coherent states, considering quadrature phase-shift keying (QPSK) constellations, as a natural generalization of the previously considered BPSK encoding.

Then, our interest turned to CVQKD. In Sec.~\ref{chap:CVQKD}, we presented the basic tools of QKD, and discussed the main characteristics of CV protocols, where coherent states with randomly modulated amplitude are transmitted from a sender (Alice) to a receiver (Bob), communicating by an untrusted noisy quantum channel.
Given this scenario, the protocol is considered secure as long as the information shared by Alice and Bob is larger than the one that can be extracted by a possible eavesdropper (Eve), leading to a nonzero key generation rate (KGR). Furthermore, we outlined the different security framework under which the analysis may be conducted, namely unconditional security, trusted-device scenario and wiretap channel. In this Section, we focused only on the unconditional security approach, and provide security proof for the GG02 protocol, based on Gaussian modulation of coherent states, as well as discrete-modulation protocols, employing both PSK and quadrature amplitude modulation (QAM), that provide a feasible alternative being easier to implement into practice. To this aim, we proved the fundamental theorem on the ``optimality of Gaussian attacks", establishing a manageable lower bound to the KGR of protocols employing Gaussian detection.

Thereafter, in Sec.~\ref{chap:RESTREAV} we studied the two remaining security frameworks, that provide examples of restricted eavesdropping, in which we pose realistic limitations to the possible attacks that Eve may launch.
In the trusted-device scenario, we included detection losses and noise in the theoretical description, assuming them to be simply lost to the environment and not intercepted by Eve, who, instead, controls both the losses and noise acquired during signal transmission. Remarkably, we extended the validity of optimality of Gaussian attacks to this scenario, thus determining a useful result to assess security in all the cases of partial lack of information at Eve's sides. Then, we addressed the wiretap channel description, that provides an example of a particular eavesdropping strategy, where Eve's action is completely characterized and specified. In both the scenarios, we computed the KGR for the QPSK protocol, comparing the results with the unconditional security approach, and obtaining an increase in the distilllable key rate.

In Sec.~\ref{chap:Amplifiers}, we investigated the potentialities of optical amplifiers to perform loss mitigation of the quantum channel and enhance CVQKD. We established the quantum limits of amplification, introducing both conventional amplifiers, phase-insensitive (PIAs) and phase-sensitive amplifiers (PSAs), as well as probabilistic noiseless linear amplifiers (NLAs), studying their application for CVQKD in the Gaussian modulation format.

Finally, in Sec.~\ref{chap:nonGauss}, we merged the acquired knowledge of the two main Parts of the thesis, and designed an optimized state-discrimination receiver, the key-rate optimized receiver (KOR), for the QPSK CVQKD protocol, providing a first step towards non-Gaussian CVQKD. We assessed security under a pure-loss wiretap channel, and obtain an enhancement of the KGR with respect to the conventional protocol in the metropolitan-network distance regime. Furthermore, we also consider the performance of displacement receivers for CVQKD, as a benchmark example of a feasible scheme, obtaining an increase in the KGR up to a maximum transmission distance.

\subsection{Future directions and outlooks}

The results obtained in the thesis provide a detailed analysis of the current state of the art in the fields of both quantum state discrimination and CVQKD, and present innovative solutions to enhance the existing communication protocols by means of improved quantum receivers and advanced stragies for transmission losses mitigation. Furthermore, they identify the limits of quantum communications in realistic conditions, e.g. imperfect detection, non-Gaussian noise, imperfect signal modulation, \ldots, paving the way for new applications, from both a theoretical and experimental point of view.
A brief overview of possible further developments is presented in the following.

\subsubsection{Novelties in coherent states discrimination}

Within quantum decision theory, one of the relevant results of this thesis is the proposal of hybrid receivers, namely the hybrid near-optimum (HYNORE) and the hybrid feed-forward (HFFRE) receivers, to enhance BPSK discrimination of coherent states \cite{Notarnicola2023:HYNORE, Notarnicola2023:FF, Notarnicola2023:PhN}.
These hybrid schemes exploit the photon-number resolving (PNR) technologies, gaining fast progresses in the latest years \cite{PNRD1,PNRD2,PNRD3}, to suitably combine the homodyne like and displacement setups, thus jointly probing the wave-like and particle-like properties of optical fields. In particular, in hybrid receivers the incoming signal is split at a beam splitter of variable transmissivity, and the reflected beam undergoes homodyne like detection, whose outcome determines a conditioned displacement operation on the transmitted beam: accordingly, we obtain a reduced error probability with respect to the standard displacement receivers.

Interestingly, the present philosophy offers a powerful approach to improve quantum receivers also for quaternary and $M$-ary phase-shift-keying discrimination \cite{Izumi2012,Izumi2013,Becerra2013,Izumi2020}, where displacement feed-forward schemes are less powerful and may benefit even more from a suitable combination with weak-field measurements. 
In particular, two possible paths can be pursued.
On the one hand, a fundamental problem is to assess the most efficient usage of PNR detectors to improve the maximum a posteriori probability (MAP) decision strategy. In the presence of multiple state discrimination, the capability to resolve individual photons promises a powerful enhancement for displacement feed-forward schemes, as the outcomes of PNR detection yield more information about the incoming signals than on-off detection. In fact,
the encoded pulses are associated with Poisson statistics with different rates, thereby the number of registered clicks gives indirect information on which was the probed signal, instead of the simple acceptance/rejection of the nulled hypothesis occurring with on-off strategies. 
In turn, a proper extension of the MAP criterion may lead to significant enhancements of the feed-forward rules, possibly closing the gap with the type II Bondurant receiver.
On the other hand, given a displacement-PNR setup, we may also claim whether or not hybrid setups combining either weak-field homodyne or double weak-field homodyne detection, being the natural extension of the HYNORE, are able to further reduce the error probability, especially in the low-energy limit where non-optimized displacement schemes do not outperform the standard quantum limit.

\subsubsection{Progresses in CVQKD}

As regards the CVQKD analysis presented in this work, different problems can be addressed to obtain innovative solutions.

First of all, in the thesis we widely discussed about the practical limitation of Gaussian modulation, and address discrete modulation protocols as a more practical solution, highlighting the tradeoff between increased practicality of the setup and reduced amount of KGR to be achieved.
In particular, PSK modulation proved itself as the simplest scheme for a realistic implementation, whereas QAM yielded a better tradeoff between the modulator complexity and the resulting KGR, allowing to close the gap with respect to GG02. 
A further solution within this topic may be offered by amplitude phase-shift keying (APSK) modulation, where both the amplitude and the phase of a carrier field are modulated to generate a multiple-ring constellation geometry in the phase space. APSK is becoming the emerging modulation format for deep-space communications, and it guarantees high information rates and energy efficiency with respect to competitive schemes, being also able to reach the Gaussian modulation capacity as the constellation size grows to infinity \cite{Thomas1974, DeGaudenzi2006, Meric2015, Oztekin2019, Inoue2024, Wang2024}. Accordingly, it candidates itself as a powerful scheme also for CVQKD applications, especially in the presence of smaller constellations with few symbols.

A second relevant issue in CVQKD is represented by channel loss mitigation, that can be partially addressed by the exploitation of optical amplifiers. In particular, we proved NLAs to provide an effective solution to achieve long-distance key distribution, being also robust against a reduced detection quantum efficiency.
In turn, these results open new perspectives for the applications of NLAs in realistic conditions for both one-way
communication and end-to-end communication over quantum repeater chains \cite{Furrer2018, Pirandola2019Commun, Dias2020}, with the ultimate goal of increasing the KGR up to the quantum channel secure-key capacity established in~\cite{Pirandola2019Commun}.
On the other hand, conventional amplifiers provide a simpler choice for large-scale applications in the framework of conditional security CVQKD. In particular, the advantage given by PSA, being a phase-sensitive operation, may be potentially further boosted by employing modulation of squeezed states \cite{Usenko2011, Usenko2015, Usenko2018, Usenko2020, Derkach2020, Gottesman2001, Schafer2016}.

Finally, in the last Section we made a first step towards the analysis of CVQKD with non-Gaussian measurements, suggesting suitable non-Gaussian receivers as a resource to increase the achievable key rate. 
Differently from conventional CVQKD protocols, this field still provides unclear and attractive open problems, fostering new research with a possibly higher potential impact.
Given this premise, our results, obtained under a (restrictive) pure-loss wiretap channel assumption, leave many points as open problems. 
At first, the extension of the present analysis to the more realistic case of a thermal-loss channel remains a challenging task. In fact, in the presence of thermal mixed states, designing the receiver achieving the minimum error probability is non-trivial. The general structure of quantum receivers exploited to the design the KOR does not hold anymore, as the $M\ge 2$ mixed states now span the whole infinite dimensional Hilbert space. Moreover, from the perspective of quantum communications, the optimum receiver achieving the minimum error probability can only be
obtained numerically via linear convex semidefinite programming \cite{Cariolaro2015}. As a consequence, the search of the KOR could only be obtained via a brute-force functional optimization over all possible POVMs, being a
nonlinear and non-convex problem.
Secondly, the sketch of an unconditional security proof may be designed, identifying which is the optimal
Eve’s attack. To do so, we should optimize over all the possible attacks compatible with Alice and Bob’s
statistics, retrieving the Devetak-Winter bound by extending the methods of \cite{Lin2020,Lin2019}. Indeed, the question whether or not protocols employing non-Gaussian measurement guarantee higher security than Gaussian ones is an
interesting open problem.
Finally, we should investigate the scalability of the present scheme with discrete modulation formats of
higher order, like PSK schemes with $M\ge 4$ states or QAM constellations, in which the GUS is not satisfied anymore \cite{Notarnicola2024:SEC, Denys2021, Roumestan2021,  Roumestan2022}.

\section*{Acknowledgements}
The work was done while the author was at Universit\`a degli Studi di Milano. The author acknowledges S. Olivares, M.~G.~A. Paris, K.~Banaszek, and M.~Jarzyna for both stimulating discussion and the contribution given to each of the research papers mentioned in the present Thesis.

\section*{ORCID}
Michele N. Notarnicola \url{https://orcid.org/0000-0002-7492-6143}


\appendix

\def\IN{{\rm in}}
\def\OUT{{\rm out}}
\def\Re{{\mathrm{Re}}}
\def\Im{{\mathrm{Im}}}
\def\id{ {\mathrm{id}} }
\def\a{ {\mathrm{QS}} }
\def\b{ {\mathrm{SPC}} }
\def\p{ {\mathrm{p}} }
\def\sigmamA{\boldsymbol\sigma^{\rm(m)}_{A}}
\def\sigmamB{\boldsymbol\sigma^{\rm(m)}_{B}}
\def\sigmamAB{\boldsymbol\sigma^{\rm(m)}_{A(B)}}
\def\sigmaz{\boldsymbol\sigma_z}
\def\aA{a_A}
\def\aB{a_B}
\def\aC{a_{B_1}}
\def\aD{a_{B_2}}
\def\bA{b_A}
\def\bB{b_B}
\def\bC{b_{B_1}}
\def\bD{b_{B_2}}
\def\alphaA{\alpha_A}
\def\alphaB{\alpha_B}
\def\alphaC{\alpha_{B_1}}
\def\alphaD{\alpha_{B_2}}
\def\betaA{\beta_A}
\def\betaB{\beta_B}
\def\betaC{\beta_{B_1}}
\def\betaD{\beta_{B_2}}\def\alphaA{\alpha_A}
\def\alphaB{\alpha_B}
\def\alphaC{\alpha_{B_1}}
\def\alphaD{\alpha_{B_2}}
\def\betaA{\beta_A}
\def\betaB{\beta_B}
\def\betaC{\beta_{B_1}}
\def\betaD{\beta_{B_2}}
\def\bmalpha{\boldsymbol\alpha}
\def\bmbeta{\boldsymbol\beta}
\def\NLA{\mathcal{T}}

\section{The maximum a posteriori probability criterion}\label{app:MAPcriterion}

The maximum a posteriori probability (MAP) criterion represents a strategy based on Bayesian inference to improve the decision rule of a displacement-photon counting discrimination scheme \cite{DiMario2018, Notarnicola2023:HYNORE}.
Here, we consider as a paradigmatic example the displacement-PNR (DPNR) receiver presented in Sec.~\ref{subsec4:Kennedy} and, for the sake of simplicity, we address the case of binary coherent-state discrimination.
That is, we discriminate between the two coherent states $|\alpha_k\rangle=|e^{i k \pi} \alpha\rangle$, $k=0,1$, $\alpha>0$, generated with equal {\it a priori} probabilities $q_k=1/2$. 

In all displacement receivers, we apply a displacement operation $D(\beta)$, $\beta>0$, to the incoming signal, mapping the states into
\begin{align}
    |\alpha_j\rangle \rightarrow |\alpha_j+\beta \rangle \, .
\end{align}
If we fix $\beta=\alpha$ we retrieve the usual ``nulling" technique, whereas when $\beta$ is considered as a free parameter to be optimized we obtain the improved receiver proposed by Takeoka and Sasaki \cite{Takeoka2008}.

Thereafter, we perform a PNR measurement on the displaced state and obtain the outcome $n$. Without loss of generality, we consider ideal photo-detection, namely with infinite resolution. The MAP criterion states that, for each $n$, we infer the state $\alpha_j$, $j=0,1$, with the largest {\it a posteriori} probability:
\begin{align}\label{eq: AP prob}
    p(\alpha_j | n) = \frac{p(n|\alpha_j) \ q_j}{p(n)} \, ,
\end{align}
where
\begin{align}
    p(n|\alpha_j) = e^{-|\alpha_j+\beta|^2} \frac{|\alpha_j+\beta|^{2n}}{n!}
\end{align}
is the probability of getting $n$ photons given $\alpha_j$ and
\begin{align}
p(n)= \sum_{j=0,1} q_j p(n|\alpha_j) = \frac{p(n|\alpha_0)+p(n|\alpha_1)}{2}
\end{align}
is the overall probability of detecting $n$ photons. For example, we infer state $\alpha_0$ if $p(\alpha_0 | n)>p( \alpha_1 | n)$, which is equivalent to condition $p(n|\alpha_0)>p(n| \alpha_1)$ since we have $q_j=1/2$.

The correct decision probability is then equal to
\begin{align}\label{eq: pc}
    {\cal P}_{\rm c} &= q_0 \sum_{n=0}^\infty p(n|\alpha_0) \chi_{0} + q_1 \sum_{n=0}^\infty p(n|\alpha_1) \chi_{1} \\
    &= \frac{1}{2} \sum_{n=0}^\infty \max  \left[p(n|\alpha_0), p(n|\alpha_1)\right] \, ,
\end{align}
where
$\chi_{0}=1$ if $p(n|\alpha_0)>p(n|\alpha_{1})$ and 0 otherwise and $\chi_{1}=1$ if $p(n|\alpha_1)>p(n|\alpha_0)$ and 0 otherwise.
The error probability is obtained immediately as $P_{\rm err} = 1- {\cal P}_{\rm c} $.

The decision rule $p(n|\alpha_0) \lessgtr p(n|\alpha_1)$ is equivalent to the definition of a threshold outcome $\nth$ such that all measurement outcomes $n\geq \nth$ are assigned to state $\alpha_1$ and all $n<\nth$ are assigned to state $\alpha_0$. The threshold number is obtained by equating $p(\bar{n}|\alpha_0)= p(\bar{n}|\alpha_1)$, $\bar{n} \in \mathbb{R}$, and considering the lowest integer greater than the obtained root $\bar{n}$, namely $\nth = \ceil{\bar{n}}$, where $\ceil{x}$ is the ceiling function, returning the smallest integer greater than $x$.
We have:
\begin{align}
    \nth = \ceil[\Bigg]{\frac{|\alpha+\beta|^2-|\alpha-\beta|^2}{\ln \bigl(|\alpha+\beta|^2\bigr) - \ln \bigl(|\alpha-\beta|^2\bigr)}} \, .
\end{align}
Thus, ${\cal P}_{\rm c}$ may be equivalently written as:
\begin{align}
    {\cal P}_{\rm c} &= \frac12  \left[ \sum_{n=0}^{\nth-1} p(n|0) + \sum_{n=\nth}^{\infty} p(n|1)\right] \, .
\end{align}
Finally, we note that for the standard Kennedy receiver, where the displacement amplitude is $\beta=\alpha$, we have $p(n|\alpha_0)= \delta_{n,0}$, therefore the correct probability of Eq.~(\ref{eq: pc}) reduces to the well known expression ${\cal P}_{\rm c}= 1- \exp(-4\alpha^2)/2$.

\section{A model for the visibility reduction at a beam splitter}\label{app:VisibilityModel}

In the framework of quantum optics, phase-sensitive operations, e.g. displacement operations and homodyne detection, are implemented via interference at a beam splitter between the signal beam and a suitable local oscillator (LO) \cite{Paris1996, Ferraro2005, Olivares2021, Serafini2017}. To obtain perfect interference in realistic implementations, it is required that the two optical modes impinging at the beam splitter are perfectly matched in both the frequency and spatial domain.
This represents a nontrivial task from a practical point of view. In fact, realistic optical beams are associated with a finite spatial linewidth, and perfect interference can be only achieved when the spatial profiles of both the signal and the LO are fully superimposed. 
On the contrary, the presence of any mode mismatch due to misalignment of the two wave-fronts, leads to reduced visibility $\xi\le 1$ of the optical interference, determined by how the intensity distributions of the signal beam and the LO overlap with each other.
Perfect mode matching corresponds to unit visibility, $\xi=1$, whilst in the case $\xi<1$ the LO partially overlaps with the spatial modes orthogonal to the signal mode, with detrimental effects for any quantum measurement performed thereafter \cite{Gupta2020}.
Here we present a model to describe the present effect, being relevant for many applications in quantum communications.

\begin{figure}
\centerline{\includegraphics[width=0.8\columnwidth]{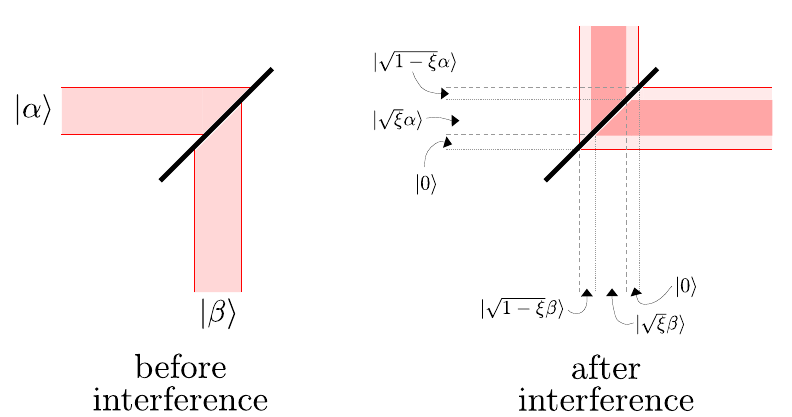}}
\centering
\caption{Scheme of imperfect interference of coherent states due to mode mismatch at the beam splitter, associated with visibility $\xi\le 1$. The two input beams before interference are misaligned, therefore only a fraction $\xi$ of each beam is effectively overlapped, leading to quantum interference, whilst the remaining portions of both the signal and the LO interferes with spatial vacuum modes.}\label{fig:01:appII.B-VisCoh}
\end{figure}

To begin with, we deal with a particular case and consider both the signal and the LO to be excited in coherent states $|\alpha\rangle$ and $|\beta\rangle$, $\alpha, \beta \in \mathbb{C}$, respectively, as schematized in Fig.~\ref{fig:01:appII.B-VisCoh}. In the presence of visibility reduction, the two input coherent beams are mismatched, and interference holds only for the beam portions that effectively overlap with each other. On the contrary, the remaining parts of both the signal and the LO overlap with spatial vacuum modes, being split into a transmitted and a reflected beam.

In turn, the presence of reduced visibility $\xi \le 1$, quantifying the spatial overlap between the input optical beams,  is equivalent to the splitting of the input coherent states into three channels, as described in Fig.~\ref{fig:01:appII.B-VisCoh}, namely:
\begin{subequations}\label{eq:3channels}
\begin{align}
|\alpha\rangle  \to |\sqrt{\xi} \alpha\rangle \otimes |\sqrt{1-\xi} \alpha\rangle \otimes |0\rangle \, ,  \\[1ex]
|\beta\rangle \to |\sqrt{\xi} \beta\rangle \otimes |0\rangle \otimes |\sqrt{1-\xi} \beta\rangle \, ,
\end{align}
\end{subequations}
where only the reduced pulses $|\sqrt{\xi} \alpha\rangle$ and $|\sqrt{\xi} \beta\rangle$ are mode-matched, while the other ones impinge with further modes prepared in the vacuum.
Thereafter, states~(\ref{eq:3channels}) interfere at the beam splitter of transmissivity $\tau\le 1$, leading to the output states on the transmitted and reflected side:
\begin{subequations}\label{eq:outbeams}
\begin{align}
|\psi^{(t)}\rangle &= |\sqrt{\xi} \left(\sqrt{\tau} \alpha + \sqrt{1-\tau} \beta \right)\rangle \nonumber \\
&\hspace{1.cm} \otimes |\sqrt{\tau(1-\xi)} \alpha\rangle \otimes |\sqrt{(1-\tau)(1-\xi)} \beta\rangle \, , \\[1.5ex]
|\psi^{(r)} \rangle &= |\sqrt{\xi} \left(\sqrt{\tau} \beta - \sqrt{1-\tau} \alpha \right)\rangle  \nonumber \\
&\hspace{1.cm} \otimes |-\sqrt{(1-\tau)(1-\xi)} \alpha\rangle \otimes |-\sqrt{\tau(1-\xi)} \beta\rangle \, ,
\end{align}
\end{subequations}
where the beam splitter operation acts independently on each channel.
The mean number of photons $\bar{n}$ on both branches is then equal to:
\begin{subequations}\label{eq:outen}
\begin{align}
\bar{n}^{(t)} &= \tau |\alpha|^2 + (1-\tau) |\beta|^2 + 2 \xi \sqrt{\tau(1-\tau)} \, {\rm Re}(\alpha \beta^{*}) \, , \\[1ex]
\bar{n}^{(r)} &= \tau |\beta|^2 + (1-\tau) |\alpha|^2 - 2 \xi \sqrt{ \tau(1-\tau)} \, {\rm Re}(\alpha \beta^{*}) \, .
\end{align}
\end{subequations}
We note that the overall effect of the mode mismatch is the reduction of the interference terms in~(\ref{eq:outen}), namely the terms proportional to ${\rm Re}(\alpha \beta^{*})$.

In particular, from Eq.~(\ref{eq:outen}) we verify that the parameter $\xi$ coincides with the inteference visibility ${\cal V}$ that can be experimentally measured. In fact, fringes visibility is measured by considering a balanced beam splitter, $\tau=1/2$, when the LO and the signal are equal in power, namely $|\alpha|^2=|\beta|^2$. We consider a single output port, e.g. the transmitted one, and evaluate the maximum and minimum output power as we change the phase difference between the two input beams, retrieving:
\begin{align}
{\cal V} = \frac{\bar{n}^{(t)}_{\rm max}- \bar{n}^{(t)}_{\rm min}}{\bar{n}^{(t)}_{\rm max} + \bar{n}^{(t)}_{\rm min}} = \frac{ 2 |\alpha|^2 \xi }{ 2 |\alpha|^2} = \xi \, .
\end{align}

Given the previous results, we now implement a quantum operation on the output beams, discussing the two relevant cases of displacement operation and homodyne detection.
In the former one, to implement the displacement $D(\zeta)$, $\zeta \in \mathbb{C}$, we should choose $\beta= \zeta/\sqrt{1-\tau}$ in Eq.~(\ref{eq:outbeams}), take the limit $\tau\to 1$, and trace out the signal on the reflected branch \cite{Paris1996}. Then, the mean number of photons on the transmitted beam becomes:
\begin{align}
\bar{n}^{(t)}_{\rm D} &= |\alpha|^2 + |\zeta|^2 + 2 \xi \, {\rm Re}(\alpha \zeta^{*}) \ne |\alpha+\zeta|^2 \, .
\end{align}
In particular, for the ``nulling" displacement amplitude $\zeta=-\alpha$, we have $\bar{n}^{(t)}_{\rm D}=2|\alpha|^2(1-\xi) \ne 0$, thus reduced visibility prevents the displacement of state $|\alpha\rangle$ into the vacuum.
Instead, in the case of homodyne detection, we adopt a balanced beam splitter, $\tau=1/2$, and the LO amplitude $\beta=z e^{i\phi}$, with $z\ge 0$ and $0\le \phi <\pi$.
In this case, the mean number of photons on the two branches reads:
\begin{align}
\bar{n}^{(t)}_{\rm HD} = \frac{|\alpha|^2 + z^2 + \xi |\alpha| z \cos\phi}{2} \quad \mbox{and} \quad
\bar{n}^{(r)}_{\rm HD} = \frac{|\alpha|^2 + z^2 - \xi |\alpha| z \cos\phi}{2} \, .
\end{align}
We, then, evaluate the difference photocurrent $\Delta$, that follows a Skellam distribution with mean value $\langle \Delta \rangle = \bar{n}^{(t)}_{\rm HD}- \bar{n}^{(r)}_{\rm HD} = \xi |\alpha| z \cos\phi$, and variance ${\rm Var}[\Delta]= \bar{n}^{(t)}_{\rm HD} + \bar{n}^{(r)}_{\rm HD} = |\alpha|^2 + z^2$. 
We conclude that reduced visibility in homodyne detection acts as a loss, playing the role of an ``effective" quantum efficiency, that only reduces the average $\langle \Delta \rangle$, without affecting the variance of the homodyne distribution.

\begin{figure}
\centerline{\includegraphics[width=0.6\columnwidth]{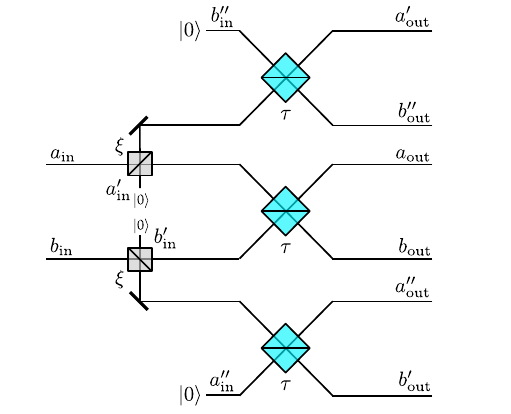}}
\centering
\caption{Modes evolution in the presence of reduced visibility $\xi\le 1$. We model the overall effect of visibility as a beam splitter of transmissivity $\xi$, acting on the input modes before the interference at the (physical) beam splitter of transmissivity $\tau$.}\label{fig:02:appII.B-VisMod}
\end{figure}

The former analysis suggests that visibility reduction may be described in terms of a loss dynamics. Therefore, we now proceed beyond coherent-states interference and provide a more general description, deriving the Heisenberg evolution of the modes impinging at a beam splitter with $\xi\le 1$. The scheme is reported in Fig.~\ref{fig:02:appII.B-VisMod}, where modes $a_\IN$ and $b_\IN$ are the signal and LO modes impinging at a (physical) beam splitter of transmissivity $\tau$.
Following the previous considerations, we model the effect of visibility on both the signal and the LO as a beam splitter with transmissivity $\xi\le 1$, in which, before interference, modes $a_\IN$ and $b_\IN$ are mixed with two ancillary modes $a'_\IN$ and $b'_\IN$, respectively, prepared in the vacuum state. 
Then, the output transmitted modes impinge at the physical beam splitter, while the reflected ones are coupled with a further pair of vacuum modes $a''_\IN$ and $b''_\IN$.
We describe the whole evolution via the input-output formalism. The input modes are ${\bf a}_\IN=(a_\IN, a'_\IN, a''_\IN,b_\IN, b'_\IN, b''_\IN)$. The evolution after the two beam splitters modeling the visibility effect is described by the unitary matrix:
\begin{align}
U_1=\begin{pmatrix}
\sqrt{\xi} & \sqrt{1-\xi} & 0 & 0 & 0 & 0 \\
-\sqrt{1-\xi} &\sqrt{\xi} & 0 & 0 & 0 & 0 \\
0 & 0 & 1 & 0 & 0 & 0 \\
0 & 0 & 0 & \sqrt{\xi} & \sqrt{1-\xi} & 0 \\
0 & 0 & 0 &-\sqrt{1-\xi} &\sqrt{\xi} & 0 \\
0 & 0 & 0 & 0 & 0 & 1 \\
\end{pmatrix} \, ,
\end{align}
while the interference at the physical beam splitter of the three output channels is associated with the unitary:
\begin{align}
U_2=\begin{pmatrix}
\sqrt{\tau} & 0 & 0 & \sqrt{1-\tau} & 0 & 0 \\
0 & \sqrt{\tau}& 0 & 0 & 0 &  \sqrt{1-\tau} \\
0 & 0 & \sqrt{\tau} & 0 & \sqrt{1-\tau} & 0 \\
-\sqrt{1-\tau} & 0 & 0 & \sqrt{\tau} & 0 & 0 \\
0 & 0 & -\sqrt{1-\tau} &0 &\sqrt{\tau} & 0 \\
0 & -\sqrt{1-\tau} & 0 & 0 & 0 & \sqrt{\tau} \\
\end{pmatrix} \, .
\end{align}
The output modes ${\bf a}_\OUT=(a_\OUT, a'_\OUT, a''_\OUT, b_\OUT, b'_\OUT, b''_\OUT)$ are then obtained as:
\begin{align}
{\bf a}_\OUT= U_2 U_1 {\bf a}_\IN \, ,
\end{align}
where
\begin{subequations}
\begin{align}
a_\OUT&= \sqrt{(1-\xi)(1-\tau)}b'_\IN + \sqrt{\xi (1-\tau)} b_\IN + \sqrt{\tau} \left(\sqrt{1-\xi} a'_\IN+ \sqrt{\xi} a_\IN\right) \, , \\[1ex]
a'_\OUT &= \sqrt{1-\tau} b''_\IN  + \sqrt{\tau }\left(\sqrt{\xi }a'_\IN  - \sqrt{1-\xi } a_\IN  \right) \, , \\[1ex]
a''_\OUT&= - \sqrt{(1-\xi)(1-\tau)} b_\IN + \sqrt{\xi (1-\tau) } b'_\IN + \sqrt{\tau } a''_\IN  \, ,
\end{align}
\begin{align}
b_\OUT&=- \sqrt{(1-\xi)(1-\tau)} a'_\IN  -\sqrt{\xi(1-\tau) } a_\IN + \sqrt{\tau} \left(\sqrt{1-\xi }b'_\IN +\sqrt{\xi }b_\IN \right) \, , \\[1ex]
b'_\OUT&=-\sqrt{1-\tau }a''_\IN + \sqrt{\tau}\left(\sqrt{\xi } b'_\IN  - \sqrt{1-\xi } b_\IN  \right) \, ,\\[1ex]
b''_\OUT&= \sqrt{(1-\xi)(1-\tau)} a_\IN - \sqrt{\xi (1-\tau) } a'_\IN  +\sqrt{\tau } b''_\IN  \, .
\end{align}
\end{subequations}
Moreover, the photon-number operators on the transmitted and reflected branches are equal to:
\begin{subequations}
\begin{align}
N^{(t)} &= a_\OUT^\dagger a_\OUT + (a'_\OUT)^\dagger a'_\OUT + (a''_\OUT)^\dagger a''_\OUT \, , \\[1ex]
N^{(r)} &= b_\OUT^\dagger b_\OUT + (b'_\OUT)^\dagger b'_\OUT + (b''_\OUT)^\dagger b''_\OUT \, ,
\end{align}
\end{subequations}
respectively.
As an example, we consider homodyne detection, where we evaluate the difference photocurrent $\hat{\Delta}= N^{(t)}- N^{(r)}$ and rescale its value by the LO amplitude.
In this scenario, all the ancillary modes are in the vacuum, while mode $b_\IN$ is excited in the coherent state $|ze^{i\phi}\rangle$. Accordingly, in the limit $z\to \infty$ we obtain:
\begin{align}
\frac{\langle z e^{i\phi}, {\bf 0} | \, \hat{\Delta} \, |z e^{i\phi}, {\bf 0} \rangle}{z} &= \xi \,  (a_\IN e^{-i\phi} + a^\dagger_\IN e^{i\phi}) =  \xi \, x_\phi \, , \\[1ex]
\frac{\langle z e^{i\phi} , {\bf 0} | \, \hat{\Delta}^2 \, |z e^{i\phi}, {\bf 0} \rangle}{z^2} &= \xi^2 \,x_\phi^2 + (1-\xi^2) \, ,
\end{align}
where quadrature operators are expressed in shot-noise units and 
\begin{align}
|z e^{i\phi}, {\bf 0} \rangle= |z e^{i\phi} \rangle_{b_\IN} |0\rangle_{a'_\IN} |0\rangle_{b'_\IN} |0\rangle_{a''_\IN} |0\rangle_{b''_\IN}\, .
\end{align}
Thus, homodyne detection still provides measurement of quadrature $x_\phi$, albeit with an ``effective" quantum efficiency $\eta=\xi^2$, proving that the impact of visibility can be modeled in terms of inefficient detection, consistently with the previous discussions \cite{Gupta2020}.

\section{Effective channel parameters in ideal NLA-assisted CVQKD}\label{app:Blandino}
\begin{figure}
\includegraphics[width=0.8\columnwidth]{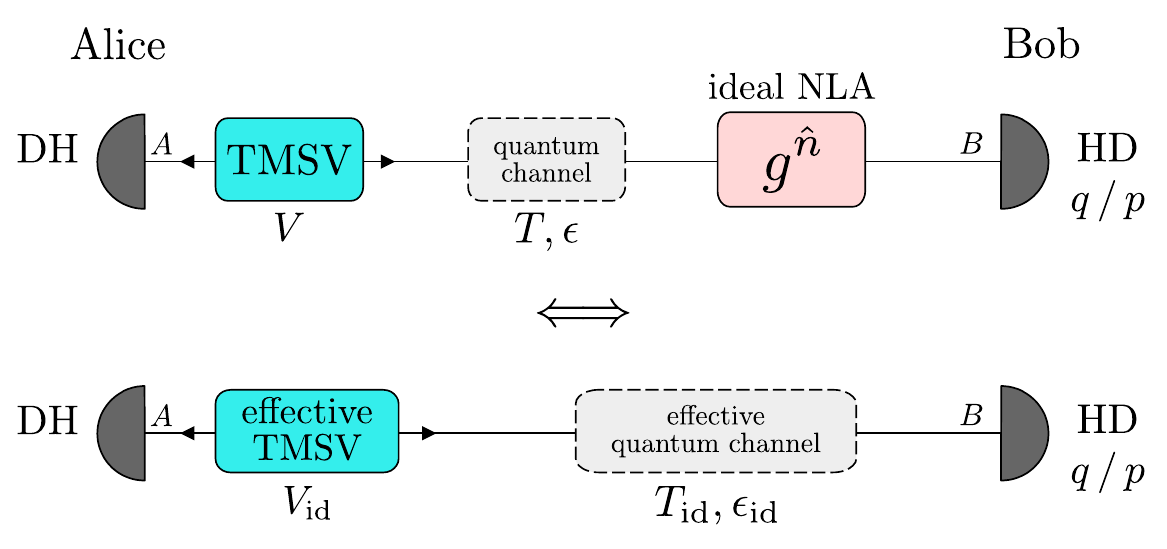}
\centering
\caption{Construction of the effective GG02 protocol associated with the ideal NLA-assisted protocol discussed in Sec.~\ref{subsec:idealNLAass}.}
\label{fig01:appIII.B_equiv}
\end{figure}

In this appendix, we perform explicit derivation of the effective channel parameters in Eq.~(\ref{eq: IdealPar}), describing the performance of the GG02 protocol assisted by an ideal NLA, associated with the unbounded operator $\NLA= g^{\hat{n}}$, where $g>1$ is the amplifier gain and $\hat{n}$ is the photon-number operator of the incoming optical mode.
As discussed in Sec.~\ref{subsec2:NLA}, the ideal NLA operation can be formally described in terms of the quantum CP map ${\cal E}_\id$, such that:
\begin{align}\label{eq:appTid}
{\cal E}_\id (\rho) = P_\id \, \frac{\NLA \, \rho \, \NLA^\dagger}{\Tr[\NLA \, \rho \, \NLA^\dagger]} + (1-P_\id) |0\rangle \langle 0| \, ,
\end{align}
$P_\id \le 1$ being the success probability of the transformation, that does not coincide with the trace of the post-selected state $\NLA \, \rho \, \NLA^\dagger$, since $\NLA$ is an unbounded operator.
Furthermore, the operator $\NLA$ can be formally written as the exponential as $\NLA=\exp(H)$, $H= (\ln g) \, \hat{n}$ being a bilinear function of the creation and annihilation operators, therefore it preserves Gaussianity. That is, if $\rho$ is Gaussian, $\NLA \, \rho \, \NLA^\dagger$ is Gaussian too. Nevertheless, due to its non-unitarity, it cannot be associated to any symplectic transformation, and the standard tools of Gaussian formalism cannot be straightforwardly applied to perform the security analysis of the CVQKD protocol. 

Given these considerations, we conclude that the ideal NLA-assisted protocol depicted in the top panel of Fig.~\ref{fig01:appIII.B_equiv}, in which Alice generates a TMSV state of variance $V>1$, namely:
\begin{align}
|{\rm TMSV} \rangle\!\rangle =
\sqrt{1-\lambda^2}\sum_{n=0}^{\infty} \lambda^n |n\rangle |n \rangle \, ,
\end{align}
where $\lambda= \sqrt{ (V-1)/(V+1) }$, whose second brach, thereafter, is injected into a thermal loss channel $(T,\epsilon)$, followed by the ideal NLA $\NLA$, is equivalent to the effective GG02 scheme depicted in the bottom panel of Fig.~\ref{fig01:appIII.B_equiv}, where Alice generates a TMSV state with effective variance $V_\id= (1+\lambda_\id^2)/(1-\lambda_\id^2)$, with $0\le \lambda_\id < 1$, which then propagates throughout thermal-loss channel with effective transmissivity $T_\id <1$ and excess noise $\epsilon_\id >0$ \cite{Blandino2012}. Both the schemes are performed only for those runs when noiseless amplification is successful, occurring with probability $P_\id$, otherwise the protocol is aborted.
To construct this equivalent protocol, we remind that the conditional state probed by Bob after Alice's DH measurement is a displaced thermal state, while the overall state is a thermal state. Therefore, we first compute the general action of the NLA operation $\NLA$ on these states and, then, specify the results to the CVQKD scheme under investigation.

\paragraph{Amplified displaced thermal states.} As discussed in Sec.~\ref{subsec2:NLA}, when a coherent state $|\alpha\rangle$, $\alpha \in \mathbb{C}$, undergoes ideal noiseless linear amplification, we have:
\begin{align}\label{eq:AppNLAcoh}
\NLA |\alpha\rangle = e^{-|\alpha|^2/2} \sum_{n=0}^{\infty} \frac{\alpha^n}{\sqrt{n!}} \, g^n |n \rangle = e^{(g^2-1)|\alpha|^2/2} \, |g \alpha\rangle \, ,
\end{align}
in which we note that the output state is not normalized. 

Now, we compute the output state of the ideal NLA when a displaced thermal state is considered as input, namely:
\begin{align}\label{eq:AppDT}
\rho_{\rm DT} = D(\beta) \nu^{\rm th} (\bar{n}) D^\dagger (\beta) \, ,
\end{align}
where $\beta \in \mathbb{C}$ is the displacement amplitude and $\bar{n}$ is the mean number of thermal photons.
As will become clearer in the following, it is useful to define $\bar{n}$ in terms of a parameter $0\le \kappa <1$ as $\bar{n}= \kappa^2/(1-\kappa^2)$, such that $\kappa^2= \bar{n}/(\bar{n}+1)$.
Eq.~(\ref{eq:AppDT}) can be also expressed in the Glauber-Sudarshan representation as $\rho_{\rm DT} = \int d^2 \alpha P_{\rm DT}(\alpha) |\alpha\rangle \langle \alpha|$, where:
\begin{align}
P_{\rm DT}(\alpha) = \frac{1-\kappa^2}{\pi \kappa^2} \exp\left( - \frac{1-\kappa^2}{\kappa^2} \, |\alpha-\beta|^2\right)
\end{align}
is the $P$-function associated with $\rho_{\rm DT}$ \cite{Blandino2012}.
Thanks to~(\ref{eq:AppNLAcoh}), the output state $\rho_{g}$ obtained after application of the NLA reads:
\begin{align}
\rho_{g} &= \frac{1}{\cal{N}} \, \NLA \, \rho_{\rm DT} \, \NLA^\dagger \nonumber  \\[1ex]
&= \frac{1}{\cal{N}} \int_\mathbb{C} d^2 \alpha P_{\rm DT}(\alpha) e^{(g^2-1) |\alpha|^2} \, |g\alpha\rangle \langle g\alpha|\, ,
\end{align}
${\cal N}= \Tr[\NLA \, \rho_{\rm DT} \, \NLA^\dagger]$ being the normalization factor.
We perform the change of variable $u=g\alpha$ and re-express the former equation as $\rho_g= \int d^2 u P_g (u) |u\rangle \langle u |$,
where:
\begin{align}
P_g(u)= \frac{1}{g^2 {\cal N}} e^{\frac{g^2-1}{g^2} |u|^2} P_{\rm DT}\left(\frac{u}{g}\right)
\end{align}
is the $P$-function associated with $\rho_g$.
Straightforward calculation leads to:
\begin{align}\label{eq:AppPgu}
P_g(u) &= \frac{1}{g^2 {\cal N}} \frac{1-\kappa^2}{\pi \kappa^2} \, \exp\left\{ - \frac{g^2-1}{g^2} |u|^2 - \frac{1-\kappa^2}{g^2\kappa^2} \, |u- g \beta|^2\right\} \nonumber \\[1ex] 
&=\frac{1}{g^2 {\cal N}} \frac{1-\kappa^2}{\pi \kappa^2} \, \exp\Bigg\{ - \frac{1- g^2 \kappa^2}{g^2\kappa^2} |u|^2  \nonumber \\
&\hspace{3.5cm} - \frac{1-\kappa^2}{\kappa^2} |\beta|^2 + 2 \frac{1-\kappa^2}{g^2\kappa^2} \Re(u^* \beta) \Bigg\} \nonumber \\[1ex] 
&=\frac{1}{g^2 {\cal N}} \frac{1-\kappa^2}{\pi \kappa^2} \,e^{\frac{(g^2-1)(1-\kappa^2)}{1-g^2 \kappa^2} |\beta|^2} \times  \nonumber \\[1ex] 
&\hspace{3.5cm} \exp\left\{ - \frac{1- g^2 \kappa^2}{g^2\kappa^2} \left|u - g \frac{1-\kappa^2}{1-g^2 \kappa^2} \beta \right|^2 \right\} \, .
\end{align}
Eq.~(\ref{eq:AppPgu}) represents a Gaussian complex function up to an irrelevant normalization factor independent of $u$, thus state $\rho_{g}$ is still a displaced thermal state in the form:
\begin{align}
\rho_{g} = D(\beta_g) \nu^{\rm th} (\bar{n}_g) D^\dagger (\beta_g) \, ,
\end{align}
with amplified displacement amplitude and thermal energy equal to:
\begin{align}\label{eq:AmplifiedParamDT}
\beta_g = g \frac{1-\kappa^2}{1-g^2 \kappa^2} \, \beta \qquad \mbox{and} \qquad \bar{n}_g = \frac{\kappa_g^2}{1-\kappa_g^2} = \frac{g^2\kappa^2}{1-g^2\kappa^2}  \, ,
\end{align}
with $\kappa_g= g \kappa$, provided that condition $g \kappa <1$ holds, which fixes a limit on the gain amplitude of the amplifier \cite{Blandino2012}.

We conclude that an ideal NLA maps a displaced thermal state into another displaced thermal state with larger displacement amplitude and mean number of thermal photons. In particular, a thermal state (retrieved by fixing $\beta=0$) is transformed into another thermal state with higher energy.

\paragraph{Derivation of the effective channel parameter for ideal-NLA assisted CVQKD.} Given the former results, we are now ready to provide derivation of the equivalent channel displayed in Fig.~\ref{fig01:appIII.B_equiv}. In particular, we derive a set of $3$ equations relating the original parameters $(\lambda, T,\epsilon)$ to the effective ones $(\lambda_\id, T_\id,\epsilon_\id)$ \cite{Blandino2012}. To this aim, we proceed as follows.

To begin with, we consider the conditional state at Bob's side. That is, when Alice performs DH detection on the first mode of the TMSV, obtaining outcomes $(x_A,y_A)$, the second branch is projected onto the coherent state $|\lambda \alpha_A\rangle$, with $\alpha_A= x_A+i y_A$. After propagation throughout the channel, the coherent pulse is transformed into a displaced thermal state $\rho_{\rm DT}(\alpha_A)$ with amplitude $\beta= \sqrt{T} \lambda \alpha_A$ and variance $1+T \epsilon$, such that the mean number of thermal photons reads $\bar{n}= T \epsilon/2$, being associated with parameter $\kappa^2=  \bar{n}/(\bar{n}+1) = T \epsilon/(2+T \epsilon)$.
Thereafter, Bob performs noiseless linear amplification on $\rho_{\rm DT}(\alpha_A)$, obtaining a displaced thermal state with $\kappa_g= g \kappa$ and the amplitude $\beta_g$ reported in Eq.~(\ref{eq:AmplifiedParamDT}). This provides us with the first two relations \cite{Blandino2012}:
\begin{align}
\sqrt{T_\id} \lambda_\id \alpha_A &= g \frac{1-\kappa^2}{1-g^2 \kappa^2} \, \sqrt{T} \lambda \alpha_A \, , \label{eq:ConditionEquiv1}\\[1ex]
\frac{T_\id \epsilon_\id}{2+T_\id \epsilon_\id}&= g^2 \frac{T \epsilon}{2+T \epsilon} \, . \label{eq:ConditionEquiv2}
\end{align} 
On the contrary, when Bob does not have access to the result of Alice's DH measurement, the overall state after propagation through the channel is a thermal state $\nu^{\rm th}(\bar{n}')$ with variance $T(V+\chi)= 1+T(V-1+\epsilon)$, and mean number of photons:
\begin{align}\label{eq:appnprime}
\bar{n}' = T \left(\frac{\lambda^2}{1-\lambda^2} + \frac{\epsilon}{2} \right) \, ,
\end{align}
associated with:
\begin{align}
(\kappa')^2 = \frac{\bar{n}'}{\bar{n}'+1}= \frac{T[\lambda^2 (2-\epsilon) + \epsilon]}{2  + T \epsilon -\lambda^2 [2- T (2-\epsilon)]} \, .
\end{align}
After the NLA, the state is converted into a thermal state with $\kappa'_g = g  \kappa'$, leading to:
\begin{align}\label{eq:ConditionEquiv3}
\frac{T_\id[\lambda_\id^2 (2-\epsilon_\id) + \epsilon_\id]}{2  + T_\id \epsilon_\id -\lambda_\id^2 [2- T_\id (2-\epsilon_\id)]} = g^2 \frac{T[\lambda^2 (2-\epsilon) + \epsilon]}{2  + T \epsilon -\lambda^2 [2- T (2-\epsilon)]}\, .
\end{align}

Eq.s~(\ref{eq:ConditionEquiv1}),~(\ref{eq:ConditionEquiv2}) and~(\ref{eq:ConditionEquiv3}) provide a system of equations for the variables $(\lambda_\id, T_\id,\epsilon_\id)$, with corresponding solutions:
 \begin{align}
\lambda_{\id}&= \lambda \, \sqrt{\frac{2+ T(g^2-1)(2-\epsilon)}{2- T \epsilon(g^2-1)}} \, ,\\[1ex]
T_{\id} &=\frac{g^2 T}{1+ T (g^2-1)[1+T \epsilon (g^2-1)(2-\epsilon)/4-\epsilon]} \, ,\\[1ex]
\epsilon_{\id} &= \epsilon + (g^2-1) \frac{T \epsilon(2-\epsilon)}{2}  \, ,
\end{align}
retrieving the results of Eq.~(\ref{eq: IdealPar}). Moreover, we obtain the expression of the effective TMSV variance as:
 \begin{align}
V_\id = \frac{1+\lambda_\id^2}{1-\lambda_\id^2} = V+ \frac{T (g^2-1) Z^2}{2-T (g^2-1)(V-1+\epsilon)} \, ,
\end{align}
with $Z=\sqrt{V^2-1}$.
These parameters can be interpreted as physical parameters of an effective channel only if they keep a physical meaning, that is if they satisfy the constraints $0 \le \lambda_\id <1$ (equivalent to $V_\id \ge V$), $0\le T_\id \le 1$ and $\epsilon_\id \ge 0$.
The first request determines a tradeoff between the amplifier gain $g$, the channel transmissivity $T$, and the modulation variance $V$, that is:
\begin{align}
T (g^2-1)(V-1+\epsilon) \le 2 \, ,
\end{align}
which ultimately leads to Eq.~(\ref{eq: Conditiong}). On the contrary, the second and third constraints impose conditions on the amplifier gain $g$ and the excess noise $\epsilon$, namely:
\begin{align}
g &\le \sqrt{\frac{\epsilon \left(T \left(\epsilon{-}4\right){+}2\right){+}4 \sqrt{\frac{T \left(\epsilon-2\right){+}2}{\epsilon}}{-}2 \sqrt{\epsilon \left(T \left(\epsilon{-}2\right){+}2\right)}{+}4
   T{-}4}{T \left(\epsilon{-}2\right){}^2}} \, ,
\end{align}
and $\epsilon \le 2$, respectively, the latter being equivalent to the constraint on the maximum tolerable excess noise associated with the PLOB bound \cite{PLOB}. 

\paragraph{Success probability of the ideal NLA.} Finally, to complete the construction of the equivalent protocol, we should determine the value of the success probability $P_\id$ associated with the NLA, when one arm of the TSMV is considered as input. However, the exact computation cannot be handled, as $\NLA$ is unbounded, therefore the trace of the amplified states does not provide the corresponding success probability. Nevertheless, we establish an upper bound to $P_\id$, which can be considered as a best-case scenario for the security analysis \cite{Blandino2012}. 

First of all, from Eq.~(\ref{eq:appTid}), we note that ${\cal E}_\id (|0\rangle \langle 0|)= |0\rangle \langle 0|$, proving the vacuum state to be a fixed point of ${\cal E}_\id$. Then, we invoke the contractivity of quantum maps: that is, any trace preserving quantum CP map cannot decrease the fidelity $\mathcal{F}$ between two quantum states \cite{NielsenChuang, Blandino2012}, thus for all states $\rho$ we have:
\begin{align}\label{eq:fiden}
{\cal F} \left( \rho, |0\rangle \langle 0 | \right) \le {\cal F} \left( {\cal E}_\id(\rho), |0\rangle \langle 0 | \right)  \, . 
\end{align}
In particular, in the CVQKD protocol under investigation, the overall state transmitted into the channel is a thermal state $\nu^{\rm th}(\bar{n}')= [1-(\kappa')^2] \sum_n (\kappa')^{2n}|n\rangle\langle n|$, with $(\kappa')^2 = \bar{n}'/(\bar{n}'+1)$, see Eq.~(\ref{eq:appnprime}), while ${\cal E}[\nu^{\rm th}(\bar{n}')]= \nu^{\rm th}(\bar{n}'_g)$, having mean energy $\bar{n}'_g = \kappa_g^2/(1-\kappa_g^2)$, with $\kappa'_g = g  \kappa'$. Then, Eq.~(\ref{eq:fiden}) becomes:
\begin{align}
\langle 0 |\nu^{\rm th}(\bar{n}') |0\rangle \le P_\id \langle 0 |\nu^{\rm th}(\bar{n}'_g) |0\rangle + (1-P_\id)  \, , 
\end{align}
being satisfied iff:
\begin{align}
P_\id\le \frac{1}{g^2} \, ,
\end{align}
providing an upper bound to the NLA success probability \cite{Blandino2012}.
%


\section{Unconditional security in physical NLA-assisted CVQKD}\label{app:CovMatNLA}

\begin{figure}
\includegraphics[width=0.8\columnwidth]{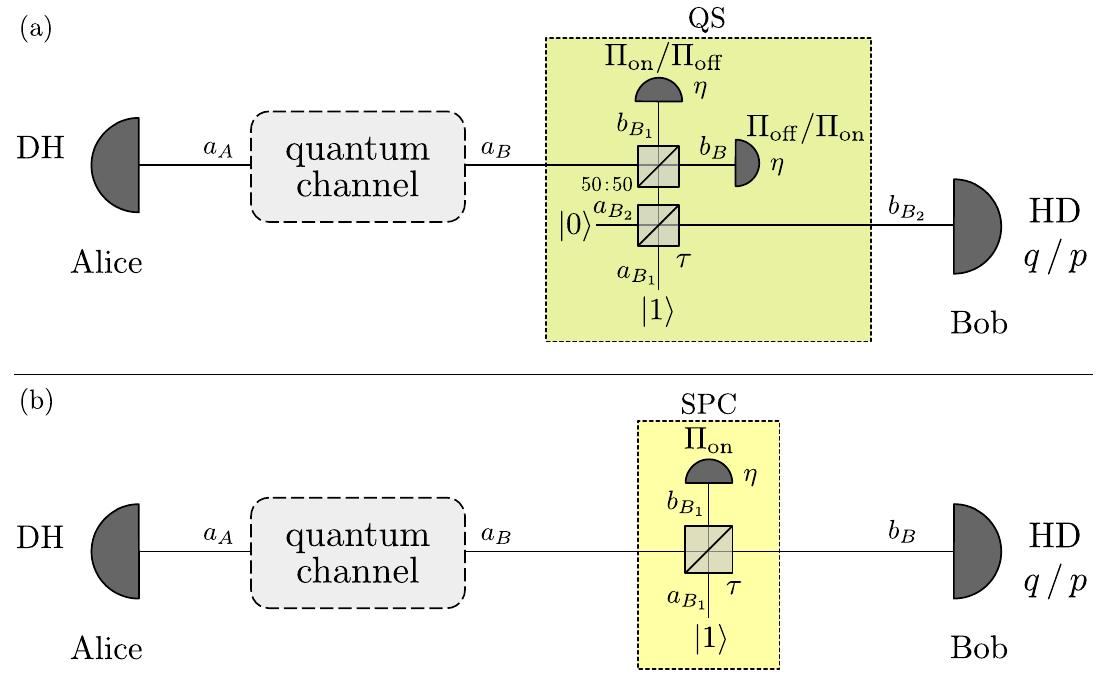}
\centering
\caption{Schematic representation of the two physical NLA-assisted protocol discussed in Sec.~\ref{sec:CVQKD-NLA}. (a) Strategy based on quantum scissors (QS); (b) strategy based on single-photon catalysis (SPC).}
\label{fig01:appIII.C_protocol}
\end{figure}

In the present appendix, we perform explicit derivation of the CM~(\ref{eq: GammaAB|p}), being necessary for the unconditional security proof of the physical NLA-assisted CVQKD discussed in Sec.~\ref{sec:CVQKD-NLA}.
As pointed out in the main text, we perform the security analysis by exploiting the optimality of Gaussian attacks. If Alice and Bob share a non-Gaussian state $\rho$, a lower bound of the exact KGR is obtained by considering a Gaussian protocol in which they share the Gaussian state $\rho_{\rm G}$ with the same CM of $\rho$. In particular, here we derive the CM for both the the quantum scissors ($\a$) and the single-photon catalysis ($\b$). To do so, we exploit the input-output formalism and the phase-space representation of quantum states.

\paragraph{Quantum scissors (QS).}
By following the notation introduced in Fig.~\ref{fig01:appIII.C_protocol}(a), the protocol employing QS works as follows \cite{Ghalaii}.
Alice prepares the TMSV and injects one mode into the thermal-loss channel, thereafter Bob performs the QS protocol on the received beam. The input modes are ${\bf a} = (\aA, \aB, \aC, \aD)^{\mathsf{T}}$, where $\aA, \aB$ are the modes shared by Alice and Bob after the channel whereas $\aC, \aD$ are the modes exploited locally by Bob for the QS. The global input state, according to Glauber's formula, reads:
\begin{align}
\rho_{\bf a} = \int \frac{d^2 \bmalpha}{\pi^4} \chi_{\bf a}(\bmalpha) D_{\bf a}(\bmalpha)\dag \, ,
\end{align}
where $\bmalpha=(\alphaA, \alphaB, \alphaC, \alphaD)^{\mathsf{T}}$ and
\begin{align}
D_{\bf a}(\bmalpha) = \bigotimes_{k} D_{a_k}(\alpha_k) \, ,
\end{align}
where $D_{a_k}(\alpha_k)$ is the displacement operator acting on mode $a_k$, namely,
\begin{align}
D_{a_k}(\alpha_k)= \exp(\alpha_k a_k\dag - \alpha_k^* a_k) \, .
\end{align}
Furthermore, in the previous expression we introduced the characteristic function:
\begin{align}
\chi_{\bf a}(\bmalpha) = \chi_{\rm G}(\alphaA, \alphaB) \times\big(1-  |\alphaC|^2\big) \, e^{- (|\alphaC|^2 + |\alphaD|^2)/2} \, ,
\end{align}
$\chi_{\rm G}(\alphaA, \alphaB)$ being the Gaussian characteristic function:
\begin{align}\label{eq: chi_Gauss}
    \chi_{\rm G}(\alphaA, \alphaB)=  \exp \bigg[ 
    - \displaystyle \frac12 \tilde{\bmalpha}^\mathsf{T}_{AB} \, \Gamma_{AB} \, \tilde{\bmalpha}_{AB}
    \bigg] \, ,
\end{align}
with null prime moments and the CM $\Gamma_{AB}$ introduced in Eq.~(\ref{eq: Gamma_GG02}), and where $\tilde{\bmalpha}_{AB}= [\Re ( \alphaA), \Im (\alphaA),\Re (\alphaB), \Im (\alphaB)]^{\sf T}$.

The output modes after the mode mixing operations performed by Bob are ${\bf b} = (\bA, \bB, \bC, \bD)^{\mathsf{T}}= {\cal M}_{\a} {\bf a}$, where
\begin{align}
{\cal M}_{\a}= 
\begin{pmatrix}
1 & 0 &  0 & 0 \\
0 & \frac{1}{\sqrt{2}} &  \sqrt{\tau/2} & -\sqrt{(1-\tau)/2} \\[1ex]
0 & -\frac{1}{\sqrt{2}} &  \sqrt{\tau/2} & -\sqrt{(1-\tau)/2} \\[1ex]
0 & 0 &  \sqrt{1-\tau} & \sqrt{\tau}
\end{pmatrix} \, ,
\end{align}
with $\tau= \tau_{\a}(g)= (1+g^2)^{-1}$.
The output state then writes:
\begin{align}
\rho_{\bf b} = \int \frac{d^2 \bmbeta}{\pi^4} \chi_{\bf b}(\bmbeta) D_{\bf b}(\bmbeta)\dag \, ,
\end{align}
where, exploiting the properties in Eq.~(\ref{eq: propertiesDisplacementOperation}), $\chi_{\bf b}(\bmbeta)= \chi_{\bf a}({\cal M}_{\a}^{\mathsf{T}} \bmalpha)$.

Finally, Bob performs on-off detection on modes $\bB, \bC$, corresponding to the positive-operator-valued measurement (POVM) $\{\Pi_{\rm off}, \Pi_{\rm on}=\Id-\Pi_{\rm off}\}$, with associated characteristic functions \cite{Ferraro2005, Olivares2005}:
\begin{align}
\chi_{\rm off} (\alpha) = \frac{1}{\eta} \exp\left(- \frac{2-\eta}{2\eta}|\alpha|^2\right)\quad \mbox{and} \quad
\chi_{\rm on} (\alpha) = \pi \delta^{(2)}(\alpha)- \chi_{\rm off} (\alpha) \, .
\end{align}
The amplification is successful if one of the two detectors gives the outcome ``on" \cite{Ghalaii, NLARalphLund}. In the following we assume to retrieve the couple (on,off), respectively for modes $\bB, \bC$. The post-selected state then equals to:
\begin{align}
\varrho_{\a} = \frac{1}{\widetilde{P}_{\a}} \int \frac{d^2 \betaA}{\pi} \frac{d^2 \betaD}{\pi} \chi_{\a}(\betaA,\betaD) D_{\bA}(\betaA)\dag D_{\bD}(\betaD)\dag  \, ,
\end{align}
where:
\begin{align}
\chi_{\a}(\betaA,\betaD)=  \int \frac{d^2 \betaB}{\pi} \frac{d^2 \betaC}{\pi} \chi_{\bf b}(\bmbeta) \chi_{\rm on} (-\betaB) \chi_{\rm off} (-\betaC) \, ,
\end{align}
and
\begin{align}
\widetilde{P}_{\a} &= \Tr \Bigg[ \int \frac{d^2 \betaA}{\pi} \frac{d^2 \betaD}{\pi} \chi_{\a}(\betaA,\betaD) D_{\bA}(\betaA)\dag D_{\bD}(\betaD)\dag \Bigg] \notag \\[1ex]
&= \chi_{\a}(0,0) 
= 2 \left[\frac{8 \eta \tau + (w-1)(3+w)(1+\eta \tau)}{(1+w)^2 (3+w)^2}\right]
\end{align}
is the success probability of this conditional operation, with $w= 1+\eta T (V+\epsilon-1)$. The same results hold if Bob gets the pair (off,on), thus the global success probability of the QS-based NLA is:
\begin{align}
P_{\a}= 2\widetilde{P}_{\a} = 4 \left[ \frac{8 \eta \tau + (w-1)(3+w)(1+\eta \tau)}{(1+w)^2 (3+w)^2} \right]\, .
\end{align}

Finally, we compute the CM associated with the state $\varrho_{\a}$, for which we should compute terms proportional to $\Tr[D_{b_k}(\beta_k) q^2_{b_k}]$.
To handle this calculation, we exploit the following property of displacement operations, derived in \cite{Ghalaii}:
\begin{align}\label{eq: CovDisp}
\Tr\big[D(\alpha) q^2\big]= e^{-(x^2+y^2)/2} \Bigg[\pi \delta^{(2)} (\alpha)+ 2 \pi y \delta(x) \frac{d}{d y} \delta(y) - \pi \delta(x) \frac{d^2}{d y^2} \delta(y)\Bigg] \, ,
\end{align}
where we consider a single single radiation mode $a$ with its corresponding quadrature $q=a+a\dag$, expressed in shot-noise units,
with $\alpha=x+i y$ and $\delta(x)$ being the Dirac delta distribution.

By exploiting Eq.~(\ref{eq: CovDisp}), we have:
\begin{subequations}
\begin{align}
V_{\a}&= \Tr\big[\varrho_{\a} q_{\bA}^2\big] = - 1 - \frac{{\cal V}_{\a}}{\widetilde{P}_{\a} } \, ,\\[1ex]
W_{\a}&= \Tr\big[\varrho_{\a} q_{\bD}^2\big] = - 1 - \frac{{\cal W}_{\a}}{\widetilde{P}_{\a} } \, , \\[1ex]
Z_{\a}&= \Tr\big[\varrho_{\a} q_{\bA} q_{\bD}\big] = -\frac{{\cal Z}_{\a}}{\widetilde{P}_{\a} } \, ,
\end{align}
\end{subequations}
where
\begin{subequations}
\begin{align}
{\cal V}_{\a}&=\Bigg[\frac{d^2}{dy^2} \Big(e^{-y^2/2}\chi_{\a}(i y,0)\Big)\Bigg]_{y=0} \notag \\[1ex]
&= 2 (V+1) \Bigg[ \frac{(2+\eta T \epsilon)(1-\eta \tau)}{(1+w)^2} \notag \\[1ex]
& \hspace{2cm} - \frac{8(3+w)+2 \eta T \epsilon(3+w-4\eta\tau)+4\eta \tau (w-5)}{(3+w)^3}\Bigg] \, , \\[2ex]
{\cal W}_{\a} &= \Bigg[\frac{d^2}{dv^2} \Big(e^{-v^2/2}\chi_{\a}(0,i v)\Big)\Bigg]_{y=0}  \notag \\[1ex]
&= -4 \, \frac{8 \eta \tau + (w-1)(3+w) [2-(1-\eta)\tau]}{(1+w)(3+w)^2}\, ,\\[2ex]
{\cal Z}_{\a}&= \Bigg[\frac{d^2}{dy dv} \Big(e^{-(y^2-v^2)/2} \chi_{\a}(i y,i v)\Big)\Bigg]_{y=0, v=0}  \notag \\[1ex]
&= \sqrt{T} Z \, \frac{8 \eta \sqrt{\tau(1-\tau)}}{(3+w)^2}\, .
\end{align}
\end{subequations}
Accordingly, the CM writes:
\begin{align}
    \Gamma^{(\a)}_{AB} =
    \begin{pmatrix} 
V_{\a} \, \Id_2 & Z_{\a} \, \sigmaz\\[1ex]
    Z_{\a} \, \sigmaz & W_{\a} \, \Id_2 
\end{pmatrix} \, .
\end{align}

\paragraph{Single-photon catalysis (SPC).}
For SPC we follow the analogous procedure of the previous subsection. The input modes depicted in Fig.~\ref{fig01:appIII.C_protocol}(b) are ${\bf a} = (\aA, \aB, \aC)^{\mathsf{T}}$, where $\aA, \aB$ are the modes shared by Alice and Bob after the channel and $\aC$ is Bob's ancillary mode. The global input state reads:
\begin{align}
\rho_{\bf a} = \int \frac{d^2 \bmalpha}{\pi^3} \chi_{\bf a}(\bmalpha) D_{\bf a}(\bmalpha)\dag \, ,
\end{align}
where $\bmalpha=(\alphaA, \alphaB, \alphaC)^{\mathsf{T}}$  and
\begin{align}
\chi_{\bf a}(\bmalpha) = \chi_{\rm G}(\alphaA, \alphaB) \times e^{- |\alphaC|^2/2} \big(1-  |\alphaC|^2\big) \, ,
\end{align}
$\chi_{\rm G}(\alphaA, \alphaB)$ being the Gaussian characteristic function in Eq.~(\ref{eq: chi_Gauss}) with null prime moments and the CM~(\ref{eq: Gamma_GG02}).
\par
The output modes after the mode mixing operation performed by Bob are ${\bf b} = (\bA, \bB, \bC)^{{\sf T}}= {\cal M}_{\b} {\bf a}$, where
\begin{align}
{\cal M}_{\b}= 
\begin{pmatrix}
1 & 0 &  0 \\[0.5ex]
0 & \sqrt{\tau} &  \sqrt{1-\tau} \\[0.5ex]
0 &  -\sqrt{1-\tau} & \sqrt{\tau}
\end{pmatrix} \, ,
\end{align}
with $\tau=\tau_{\b}(g)= \left( 4 + g^2 - g \sqrt{8 + g^2}\right)/8$.
The output state then writes:
\begin{align}
\rho_{\bf b} = \int \frac{d^2 \bmbeta}{\pi^3} \chi_{\bf b}(\bmbeta) D_{\bf b}(\bmbeta)\dag \, ,
\end{align}
where
\begin{align}
\chi_{\bf b}(\bmbeta)= \chi_{\bf a}({\cal M}_{\b}^{\mathsf{T}} \bmalpha) \, .
\end{align}

After the conditional on-off detection on mode $\bC$, the post-selected state reads:
\begin{align}
\varrho_{\b} = \frac{1}{P_{\b}} \int \frac{d^2 \betaA}{\pi} \frac{d^2 \betaB}{\pi} \chi_{\b}(\betaA,\betaB) D_{\bA}(\betaA)\dag D_{\bB}(\betaB)\dag  \, ,
\end{align}
where:
\begin{align}
\chi_{\b}(\betaA,\betaB)=  \int \frac{d^2 \betaC}{\pi} \chi_{\bf b}(\bmbeta) \chi_{\rm on} (-\betaC) \, ,
\end{align}
and
\begin{align}
P_{\b} &= \Tr \Bigg[ \int \frac{d^2 \betaA}{\pi} \frac{d^2 \betaB}{\pi} \chi_{\b}(\betaA,\betaB) D_{\bA}(\betaA)\dag D_{\bB}(\betaB)\dag \Bigg]\notag \\[1ex]
&= \chi_{\b}(0,0) = 1-\frac{4(1-\eta \tau) + 2(w-1)(1- \tau)}{[2+(w-1)(1-\tau)]^2}
\end{align}
is the success probability of the SPC, and we introduced the quantity $w= 1+\eta T (V+\epsilon-1)$.

The CM associated with the state $\varrho_{\b}$ reads:
\begin{align}
\Gamma^{(\b)}_{AB} =
\begin{pmatrix} 
V_{\b} \, \Id_2 & Z_{\b} \, \sigmaz\\[1ex]
    Z_{\b} \, \sigmaz & W_{\b} \, \Id_2 
\end{pmatrix} \, .
\end{align}
As for QS, we have:
\begin{align}
V_{\b}&= \Tr\big[\varrho_{\a} q_{\bA}^2\big] = - 1 - \frac{{\cal V}_{\b}}{P_{\b} } \, , \\[1ex]
W_{\b}&= \Tr\big[\varrho_{\a} q_{\bB}^2\big] = - 1 - \frac{{\cal W}_{\b}}{P_{\b} } \, , \\[1ex]
Z_{\b}&= \Tr\big[\varrho_{\a} q_{\bA} q_{\bB}\big] = -\frac{{\cal Z}_{\b}}{P_{\b} } \, ,
\end{align}
and
\begin{subequations}
\begin{align}
{\cal V}_{\b}&=\Bigg[\frac{d^2}{dy^2} \Big(e^{-y^2/2}\chi_{\b}(i y,0)\Big)\Bigg]_{y=0} \notag \\[1ex]
&= - 2(V+1) \Bigg[\frac12- \frac{4+\eta T \epsilon (1-\tau)(1+q-4\eta\tau)}{(1+q)^3} \notag \\[1ex]
& \hspace{5cm} +\frac{2(1+\eta\tau)(q-1)-4\eta\tau}{(1+q)^3}\Bigg] \, ,\\[2ex]
{\cal W}_{\b} &= \Bigg[\frac{d^2}{dv^2} \Big(e^{-v^2/2}\chi_{\b}(0,i v)\Big)\Bigg]_{y=0} \notag \\[1ex]
&= -4 -\tau(r-3) + 4\, \frac{(q-1)^2 +(r-1)(q-1)(\eta+\tau) +2\tau (r-1)-2 \eta\tau (q-1)}{(1+q)^3}\notag \\[1ex]
& \hspace{2cm}+ 4\, \frac{2(w-1)(4-4\tau-\tau^2)+4 (2 - \tau (1 + \eta))}{(1+q)^3}\, ,\\[2ex]
{\cal Z}_{\b}&= \Bigg[\frac{d^2}{dy dv} \Big(e^{-(y^2-v^2)/2} \chi_{\b}(i y,i v)\Big)\Bigg]_{y=0, v=0} \notag \\[1ex]
&= \sqrt{\tau T}Z \Bigg[1-4\, \frac{2+(1+\eta)(q-1)+2\eta(1-2\tau)}{(1+q)^3} \Bigg] \, ,
\end{align}
\end{subequations}
with $q= 1+ \eta T(1-\tau) (V+\epsilon-1)$ and $r= 1+ T (V+\epsilon-1)$.


\end{document}